**Resistance Switching Devices based on Amorphous Insulator-metal Thin Films**

Xiang Yang

A DISSERTATION

in

Materials Science and Engineering

Presented to the Faculties of the University of Pennsylvania

in

Partial Fulfillment of the Requirements for the

Degree of Doctor of Philosophy

2014

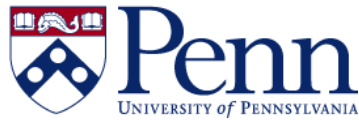

Supervisor of Dissertation

I-Wei Chen

Professor of Material Science and Engineering

Graduate Group Chairperson

Shu Yang, Professor of Material Science and Engineering

Dissertation Committee

Ritesh Agarwal, Professor of Materials Science and Engineering

Cherie R. Kagan, Professor of Materials Science and Engineering

James M. Kikkawa, Professor of Physics and Astronomy

To Mom, Dad, and my loving and ever-supportive wife, Jie.

# ACKNOWLEDGMENTS

First and foremost, I would like to thank my advisor, Prof. I-Wei Chen, for all his guidance, inspiration, support, and patience throughout the past five years. I am very grateful to have been introduced to the fascinating research, asked to always try to solve all problems independently, and rigorously trained to be a good scientist and engineer. I sincerely appreciate that he devoted so much time to shed light on my research, answer my endlessly naive and crazy questions, encourage me to question the authority bravely. This dissertation would not be possible without his advice.

Secondly, I would like to express my deepest love to all my family members, especially my father and mother, my wife's father and mother. Their unconditional love and trust are the strongest supports for me to fearlessly pursue my dream. I would also express my special gratitude to my beloved wife, Jie Li, for always being there for me. This work would not have been done without her consistent understanding and encouragement. I truly appreciate she trusted me, married me and sacrificed so much for me, for our family. I sincerely thank her for tolerating my stubbornness, "BU KAO PU", stupid ideas and decisions from time to time. I am proud of being her husband, and will take great pride in giving back everything I have and will have to her, to my family, in every possible way.

I want to acknowledge my thesis committee members, Prof. James M. Kikkawa, Prof. Ritesh Agarwal, Prof. Cherie R. Kagan, and Prof. A. T. Charlie Johnson, for their valuable time, effort and feedback to me for the research proposal, the annual progress reports and the final dissertation. Their insightful comments and suggestions are of great importance to me to make this thesis more thorough and correct. My special thanks go to Prof. James M. Kikkawa, who is a fantastic experimental physicist. He had spent a lot of time helping me set up PPMS experiments, and teaching me useful techniques and hand-on experiences. Chapter V would never be done without his generous supports. In addition, I appreciate Prof. Nadar Engheta's excellent suggestions on interpretation of electron-bunch experiment, which indeed helped me and encouraged me a lot to touch an unknown area.

I've had the good fortune to be able to work with excellent colleagues and friends at Penn, and I would like to express my appreciation to all of them. I would especially thank Dr. Albert B. K. Chen and Dr. Byung Joon Choi, for guiding me to the research field and generously sharing valuable experience; Dr. Yudi Wang and Dr. Soo Gil Kim, for their pioneering works which provide me the best references ever. I would also thank Dr. Ioan Tudosa, who is a brilliant scientist at SLAC, for his crucial support to experiment, simulation and theory in ebunch project. I have enjoyed the inspiring discussions with him on electromagnetics and high energy physics. Special thanks go to my project members: Dr. Jongho Lee, Yang Lu, Wei-Shao Tung, for their friendship and technical supports; other group members: Yanhao Dong, Dr. Jungdeok Park, Dr. Hoon Choi, for their friendship and mental supports; Dr. Lauren Willis in the physics department, for training me on e-beam lithography and sharing useful experiences; Annemarie Exarhos and Michael E Turk in Prof. Kikkawa's group, for assisting me on PPMS project; Pavan Nukala in Prof. Agarwal's group, for helping me debug problems in cryogenic station.

I also want to acknowledge many people for all kinds of assistance. I thank Prof. Shu Yang for granting me access to FTIR. I am grateful to Nanoscale Characterization Facility (NCF), formerly PRNF, for access to SEM, FIB, EDX, TEM, especially to Dr. Lolita Rotkina, Dr. Douglas Yates, and Dr. Jamie Ford with the training and support on these instruments. I acknowledge Wolf Nanofabrication Facility (WNF) for access to various lithography, film deposition and etching tools, especially to Iulian Codreanu, Kyle Keenan and Hiromichi Yamamoto for all their kind help with my experiments. I thank Nano- Bio Interface Center (NBIC) for access to AFM and Cryogenic Probe Station, especially to Dr. Matthew Brukman for all trainings. I am also grateful to Steve Szewczyk, for training me on XRD, optical microscope, hot press and furnaces, and offering me other countless technical/nontechnical supports.

In addition, I would express my gratitude to our kind LRSM staff, especially Vicky Lee Truei, Enrique Vargas, Fred Helmig, Raymond Hsiao, Irene Clements and Pat Overend, for being so nice to me and offering me help all the time, including lending me MSE parking permit, which turned out the biggest motivation for my weekend research!

My Ph.D. project was supported by the US National Science Foundation (Grant No. DMR-11-04530 and DMR-14-09114, primarily, and DMR-09-07523 and DMR-11-20901, in part). I would like to acknowledge them all.

# ABSTRACT

**Resistance Switching Devices based on Amorphous Insulator-metal Thin Films**

**Xiang Yang**

**Advisor: I-Wei Chen**

Nanometallic resistance switching devices based on amorphous insulator-metal thin films are developed to provide a novel non-volatile resistance-switching random-access memory (RRAM) that is CMOS-compatible and meeting technological demand. In these devices, data recording/converting is controlled by a bipolar voltage, which tunes electron localization lengths, hence resistivity, through electron trapping and detrapping. The low-resistance state is a metallic state while the high-resistance state is an insulating state, as established by conductivity studies from 2K to 300K.

The material is exemplified by a $Si_3N_4$ thin film with randomly dispersed Pt or Cr. It has been extended to other materials, spanning a large library of oxide and nitride insulator films, dispersed with transition and main-group metal atoms. Metallic nanoparticles, which form at metal levels greater than 10 atomic percent, are nonessential for resistance switching: nanometallicity and resistance switching in nanometer thin films start at levels well below the metal percolation threshold.

Nanometallic RRAMs have superior properties that set them apart from other RRAMs. The critical switching voltage is independent of the film thickness, device area, operating temperature and switching speed. Trapped electrons are relaxed by electron-phonon interaction, adding stability which enables long-term memory retention despite a low switching voltage. As electron-phonon interaction is mechanically altered, trapped electron can be destabilized, and sub-picosecond switching has been demonstrated using an electromagnetically generated stress pulse. The resistance state is finely tunable throughout the entire continuum between the fully metallic state and the fully insulating state, by voltage, thickness and composition. AC impedance spectroscopy confirms the resistance state is spatially uniform, providing a capacitance that linearly scales with area and inversely scales with thickness. The spatial uniformity is also manifested in outstanding uniformity of switching properties. Device degradation, due to moisture, electrode oxidation and dielectrophoresis, is minimal when dense thin films are used or when a hermetic seal is provided. The potential for low power operation, multi-bit storage and complementary stacking have been demonstrated in various RRAM configurations.

These studies furnish a firmer understanding of nanometallicity and nanometallic switching. They also establish nanometallic RRAM as a viable candidate for emerging memory.

# TABLE OF CONTENTS

# LIST OF TABLES

# LIST OF ILLUSTRATIONS

xvi

xxii

xxiii

xxix

resistance increases as negative voltage limit reduces, causing off-switching voltage to decrease. (d) Simulated $R$-$V$ curves under different $-V_{max}$ using parallel circuit model in **Figure 9.2b**. Percentage in the bracket shows different $F$ at plateau resistance. Simulation parameters: $V_c*(V)=\pm(1.2\pm0.2)$, $R_l(\Omega)=330$, $r_L/A(\Omega)=90$, .400

**Part A: Nanometallic memory: materials, transitions, and mechanisms**

# Chapter I. Introduction

## 1.1 Background

### 1.1.1 Traditional Memory and Emerging Memory

The increasing demand for electronic memory has motivated intensive research in academia and industry over several decades. Such demand never saturates, instead it has accelerated in this era of big data and cloud computing/storage. An ideal memory should feature fast read and write access, high density and low cost, low voltage/power operation, robust endurance and retention during the product cycle. Clearly, these features are difficult to realize within any one memory today, hence the need for new product. For example, static RAM (SRAM), serving as CPU on-chip cache for temporarily storing computed results within nanosecond, has a fast access time ~100 ps but suffers from high cost, large cell size and extremely poor retention (volatile). On the other hand, flash memory, which has non-volatility (>10 years), high density, and a steadily more competitive cost, sadly needs hundreds of microsecond and high power to drive. To cope with this issue, data and instructions in computer architecture follow a hierarchical arrangement referred to as memory hierarchy. Instead of one "universal" memory solution, data and instructions are implemented and stored in different memory levels, depending on their priority and system performance trade-off. A memory hierarchy design typically consists of embedded memory (SRAM, eDRAM) as on-chip caches, commodity DRAM as main memory, and a peripheral drive (HDD or SSD) as storage. Generally, the closer the memory is to the microprocessor, the faster and higher bandwidth it must present, but at the cost of a lower density or high expense (**Figure 1.1**). Within a memory module, functional blocks regardless of specific memory type essentially share similar structures as shown in **Figure 1.2**. Memory inputs typically consist of data, address and control signals, while memory outputs consist of data and status signals. A binary address of some prescribed length and structure is sent to a decoder which generates a memory readable address (*e.g.* one-hot) to select a specific memory unit. Input data are then written to such unit or the content of such unit is read out through write/read circuits.

Following Moore's law over the past several decades, CMOS technology has eventually approached the size of ~10 nm today. At this size, the increasing leakage current and the dynamic power density for SRAM/DRAM pose the greatest challenge for circuit/architecture designers. In addition, the slow speed of hard disk has long become a bottleneck in computer systems with increasingly faster speed. These needs have motivated the pursuit of disruptive technologies for future memory hierarchy designs.

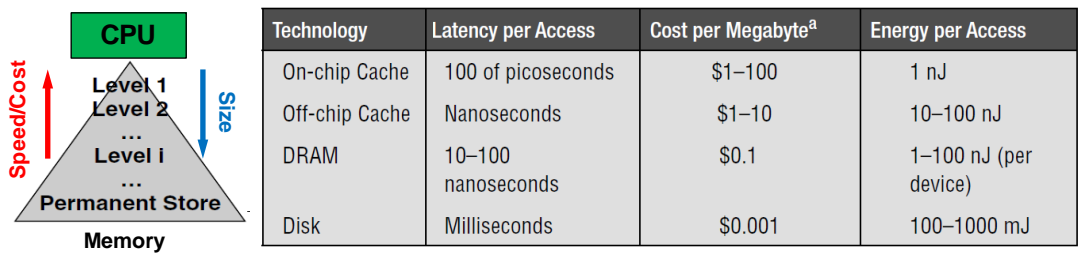

| Technology | Latency per Access | Cost per Megabyte[a] | Energy per Access |
|---|---|---|---|
| On-chip Cache | 100 of picoseconds | $1–100 | 1 nJ |
| Off-chip Cache | Nanoseconds | $1–10 | 10–100 nJ |
| DRAM | 10–100 nanoseconds | $0.1 | 1–100 nJ (per device) |
| Disk | Milliseconds | $0.001 | 100–1000 mJ |

**Figure 1.1.** A memory hierarchy: as the distance from the processor increases, so does the size, but opposite for speed and cost (adapted from ref.[1]).

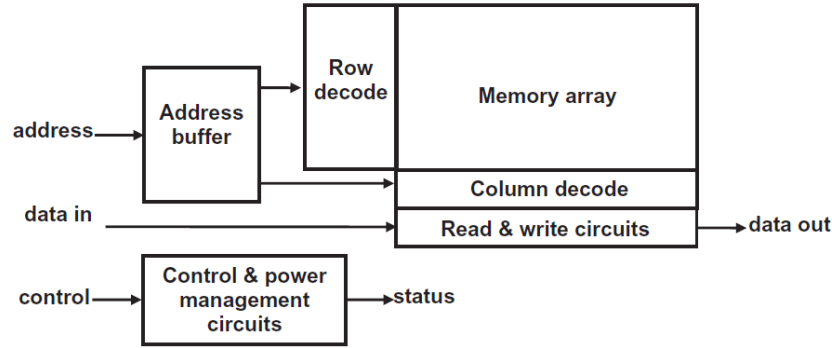

**Figure 1.2.** Memory functional block diagram (adapted from ref.[2]).

Since 1990s, a variety of emerging memory technologies have been proposed, including magnetoresistance RAM (MRAM), ferroelectric RAM (FeRAM), phase-change RAM (PRAM), resistive RAM (RRAM), nanomechanical memory, Mott memory, molecular memory, *etc*. A common goal of such emerging memory technologies is to combine the speed of SRAM, the density of DRAM, and the non-volatility and low cost of hard drive. Among them, MRAM, PRAM, RRAM and FeRAM are considered promising and thus have attracted much attention. **Table 1.1** lists demonstrated properties of these emerging memories in comparison with commercialized state-of-art memory in 2013.

| | RRAM | PRAM | MRAM | FeRAM | | SRAM | DRAM | Flash | HDD* |
|---|---|---|---|---|---|---|---|---|---|
| | Prototypes | | | | | Commercialized technologies | | | |
| W/E time (ns) | 1 | 100 | 35 | 65 | | 0.2 | 10 | $10^5$ | $10^6$ |
| Read time (ns) | 1 | 12 | 35 | 40 | | 0.2 | 10 | $10^5$ | $10^6$ |
| Retention | 10 yr | 10 yr | 10 yr | 10 yr | | [D] | 64 ms | 10 yr | 10 yr |
| Endurance (#) | $10^{12}$ | $10^9$ | $10^{12}$ | $10^{14}$ | | $10^{16}$ | $10^{16}$ | $10^5$ | $10^4$ |
| *F*-Size (nm) | 5 | 45 | 65 | 180 | | 45 | 36 | 16 | |
| Density ($F^2$) | $4F^2$ | $4F^2$ | $20F^2$ | $22F^2$ | | $140F^2$ | $6F^2$ | $4F^2$ | $2/3F^2$ |
| W/E voltage (V) | 0.6 | 3 | 1.8 | 1.3-3.3 | | 1 | 2.5 | 15 | --- |
| Read *V.* (V) | 0.1 | 1.2 | 1.8 | 1.3-3.3 | | 1 | 1.8 | 4.5 | --- |
| Energy/bit (pJ) | 0.1 | 6 | 2.5 | 0.03 | | 0.0005 | 0.004 | 10 | $10^9$ |
| Cell elements | 1T(D)1R | 1T(D)1R | 1(2)T1R | 1T1C | | 6T | 1T1C | 1T | --- |

**Table 1.1.** Comparison of state-of-art memory/storage technology from **2013 *International Technology Roadmap for Semiconductors (ITRS)*[3]**. All parameters represents stand-alone device. RRAM: resistive switching RAM; PRAM: phase change RAM; MRAM: magnetoresistance RAM; FeRAM: ferroelectric RAM; SRAM: static RAM; DRAM: dynamic RRAM; Flash: NAND stand-alone flash cell; HDD: hard disk drive. *Adapted from ref.[4]

Although there is a desire for "universal" memory, at least for now this is not possible by direct replacement of existing hierarchy with emerging memories. This is at least for the following reasons. (1) Although MRAM can reduce the cache miss rate with a larger capacity, using high density MRAM to replace SRAM as on-chip cache causes longer write latency and thus degrade the performance for write-intensive applications. (2) Likewise, although high density on-chip memory will reduce CPU requests to the off-package DRAM, hence decrease the average access time, more extra space needed on CPU must be taken up by tags and logics, which would have been better utilized as the next level cache. (3) It is common for emerging memory to be non-volatile, but it generically takes a longer time and more energy for write operation. (4) Some emerging memory such as PRAM has an issue with lifetime reliability, which could be a major obstacle for using it as storage class memory even though it is suitable for working memory[5]. A better interim solution may be to leverage the benefits of traditional SRAM/DRAM and the emerging memory in a hybrid memory architecture, such as MRAM/SRAM hybrid on-chip cache[6] or PRAM/DRAM hybrid main memory[7]. The idea is to keep most write-intensive data within SRAM/DRAM while use high-density emerging memory as fast local storage.

### 1.1.2 Resistive Random Access Memory (RRAM)

As shown in 2013 ITRS (**Table 1.1**), RRAM exhibits many desired features of "good memory": fast speed, low energy and high endurance of SRAM and low cost, high density and non-volatility of flash memories. The basic structure of RRAM is a MIM "capacitor" with an active layer (I) sandwiched between two electrodes (M), as shown in **Figure 1.3**. The top electrode (TE) and the bottom electrode (BE) can be the same or different metals, which may participate in the switching process in some devices. Usually, the active layer is the key component, where switching occurs. The MIM structure must hold at least two stable resistance states: a high resistance state (HRS) and a low resistance state (LRS), which can be repeatedly converted between them by electrical stimuli (*e.g.*, voltage, current).

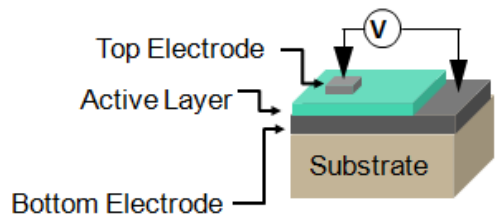

**Figure 1.3.** Schematic of a typical RRAM structure.

Ultrathin metal/oxide/metal films showing pronounced resistive switching began to attract interest in 1960s[8-9]. However, the rise of silicon integrated circuit technology put off the progress of resistive switching devices, despite some theoretical interest expressed by Chua and coworkers[10,11]. Intensive studies of RRAM reappeared in late 1990s in view of the seemingly imminent end-of-the-road-map for silicon technology. In the past two decades, RRAM has advanced at a remarkably pace and single-device performance has apparently achieved a satisfactory status for memory applications (**Figure 1.4**). Industry has also engaged in RRAM technology, notably at Fujitsu, Sharp, Samsung Electronics, Hynix Semiconductor, Micron Technology, HP, Adesto Technology Inc. (a spin off company from AMD), Crossbar Inc., 4DS, Elpida (acquired by Micron) and Unity Semiconductor (acquired by Rambus)[5]. Although the majority of efforts are on materials and devices, some circuit/architecture-level issues have also been addressed. For example, in 2013 ISSCC, Sandisk and Toshiba successfully demonstrated a metal-oxide based 32 Gb RRAM prototype developed in 24 nm technology[12].

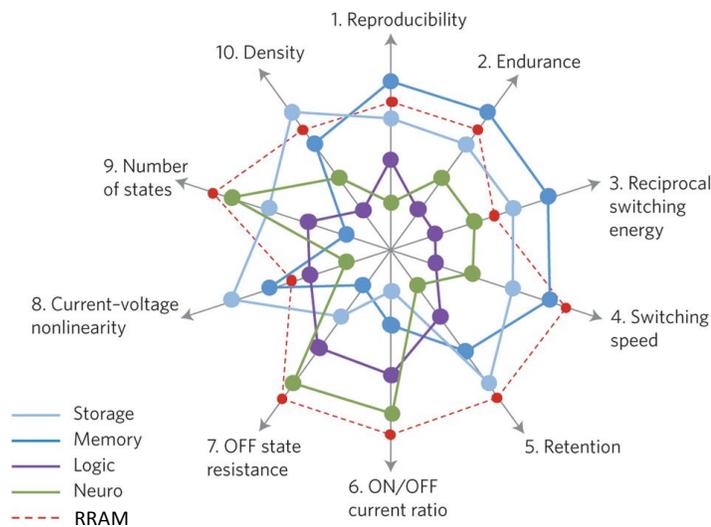

**Figure 1.4.** Device requirements for main applications (adapted from ref.[4]). A higher point on the axis represents a higher required value of the specific property. The dash line is the best reported RRAM data (from different

devices). Quantitative data is summarized in Supplementary Information in ref [4].

### 1.1.3  RRAM Classification

There are several classifications for RRAM. A phenomenological one is based on the apparent electrical polarity required for resistive switching, namely unipolar and bipolar[13]. In a "unipolar" RRAM, the switching procedure is independent of the polarity of voltage and current (**Figure 1.5a**). The current for the OFF-to-ON ("set") transition is usually limited by the compliance current (CC), which is always smaller than the current required for the ON-to-OFF ("reset") transition; meanwhile, the set voltage is always higher than the reset voltage. A unipolar RRAM can be used in either one-polarity (+ or -) or two-polarity (+ and -) mode. In contrast, a bipolar RRAM requires two opposite voltage polarities to trigger set and reset switching with or without CC (**Figure 1.5b**). Certain asymmetry elements (materials, electrodes or geometries) need to be introduced to bipolar systems to realize such polarity preference Unipolar and bipolar switching can coexist in some materials such as $TiO_2$ (ref.[14]), NiO (ref.[15]) and $SrTiO_3$ (ref.[16]), and depends on the compliance control (CC)—a high CC typically results in unipolar switching while a low CC leads to bipolar switching. Even in the same bipolar category, opposite polarities (clockwise or counterclockwise switching directions in the loops of **Figure 1.5b**) can coexist, as seen in $TiO_2$ (ref.[17]) and $SrTiO_3$ (ref.[18]). This may be caused by the competition of the top and bottom electrode interfaces which see opposite electrical fields.

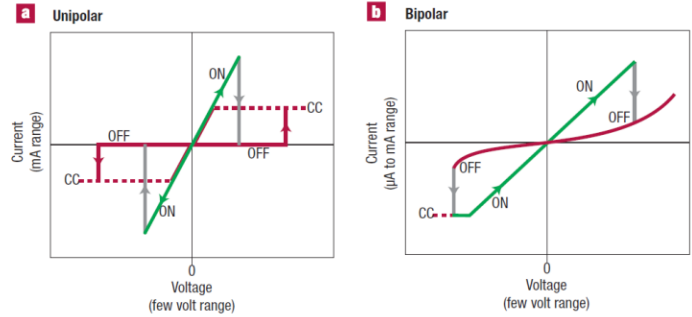

**Figure 1.5.** Classification of resistive switching (adapted from ref.[13]): (a) Unipolar switching. (b) Bipolar switching.

One can also classify RRAMs by active-layer material. For example, binary oxide RRAMs include transition-metal oxides such as $TiO_x$[19-21], $NiO$[22-25], $HfO_x$[26-27], $TaO_x$[28-29], $ZrO_x$[30] and $MoO_x$[31], as well as simple oxides such as $SiO_x$[32-33], $AlO_x$[34], $MgO$[35] and their oxynitrides or nitrides such as $SiO_xN_y$[36] and $Si_3N_4$[37-38]. Perovskite RRAMs include $SrZrO_3$[39], $SrTiO_3$[40-41] and $(Ba, Sr)TiO_3$[42] with or without doping (*e.g.*, by Cr, Nb, V, Mo), some being colossal magnetoresistance perovskites such as $Pr_{0.3}Ca_{0.7}MnO_3$ (PCMO)[43-44], $La_{1-x}Sr_xMnO_3$ (LSMO)[45-46] and $La_{1-x}Ca_xMnO_3$ (LCMO)[47]. Polymer RRAMs include polystyrene[48] or other organics[49-50], often with conducting metallic nanoparticles[50] or nanowires[48]. Ionic conductor RRAMs include Ag-Ge-S(Se)[51-53], Cu-Ge-S(Se)[53-54], *etc*. Semicondutor RRAMs include amorphous Si[55], *etc.*. Indeed, one is inclined to believe that *every dielectric material can be made into an RRAM*. Therefore, such material-based classification seems not informative and may even be misleading since sometimes the most critical switching mechanisms may occur at the interface and not in the "dielectric" layer.

Classification based on the dominant transport mechanism divides RRAMs into two categories: ionic RRAMs and electronic RRAMs. In addition, ionic RRAMs typically involve generation/dissolution of certain conducting filaments populated by cations (*e.g.*, metal $M^{n+}$) or anions (*e.g.*, oxygen $O^{2-}$) that can relatively easily migrate in these conduits, and electronic RRAMs typically involve electronic trapping/detrapping, space-charge limited current (SCLC), or strongly correlated electron effects. By and large, ionic RRAMs have been the major focus of recent research, which is briefly described below.

### 1.1.3.1 Ionic RRAM

Ionic RRAMs can be further divided into two sub-categories: cation devices and anion devices.

**Cation devices**

In the literature, cation devices are sometimes referred to as electrochemical metallization (ECM) memory, programmable cells, conductive bridging RAM (CBRAM) or atomic switches. In 1976, Hirose *et al.* first reported a polarity-dependent memory effect demonstrating conducting Ag dendrite growth in amorphous $As_2S_3$ films under an electrical field[56]. The mobile species here are believed to be metal cations. After three decades' studies, it is widely accepted that off→on resistive switching occurs through the following steps, which occur in localized filaments (see **Figure 1.6** from ref.[57]):

(i) Anodic dissolution of metal M: $M \longrightarrow M^{n+} + ne^-$;

(ii) $M^{n+}$ migration across the solid-electrolyte towards cathode under the electrical

field;

(iii) Reduction and electrocrystallization on the surface of cathode: $M^{n+} + ne^{-} \longrightarrow M$.

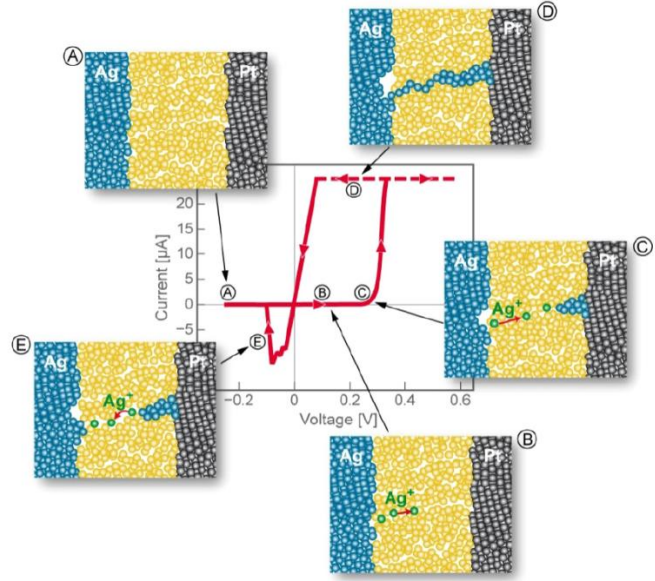

**Figure 1.6.** Schematic of operation of an ECM memory (adapted from ref.[57]).

If a reverse voltage is applied, on⟶off switching occurs during which the previous cathode with deposited M becomes anode and will dissolve. The dissolution will bring the device back to the original insulator state. Any of these steps could be rate-limiting for the event of resistive switching, depending on the specific material system. For example, cation migration (step ii) rate is usually high in chalcogenides and thus redox process (step i and iii) are the rate limiting factors. In the other extreme, in dense dielectrics (*e.g.* SiO$_2$ or amorphous Si), cation migration is very low and thus rate limiting. Different rate limiting processes can lead to very different filament shapes. In fast ion conductors (*e.g.*, chalcogenides), the large population of cations and the low reduction rate dictate that cations can easily arrive at the inert electrode long before reduction occurs, hence forming a cone-shaped filament growing from the inert electrode to the active electrode. However in a sluggish conductor ($SiO_2$ or amorphous Si), the reduction rate is high compared to the limited ion solubility and mobility. Therefore, cations prefer to migrate for a shorter distance and become reduced at the end of the existing filament. This leads to a reversed cone-shaped filament.

This switching mechanism has been confirmed by several (*in-situ*/*ex-situ*) TEM experiments[58-59]. An example (in $SiO_2$) is shown in **Figure 1.7**, in which Ag dendrites eventually form a reversed cone-shaped filament initiated from the Pt electrode. It is worth noting that although *in-situ* TEM seems to be a powerful technique to study filament dynamics, it might also easily introduce "filament growth" because of its high energy electron beam. In this sense, *ex-situ* TEM examination may provide more reliable information with fewer artifacts.

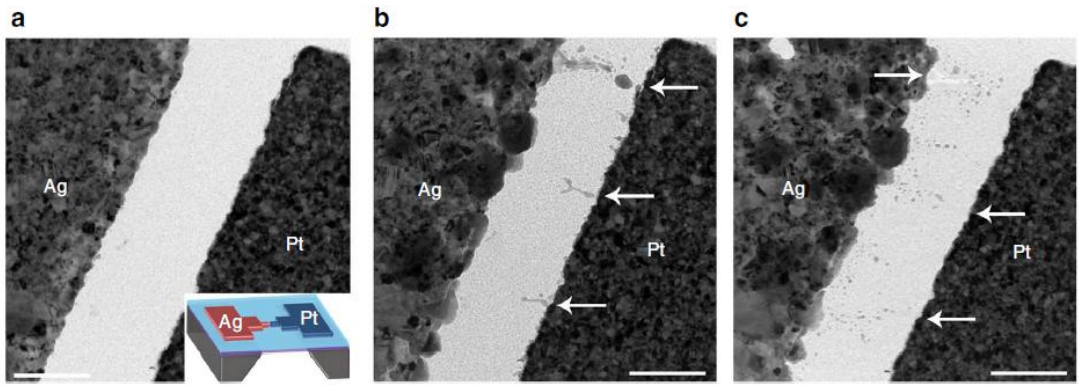

**Figure 1.7.** Dynamic filament growth in $SiO_2$. (a) TEM image of an as-fabricated $SiO_2$-based planar device. (b) TEM image after the forming process. (c) TEM image after erasing. Scale bar: 200 nm (adapted from ref.[58]).

A cation-based device must use an electrochemically active material as anode, which may be Cu, Ag or an alloy of these metals. In contrast, the cathode is typically an electrochemically inert metal, such as Pt, Au, W, Mo, Ir, Ru, TiW, or TaN (ref.[57]). The dielectric material can be a traditional electrolyte chosen from sulphides, iodides, selenides, tellurides, ternary chalcogenides and even water[60]. Other materials such as insulating oxides, nitrides, amorphous Si, C, doped organic semiconductors and vacuum gaps[61-62] can also serve the same purpose. One noticeable effect of evolving from traditional electrolytes to oxide materials is on the switching voltage, which can increase from below 0.3 V to above the operating voltage of CMOS devices[4].

In the above mechanism, the electrical field provides the major driving force; its thermodynamics also guarantees bipolar switching. However, Joule heating is not always negligible, especially poor electrolyte materials, such as oxides or amorphous Si, are used because of the higher switching voltage required (thus substantial heating). Thermally assisted diffusion has been demonstrated to be critical in $Cu/Ta_2O_5/Pt$ during reset switching[60]. Under extreme circumstances where the ON-state has a very low resistance ($\sim 1\ \Omega$), on$\longrightarrow$off switching could be triggered by heating-induced filament breakdown. In this scenario, the non-polar thermal effect overwhelms the directional electrical field drift, leading to unipolar switching. Because of the substantial thermal energy involved, unipolar RRAMs typically exhibit worse reliability than bipolar RRAMs.

From the device perspectives, key advantages of ECM are its good endurance ($>10^{10}$ cycles demonstrated[63]), scalability ($<20$ nm demonstrated[64]), fast speed ($<1$ ns

demonstrated[65]) and low energy operation (1 pJ (W) and 8 pJ (E) demonstrated[66]). Current research challenges for ECM lie in obtaining assurance of robust operation, developing reliability models, reducing random telegraph noise (RTN), improving memory controller design and CMOS compatibility, and finding compatible reliable selectors[3].

**Anion devices**

Compared to cation devices, the study of anion devices are less mature in the sense that switching mechanisms are still under debate. In general, these devices are based on anion migration (*e.g.*, oxygen), which can lead to two effects. First, the accumulation/depletion of oxygen ions (or its counterpart "oxygen vacancies") at the electrode/oxide interface may modify the energy profile of the Schottky barrier and thus induce interface-type resistive switching (non-filament)[67]. Second, the migration and redistribution of oxygen vacancies can induce a localized valence change and thus form a new phase (filament). This RRAM category is sometimes referred to as valence change memory (VCM), covering oxide insulators (transition metal oxides, complex oxides), nitrides and chalcogenides. Similar to cation devices, most anion devices can switch in both unipolar and bipolar modes depending on the details of device fabrication and electrical operation conditions.

Starting from early exploration of resistive switching 50 years ago[8-9] (referred as negative resistance at that time), a rich family of simple oxides have been extensively investigated, including $MgO_x$, $TiO_x$, $ZrO_x$, $HfO_x$, $VO_x$, $NbO_x$, $TaO_x$, $CrO_x$, $MoO_x$, $WO_x$, $MnO_x$, $FeO_x$,

$CoO_x$, $NiO_x$, $CuO_x$, $ZnO_x$, $AlO_x$, $GaO_x$, $SiO_x$, $GeO_x$, $SnO_x$, $BiO_x$, and $SbO_x$ (ref.[4]). Their detailed switching mechanism varies from system to system, but basic principles are quite similar. Before the material becomes repeatedly switchable, a high voltage electroforming process is usually required to define a highly localized new phase region, *i.e.*, a filament. During electroforming, electro-reduction and defect creation (*e.g.*, oxygen vacancies) is triggered by the high electrical field, possibly enhanced by Joule heating. Next, $O^{2-}$ ions drift towards the anode and are discharged there, creating $O_2$ gas and possibly causing physical deformation[68]. After electroforming, subsequent switching occurs locally within the newly formed phase region and resistance states are determined by its various oxidation/structural states.

$TiO_2$ is one of the best understood simple oxides. Resistive switching in $TiO_2$ was believed to occur by locally reducing the stoichiometric $TiO_2$ phase to a more conducting $TiO_{2-x}$ phase under the electrical field[19-20,69]. The maximum value of $x$ to maintain statistically non-associated point defects is on the order of $10^{-4}$ (ref.[70-71]), above which extended defects (*e.g.*, vacancy chains, Wadsley defect[71]) form easily. Recent studies aided by *in-situ*/*ex-situ* TEM technique confirmed the conducting channels to be made of $Ti_nO_{2n-1}$, a Magnéli phase, a product of local oxygen deficiency[21]. Specifically, a locally reduced $Ti_4O_7$ filament can be observed in the on-state of a $Pt/TiO_2/Pt$ device, which is then oxidized to $TiO_2$ phase in the off-state (**Figure 1.8**). (The same conclusion was also reached in other studies[72].) Convincing TEM evidence for VCM was too seen in the ZnO system, in which Zn-rich $ZnO_{1-x}$ and $ZnO$ phases were demonstrated to be responsible for switching[73].

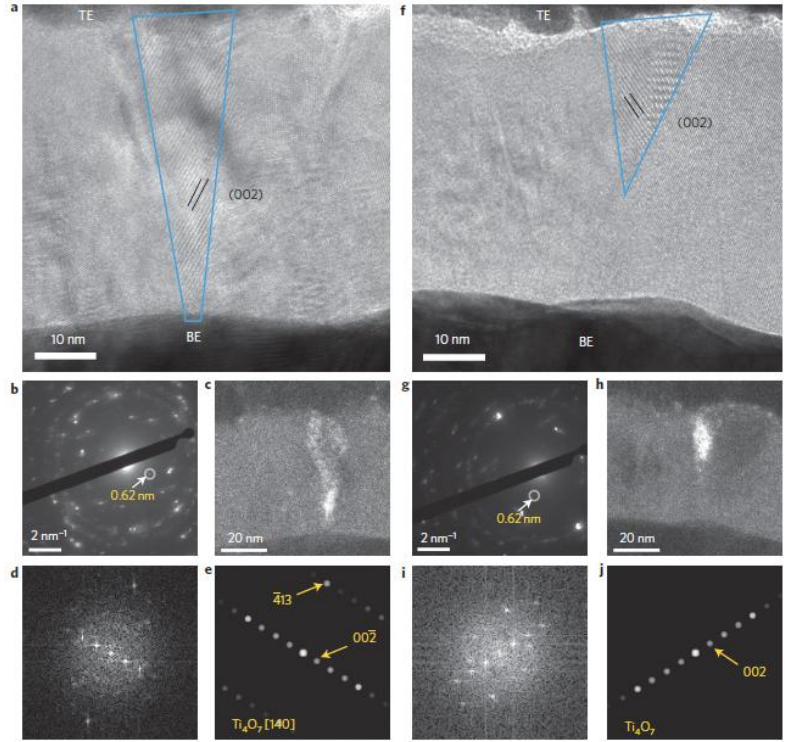

**Figure 1.8.** Magnéli phase in the switched $TiO_2$ device[21]. (a) High-resolution TEM image of a $Ti_4O_7$ filament. (b) Selected-area diffraction pattern. (c) Dark-field image obtained from the diffraction spot marked as a circle in (b), (d) FFT image of the high resolution image of $Ti_4O_7$. (e) Simulated diffraction pattern. (f)–(j) Disconnected $Ti_4O_7$ structure in the conical shape. The images are presented in the similar manner as for the connected filament in (a)–(e). (Adapted from ref. ref. [21]).

$HfO_2$ is another model material, which is CMOS compatible and already suitable for gate oxide in MOSFET. As a resistive switching material, it features outstanding endurance ($>10^{10}$ cycles), retention ($>10$ years) and fast switching speed ($<1$ ns)[26]. Like other VCMs, oxygen vacancy migration driven by the electrical field is believed to induce local stoichiometry changes between insulating $HfO_2$ and conducting $HfO_{2-x}$. Similar valence change probably also applies to $Ta_2O_5$ RRAM, which also features $>10^{10}$ cycles of switching[29]. Noting the significant performance (*e.g.*, endurance) difference between $HfO_x$, $TaO_x$ and other transition metal oxides such as $TiO_x$. Yang *et al.*[29] argued that $HfO_x$ and $TaO_x$ are unique because their phase diagrams contain only two stable phases (one being insulating and nearly stoichiometric, the other conducting and oxygen deficient); no other intermediate phase exists unlike the case of $TiO_x$ containing many Magnéli phases. This implies the phases in $HfO_x$ and $TaO_x$ are rather stable with a reasonably large barrier to any other metastable phases, whereas many if not all the phases $TiO_x$ are easily convertible making them susceptible to degradation at the elevated temperature caused by Joule heating in the device. This idea suggests reliable VCM materials should be associated with a simple phase diagram such as the one shown in **Figure 1.9**: the two compatible phases are stable and do not form any intermediate phase even at high temperature[4]. According to this proposal, the $MO_x$ phase in the phase diagram is the insulating phase and the M phase is the conducting phase in the filament. Since the latter has a large solubility for oxygen, the filament can easily accommodate mobile oxygen ions.

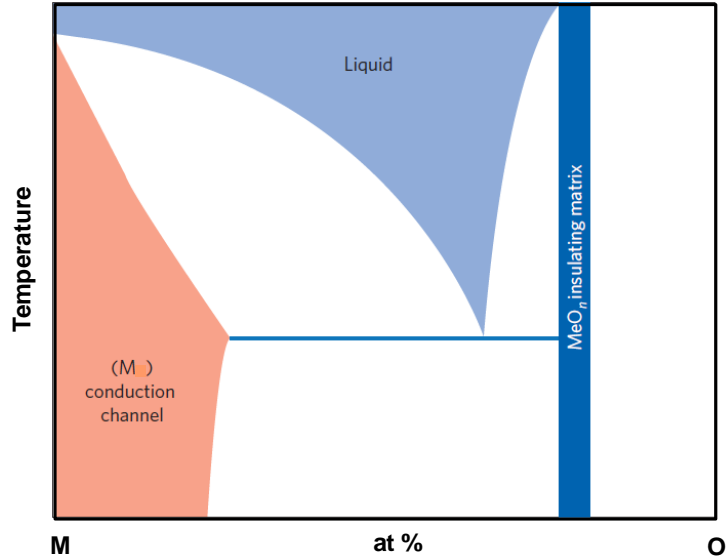

**Figure 1.9.** Proposed phase diagram of a metal-oxygen (M-O) system by Joshua Yang, *etc.*, with two solid-state phases at low temperature. $MO_x$ phase is a stoichiometric (insulating) phase. M phase is a metal-oxygen solid solution (conduction channel). The two phases are thermodynamically compatible with each other (adapted from ref.[4]).

Resistive switching performance can be further improved by engineering electrode/active layer interface. For example, electrical uniformity of $HfO_2$ RRAM is greatly enhanced by stacking other oxide thin layers such as $AlO_x$ (ref.[27]), $ZrO_x$ (ref.[74]) and Ge (ref.[75]) into the $HfO_2$ layers. Such embedded layer technique is thought to help better control of the generation of conductive filaments through ionic diffusion. A significant improvement was also found in the $Ta_2O_5$ system by using $TaO_{2-x}$ as a base layer to provide a better control of filament growth and to reduce the current and power consumption. Using this technique, a $Ta_2O_{5-x}/TaO_{2-x}$ bilayer RRAM can achieve $10^{12}$ cycles (a record to-date), 10 ns and <60 μW switching (ref.[28]). Another technique is to introduce metallic nanoparticles (*e.g.*, Ru nanodots) at the interface, probably for field concentration, which can more effectively define the formation of filaments in the device (*e.g.*, using TiO$_2$)[76].

NiO is another intensively studied model material for RRAM. Using a Hg drop as a temporary electrode on NiO, Son and Shin[23] were able to observe multiple conducting filaments (instead of one single filament) in the LRS under conducting AFM (**Figure 1.10**). To provide a direct proof of ion migration in response to different voltages, Yoshida *et al.* incorporated $^{18}$O tracer into NiO film and examined the $^{16}$O and $^{18}$O profiles using time-of-flight secondary ion mass spectroscopy. Clear evidence of oxygen migration led them to conclude that resistance changes originated from the voltage-induced $\delta$ variation in Ni$_{1-\delta}$O ($\delta$=+/- or 0) [24]. Although NiO is antiferromagnetic and this was observed by Son *et al.* in the magnetization hysteresis loops (*M-H*) at the OFF state, they found a ferromagnetic hysteresis loop with exchange coupling at the ON state, indicating the formation of ferromagnetic Ni during resistive switching [25].

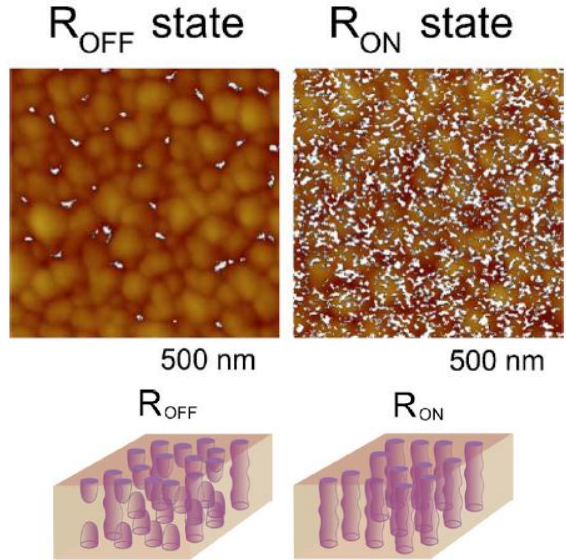

**Figure 1.10.** (a) The CAFM image of the NiO at different states. White dot indicates conducting channels. At OFF-state, only leaky current exists near grain boundary. Pictorial view of the filament model reveals a "local break" nature. (adapted from ref.[23])

From the device perspective, the key advantages of VCM are its good endurance ($>10^{12}$ cycles demonstrated[28,77]), scalability (<5 nm demonstrated[78]), fast speed (<1 ns demonstrated[79]), low energy operation (115 fJ (W)[80], 1 pJ (E)[81]) and CMOS compatibility. Current key research challenges for ECM lie in obtaining a better understanding of the physical mechanisms, improving reliability and uniformity, reducing random telegraph noise (RTN) and finding compatible and reliable selectors[3].

### 1.1.3.2 Electronic RRAM

As early as 1960s, an electronic mechanism for resistive switching was reported by Simmons and Verderber in the Au:SiO$_x$ system[82]. It was postulated that Au atoms introduce a broad band of localized impurity levels within the band gap of the insulator. Such impurity band allows electrons tunneling through adjacent sites, but the associated barrier can be repeatedly modified by trapping/detrapping events under a high field, resembling gate-channel modulation in flash FET. In 2000s, researchers extended the idea to other systems, *e.g.*, using Co, Ni (ref.[83]), Ge (ref.[84]), Mg, Ag, Al, Cr, CuPc (ref.[50]) to replace Au, and using complex oxide[85] or polymers[49-50] to replace $SiO_x$. Meanwhile, another electronic switching mechanism was also reported[40,44,86]. This mechanism occurs at the interface between a conducting oxide and a metal, by way of modulation of the Schottky barrier by the trapped electrons near the interface. As the population of the trapped electrons may be altered by the applied field, it can also cause a large resistance change for the electronic transport across the Schottky barrier. Another electronic switching mechanism is proposed for strongly correlated electron systems, which are typically based on transition-metal complex oxides (*e.g.*, perovskite). Here, electron injection/removal under an electrical field acts like doping to change the charging state, therefore inducing electron localization/delocalization. Such metal-insulator transition has been demonstrated in $La_{0.67}Ca_{0.33}MnO_3$ (ref.[87]) and $Pr_{0.7}Ca_{0.3}MnO_3$ (ref.[88]). In addition, nanometallic RRAM[89], the focus of this thesis, operates on an electronic mechanism as will be extensively discussed in several later chapters.

### 1.1.3.3 Summary of Reported RRAM Systems

**Table 1.2** summarizes reported RRAM systems in the literature. The data of ionic devices are largely based on Yang's review paper (see the section of Supporting Information of ref. [4]).

**Table 1.2A**. Cation devices (adapted from ref.[4])

| Active Layer | BE | TE | Mode |
|---|---|---|---|
| $Ta_2O_5$ | Pt | Cu | Bipolar |
| $SiO_2$ | W | Cu | Uni/Bipolar |
| $HfO_2$ | Pt | Cu | Bipolar |
| $WO_3$ | Pt | Cu | Bipolar |
| $ZrO_2$ | Ag | Au | Bipolar |
| $SrTiO_3$ | Pt | Ag | Bipolar |
| $TiO_2$ | Pt | Ag | Bipolar |
| $CuO_x$ | Cu | Al | Unipolar |
| ZnO | (Pt, Al):ZnO | Cu | Bipolar |
| $Al_2O_3$ | Al | Cu | Bipolar |
| $MoO_x$ | Cu | Pt | Bipolar |
| $GdO_x$ | Pt | $Cu:MoO_x$ | Bipolar |
| $Ge_xS_x$ | W | Ag | Bipolar |
| $As_2S_3$ | Au | Ag | Bipolar |
| $Cu_2S$ | Cu | Pt | Bipolar |
| $Zn_xCd_{1-x}S$ | Pt | Ag | Bipolar |
| AgI | Pt | Ag | Bipolar |
| $RbAg_4I_5$ | Pt | Ag | Bipolar |
| $Ge_xSe_y$ | W | Ag, Cu | Bipolar |
| $Ge_xTe_y$ | TiW | Ag | Bipolar |
| Ge-Sb-Te | Mo | Au, Ag | Bipolar |
| MSQ | Pt | Ag | Bipolar |
| Doped organic semiconductors | Pt | Cu | Bipolar |
| Nitrides | Pt | Cu | Bipolar |
| Amorphous Si | $p^+$-Si | Ag | Bipolar |
| Carbon | Pt | Cu | Bipolar |
| Vacuum gaps | $RbAg_4I_5$/Ag, $Ag_2S$/Ag | W, Pt | Bipolar |

**Table 1.2B**. Anion devices (adapted from ref.[4])

| Active Layer | BE | TE | Mode |
|---|---|---|---|
| MgO | Pt | Pt | Unipolar |
| $TiO_x$ | Ru, Pt | Al, Pt | Uni/Bipolar |
| $ZrO_x$ | $p^+$-Si, $n^+$-Si | Pt, Cr | Uni/Bipolar |
| $HfO_x$ | TiN, Ti | TiN, Pt | Bipolar |
| $NbO_x$ | $p^+$-Si | Pt | Unipolar |
| $TaO_x$ | Ta, Pt | Pt, Ta | Bipolar |
| $CrO_x$ | TiN | Pt | Bipolar |
| $MoO_x$ | Pt | Pt-Ir | Uni/Bipolar |
| $WO_x$ | W, FTO | TiN, Au | Bipolar |
| $MnO_x$ | Pt | Al, TiN | Bipolar |
| $FeO_x$ | Pt | Pt | Uni/Bipolar |
| $CoO_x$ | Pt | Pt | Unipolar |
| $NiO_x$ | Pt | Pt, Hg | Uni/Bipolar |
| $CuO_x$ | TiN, TaN, SRO, Pt | Pt | Bipolar |
| $ZnO_x$ | Pt, Au | TiN, Ag | Bipolar |
| $AlO_x$ | Ru, Pt | Pt, Ti | Uni/Bipolar |
| $GaO_x$ | ITO | Pt, Ti | Bipolar |
| $SiO_x$ | Poly-Si, TiW | Poly-Si, TiW | Unipolar |
| $SiO_xN_y$ | W | Cu | Bipolar |
| $GeO_x$ | ITO, TaN | Pt, Ni | Bipolar |
| $SnO_2$ | Pt | Pt | Unipolar |
| $BiO_x$ | Bi | W, Re, Ag, Cu | Bipolar |
| $SbO_x$ | Pt | Sb | Uni/Bipolar |
| $SmO_x$ | TiN | Pt | Bipolar |
| $GdO_x$ | Pt | Pt | Unipolar |
| $YO_x$ | Al | Al | Unipolar |
| $CeO_x$ | Pt | Al | Bipolar |
| $EuO_x$ | TaN | Ru | Uni/Bipolar |
| $PrO_x$ | TaN | Ru | Bipolar |
| $ErO_x$ | TaN | Ru | Unipolar |
| $DyO_x$ | TaN | Ru | Unipolar |
| $NdO_x$ | TaN | Ru | Unipolar |
| $Ba_{0.7}Sr_{0.3}TiO_3$ | $SrRuO_3$ | Pt, W | Bipolar |
| $SrTiO_3$ | $SrRuO_3$, Au, Pt | Au, Pt | Bipolar |
| $SrZrO_3$ | $SrRuO_3$ | Au | Bipolar |
| $BiFeO_3$ | $LaNiO_3$ | Pt | Bipolar |
| $Pr_{0.7}Ca_{0.3}MnO_3$ | YBCO, Pt, $LaAlO_3$ | Ag | Bipolar |
| $La_{0.33}Sr_{0.67}FeO_3$ | Au | Al | Bipolar |
| $Pr_yLa_{0.625-y}Ca_{0.375}MnO_3$ | Ag | Ag | Bipolar |
| $AlN$ | Al, TiN, Pt | Al, TiN, Pt | Bipolar |
| $ZnTe$ | Si | Au | Bipolar |
| $ZnSe$ | $p^+$-Ge | In, In-Zn | Bipolar |
| Polymers | Al, ITO, Cu | Al, ITO, Cu | Bipolar |

**Table 1.2C**. Electronic devices (adapted from ref.[4])

| Active Layer | BE | TE | Mode |
|---|---|---|---|
| $Zr^+$:$ZrO_2$[90] | $n^+$-Si | Au | Bipolar |
| $SiO_x$[82] | Al | Au | Bipolar |
| $Pr_{0.7}Ca_{0.3}MnO_3$[88] | SRO, Pt | Ti, Au | Bipolar |
| $La_{0.67}Sr_{0.33}MnO_3$[46] | Ag | N/A | Bipolar |
| $La_{0.7}Ca_{0.3}MnO_3$[47] | Pt | Ag | Bipolar |
| $Sm_{0.7}Ca_{0.3}MnO_3$[86] | Ti | N/A | Bipolar |
| $Nb:SrTiO_3$[40] | N/A | SRO | Bipolar |
| $Cr:SrZrO_3$[39] | SRO | Pt | Bipolar |
| (Au, Co, Ni): $SiO_2$[83] | $p^+$-Si | Al | Bipolar |
| Polymers[50] | Al, Cr, Cu, ITO, Au, Ni | Al | Bipolar |
| $LaNiO_3:LaAlO_3$[89,91] | SRO | Pt | Bipolar |
| $SrRuO_3:LaAlO_3$[89,91] | SRO | Pt | Bipolar |
| $LaNiO_3:CaZrO_3$[89,91] | SRO | Pt | Bipolar |
| $SrRuO_3:CaZrO_3$[89,91] | SRO | Pt | Bipolar |
| $Pt:SiO_2$[33,89,92] | SRO, Mo | Pt | Bipolar |
| *(Al, Cr, Cu, Ta, Pt):$Si_3N_4$ | SRO, Ti, TiN, Mo, Ta | Pt | Bipolar |
| *$Pt:SiO_xN_y$ | Mo | Pt | Bipolar |
| *Pt: ($MgO$, $Al_2O_3$, $Y_2O_3$, $HfO_2$, $Ta_2O_5$) | Mo | Pt | Bipolar |

* will be covered in this thesis

## 1.1.4 The Need for A New RRAM

Ionic RRAM devices, which rely upon physical migration of cations (metal$^+$) or anions ($O^{2-}$), come with some intrinsic problems that may affect their performance. One is the physical damage incurred during operation because of excess Joule heating or high-field breakdown. For example, Joule heating in unipolar devices causes irreversible unidirectional atom migration; these devices typically only lasts ~$10^2$ cycles[93]. As another example, electrodes can be easily blown off in $TiO_2$ devices during electrical switching [21,72]. The stochastic nature of localized filament development is another origin of poor device performance. To mitigate this problem, uniform dopants in the dielectric layers (to provide conduction bridges or seed crystals[94-95]) and nanodot "dopants" near one electrode (to serve as field concentrators) have been attempted[76]. A suboxide layer[96-97], a multilayer architecture[98], and a selection switch[99] have also been introduced to the RRAM stack to improve its performance. These reliability issues are clearly important for RRAM and their uncertain nature is intrinsic to ionic RRAM stemming from its thermoionic transport mechanism and filamentary paths. Moreover, some ionic devices do have outstanding endurance: a $TaO_x$ ionic-RRAM lasting ~$10^{12}$ cycles was reported by the Samsung group[28], which much exceeds the requirement for a typical non-volatile storage (*e.g.*, flash memory[100]) of $10^5$-$10^6$ cycles. Such large disparity in the endurance limit of ionic RRAM devices is not understood, adding to the uncertainty of the state of affairs.

Another problem of ionic RRAM is the switching speed, which is strongly dependent on the applied voltage[101-102]. Again, this is intrinsic to the ionic RRAM since physically, a larger voltage injects more energy into the filaments and generates a higher temperature, thus facilitating ion migration to achieve switching within a shorter time. Typically, a few fold increase of the switching voltage is required to increase the switching speed from 100 s to 100 ns. Since a low voltage/power circuit is desired in modern digital electronics, the large switching voltage demanded by a fast RRAM is certainly a drawback for future applications.

In view of this background, we believe there is a need to develop another RRAM based on electronic switching mechanisms and not relying on filamentary-type of switching paths. Nanometallic materials described next contain atomically dispersed metal atoms in a uniform amorphous dielectric matrix. Such a uniform composition with dispersed metal atoms may have some advantages. It may effectively spread the electric field across the entire film, it may make the film immune to local Joule heating problems, and it may provide electronic conduction thus obviating the need for ionic transport. These features may in turn provide better RRAM characteristics, leading to superior device performance.

## 1.2 Nanometallicity in Random-Electron Materials

Random-electron materials are ones that do not provide a periodic structure for electrons. They include amorphous materials, but they may also include solid solutions of an insulator and a conductor set in a crystalline framework. Amorphous materials have a non-periodic structure containing "randomness". They lack long-range order but, typically, some short-range order still exists at the local, atomic level due to the specific nature of chemical bonding. Amorphous materials cover a wide variety of materials classes including polymers, oxides, nitrides, borides, chalcogenides, semiconductors and even metals. In fact, it was realized as early as 1960s that "*nearly all materials can, if cooled fast enough and far enough, be prepared as amorphous solids*"[103].

Due to non-periodicity, amorphous materials cannot be described within the framework of traditional band theory. However, one can still use the term *extended states* referring to electronic states which have wave functions with appreciable amplitude throughout the solid. They may be contrasted with *localized states* that have wave functions with decaying amplitudes. In random materials, lacking periodic boundary conditions, there is no well-defined band gap; instead states are allowed to continuously distribute within the "band gap". But these states have relatively low density of states, meaning their available sites are sparse and they are spread far apart in space. That is, they are localized states through them "percolation" is difficult. By replacing band edge with mobility edge as shown in (see **Figure 1.11**)[103], band pictures are still applicable in amorphous materials to some extent, which can facilitate our understanding.

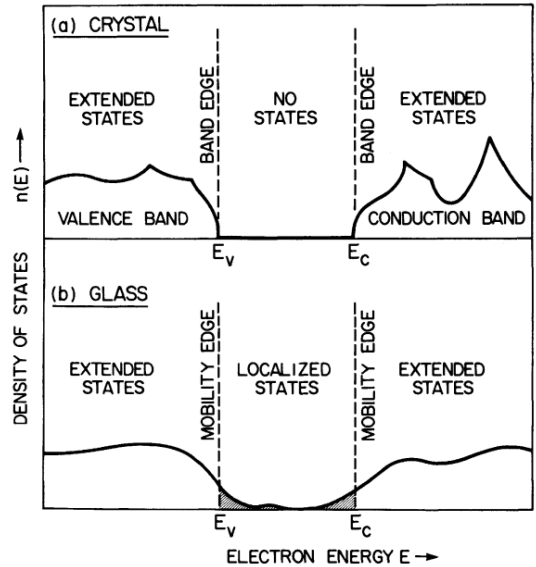

**Figure 1.11**. Schematic density of states for a crystalline and amorphous semiconductor (adapted from ref.[103]).

The wave function of states can be formally described as $|\Psi|^2 = P \sim \exp(-r/\xi)$, which is the probability $P$ of finding electrons. A localized state has a finite $\xi$, and its wave function falls off exponentially vanishing at distance $r$ much larger than $\xi$ from the localization center. This localization length $\xi$ is the "diffusion distance" $\zeta$ of Anderson[104]. It provides a straightforward criterion to distinguish a random insulator from a conductor: $\zeta$ describes how far a free electron can tunnel elastically at 0 K, thus an insulator has a finite $\zeta$, while a conductor has an infinite $\zeta$.

Our concept of nanometallicty stems from the realization that, even for localized states, their electrons can be visualized as "free" when the relevant spatial extent is less or on the order of $\xi$. The concept implies a size-triggered metal-insulator transition: a material is insulating if the sample size is much larger than $\xi$, but conducting if the size is much smaller than $\xi$. This size-dependent idea was experimentally confirmed by my colleagues; they realized "nanometallicity" by either reducing the sample size of an insulator to fall below the diffusion length; or by increasing the diffusion length through addition of metallic content to an insulator[89,91-92]. Remarkably and technologically importantly, they also discovered that the diffusion length in a given sample can be manipulated by a voltage in a reversible manner: for example, a positive critical voltage apparently causes $\xi$ to decrease, and a negative critical voltage apparently returns $\xi$ to the original value. This latter finding enables a new class of RRAM, the nanometallic RRAM. This thesis seeks to explore the generality of nanometallicity and nanometallic RRAM in various random materials, and provide an experimental foundation to understand and to guide future nanometallic materials selection and design. This will hopefully contribute to the development of novel and superior materials and devices for the emerging memory technology. In the next section, some background for the underlying amorphous materials is provided.

## 1.3   Structures and Properties of $Si_3N_4$

Silicon nitride is one of the most studied materials, as it exhibits excellent mechanical properties, high tolerance to harsh chemical/thermal environment, superior creep and tribological/wear resistance, and desirable dielectric responses including a high dielectric strength. Bulk, crystalline silicon nitride and its modifications are deployed in a variety of industrial applications such as automobile engine components, pump seal parts, heat exchangers, bearings and cutting tools, while amorphous silicon nitride films are widely used as dielectrics in modern electronics[105]. These outstanding properties and knowledge base make silicon nitride an excellent base and benchmark material for exploring amorphous nanometallic RRAM in this thesis.

There are three crystalline phases for $Si_3N_4$, known as $\alpha$, $\beta$ and $\gamma$ phase. The $\gamma$ phase (a cubic spinel structure with a space group $Fd$-$3m$, $a$=7.7339 $\pm$ 0.0001Å)[106] is synthesized under a high pressure and high temperature although it can be retained at ambient conditions afterwards, but both $\alpha$ and $\beta$ phases are commonly produced under normal or slightly elevated processing pressure condition. The latter phases in their pure form exhibit the same chemical composition and essentially identical measured densities. They are present in a hexagonal structure but with different stacking sequences: ABCD in $\alpha$ (Space group: $P31c$, $a$=7.748 $\pm$ 0.001Å, $c$=5.617 $\pm$ 0.001Å) and ABAB in $\beta$ (Space group: $P6_3/m$, $a$=7.608 $\pm$ 0.001Å, $c$=2.9107 $\pm$ 0.0005Å)[107]. As a result of the longer stacking sequence, the $\alpha$-phase has a higher hardness than the $\beta$-phase because of the larger Burgers vector required for slip dislocations[108]. In its pure $Si_3N_4$ form, the $\alpha$-phase is thermodynamically metastable compared to the $\beta$-phase, which means at high temperature the $\alpha$-phase always converts to the $\beta$-phase[109]. Under typical processing/synthesis conditions (sintering, hot pressing, $etc$.), during the $\alpha$-$\beta$ phase transformation, the newly grown $\beta$ crystals consume the unstable matrix and tend to form whiskers, resulting in a self-reinforced composite. This long-rod microstructure endows high toughness and strength to $\beta$-Si$_3$N$_4$. However, with the addition of stabilizing cations and tailoring of powders, it is also possible to form rod-containing, self-reinforced $\alpha$-Si$_3$N$_4$ (more precisely, its cation-stabilized oxynitride) which has high toughness and strength in addition to superior hardness[110]. Amorphous silicon nitride can be easily obtained through low temperature processing such as sputtering and low temperature CVD. X-ray diffraction radial distribution studies suggest the short-range order in amorphous silicon nitride is similar to that in $\beta$-Si$_3$N$_4$ (ref.[111]). Basic Si$_3$N$_4$ properties are summarized in **Table 1.3**.

| Property | Value |
|---|---|
| Mass density | 3.169 kg/m$^3$ (Ref.[112]) |
| Melting point | 1900$^\circ$C (Ref.[113]) |
| Energy band gap | 4.6 eV (Ref.[114]) |
| Young's modulus | Dense ~300 GPa<br>Reaction-bonded ~170 GPa (Ref.[113]) |
| Poisson ratio | 0.28 (Ref.[115]) |
| Tensile or fracture strength | Dense ~400-950 GPa<br>Reaction-bonded ~120-220 GPa (Ref.[113]) |
| Residual stress on silicon | 600 MPa (compressive) (Ref.[116]) |
| Thermal conductivity | Dense ~15-50 Wm$^{-1}$K$^{-1}$<br>Reaction-bonded ~4-30 Wm$^{-1}$K$^{-1}$ (Ref.[113]) |
| Thermal expansion coefficient | 2.9-3.6×10$^{-6}$ $^\circ$C$^{-1}$ (Ref.[113]) |
| Dielectric constant | 7 (Ref.[117]) |
| Refractive index | 2.016 (Ref.[118]) |
| Electrical conductivity | 10$^{12}$ $\Omega$ cm (Ref.[119]) |
| Wet etching | BHF/Phosphoric acid (Ref.[120]) |
| Plasma etching | SF$_6$/O$_2$/N$_2$ (Ref.[121]) |
| Adhesion to SiO$_2$ | Good (Ref.[122]) |
| Hydrophobicity | No, but achievable by chemical surface modification (Ref.[123]) |

**Table 1.3.** Basic parameters of $Si_3N_4$. Data are extracted from cited literature which may be valid for a specific phase (*e.g.*, $\beta$-phase).

Silicon nitride can be extended to silicon oxynitride $SiO_xN_y$ by substituting O for N in the structure. This is especially straightforward in an amorphous structure since it is based on silicon tetrahedra connected by corner N or O. (Each Si atom is coordinated by four N/O atoms, while N/O atoms are coordinated by three/two Si atoms, respectively[124].) Regarding the bonding states, ideal silicon oxynitride can be quantitatively described by the random bonding (RB) model: $SiO_xN_y$ is composed of five types of tetrahedra $SiO_vN_{4-v}$, where $v$=(0, 1, 2, 3, 4). The distribution function of $SiO_vN_{4-v}$ tetrahedra with a particular composition $x$ and $y$ is formulated as[124-125]:

$$W(v, x, y) = \left(\frac{2x}{2x+3y}\right)\left(\frac{3y}{2x+3y}\right)\frac{4!}{v!(4-v)!}$$

Such formula presumes that $SiO_xN_y$ does not contain intrinsic defects (*e.g.*, Si-Si, Si-O-O-Si, or N-O bonds). The experimental Si $2p$ XPS spectra show perfect agreement with the above formula (**Figure 1.12**)[125].

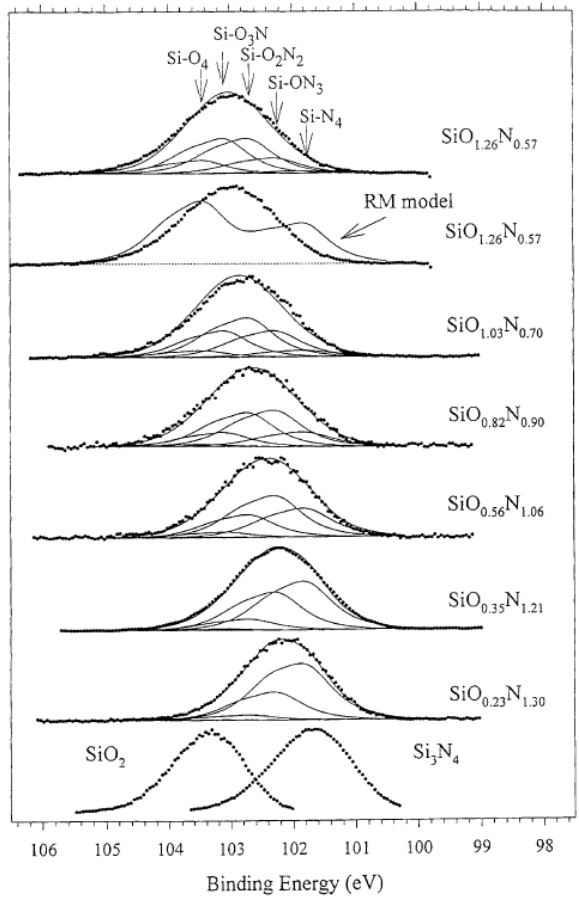

**Figure 1.12.** Si 2$p$ XPS spectra of SiO$_x$N$_y$ films with different compositions. Dotted lines: experimental data. Solid lines: simulation from RB model (adapted from ref.[125]).

Despite the agreement, chemical shifts of Si 2$p$ (also N 1$s$) were observed in some SiN$_y$ films, which cannot be simply predicted by the RB model[126]. The underlying reason lies in the fact that random mixing of Si and Si$_3$N$_4$ do exist to some extent and the RB model need to be modified, *i.e.*, the intermediate mixing (IM) model is needed[114]. The IM model incorporates the presence of separate "phases" of Si and Si$_3$N$_4$ in the film with sub-nitrides on the interfaces. The associated band diagram can be visualized as **Figure 1.13**, where the insulating large band gap of $Si_3N_4$ is randomly disturbed by the semiconducting small band gap of Si.

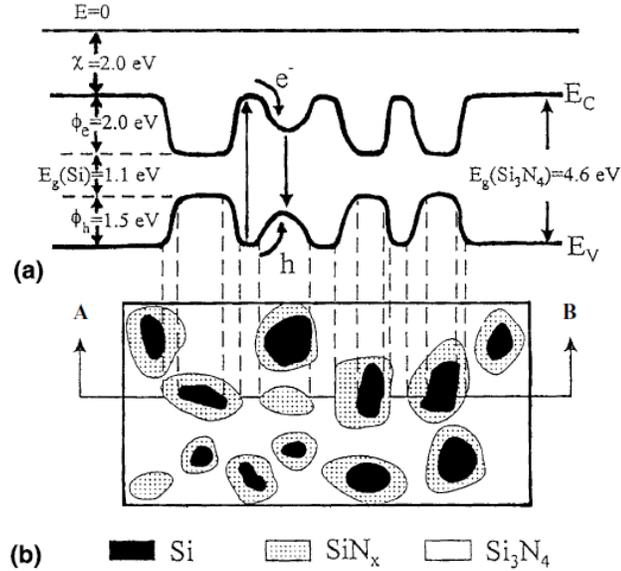

**Figure 1.13**. (a) 1-D energy band diagram for $SiN_x$; (b) the corresponding spatial chemical composition fluctuation (adapted from ref.[126]).

## 1.4  Thesis Outline

Nanometallic RRAM research was initiated by my colleagues Yudi Wang and Soo Gil Kim (ref.[91]). Nanometallicity and nanometallic RRAM were documented, for the first time in the PhD thesis of Yudi Wang[91] using crystalline thin films of four perovskites ($LaAlO_3$:$LaNiO_3$, $LaAlO_3$:$SrRuO_3$, $CaZrO_3$:$LaNiO_3$, $CaZrO_3$:$SrRuO_3$) expitaxilly grown (by pulse laser deposition) on $SrTiO_3$ single crystals of 110 and 111 orientations. Despite their crystalline lattice framework, these materials are random-electron systems since cations of the conducting component perovskites ($LaNiO_3$ and $SrRuO_3$) are randomly mixed with cations of the insulating component perovskites (LaAlO$_3$ and CaZrO$_3$). Their pioneering work was followed by my colleagues Albert B. K. Chen and Byung Joon Choi (ref.[92]) who extended the material category for nanometallicity and nanometallic RRAM from crystalline to amorphous materials (SiO$_2$:Pt and Si$_3$N$_4$:Pt), which have the benefit of not requiring a special substrate, being CMOS compatible, and using a much more manufacturable process such as sputtering. (The work on SiO$_2$:Pt was documented in the PhD thesis of Albert B. K. Chen[92].)

In this thesis, my first aim is to explore the generality of the "random mixture" idea by extending the SiO$_2$:Pt work to other insulators (various oxides, nitrides, and oxynitrides) and to other metals beyond Pt (non-noble metals including main-group metals.) This work is successful as documented in **Chapter II**, which provides some of the very best RRAMs available today. (This chapter also includes a preview of our best understanding of the nanometallic RRAM mechanism to date.) Such extension adds considerable flexibility to nanometallic RRAM design and applications.

The above work is followed by a detailed study of three topics of fundamental importance to the understanding of nanometallicity and nanometallic RRAM: a non-electrical, mechanical stimulus—both static ones and dynamic, sub-picosecond ones—for electronic switching in **Chapter III**, dielectric properties in **Chapter IV** and electron conduction mechanisms in **Chapter V**. In addition, the environmental effect is described in **Chapter VI** to provide a practical view of additional transport (including ionic current) and degradation mechanisms at ambient temperature in the presence of moisture. This research goes much beyond the level and depth of past RRAM research, and will hopefully aid the understanding and modeling of not only our but all RRAM materials and devices.

The last part of the thesis explores several means to model and build practical devices that are of technological relevance to RRAM applications: a phenomenological (parallel circuit) model in **Chapter VII**, a multibit memory in **Chapter VIII**, a low-power memory in **Chapter IX**, and a complementary memory in **Chapter X**. Additional information of documentary interest but too long or too numerate to include in the main chapters of the theses are provided in **Appendices**.

# Chapter II. Nanometallic Thin Films: Material Characteristics and Voltage Induced Metal-Insulator Transition

## 2.1 Introduction

An electron in random materials does not propagate as plane wave; instead it is repeatedly scattered and redirected to random directions in a way akin to random walk, termed *diffusion* by P. W. Anderson (1958)[1]. Along the way, the amplitude of the random-walk wave function may decay: such decay signifies electron localization and the decay distance $\zeta$ defines the localization length. A conventional random insulator has a finite $\zeta$, whereas a conventional random metal has an infinite $\zeta$. But if the sample size $\delta$ falls below $\zeta$, then even a bulk-insulating sample is effectively metallic. This simple idea of size-defined metal insulator transition (MIT) in random materials was not revealed until our group's recent work[2-3]: by randomly inserting atomically dispersed conducting components into an insulating matrix (*e.g.*, Pt to $SiO_2$), $\zeta$ can be raised above $\delta$ (5-40 nm, the film thickness) and a size-dependent MIT is demonstrated. Importantly, a sudden decrease of $\zeta$ can too be rendered at a critical voltage $V_c$ by injected charge that alters the landscape of the Coulomb repulsion, which curtails electron "diffusion". After voltage removal, trapped charge still remains, apparently indefinitely, unless it is released by an opposite voltage. Such voltage-triggered MIT forms the basis of a new type of nonvolatile resistive switching random access memory (RRAM).

In this chapter, we will explore the basic characteristics of such electronic memory to set the stage of a more detailed investigation in later chapters on several new aspects not yet fully explored in nanometallic memory. The data on electrical switching properties presented here were primarily obtained from $Si_3N_4$:Cr films, namely amorphous $Si_3N_4$ films with atomically dispersed Cr of various compositions. But the same observations and picture were also found in other CMOS-compatible compositions. We will demonstrate RRAMs from these materials feature fast switching speed, long retention, high endurance, easy scaling and superior statistical uniformity. On the physics side, since the high resistance state (HRS) falls into the localization regime, its resistance exponentially increases with thickness in agreement with the random-walk quantum physics but in gross violation of Ohm's law. Such thickness-sensitivity, along with thickness-composition map that defines the size-dependent MIT, offers new degrees of freedom to engineer material resistance. In addition, the voltage-time dilemma commonly encountered by electronic memory will be addressed and a solution based on lattice relaxation will be proposed. This is fundamentally important because the resolution of the dilemma assures that the memory can be simultaneously fast, robust, with a long retention yet a low switching voltage. To explain all the experimental evidence, we will provide a detailed switching model at the end of the chapter.

## 2.2  Sample Preparation

Magnetron co-sputtering is the main tool for thin film fabrication in this thesis. As introduced in the previous chapter, sputtering can deposit almost any amorphous material composition using proper targets at low temperature. The stoichiometry depends on the DC/RF power, working gas pressure, target and substrate geometry, temperature, *etc*. For co-sputtering, separate control of sputtering conditions for different materials is of great importance; these conditions can be quite different for different materials. A precise determination of the stoichiometry must rely upon proper post-deposition characterization, which in turn informs the best deposition conditions for optimal film composition and quality.

There are several choices for substrates in this work. The most commonly used one is a moderately doped p-type (100) silicon wafer (resistivity~10 ohm-cm) covered with 200 nm thermal oxide film. Others are highly doped silicon wafers and fused silica. The bottom electrode (BE), ~20 nm in thickness, was first deposited by DC sputtering. An "active" mixture layer containing an insulator ($I=Si_3N_4$, $SiO_2$, *etc.*) and a metal ($M=$ Pt or non-noble metals) component were next co-sputtered by RF magnetron sputtering using appropriate targets. The metal composition was controlled by varying the RF power of the metal target. Before deposition, the deposition chamber was evacuated to reach a typical vacuum $<10^{-6}$ Torr. Finally, a top electrode (TE) was deposited, again by sputtering, through a shadow mask; or it was deposited without a mask but photolithographically patterned later. The deposition geometry is shown in **Figure 2.1a**. The final device geometry is shown in **Figure 2.1b**. In the following, devices or samples will be referred to by their composition of the bottom electrode, nanometallic film, and top electrode, such as $Mo/Si_3N_4$:20%Cr/Pt. Wherever appropriate, the thickness of the nanometallic film (*e.g.*, 10 nm) and the size of the top electrode (*e.g.*, 100 µm x 100 µm) are also specified.

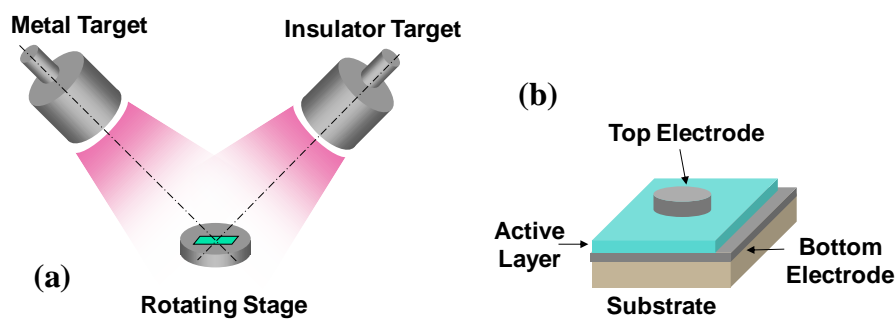

**Figure 2.1.** (a) Sputter deposition configuration, (b) Device configuration.

## 2.3 Chemical Bonding and Stoichiometry

### 2.3.1 Infrared (IR) Spectroscopy

IR is suitable for examining chemical bonding and stoichiometry information. Transmission spectra in the IR range were performed by a Fourier transform infrared spectrophotometer (FT-IR, Nexus 470, Thermo Nicolet, International Equipment Trading Ltd. Vernon Hills, Illinois, USA). To minimize background signal, two-side-polished IR-transparent KBr crystals (International Crystal Laboratories, Garfield, NJ) were used as substrates. Thicker films on the order of 100 nm thick, much thicker than typical device ones, were used to maximize the IR signal. The data were analyzed and fitted using a self-developed Mathematica program.

Generally speaking, oxygen introduction was unavoidable during nitride sputtering under the deposition conditions in this work. Oxygen contamination was more severe when the argon pressure was high or when the deposition rate was slow. For example, in **Figure 2.2b**, a film deposited under a high argon pressure using a $Si_3N_4$ (Reactive bonded $Si_3N_4$, free of sintering aids) target is mostly $SiO_2$ according to the strong Si-O peak at 1060/cm (ref.[4]). However, when the argon pressure is sufficiently low, the characteristic Si-N bonding at 860/cm becomes prominent (see **Figure 2.2**)[5]. This is understood as follows: a higher Ar working pressure yields more collisions, so the sputtered Si atom has a larger probability to be oxidized. The optimized deposition condition of Ar was 5 sccm (**Figure 2.2e**), which was used for all subsequent work described below unless otherwise noted.

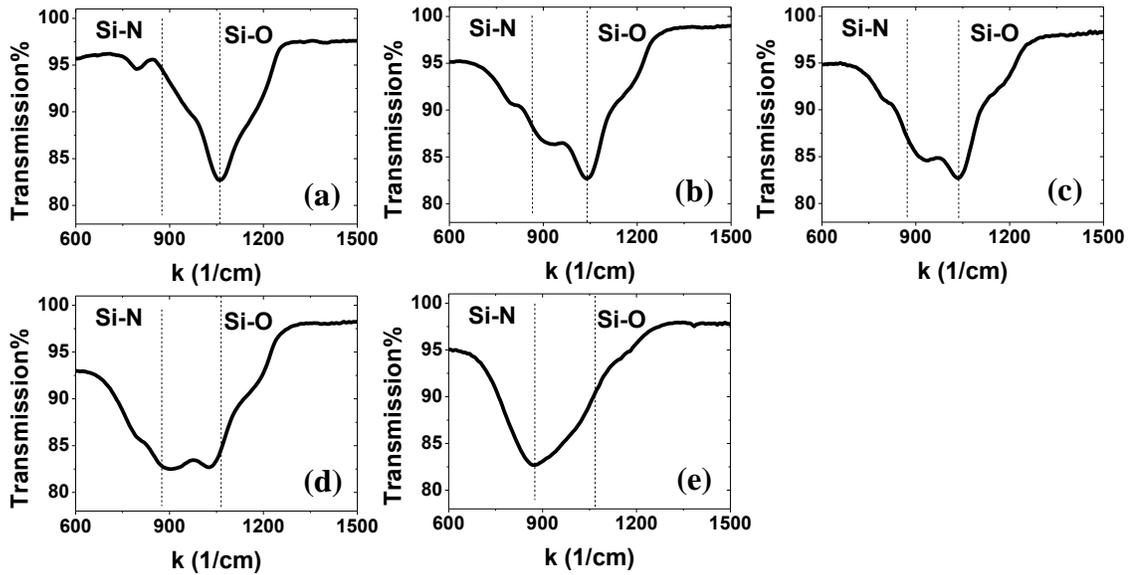

**Figure 2.2.** IR transmission data of films deposited with different Ar flow rate (a) 45 sccm ($SiO_2$ target), (b) 45 sccm (Ar 45 sample, $Si_3N_4$ target), (c) 32 sccm (Ar 32 sample, $Si_3N_4$ target), (d) 20 sccm (Ar 20 sample, $Si_3N_4$ target), and (e) 5 sccm (Ar 5 sample, $Si_3N_4$ target).

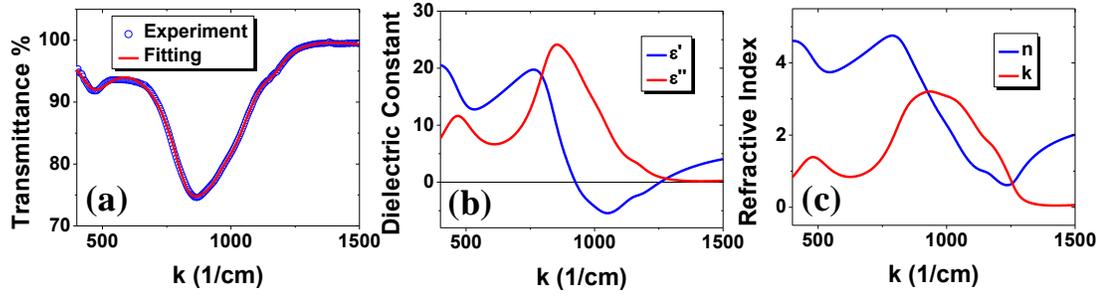

**Figure 2.3.** IR spectroscopy data fitting. (a) Experiment *vs.* fitting of transmission data based on Drude-Lorentz model. (b) Calculated dielectric constant. (c) Calculated refractive index.

Simulation was performed to fit the spectrum of the Ar5 sample using the Drude-Lorentz model and to extract complex dielectric constant ($\varepsilon$'+i$\varepsilon$'') and refractive index ($n$+i$k$) as a function of wavenumber. In the Drude-Lorentz model, dielectric response arises from harmonic oscillators with (Drude) or without damping (Lorentz),

$$\varepsilon(\omega) = \varepsilon_\infty + \sum_m \frac{\omega_{pm}^2}{\omega_{0m}^2 + i\Gamma_m\omega - \omega^2}$$

Here, $\omega_{0m}$ is the resonant frequency, $\omega_{pm}$ is the oscillator strength at $\omega$=0, $\Gamma_m$ is the damping coefficient and $\varepsilon_\infty$ is the permittivity at infinite frequency. Since our study indicated the existence of both Si-N and SiO bonding, and each distinct chemical bonding has three normal modes (stretching, bending, and rocking[4]), six oscillators (see **Figure 2.3a**) are used for fitting. This model gives very good fitting with the experimental data, as shown in the red curve of **Figure 2.3a**. With the fitting parameters listed in **Table 2.1**, we then compute the dielectric constant and refractive index in **Figure 2.3b** and **Figure 2.3c**. The minimum in transmission at ~860/cm is now seen to be related to the peak of the imaginary part of dielectric constant and refractive index, *i.e.*, poor transmission is due to strong resonant absorption. Our calculated optical parameters of the Ar 5 sample are consistent with Engheta *et al.*'s for $Si_3N_4$ covering the same energy range. This provides some confidence that we have indeed obtained a $Si_3N_4$ film.

| Parameter | Value | | Parameter | Value |
|---|---|---|---|---|
| $\varepsilon_\infty$ | 7 | | | |
| $\omega_{01}$ | 471.77 | | $\omega_{04}$ | 935.16 |
| $\Gamma_1$ | 168.64 | | $\Gamma_4$ | 193.11 |
| $\omega_{p1}$ | 859.351 | | $\omega_{p4}$ | 1439.00 |
| $\omega_{02}$ | 838.14 | | $\omega_{05}$ | 1167.36 |
| $\Gamma_2$ | 162.79 | | $\Gamma_5$ | 100.21 |
| $\omega_{p2}$ | 1336.52 | | $\omega_{p5}$ | 367.53 |
| $\omega_{03}$ | 1015.33 | | $\omega_{06}$ | 1203.3 |
| $\Gamma_3$ | 203.20 | | $\Gamma_6$ | 1056.57 |
| $\omega_{p3}$ | 1365.85 | | $\omega_{p6}$ | 1547.9 |

**Table 2.1.** Fitting parameters in **Figure 2.3a**.

### 2.3.2 Electron Energy Loss Spectroscopy (EELS)

The influence of deposition condition on O/N contents was confirmed using EELS, which is especially suited for light element mapping. Thin films ($\delta$=10 nm) were directly deposited on regular TEM grids and EELS spectra were collected in a transmission electron microscopy (TEM, 200 kV JEOL 2010F). **Figure 2.4** shows that as the working pressure decreases, the nitrogen to oxygen ratio increases. The residue oxygen in the Ar 2 sample is the lowest and may be attributed to the contamination (or surface oxygen/hydroxyl/water) of the TEM grid because a bare grid also shows a similar, small oxygen edge.

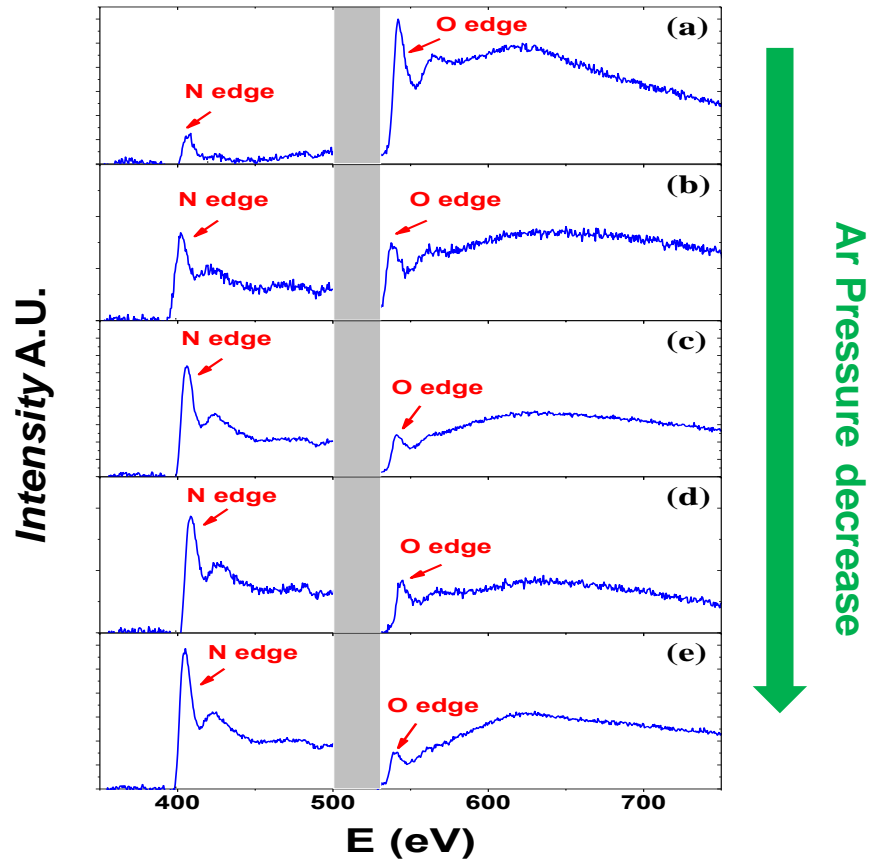

**Figure 2.4.** Electron energy-loss due to N and O for $Si_3N_4$ samples deposited under different Ar pressure. As Ar pressure decreases, N to O ratio greatly increases corresponding to better nitride film. (a) Ar 45, (b) Ar 12, (c) Ar 5, (d) Ar 3, and (e) Ar 2.

Film stoichiometry was further studied by energy-loss near edge structure (ELNES) of silicon *L*-edge, as shown in **Figure 2.5**. The fine structures of $SiO_2$, SiON and $Si_3N_4$ in the literature[6] (**Figure 2.5e**) were used as benchmark fingerprints for $SiO_2$, SiON and $Si_3N_4$ bonding. As Ar pressure decreases, the Si *L*-edge moves from 108 eV to 105 eV and the post-edge shoulder at ~ 115 eV gradually disappears, indicating a change from SiON to $Si_3N_4$. These results are consistent with the IR results where Si-N ionic bonding dominates at Ar 5. We thus conclude good silicon nitride films were obtained under Ar 5 sccm (3.5 mtorr) condition.

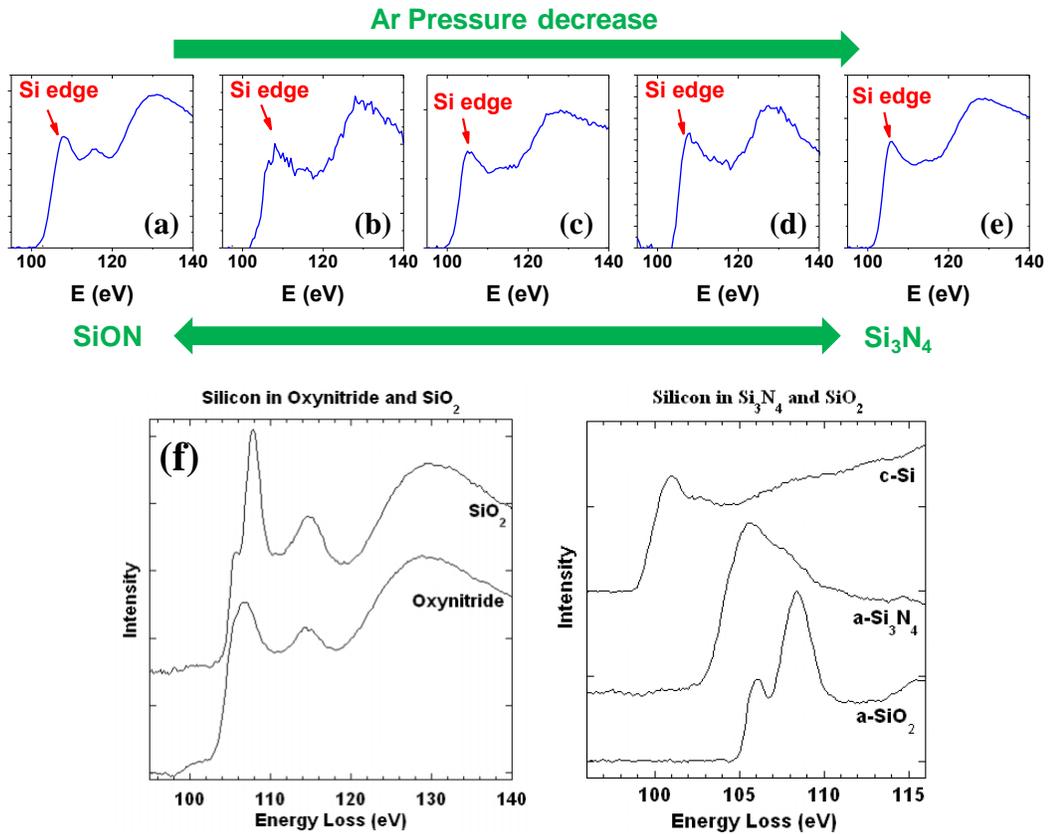

**Figure 2.5.** Silicon *L*-edge near edge fine structure for various Ar pressure. (a) Ar 45 sccm, (b) Ar 12 sccm, (c) Ar 5 sccm, (d) Ar 3 sccm, (e) Ar 2 sccm. (f) Standard silicon *L*-edge near edge fine structure for various silicon compounds (ref.[6]).

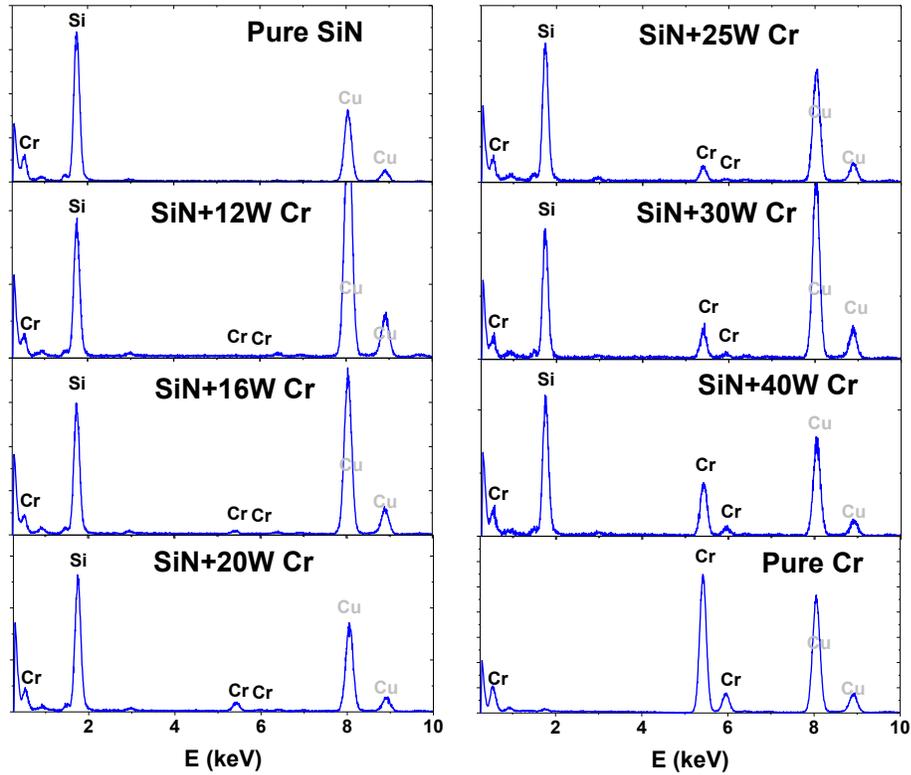

**Figure 2.6.** EDX data showing Si and Cr peaks at 0-10 keV energy range. Cu peak comes from TEM grid.

### 2.3.3 Energy-dispersive X-ray Spectroscopy (EDX)

Energy-dispersive X-ray spectroscopy (EDX) cannot differentiate N and O because the energy resolution is too coarse (~keV) to resolve the C/N/O peaks. However, it can be used to verify the amount of metal elements, which increases with the sputtering power. **Figure 2.6** shows the EDX spectra of a representative set of $Si_3N_4$:Cr films. The Cr peaks increase as the sputtering power of the Cr target increases. Using the integrated peak intensity for silicon and chromium, calibrated by the EELS data, we find the Cr concentration can be fitted to the power $P$ (in Watt) by:

$$Cr(atom\%) = 9.93 \times 10^{-2} P + 2.40 \times 10^{-2} P^2$$

## 2.4 Microstructure

### 2.4.1 Film Density

X-ray-reflectometry (XRR) is a non-destructive method for thin film characterization. It is useful to determine the film density, roughness and thickness. As shown in **Figure 2.7**, pure $Si_3N_4$ film (*i.e.*, film deposited without any co-sputtering metal) has a density ~3.5 g/cm$^3$, close to the ideal density. Comparing the XRR-determined film density with the following theoretical density, which assumes no porosity and no density (deviation) of mixing,

$$\rho_{Theoretical} = f_{Cr}\rho_{Cr} + (1 - f_{Cr})\rho_{SiN_{4/3}}$$

we find a reasonably good comparison only for 93%$SiN_{4/3}$:7%Cr, which is 3% less than the above estimate, indicating it is 97% dense. Here $f$ is the atomic fraction of metal, and $\rho_{SiN_{4/3}}$ =3.44g/cm$^3$ and $\rho_{Cr}$ =7.19g/cm$^3$ are theoretical densities of pure $Si_3N_4$ and Cr, respectively. In all other cases, a negative density of mixing is seen suggesting nano pores are generated during deposition. The relative density and nanoporosity have a profound influence on microstructure and electrical properties, as will be discussed in **Chapter VI**.

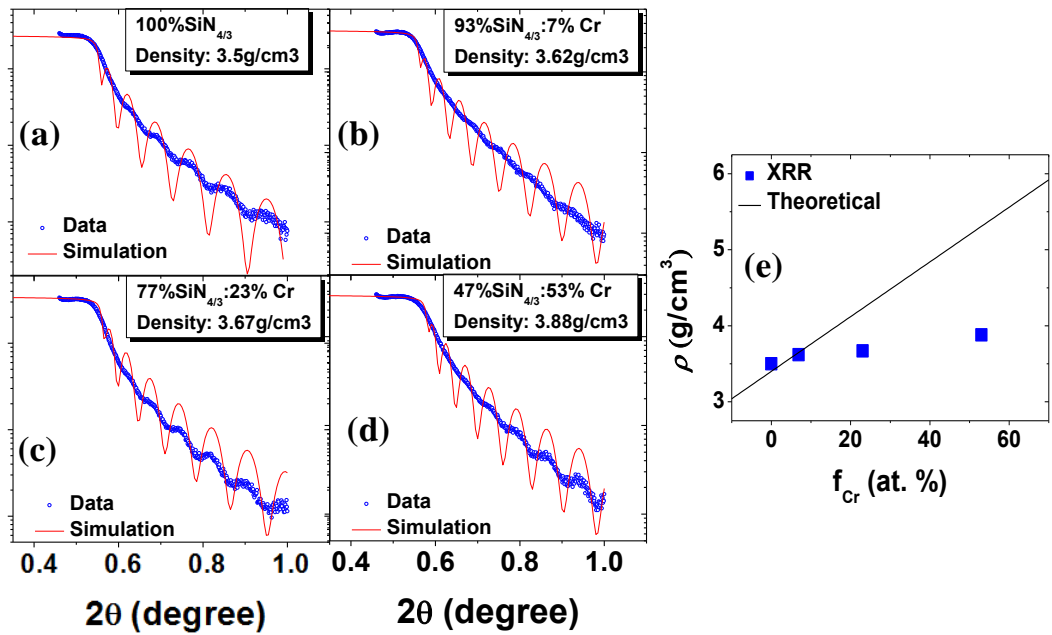

**Figure 2.7.** XRR fitting for (a) $SiN_{4/3}$, (b) 93%$SiN_{4/3}$:7%Cr, (c) 77%$SiN_{4/3}$:23%Cr, and (d) 47%$SiN_{4/3}$:53%Cr. (e) Experimental data *vs.* theoretical prediction.

### 2.4.2 Crystallinity

The crystallinity of nanometallic films was investigated by x-ray diffraction (XRD) using Cu Kα radiation. As shown in **Figure 2.8**, the Si/SiO$_2$ substrate exhibits one strong peak only: it is located at 69$^o$ corresponding to Si (100) of the substrate. This peak is present in all other samples since the same substrate was used. For pure Si$_3$N$_4$ film, there is no obvious difference between its XRD spectrum and that of the bare substrate, suggesting the amorphous nature of nitride film. Sputtered Cr film, on the other hand, exhibits two peaks, at 45$^o$ (100) and 82$^o$ (211), indicating it is polycrystalline. However, these peaks are not seen in **Figure 2.8** in 88%$SiN_{4/3}$:12%Cr and 76%$SiN_{4/3}$:24%Cr films (both are typical compositions for device). Therefore, we conclude that metal atoms (*e.g.*, Cr) mostly assume the same amorphous structure of the matrix $Si_3N_4$ films. That is, metal atoms are atomically dispersed in the amorphous matrix for the most part, and they do not crystallize even though some of them may have clustered, most likely by chance. This conclusion is further supported by high resolution TEM image as will be shown in a later section.

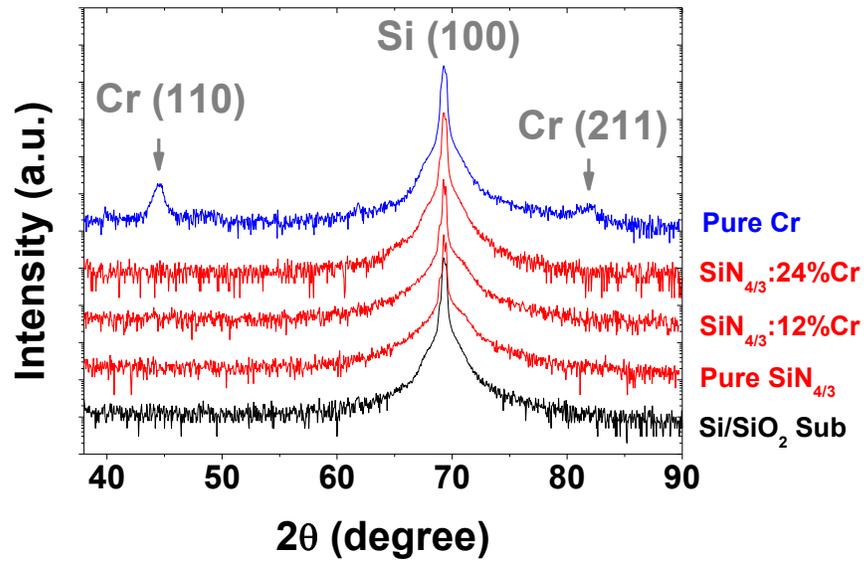

**Figure 2.8**. X-ray diffraction (*θ-2θ*) patterns of $Si_3N_4$:Cr films on Si substrate (film thickness: 100 nm). Also displayed are patterns of polycrystalline Cr film (showing (110) and (211) peaks, as labeled, Cr (200) is located at $64^o$ but too weak to be observed) on same substrate, and substrate itself (Si with native $SiO_2$ oxide). $Si_3N_4$:Cr nanometallic films show no obvious Cr crystal reflection.

### 2.4.3 Surface Morphology

Since our film is only ~10 nm thick, whose electrical integrity is critical, surface roughness of the bottom electrode and the active film is extremely important and must be kept well below 10 nm. In general, a larger sputtering power results in a larger roughness because surface atoms do not have sufficient diffusion time before the arrival of the next atomic layer. This would yield more porosity or poorer density. Roughness also accumulates as the film becomes thicker.

Surface morphology of some optimized films is shown in **Figure 2.9**. The uniform dark region is covered by a $Si_3N_4$:Pt hybrid film with 0.29 nm rms roughness (**Figure 2.9b**). Bright regions are covered by 40 nm thick Pt top electrodes, which have a similar roughness. This indicates both the active film and the electrode-covered devices are uniform in thickness.

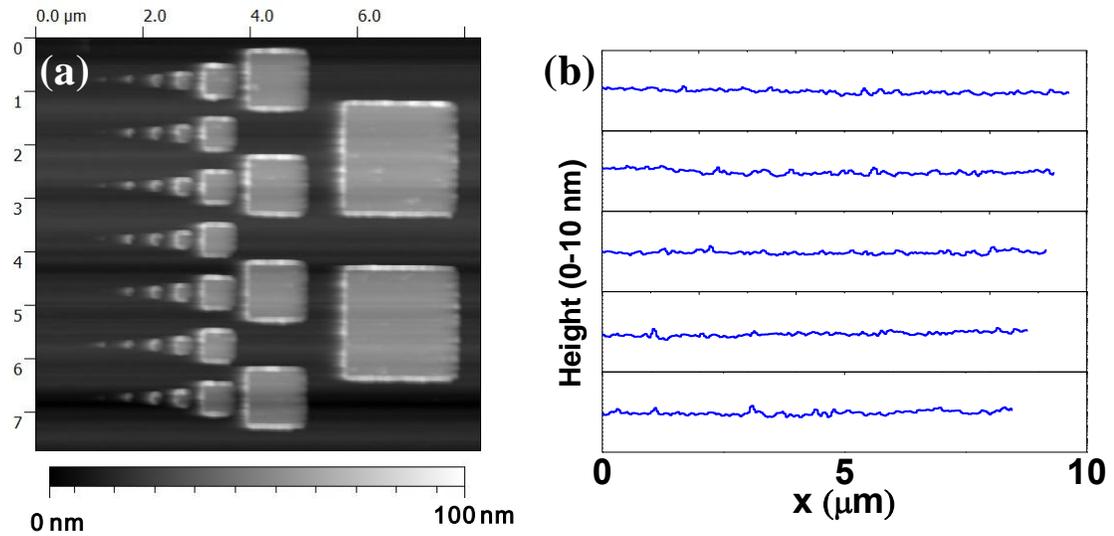

**Figure 2.9**. (a) AFM image of surface morphology of Mo/Si$_3$N$_4$:Pt/Pt device on Si/SiO$_2$ substrate. (b) Line scan profiles at 5 randomly selected locations.

Surface conductivity of as-fabricated devices was also examined by employing a conducting atomic force microscope (CAFM) technique. **Figure 2.10** shows a uniform black color (indicating low current) in the exposed region of the hybrid film regions, indicating an extremely resistive circuit when the contact area is limited to that of the AFM tip. No pinhole for large current passage is apparent. On the other hand, electrode regions show a much higher current (reaching 2 nA, which is the current compliance limit set by Asylum AFM) because there is a large equivalent measurement area as defined by the electrode size. A further check using a smaller voltage to stay within the compliance confirmed that the current is indeed proportional to the electrode area obeying Ohm's law.

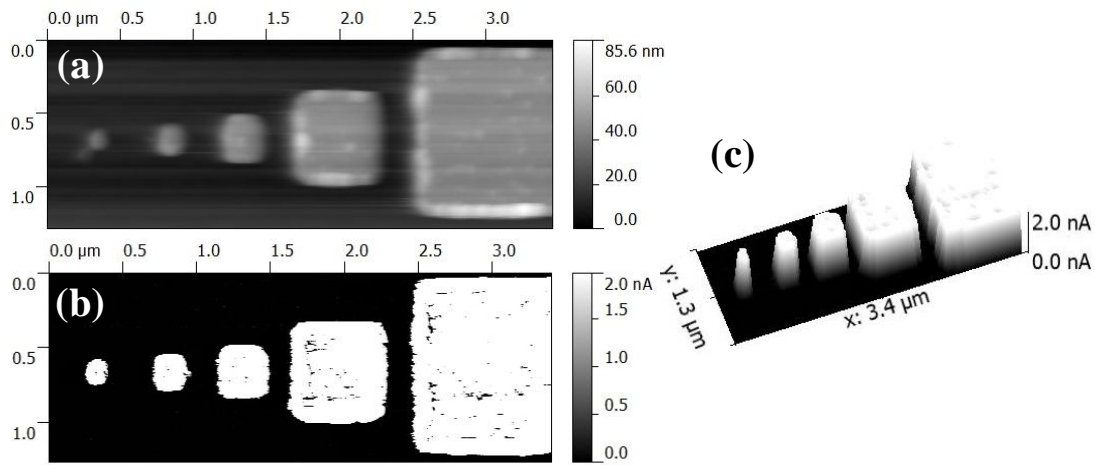

**Figure 2.10.** (a) Surface morphology *vs.* (b) electrical current map. (c) 3D current map.

### 2.4.4 Nanostructure

Nanostructure of the hybrid films was examined using TEM. To simplify the comparison, we show below only films of the same thickness (10 nm) but with different concentration. The results vary with material systems, which can be summarized into two cases:

1) In the $Si_3N_4$-based system, at low or moderate concentrations, metal-rich regions are not crystallized. As shown in **Figure 2.11**, $Si_3N_4$ and $Si_3N_4$:Cr films up to 17% Cr (**Figure 2.11a-e**) exhibit similarly uniformly amorphous morphology with a worm-like structure. This suggests that Cr atoms are atomically dispersed within the $Si_3N_4$ matrix. Non-uniformity occurs when Cr concentration reaches ~26%. As shown in **Figure 2.11 f-h**, the metal-rich clusters have different bright and light contrast which is likely due to different diffraction conditions of differently oriented crystals (Moire fringes seem unlikely because the thickness within the distance between clusters is probably the same) since high resolution images and electron diffraction patterns still suggest an amorphous structure without noticeable lattice fringes or diffraction spots. In other words, conventional metallic nanocrystals have not formed in these films; instead, there are amorphous Cr-rich clusters. Difficulty in crystallization may lie in the low atomic mobility of Cr, but it may also be caused by the (relatively) high density of the $Si_3N_4$ matrix, which provides little room for Cr to move. Therefore, any extra Cr atoms are forced to disperse atomically within the matrix. Interestingly, the relative density starts to decrease as the Cr concentration increases according to **Figure 2.7**. Eventually, it leads to nanopore generation and Cr crystallization, *i.e.*, a two-phase structure containing Cr nanocrystals and a nanoporous amorphous $Si_3N_4$ matrix.

2)     In the $SiO_2$ and the $SiO_xN_y$ system, metal inclusions form at moderate concentrations. As shown in **Figure 2.12**, Pt atoms in $SiO_xN_y$ matrices quickly cluster into Pt rich (black) regions. Differently from the $Si_3N_4$:Cr case, even at low metal concentrations, these black regions already exhibit lattice fringes (**Figure 2.13**) indicating Pt atoms are indeed crystallized. The reason for such difference becomes transparent if we recall that the relative density of $SiO_xN_y$:Pt film is rather low and a large amount of nanopores exist. Therefore, Pt atoms which are immiscible in $SiO_xN_y$ prefer to settle in the porous region and not enter the amorphous matrix. This will in turn allow fast surface diffusion/reorganization and eventual crystallization.

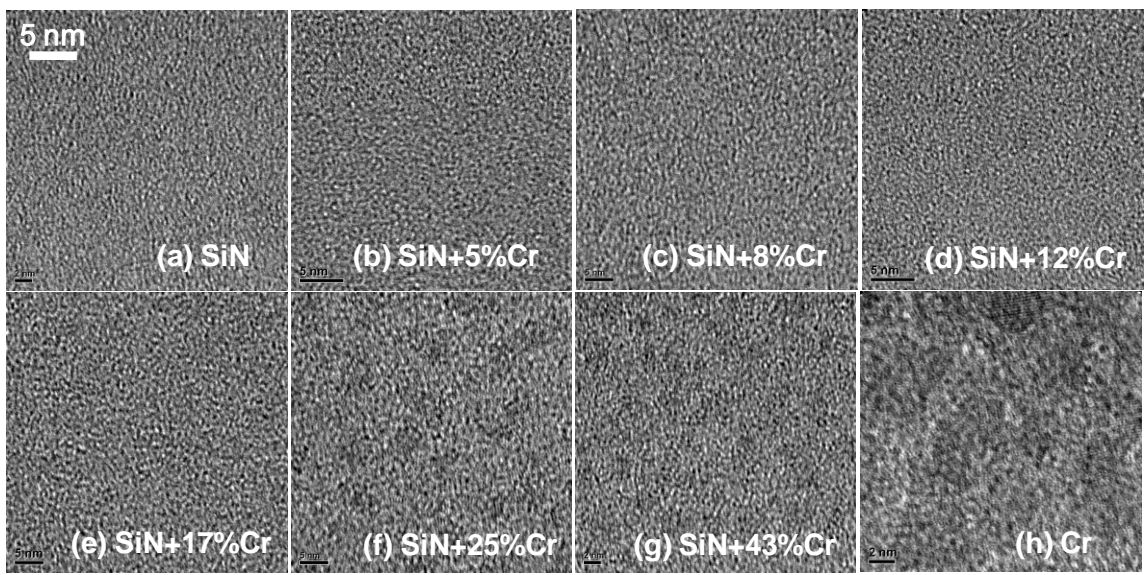

**Figure 2.11**. TEM image for $Si_3N_4$:Cr mixture film. Non-uniformity (still amorphous) starts to form at 25% Cr.

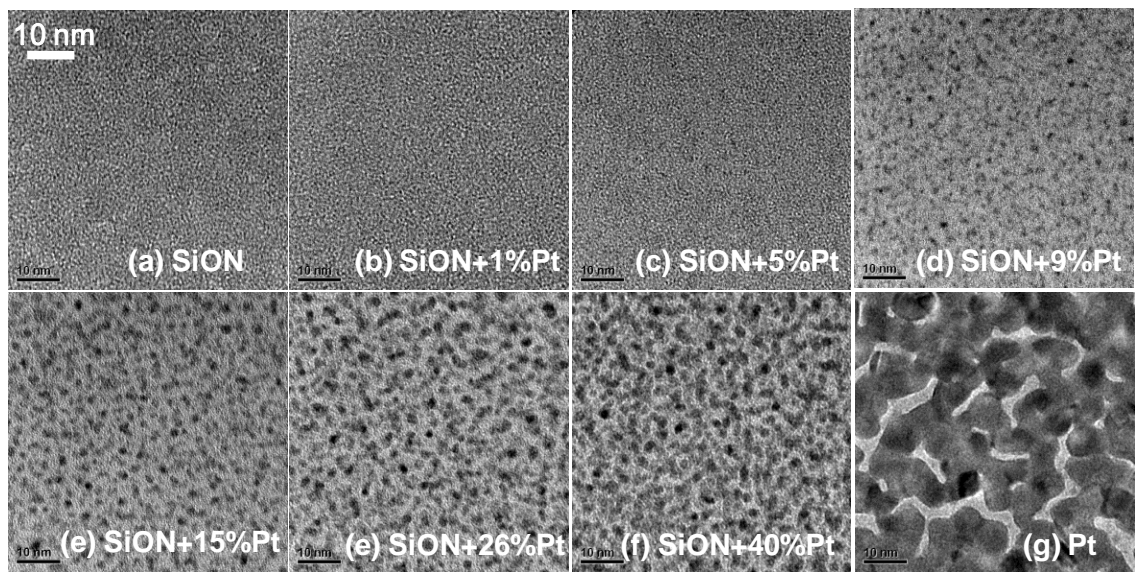

**Figure 2.12**. TEM image for $SiO_xN_y$:Pt mixture film. Non-uniformity (crystallized) starts to form near 9% Pt.

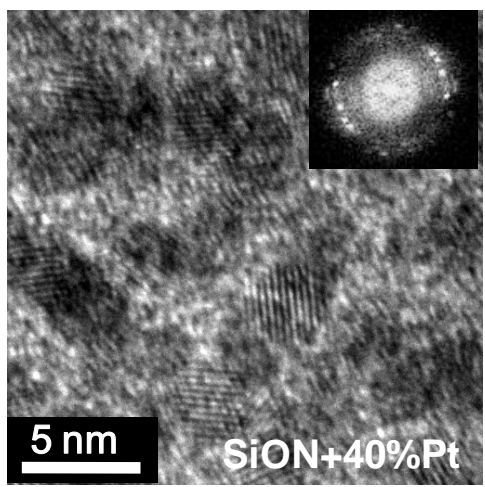

**Figure 2.13**. High resolution TEM image for $60\%SiO_xN_y{:}40\%Pt$ mixture film.

From these observations, we may summarize the hybrid film growth kinetics as in **Figure 2.14**. (a) In a fully dense amorphous dielectric matrix, metal atoms are dispersed randomly except for chance coalescence into dimers, trimmers and oligomers, thus an amorphous hybrid structure forms, and (b) in a less dense, nanoporous amorphous dielectric film, some metal atoms are randomly dispersed in the matrix but many are deposited onto surfaces of nanopores of almost atomic size or (c) nanometer size. In (c), they tend to crystallize into metal nanocrystals.

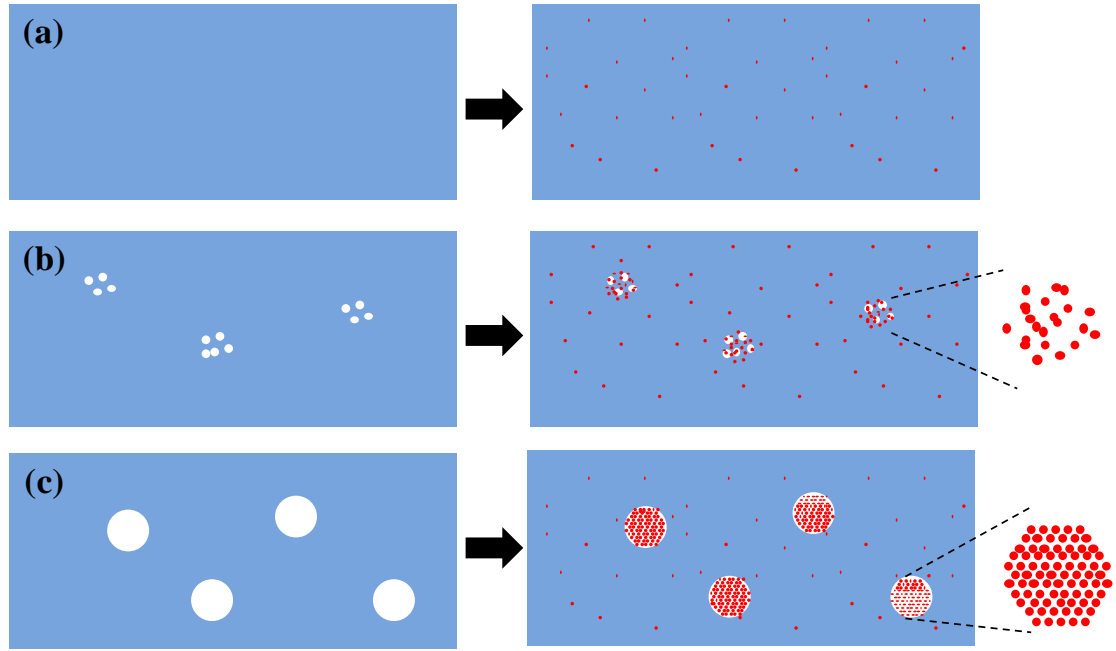

**Figure 2.14.** Schematic nanostructures of nanometallic films. (a) Dense film. (b) Less dense film with atomic-sized porosity. (c) Porous film with larger porosity. White regions are pores; red dots are metal atoms.

## 2.5   Voltage Induced Metal Insulator Transition

### 2.5.1   DC & AC Measurement

We next describe electrical characterization using voltage as the main control variable. First, (Quasi) DC measurements used digitized DC voltage segments, each lasing 20 ms before changing to the next segment over a transient (rise/fall) time of <1 ms. The increment or decrement between segments is controllable (typically from 0.001 V to 1 V). As shown in **Figure 2.15a**, at a critical forward voltage bias (defined as having a positive voltage, ~4 V on the TE), the current suddenly drops, which transitions the cell from the low resistance state (LRS, or "on" state) to the high resistance state (HRS, or "off" state). The state is "non-volatile" and "bipolar" in that the high resistance is maintained at 0 V. In fact, it is typically orders of magnitude higher than the resistance at the critical voltage because of the non-Ohmic resistance-voltage dependence. Only a reverse bias of a critical voltage can switch it back to the low resistance state (LRS, or "on" state). The latter switching typically involves increasing current going through multiple stages. Having a Pt TE and a Mo BE, the switching direction in **Figure 2.15** is counterclockwise in the *R-V* hysteresis, changing from the LRS to the HRS at a positive bias and vice versa at a negative bias.

Switching is quite robust as evidenced by the cycling/fatigue test. In **Figure 2.15b**, consecutive 100 DC sweeping cycles were performed on a Mo/Si$_3$N$_4$:Cr/Pt device (the voltage from −5 V to +6 V with 0.2 V increment), showing overlapping *R-V* curves without noticeable change (degradation). To test over more cycles, AC measurements used 1 μs square shaped voltage pulses (between −5 V and +7 V) for excitation, followed by DC measurements of resistance (at 0.1 V) after each pulse, and the process was repeated over many cycles. These cyclic tests found both HRS and LRS are stable (**Figure 2.15c**). (The device performance of SiO$_2$:Pt is worse due to moisture effects on nanoporous devices, as will be described in **Chapter VI**.)

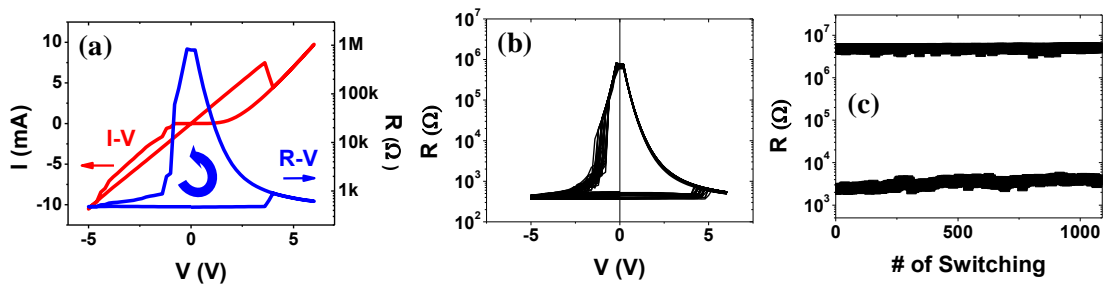

**Figure 2.15**. (a) Typical *I-V* and *R-V* curve for resistive switching device. (b) Consecutive 100 cycles of DC *R-V* curves for Mo/Si$_3$N$_4$:Cr/Pt device. (c) AC pulse tests for Mo/Si$_3$N$_4$:Cr/Pt device. ($f_{Cr}$=7%, $\delta$=10 nm).

## 2.5.2 Thickness-Composition Map

Devices (100×100 μm$^2$) of Mo/Si$_3$N$_4$:Cr/Pt with various Si$_3$N$_4$:Cr film thickness and composition were tested to determine their characteristics of nanometallic transition. Nanometallicity occurs at a nominal metal fraction $f$ well below the bulk percolation limit $f_p$. This is demonstrated in **Figure 2.16a**, which depicts a thickness-composition $\delta$-$f$ map for the electrical conduction. The map at large $\delta$ follows the conventional percolation transition: above $f_p$ the conducting region reigns; below $f_p$ it is the insulating region. Here $f_p$~0.45, which is comparable to the value determined for thin films of (1-$f$) SiO$_2$:$f$ Pt (ref.[2-3]) and several granular mixtures of oxides and metals[7]. However, below $f_p$, there is still a conducting region at small $\delta$, which we call the nanometallic region. Moreover, below $f_p$, there is also a (voltage-triggered) switching region between the nanometallic region and the conventional insulating region. Both the nanometallic region and the switching region apparently can extend to $f$→0 when $\delta$ approaches 0. This means that: if $\delta$ is small enough, there is always nanometallicity. Traversing vertically in the map, we can see that the nanometallic MIT is a transition defined by the size. **Figure 2.16b** illustrates that at $f_{Cr}$=7%, as the thickness decreases from 35 nm, to 17 nm, to 10 nm, to 5 nm, the device characteristics changes from insulating, to switching to conducting. Traversing horizontally in the map, we can see that the nanometallic MIT is a transition defined by the composition. **Figure 2.16c** illustrates that at 10 nm film thickness, as the Cr concentration increases from 2%, to 7%, to 30%, the device characteristics changes from insulating, to switching to conducting.

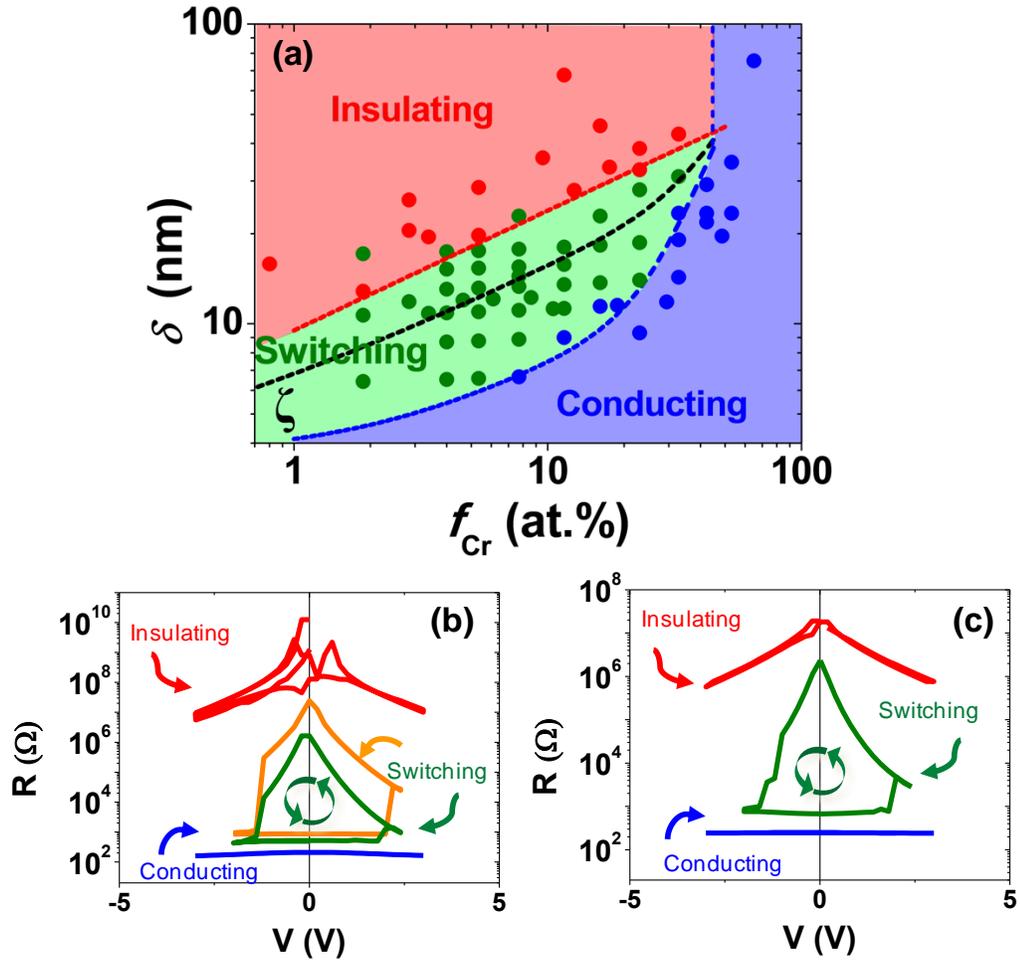

**Figure 2.16.** (a) $\delta$-$f$ map for Si$_3$N$_4$:Cr system delineating data points for insulating (I), conducting (C) and switching (S) behaviors, showing percolation at metal composition $f$=0.45, nanometallicity at thinner thickness from $f$~0 to $f$=0.45, and switching when thickness $\delta \approx \zeta$, the latter in black dot line. (b) $R$-$V$ curves for insulating (non-Ohmic, 93%SiN$_{4/3}$:7%Cr, 35 nm), conducting

(Ohmic, 93%SiN$_{4/3}$:7%Cr, 5 nm) and switching (hysteretic, 93%SiN$_{4/3}$:7%Cr, 10 nm (green) and 17 nm (orange)) films sandwiched between Pt and Mo. (c) *R-V* curves for insulating (non-Ohmic, 98%SiN$_{4/3}$:2%Cr, 10 nm), conducting (Ohmic, 70%SiN$_{4/3}$:30%Cr, 10 nm) and switching (hysteretic, 93%SiN$_{4/3}$:7%Cr, 10 nm) films sandwiched between Pt and Mo. During switching, an initially conducting film transitions to high-resistance insulating state at +2V and returns to low-resistance metallic state at −1.5 V.

### 2.5.3   Scaling Laws

The area dependence of HRS resistance follows Ohm's law, $R \propto 1/A$, as shown in **Figure 2.17**. This is an indication of the overall uniformity nature of the HRS: there is no variation over the length scale of the smallest size in **Figure 2.17** (20 μm). On the other hand, the value of LRS seems to be area independent. Such independence comes from an artifact and the metallic nature of the LRS. In our film, the LRS is not a unique metallic state in that the resistance value of the film is affected by the history of voltage, current compliance, *etc.*, and it may be so low that it falls below the resistance of the electrodes. (The bottom electrode, which is very thin, is especially resistive, typically ~ 100-1000 Ω. Most of this resistance is due to the spreading resistance from the bottom of the cell to the edge of the chip where the probe electrode makes contact.) When this happens, which is common, the LRS resistance primarily reflects the resistance of the spreading resistance, which has a very weak cell area dependence. We will formally deal with the LRS area scaling after developing a circuit model in **Chapter VII**.

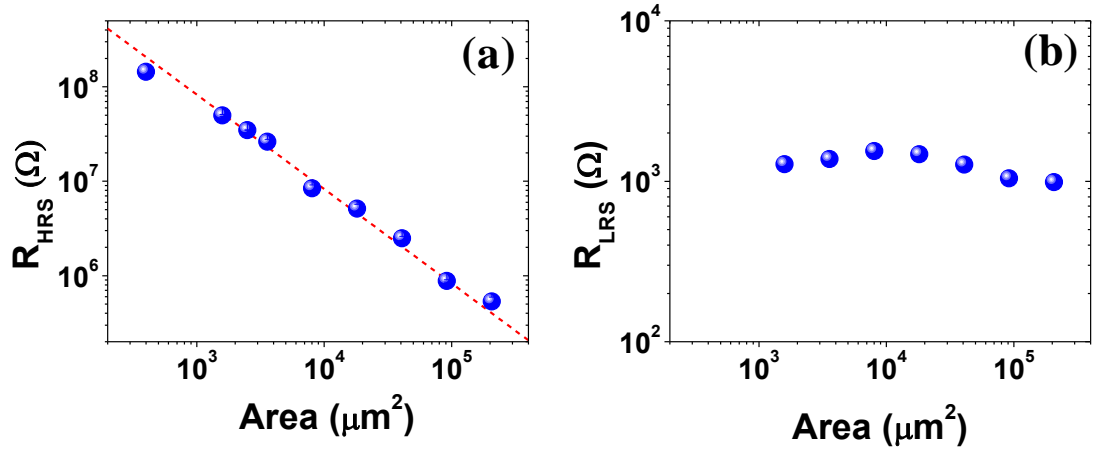

**Figure 2.17.** (a) Area scaling of HRS. Blue dots are experimental data and red dash line is fitted straight line with slope=1. (b) Area scaling of LRS. ($f_{Cr}$=7%, $\delta$=10 nm).

The resistance of the HRS reveals an unconventional (non-Ohmic) thickness dependence, as shown in **Figure 2.18a**. It exponentially increases with thickness, apparently following Anderson's scaling theory $R = R_0 \exp(\delta / \zeta)$. (We will revisit this subject in a later chapter, since strictly speaking Anderson's picture of random wave physics only holds at 0 K where electron movement is not thermally assisted. This is not the case at the room temperature.) By fitting the HRS resistance with this form, we can obtain the localization length for various metal concentrations. As shown in **Figure 2.18b**, the localization (diffusion) length of the HRS increases with metal concentration, indicating electrons can reach further as more metals are incorporated into the film. Meanwhile, the localization length of the LRS can be obtained from the switching region in **Figure 2.16**, by assigning it to the value halfway between the insulating boundary and the conducting (nanometallic) boundary. The localization lengths of the two states are compared in **Figure 2.18b** illustrating a large increase from the LRS to the HRS. It is this increase, caused by a voltage trigger, that is responsible for the metal insulator transition in RRAM. These results are consistent with the ones previous seen in the $SiO_2$ system[2-3,8]. From a practical viewpoint, the above exponential dependence offers new freedom to engineer the device resistance without affecting other switching characteristics. An example of utilizing such highly tunable HRS is discussed in **Appendix**.

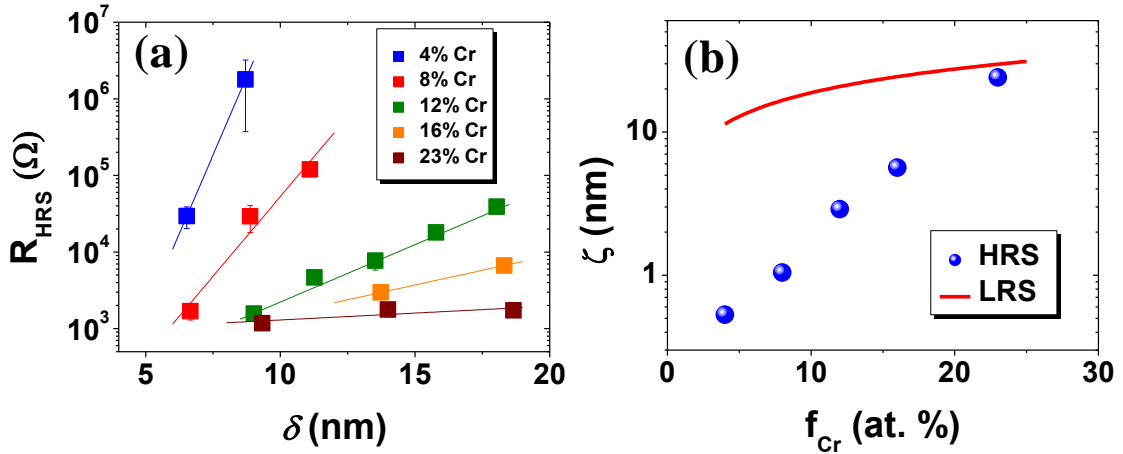

**Figure 2.18**. (a) Thickness dependence of HRS. (b) Diffusion length ($\zeta$) of HRS and LRS.

The switching voltage is found to be device independent (**Figure 2.19**), always ~1 V irrespective of the TE dimension and the film thickness. Therefore, switching is not controlled by the electrical field; instead, it is an energy (voltage) controlled process. This finding has a direct bearing on the candidate mechanisms that may explain the transition.

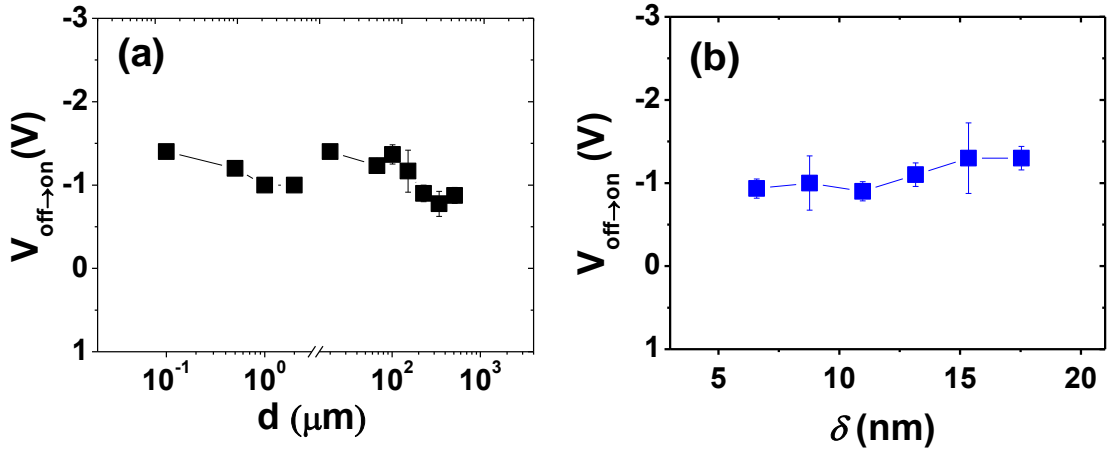

**Figure 2.19**. (a) Area and (b) thickness dependence of off→on switching voltage. ($f_{Pt}$=5%).

## 2.6  UV Induced Metal Insulator Transition

Although a relatively small voltage is used in our testing, even a small voltage can easily lead to a large field in the thin film (*e.g.*, 1 V across a 10 nm film results in a field of ~$10^6$ V/cm, which is quite large). Under such high electrical field, ionic and (metallic) atomic migration is possible, especially in a nanoporous films. To determine whether ion migration is involved during switching or not, an alternative non-electrical excitation is required. In a later chapter, we will describe how a mechanical stress can provide such excitation. Here we borrowed the idea in standard Erasable Programmable ROM (EPROM), where exposure to UV light is used to remove trapped charge on the floating gate[9]. For the following experiment, we used a UV light source (ELC-403, Electro-Lite Corp., Bethel, CT), which provide a wavelength of 300-420 nm (4.2-3.0 eV) with an output power of 70 mW/cm$^2$ at 3.4 eV. In order to minimize substrate absorption, UV-transparent fused silica (two-side polished) was used as the substrate and the radiation was shone from below (**Figure 2.20a**). The resistance was examined using a small reading voltage of 0.2 V, which is insufficient to cause switching according to our previous tests.

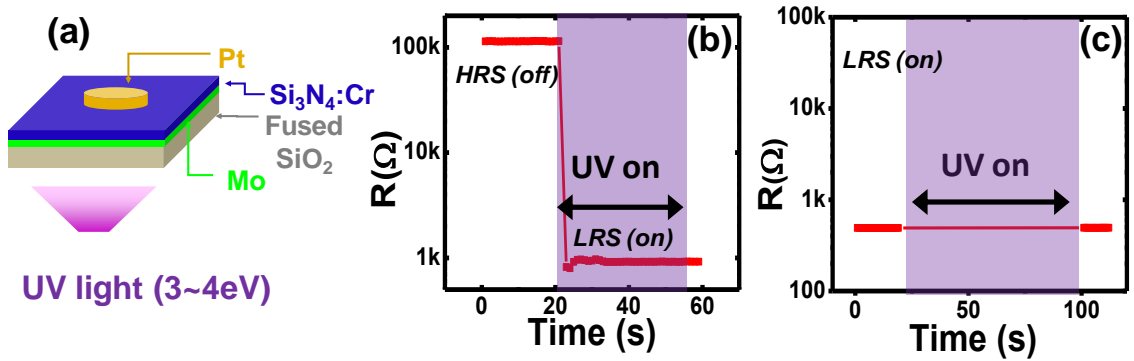

**Figure 2.20**. (a) Schematic of set up for UV test of Mo/Si$_3$N$_4$:Cr/Pt device. (b) Initially HR state. (c) Initially LR state. TE/BE electrodes are in series with external circuit. ($f_{Pt}$=5%, $\delta$=10 nm).

As shown in **Figure 2.20b**, a device initially at the HRS immediately drops to the LRS as UV is turned on, and it remains in the LRS even after UV is shut off. (This excludes the possibility of enhanced current being a photocurrent). On the other hand, UV has no effect on the LRS as shown in **Figure 2.20c**. Since the nanometallic film is only 10 nm thick, it is unlikely to cause any significant attenuation of the UV radiation. Therefore, the UV irradiation does not impose any directional field, which rules out any possibility of ion/atom transport. Since ion/atomic transport cannot be the mechanism of nanometallic switching, electronic transport is the only mechanism that may be responsible for nanometallic switching. Very likely, the HRS corresponds to an energetically metastable state with trapped electrons, which can be detrapped with the excitation of photons of an appropriate energy. The hypothesis of HRS being a metastable state will be further tested in **Chapter III.**

## 2.7    Device Performance

### 2.7.1    Retention

Retention test was performed by switching a device to the HRS or LRS and monitoring its resistance change (periodically read at 0.1 V) as a function of time. According to these tests, both HRS and LRS are stable, see **Figure 2.21**. If the small resistance change of the HRS is extrapolated to 10 years, it will amount to a $3\times$ increase in resistance. The origin of the resistance increase will be explored in **Chapter VI**.

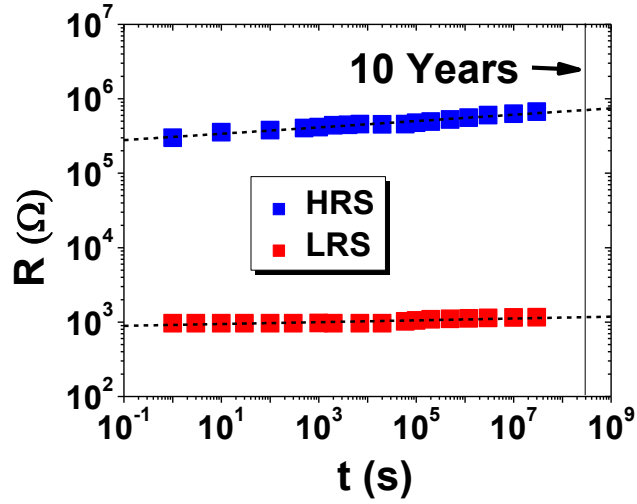

**Figure 2.21**. Resistance retention (Mo/93% $Si_3N_4$: 7%Cr/Pt) at room temperature. ($\delta$=10 nm).

### 2.7.2 Uniformity

As already shown in **Figure 2.15b-c**, switching curves of the Mo/$Si_3N_4$:Cr/Pt device are highly reproducible and the resistance values of the HRS and the LRS remain constant over many cycles. The Weibull plot provides a quantitative comparison between our data and those from the literature, on ionic-type RRAM. The Weibull distribution for the cumulative probability ($F$) of a random variable $x$ is prescribed by a shape parameter $k$ (or Weibull exponent):

$$F(x) = 1 - \exp\left(-\left(x/x_0\right)^k\right)$$

where $x_0$ is a scaling constant for $x$. It is apparent that a higher $k$ indicates a narrower distribution and thus higher uniformity. This $k$ value is related to the ratio of standard deviation ($\Delta$) to mean ($\mu$), $\Delta/\mu$, of the distribution, through the following analytical expression:

$$\Delta/\mu = \frac{\left[\Gamma\left(1+2/k\right) - \Gamma^2\left(1+1/k\right)\right]^{1/2}}{\Gamma\left(1+1/k\right)}$$

where $\Gamma$ is the Gamma function. This is verified in **Figure 2.22** using our data (from **Figure 2.15b-c**) and those in the literature, which shows a tight correlation between our $\Delta/\mu$ and $k$ data and those in the literature. (We analyzed the statistics of the data of ionic devices reported in the literature, from which $k$ and $\Delta/\mu$ values were computed.) Along with the previous data on another nanometallic RRAM (SiO$_2$:Pt), our Si$_3$N$_4$:Cr device clearly have outstanding uniformity for all switching parameters (resistance values of HRS and LRS, as well as on/off switching voltages.) Such uniformity would benefit multi-bit storage for which multiple resistance states need to be distinguished within a relatively narrow switching window (we will describe a multi-bit storage in **Chapter VIII**).

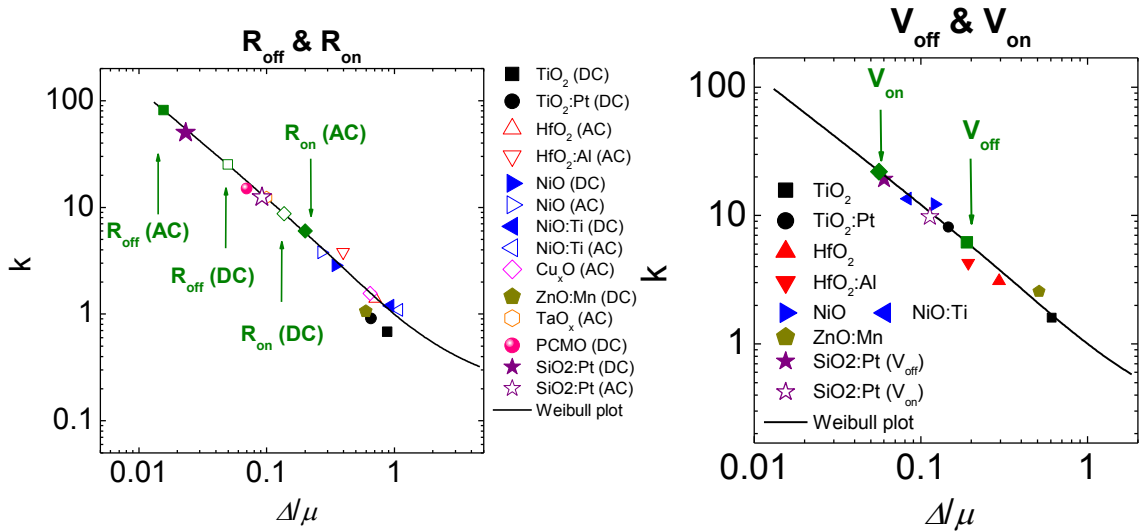

**Figure 2.22.** *k vs.* Δ/μ plot of (a) resistance value and (b) switching voltage of Mo/Si₃N₄:Cr/Pt device (green points), compared with reported data of other devices in the literature. The literature data can be found in Ref. [2,8].

### 2.7.3 Endurance

Endurance tests beyond the one shown in **Figure 2.15b** were performed for Si₃N₄:Cr devices. They can maintain its switching capability up to ~10⁶ cycles (**Figure 2.23a**). The endurance is expected to be improved if (a) mechanical contact is avoided during testing (see **Chapter III**), (b) the device is isolated from ambient moisture (see **Chapter VI**) and (c) the quality of electrodes is improved to avoid wear, decohesion or oxidation. For example, our former colleague Dr. B-J. Choi fabricated a crossbar nanometallic device made of SiO₂:Pt, which allows electrical testing without direct mechanical contact of the cell. He reported an improved endurance of >3×10⁷ cycles (**Figure 2.23b**)[10].

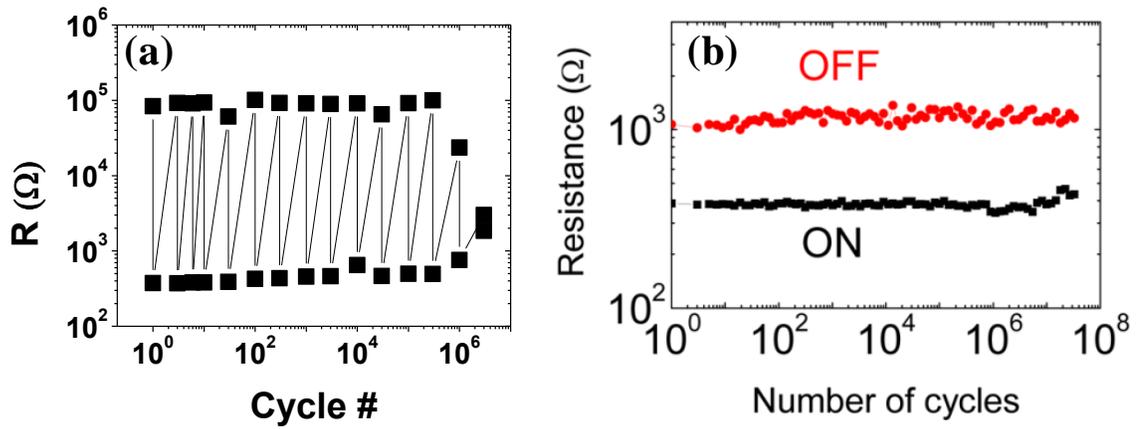

**Figure 2.23.** Endurance test for (a) Si$_3$N$_4$:Cr device, ($f_{Cr}$=7%, $\delta$=10 nm) and (b) SiO$_2$:Pt device (data from ref.[10]).

### 2.7.4 Switching Speed

To test the switching speed, we again employed a pump-probe method (previously described as the AC test in **Section 2.5.1**): an square-shaped excitation voltage pulse with a certain width (from 20 ns to 1s) and height (from 0 V to the switching voltage) was first provided to the device, then a small probe DC voltage (0.2 V) was used to read the resistance to determine the current resistance state. A switch box was employed to isolate the two electrical sources used in the two steps (**Figure 2.24**).

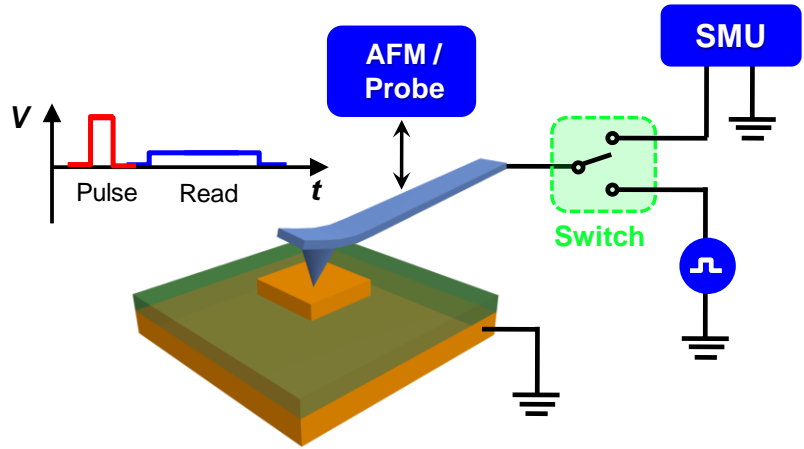

**Figure 2.24**. Schematic of measurement system.

**Figure 2.25a** shows the on→off switching voltage for different pulse widths for a 400×400 µm² cell, which was initially set to the LRS. (After the resistance change reached the ~MΩ range, taken to be already in the HRS, the cell was reset to the LRS for additional testing.) It is clear that the switching voltage is independent of the pulse width over the range from 1 s to 10 µs. However, at shorter pulse widths, the on→off switching voltage rapidly increases, reaching 10 V for a pulse width of 100 ns. A similar observation was found for off→on switching (**Figure 2.25b**).

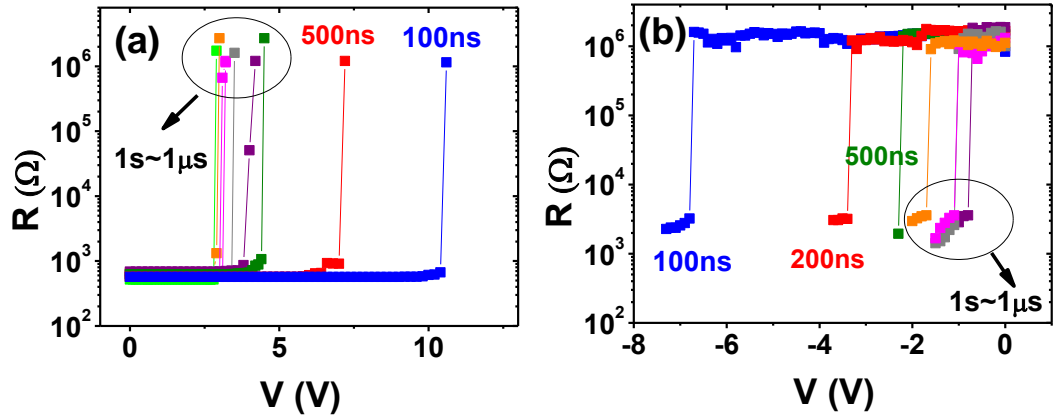

**Figure 2.25.** Switching speed testing for (a) on→off (b) off →on of 400×400 μm² cell. ($f_{Pt}$=5%, $\delta$=10 nm).

Similar observations were made with other cell sizes. As shown in **Figure 2.26a & b**, the threshold pulse width below which the switching voltage significantly increases varies with the device size: a smaller size exhibits a smaller threshold pulse width. For the smallest cell size tested, having a 2 μm cell, the threshold pulse width apparently falls below our instrumental capability (from 20 ns to 1 s), as both the on→off and off→on switching voltages are constant within the testing range. This is counter to the observation of ionic-RRAM for which there is typically a 4 fold increase of the switching voltage as the pulse width decreases from 1 s to 100 ns. (See **Chapter I**) It is also counter to the expectation that smaller cells are more difficult to switch because of the less favorable statistics of finding a switching filament.

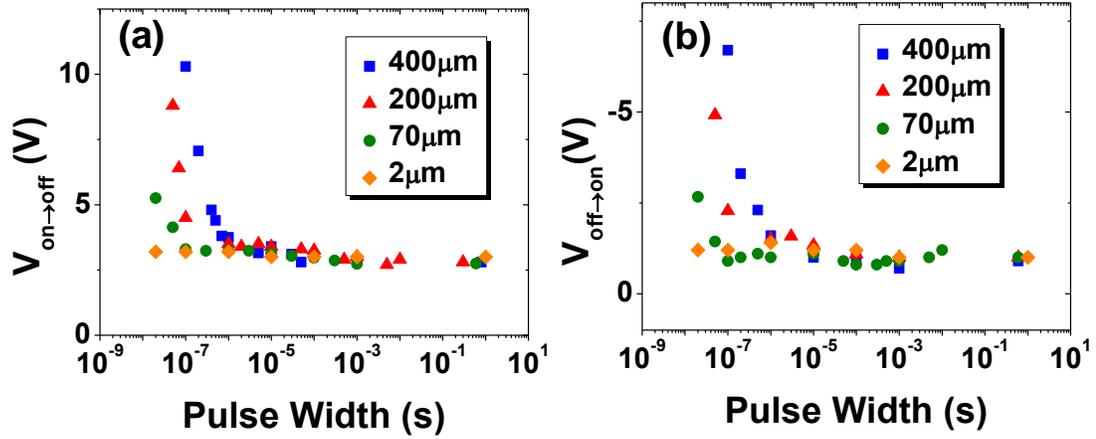

**Figure 2.26.** Threshold voltage of square-shaped voltage pulse required for switching in devices of various lateral sizes. (a) $V_{\text{on}\rightarrow\text{off}}$, *vs.* pulse width. (b) $V_{\text{off}\rightarrow\text{on}}$ *vs.* pulse width. ($f_{\text{Pt}}$=5%, $\delta$=10 nm).

The size dependence of the threshold voltage in our devices turns out to be a circuit artifact related to its *RC* delay as illustrated in **Figure 2.27a**. The delay is controlled by the product of device capacitance and the load resistance. In our device, the dominant load resistance is identified as $R_{\text{BE}}$ originating from the bottom electrode resistance, interface resistance or any parasitic line resistance. For a square-wave pulse of applied voltage $V$, the steady state voltage $V*_{\text{cell}}=(R_{\text{cell}}/(R_{\text{cell}}+R_{\text{BE}}))V$ for the cell is not reached until a transient time ~$R_{\text{BE}}C$ has passed. Therefore, a pulse width shorter than this transient time is not able to provide the steady state voltage to the cell, thus necessitating a higher applied voltage to trigger switching. The analytical solution to the cell voltage problem

$$i = \frac{V - V_{cell}}{R_{BE}} = \frac{V_{cell}}{R_{cell}} + C \frac{dV_{cell}}{dt}$$

is trivial, given by

$$V_{cell} = \frac{R_{cell}}{R_{cell} + R_{BE}} \left[ 1 - \exp\left( -\frac{R_{cell} + R_{BE}}{CR_{cell}R_{BE}} t \right) \right] \times V$$

in which the $RC$-delay time is $CR_{BE}R_{cell}/(R_{cell} + R_{BE})$. Assuming a linear capacitor for the cell, for which $C = \varepsilon d^2/\delta$ ($\varepsilon$: dielectric constant, $d$: cell lateral length, $\delta$: film thickness), and a size-independent $R_{BE}$ which is dominated by the size-insensitive spreading resistance, we can expect the delay time $\sim R_{BE}C$ to exhibit a strong size dependence, proportional to the cell area. For a 400 μm cell, we let $C$=700 pF, $R_{BE}$=800 Ω, $R_{cell,\ HRS}$=1 MΩ and $R_{cell,LRS}$=400 Ω, $V_{cell}^{on \to off}$ =3 V and $V_{cell}^{off \to on}$ =-1 V (from DC switching data), the computed switching voltages for on→off and off→on switching are shown in **Figure 2.27b** and **c**. These plots capture the experimentally observed switching characteristic; in particular, for a cell size of 2 μm or less, it shows $R_{BE}C$=800 Ω×18 fF=15 ps, so there is no obvious increase in the switching voltage within the 1 s to 1 ns test range in **Figure 2.26**. Therefore, the size dependence of the switching voltage of our devices is a purely a circuit artifact.

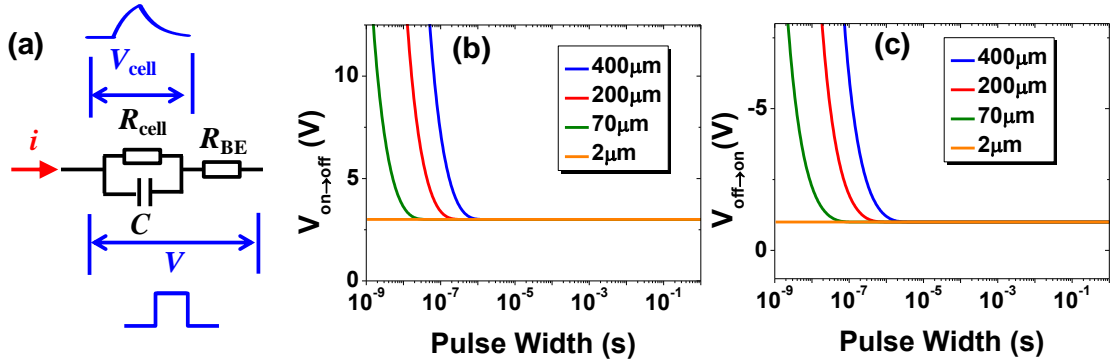

**Figure 2.27.** (a) Equivalent circuit. Simulated data for (b) $V_{\text{on}\rightarrow\text{off}}$ *vs.* pulse width (c) $V_{\text{off}\rightarrow\text{on}}$ *vs.* pulse width for various lateral sizes.

The above interpretation is consistent with the observation of time-independence of the switching voltage in other nanometallic RRAM. For example, both SiO$_2$:Pt (**Figure 2.28a**, data from ref.[2]) and perovskite nanometallic film LaAlO$_3$:LaNiO$_3$ (**Figure 2.28b**, data from ref.[11]) exhibit a similar switching-time -independence. In both cases, the apparent switching voltage only rises when the pulse widths are less than 100 ns, which is approximately the estimated *RC* delay.

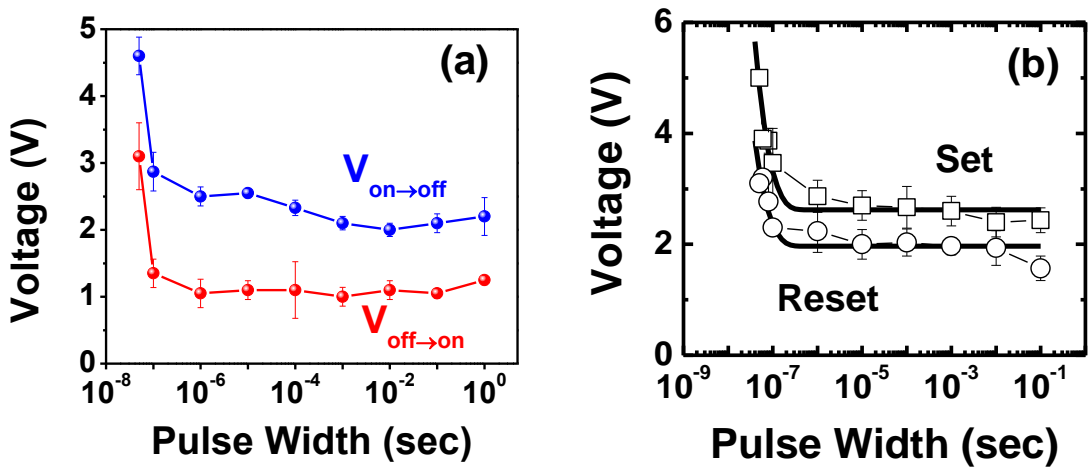

**Figure 2.28.** Threshold voltage *vs.* pulse width in (a) $SiO_2$:Pt (adapted from ref. [2]) and (b) $LaAlO_3$:$LaNiO_3$ (adapted from ref. [11]) devices.

To estimate the threshold pulse width in practical devices, we consider the scaling laws of IC circuits in CMOS technology. The interconnect line resistance is expected to follow a linear relation $R_l \propto N$, where $N \times N$ is the memory size, and the capacitance to follow $C \propto N^2$, giving $RC \propto N^1$. Thus, the $RC$ delay rapidly decreases with $N$. To provide a rough numerical estimation for a state-of-the-art 1 Tbits ($10^6 \times 10^6$) memory CMOS technology, we use a typical sheet resistance of 0.05 $\Omega$/sq for the metal conductor layer to estimate the interconnect line resistance, which is 0.05 $\Omega$/sq $\times 10^6$ sq or 50 k$\Omega$. We also estimate the capacitance of a $100 \times 100$ nm$^2$ cell to be of 0.1 fF based on the measured $C$=100 pF for a $100 \times 100$ $\mu$m$^2$ cell. Therefore, the estimated $RC$ delay for the 1 Tbit nanometallic memory is ~5 ps. This is fast enough to easily support any nanosecond or even 25 ps memory device.

Our data above provides an upper limit of the intrinsic switching time to be 20 ns, which is the end of the test range which did not cause any voltage rise in the 2 $\mu$m device in **Figure 2.26**. Our former colleague Dr. B. J. Choi tested a $SiO_2$:Pt nanometallic device in a circuit that has a much smaller $RC$ delay; he demonstrated a switching time <100 ps (**Figure 2.29**)[10]. In the above test and even in the state-of-art electrical circuit ($RC$~1 ps) the capacitance is ultimately dominated by the parasitic capacitance, whereas we believe the intrinsic switching time is ultimately determined by electron's trapping/detrapping time within a nanometer trap barrier, which could be as short as ~femtosecond. Therefore, probing the intrinsic switching time is far beyond the capability of any electrical testing method today. In **Chapter III**, I will describe an entirely different setup using a  stress-trigger to demonstrate the HRS to LRS switching that apparently occurs within 0.1 ps.

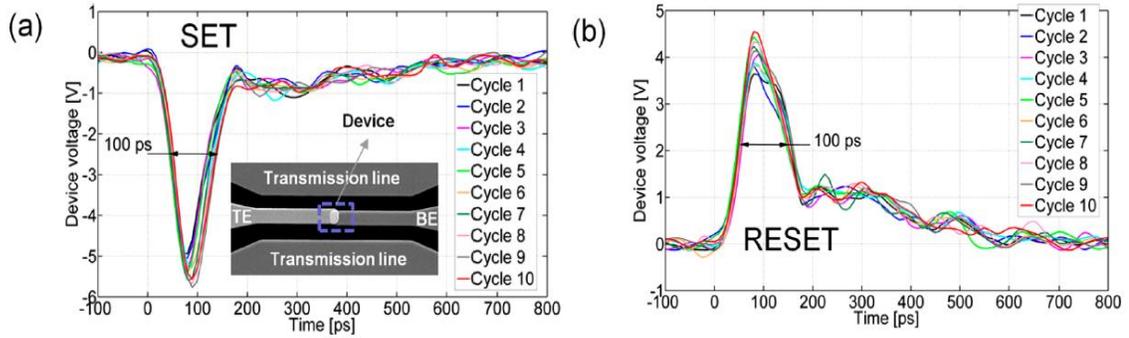

**Figure 2.29.** SiO$_2$:Pt device (5×5 µm$^2$) during (a) set and (b) reset operations with 100 ps (FWHM) pulses (adapted from ref.[10]).

### 2.7.5  Size Limitation

To investigate the size limitation of nanometallic devices, nano-sized devices were fabricated using electron beam lithography. The fabrication procedure is summarized as follows. 1) A Mo bottom electrode (~20 nm) was DC sputter-deposited, followed by nanometallic film (Si$_3$N$_4$:Pt) deposition. This step is identical to the one used in fabricating micron-sized devices described above. 2) PMMA (A4 495) was spin-coated onto the nanometallic film at 4000 rpm, then baked for 10 min at 180 ℃. After that, a second PMMA (A2 950) layer was spin-coated and baked in an identical manner. This two-layered e-beam resist containing PMMA of different molecular weights was used to facilitate the lift-off process in later steps. 3) The e-beam resist was exposed to e-beam in an Elionix ELS-7500EX (current: 100 pA, dose: 500-800 $\mu C/cm^2$). The CAD file used included square shaped patterns from $20 \times 20$ $nm^2$ to $2 \times 2$ $\mu m^2$. After exposure, it was developed in MIBK(1):IPA(3) developer for 1 min (ultrasonication was not used). 4) A Pt top electrode (~40 nm) was RF sputter-deposited, followed by 2 min ultrasonication in acetone. The obtained pattern examined by SEM (**Figure 2.30a** and **b**) showed the desired square shape for the larger patterns (from $200 \times 200$ $nm^2$ to $2 \times 2$ $\mu m^2$), but below $100 \times 100$ $nm^2$ devices had rounded corners due to imperfect exposure. The pattern was also checked by AFM (**Figure 2.30c** and **d**), confirming the electrode thickness and surface flatness.

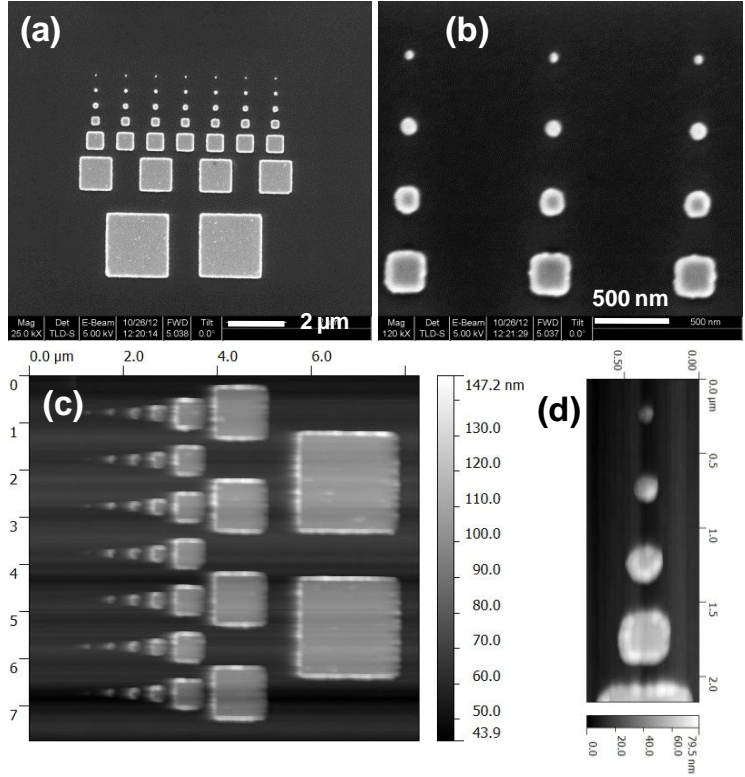

**Figure 2.30**. Nano-devices fabricated by e-beam lithography. (a) SEM image of nano-devices from $20\times20$ nm$^2$ to $2\times2$ μm$^2$; higher magnification image shown in (b). (c) & (d) AFM images of fabricated nano-devices. ($f_{Pt}$=5%, $\delta$=10 nm).

Electrical characterization was performed by a customized conducting AFM (C-AFM). By rerouting connections of Asylum MFP-3D (equipped with a Pt/Ir-coated CAFM tip with a diameter <20 nm) to an external circuit board, we obtained a larger range of current compliance for our devices. (The built-in circuit in Asylum MFP-3D ORCA has a 2 nA current compliance, compared to our customized compliance, up to 100 mA, provided by our own source meter.) **Figure 2.31** shows a series of *R-V* curves for various cell sizes ranging from $2\times2$ μm$^2$ to $100\times100$ nm$^2$. All devices present resistive switching with a similar switching voltage undistinguishable from that of the micron-devices. This provides evidence that the nanometallic RRAM has no size limit down to ~100 nm, which can support a ∼10 Gbit/cm² memory.

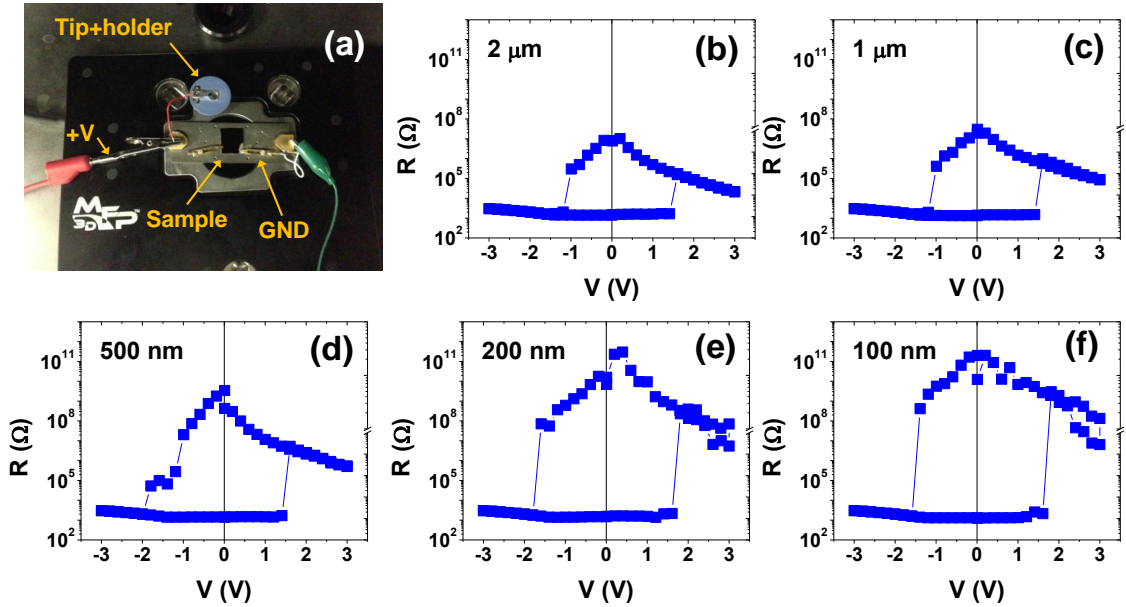

**Figure 2.31**. (a) Measurement set up. (b)-(f) *R-V* characteristics of nano-device from 2×2 μm² to 100×100 nm². ($f_{Pt}$=5%, $\delta$=2 nm).

### 2.7.6   Crossbar Feasibility

On an application level, individual RRAM devices need to be integrated to RRAM arrays. The most straightforward scheme is a 2-D crossbar structure, in which top and bottom electrodes are accessed by word and bit lines controlled by peripheral circuits. In the literature, RRAM crossbar structures have been reported. For example, Kim *et al.* fabricated 32×32 arrays for a 1 kbit TiO₂ RRAM without cross-talk problems[12]. A 3D cross-point architecture is also possible[13-14] and may have some advantage, but it  must utilize a conformal deposition method (*e.g.*, ALD), and its read/write feasibility (including demonstrating the absence of crosstalk) remains to be verified experimentally. We have fabricated a crossbar structure for the nanometallic RRAM. A set of 40 nm Pt (with a thin Mo adhesion layer) bottom "word lines" was first obtained by conventional photolithography. A nanometallic $Si_3N_4$:Pt film was next deposited on these word lines, fully covering the metal line and beyond in order to avoid short circuit during later cross-bridging. Finally, a set of top Mo "bit lines" were fabricated by conventional photolithography with appropriate alignment. The fabricated structure and associated electrical characteristics are shown in **Figure 2.32** to demonstrate the same switching behavior as our standard micron-size devices. (Crosstalk was not addressed here since our experiment only involved one individual crossbar structure, which is the same status as that for the 3D crossbar architecture mentioned above.)

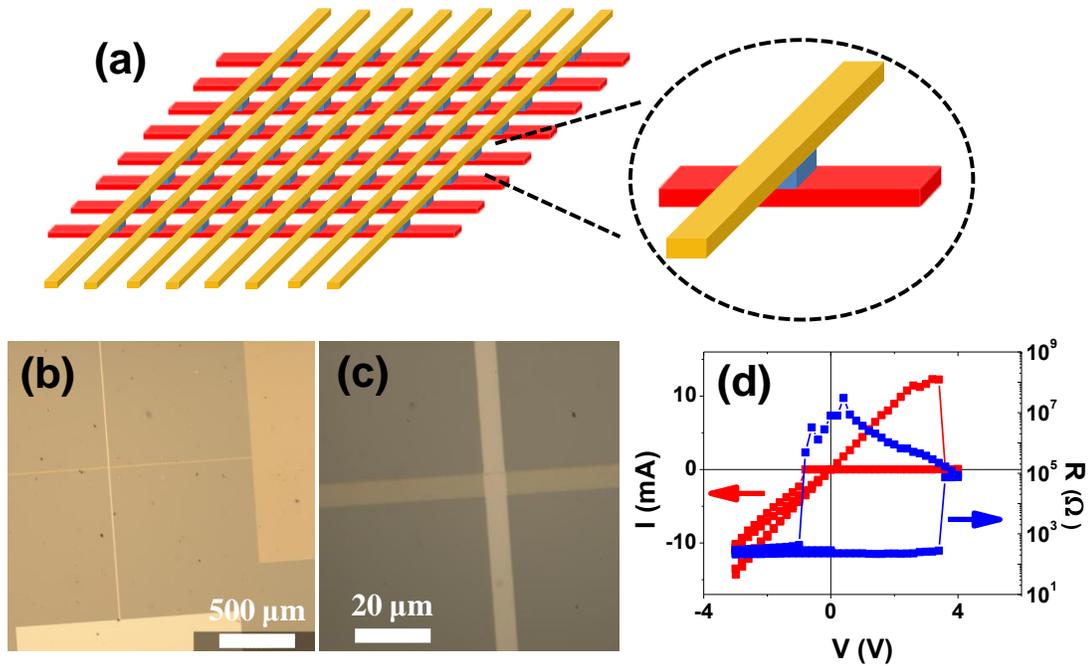

**Figure 2.32.** A crossbar structure of nanometallic RRAM. (a) Schematic of 2-D crossbar structure. (b) Top view of photolithographically fabricated crossbar nanometallic RRAM and its high magnification picture (c). (d) Characteristic *I-V* (*R-V*) curve. ($f_{Pt}$=5%, $\delta$=10 nm).

## 2.8  Other Nanometallic RRAMs

### 2.8.1  Nanometallic Films

Before addressing the switching mechanism of nanometallic RRAM at the end of this chapter, we first take a broader view of the nanometallic materials universe to establish the generality of the observed phenomena. Nanometallic resistive switching turns out to be a quite general phenomenon，observed in a large set of materials, listed in **Table 2.2**. The insulator matrices cover oxides/oxynitrides/nitrides of group II, III, IV and transition metals, and the metal atoms cover both main group metals (*e.g.*, Al) and transition metals (not limited to noble metals). Surprisingly, films of all combinations exhibit resistive switching phenomenon with almost identical switching characteristics (**Figure 2.33**). Such remarkable universality confers simplicity and flexibility to CMOS design and manufacturing.

| Insulator : Metal | Insulator : Metal |
|---|---|
| $SiO_2$:Pt | AlN:Pt |
| MgO:Pt | $Si_3N_4$: Al |
| $Al_2O_3$:Pt | $Si_3N_4$: Cr |
| $Y_2O_3$:Pt | $Si_3N_4$: Cu |
| $HfO_2$:Pt | $Si_3N_4$: Ta |
| $Ta_2O_5$:Pt | $Si_3N_4$: Pt |
| $SiO_xN_y$: Pt | |

**Table 2.2.** Atomic insulator:metal hybrids exhibiting nanometallic transitions and switching behavior.

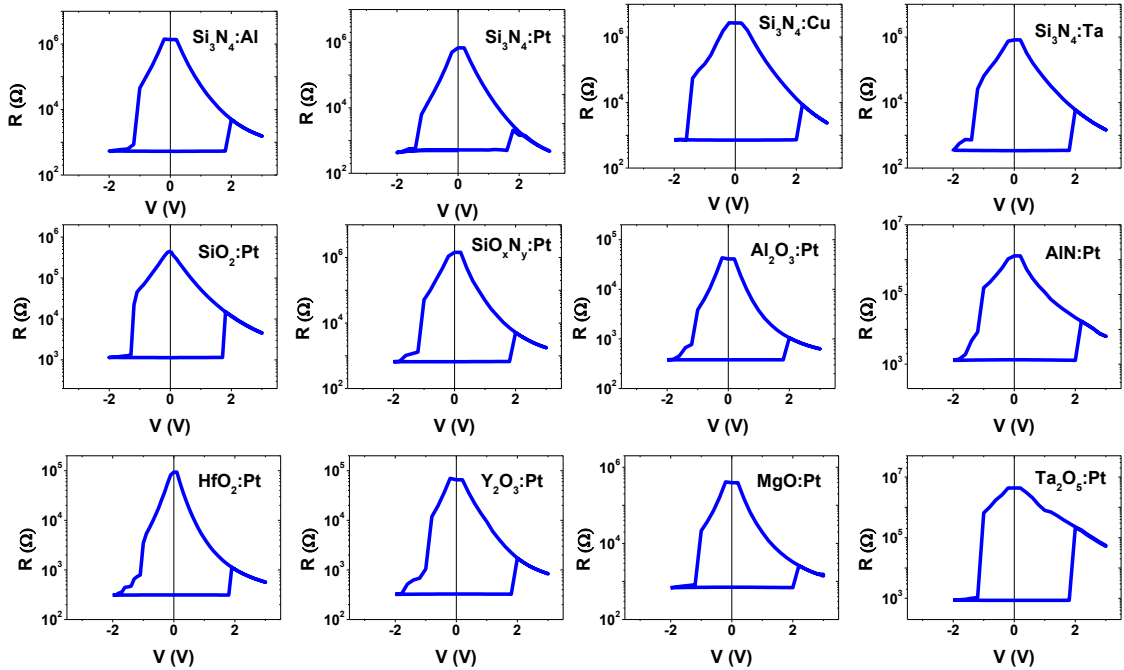

**Figure 2.33.** Characteristic switching *R-V* curves. For several other insulator : metal systems: Si$_3$N$_4$:Al, Si$_3$N$_4$:Cu, Si$_3$N$_4$:Ta, Si$_3$N$_4$:Pt, SiO$_2$:Pt, SiO$_x$N$_y$:Pt, Al$_2$O$_3$:Pt, AlN:Pt, HfO$_2$:Pt, Y$_2$O$_3$:Pt, MgO:Pt, Ta$_2$O$_5$:Pt. Bottom electrode: Mo; top electrode: Pt.

As we have shown earlier in **Figure 2.8**, all switchable films appeared amorphous to conventional X-ray diffraction, but their TEM nanostructures ranged from ones with a worm-like contrast (typical for amorphous networks) without any resolvable inclusion to ones with metal-rich inclusions embedded in an amorphous background (**Figure 2.11** and **Figure 2.12**). As already discussed earlier, the interesting questions here are: how are these different nanostructures correlated to device characteristics, and what is inclusion's role in nanometallicity and switching?

To answer these questions, we performed $R$-$V$ tests on devices fabricated with various compositions. The insulating matrices of all the films are amorphous as evident from the worm-like appearance in the plan-view micrographs shown in **Figure 2.34**. At the highest $f$ (right column, **Figure 2.34** (**g**, **h**, **i**)), dark-colored inclusions associated with metal-rich compositions, which derive the contrast from atomic number ($Z_{Pt,Cr,Si,Al,O,N}$ =78, 24, 14, 13, 8, and 7, respectively) differences, are readily visible. Some clusters appear to be in contact with each other; not surprisingly, these films are always conducting regardless of thickness (hence above the percolation threshold, $f_p$), which was confirmed by the $R$-$V$ curves in the insets. In contrast, at the lowest $f$ (left column, **Figure 2.34** (**a**, **b**, **c**)), no such inclusion is visible. Yet with a suitable δ these *structureless* films are switchable featuring "generic" $R$-$V$ curves (see insets) for the voltage-controlled MIT. At an intermediate $f$ (middle column, **Figure 2.34** (**d**, **e**, **f**)), dark inclusions are also visible but obviously separated from each other. Again, with a suitable δ these films are switchable featuring $R$-$V$ curves (see insets) identical to their lower-$f$ counterparts but for a lower insulator resistance.

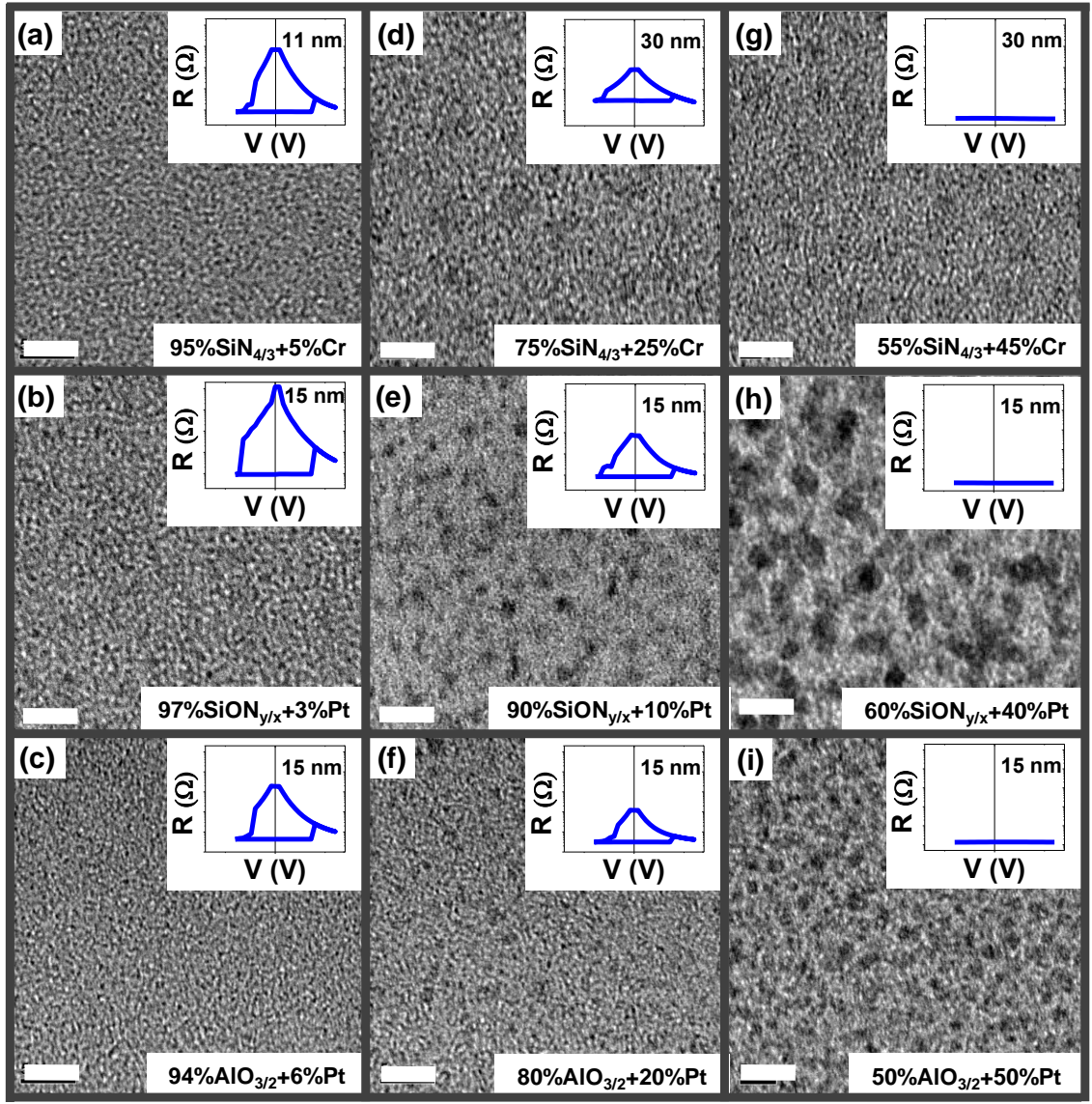

**Figure 2.34**. Nanostructure and switching. Plan-view TEM images (bright field) of $Si_3N_4$:Cr (a, d, g), $SiO_xN_y$:Pt (b, e, h) and $Al_2O_3$: Pt (c, f, i) systems. Scale bar: 5 nm. Inset within each image is the characteristic *R-V* curve for film of the corresponding composition at the thickness specified, under identical voltage (*V*) scan between -2 V to +3 V, drawn with identical resistance (*R*) range from 80 Ω to 20 MΩ for ease of comparison. (a, b, c): switching films without metal-rich clusters; (d, e, f): switching

films with metal-rich clusters; (g, h, i): conducting films with percolating metal-rich clusters.

These observations verified that contacting metal-rich clusters are indeed needed for bulk percolation, but whether any cluster exists or not is immaterial for nanometallicity and switching. Since nanoclusters are ruled out as the cause for nanometallicity, we must explain conduction by Anderson's picture: in a random material some electrons of atomically dispersed metal atoms can tunnel to span a sufficiently large distance $\zeta$, thus rendering nanometallicity. Likewise, since nanoclusters are also ruled out as the cause for switching, we suggest that localization must arise from the lack of tunneling paths and/or the presence of Coulomb barriers (due to trapped charge injected during voltage-triggered MIT), thereby decreasing $\zeta$. Such localization effect decreases at a higher $f$, where a higher electron concentration is expected in Anderson's random lattice together with the contribution from metal-rich clusters. Thus, unlike metal-nanocrystal memory, which stores electrons in discrete metal nanodots (*e.g.*, Au and Pt in $SiO_2$, ref.[15-17]), isolated metal-rich clusters are merely spectators of nanometallic MIT. Random-wave electrons must have come from the random network itself, most likely from the electron-rich metal atoms/oligomers dispersed on the three-dimensional insulator scaffold.

## 2.8.2 Electrode Combinations

Electrode is a critical factor for nanometallic switching. Electrodes in contact with the nanometallic film provide definite work functions, which decisively control the ease and direction of electron injection, removal, trapping and detrapping. We have investigated various electrode combinations and found the following correlations.

(1)     If identical electrodes are used in a cell, no bipolar switching can be observed. From symmetry, this is completely expected. More specifically, a cell with identical low work-function electrodes (*e.g.*, Mo-Mo) is in a non-switching conducting state while a cell with identical high work-function electrodes (*e.g.*, Pt-Pt) results in a non-switching insulating state.

(2)     Devices with dissimilar electrodes are switchable and their switching polarity depends on the relative work function of the electrodes. If the top electrode has a higher work function, then switching is counterclockwise (**Figure 2.15**), which is our standard configuration. Conversely, if the bottom electrode has a higher work function, then switching is clockwise. This correlation is summarized in **Figure 2.35**. Asymmetric electrodes are likely to provide directional electron injection/removal/trapping/detrapping and therefore bipolar switching.

These correlations will be fully addressed in **Section 2.9.4**.

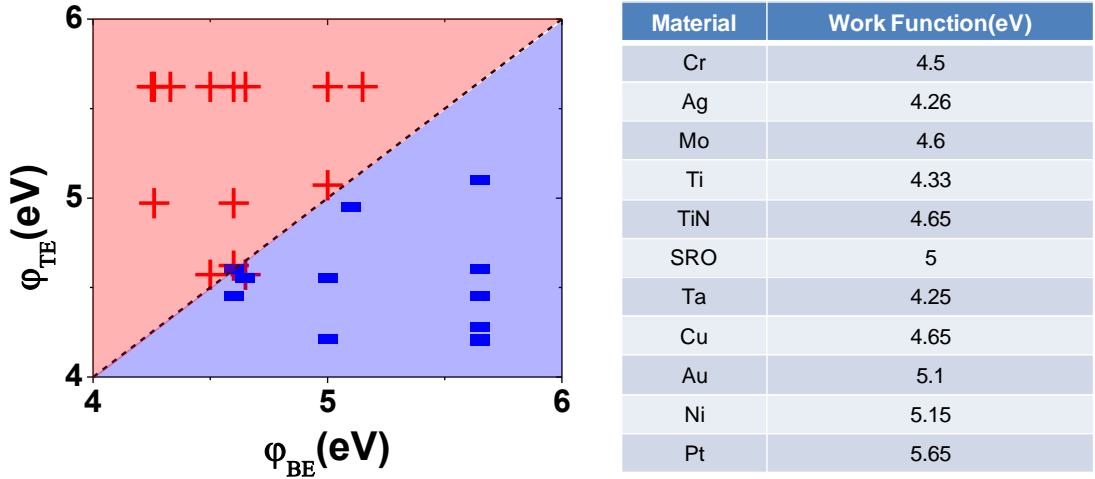

**Figure 2.35.** Electrode effect on switching polarity. "+" represents counterclockwise switching, "-" represents clockwise switching. (Polarity convention was previously defined in **Figure 2.15**)

## 2.9 Mechanism and Discussion

### 2.9.1 Localization Length

Resistance change between different states can be attributed to a corresponding change in localization length $(\zeta)$, *i.e.* $\zeta = \zeta(\Omega)$, where $\Omega$ is a state variable depending on the resistance state, doping level, field, *etc.* The LRS enjoys a large localization length, $\zeta_{LRS} > \delta$, allowing "free" electron diffusion across the film thickness. In this case, $R = R_0 \exp(\delta/\zeta) \approx R_0(1 + \delta/\zeta)$, so there is a weak thickness dependence for resistance. As electrons gain energies from the external voltage, at a critical voltage $V_c$ some electrons can overcome certain barriers and land on certain metastable sites that are local minima in the energy landscape. If these states are separated from the conduction patch, then the electrons there are trapped and become "permanent". Their Coulombic repulsion will turn away other (itinerant) electrons and curtail their localization lengths to $\zeta_{HRS} < \delta$, which can be a drastic reduction (**Figure 2.18b**). This change is non-volatile until a negative voltage $-V_c$ is applied to empty the trapped charge.

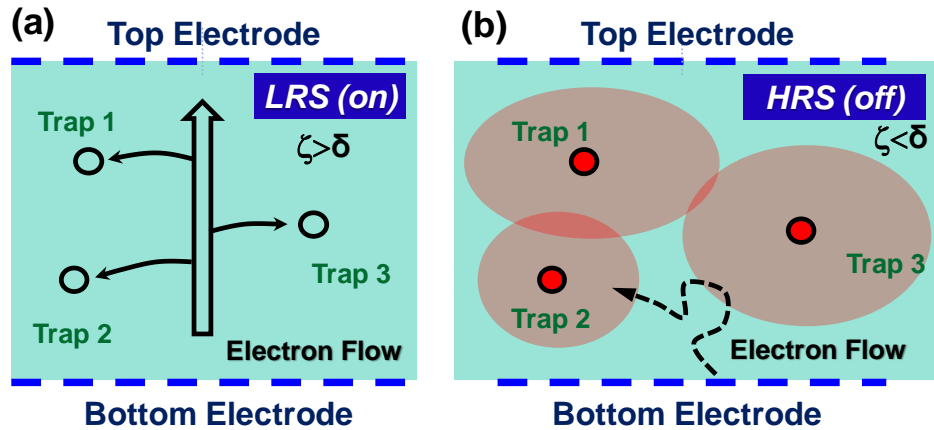

**Figure 2.36**. Schematic of electron conduction in (a) LRS and (b) HRS.

### 2.9.2 Trapping Centers and Negative-*U* centers

Where are these traps? In crystalline solids, likely locations for electron trapping sites are point defects. However, nanometallic films are amorphous, which have no well-defined defects. We believe trapping sites are associated with dangling bonds, which are common in amorphous compounds. With the insertion of metal atoms, ionic bonds (*e.g.*, O-Si-O) may be interrupted and dangling bonds (*e.g.* O-Si-) may form. These unpaired dangling bonds are energetically favorable sites for acquiring one additional electron; thus, they may provide the trapping centers (**Figure 2.37**).

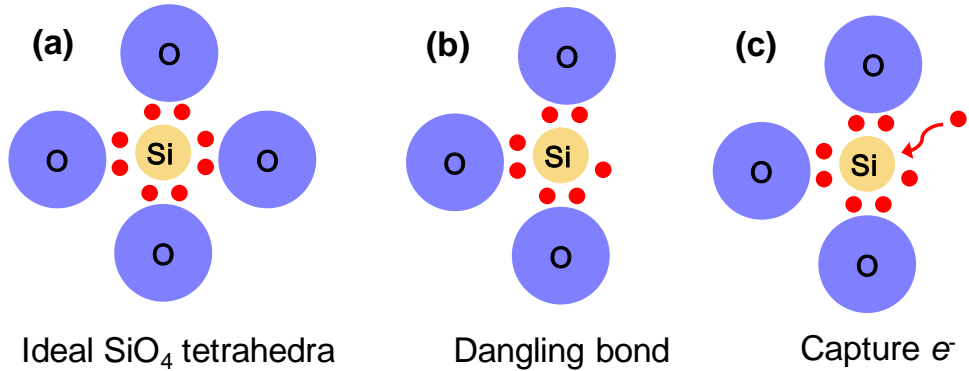

**Figure 2.37**. Schematic of electron trapping site comparing (a) ideal Si-O bonding, (b) Si with dangling bond and (c) dangling bond capturing one extra electron.

Existence of dangling bonds implies the possibility of the so-called negative-$U$ centers[8]. Electronically, $U$ refers to the on-site electron energy, which is positive under normal circumstances mostly because of the on-site Coulomb repulsion. A negative-$U$ center nevertheless arises when local bond/atomic distortions lower the system energy so much that the net on-site energy may be regarded negative despite a positive Coulomb contribution[18]. The prerequisite for such negative-$U$ states is (a) the addition of a second electron, and (b) a strong electron-phonon interaction that leads to a sufficiently large local structural distortion. In this sense, it is a polaron albeit a highly localized one, at the electron trapping site and not found everywhere in the material. The above bond relaxation occurs shortly after (over a time of $10^{-13}$ to $10^{-12}$ s) electron filling (happening in $\sim 10^{-15}$ s) and lasts well within the residence time of the electron at the defect/center (ranging from $10^{-9}$ to $10^{-4}$ s, which would be the natural retention time of an electronic memory)[19]. Naturally, the need for easy structural distortions dictates that these centers tend to be situated near internal defects (*e.g.* vacancies)[19-22] and surfaces (*e.g.* internal voids)[19] in a crystalline solid, and they are especially common in amorphous materials in which flexible (cation/anion) polyhedra with dangling bonds are commonplace[18-20]. Indeed, the negative-$U$ center was first proposed for *amorphous* chalcogenides because chalcogenides are relatively flexible and covalent, thus accentuating the electron-phonon interactions[18]. (Incidentally but not surprisingly, we also note that the negative-$U$ centers have been studied in phase change memory materials that use similar chalcogenides[23].) In addition to amorphous chalcogenides and $SiO_2$, negative-$U$ centers have been studied in a broad range of materials. A cursory literature search found negative-$U$ centers have been investigated in the materials listed in the following table.

**Materials with experimentally verified negative-$U$:**

Semiconductor: $Si$[24-25], $GaAs$[26-27], $Al_xGa_{1-x}As$[28], $CdTe$[29], $GaSb$[30], $HgCdTe$[31].

Oxides: $SiO_2$[19-20], $HfO_2$[32], $BaTiO_3$[33].

Nitrides: $Si_3N_4$[34-35], $GaN$[36].

Other Chalcogenide: $(Ge_2Sb_2Te_5)$[23]. Additionally, it was for elementary chalcogenides (S and Se)[37] where Anderson[18] first proposed the negative-$U$ mechanism.

**Theoretically predicted negative-$U$:**

Semiconductor: $SiC$[38], $Ge$[39], $ZnSe$[40].

Oxides: $ZnO$[22], $HfO_2$[41], $ZrO_2$[41], $FeO$[42].

Nitrides: $AlN$[43], $GaN$[44].

All strongly ionic Halides[45].

We believe such negative-$U$ centers broadly exist in nanometallic films, given its rich dangling bonds. The metal-rich nanometallic pathways can then supply mobile electrons to these negative-$U$ centers, and when two-electron filling is made at a center—which occurs when the Fermi level is sufficiently lifted by $V_{off}$ to counter the internal bias caused by the work function differential between the electrodes—*off-switching* (electron trapping/localization) takes place at the site. Since the spontaneous conversion of mobile electrons to localized electrons at the negative-$U$ centers is energetically irreversible after $V_{off}$'s removal, the trapped electrons and the local Coulomb barriers they erect will remain. They can "choke off" the electron passage in the nearby nanometallic paths, making non-volatile memory possible. Later, from the fact that a critical opposite bias, $-V_{on}$, can reverse the process and cause delocalization again, we estimate the stabilization of the negative $U$ contribution is about $V_{on}+V_{off/2}$. We will provide more experimental evidence on negative-$U$ centers in **Chapter III**.

### 2.9.3 Voltage-time Dilemma

As we have shown in **Section 2.7**, fast switching (<ns or <ps) and long retention (>10 years) can be achieved simultaneously, which seems contrary to the "voltage-time dilemma" that is believed to prevail in all electronic systems. The rationale behind this dilemma is that the same barrier that need to be overcome, by tunneling, thermal emission, or their combination, with and without a field assistance cannot be too difficult to overcome (thus good for retaining data) on one hand, yet too easy to overcome (thus good for fast programming at a low voltage) on the other hand. Schroeder *et al.* provided a quantitative assessment for a metal/insulator/metal thin film stack[46]. Their calculation confirms the dilemma: there is an incompatibility between the requisite long retention time (10 years) and the desired short (≤100 ns) read/write current pulses at low voltages (≤1 V). However, negative-$U$ centers provide a solution to the dilemma by offering a variable energy barrier, which dynamically adjusts the barrier for during programming and for electron storage. We will revisit this topic in **Chapter III**.

### 2.9.4  Mechanism for Nanometallic Switching

Key experimental observations that must be explained by the model are summarized as follows.

(1) Switching voltage is insensitive to temperature and film thickness; device resistance is polarity-symmetric before switching occurs.

(2) The initial state is Ohmic conducting down to <2K.

(3) Switching is bipolar: on-switching at a negative voltage, off-switching at a positive voltage.

(4) After on-switching there is no further unipolar switching at more negative voltages; after off-switching there is no further unipolar switching at more positive voltages.

(5) Both the HRS and LRS are non-volatile at zero voltage.

(6) When a low-work-function metal is used as both top and bottom electrodes, the device is always at the LRS.

(7) When a high-work-function metal is used as both top and bottom electrodes, the device is initially conducting but permanently switches to the HRS at either a positive or a negative voltage.

### 2.9.4.1 Electron energy states in an insulator:metal film

Nanometallic film contains metal atoms randomly separated at various spacing. As shown in **Figure 2.38a**, when metal atoms are closely spaced, their outer-electron orbitals overlap forming band structures of the conventional type. For the simplest case of metal atoms with only one outer-electron per atom, the conduction band is half filled, up to the level of $E_f$, the Fermi energy. Adding an extra electron to the metallic state requires $E_f$, the same energy as filling the last electron. Nanometallic film also contains metal atoms of a larger spacing, with less overlapping of outer-electron orbitals, for which the band structure progressively deforms as shown in **Figure 2.38b-d**. The deformation is caused by the (positive ) on-site Coulomb potential: adding a second electron to the same orbital of an already highly *localized* electron entails Coulomb repulsion between the two electrons; in effect, the second electron must occupy a higher energy state as shown in **Figure 2.38d**. (This effect, the so-called correlated-electron effect, is absent in the conventional band structure, in which the Coulomb potential is small because of delocalization.) Therefore, with close atomic spacing, a delocalized-electron system of one-electron per metal atom is a conductor (**Figure 2.38a-b**), whereas with large atomic spacing a partially-localized-electron system of the same is an insulator (**Figure 2.38c-d**). In nanometallic films, negative-$U$ centers associated with **Figure 2.38d** ensures the following features for the trap: (i) adding an extra electron to the trap requires a critical energy, (ii) the added electron is localized, (iii) with the extra electron the trap becomes negatively charged, and (iv) the trapped charge is stable after bond relaxation.

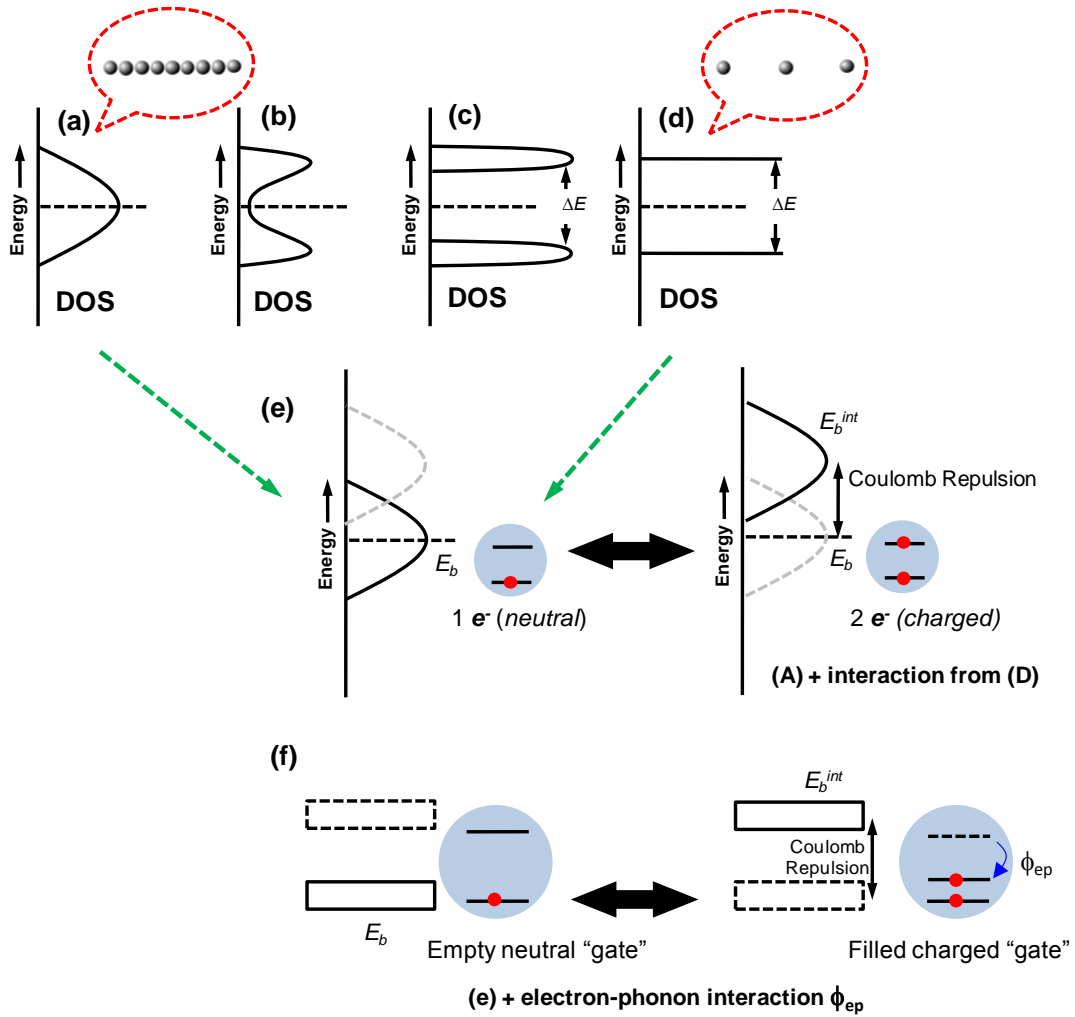

**Figure 2.38**. Schematic energy/density of states for one-electron metal atoms placed at various spacing (a-d). Correlated-electron effects increase from (a) to (d). (e) The energy level of (a) depends on whether a second electron (the second <span style="color:red">red dot</span>) occupies the (d) state (the right panel) or not (the left panel). (f) Simplified representation of (e) with electron-phonon interaction included,

which lowers the energy level of the second electron. In (e) and (f), the states derived from (a) form the conduction channel, the states derived from (d) form the gate channel.

### 2.9.4.2  Two channel model: Conduction channel and gate channel

As-fabricated device is designed to contain enough metal atoms to form enough metallic states (**Figure 2.38a**), so that there exist some continuous conducting paths between the two electrodes. They will be referred to as the *conduction channel*. An isolated conduction channel has a characteristic energy $E_b$. Since the conduction channel coexists with traps, its energy can be influenced by the traps. When a trap is occupied by an extra electron, it becomes negatively charged and creates a long-range (inter-site) Coulomb repulsion. As a result, it raises the energy of the nearby conduction channel from $E_b$ to $E_b^{int}$ as shown in **Figure 2.38e**. This may choke off conduction. Therefore, the traps will be referred to as the *gate channel.* Note that there is a fundamental difference between the two channels. The correlated-electron effect is absent or weak for the conduction channel, so additional electrons can be accommodated at the same (Fermi) energy $E_f$. In contrast, the effect is strong for the gate channel, so there needs an extra energy to add an electron to the gate channel.

In the following, we will use the simplified drawing of **Figure 2.38f** to represent **Figure 2.38e**. In **Figure 2.38f**, we also introduce the electron-phonon interaction $\phi_{ep}$, which lowers the energy of the second electron to counter the correlated-electron effect. The conduction channel is represented by a rectangular box: the operating one reflecting the current status of the gate channel (and its interaction) is drawn in solid lines, and the hypothetical one reflecting a past/future status of the gate channel is drawn in broken lines. The gate channel is represented by a blue circle with up to three states occupied by up to two electrons (as <span style="color:red">red dots</span>). The lowest level is the one-electron state, the highest level is the two-electron state, and the middle level is the two-electron state after partial stabilization by the electron-phonon interaction $\phi_{ep}$. The exact magnitude of $\phi_{ep}$ is unimportant. If it is large enough, then the energy state of the second electron becomes lower than that of the first electron, which is the case of negative-$U$. In the figures shown below, for simplicity a smaller $\phi_{ep}$ is used. But the essential results below are all preserved if a larger $\phi_{ep}$ is used instead. Different cell configurations and resistance states will be illustrated to show the model can explain all the observations listed at the start of this section.

**Different material electrodes**

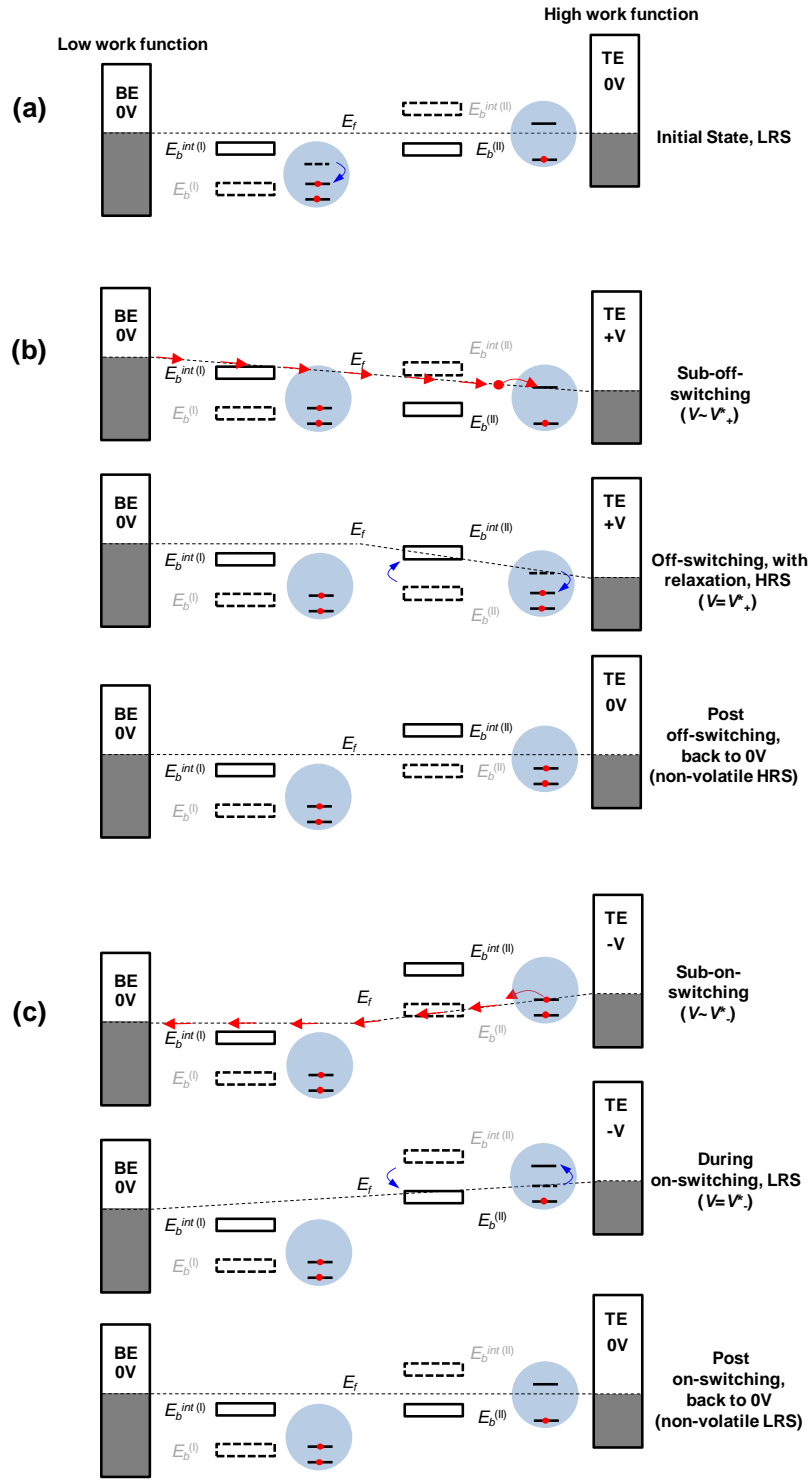

**Figure 2.39**. Device with top electrode made of higher-work-function metal and bottom electrode made of lower-work-function metal. (a) Metal concentration rich enough to have conducting device at $V$=0. (b) Positive $V$ applied to top electrode, causing off-switching, then back to $V$=0, with device in insulating state. (c) Negative $V$ applied to top electrode, causing on-switching, then back to $V$=0.

### 2.9.4.3  Device with electrodes of different metals: the initial state

Consider a film in contact with a bottom electrode I (*e.g.*, Mo, with a lower work function) and a top electrode II (*e.g.*, Pt, with a higher work function). After the conduction channel redistributes electrons between the two electrodes to establish a constant $E_{\mathrm{f}}$, the accumulated charge on the two electrodes causes all energy levels to shift, hence asymmetry arises. Suitable electron doping (*via* adjusting the metal concentration during fabrication) can create an as-fabricated state as shown in **Figure 2.39a**, with side I having two electrons in the gate channel and side II having one electron in the gate channel, plus an $E_{\mathrm{f}}$ high enough to allow electron flow through $E_{\mathrm{b}}^{\mathrm{int(I)}}$ and $E_{\mathrm{b}}^{\mathrm{(II)}}$. This is the low-resistance state (LRS): the electron can flow in either direction.

### 2.9.4.4  Off-switching

At a critical positive voltage, $+V^{*+}$, electron filling (by an electron from the bottom electrode) of the gate channel on side II becomes energetically feasible (top panel of **Figure 2.39b**). As this proceeds to completion, shown in the second panel of **Figure 2.39b**, four effects set in. (i) The energy of the two-electron state in the gate channel is lowered by electron-phonon interaction $\phi_{ep}$; (ii) the energy of the conduction channel on side II is raised to $E_b^{\text{int(II)}}$ by long-range Coulomb repulsion; (iii) if $E_f$ now lies below $E_b^{\text{int(II)}}$, then there is no electron flow to $E_b^{\text{int(II)}}$, so the device is switched to the high-resistance state (HRS), and (iv) the voltage in the film next readjusts to reflect the fact that side II is insulating while side I remains conducting. It is worth noting that without $\phi_{ep}$, non-volatility cannot be achieved in the model. This is shown in the third panel in **Figure 2.39b**. A sufficiently large $\phi_{ep}$ and a sufficiently high Fermi energy (controlled by metal concentration) will ensure the HRS is stable at zero voltage, *i.e.*, it is non-volatile.

### 2.9.4.5 On-switching

At a critical negative voltage, $-V^{*-}$, the release of the second electron (returning to the bottom electrode) in the gate channel on side II becomes energetically feasible (top panel of **Figure 2.39c**). As this proceeds to completion in the second panel of **Figure 2.39c**, four effects set in. (i) The energy of the (empty) two-electron state in the gate channel recovers to the previous level prior to $\phi_{ep}$ stabilization; (ii) the energy of the conduction channel on side II is lowered to $E_b^{\text{(II)}}$ as the long-range Coulomb repulsion disappears; (iii) if $E_f$ now lies above $E_b^{\text{(II)}}$, then the electron flow in $E_b^{\text{(II)}}$ resumes, the electron flows back to the bottom electrode, and the device is switched to the low-resistance state (LRS), and (iv) the voltage in the film next readjusts to reflect the fact that both side II and side I are conducting; indeed, the *actual* voltage across the electrodes through the film must drop if the applied voltage also passes through an external load. Moreover, the LRS is stable at zero voltage, *i.e.*, it is non-volatile, as shown in the third panel of **Figure 2.39c**.

### 2.9.4.6 Device with electrodes of identical low work function metal

As shown in **Figure 2.40**, when a low-work-function metal is used for both electrodes, the Fermi energy $E_f$ lies above the (repulsion-elevated) conduction band edge $E_b^{int}$, so there is always conduction. This is despite the fact that there is trapped charge in the film, which elevates the conduction band energy from $E_b$ to $E_b^{int}$.

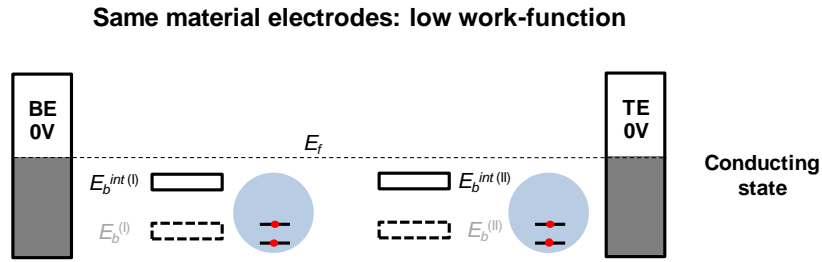

**Figure 2.40**. Device with top and bottom electrodes made of same low-work-function metal. Metal concentration rich enough to have conducting device initially.

### 2.9.4.7 Device with electrodes of identical high work function metal

As shown in **Figure 2.41a**, when a high-work-function metal is used for both electrodes, $E_f$ lies above the conduction band edge $E_b$, so there is also conduction, initially. However, this conducting state is unstable: once a critical voltage is applied, electron trapping on one side (side II, if the voltage is positive) of the gate channel causes $E_b^{(II)}$ to rise to $E_b^{int(II)}$, which now lies above $E_f$, as shown in the first two panels of **Figure 2.41b**. Thus the device is switched off. This is despite the fact that there is no trapped charge on side I of the gate channel. At the opposite voltage polarity, trapped electron on side II are detrapped but they are immediately trapped again on side I. Therefore, the device will always remain in a HRS after initial trapping, although microscopically the trapped electrons are allowed to internally exchange within the nanometallic film.

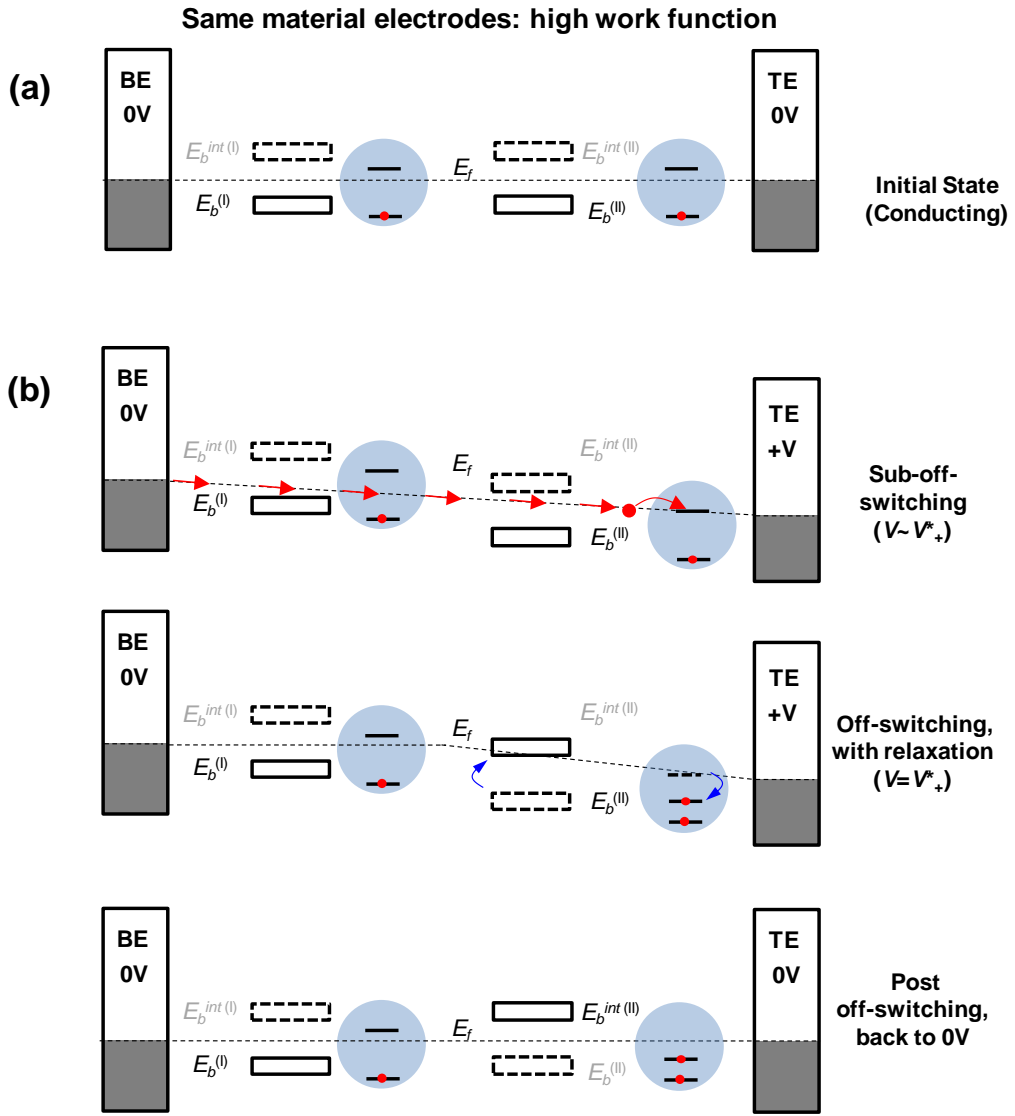

**Figure 2.41**. Device with top and bottom electrodes made of same high-work-function metal. (a) Metal concentration rich enough to have conducting device initially. (b) Positive *V* applied to top electrode, causing off-switching, then back to *V*=0, with device remaining in insulating state.

## 2.10 Conclusions

1. Nanometallicity have been found in thin films of a wide variety of oxide/nitride/oxynitride:metal combinations, with metal constituents atomically dispersed in an amorphous hybrid. Thickness ($\delta$)-composition ($f$) maps constructed for these thin films demonstrated nanometallicity and nanometallic switching regardless of the presence/absence of metallic clusters, even at compositions far below the cluster percolation threshold.

2. Reproducible bipolar resistive switching phenomenon triggered by a set of characteristic switching voltages (~ 1 V) independent of device thickness, area (100 nm × 100 nm), composition, and switching speed is found for nanometallic films sandwiched between two electrodes of different work functions. They provide an RRAM with outstanding performance: fast speed (<100 ps), long retention (>10 years), high endurance, scalability, and excellent reproducibility and uniformity.

3. Nanometallic switching is purely electronic: the insulator to metal transition can be alternatively triggered by UV irradiation. The insulating state (the high resistance state or HRS) is a metastable state, with a resistance that scales with thickness and composition exponentially. The low resistance state is stable with a resistance that can be arbitrarily lowered by depleting residual trapped charge.

4. Electrons in nanometallic films can be trapped at negative-$U$ centers. The negative-$U$ effect, which involves lattice distortion (electron-phonon interaction) to stabilize trapped electrons and provide resistance-state-dependent adjustable energy barriers, enables nanometallic RRAM to simultaneously satisfy the requirements of fast switching speed at low voltage and good memory retention. Thus, the voltage-time dilemma that plagues electronic memory is fully resolved in nanometallic RRAM.

5. A two-channel model that assigns metal atoms of different spacing, thus different correlated electron effects, to "gate" and "conduction" channels is proposed. When aided by electron-phonon interaction that stabilizes trapped electrons in the gate channel, the model can explain all the experimental observations including nonvolatile switching and electrode effects.

# Chapter III. Stress Induced Metal-Insulator Transition

## 3.1 Introduction

In **Chapter II**, we demonstrated nanometallic RRAM has an electronic nature，free from the "voltage-time dilemma" that troubles conventional electronic memories. Theoretically, no energy barrier (having *two* independent characteristics: barrier height and width) that separates a trapped-charge-state from a free-charge-state can simultaneously satisfy *three* specifications: low programming voltage (<1 V), fast programming time/speed (<100 ns) and long retention time (>10 years)[1-5]. Moreover, since the trapped-electron-state often experiences an on-site Coulomb repulsion (the Hubbard $U$), the positive-$U$ electron sees a lowered barrier making the state even less capable of retaining memory. However, such dilemma can be lessened or eliminated by converting a positive-$U$ state to a "negative-$U$" state through strong electron-phonon interaction, which lowers the energy of a trapped electron by $\phi_{ep}$, as schematically illustrated in **Figure 3.1**. Atomically, this requires a local bond distortion (relaxation) around the newly trapped electron, stabilizing both the electron and the bond. Since the electron trapping/escape times are short (~1 fs)[6], the rate-limiting step for memory switching in a nanometallic RRAM is that of bond distortion, which takes about the time for one atomic vibration, or 0.1-1 ps.

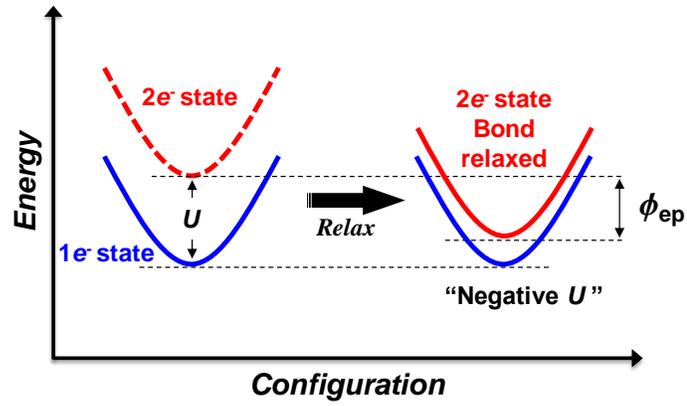

**Figure 3.1.** Energy of state of freshly captured electron (upper) and the relaxed state (lower) in configuration coordinate.

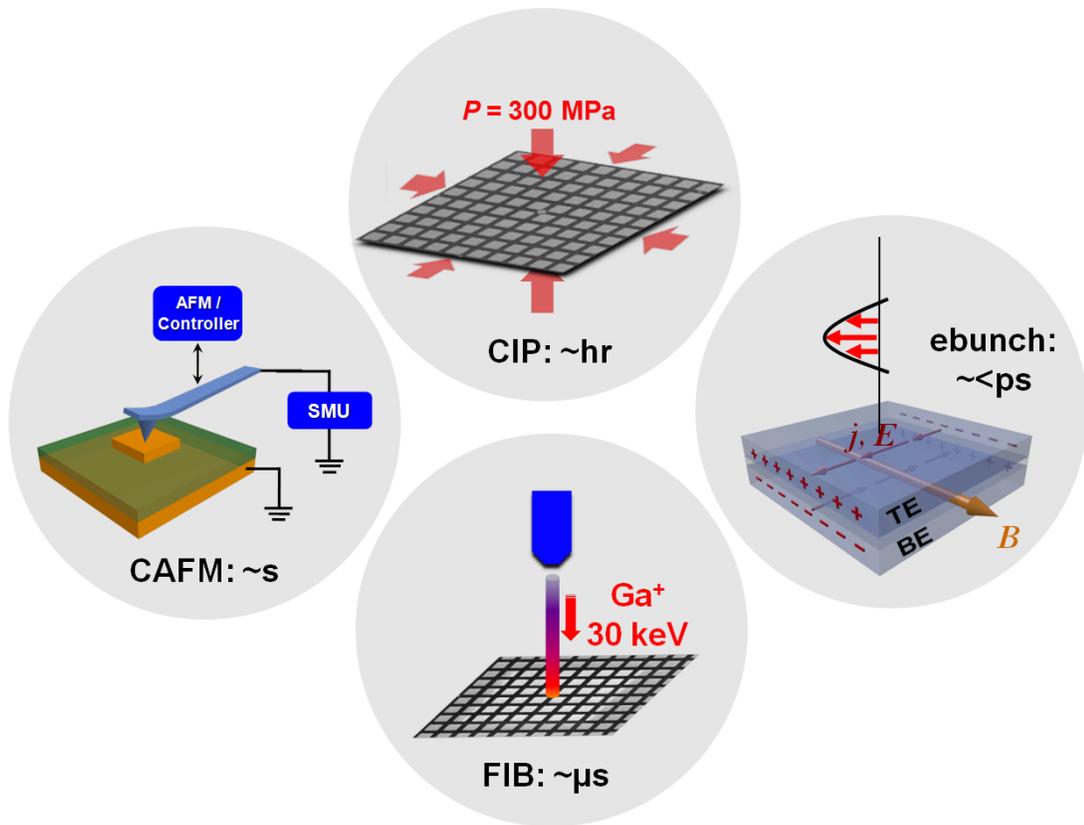

**Figure 3.2.** Four experiments in this chapter: CIP, CAFM, FIB, ebunch.

In this chapter, we provide four pieces of experimental evidence revealing the existence of electron-phonon interaction and its ability to directly influence switching (**Figure 3.2**). These experiments demonstrate another degree of freedom to trigger the metal-insulator transition: *mechanical stress*. A surprisingly modest mechanical excitation of the orders of 20 to 300 MPa covering a very wide time domain from $10^{-13}$ s to $10^3$ s can destabilize a trapped-charge state, converting it to a free-charge state, thus leading to a one-way resistance switching. Beyond RRAM, the ultra-fast test methodology developed in this chapter could also be of use for probing mechanical/ electrical properties of other materials.

## 3.2 Pressure Induced MIT at $10^3$ s

Our first mechanical test was performed using a uniform isostatic pressure of 300 MPa applied to all the devices on a chip inside a Cold Isostatic Press (CIP). The system uses oil as working fluid, which guarantees an isotropic compressive stress and thus avoids tensile or shear deformation, cracking and other anisotropic damage. Device was mounted, with silver paste, on a rounded metal plate and covered by an aluminum foil (as charge sink), then placed inside an evacuated rubber glove before inserted into the pressure vessel.

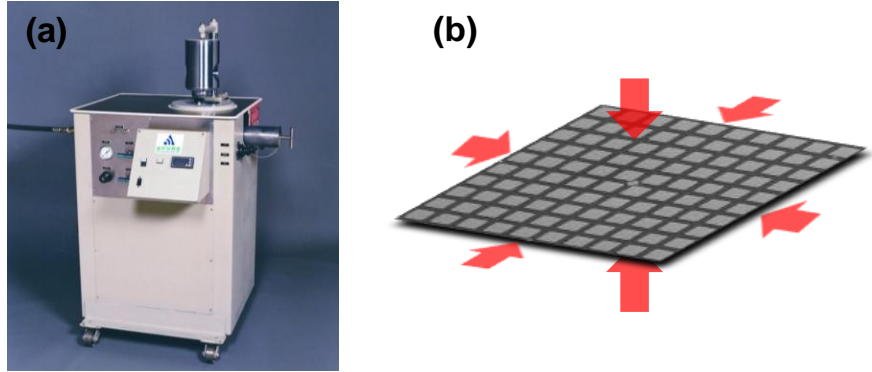

**Figure 3.3.** (a) Cold Isostatic Press (CIP) system. (b) Schematic of RRAM arrays under isotropic pressure.

### 3.2.1  Results

Before pressurization, cells in an RRAM array were first checked for their initial resistance states (HRS or LRS). As shown in **Figure 3.4a**, as-fabricated cells are in either state, with the fraction of the LRS being higher in larger cells. (Each array contains many cells of different sizes.) This is common in our experience, and the statistical presence of the HRS despite the fact that all cells have the same thickness and composition is attributed to processing-introduced trapped charge (*e.g.*, excess charges during sputtering deposition). Then samples were loaded into a CIP and isotropic pressure of 300 MPa was applied at room temperature (**Figure 3.3b**). A range of loading time (typically between 10 min and 1 hour) was used, but it is not expected to make any difference. After depressurization, samples were removed and electrically tested again.

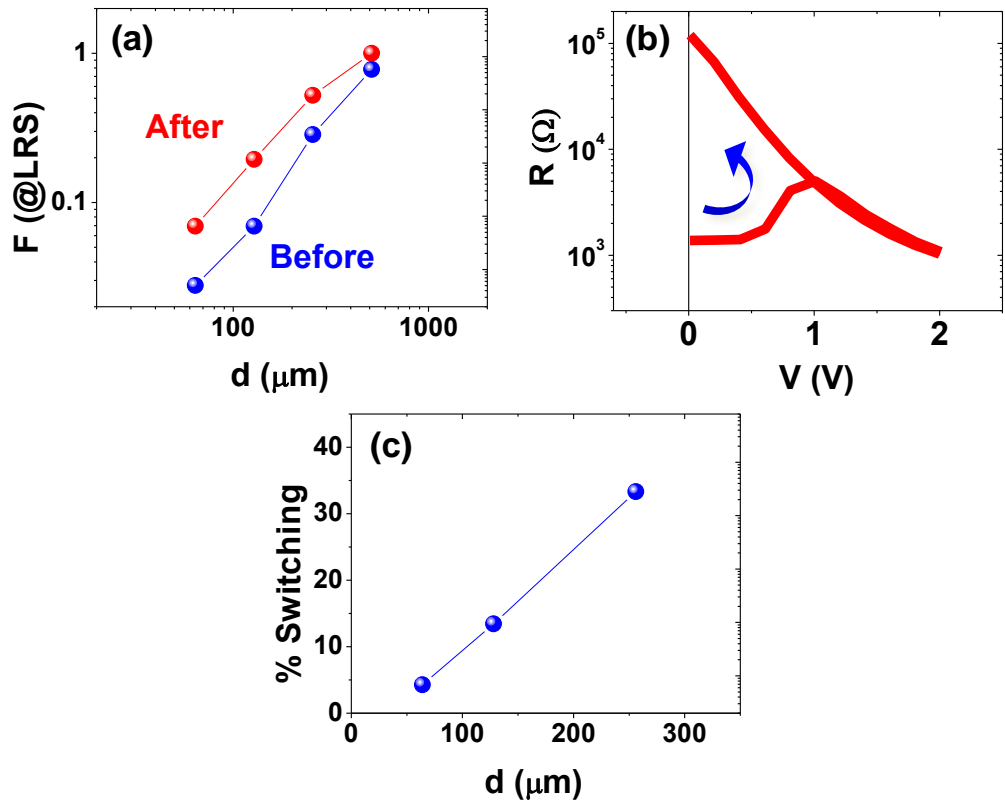

**Figure 3.4.** (a) Fraction of LRS before and after 300 MPa pressure treatment (the tested cells are all in their fresh states without preset). (b) *R-V* characteristics of pressure-switched device being switched back to HRS under positive voltage. (c) Calculated percentage change (HRS to LRS) caused by pressure treatment in (a) (largest size data is not shown because it is 100% after pressure treatment). Device: Mo/$Si_3N_4$:5%Pt/Pt, $\delta$=10 nm.

Post-pressurization measurements found that, with increasing device sizes, an increasing percentage of HRS devices had switched to LRS (**Figure 3.4a & c**). We also found the pressure-induced switching was non-damaging: in subsequent testing the pressurized devices could repeatedly switch in either direction with the same characteristic *R-V* curves as before (**Figure 3.4b**). But unlike an electrical voltage that can trigger two-way switching, pressure only induced one-way switching: the transitions only occur in HRS cells (HRS→LRS) while all LRS cell remains unchanged. We can thus conclude that the HRS is metastable from an energy perspective, which irreversibly transitions to the LRS, the ground state, under a mechanical perturbation. The above experiment was also repeated using arrays with preset cells, some to the HRS, and some to the LRS. The same observations were made, so the same conclusion remains. In addition, HRS→LRS switching percentage increases with applied pressure. As shown in **Figure 3.5**, switching percentage increases from ~20% to ~50% as the applied stress varies from 200 MPa to 1 GPa.

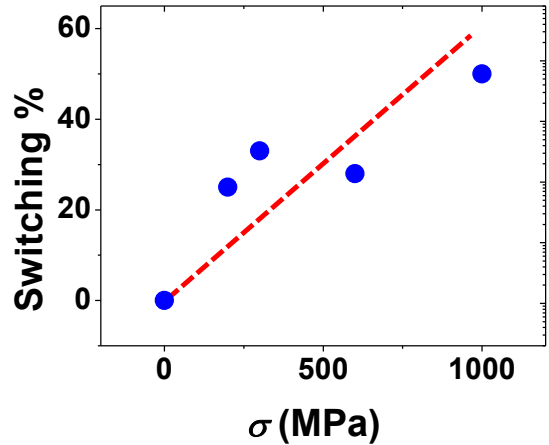

**Figure 3.5.** Switching percentage as a function of stress/pressure level (Device: Mo/$Si_3N_4$:5%Pt/Pt, $\delta$=10 nm, size: 256×256 μm$^2$).

### 3.2.2 Discussion

The key feature of the stress induced switching involved in mechanically triggered switching that differs from electrically triggered switching is the absence of ion migration. Since hydrostatic pressure is spatially uniform, there was no biased field to drive ion motion. Therefore, pressure-induced switching must be due to electron de-trapping. But can electron de-trapping be induced by mere 300 MPa—well below the critical pressure (~3-30 GPa) typical for electronic transitions? Indeed, the estimated strain is <0.002 if a typical Young's modulus of 100 GPa (ref.[7]) and a Poisson's ratio of 0.2 (ref.[8]) are assumed for the amorphous 95% $Si_3N_4$:5% Pt film used. We believe the answer lies in the fact that amorphous materials are not elastically uniform: they contain locally soft atomic spots[9] which may buckle under a modest pressure. These spots may have provided the sites for electron trapping and de-trapping. More importantly, this reveals electron-phonon interaction plays an active role during switching, which is a critical evidence for negative-$U$ sites: the negative-$U$ state is able to drive bond distortion to reduce the energy of a freshly trapped electron by $\phi_{ep}$, turning the state into a stable negative-$U$ configuration; conversely, once an opposite bond distortion unravels $\phi_{ep}$, restoring the state to the positive-$U$ configuration, it must prompt electron de-trapping.

In our material, these negative-$U$ sites are localized to some soft atomic sites. The pressure experiment unequivocally confirmed two key elements expected of the above mechanism: the action of electron-phonon interaction must (a) entail a strain *via* bond distortion (thus a stress can unravel $\phi_{ep}$), and (b) be predicated on electron occupancy (thus a stress can induce de-trapping but not trapping). Moreover, at least some trap sites must coincide with the locally soft atomic spots, which have locally low atomic density[9], so that even a modest pressure can induce switching.

### 3.3 Contact Stress Induced MIT at $10^2$ s

To repeat the above experiment under a different (and more complicated) stress state, CAFM technique was employed. The conducting AFM was a conventional non-conducting AFM module with an external circuit connecting the AFM tip to a semiconductor measurement unit (SMU), as shown in **Figure 3.6**. This homemade circuit allows a more flexible signal control (*e.g.*, a current compliance: 100 mA) than the build-in CAFM mode (*e.g.*, current compliance: 2 nA). It is suitable for DC measurement but not optimal for AC measurement (>1 MHz) because of the use of extraneous BNC and unshielded lines. A Pt/Ir coated AFM tip was used to provide a mechanical force and electrical contact. The standard contact mode was used during testing.

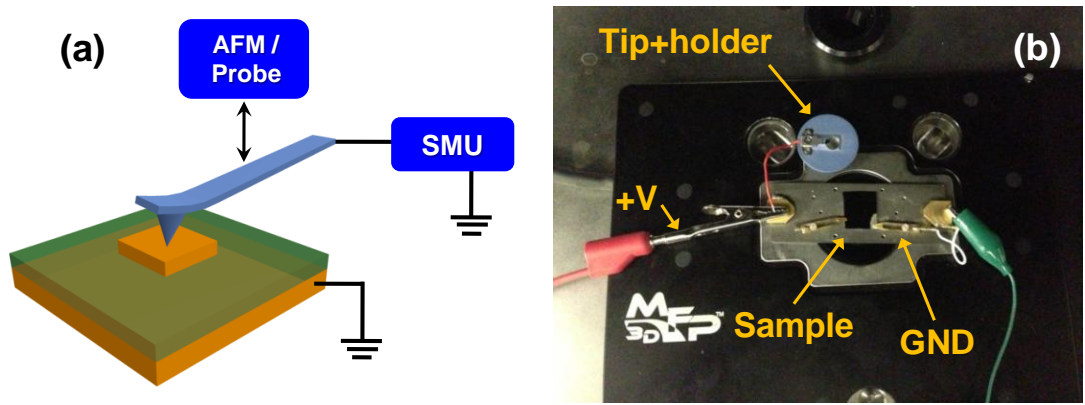

**Figure 3.6.** (a) Schematic of CAFM circuits. (b) Sample connections.

### 3.3.1  Results

The testing procedure is shown in **Figure 3.7**. First, a topographical profile was obtained using the tapping mode, which minimized tip wear. Second, a cell was selected and set to the HRS or LRS using a positive (HRS) or negative (LRS) voltage. Third, a random point on the top electrode of the cell was selected and engaged with a certain force set by the AFM controller. Fourth, the resistance ($R$) was monitored continuously by using a small reading voltage (+0.1 V), while the applied force was kept constant (made possible by the feedback mechanism of the AFM controller.) This continued for certain time or until the cell resistance showed an abrupt change to the value expected for the other state. After testing, the cell was electrically cycled again to check its switchability, and to examine whether there was any permanent damage to the cell.

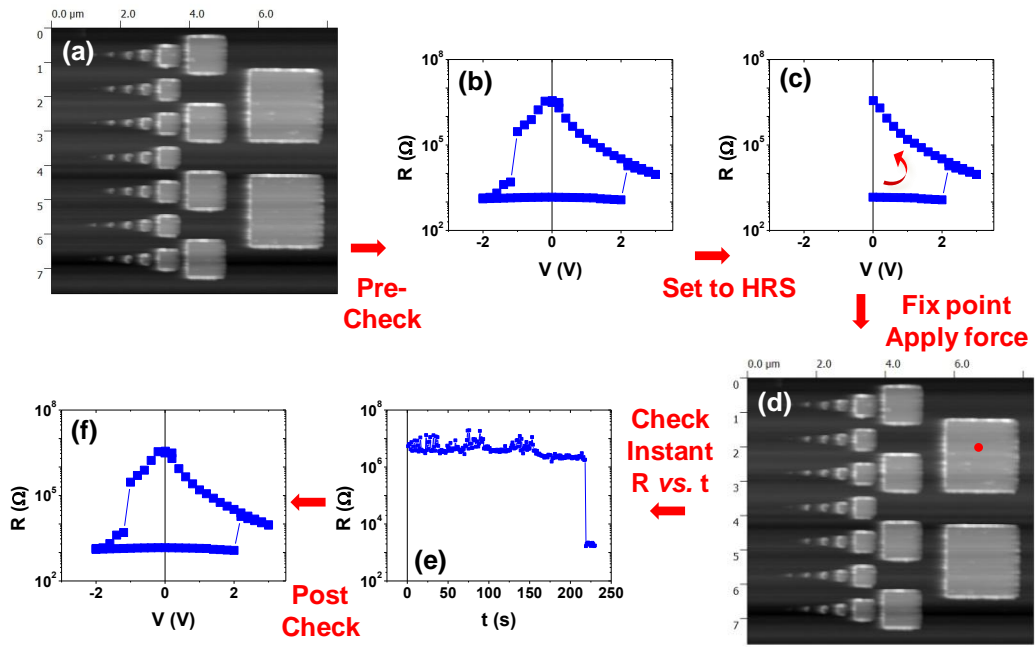

**Figure 3.7.** CAFM testing procedure.

**Figure 3.8a** shows the *in-situ* resistance variation of an HRS cell under a 100 nN compressive force. The resistance monitored at 0.1 V experienced a sudden transition to a LRS-"1" at ~100 s, followed by a second transition to LRS-"2" ~100 s later. The transition was non-transient: the low-resistance remained after removing the tip and voltage. It was also non-damaging: the device switched normally in subsequent tests (**Figure 3.8b**). Note that although the small reading voltage was *positive*, its electro-mechanical interaction cannot aid switching because the stress-free HRS→LRS transition in the device normally requires a *negative* voltage. Again, force only induced one-way switching: no LRS device identically tested switched.

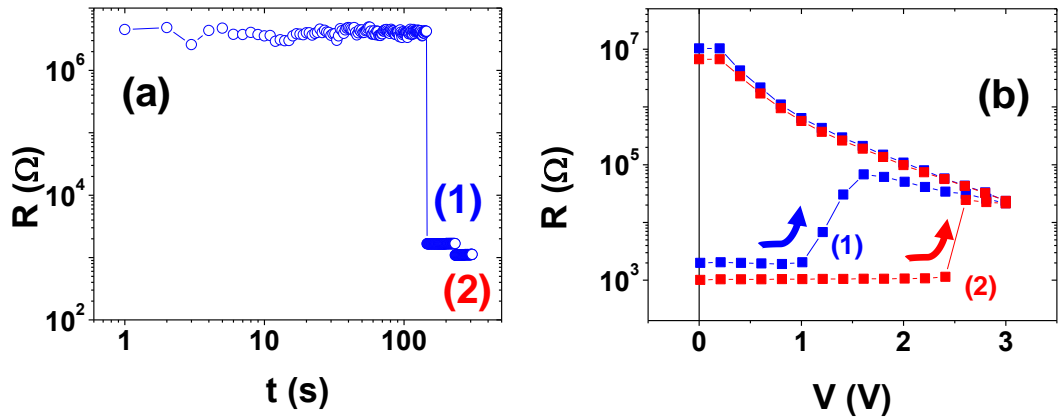

**Figure 3.8.** (a) *In-situ* resistance under 100 nN compression by CAFM tip (tip radius ~20 nm). HRS transitions to LRS (1) at ~100 s and to LRS (2) at ~200 s. (b) *R-V* curves of state (1) and (2), both switching to HRS under a positive voltage. (Device: Mo/Si$_3$N$_4$:5%Pt/Pt, $\delta$=2 nm, size: 2×2 μm$^2$).

As shown in **Figure 3.9**, the waiting time for the HRS→LRS transition decreases as the applied force or stress increases. A 50 nN force requires a significantly longer time (~$10^3$ s) than a 200 nN force (~$10^2$ s). This suggests that a larger local stress, which induces a large lattice distortion, can facilitate electron-phonon interaction. However, my experience with the experiment was that it was difficult to reproduce, *i.e.*, different tests yielded different waiting time.

Stress also affects switching voltage. This was demonstrated in another experiment in which the *R-V* curve was measured while a constant CAFM force was applied. As shown in **Figure 3.11** and **Figure 3.12**, the switching voltage $V_{on\rightarrow off}$ varies from 1.6 V to 2.4 V as the force/stress increases from 25 nN to 250 nN. However, the $V_{off\rightarrow on}$ remains unchanged under different stresses. Experiments are reproducible.

### 3.3.2 Discussion

One reason for the poor reproducibility may lie on the AFM tip which was worn during testing: it is hard to maintain an identical tip shape and contact radius between tests (see **Figure 3.10**). It is also impractical to check tip wearing *in-situ*, which would introduce uncertainty in stress estimation. For example, for the same 100 nN, an unworn tip with a tip diameter 20 nm results in a nominal stress $\sigma$=*F*/*A*=318 MPa, a slightly worn tip in **Figure 3.10** (left, with a diameter~100 nm) would lead to $\sigma$=13 MPa, and a severely worn tip in **Figure 3.10** (right, with a tip diameter ~2 μm) would lead to $\sigma$=32 kPa. Such wide range of tip size makes true stress determination very difficult if not impossible.

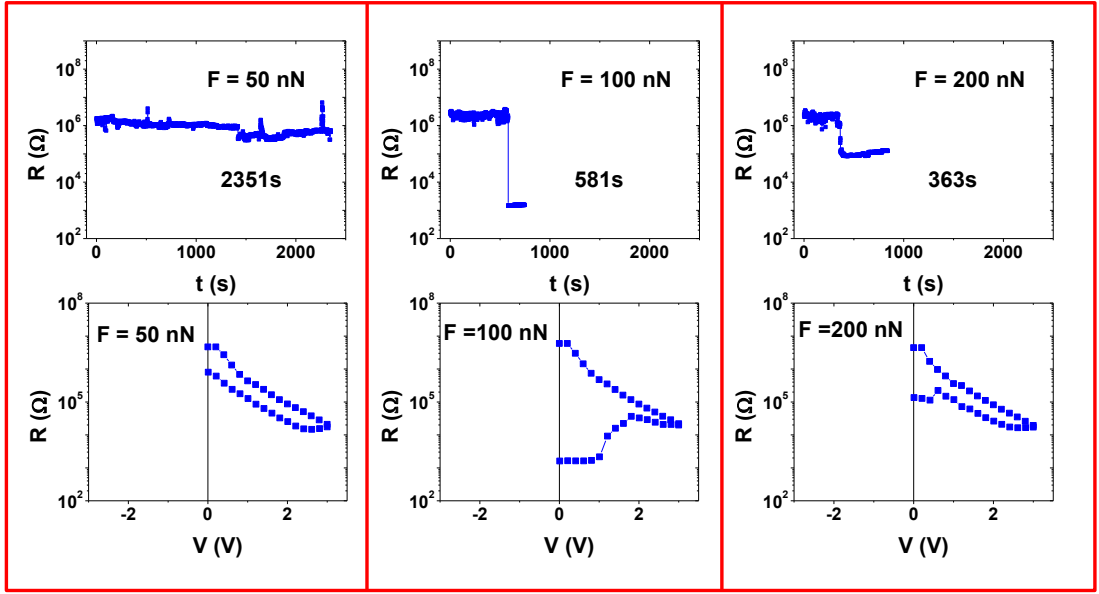

**Figure 3.9.** *In-situ* resistance under various AFM forces $F$ and post $R$-$V$ check. (Device: Mo/Si$_3$N$_4$:5%Pt/Pt, $\delta$=2 nm, size: 2×2 μm$^2$).

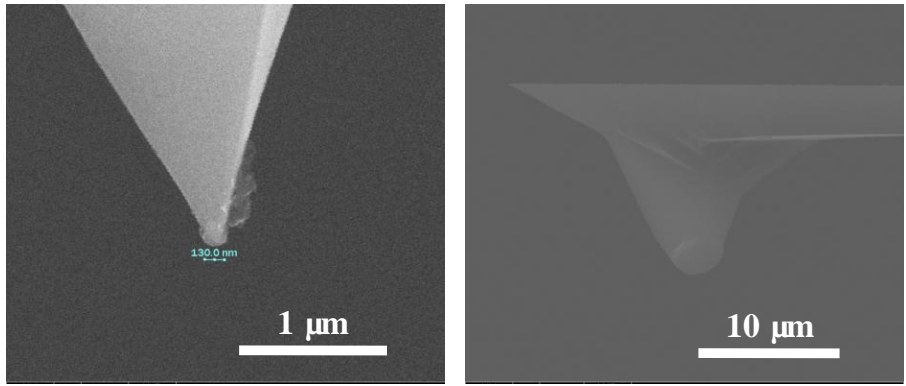

**Figure 3.10.** Two randomly picked AFM tips, after testing.

The stress effect on the $R$-$V$ curve and switching voltages is also understandable. As shown in **Chapter VII**, switching voltage $V_{on\rightarrow off}$ is positively correlated with the maximum negative voltage used during the off→on switching, but $V_{off\rightarrow on}$ is history independent. Since stress facilitates electron-detrapping process, a higher stress allows more electrons detrapped thus reaching a more conductive state, which is physically equivalent to applying a more negative voltage during the HRS→LRS transition. Therefore, a larger $V_{on \to off}$ is expected. On the other hand, since $V_{off \to on}$ is history independent, it should not be affected by stressed-induced detrapping. Therefore, $V_{off \to on}$ is the same under different stresses.

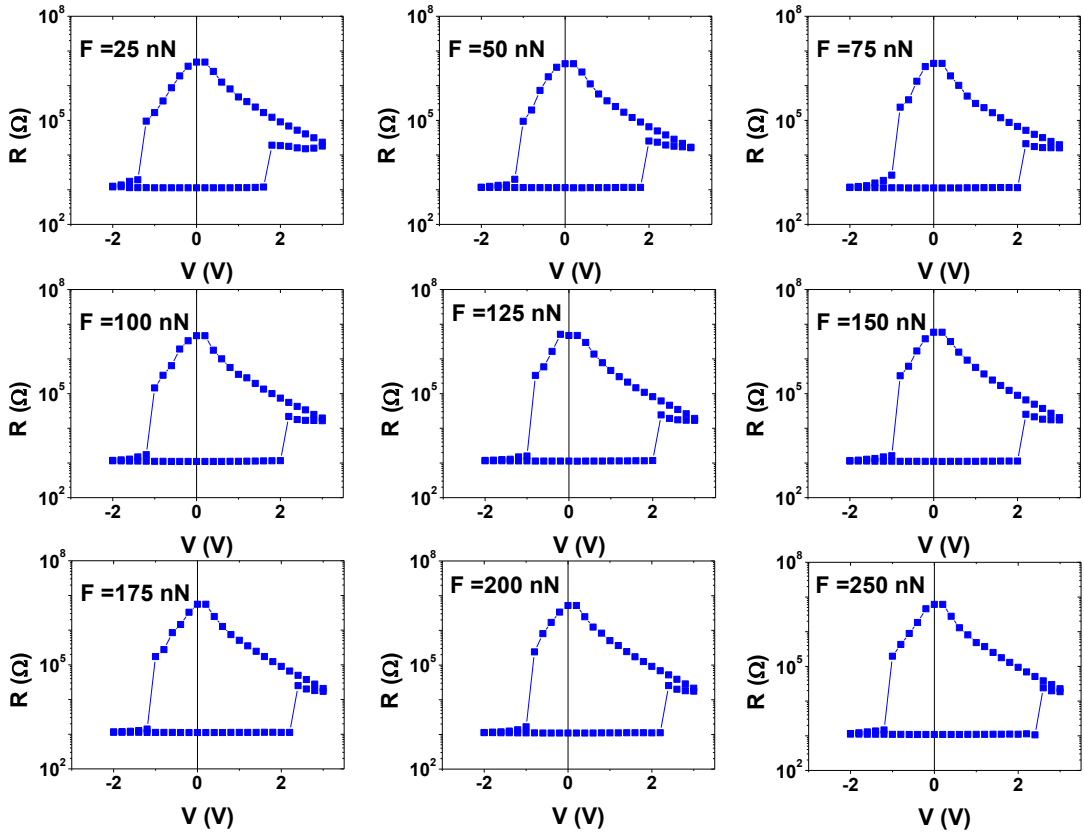

**Figure 3.11.** *R-V* curves for switching cycle under different AFM forces *F*. (Device: Mo/Si$_3$N$_4$:5%Pt/Pt, $\delta$=2 nm, size: 2×2 µm$^2$).

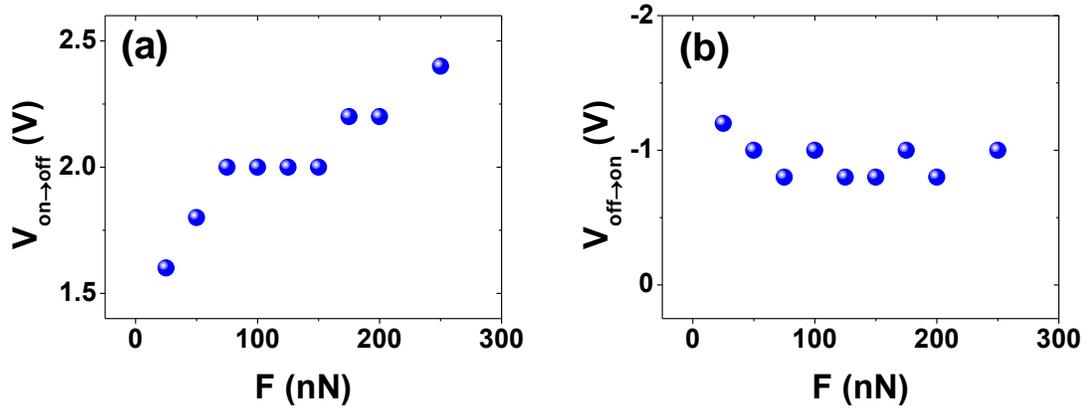

**Figure 3.12.** (a) $V_{on \to off}$ *vs.* $F$ and (b) $V_{off \to on}$ *vs.* $F$ in **Figure 3.11**. (Device: Mo/Si$_3$N$_4$:5%Pt/Pt, $\delta$=2 nm, size: 2×2 μm$^2$).

### 3.4  Bombardment Stress Induced MIT at $10^{-6}$ s

A FIB system, which provides energetic Ga$^+$ ions (30 keV) to bombard the device surface (**Figure 3.13**), was employed to provide a compressive force on the sample. As the momentum of Ga$^+$ ions is transferred to the sample, an impact is made according to $\Delta p / \Delta t$. A beam with a Gaussian size of ~1 nm scanning a ~100 μm$^2$ area during a scanning time ~10 s spends ~0.1 μs at every point. Therefore, this offers a stress testing method on the order of 1 microsecond.

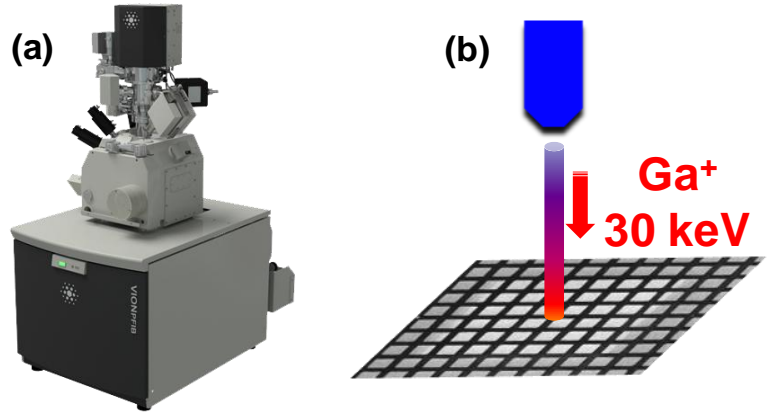

**Figure 3.13.** (a) Focus ion beam (FIB) system. (b) Schematic of Ga$^+$ bombardment inside FIB.

### 3.4.1 Results

As shown in **Figure 3.14**, device exhibits visible physical change after Ga$^+$ ion bombardment over the entire scanned region (15×12 μm$^2$). This change increased with current and was apparent even at the smallest current used. However, the electrodes were still intact and thus allowed post electrical characterization. Resistance measured *ex-situ* (**Figure 3.15**) shows a drop beginning with 50 pA bombardment; 300 pA bombardment caused a drop comparable to that of a typical voltage-triggered HRS→LRS transition. Similar to the previous cases, beam-induced resistance transition was non-transient, non-damaging as shown in **Figure 3.16**. In addition, such transition is one-way only: LRS devices treated identically did not switch. This confirms that the HRS is a metastable state.

A Ga$^+$ carrying energy $E$ with an effective mass $m$ has a velocity $v = \sqrt{2E/m}$. Since the momentum is entirely transferred to the sample when Ga$^+$ stops there, the momentum transfer is $mv = \sqrt{2mE}$ (if Ga$^+$ bounces, the momentum transfer is even larger). The number of Ga$^+$ ions per unit time, delivered by a beam current $I$, is $N = I/e$. Therefore, the total force as a result of momentum transfer per unit time is: $F = N \times mv$. Furthermore, considering a beam size ~1 nm or area $A$ ~1 nm$^2$, the normal stress or pressure over the beam cross section is:

$$\sigma = \frac{F}{A} = \frac{I\sqrt{2mE}}{eA}$$

Using $m$=1.2×10$^{-25}$ kg, $\sqrt{2mE}$=3.4×10$^{-20}$ kg·m/s, $e$=1.6×10$^{-19}$C, $A$=1 nm$^2$, we get $\sigma$=21.3 MPa for $I$=100 pA, or $\sigma$=213 MPa for $I$=1 nA. With these estimate, we find the results in **Figure 3.15** sets a threshold stress (at 50 pA) of ~10 MPa, which is even lower than the already low pressure used earlier in the hydrostatic pressure experiment.

Lastly, since any Ga$^+$ accumulation on the top electrode would produce a *positive* bias whereas HRS→LRS switching under the stress-free condition requires a *negative* bias, we can safely exclude the possibility of charge accumulation induced switching. To further corroborate this, parallel experiments were performed with 30 keV electrons, which would produce an opposite charge accumulation. None of the devices experienced any resistance change, and all remained switchable after bombardment (**Figure 3.17**). The lack of response was expected because the impact pressure of electrons (~100 kPa) was a factor of $(m_{Ga}/m_e)^{1/2}$, or 356×, significantly smaller than that of Ga$^+$.

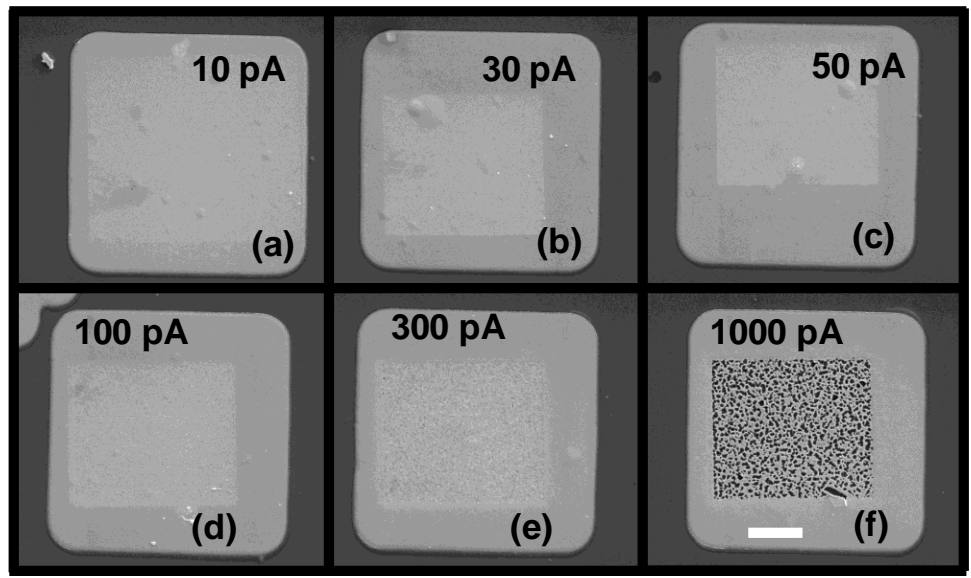

**Figure 3.14.** SEM images of 20×20 μm² devices after 20 s bombardment of Ga⁺ ions at various beam current (a) 10 pA, (b) 30 pA, (c) 50 pA, (d) 100 pA, (e) 300 pA, (f) 1000 pA. Bombarded regions (15×12 μm²) show a brighter contrast than the remainder of the square-shaped electrodes. Scale bar: 5 μm. (Device: Mo/Si$_3$N$_4$:5%Pt/Pt, $\delta$=10 nm).

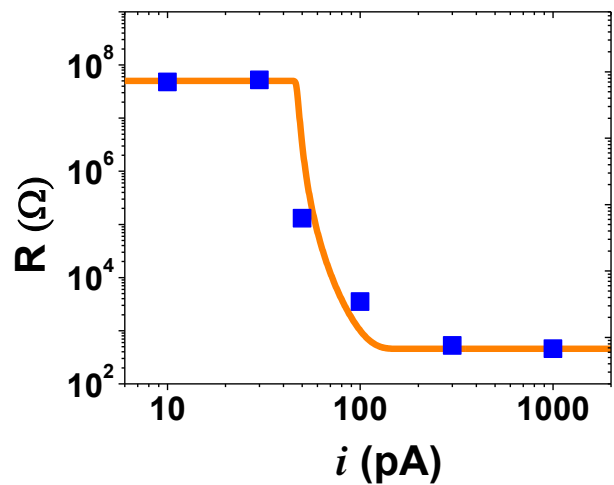

**Figure 3.15.** HRS resistance after 20 s bombardment by 30 keV Ga$^+$ ions of various currents. Switching to intermediate states starts at >50 pA, becoming nearly complete at 300 pA. (Device: Mo/Si$_3$N$_4$:5%Pt/Pt, $\delta$=10 nm, size: 20×20 μm$^2$).

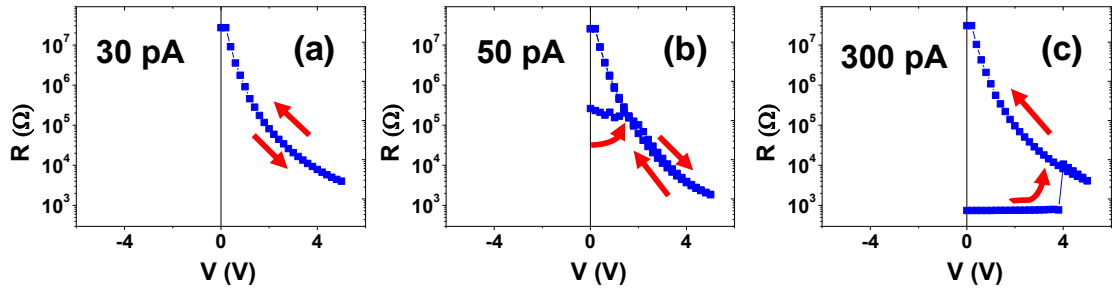

**Figure 3.16.** Device *R-V* curves after bombardment started at (a) HRS after 30 pA bombardment; (b) intermediate state after 50 pA bombardment, which switched to HRS at ~+1 V; (c) LRS after 300 pA bombardment, which switched to HRS at ~+4 V. (Device: Mo/Si$_3$N$_4$:5%Pt/Pt, $\delta$=10 nm, size: 20×20 μm$^2$).

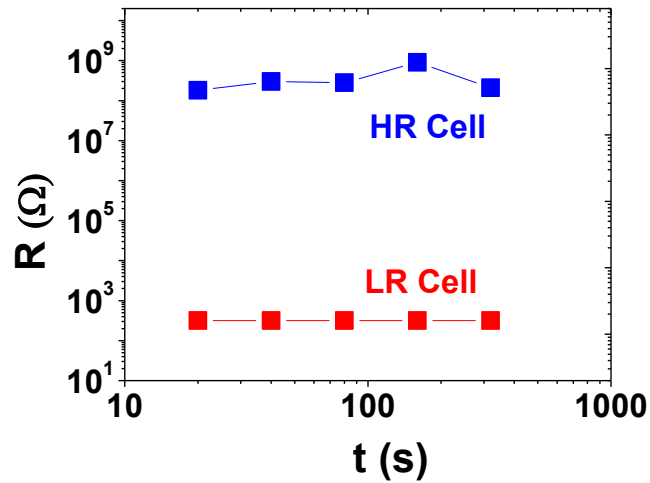

**Figure 3.17.** Resistance after bombardment by 30 keV electrons of various duration. No resistance change is observed. (Device: Mo/Si$_3$N$_4$:5%Pt/Pt, $\delta$=10 nm, size: 20×20 $\mu$m$^2$).

### 3.4.2 Discussion

Comparing with isotropic pressure, FIB requires a lower stress to trigger resistive switching. This is reasonable because bond distortion requires a shear stress, which cannot be provided by a pressure in an elastically homogeneous continuum (our amorphous films are not elastically homogeneous), but can be efficiently generated by a uniaxial compression/tension.

The gradual resistance transition in **Figure 3.15** may be rationalized using the concept of a "parallel circuit" in **Chapter VII**. The total resistance $R$ of the device of a parallel connection of an HRS component and an LRS component depends on the weights of the components, which change according to the stress once the stress exceeds a threshold value $\sigma^*$. The weight of the HRS is initially unity, and it subsequently decreases with stress; the weight of the LRS is initially zero, and it subsequently increases with stress. One representation which may account for the observed transition is

$$\frac{1}{R} = \frac{exp\left[-\left(\frac{\sigma - \sigma^*}{\Delta}\right)^n_+\right]}{R_{HRS}} + \frac{1 - exp\left[-\left(\frac{\sigma - \sigma^*}{\Delta}\right)^n_+\right]}{R_{LRS}}$$

in which the numerators are the weights. Here $(\ldots)_+$ represents the positive part of $(\ldots)$: when it negative, it is set at zero; the exponent $n$ prescribes the steepness of the transition and $\Delta$ is a scaling factor for stress. In the Ga$^+$ ion bombardment, $\sigma = Ai$, where $A$ is a proportionality constant (see **Section 3.4.1**). So we can rewrite the expression in terms of current $i$ (in pA). Therefore, we have

$$\frac{1}{R} = \frac{exp\left[-\left(\frac{Ai - \sigma^*}{\Delta}\right)_+^n\right]}{R_{HRS}} + \frac{1 - exp\left[-\left(\frac{Ai - \sigma^*}{\Delta}\right)_+^n\right]}{R_{LRS}}$$

$$= \frac{exp\left[-\left(C\left(\frac{i}{i_0} - 1\right)\right)_+^n\right]}{R_{HRS}} + \frac{1 - exp\left[-\left(C\left(\frac{i}{i_0} - 1\right)\right)_+^n\right]}{R_{LRS}}$$

This equation gives a reasonable fit to the data, as shown by the solid curve in **Figure 3.15** with the following fitting parameters:

$$\frac{1}{R(\Omega)} = \frac{1}{50\text{M}\Omega} \times exp\left[-\left(0.71 \times \left(\frac{i(pA)}{45pA} - 1\right)\right)_+^{3.5}\right] + \frac{1}{456\Omega}$$

$$\times \left(1 - exp\left[-\left(0.71 \times \left(\frac{i(pA)}{45pA} - 1\right)\right)_+^{3.5}\right]\right)$$

Later, we will revisit this model and check its consistency with the electron-bunch experiment.

To verify whether bombardment ions penetrate through top electrode and reach nanometallic film, we performed the range calculation for 30 keV Ga$^+$ in Pt using SRIM software ("the Stopping and Range of Ions in Matter", http://www.srim.org/). The results show the projected range is 6.9 nm with 6.3 nm longitudinal straggling (**Figure 3.18**) and 4.9 nm lateral straggling. Therefore, the farthest the ion will go is 13.2 nm, which is much less than our top electrode thickness of 40 nm. We can thus safely conclude that momentum transfer to the Pt top electrode is complete, hence the force on the device.

Internal dissipation, including atomic rearrangement, does not affect this calculation: momentum conservation will ensure that eventually all momentum will be transferred to the Pt electrode.

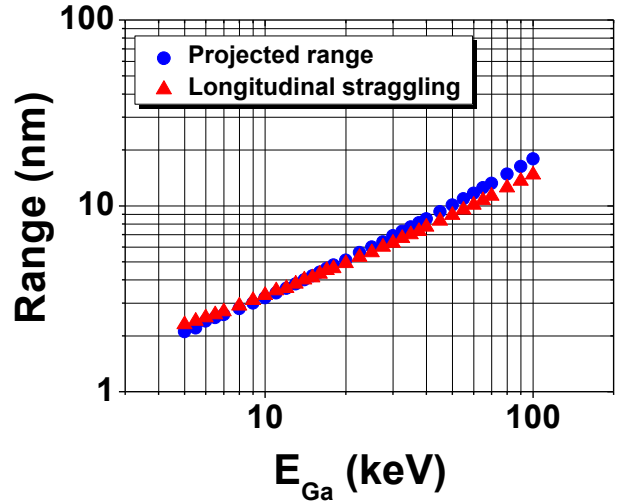

**Figure 3.18.** Projected Range and Longitudinal Straggling. Ga⁺ (5 keV to 100 keV) in platinum calculated by SRIM software ("the Stopping and Range of Ions in Matter", http://www.srim.org/). A 30 keV Ga⁺ is fully stopped by a 40 nm Pt top electrode, enabling complete momentum transfer.

To further support the above calculation, we refer to **Table 1.1** in the book _Focused Ion Beam Systems: Basics and Applications_ (Nan Yao, Cambridge University Press 2007), which lists the penetration depth of 30 keV Ga⁺ in Fe to be 20 nm. Since Pt has a higher density than Fe, the penetration depth in Pt should be much less than 20 nm. This is consistent with our calculation shown above.

## 3.5   Lorentz Stress Induced MIT at 10$^{-13}$ s

Having established the mechanical and electron-occupancy nature of the electron-phonon interaction by static and quasi-static perturbation experiments, we next address its dynamics by probing the effect of mechanical excitation in a faster time domain. The time window to form the negative-$U$ state commences after electron filling of the trap-site (happening in ~1 fs) and lasts for the duration of a strong electron-phonon interaction (about one period of atomic vibration, 0.1-1 ps)[6]. Therefore, if a sub-picosecond force can mechanically unravel $\phi_{ep}$, it will unequivocally confirm the operation of the mechanism. To provide such force, we designed a sub-picosecond experiment using the concept of the Lorentz force (magnetic pressure), which is a self-force acting on any circuit loop with a circulating current.

### 3.5.1   Lorentz Force for a Current Loop

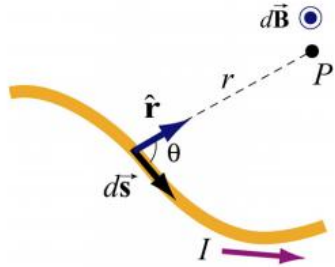

**Figure 3.19.** Magnetic field d$\vec{B}$ at point $P$ due to a current element $I$d$\vec{s}$ (ref.[10])

According to the Biot-Savart law, an electrical current $I$ flowing around a closed loop generates an induced magnetic field at the point $P$, given by:

$$\vec{B} = \oint d\vec{B} = \oint \frac{\mu}{4\pi} \frac{I d\vec{s} \times \hat{r}}{r^2}$$

Here $\mu$ is permeability ($\mu = \mu_0 = 4\pi \times 10^{-7}$ T·m/A for free space), and $r$ denotes the distance from the current source to $P$ (**Figure 3.19**)[10]. For a circular loop, the expression simplifies to: $\vec{B} = \frac{1}{2}\mu I \hat{l} \times \hat{r}$. The magnetic field in turn interacts with the current loop itself and exerts a Lorentz force on the loop, giving $\vec{f} = I\hat{l} \times \vec{B}$ per unit length, where $\hat{l}$ is the unit vector towards the current direction. Therefore, for a circular loop, the self force is

$$\vec{f} = \frac{1}{2}\mu I^2 \hat{r}$$

As illustrated in **Figure 3.20**, the self-force is always tensile (radially outwards along $\hat{r}$), irrespective of the current direction (clockwise or counterclockwise). This electromagnetic field induced mechanical force is actuated by an electrical/magnetic signal. Therefore, it can be extremely fast with the same duration of the electromagnetic excitation itself.

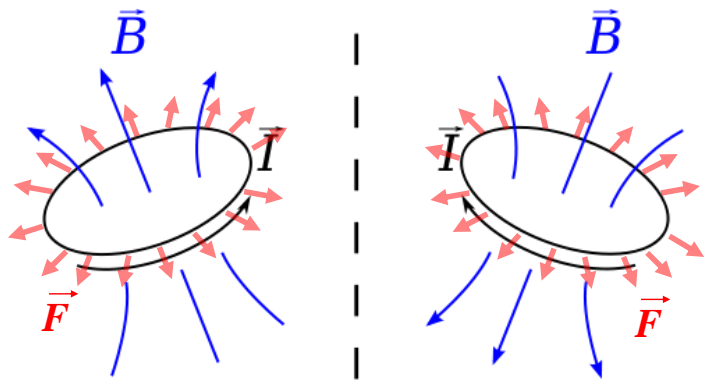

**Figure 3.20.** Lorentz force induced by a current loop. It is always tensile (radially outward) regardless of the current direction (cp. left and right).

### 3.5.2 Electron-bunch in Stanford Linear Accelerator Center (SLAC)

To induce an ultra-fast mechanical force (the Lorentz force) through an electromagnetic excitation, a short-pulse current is needed. Traditional circuit methods find it difficult to go faster than ~nanosecond because of the *LCR* delay of the circuit, including its wiring. Fortunately, this can be accomplished by use of the 3 km long Stanford linear accelerator (linac), which generates an ultra-relativistic electron beam of 20 GeV electrons with $2 \times 10^{10}$ electrons "bunched" into a short packet of 24 μm in the laboratory frame. The charged beam, *i.e.*, each electron bunch, travels at the speed of light. Therefore, such short bunch length translates into a 80 femtosecond (fs) pulse with a peak current of $3 \times 10^4$A. In fact, the manipulation of the characteristics of electron bunches (shape, size, length, *etc.*) has become a science in itself (ref.[11]).

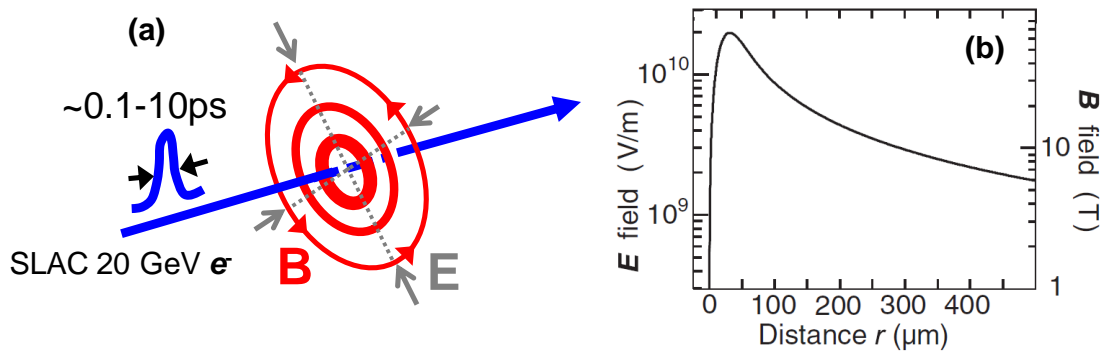

**Figure 3.21.** (a) Schematic of a single shot of relativistic electron bunch. The *E* and *B* fields are perpendicular to the beam. (b) Plot of maximum *E*-

and **B**-field amplitude *vs.* distance *r* from the beam center (adapted from ref.[12]).

A relativistic beam carrying a large charge resembles an electrical current passing through space, it thus induces a large electrical/magnetic field nearby. A simple analogy to an infinite long current results in $E{\sim}cB{\sim}1/r$ distribution, which is exactly the asymptote expression for the far field at $r{\to}\infty$. The field near the bunch needs explicit consideration of the spatial distribution of the electron package ($\sigma_r{\sim}20$ μm). This eventually leads to the following form[11]:

$$E(r,t) = \frac{Q}{r(2\pi)^{3/2}\varepsilon_0 c\tau}\left(1 - \exp\left[-\frac{r^2}{2\sigma_r^2}\right]\right)\exp\left[-\frac{t^2}{2\tau^2}\right]$$

$$B(r,t) = \frac{\mu_0 Q}{r(2\pi)^{3/2}\tau}\left(1 - \exp\left[-\frac{r^2}{2\sigma_r^2}\right]\right)\exp\left[-\frac{t^2}{2\tau^2}\right]$$

Where $Q=N_e e$ is the total charge of the bunch and $\tau$ is the bunch duration (~0.1 ps-1 ps). This relation is plotted in **Figure 3.21b**. Both **E**- and **B**-fields initially rise up to the edge of bunch, then decay as $1/r$. At the beam center $r=0$ where electrons are most concentrated, **E**- and **B**-fields are zero which is also expected from the symmetry point of view. This property will prove critical to differentiating different effects of the electron bunch in later sections.

In the following, we will describe a series of experiments using the following bunch durations: 0.1 ps or 1 ps. This time $\tau$ is the Gaussian sigma time of the electron bunch, which corresponds to a FWHM of 2.35 $\tau$, which may be taken as the time of the half-cycle field. The corresponding spatial extent of the field is 2.35 c$\tau$, where c is the speed of light. (The corresponding period is 4.7 τ, which may be used for estimating the Fourier components or wavelength when needed.) Using the above relation, the duration of the following experiments are 0.235 ps or 2.35 ps, and the length scales are 71 μm or 710 μm. (For convenience, the duration time in this chapter refers to Gaussian sigma time τ instead of FWHM).

### 3.5.3  Experimental Design

To harness the fast half-cycle electromagnetic field of the electron bunch and convert it into a Lorentz force, we invented the following method which conveniently utilizes the MIM structure already provided by RRAM. As shown in **Figure 3.22**, there is a current/circuit loop in our standard metal-insulator-metal (MIM) structure shown in **Figure 3.22**, which may be used to induce a force. When an electron bunch is normally incident on the top electrode, a transient current/electrical-field in the circuit loop is induced from the transient magnetic field. Meanwhile, the direct electrical field can be easily shielded inside the top electrode as long as the electrode thickness exceeds the skin depth (~1 nm). This current flow causes a repulsive Lorentz force between the electrodes. Therefore, our standard MIM device structure is immediately suitable for ultra-fast mechanical test.

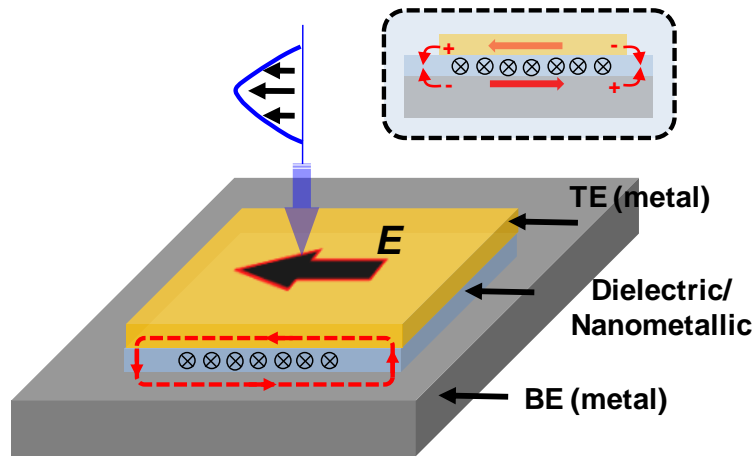

**Figure 3.22.** MIM structure drawn with a circuit loop of TE → insulator-gap → BE → insulator-gap → TE.

Other considerations are important to optimize the test condition. First, electrons in an e-bunch are laterally distributed within $\sigma_r \sim 20$ μm, so the device size need to be small enough to capture the details of the field distribution inside the bunch; Second, both **E**- and **B**-fields decay rapidly from $10^{10}$ V/m to $10^9$ V/m within ~500 μm (**Figure 3.21b**), so individual devices need to be close enough for sufficient sampling of the field within the above range; Third, the device size need to be large enough for easy pre- and post-electrical characterization (probe-tip contact typically requires >10 μm); Fourth, it will become clear that there is a "resonance" effect at play in the field-electrode interaction, yielding a peak coupling when the device size is commensurate with the effective wavelength of the field. The latter is about the bunch size, *i.e.*, ~30 μm. Therefore, devices of about this size are expected to experience the maximum Lorentz force. Based on these considerations, the array was designed to consist of 20×20 μm² squares (TE) with 10 μm spacing. **Figure 3.23** shows the schematic of fields carried by e-bunch and a device array of 20×20 μm² squares.

In the following, the main observations of the electron bunch experiments are first summarized, followed by a systematic parametric study to interrogate the mechanism for the observed effect. This is followed by evaluation of radiation and collisional damage of electrons and simulation of the field/MIM interactions.

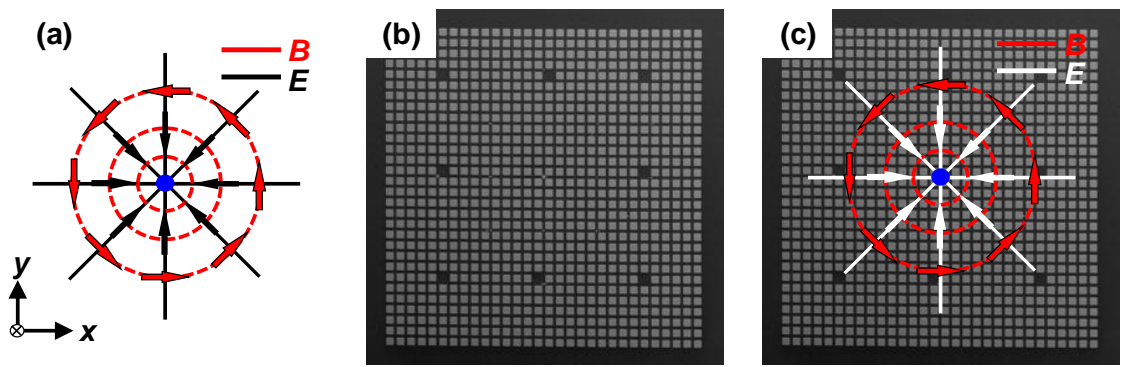

**Figure 3.23.** (a) **E**- and **B**-fields from electron bunch. (b) 20×20 μm² device array. (c) Overlap (a) & (b).

### 3.5.4 Physical Appearance After Electron-bunch Shot

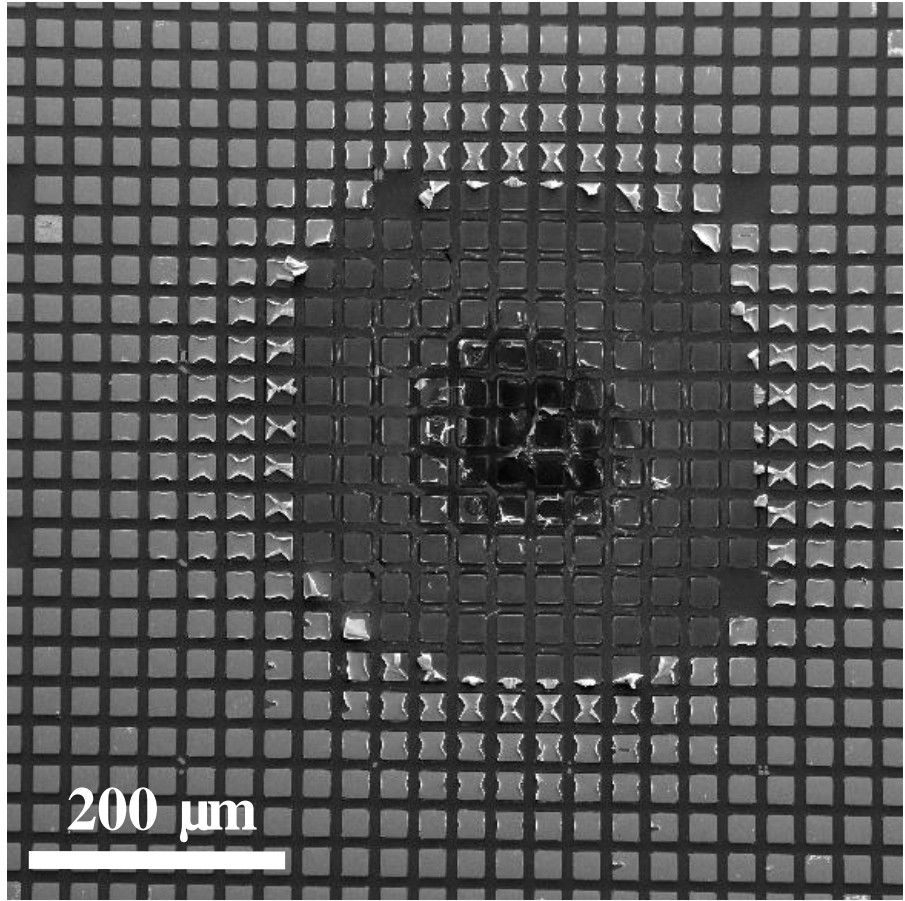

**Figure 3.24.** SEM image of square shaped device. White features are distorted remnant electrodes, between them are gray regions where top electrodes are completely torn away. Top electrode thickness: 40 nm Pt. Beam condition: 3 nC, 0.1 ps.

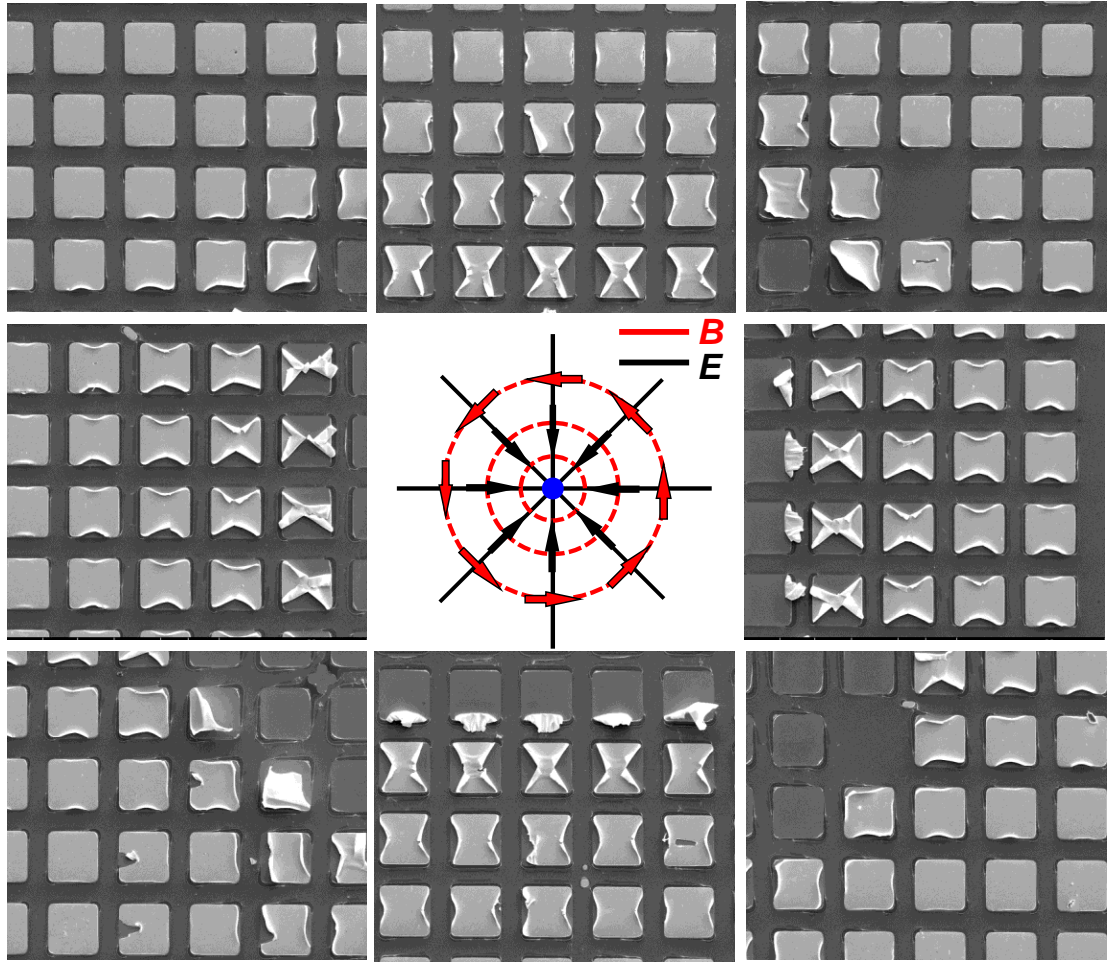

**Figure 3.25.** SEM higher resolution image of device in **Figure 3.24**. The damaged top electrodes show bow-tie shape symmetrically distributed around the bunch center. Top electrode thickness: 40 nm Pt. Beam condition: 3 nC, 0.1 ps.

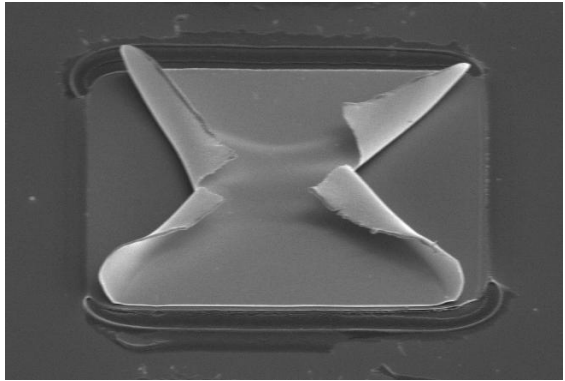

**Figure 3.26.** SEM higher resolution image of device (tilted by 50 degree) in **Figure 3.25**. Top electrode is partially torn off, showing bow-tie shape with two clamped edges that are perpendicular to the radial direction from the electron bunch. The torn free edges are along ***B***-field direction. Top electrode thickness: 40 nm Pt. Beam condition: 3 nC, 0.1 ps.

After exposed to a single shot of an electron bunch, the device array was severely damaged at and around the bombarded site (**Figure 3.24**) within a radius of 180 μm. The damage was most severe within the bunch radius ($r \sim \sigma_r$), showing "scorching" and "trench" formation around the edges of the top electrodes. Outside the bunch center, top electrodes were also completely torn off within $r \sim 200$ μm. Further away from the bunch center, however, the top electrodes remained intact. Interestingly, the transition region, near $r \sim 200$ μm, showed directional tearing which has central symmetry around the bunch center (**Figure 3.25**). This transition region carried key information regarding the nature of the mechanical forces. A closer look at the device (**Figure 3.26**) reveals a beautiful pattern: the top electrode (40 nm Pt film) was partially peeled off from the underneath dielectric film and formed a "bow-tie" shape with two edges (up & down) firmly

"clamped" to the dielectric layer while the other two edges (left & right) detached. On the clamped edges, a deep trench developed along each edge. Directional top electrode tearing and trenching clearly reveals a strongly mechanical nature of the e-bunch effect on the MIM structure. Meanwhile, trenching—most likely a result of ionization and arcing, provides evidence of electrical field concentrations.

### 3.5.5   EDX Study of Physical Damage

To further identify the nature of damage and understand the electrode detachment mechanism, a detailed EDX study was performed. The results can be summarized as follows:

1.  The "blue" region outside the cells in **Figure 3.27** shows the characteristic peaks from O ($E_{O, K\alpha}$=0.525 keV), Si ($E_{Si, K\alpha}$=1.739 keV), and Mo ($E_{Mo, L\alpha}$=2.293 keV). This signal is interpreted as coming from the $Si/SiO_2/Mo/Si_3N_4$:Pt stack, in which the nanometallic film ($Si_3N_4$:Pt) is only 10 nm thick, thus not contributing to the signal.

2.  The "red" region in **Figure 3.27** on an intact portion of the top (Pt) electrode shows a strong Pt peak ($E_{Pt, M}$=2.048 keV) while other peaks are suppressed. This is expected because Pt electrode can shield the signal from beneath. The "yellow" region in **Figure 3.27** shows only O, Si and Mo peaks from the $Si/SiO_2/Mo/Si_3N_4$:Pt stack, very similar to the spectrum obtained in the "blue" region. This indicates that Pt electrode is completely torn off.

3.  Next to the "clamped" edge, there is a trench (the "green" region in **Figure 3.27**) which shows only Si and O peaks but no Mo peak ($E_{Mo, L\alpha}$=2.293 keV). This indicates the Mo bottom electrode was also missing. However, as the distance to the path of the e-bunch increases, Mo peak is gradually recovered (the "purple" region in **Figure 3.27**), indicating the Mo layer survived if the device is far away from the bombardment center, even though some top electrode tearing is still evident.

4. Near the path of the e-bunch, devices were severely damaged and the Pt top electrode was missing. This is evident from all the spectra collected in various regions shown in **Figure 3.28**. The remnant has a layer-like appearance, although the layering is due to the Mo bottom electrode, which is initially a continuous film beneath both the cell regions and the rim (between cells) region. For example, in the cell region, closer to the e-bunch the "blue" spot has no Mo signal, but further away at the "red" spot some Mo signal appears. Likewise, in the rim region, there is a varying amount of Mo remaining, more at the "purple" spot than the "yellow" spot as indicated by their spectra. Interestingly, in the trench region (the "green" region) between the above two regions, Mo is completely gone. (The trench here is very wide.) Since this is the same edge that would have been clamped had the top electrode remained, this suggests the very severe local damage is related to the electrical field which is expected to be concentrated here.

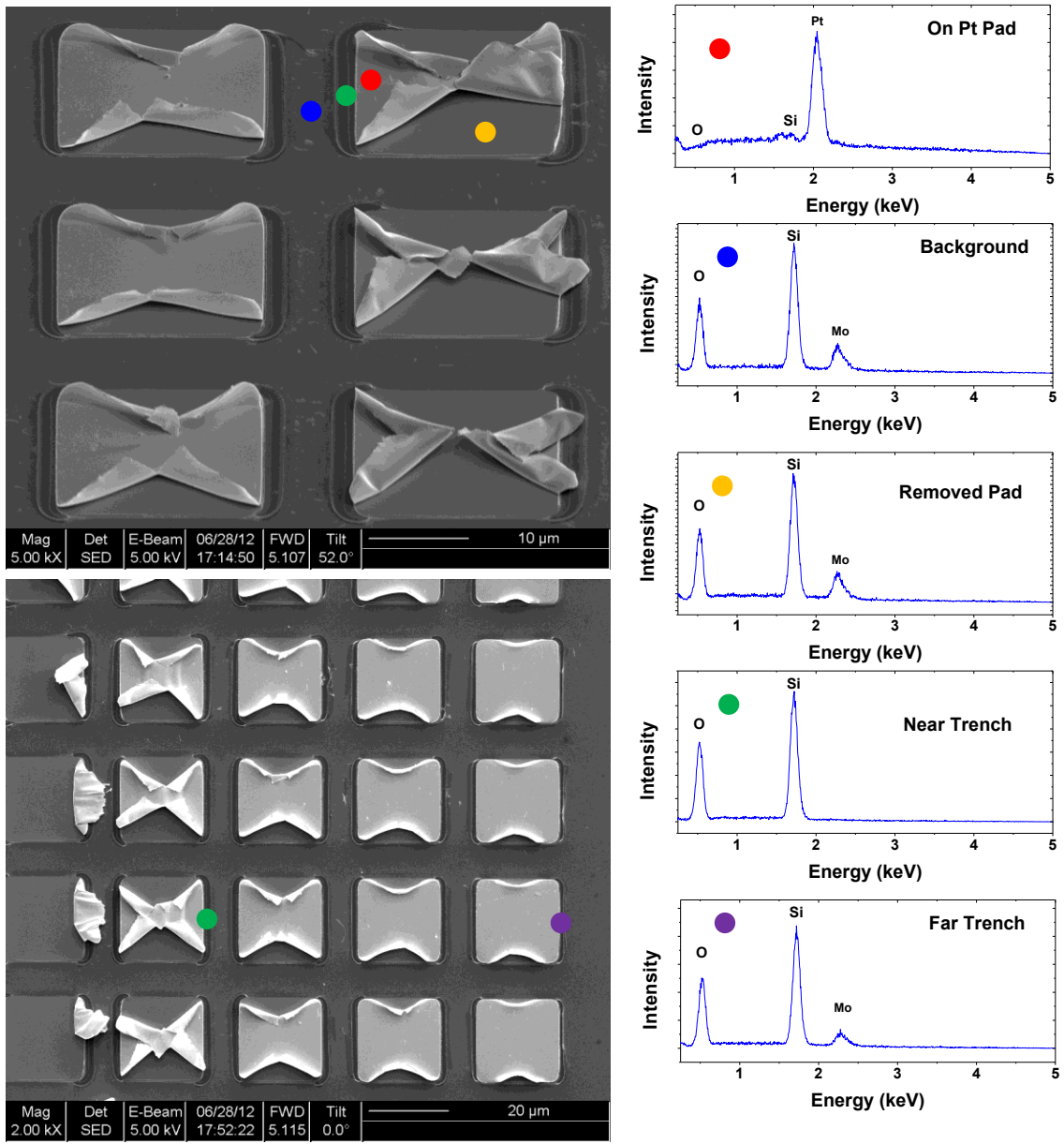

**Figure 3.27.** SEM image and EDX analysis at different regions outside e-bunch. (Device: Mo/Si$_3$N$_4$:5%Pt/Pt, $\delta$=10 nm, top electrode: 40 nm Pt. bottom electrode: 20 nm Mo. Size: 20×20 μm$^2$. Beam condition: 3 nC, 0.1 ps).

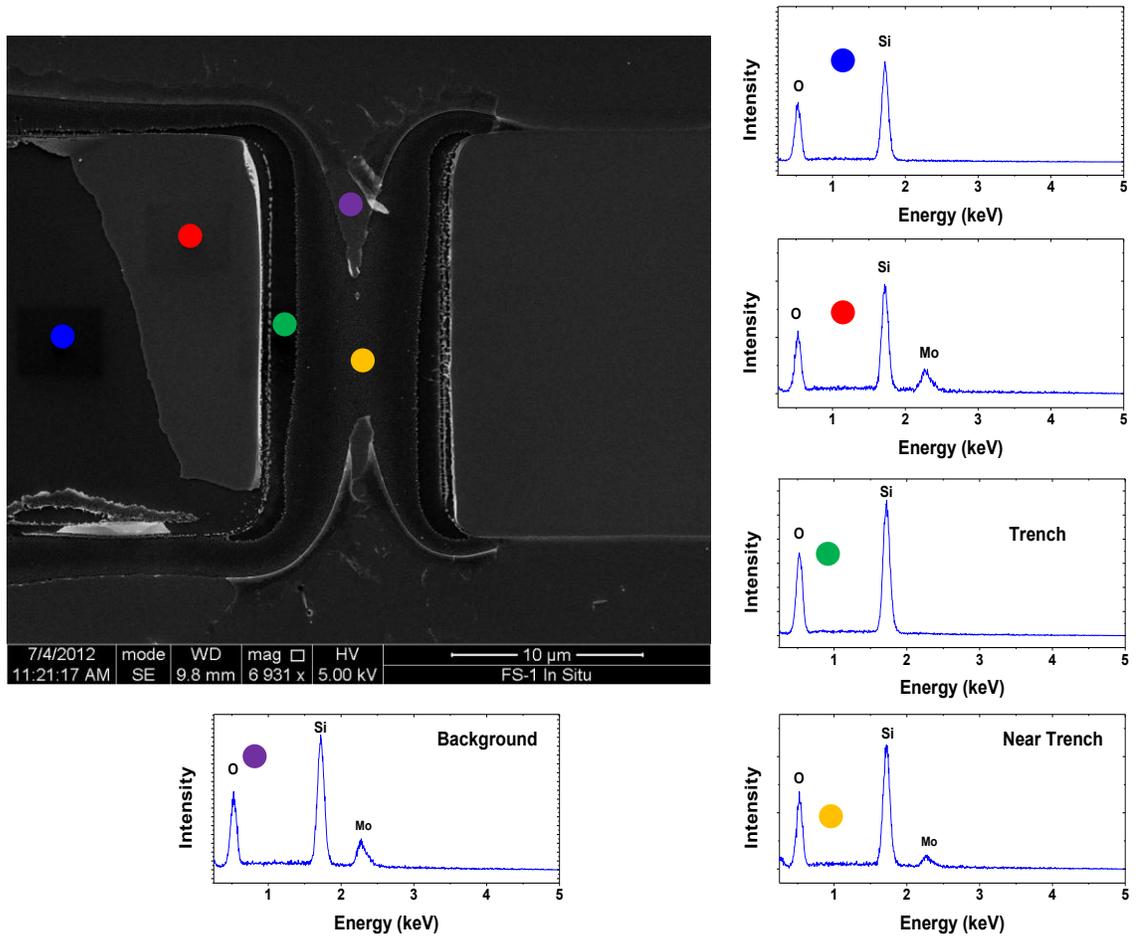

**Figure 3.28.** SEM image and EDX analysis at different regions near e-bunch, which landed on the left in the SEM image. (Device: Mo/Si$_3$N$_4$:5%Pt/Pt, $\delta$=10 nm, top electrode: 40 nm Pt, bottom electrode: 20 nm Mo. Size: 20×20 μm$^2$. Beam condition: 3 nC, 0.1 ps)

### 3.5.6 Resistance Change After Electron-bunch Shot

With a repulsive Lorentz force between electrodes, the nanometallic film experienced a (normal) tensile stress, which induced resistive switching. **Figure 3.29** shows resistance-state distribution for a circular shaped device array made of Mo (20 nm) / Si$_3$N$_4$:Pt (10 nm) / Pt (100 nm) after one single shot of electron bunch bombardment. Before the experiment, all cells were preset to the HRS. In this sample, the top electrode is thicker, which proved to be effective for suppressing tearing. This indicates that a thicker Pt film has a higher bending rigidity thus more resistance to tearing, which is convenient for electrical study. Different electrical states were found in different regions. 1. Very near the center, which suffered the highest direct mechanical impact of electrons, there were some undamaged, un-switched HRS devices (inside the small white dotted circle.) 2. Within $r{\sim}150$ μm from the center, all electrodes were torn away, so no electrical testing was performed in this region. 3. Between $r{\sim}150$ μm and $r{\sim}400$ μm, the preset HRS devices had mostly transitioned to the LRS. 4. Outside $r{\sim}400$ μm, the HRS devices were unchanged. Within the switching zone (150 μm$<r<$400 μm), electron-bunch induced switching was non-transient and non-damaging: all the switched devices—even those bow-tie ones—could be electrically switched back and forth.

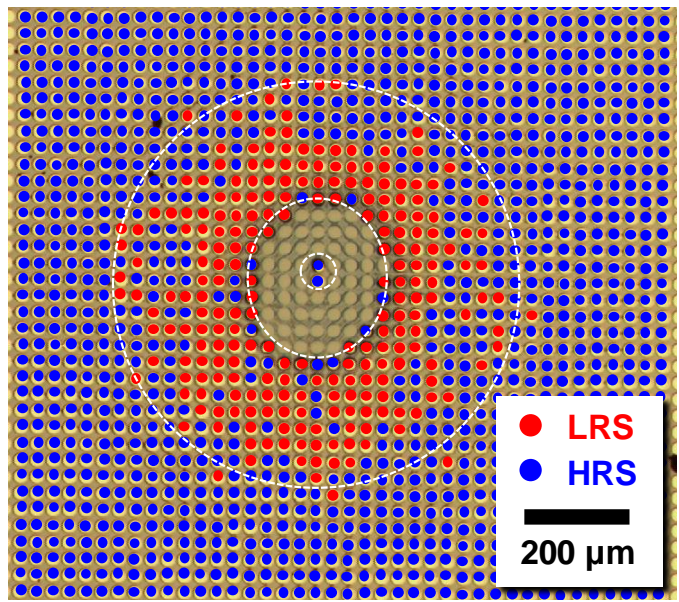

**Figure 3.29.** Optical image of circular (20 µm) devices with thicker (100 nm) TE preset to HRS. Red and blue colors were added to the cells to indicate different states after experiment, blue: HRS; red: LRS. Yellow cells near the center have lost top electrodes. Drawn circles going outward indicate four zones of different device states, starting from (no damage & HRS)→ (top electrode torn away)→ LRS→ HRS. (Device: Mo/Si$_3$N$_4$:5%Pt/Pt, $\delta$=10 nm, top electrode: 40 nm Pt, bottom electrode: 20 nm Mo. Size: $d$=20 µm. Beam condition: 3 nC, 0.1 ps)

### 3.5.7  A Parametric Study

#### 3.5.7.1  MIM *vs.* MI structure: "circuit loop"

We first verify that the MIM structure, which allows a transient current to form, is essential for stress generation. This was checked by removing the bottom electrode and replacing the highly doped conducting Si substrate with a lightly doped insulating Si substrate. A much smaller damage profile was observed (**Figure 3.30**): $r$~200 µm for the MIM structure and $r$~100 µm for the latter, MI structure. Apparently, without a bottom electrode to provide a sufficient current, the Lorentz force becomes much weaker for tearing. However, since even an insulator would behave like a conductor in fast transients when the impedance $Z$~$1/j\omega C$ is low, an insulator "counter electrode" does conduct to a certain degree, thus creating some tearing at very high field (near the center).

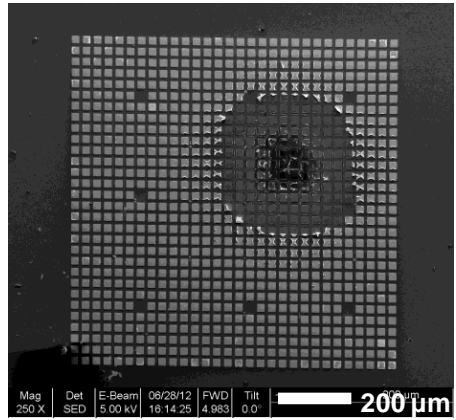
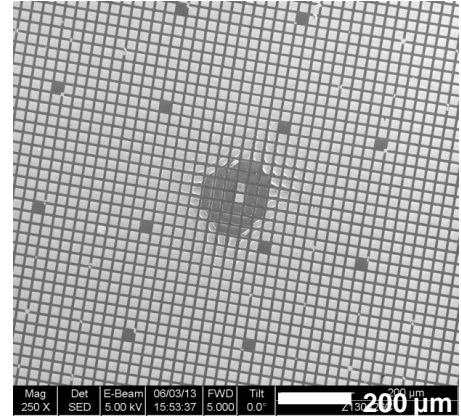

**Highly doped Si sub**  **Lightly doped Si sub**

**Figure 3.30.** Decreased damage when image charge is suppressed. Damage patterns in samples made on (**Left**) highly doped Si (conducting) substrate, with a bottom electrode, and (**Right**) lightly doped Si (insulating) substrate, without a bottom electrode. The latter sample has little image charge and shows much less tearing of top electrode. (Device: Mo/Si$_3$N$_4$:5%Pt/Pt, $\delta$=10 nm, top electrode: 40 nm Pt. Size: 20×20 μm$^2$. Beam condition: 3 nC, 0.1 ps)

#### 3.5.7.2  Charge dependence

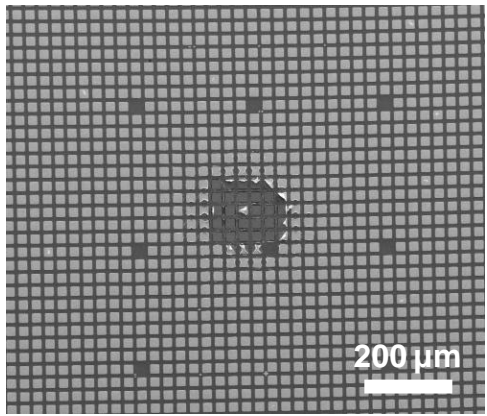
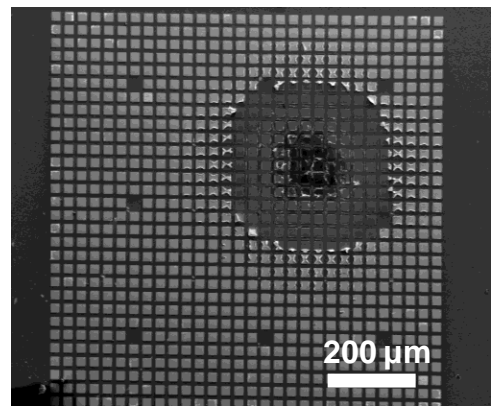

**Low charge 10$^{10}$ $e$**  **High charge 2×10$^{10}$ $e$**

**Figure 3.31.** Two identical array exposed under 20 GeV electron bunch. One used $10^{10}$ electrons (left), the other used $2 \times 10^{10}$ electrons (right). (Device: Mo/Si$_3$N$_4$:5%Pt/Pt, $\delta$=10 nm, top electrode: 40 nm Pt, bottom electrode: 20 nm Mo. Size: 20×20 µm$^2$. Beam condition: 0.1 ps)

The incident $E$ and $B$ fields scale with $Q/r$ at large $r$ ($Q$ = total electron charge of the electron-bunch, $r$ = radial distance from the bunch) and the same relation holds for the induced fields since the Maxwell equations are linear. Therefore, any field effect is $Q$ dependent. This was verified in **Figure 3.31**, where the size of the damage zone is highly dependent on the charge ($r_{damage}$~100 µm for the low charge case ($10^{10}$ electrons) *vs.* $r_{damage}$~200 µm for the high charge case ($2 \times 10^{10}$ electrons)). Since the damage contour may be described as:

$$f(E, B) = f\left(\frac{Q}{r}\right) = const.$$

it follows that $Q/r$ is constant at the same degree of damage/effect (*e.g.*, the border of the damage zone.) Therefore, the radius of the damage zone should be proportional to the charge $Q$, which is consistent with the above observation.

### 3.5.7.3  Pulse duration dependence

Since the electromagnetic field, especially the $B$ field, arises from the current of the electron bunch, vary the duration of the pulse has the same effect as varying the current. **Figure 3.32** shows two identical samples each bombarded by one shot of a 20 GeV electron bunch of the same total charge ($10^{10}$ electrons) but different pulse durations (the short pulse: 0.1 ps *vs.* the long pulse: 1 ps). The shorter pulse caused damage as before, while the longer one did not. Because identical collision/ energy-transfer/ momentum-transfer was created by the electrons in these two experiments, they clearly demonstrated that the damage was caused by the **E**- or **B**-field rather than the direct energy/momentum transfer. It is the time/transient/rise-time effect, namely the one associated with the induced magnetic field that arises from the Faraday law, that is the cause of the phenomena probed in our study. This magnetic field transient is harnessed by our MIM cell, which serves as a patch antenna (see below).

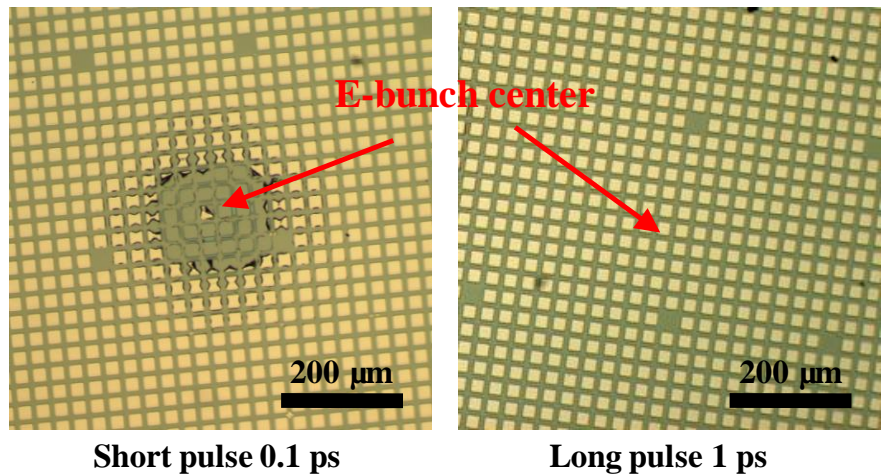

**Figure 3.32.** Two identical samples exposed to 20 GeV electron bunch. One used 0.1 ps short pulse (left). The other used 1 ps long pulse (right). (In these two samples, the damage region is smaller than previous cases because charge carried by electron bunch is $10^{10}$ electrons, half of the standard case.) (Device: Mo/Si$_3$N$_4$:5%Pt/Pt, $\delta$=10 nm, top electrode: 40 nm Pt, bottom electrode: 20 nm Mo. Size: 20×20 μm$^2$.)

#### 3.5.7.4 Electrode shape dependence

To examine whether damage was related to sharp corners, which may be associated with field concentrations, we compared square and circular cells of comparable area/size. As shown in **Figure 3.33**, square shaped top electrodes form standard "bow-ties" near the edge of damage region, while circular shaped ones also form distorted "bow-ties". The detailed shape suggests that tearing in both cases initiates from the center of the "longitudinal" edge, defined as the one pointing towards the center of the e-bunch, but is clamped on the "transverse" edges, defined as the one perpendicular to the longitudinal edge. Since circular cells have shorter transverse edges, there is less resistance to tearing in circular cells appears and the "bow-tie" shape is less prominent for them.

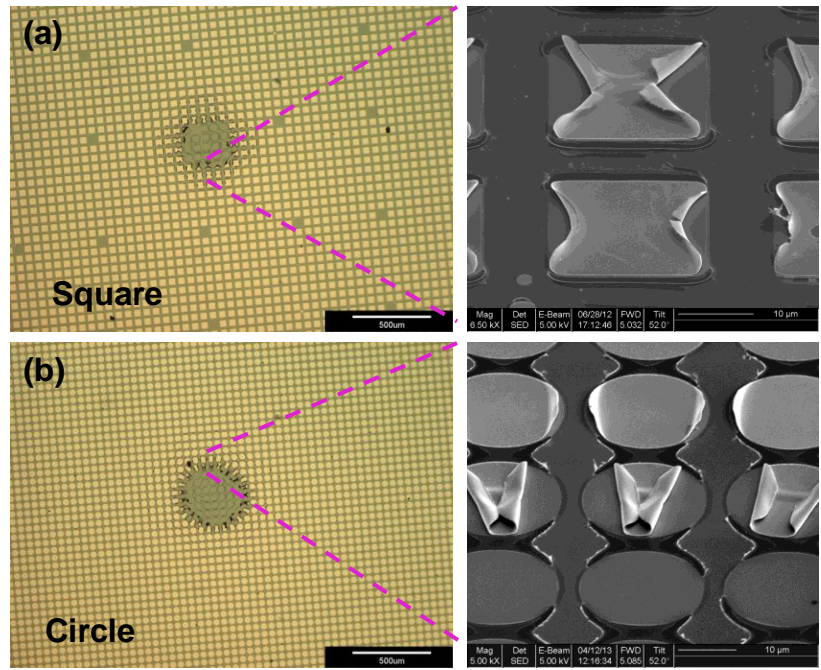

**Figure 3.33.** (a) Square *vs.* (b) circular cells exposed to identical electron bunch (20 GeV, 0.1 ps duration, $10^{10}$

electrons). SEM image on the right is taken by tilting the stage by $52^{\circ}$. (Device: Mo/Si$_3$N$_4$:5%Pt/Pt, $\delta$=10 nm, top electrode: 40 nm Pt, bottom electrode: 20 nm Mo. Size: $20\times20$ μm$^2$.)

### 3.5.7.5   Electrode aspect-ratio dependence

To further investigate the effect of cell shape, we compared rectangular patterns of various aspect ratios. The damage zones are shown in **Figure 3.34** for the ratios of 1:1, 1:2, 1:4 and 1:8. The 1:1 sample exhibits a circular shaped damage zone, while the 1:2 sample shows an elliptical zone. As the aspect ratio increases to 1:4 and 1:8, the damage zones become kidney-shaped, as if resembling the **E**-field distribution of a "dipole": damage is locally minimized along the short edges but maximized along the long edges. The asymmetry in **Figure 3.34** again excludes the possibility of direct impact–identical in all cases–to be the main cause of the damage. It is consistent with a magnetic flux mechanism augmented by the clamping force as illustrated in the right panel of **Figure 3.35**. (The clamping force is exerted on the edges where there is a current discontinuity.) Regarding the magnetic flux, the field incident on a long edge has a larger flux $\Phi=B\delta l_{\text{long}}$ than the field incident a short edge $\Phi=B\delta l_{\text{short}}$, thus generating a stronger emf in the former case. Since the circuit impedance is dominated by that of the dielectric gap, which is the same for both cases, there is a larger current (along the long side) when the magnetic field is incident on a long edge, thus a larger Lorentz force in this case. Referring to **Figure 3.35** for the high aspect ratio case, in which A and B are about the same distance from the center C, the SEM image shows the top Pt electrode of a rectangular device deformed into a "bow-tie" aligned in the "long" direction, when the long direction is more or less along the radial direction from the center. This is most evident around A. In contrast, the device around B suffered little damage apparently because along their boundaries, most length of the edges is clamped (along the long edges) and very little magnetic flux passes (through the short edge cross-section). This compares the devices around A, along their boundaries only a small length of the edges is clamped (at the short edges) and a more substantial magnetic flux passes (through the long edge "circuit loop"), thus exhibiting more tearing.

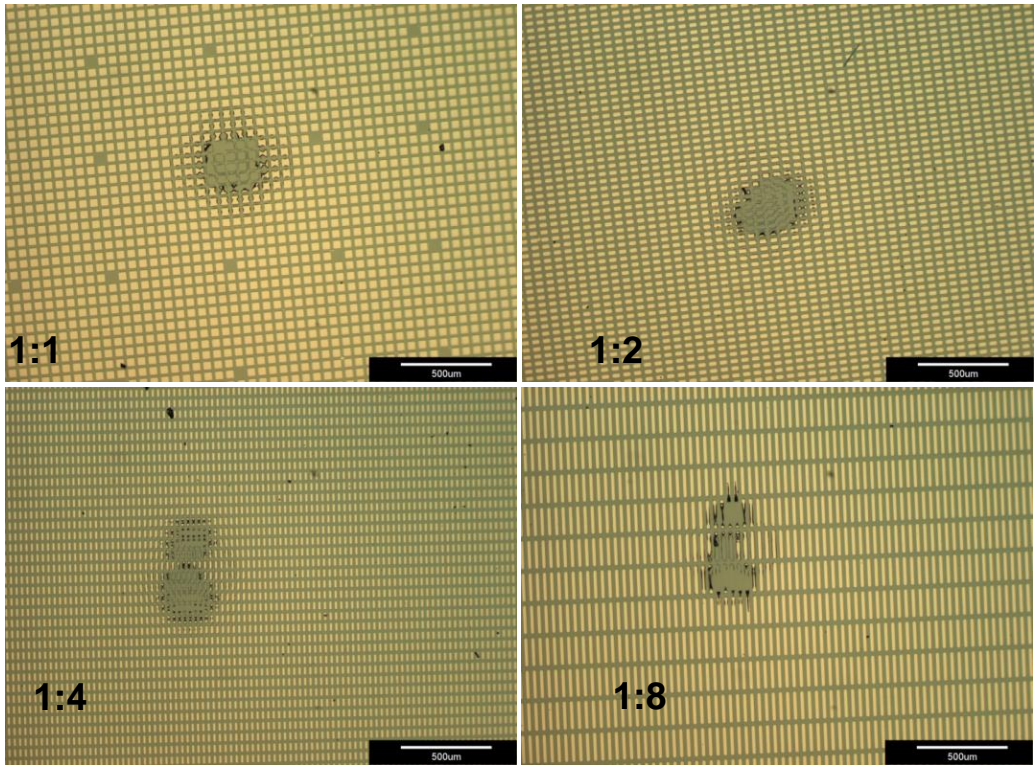

**Figure 3.34.** Damage pattern for various aspect ratio (1:1, 1:2, 1:4, 1:8). (Device: Mo/Si$_3$N$_4$:5%Pt/Pt, $\delta$=10 nm, top electrode: 40 nm Pt, bottom electrode: 20 nm Mo. Beam condition: 3 nC, 0.1 ps)

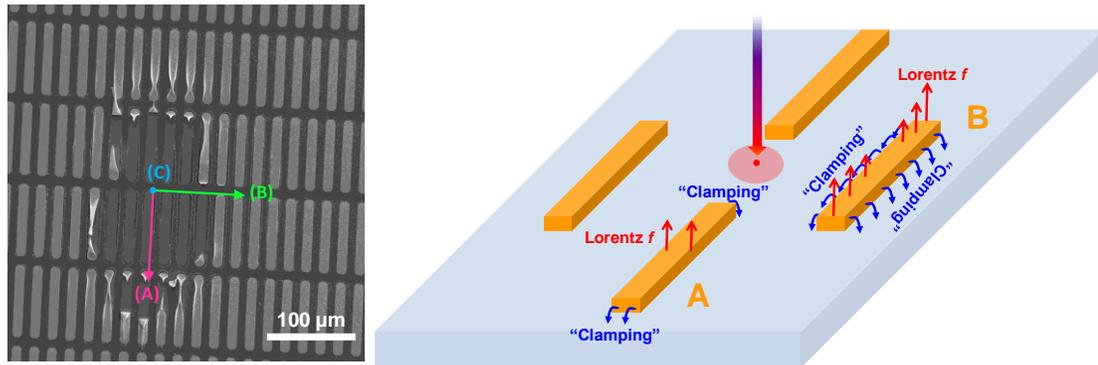

**Figure 3.35.** Damage pattern varying with aspect ratios of devices. (left) High aspect ratio (1:8) devices. C: center; A and B about same distance from C. (right) Schematic of field distribution, clamping forces, and tearing (Lorentz) forces, their magnitude indicated by the length of the red arrows. Top electrode thickness: 40 nm. (Device: Mo/Si$_3$N$_4$:5%Pt/Pt, $\delta$=10 nm, top electrode: 40 nm Pt, bottom electrode: 20 nm Mo. Beam condition: 3 nC, 0.1 ps)

### 3.5.7.6 Electrode thickness dependence

Magnetic-flux-induced Lorentz force depends on the area of the "circuit loop" but not the top electrode thickness $t$. Therefore, a weak thickness dependence is expected for induced switching. This is indeed verified in **Figure 3.36** for samples of different top electrode thickness from 15 nm to 100 nm, in which the radius of switching is similar, $r$~350 μm,.

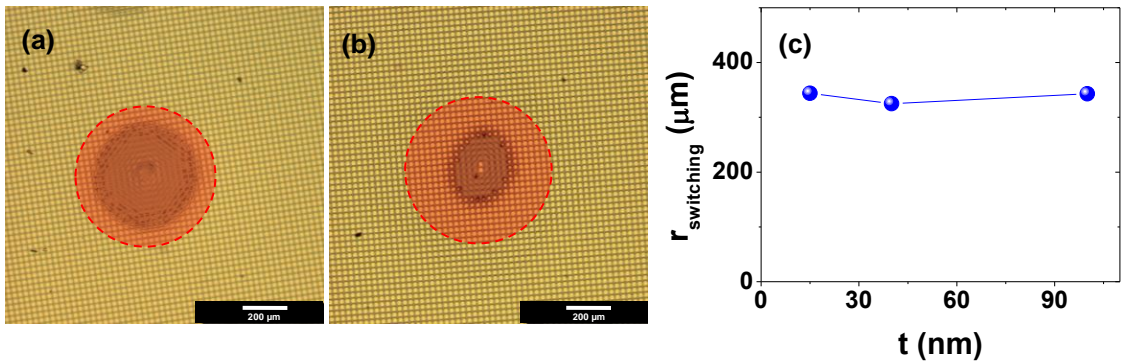

**Figure 3.36.** Top electrode thickness $t$ has no effect on switching zone (indicated in red), but the size of the damage zone in which top electrode was torn off is differet. (a) $t$=15 nm, with larger damage zone, and (b) $t$=100 nm, with smaller damage zone. (c) Radius of switching zone *vs.* thickness (Beam condition: $2\times10^{10}$ electrons, 0.1 ps). (Device: Mo/Si$_3$N$_4$:5%Pt/Pt, $\delta$=10 nm, top electrode: 40 nm Pt, bottom electrode: 20 nm Mo.)

Although resistance switching is the same, there is less degree of the physical damage when the top electrode is thicker. This is shown by the size of the inner zone in which all top electrodes are torn off: this zone is smaller in **Figure 3.36** when the top electrode is thicker. It is further verified by **Figure 3.37**: the $t$=10 nm sample shows a damage zone $r\sim500$ μm, but the $t$=100 nm sample exhibits little damage. Therefore, the electrode thickness has a strong influence on the resistance to mechanical tearing but not on switching. The resistance to tearing is easy to understand: the bending rigidity of a metal electrode follows a $t^3$ dependence, so a thicker top electrode can better resist tearing given the same Lorentz force. It is also clear that switching is caused by the Lorentz force, which is thickness independent, and not by tearing, which is strongly thickness dependent.

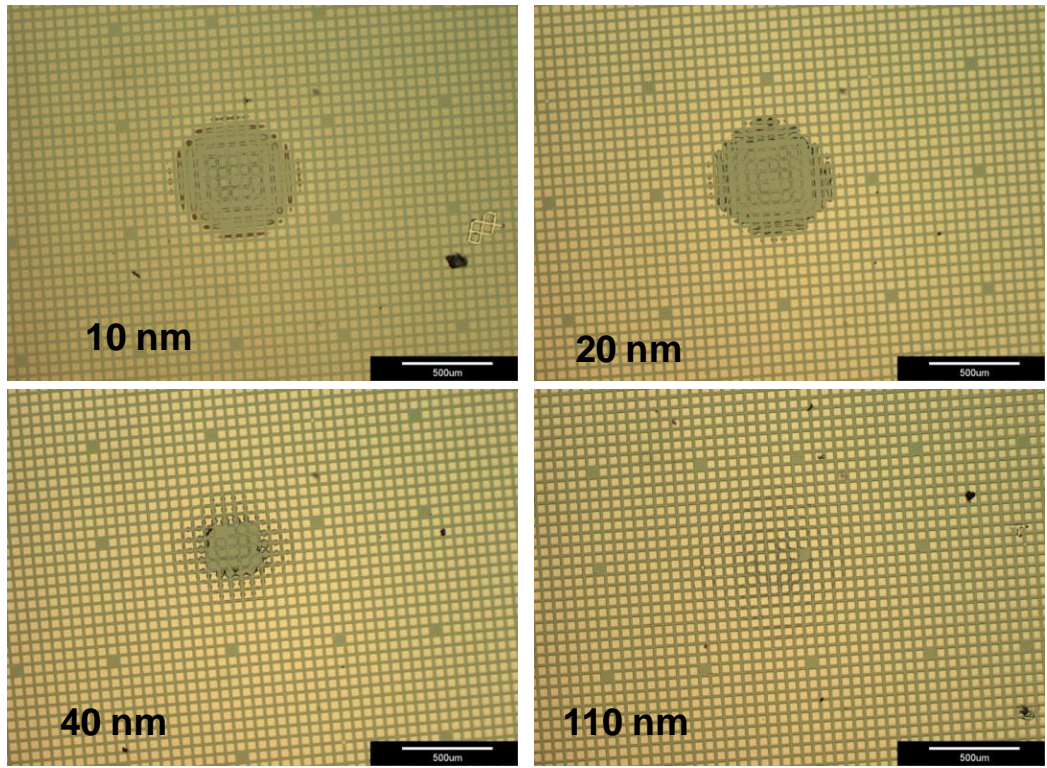

**Figure 3.37.** Top electrode (TE) thickness ($t$=10 nm, 20 nm, 40 nm, 110 nm) has a strong effect on damage zone. Damage zone shrinks for a thicker TE. (Beam condition: $10^{10}$ electrons, 0.1 ps). (Device: Mo/Si$_3$N$_4$:5%Pt/Pt, $\delta$=10 nm, top electrode: 40 nm Pt, bottom electrode: 20 nm Mo)

### 3.5.7.7 Electrode size dependence

The size of the top electrode should have no effect on the direct $\boldsymbol{E}$- or $\boldsymbol{B}$-field traveling with the electron bunch. However, a size effect is expected for the induced field. A larger electrode size allows a larger magnetic flux $\Phi = Bl\delta$ ($l$: electrode length, $\delta$: gap size) and therefore a larger Faraday emf or current $\propto \partial\Phi/\partial t \sim l$. Since the dominant impedance due to the dielectric gap remains the same, a larger Lorentz force is expected. This size effect was observed in **Figure 3.38** using devices of different sizes fabricated on the same sample: the extent of top-electrode tearing initially increased with the size but later saturated, confirming the importance of induced field in its pivotal role in causing mechanical damage. The saturation may be related to a "resonance" effect. The electron bunch of a 0.1 ps Gaussian (sigma) time $\tau$ has a FWHM of 2.35 sigma or $2.35\tau$, which is the corresponding half period, so the equivalent "period" of the half-cycle bunch is $4.7\tau$ or 0.47 ps. This half-cycle "oscillation" has a Fourier spectrum containing all components from DC to AC at a high-frequency cut-off corresponding to the wavelength of the bunch length of 140 $\mu$m (half of the wavelength is equivalent to "physical" bunch length 70 $\mu$m). Therefore, the induced field may initially increases linearly with the electrode size, but a maximum is reached when the electrode size is about 1/2 of the wavelength, after that there is no further increase.

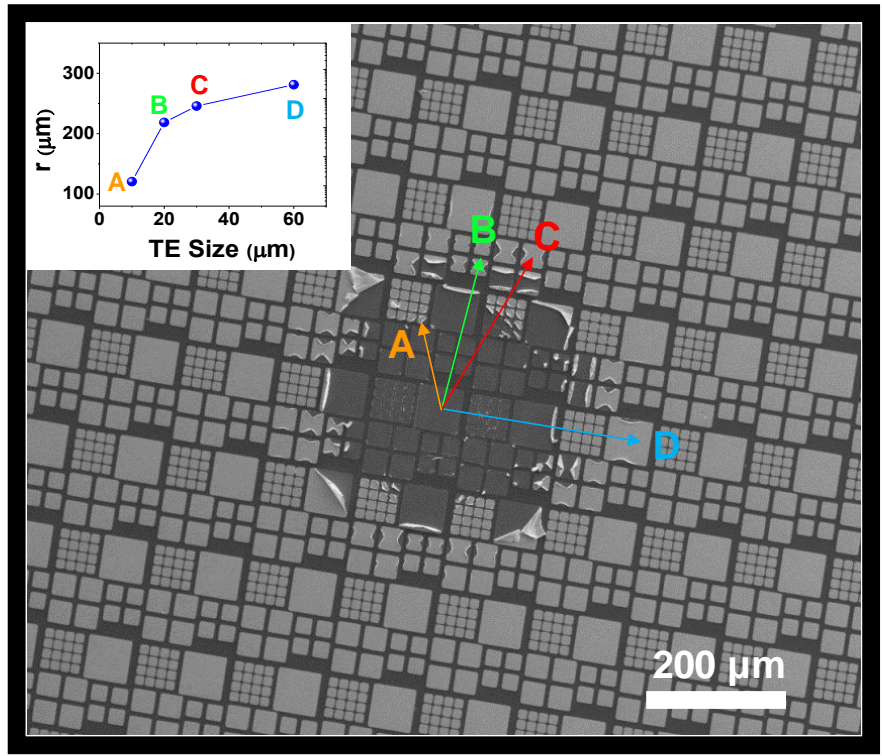

**Figure 3.38.** Decreased damage when the electrode size is smaller. Damage patterns in sample with same shaped devices of different sizes. Beam center is located at the center of the damage pattern. A-D indicates critical radial distance from beam center where damage first becomes detectable. **Inset**: critical radius (*r*) *vs.* size of top electrode. Electrode size: A: 10 μm; B: 20 μm; C: 30 μm; D: 60 μm. Electrode thickness: 40 nm. (Device: Mo/Si₃N₄:5%Pt/Pt, $\delta$=10 nm)

### 3.5.7.8 Dielectric thickness dependence

Dielectric thickness dependence was also checked. Identical devices (except the thickness of nanometallic film, 10 nm *vs.* 80 nm) were fabricated and exposed to identical electron bunch. As shown in **Figure 3.39**, the sizes of damage region exhibit very weak dependence on dielectric thickness (10 nm sample shows slightly larger affected region than 80 nm sample but such difference is far from 1:8 difference). This provides direct evidence that the gap impedance is the dominating impedance: magnetic flux $\Phi=Bl\delta$ (thus emf) increases with thickness $\delta$. On the other hand, gap impedance $Z\propto\delta$. Therefore, induced current $I\propto$emf$/Z$ is expected for a weak dielectric thickness dependence.

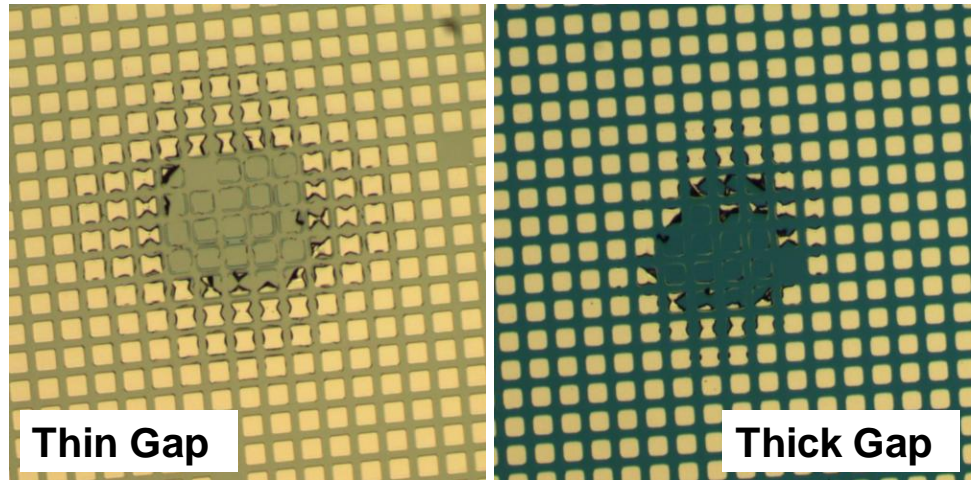

**Figure 3.39.** (a) Thin dielectric (10 nm) *vs.* thick dielectric (80 nm) exposed to identical electron bunch (20 GeV, 0.1 ps duration, $10^{10}$ electrons). (Device: Mo/Si$_3$N$_4$:5%Pt/Pt, top electrode: 40 nm Pt, bottom electrode: 20 nm Mo.)

### 3.5.7.9  Beam angle dependence

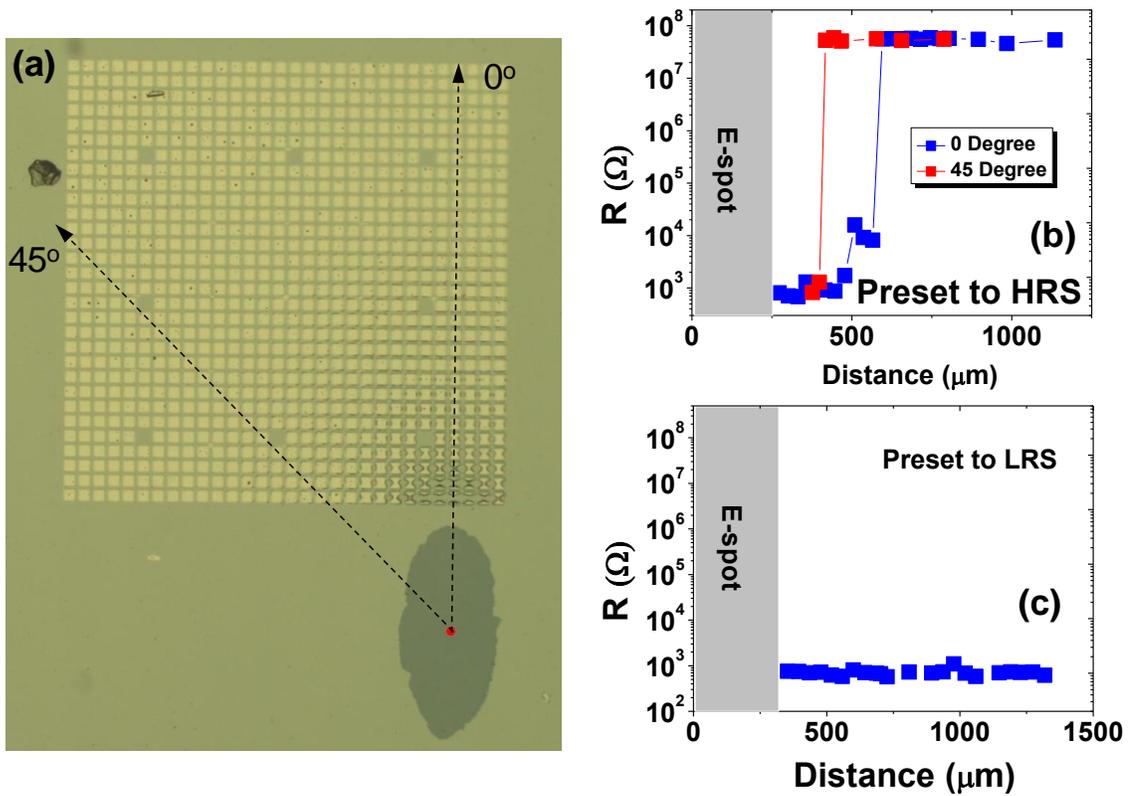

**Figure 3.40.** (a) Memory array after exposure to 45° incident electron bunch. Red dot indicates the position of beam center. (b) Resistance (preset to HRS) *vs.* distance from beam center along 0° and 45° directions in (a). (c) Resistance (preset to LRS) *vs.* distance from beam center along 0° direction in (a). (Beam condition: 20 GeV, 0.1 ps duration, $2 \times 10^{10}$ electrons, 45° incidence). (Device: Mo/Si$_3$N$_4$:5%Pt/Pt, $\delta$=10 nm, top electrode: 40 nm Pt, bottom electrode: 20 nm Mo.)

If the incident electron bunch takes on an inclined angle, asymmetry is introduced to the problem. **Figure 3.40a** shows an optical image of a memory array after exposure to a 45° incident electron bunch, in which the damage and switching regions were both distorted to an elliptical shape. For the preset HRS cells, the switching distance extends to ~565 μm along $0^o$ but to ~398 μm along $45^o$ (**Figure 3.40b**). A simple analysis of the direct field is provided below, referring to **Figure 3.41**. For a sample lying on the *x-y* plane receiving an electron beam traveling in the (*x*, -*z*) direction, the in-plane $E_{//}$ component is larger in the (0, ±*y*) region than in the (±*x*, 0) region, while the in-plane $B_{//}$ component has the opposite strength distribution. Meanwhile, the field components normal to the plane have the opposite distributions to the above trends. Comparing the field-strength distributions in **Figure 3.41** with the observed damage pattern in **Figure 3.40**, it becomes clear that the damage pattern is consistent with the pattern of in-plane magnetic field $B_{//}$. This is in support of our proposed mechanism which relies on the magnetic flux that enters the gap between the top and bottom electrodes.

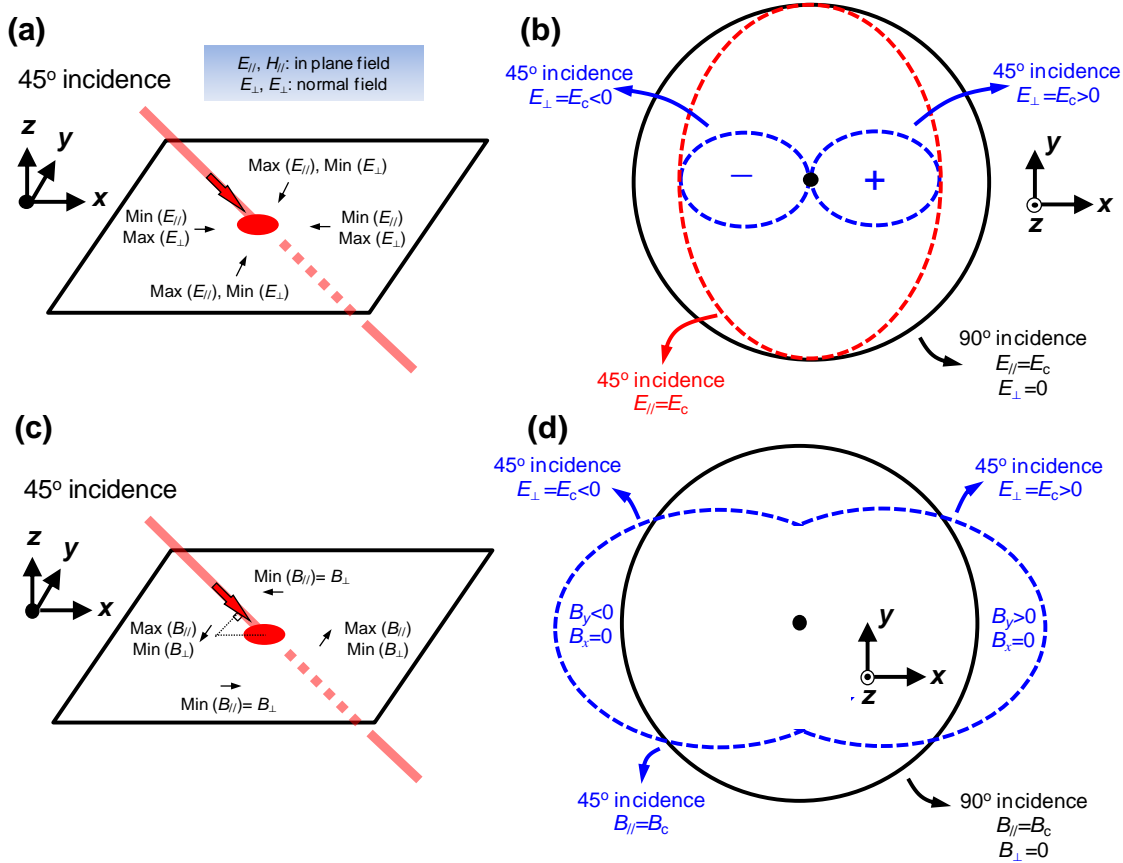

**Figure 3.41.** (a) **E**-field associated with a 45° incident electron beam. (b) **E**-field components on *x-y* plane. (c) **B**-field associated with a 45° incident electron beam. (b) **B**-field components on *x-y* plane.

**A quick derivation of angle dependence**

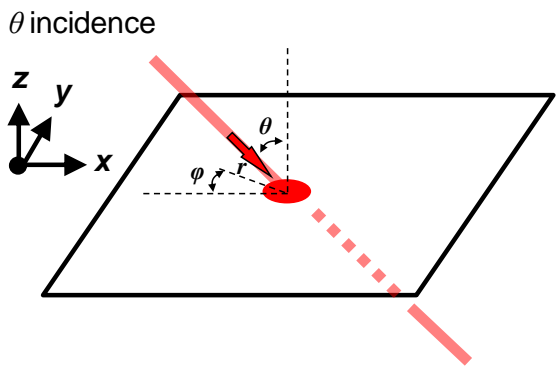

**Figure 3.42.** Schematic of spherical coordinate system.

If we set up a regular spherical coordinate system $(r, \theta, \varphi)$ as shown in **Figure 3.42**, the distance between the beam line and any given point $(r\cos\theta, r\sin\theta, 0)$ on the plane can be easily obtained as:

$$R = r\sqrt{\cos^2\theta + \sin^2\varphi\sin^2\theta}$$

Therefore, electrical field and magnetic field can be written as:

$$E = \frac{C_E}{R} = \frac{C_E}{r\sqrt{\cos^2\theta + \sin^2\varphi\sin^2\theta}}$$

$$B = \frac{C_B}{R} = \frac{C_B}{r\sqrt{\cos^2\theta + \sin^2\varphi\sin^2\theta}}$$

Where $C_E = c \times C_B$ and $\boldsymbol{E} \perp \boldsymbol{B}$. To further project $\boldsymbol{E}$ and $\boldsymbol{B}$ onto $\boldsymbol{z}$ axis ($\perp$ component) and $\boldsymbol{x}$-$\boldsymbol{y}$ plane (// component), we firstly derived (unit) directional vector as:

$$e_\perp = \frac{\sin\theta\cos\theta\cos\varphi}{\sqrt{\cos^2\theta + \sin^2\varphi\sin^2\theta}}, \; e_{//} = \frac{\sqrt{\cos^2\varphi\cos^4\theta + \sin^2\varphi}}{\sqrt{\cos^2\theta + \sin^2\varphi\sin^2\theta}}$$

$$b_\perp = \frac{\sin\theta\sin\varphi}{\sqrt{\cos^2\theta + \sin^2\varphi\sin^2\theta}}, \; b_{//} = \frac{\cos\theta}{\sqrt{\cos^2\theta + \sin^2\varphi\sin^2\theta}}$$

Therefore, we can eventually obtain the general form of $\boldsymbol{E}$ and $\boldsymbol{B}$ components:

$$E_\perp = Ee_\perp = \frac{C_E}{r}\frac{\sin\theta\cos\theta\cos\varphi}{\cos^2\theta + \sin^2\varphi\sin^2\theta},$$

$$E_{//} = Ee_{//} = \frac{C_E}{r}\frac{\sqrt{\cos^2\varphi\cos^4\theta + \sin^2\varphi}}{\cos^2\theta + \sin^2\varphi\sin^2\theta}$$

$$B_{\perp} = Bb_{\perp} = \frac{C_B}{r} \frac{\sin\theta \sin\varphi}{\cos^2\theta + \sin^2\varphi \sin^2\theta},$$

$$B_{//} = Bb_{//} = \frac{C_B}{r} \frac{\cos\theta}{\cos^2\theta + \sin^2\varphi \sin^2\theta}$$

### 3.5.8 Momentum and Energy Transfer of Electron-bunch

#### 3.5.8.1 Direct impact from momentum transfer

A high energy electron bunch has a very small (collision, energy/momentum transfer) cross section because the time available for momentum transfer is very short. Typically, a 20 GeV electron only loses 10-100 eV (see later calculation of collisional energy loss) as it passes through a 10 nm sample, which implies the direct damage of electron bunch in our experiment is rather miniscule. Quantitatively, the energy of a relativistic electron is

$$E = \sqrt{(m_0 c)^2 + (pc)^2} \approx pc$$

The momentum loss of the electron is thus:

$$\Delta p = \frac{\Delta E}{c}$$

For a bunch of a size of $A$=40 μm×40 μm, with a pulse duration of $t$=100 fs and a bunch charge of $N$=21×10$^{10}$, the direct (compressive) stress is

$$\sigma = \frac{N\Delta p}{At} = \frac{N\Delta E}{Act}$$

giving $\sigma$ =0.67 MPa (for $\Delta E$=10 eV) or $\sigma$ =6.7 MPa (for $\Delta E$=10 eV). These values are far below the critical stress estimated in previous sections. This explains why **Figure 3.29**, as well as **Figure 3.30**, **Figure 3.31**, **Figure 3.32** & **Figure 3.36** and other micrographs not shown here, always contains several undamaged cells at the very center of the collision. These cells, preset in the HRS, were still in the HRS and apparently not disturbed by the electron bunch at all. Yet cells slightly outside the center, up to a radius of ~180 μm in the same figure, were irreparably damaged with their top electrodes all gone. Since the collision impact/energy-transfer of the 20 GeV electrons is the most severe at the collision center, this observation is a clear indication that it cannot be the cause of the phenomena probed in our study. The lack of damage is consistent with the field distribution: from a symmetry perspective, there is no induced field in the beam center.

In the following, we will consider the energy dissipation in the sample of electron bunch. The loss mechanisms are first summarized, followed by more detailed evaluation of radiation energy profile and deposited energy density.

### 3.5.8.2 Energy loss mechanisms

An electron bunch generates radiation through both deceleration and fluorescence of charged particles as the bunch collides with atoms (and their electrons). Low energy radiation also arises through excitation of phonons by the electron bunch. We thus examine whether electron irradiation can cause memory change. It is known that energetic photons and charged particles can generate damage on computer memories. For example, α-particles (in the MeV range) emitted by radioactive decay of uranium or even cosmic rays can create enough electron-hole pairs near a storage node to cause a random soft error[13-14]. However, the following calculation suggests such possibility can be excluded in our experiments involving relativistic electrons.

As a high speed electron hits the sample, their energy loss is quantified by the stopping power, defined as the energy loss (through both collisions and radiation, of all causes) per unit path length traveled:

$$S(E) = -dE/dx$$

There are three components in the stopping power[15]:

**Collision stopping power**: This is the energy loss due to Coulomb collisions that result in the ionization and excitation of the target and secondary atoms. For heavily charged particles, collision stopping power is often called electronic stopping power.

**Radiative stopping power**: This is the energy loss due to collisions with atoms and atomic electrons in which bremsstrahlung quanta are emitted from the decelerating incident charged particle. It is important only for electrons.

**Nuclear stopping power**: This is the energy loss due to the energy transfer to recoiling atoms in elastic collisions. It is most important when the mass ratio is close to one.

**Total stopping power**: While for protons and helium ions, the collision and nuclear stopping powers are important, for electrons, we only need to consider collision and radiative stopping power. From NIST data[16] (**Figure 3.43**), we can see that the total stopping power is dominated by collision stopping for low energy (<1 MeV) electrons but by radiative stopping power for high energy (>1 MeV) electrons.

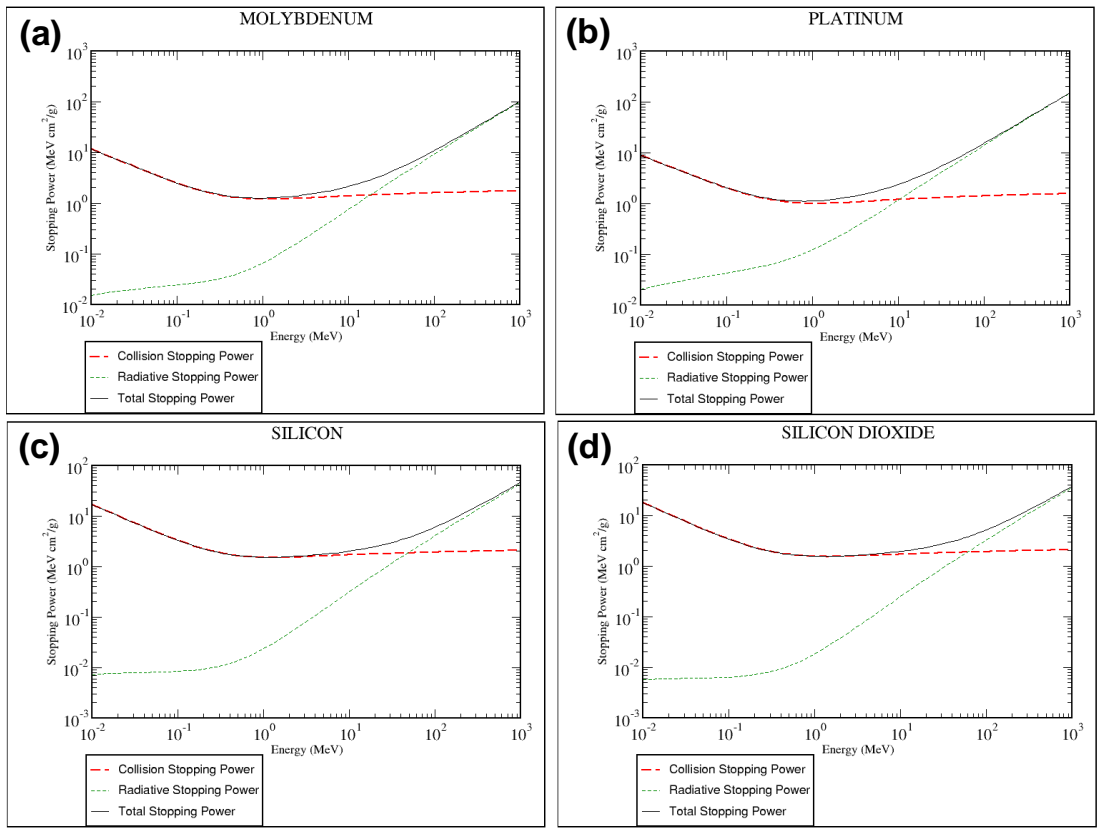

**Figure 3.43.** Stopping power of electrons up to 1 GeV; from NIST website[16].

### 3.5.8.3 Radiation energy loss of 20 GeV electron

Although radiative energy loss is several orders larger than collisional energy loss, it is of little interest for us. The reason lies in the relativistic effect of ultrahigh energy electrons.

The importance of the relativistic effect may be estimated using the ratio $\gamma = 1/\sqrt{1 - v^2/c^2}$, which is the ratio of electron energy to that of a rest mass (0.51 MeV). For a 20 GeV electron,

$$E = \frac{m_0 c^2}{\sqrt{1 - v^2/c^2}} = \gamma m_0 c^2$$

with $\gamma=40000 \gg 1$, meaning that the relativistic effects will dominate. (Since the total energy loss during the electron bunch experiment is only 10-100 eV, as shown later in the collisional-energy-loss calculation, almost identical $\gamma \approx 40000$ is maintained through the entire journey. For a relativistic electron, the radiation (gamma rays and x-rays) emitted by a decelerating electron is strongly forward-focusing along the beam direction. Therefore, as soon as the electron leaves the mixture film, there is no more radiation directed to the film. Since the energy loss in the course of passing through the top electrode and the film (50-60 nm in total) is very small due to the small thickness, and the total backward radiation when the electron is passing through the much thicker silicon substrate is also small due to forward focusing, the total radiation damage received by the mixture film is negligible for a 20 GeV electron.

The above holds along the path of the election; along the lateral dimension away from the electron path, it is even smaller. Specifically, the radiation field of an electron traveling at a speed v but decelerating at a rate a along the same direction can be expressed as [17]:

$$E = \frac{1}{4\pi\varepsilon_0} \frac{ea\sin\theta}{c^2 R\left(1 - \frac{v}{c}\cos\theta\right)^3}$$

The angular distribution of power is thus:

$$\frac{dP}{d\Omega} = \frac{e^2}{16\pi^2\varepsilon_0 c^3} \frac{a^2\sin^2\theta}{\left(1 - \frac{v}{c}\cos\theta\right)^5}$$

Under the relativistic limit $\gamma \gg 1$, one can treat $\varepsilon = 1/\gamma^2$ as infinitesimal and perform the Taylor expansion. Therefore, the maximum power intensity occurs at:

$$\cos\theta_m \approx 1 - \frac{\varepsilon}{8}, \text{ or } \theta_m \approx \frac{1}{2\gamma}$$

For $\gamma$=40000, it gives $\theta_m \approx 1.25 \times 10^{-5}$ rad. In short, almost all radiation energies are confined within a narrow range along the beam direction, as illustrated in the polar plot of energy distribution in **Figure 3.44** and in the $\theta$ plot in **Figure 3.45**. (Note that radiation energies are mostly concentrated near $10^{-5}$ rad and rapidly decay by 12 orders of magnitude at $10^{-3}$ rad.) Therefore, we may safely conclude that radiative energy loss has little effect on our devices, especially those that are more than one $\sigma_r$ (20-30 µm) away from the beam center. Indeed, even near the center of beam which should experience the largest radiation, we always found several devices survived without any physical damage or change in the resistance state. (This was shown in **Figure 3.29**, **Figure 3.30**, **Figure 3.31**, **Figure 3.32** & **Figure 3.36**.) This is the most direct experimental proof that radiation induced effect is unimportant in our experiments: it has no significant effect on the mechanical deformation or the resistance state change.

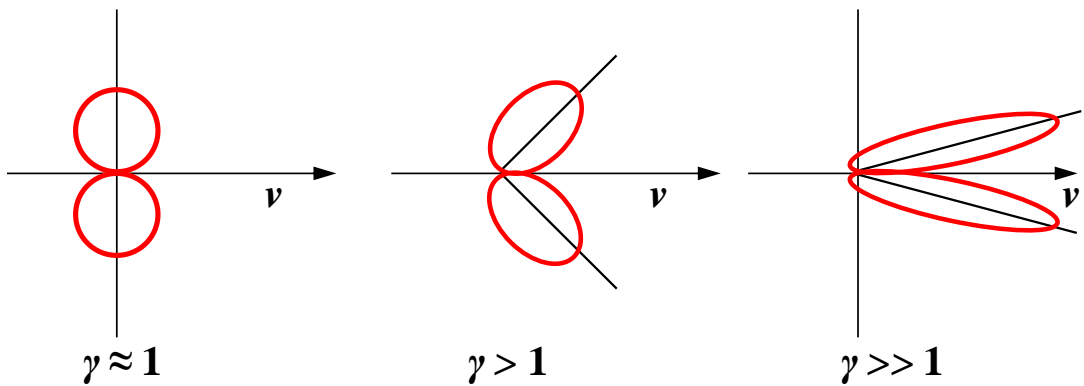

**Figure 3.44.** Schematics of radiation power distribution for the case of linear deceleration ($a//v$).

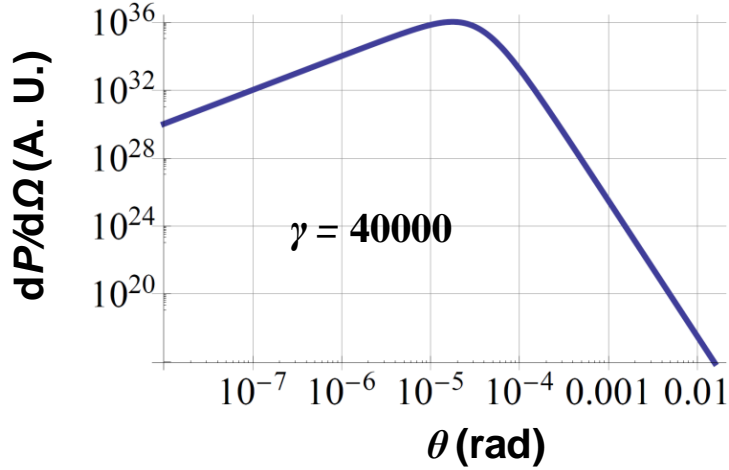

**Figure 3.45.** Angle dependence of radiation power for γ=40000.

### 3.5.8.4 Collisional energy loss and deposited energy density

We now calculate the collisional energy loss and thus deposited energy into the MIM structure. From **Figure 3.43**, we can estimate such energy by extrapolating collision stopping power to 20 GeV. As shown in **Table 3.1**, the total collisional energy loss per electron is of the order of 200 eV, the majority of that occurring in the Pt top electrode. For an electron bunch (of $2\times10^{10}$ charge) with a diameter of 30 μm passing through a 80 nm thick Mo/Si$_3$N$_4$:Pt/Pt stack, the total deposited energy density is 11.3 kJ/cm$^3$. If we further assume all the lost kinetic energies are locally converted to heat within the same 30 μm lateral diameter, we can estimate the temperature rise in each layer as $\Delta T_{Pt}$=5712K, $\Delta T_{SiN}$=4989K, $\Delta T_{Mo}$=3303K using the following heat capacity[18]: $c_{Pt}$=25.86 J·mol$^{-1}$·K$^{-1}$, $c_{SiN}$=0.17 J·g$^{-1}$·K$^{-1}$, and $c_{Mo}$=24.06 J·mol$^{-1}$·K$^{-1}$. This means that the collision energy loss should induce local melting or even evaporating, provided heat is not dissipated away by radiation or conduction heat transfer. However, given the fact that the devices at the beam center survived (**Figure 3.29**, **Figure 3.30**, **Figure 3.31**, **Figure 3.32** & **Figure 3.36**), and the fact that the sample with thick Pt electrodes (**Figure 3.37**) does not show any damage even within the beam diameter, we must conclude that only a small portion of the collision energy loss is actually used for local heating, so small that no melting/evaporating occurs. One explanation for not observing the heating effect predicted by the collisional energy deposition might be that this collisional loss gets dissipated to the Si substrate. This is because knocked-off electrons and X-rays of hundreds of eV can travel tens of nanometers to deposit their energy in the Si, thus lowering the energy density in Pt and in the nanometallic film. (The Si substrate itself has little collisional loss, at most enough to give a temperature rise of ~200K). Another explanation is that the collisional loss is overestimated by the collisional stopping power formula in the NIST data base, which is valid for thick substrates. For thin substrates the Landau distribution[41] for statistical fluctuations of energy loss needs to be used. The collisional loss is indeed smaller, though only by a factor of 2 or so. Yet another explanation for the lower value is that the $K$ shell of electrons is not excited, as hinted in ref.[42]. This reference is for heavy particles where the energy loss is by collision mainly. (We could consider the formulas for muon and adjust the mass, although in this form the method is only approximate.) Lastly, a collection of the formulas and brief descriptions of energy loss may be found in PDG booklet[43] (starting at page 243). It also has figures for photon radiation length vs energy. A longer review is available in ref.[44]. (The other possibility is that the extrapolation of the NIST data base to high energy electron is incorrect.)

Away from the beam path, the heat received is small even if the collisional energy loss is as high as calculated above. Assuming the local heat is dissipated radially in all directions in the thin sample, we can estimate the energy density when the radius of the heated material is increased from that of the beam (beam radius=15 μm) to $r$=150 μm. The energy density, which scales as ~$1/r^2$, is then reduced by 2 orders of magnitude, so at most increasing the temperature by ~50 K. This increase is rather minimum and not expected to induce any mechanical nor electronic effect on the device.

| | $d$ (cm) | Collision stopping power at 20 GeV (MeVcm$^2$/g) | Energy loss per electron, $\Delta E$ | | Total deposited energy, $E_{tot}$ |
| | | | eV | J | J |
|---|---|---|---|---|---|
| Pt (TE) | $4\times10^{-6}$ | 1.7 | 150 | $2.4\times10^{-17}$ | $4.5\times10^{-7}$ |
| Si$_3$N$_4$:Pt | $1\times10^{-6}$ | 2 | 6.3 | $1\times10^{-18}$ | $1.9\times10^{-8}$ |
| Mo (BE) | $3\times10^{-6}$ | 1.91 | 59 | $9.4\times10^{-18}$ | $1.8\times10^{-7}$ |
| Total | | | 215.3 | $3.4\times10^{-17}$ | $6.4\times10^{-7}$ |

**Table 3.1.** Collision energy loss for stacks of Mo/Si$_3$N$_4$:Pt/Pt assuming 3 nC charge was used.

Combining all the information above, we can safely conclude that neither radiation energy loss nor collision energy loss can cause the physical and electrical change of devices observed in our experiments.

### 3.5.9 Stress and Field Analysis

#### 3.5.9.1 Stress analysis

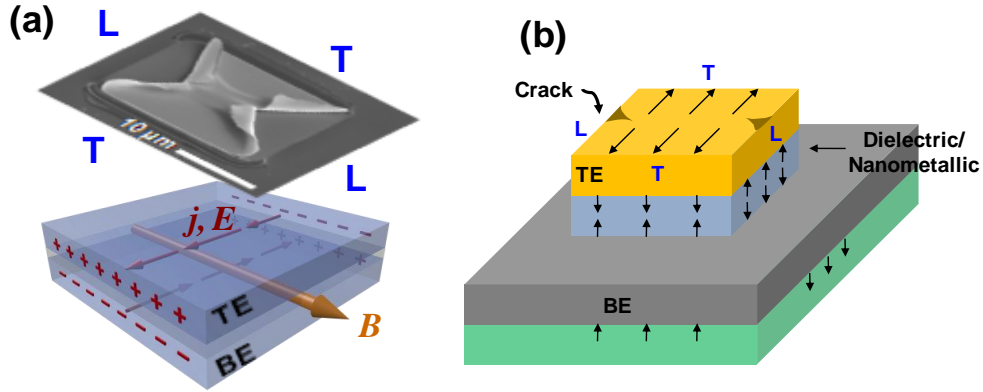

**Figure 3.46.** (a) Top panel: top Pt electrode of square device deformed into "bow-tie"; Bottom: incoming magnetic field *B* and induced current *j*, electric field *E*, and charge in schematic device with top electrode (TE) and bottom electrode (BE) separated by an insulator gap. (b) Stress states in three regions with fracture in top electrode.

Having excluded direct impact and radiation/collision energy transfer from further consideration, we find the electromagnetic field associated with an electron bunch to be the only possible mechanism that may account for the phenomena described above. Here we first give a qualitative analysis of the field and stress in the MIM structure, to be followed by a more detailed analysis of the field in the next subsection. Since the *E*-field is largely screened by the top electrode but the *B*-field is not, the main effect on the device is exerted by a transient magnetic field that crosses the "circuit loop" of TE → film-gap → BE → film-gap → TE (**Figure 3.46a**, bottom panel). For this configuration, an induced current (*j*) in the top electrode (TE) is accentuated at the longitudinal edges (L), and it has an opposite image current in the bottom electrode (BE). This circuit loop activates a 0.1 ps (sigma) Lorentz-force repulsion (*i.e.*, a magnetic pressure) to push the two electrodes apart. However, because the loop is not continuous, the current discontinuity at the transverse edges (T) in the top electrode produces a set of opposite charges at T, matched by another set of opposite image charges in the bottom electrode/substrate (**Figure 3.46a**, bottom panel). Thus the overall magnetic pressure is countered by the image-charge attractions at T. As the pressure forces the top electrode to tear, along the electrode-film interface, starting at the places where the Lorentz force is strongest (around L, especially at its center), the clamping forces counteract to turn the partially torn top electrode into an elegant platinum "bow-tie".

Several other stresses also exist in the device stack, as shown in **Figure 3.46b**: (a) an in-plane tension (especially at the crack-tip) in the TE, caused by the clamping forces that constrains the TE from being repelled, (b) a normal tension in the insulator (or mixture layer) along the L side (associated with the tendency of TE bulging up), and (c) normal compression in the insulator along the T side (a reaction to clamping) and (d) normal compression (L side) and tension (T side) beneath the BE. For switching, the stress (b) and stress (c) are relevant. Their importance is supported by the parametric studies described above. The following analysis will provide a more quantitative description of the origin of the induced fields in the MIM device, from which the Lorentz force and stresses originate.

### 3.5.9.2  Current and voltage simulation: "patch antenna"

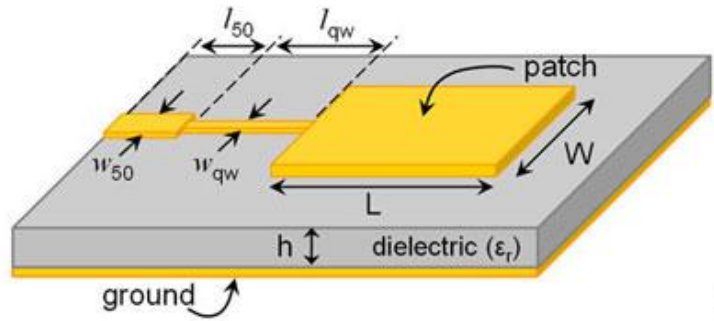

**Figure 3.47.** A schematic of conventional patch antenna structure (adapted from ref.[19]).

The MIM structure in the RRAM array finds its physical correspondence in a radio frequency (RF) circuit, called patch antenna, **Figure 3.47**. It consists of a flat rectangular / circular / elliptical "patch" of metal (with a certain feed-in transmission line), placed over a sheet of grounding metal, and the two are typically separated by a dielectric material. The discontinuities at each truncated edge allow such structure to generate particular radiation characteristics, governed by the Maxwell equation. This analogy allows us to use the well developed patch antenna theory to understand how current and voltage are distributed inside the MIM structure by a THz electromagnetic wave excitation.

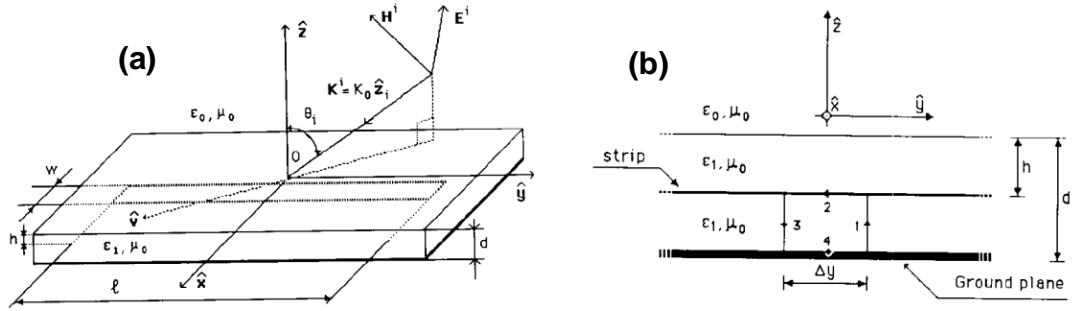

**Figure 3.48.** (a) Microstrip structure illuminated by a linearly polarized plane wave and associated coordination setup. The $y$ axis is along the direction of the strip. (b) Side view of microstrip structure (adapted from ref.[20]).

Reference[20] provides a detailed mathematical description of the patch antenna effect. A generalized structure, with a microstrip line of a length $l$ and width $w$, embedded at $z{=}{-}h$, is illuminated by a linearly polarized uniform plane wave $\boldsymbol{E}_i$, $\boldsymbol{H}_i$, as shown in **Figure 3.48**. The ground plane and the metal strip are assumed to be a perfect conductor, and the dielectric having a dielectric constant ($\varepsilon_r$) is assumed lossless. By integrating the Maxwell equations along the $z$ direction, a canonical set of voltage and current distributions can be derived as:

$$\frac{dV}{dy} = -j\omega \frac{Z_w}{v_f} I - v$$

$$\frac{dI}{dy} = -j\omega \frac{1}{Z_w v_f} V - i$$

where $v_f = c/\sqrt{\varepsilon_{eff}}$ is the phase velocity, $Z_w = Z_0/\sqrt{\varepsilon_{eff}}$ is the characteristic impedance ( $\sqrt{L/C}$ ) and $\varepsilon_{eff}$ is an effective dielectric constant taking account of the fringing field near capacitor edges. The excitation terms $v(y)$ and $i(y)$ can be explicitly expressed as:

$$v(y) = j2\omega\mu_0 \frac{\sin\left[K_{Z1}(d-h)\right]}{K_{Z1}} \cdot \left\{ \begin{array}{c} \dfrac{\vec{E}_0 \cdot \hat{v}}{jZ_{w1}^{TM}\sin(K_{Z1}d) + Z_{w2}^{TM}\cos(K_{Z1}d)}\,\vec{x}\cdot\hat{u} + \\[2ex] \dfrac{\vec{E}_0 \cdot \hat{u}}{jZ_{w1}^{TM}\sin(K_{Z1}d) + Z_{w2}^{TM}\cos(K_{Z1}d)}\,\vec{x}\cdot\hat{v} \end{array} \right\} \cdot e^{-jK_y y}$$

$$i(y) = -j2K_t \frac{\vec{E}_0 \cdot \hat{v}}{jZ_{w1}^{TM}\sin(K_{Z1}d) + Z_{w2}^{TM}\cos(K_{Z1}d)} \cdot \left\{ \begin{array}{c} 2\cos\left[K_{z1}(d-h)\right]\dfrac{\sin\left(K_x\dfrac{w}{2}\right)}{K_x} + \\[2ex] \dfrac{C}{\varepsilon_1}\dfrac{\sin\left[K_{z1}(d-h)\right]}{K_{z1}} \end{array} \right\} \cdot e^{-jK_y y}$$

Reference[20] provided a nice but lengthy analytical solution for voltage $V(y)$ and current $I(y)$ distribution. Here we will only use their numerical results which cover several configurations that can be further extended to other dimensions using scaling arguments. This is first shown for the case of a dielectric ($\varepsilon_r$=10, corresponding to $\varepsilon_{eff}$=6.66) of a thickness of 1.57 mm, a characteristic impedance of 50 $\Omega$, metal lines of a width $w$=1.3 mm and length $l$=15 cm, and an incident uniform plane wave with an electric field intensity of 1 V/m and a frequency of 3 GHz ($\lambda$=0.1 m), at normal incidence.

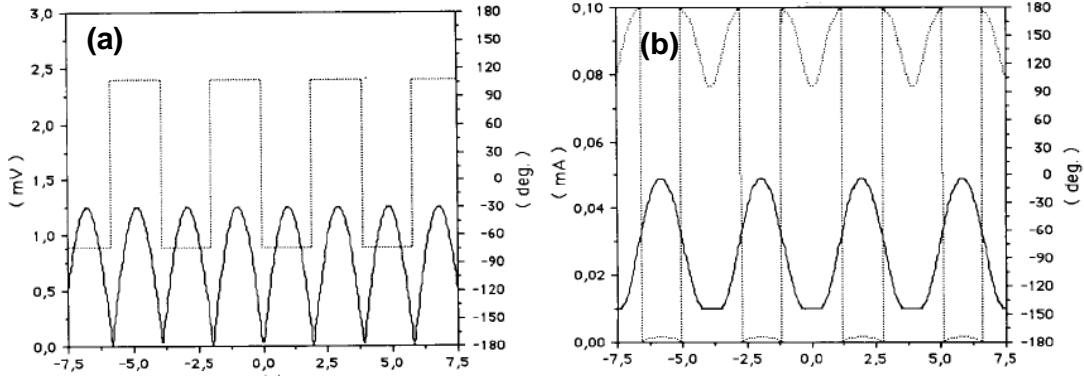

**Figure 3.49.** (a) Voltage magnitude (solid line) and phase (dotted line) and (b) current magnitude (solid line) and phase (dashed line) induced along the microstrip line ($\varepsilon_r$=10, $w$=1.3 mm, $h$=0) with an incidence angle $\theta$=0 and incidence field $E_0$=1 V/m at $f$=3 GHz (adapted from ref.[20]).

The simulation results are shown in **Figure 3.49**. It gives $V(y)$ & $I(y)$ [20] with a peak voltage of ~1.25 mV and peak current of ~0.05 mA. At two opposite edges, voltage amplitudes are identical but the phases differ by 180°, which confirms that opposite charges accumulate at opposing edges. The number of nodes is expected from the general resonance criterion: $m \times \lambda / 2 \sqrt{\varepsilon_{eff}} = l$, which leads to the node number $m \approx 8$ consistent with the simulation results. In contrast, the current amplitudes fall to the minimum near the two edges, which is required by the boundary condition at the metal-insulator interface.

### 3.5.9.3   Estimate of fields and stresses in the MIM structure

To apply these results to our experiment, we first note that our half cycle (with a Gaussian width $\sigma_t$=0.1 ps, giving a FWHM of 0.235 ps, or an equivalent full cycle period of 0.47 ps) pulse corresponds to ~2.1 THz ((1/2.35$\sigma_t$)/2) wave or a $\lambda$=141 µm wave. Since Maxwell's equation is linear, identical excitation terms $v(y)$ and $i(y)$ obtain if we simply scale all lengths by a factor of $\lambda$(0.1 m)/ $\lambda$(141 µm) =709 to keep factors like $K_{z1}$d unchanged in the above equation. Therefore, a stripline with a dielectric constant of 10, a dielectric gap $\delta$=2.21 µm, a width 1.83 µm, and a length 212 µm will also develop a voltage of 1.76 µV at the ends in a field of 1 V/m at 2.1 THz. In the case of an electrical field $E$~3$\times$10$^9$ V/m, which is representative of the field around an electron bunch, the voltage at the edges will be 1.76$\times$3$\times$l0$^9$ µV=5.28 kV across a 2.21 µm gap, or 24 V across a 10 nm gap. Such a high voltage, even a transient one, may cause dielectric breakdown or ionization, thus creating "trenches" at the two opposing edges.

In our simulations we assumed a more realistic excitation with a Gaussian pulse profile of a linearly polarized plane wave at normal incidence to the sample plane. **Figure 3.50** shows a snapshot (after 2 ps) of the field dynamics, which confirmed the electrical field near the two counter edges are maximum but in opposite directions. According to the Maxwell equation (the Gauss law), the surface charge density $\sigma_q$ is directly related to the normal field, $\sigma_q$=2$\varepsilon E$. Therefore, we can immediately conclude that charges are accumulated on the two transverse edges, which confirms our earlier qualitative speculation in **Figure 3.46a**. The simulation was done with an excitation plane wave field of 1 V/m, so scaling it linearly to 3$\times$10$^9$ V/m will yield 3$\times$10$^8$ V/m at the edges for this time snapshot (corresponding to 6 V across a 20 nm gap). The current distribution on the top and bottom electrodes have opposite signs indicating the magnetic field is contained between the electrodes. The current distribution is consistent with the "bow-tie" shape of top electrodes in the affected zone, which serves as strong evidence supporting the crucial role played by the induced **E-B** fields and their interactions.

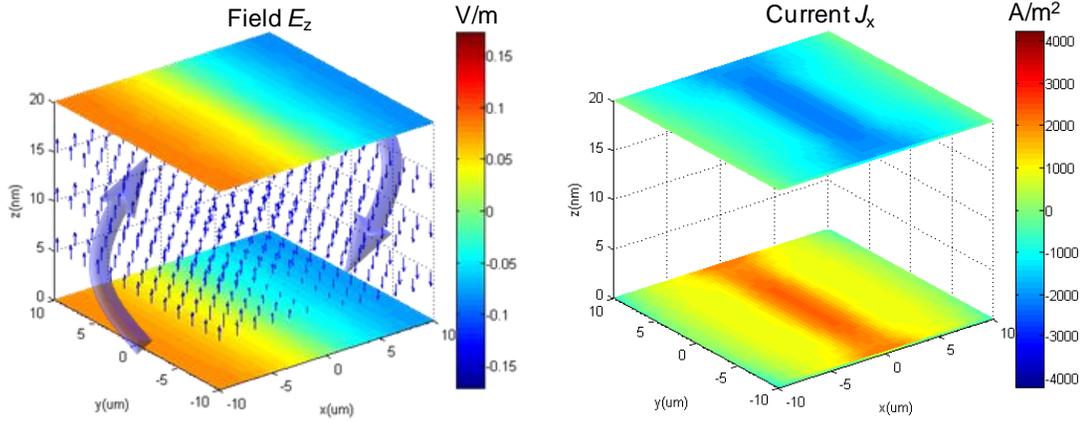

**Figure 3.50.** Simulation of (a) field $E_z$ and (b) current $J_x$.

A quantitative estimation of the magnitude of the stress can be conveniently done using the concept of magnetic pressure $P_B = B^2/2\mu_0$, which is an equivalent energy density associated with a magnetic field[21], exerting a Lorentz force on the electrodes containing it. Since the field inside the patch is similar in magnitude to the external incident field, we obtain $P_{B, max} = B^2/2\mu_0 = 1.4$ GPa ($B_{max} \approx 60$ T, see **Figure 3.21b**) near the bunch center and $P_B = B^2/2\mu_0 = 40$ MPa ($B \approx 10$ T, see **Figure 3.21b**) near the edge of the HRS/LRS-switching region. This range, 40 MPa~1 GPa, is not unreasonable, comparable to our earlier estimates: $P_{CIP} = 300$ MPa, $P_{CAFM} = 1~100$ MPa, $P_{FIB} = 40$ MPa. In particular, the stress threshold of 40 MPa at the edge of the switching zone is comparable to the corresponding stress estimate of CAFM and FIB experiments, since in all three cases a uniaxial stress is involved.

### 3.5.9.4  Statistics of resistance & stress distribution

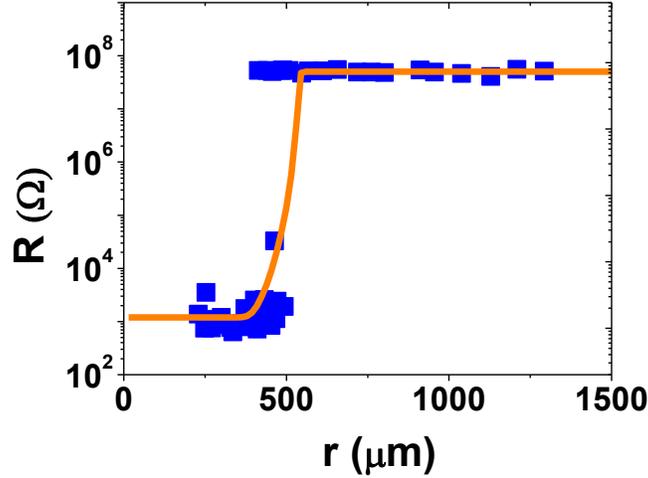

**Figure 3.51.** Devices (TE thickness: 40 nm) preset to HRS switched to LRS if located within ~450 μm from hit-spot; devices preset to LRS remained in LRS at all distance. Devices within 200 μm lost the top electrodes completely and were not tested.

**Figure 3.51** shows the resistance distribution for a device array consisting of Mo (20 nm) / Si$_3$N$_4$:Pt (10 nm) / Pt (40 nm). With a thinner TE (40 nm), the damage region is larger than the previous case (TE torn away for $r$<250 μm) due to a lower bending rigidity. However, switching (HRS→LRS) occurs from $r$~400 μm to $r$~500 μm, similar to the case of **Figure 3.29**. A quantitative analysis can be made following the same method as in **Section 3.4.2**, with the additional consideration of field dependence. As shown in **Section 3.4.2**, the **B**-field follows an $r^{-1}$ dependence, so the induced Lorentz force has an $r^{-2}$ dependence. Hence, stress has the same $r$ dependence, namely $\sigma=A'/r^2$, where $A'$ is a proportionality constant). So we can rewrite the expression of resistances in terms of radius $r$ as:

$$\frac{1}{R} = \frac{exp\left[-\left(\frac{A'/r^2 - \sigma^*}{\Delta}\right)_+^n\right]}{R_{HRS}} + \frac{1 - exp\left[-\left(\frac{A'/r^2 - \sigma^*}{\Delta}\right)_+^n\right]}{R_{LRS}}$$

$$= \frac{exp\left[-\left(C\left(\frac{r_0^2}{r^2} - 1\right)\right)_+^n\right]}{R_{HRS}} + \frac{1 - exp\left[-\left(C\left(\frac{r_0^2}{r^2} - 1\right)\right)_+^n\right]}{R_{LRS}}$$

The pre-factor $C$ and the shape factor $n$ should be the same as those found in **Section 0**, since the switching condition (being dependent on stress) and statistics are the same. The best fitting shown as the solid curve in **Figure 3.51** has the following form with an identical $n$=3.5 as in the FIB experiment but a slightly larger $C$=1.2 compared to $C$=0.71 in the FIB ion experiment. This verifies our model is self-consistent.

$$\frac{1}{R(\Omega)} = \frac{1}{50\text{M}\Omega} \times exp\left[-\left(1.2 \times \left(\left(\frac{550\mu m}{r(\mu m)}\right)^2 - 1\right)\right)_+^{3.5}\right] + \frac{1}{1200\Omega}$$

$$\times \left(1 - exp\left[-\left(1.2 \times \left(\left(\frac{550\mu m}{r(\mu m)}\right)^2 - 1\right)\right)_+^{3.5}\right]\right)$$

We note that the data of the electron bunch experiment appear to undergo a more abrupt transition than those collected in the Ga$^+$ ion bombardment experiment (**Figure 3.51** *vs.* **Figure 3.15**). Such difference might arise for the following reasons. (i) The electron bunch experiment follows an $r^{-2}$ dependence compared to the linear $I$ dependence. Therefore, the transition in the $R$-$r$ plot appears more abrupt than in the $R$-$I$ plot. (ii) In the electron bunch experiment, the resistance data are bimodal, segregated to the two binary states, HRS and LRS, even though the transition window extends over 150 μm: approximately 34% of the devices in the transition zones are in the HRS, *vs.* 66% in the LRS. This "binary" feature may have contributed to the impression that the transition is more abrupt. The binary feature may be caused by the subsidiary oscillation of the induced field: the initial induced field may have left some devices at an intermediate resistance, which may continue to shake down to the lower-resistance states when perturbed by the subsequent oscillations (of decreasing amplitudes) of the induced electromagnetic fields in the samples. As a result, the binary feature of the resistance transition in the electron bunch experiment becomes manifest.

### 3.6 Discussion

The above four experiments at different time domains leave little doubt that electron-phonon interaction is actively involved in nanometallic MIT, and the transition is fundamentally different from all other stimuli-driven reversible MIT in the literature[22-28]. A mechanistic picture of memory switching that emerges may be depicted as follows. During voltage-controlled MIT, electron filling occurs at a threshold forward bias $V^*$ when itinerant electrons are energetic enough to enter a prospective negative-$U$ trapping site, which then undergoes energy-lowering local bond distortion. This is the LRS→HRS transition. Under a reverse threshold bias, the distorted bond is inelastically restored to the original configuration by the electric force, thereby removing $\phi_{ep}$ stabilization, prompting electron de-trapping. This is the HRS→LRS transition. What we witnessed in the electron-bunch experiment was the unraveling of the negative-$U$ state: the distorted bond was mechanically restored—hence $\phi_{ep}$ removed—in about 0.1 ps, thereby converting the electron state from a stable negative-$U$ state to an unstable positive-$U$ state, which necessitated immediate (in ~1 fs) electron release and the HRS→LRS transition. Unlike the electron-phonon interaction in the global Hamiltonian of conventional MIT[29-32], in amorphous films the interaction is localized to the vicinity of trapped-electrons and their surrounding over/under-bonded bonds, which correspond to a local environment that is non-compact and soft[9]. These sites are especially susceptible to the applied force/stress/strain, allowing pivotal, localized electron-phonon interaction to turn on/off to effect both voltage and force-triggered MIT. In the past, non-electrically induced electronic transitions were seen in phase-change chalcogenide (Ge-Sb-Te) memory[33-35], by photo-induced transition[36] or dislocation jamming-triggered non-melting crystalline-to-amorphous transition[37]. Compared to these electronic transitions mediated by a bulk structural transition, nanometallic MIT involves only the local electronic/structural state and not the bulk structural state. Therefore, it is far more robust and can be triggered by an atomic-scale lever of localized electron-phonon interaction. In particular, since the sub-picosecond on/off of electron-phonon interactions controls nanometallic MIT, the switching voltage $V^*$ should be independent of switching time down to ~1 ps. This was directly confirmed in our electrical testing from 1 s to 20 ns as shown in **Chapter II**.

The idea of electron trap through the electron-phonon coupling and negative-$U$ center is similar to the mechanism of polarons. Indeed, when Anderson first proposed the negative-$U$ concept, he already recognized its close connection to bipolarons[38]. However, there is an important distinction between switchable nanometallic material and polaronic material. Polaron arises from electron-phonon interaction, which is a generic term in the global Hamiltonian: it becomes important when the coupling is strong. Naturally, this occurs in a structure that is globally susceptible to polarization. Unlike some of Anderson's materials which include elemental amorphous chalcogenides (*e.g.*, S and Se) that are highly polarizable, the unique and remarkable feature of our switchable nanometallic materials is that, globally, they are not very susceptible to polarization at all—their dielectric constant ($\varepsilon_r \sim 10$) is actually quite low (**Chapter IV**). However, being amorphous these materials do contain bonding/structure variations, which provide locally soft spots that are highly polarizable, especially when doped electrons are localized thereon. This contributes an "impurity" term to the local Hamiltonian, instead of a generic term to the global Hamiltonian. In practice, to find switchable nanometallicity with negative-$U$ centers or local polarons, one need not search the sub-universe of highly polarizable materials. Instead, one can simply introduce defects to relatively non-polar materials, or make amorphous forms of relatively non-polar materials, then dope them with electrons/holes to bond with some of the locally polarizable sites. This explains why the switchable nanometallicity phenomenon is so ubiquitously found in so many materials, which seem to share little commonality.

Our findings on electron-phonon interaction and its manifestation in nanometallic memory also shed light on resolving the long-standing dilemma in electronic memory as we discussed in **Chapter II**: electron-trapping memory that is easily programmable (at low voltage and fast speed) cannot retain memory for a long time. Indeed, almost all memory structures encounter such dilemma, and it is usually engineered around by designing a separate high-temperature (transient) route for the program/erase step— thermally assisted switching in magnetic memory[39], melting in phase-change memory[36], hot-filament formation in resistance memory[40], and hot-carrier injection in flash memory[3]—albeit always at the cost of excess power consumption or damage. Electron-phonon interaction, which works as an atomic-scale lever that can be externally triggered to readjust the local barrier height, offers a fundamentally elegant solution—in principle, such interaction can be embedded into any electronic memory structure. It already enables purely electronic nanometallic memory with wide-ranging sizes, compositions and random nanostructures. Harnessing such sub-picosecond atomic levers may further enable nanoscale electron storage and gating to usher in new nanodevices with unconventional functionalities.

### 3.7 Conclusions

(1) HRS is an energetically metastable state, while LRS is the more stable state. A one-way HRS→LRS conversion can be triggered by a purely mechanical stress, which destabilizes trapped electrons in the HRS.

(2) An isotropic compressive stress ~300 MPa as provided by a CIP can induce one-way HRS→LRS transition in 1 s to hours.

(3) A uniaxial compressive stress ~1-100 MPa as provided by an AFM tip can induce one-way HRS→LRS transition in ~1 s.

(4) A uniaxial compressive stress ~20 MPa as provided by an FIB can induce one-way HRS→LRS transition in ~1 μs.

(5) An electromagnetic-field-induced tensile stress as provided by an electron bunch can

induce one-way HRS→LRS transition in ~0.1 ps.

(6) Strong electron-phonon interactions (negative-$U$), operating at a time scale of 0.1-1 ps, exist in nanometallic devices. They stabilize the metastable state (HRS) and resolve the voltage-time dilemma in nanometallic RRAM. Properly engineered, they may also benefit other electronic designs.

Endnote: Part of this chapter is available at Ref.[45].

# Chapter IV. AC Response of Nanometallic RRAM

## 4.1 Introduction

Practical electronic devices are preferably operated at high frequency to achieve high throughput. One of the potential limiting factors for high speed devices is the capacitive and inductive components in the circuit. These components become increasingly important at higher frequencies. For example, charging and discharging of a capacitor require a certain time $\tau_c$; likewise, current build up in an inductor also need a certain time $\tau_L$. If the system clock cycle time $\tau$ approaches or even falls below $\tau_c$ or $\tau_L$, severe signal distortion and delay will occur, which may lead to malfunction. Capacitors also consume power, $\sim CV^2 f$, which becomes significant at high frequencies. For these reasons, data of the AC response are needed for predicting the ultimate speed and power of electronic devices.

Understanding and engineering the AC response is critical for modern computer and communication electronics. In VLSI design, by analyzing the causes of the AC response of MOSFET ($C_{ox}$, $C_{junction}$, *etc.*) and interconnect parasitics, circuit designers have found it possible to achieve high speed, without significant delay, by scaling down the integrated circuit. Scaling analysis also revealed that the delay issue is less important than the power issue, which becomes the "show-stopper". This led to the development of multi-core CPU in recent years, which consumes less power than single-core CPU of a higher frequency. Another example is in the RFIC world, where "simple" *LCR* elements are used to artificially engineer the input and output impedance for the best use of signal and power.

Impedance spectroscopy (IS) offers a powerful tool to explore electrical properties of materials. By superposing an AC oscillation signal of various frequencies on a DC signal, one can extract resistance, inductance, and capacitance values of the corresponding elements of a circuit. In many cases, with the aid of physical modeling, circuit elements can be individually attributed to a physical or chemical process such as conduction, dielectric relaxation, diffusion or even chemical reactions. This, for example, can differentiate device characteristics from parasitic contributors such as testing lines, electrodes, boundaries and interfaces. Impedance spectroscopy has been employed to characterize RRAM made of $Pr_{0.7}Ca_{0.3}MnO_3$[1], $NiO$[2-3], $TiO_2$[3-5] and $HfO_x$[6]. These RRAM all rely upon ionic transport, mostly in highly localized filaments which change from highly conductive ones to insulating ones depending on electric loading. These different RRAM states have been found to exhibit essentially identical capacitive behavior. This chapter describes an impedance spectroscopy study of nanometallic RRAM materials and their devices. The result reveals some important features not seen in filamentary RRAM; such features are presumably related to the creation/elimination of additional charge storage sites and conducting pathways, and as such they help provide a better understanding of the two families of RRAM. Lastly, the findings of impedance spectroscopy will be used to supplement the understanding from DC responses ("resistance" only) to provide input to establish an equivalent circuit model (**Chapter VII**) for our device.

## 4.2 Theory of Impedance Spectroscopy

### 4.2.1 Impedance Spectroscopy

Impedance spectroscopy (IS) refers to the impedance data in the frequency domain, although actual measurements are typically carried out in the time domain and the data are next Fourier transformed to the frequency domain[7]. (The transform is typically carried out by the measurement instrument itself.) In general, each circuit (or circuit element) has a generalized "impedance", which is a complex number. The impedance of standard linear circuit elements is known: for a resistive element, it is $Z=R$; an inductive element $Z=j\omega L$; and a capacitive element $Z=1/j\omega C$. The impedance of a more complicated circuit with multiple linear elements can then be obtained by simple algebra using the above. The fact that different circuit elements have distinctly different frequency responses makes it easy to interpret the experimental data sometimes: the influences of different elements are well segregated into different frequency regimes.

Several equivalent representations of impedance spectroscopy have been developed. In 1920s and 1930s, C. W. Carter introduced the circle diagram[8] and P. H. Smith introduced the Smith-Chart impedance diagram[9]. These methods were followed by the Cole-Cole plot in 1941: a plot of $Z''$ or $\varepsilon''$ on the $y$ (imaginary) axis $vs.$ $Z'$ or $\varepsilon'$ on the $x$ (real) axis. The Cole-Cole plot offers a straightforward two-dimensional representation of the conduction properties but the frequency information is made implicit. Another approach is the Bode diagram, which plots $|Z|$ (modulus) and $\phi$ (phase) $vs.$ frequency $f$. (Alternatively, one can plot $Z'$ and $Z''$ $vs.$ $f$). This can also be represented using a three-dimensional plot, introduced by J. R. Macdonald in 1981, which provides complete information including the Cole-Cole plot and the $Z'(f)$ and $Z''(f)$ plots in the same diagram[10].

The simplest physical model to employ to interpret the impedance spectroscopy is from original Debye model, which describes the frequency-dependent dielectric response in terms of a single time constant $\tau$:

$$\varepsilon = \varepsilon_\infty + \frac{\varepsilon_0 - \varepsilon_\infty}{1 + i\omega\tau}$$

In circuit terms, the model can be realized using a linear capacitor $\varepsilon_\infty C_c$ in parallel with a serial combination of a linear resistor $R$ (causing dissipative effects) and another linear capacitor $C=(\varepsilon_0-\varepsilon_\infty)C_c$, in which $R$ and $C$ are related to each other by $RC=\tau$. Its Cole-Cole plot features a half circle, shown in **Figure 4.1a** with the arrow indicating the direction of increasing frequency[7]. For a conducting system, Debye response is typically abstracted as a resistor ($R_0$) in series with a parallel $RC$ (**Figure 4.1b inset**). This leads to circuit impedance:

$$Z = R_0 + \frac{R}{1 + i\omega CR}$$

Using the real part and the imaginary part of the impedance

$$Z' = R_0 + \frac{R}{1 + (\omega CR)^2} \text{ , } Z'' = -\frac{\omega CR^2}{1 + (\omega CR)^2}$$

we obtain the following Cole-Cole form by eliminating the frequency from the above expressions

$$\left(Z' - \frac{R}{2} - R_0\right)^2 + Z''^2 = R^2$$

This is a perfect semicircle, with a radius $R$, center at $(R/2+R_0, 0)$. **Figure 4.1b** shows a 3-D representation of such Debye system.

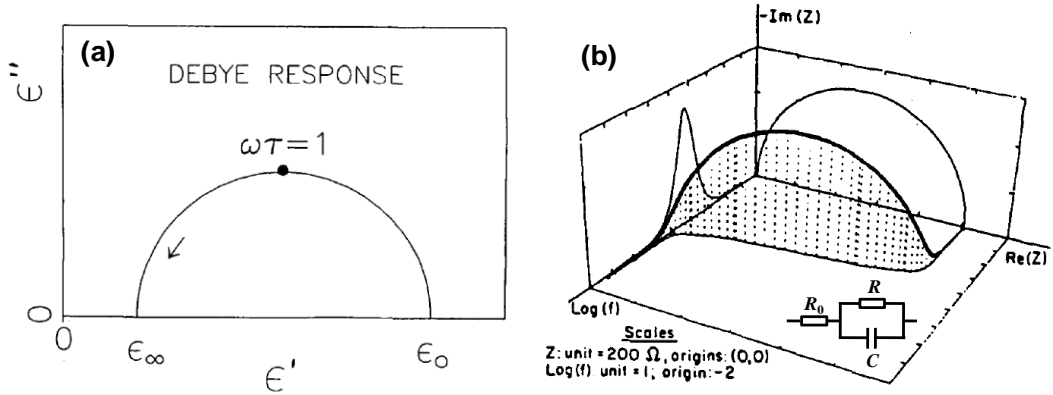

**Figure 4.1.** (a) Complex-plane plot of the complex dielectric constant for Debye frequency response. (b) 3-D perspective plot of $R_0$-$R//C$ circuit. (adapted from ref.[7]).

In practice, two complementary methods can be used to interrogate the circuit structure of an RRAM using impedance spectroscopy. The first method (**Figure 4.2a**) presumes an equivalent circuit, such as a serial $R$-$C$, and directly returns the values of "resistance" and "capacitance" by fitting the data with the equivalent circuit model. This method provides a fast measurement of $R$ and $C$. For example, it is suitable for the case of an ideal capacitor $C$ or a capacitor that is in parallel with a large resistor $R_{//}$. The method is useful because most RRAM can be thought of as a capacitor with a serially connected resistance $R_0$ due to the electrode resistance, spreading resistance, and/or load resistance. Using the above method, one can find the serial resistance $R_0$ and the parallel resistance $R_{//}$, following $Z = R_0 + R/(1 + j\omega CR)$. At large frequency ($\omega \gg RC$), the impedance of the capacitor becomes extremely small ($Z = 1/j\omega C$), which "shorts" the resistor $R$, so the total impedance can be approximated as:

$$Z \approx R_0 + \frac{1}{j\omega C} \text{, at } \omega \gg RC$$

which is like a serial $R_0$-$C$ circuit. We can conveniently use this method to investigate capacitance information. However, since it approximates $C = -1/\omega Z''$, it underestimates the actual capacitance somewhat ($Z''_{R0-R//C} < Z''_{R0-C}$).

The second method (**Figure 4.2b**) is a more rigorous analysis, which treats the circuit as a "black box". The measured impedance data ($Z'$ and $Z''$) in the frequency domain are fit with various equivalent circuits of increasing complexity until a satisfactory fit is obtained. This method is limited by the capability of the instrument to measure current. In particular, if the impedance is too high, then this method is not applicable because the current is too low to provide any useful data over a large range of frequency. In the case of HP 4192 A used in this thesis study, the instrument resolution requires $|Z| < 2$ M$\Omega$. That is, the input impedance of the instrument sets the upper limit of the circuit impedance.

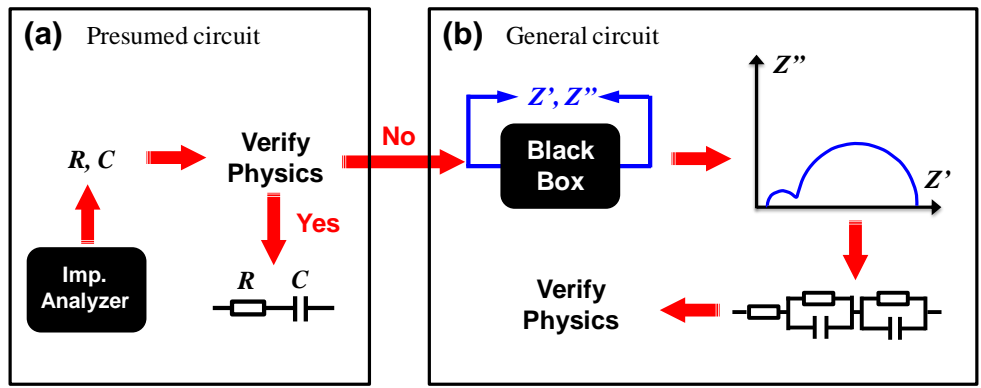

**Figure 4.2.** Two analyzing methods used in my work. (a) Presumed *R*-C serial circuit model at a fixed frequency. (b) Full impedance analysis over frequency domain.

### 4.2.2   Constant Phase Element (CPE)

A linear capacitance should be frequency independent reflecting the fact that the ability of the dielectric material (in the capacitor) to store electrical charges, by polarization, is not dependent on the speed the charge is provided or extracted. However, in some cases, this assumption is not valid. In a simple parallel *RC* circuit, such nonlinearity or frequency dispersion leads to a Cole-Cole plot of a "depressed semicircle", see **Figure 4.3**, which is a semicircle with a center some distance below the real axis. The depressed semicircle has been attributed to a number of physical causes, which share the common origin that certain material/system properties are not homogeneous. Specifically, surface (*e.g.*, electrode) roughness[11], inhomogenous composition and thickness[12], non-uniform current distribution[13] and distribution of reaction rates[14] can all contribute to the above nonlinear behavior.

The type of nonlinear capacitor associated with a depressed semicircle in the Cole-Cole plot can be mathematically represented using the following equivalent impedance [15]:

$$\frac{1}{Z_{CPE}} = (j\omega)^n Q$$

Here, *n* is an ideality factor ($n \leq 1$, with $n=1$ being ideal), *Q* has the numerical value of the admittance at $\omega=1$ rad/s, and the subscript CPE stands for constant phase element, which is the name given to such capacitor. By replacing an ideal capacitance with a CPE, the impedance of a parallel $R$-CPE (**Figure 4.3**) can be written as:

$$\frac{1}{Z} = \frac{1}{R} + (j\omega)^n Q$$

It can be further reduced into the Cole-Cole form as:

$$\left(Z' - \frac{R}{2}\right)^2 + \left(Z'' + \frac{R}{2}\cot n\frac{\pi}{2}\right)^2 = \left(\frac{R}{2\sin n\frac{\pi}{2}}\right)^2$$

by eliminating the frequency dependence between the real part and the imaginary part. The above result corresponds to a new capacitive circle with a shifted center at ($\frac{R}{2}$, $-\frac{R}{2}\cot n\frac{\pi}{2}$) and a radius $R/(2\sin n\pi/2)$. Because of rotational symmetry, we may have two equivalent views of the new circle. 1. The circle in a normal Cole-Cole plot for a linear $RC$ circuit is rotated by $(1-n)\cdot 90^o$ (as shown in **Figure 4.3**). 2. The circle is translated along $+Z''$ axis by $\frac{R}{2}\cot n\frac{\pi}{2}$ (a "depressed circle"). If a serial resistor $R_0$ is further added, the depressed semicircle is further shifted along $+Z'$ by $R_0$:

$$\left(Z' - \frac{R}{2} - R_0\right)^2 + \left(Z'' + \frac{R}{2}\cot n\frac{\pi}{2}\right)^2 = \left(\frac{R}{2\sin n\frac{\pi}{2}}\right)^2$$

By fitting the experimental data using the above formula, one can obtain $R$ and $n$. The "real" capacitance can then be taken as:

$$C = Q \times \omega_{max}^{n-1} = \frac{1}{\omega_{max}R} = \frac{(QR)^{1/n}}{R}.$$

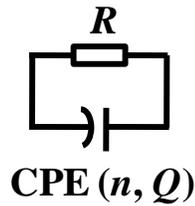
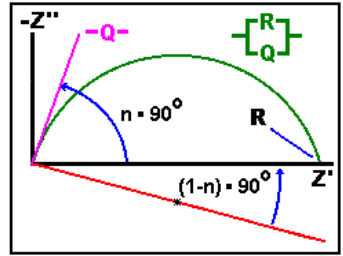

**Figure 4.3.** (left) Circuit representation of a parallel *R*-CPE. (right) Cole-Cole plot of depressed semicircle of a parallel *R*-CPE (adapted from ref. [15]).

Note that the quality of fit is typically better when the capacitor is nearly ideal, *i.e.*, *n* is close to unity.

Another way to describe a non-ideal capacitance is to generalize *C* into a complex number which is itself frequency-dependent: $C(f)=C'(f)-jC''(f)$. This representation has a transparent physical meaning: it corresponds to the real part and the imaginary part of the dielectric response, which is known to be frequency dependent. In particular, $C''(f)$ is attributed to dielectric damping, due to leakage current or internal dissipation, *e.g.*, caused by conduction. The frequency dependence of dielectric "constant" and dielectric damping is also known as dielectric relaxation, often modeled by a set of oscillators of different characteristic frequencies, thus each dominating in a different frequency regime[4]. Since the CPE impedance can be represented as a complex capacitance, the two descriptions are mathematically equivalent. This correspondence is summarized in **Table 4.1**.

| | CPE | General Capacitance |
|---|---|---|
| **Model** | 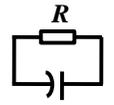 $R$ <br> **CPE** $(n, Q)$ | 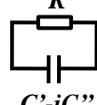 $R$ <br> $C'\text{-}jC''$ |
| **Cole-Cole EQ.** | $\left(x-\dfrac{R}{2}\right)^2 + \left(y+\dfrac{R}{2}\cot n\dfrac{\pi}{2}\right)^2$ <br><br> $= \left(\dfrac{R}{2\sin n\dfrac{\pi}{2}}\right)^2$ | $\left(x-\dfrac{R}{2}\right)^2 + \left(y-\dfrac{C''}{2C'}R\right)^2$ <br><br> $= \left(\dfrac{R}{2}\right)^2 + \left(\dfrac{C''}{2C'}R\right)^2$ |
| **Equivalency** | $C'=\omega^{n-1}Q\sin n\dfrac{\pi}{2}$ , $C''=\omega^{n-1}Q\cos n\dfrac{\pi}{2}$ | |

**Table 4.1.** CPE method and general capacitance method.

## 4.3 Experimental Procedure and Setup

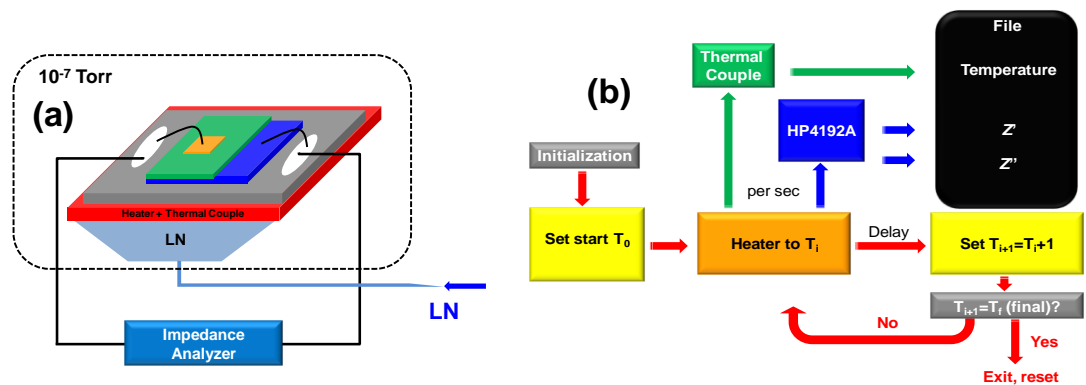

**Figure 4.4.** (a) Schematic of experiment setup and sample connections.
(b) Flow chart of computer automation program (written by LabView).

Our room temperature impedance measurements used a HP 4192A impedance analyzer. Samples were placed on a probe station (Signatone S1160) and a voltage was applied between the top and the bottom electrodes. (Current flowing from the top electrode to the bottom electrode is considered positively biased, similar to the DC configuration in earlier chapters.) Temperature spectra were investigated in a Lakeshore cryogenic probe station. The system provided a heating/cooling stage from 77 K to 400 K in a $10^{-7}$ Torr level vacuum. Cooling was provided by purging the sample stage with liquid nitrogen (LN) while heating was provided by a resistance heater (**Figure 4.4a & Figure 4.5c**). Built-in probe tips are available for small device measurement but wire-bonded connections (**Figure 4.4a**, **Figure 4.5a & b**) are more mechanically reliable. Testing protocol in **Figure 4.4b** includes a customized LabView program developed for controlling/ monitoring temperature and performing impedance analysis.

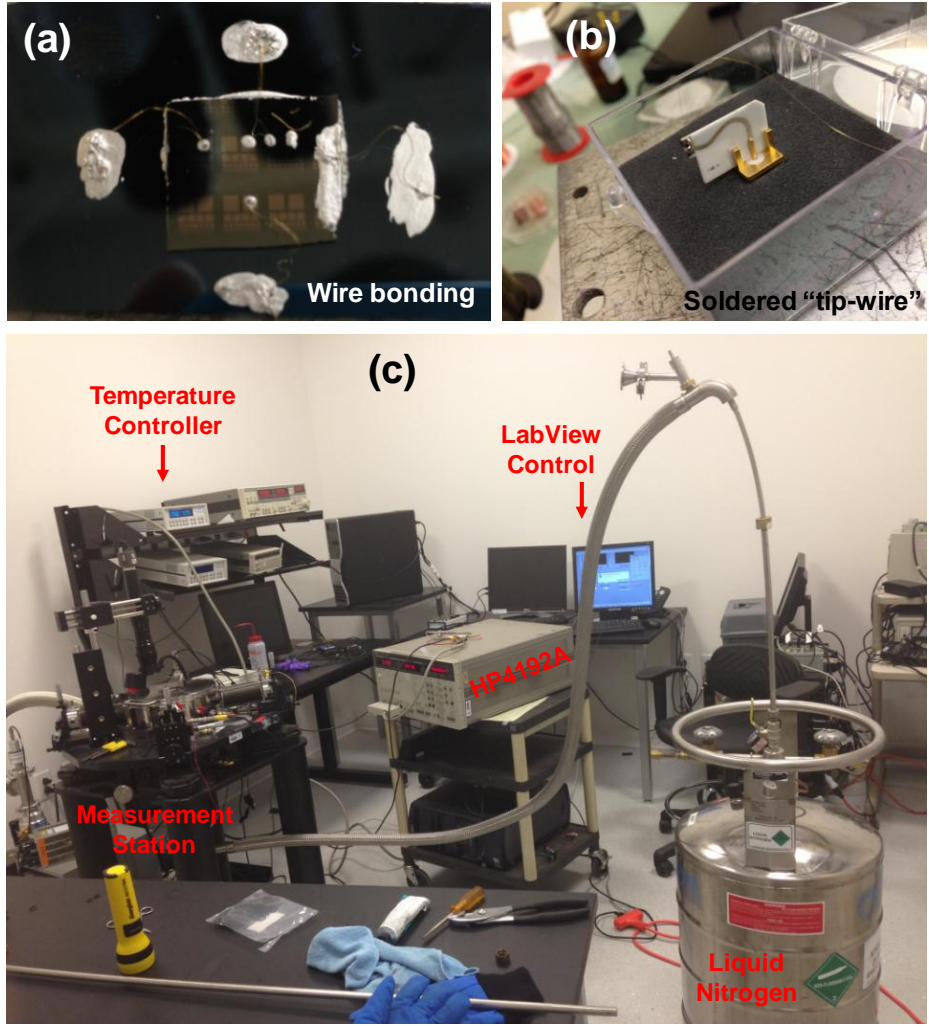

**Figure 4.5.** (a) Optical image of wire-bonded sample. (b) Reconfigured probe tip with connected gold wire (2 mils). (c) Entire test bench.

## 4.4 Results

### 4.4.1 Resistance State Dependence

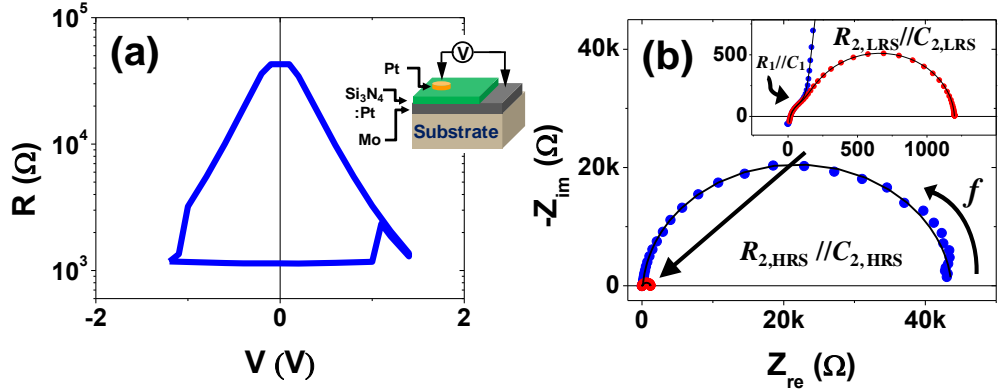

**Figure 4.6.** (a) *R-V* switching curve. (b) Cole-Cole plot for HRS and LRS at 0 V DC bias. (Device: 90% Si$_3$N$_4$:10% Pt, $\delta$=10 nm, $d$=1100 μm).

Data are first presented to illustrate the general information obtainable from impedance spectroscopy and standard model fitting. **Figure 4.6a** is a typical bipolar switching curve of a device following the voltage sweep sequence: 0 V, to 1.4 V, to -1.2 V, and to 0 V, where a positive voltage indicates current flowing from the top to the bottom electrode. Impedance (*Z*) spectroscopy of this device at 0 V under a 100 mV AC excitation (10 Hz to 5 MHz) are represented in the Cole-Cole plots (Nyquist plots) in **Figure 4.6b** (for the HRS) and its inset (for the LRS). For both states, the plots mainly consist of two semi-circles. The low frequency data of the HRS lie on a larger semi-circle with a diameter ~42 kΩ, having a "peak" at the resonance frequency $f_{max}$~330 Hz, while the LRS data fall on a much smaller semi-circle with a diameter ~1100 Ω and a resonance frequency $f_{max} \sim 17$ kHz. Since the imaginary part of impedance $Z_{im}$ is negative, these semi-circles are representative of a capacitive parallel $RC$ element. At the high frequency limit of our tests, the data of the HRS and LRS converge to a common capacitive semi-circle (**Figure 4.6b** inset), which eventually crosses over to an inductive behavior ($Z_{im} > 0$) at very high frequencies (>1 MHz). This convergence at high frequency suggests that there is a common resistive element (with unchanged resistance between the two states) in series with the two distinct capacitive $RC$ elements, one for each state.

Guided by the common practice of impedance spectroscopy, we use the following equivalent circuit (**Figure 4.7**) to extract the component characteristics from the impedance data:

$$Z = j\omega L + R_0 + R_1//C_1 + R_2//C_2 = j\omega L + R_0 + \frac{R_1}{1 + (j\omega)^{n1} Q_1} + \frac{R_2}{1 + (j\omega)^{n2} Q_2}$$

Here, the $j\omega L + R_0$ terms are responsible for the behavior at the high frequency limit, arising from the inductance of the testing circuit (line) and other non-capacitive parasitic resistive components serially connected to the memory cell, which include the geometric spreading resistance of the planar bottom electrode and any load/line resistance. (The parasitic inductance of the testing circuit is probably from the probe-station line made of a "twisted" metal line with an inductance $L \sim 3$ μH.) The term $R_1//C_1 = \frac{R_1}{1 + (j\omega)^{n1} Q_1}$ is used to describe the high frequency semi-circle that is also common to the LRS and the HRS; it may arise from the electrodes, electrode/film interfaces and other serial elements. (Our model in **Chapter II** suggests that the nanometallic film is divided into two parts, the part closer to the lower work function electrode is always conducting—thus unaffected by switching, while the other part closer to the higher work function electrode undergoes resistance switching.) Here, a non-ideality factor $n_1$ and a constant phase element $Q_1$ have been introduced to account for the possible time constant ($RC$) distribution so that the capacitance follows $C_1 = (Q_1R_1)^{1/n1}/R_1$. A similar fourth term $R_2//C_2 = \frac{R_2}{1+(j\omega)^{n2}Q_2}$ is used to describe the low frequency semi-circle, which is apparently closely related to the switchable serial part of the nanometallic film. It again is described by a non-ideality factor $n_2$ and a constant phase element $Q_2$ to allow the capacitance to vary as $C_2 = (Q_2R_2)^{1/n2}/R_2$. (It contains a parallel capacitor since, after all, the MIM device, physically, may be regarded as a capacitor to the first order approximation.) This model gives good fitting to the data as shown by the solid fitting curves in **Figure 4.6b**. Here, the parameters used are: $R_1$=86 $\Omega$, $C_1$=3.8 nF, $R_{2,\text{HRS}}$= 42.6 k$\Omega$, $R_{2,\text{LRS}}$=1047 $\Omega$, $C_{2,\text{HRS}}$=11.3 nF, $C_{2,\text{LRS}}$=9.3 nF, $n_1$=1, $n_{2,\text{HRS}}$=0.95 and $n_{2,\text{LRS}}$=0.97.

In the above, the near unity values of $n_2$ indicate that the capacitor in the switchable part of the film is apparently a linear capacitor, which changes its capacitance depending on the resistance state. This indicates that the nanometallic capacitors of both states are uniform with a narrow time constant distribution around $\tau$=1/$f_{\text{max}}$, as in an ideal capacitor with $n$=1, and that both the HRS and the LRS can be considered as a set of parallel, linear RC circuits. The 20% difference between the HRS and LRS capacitance at zero bias is rather large and is well beyond the uncertainty of the data and fitting (typically <5%).

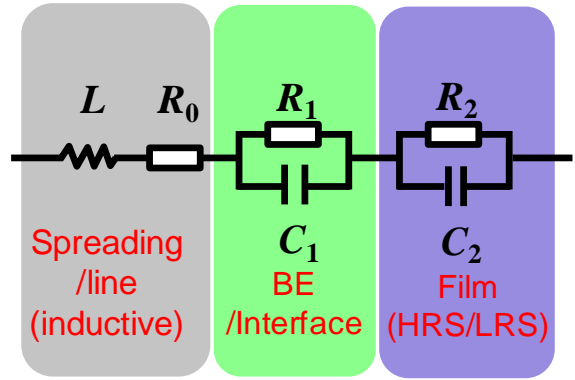

**Figure 4.7.** Equivalent circuit.

### 4.4.2   Area and Thickness Dependence

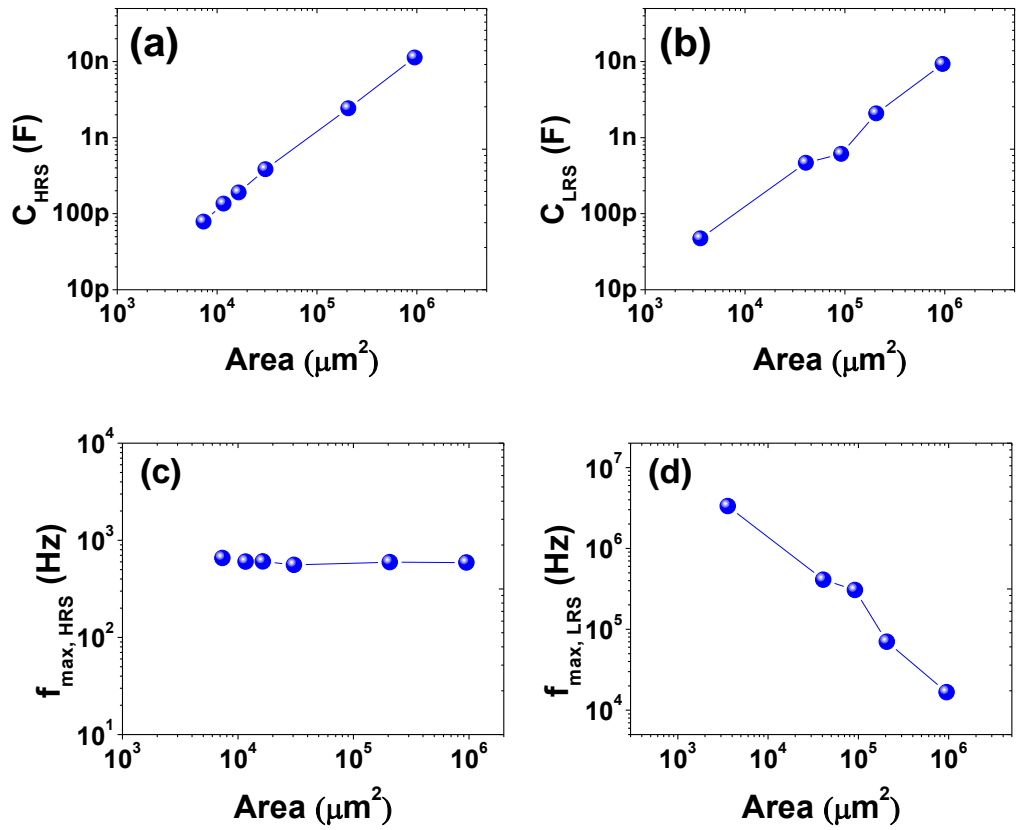

**Figure 4.8.** Scaling behavior of capacitance following $C \sim A$, where $A$ is cell area. (a) HRS and (b) LRS capacitance linearly increases with cell area. Resonance frequency as a function of device area for (c) HRS and (d) LRS. (nanometallic film: $Si_3N_4$:6%Cr, $\delta$=10 nm).

Having established the validity of the parallel RC circuit, we next investigated the area and thickness dependence to verify that the capacitance is indeed made of a uniform linear dielectric. Size dependence was studied in $Si_3N_4$:6%Cr cells with a diameter ranging from 20 μm to 1100 μm. As shown in **Figure 4.8a & b**, both the HRS and LRS capacitances scale linearly with the device area: $C \propto A$. This confirms that the device is made of a uniform dielectric, at the length scale of the order of the cell dimension. Since resistance (HRS) is inversely proportional to $A$, we conclude that the $RC$ product should remain identical ($\sim \rho\varepsilon$) for devices of different size. This immediately implies a size-independent resonance frequency ($f_{max,HRS}$ or $\omega_{max,HRS}$), verified in **Figure 4.8c**. On the other hand, resistance (LRS) is weakly dependent on $A$, thus resonance frequency ($f_{max,LRS}$ or $\omega_{max,LRS}$) is inversely proportional to device area.

HRS thickness dependence was investigated in a set of devices made of Mo/$Si_3N_4$:Cr/Pt with various Cr content and with various thicknesses but a fixed lateral size (diameter: 200 μm). Because of the limited bandwidth of the instrument, $R$-$C$ measurements (**Figure 4.2a**) at a fixed frequency of 100 kHz were used to extract the capacitance value of the HRS assuming it is a parallel RC circuit. (We ignored the line/spreading/BE resistance, which is relatively insignificant for the HRS.) As shown in **Figure 4.9**, capacitances of various compositions follow a linear scaling law $C \propto 1/\delta$, indicating that the capacitance is a uniform linear dielectric ($C=\varepsilon_r\varepsilon_0 A/\delta \propto 1/\delta$) over the length scale of a few nm, and is not one in series with another (interface) capacitance. However, because the HRS resistance is exponentially dependent on thickness ($\delta$) but capacitance is merely linearly dependent on $\delta$, the $RC$ product and the resonance frequency ($f_{max}=1/2\pi RC$) is exponentially thickness dependent (**Figure 4.10**).

LRS thickness dependence was also investigated in a set of devices made of Mo/Si$_3$N$_4$:4%Pt/Pt with various thicknesses but a fixed lateral size (512 μm). Identical (negative) voltage procedure (0V⟶ -2V⟶ 0V) was used to obtain LRS in order to provide a fair comparison and Cole-Cole plots were used to extract $R$ and $C$, respectively. As shown in **Figure 4.11a**, LRS capacitances also follow a linear scaling law $C \propto 1/\delta$, indicating that the capacitance is a uniform linear dielectric ($C=\varepsilon_r\varepsilon_0 A/\delta \propto 1/\delta$). However, because LRS resistance is load/compliance controlled, $R$ only shows a weak thickness dependence (**Figure 4.11b**). The resonance frequency ($f_{max}=1/2\pi RC=\delta/2\pi R\varepsilon_r\varepsilon_0 A$) is linearly dependent on thickness (**Figure 4.11c**).

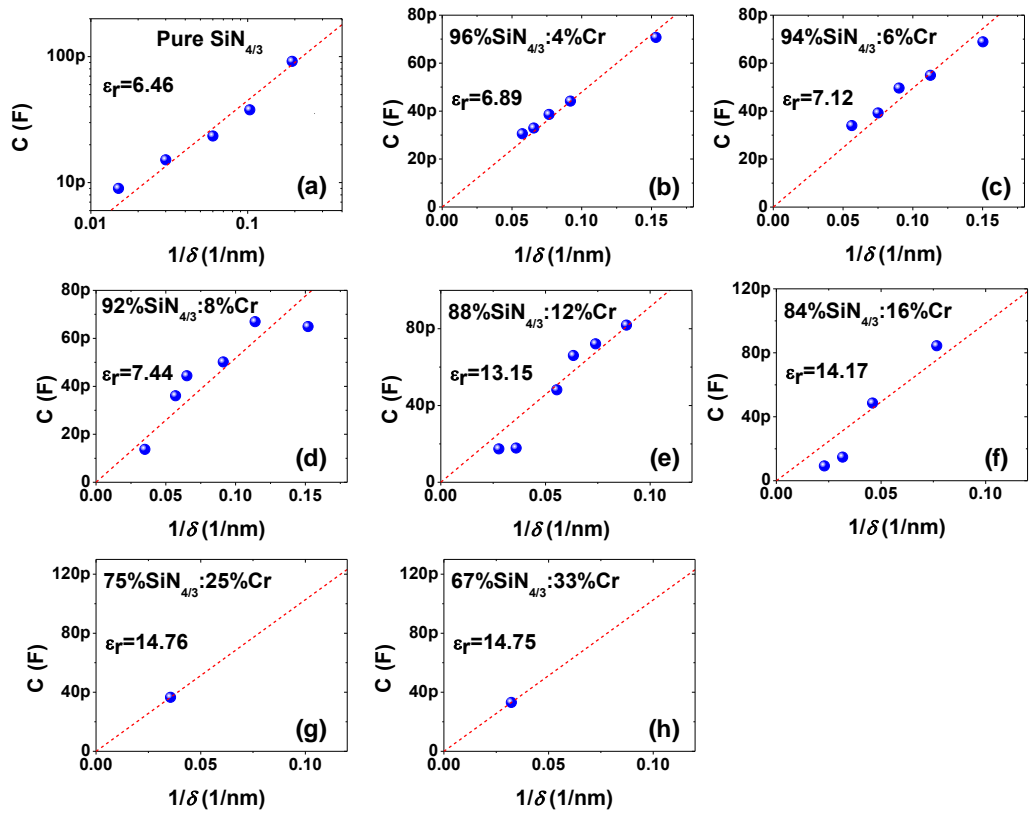

**Figure 4.9.** HRS capacitance (*C*) *vs.* reciprocal thickness (1/δ) for various metal concentration. (a) SiN$_{4/3}$, (b) 96% SiN$_{4/3}$:4%Cr, (c) 94% SiN$_{4/3}$:6%Cr, (d) 92% SiN$_{4/3}$:8%Cr, (e) 88% SiN$_{4/3}$:12%Cr, (f) 84% SiN$_{4/3}$:16%Cr, (g) 75% SiN$_{4/3}$:25%Cr, (b) 67% SiN$_{4/3}$:33%Cr. (*d*=200 μm)

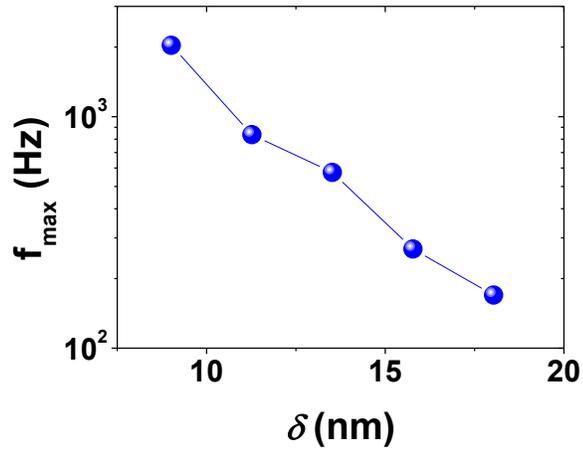

**Figure 4.10.** HRS resonance frequency as a function of thickness (nanometallic film: $Si_3N_4$:12%Cr).

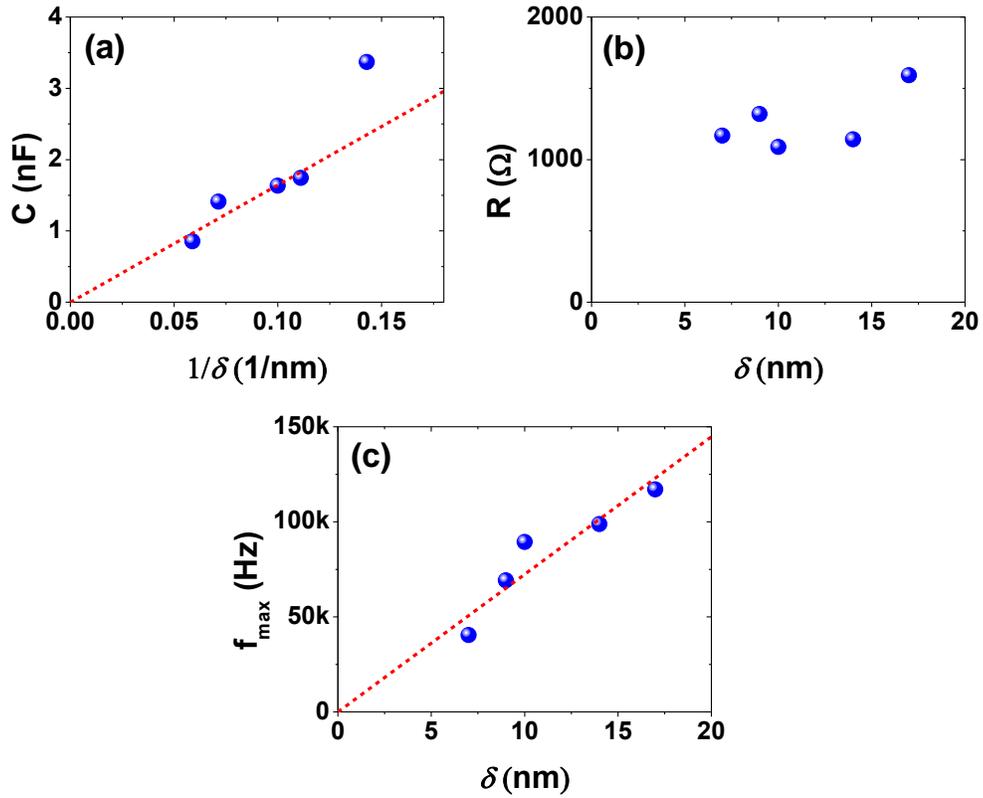

**Figure 4.11.** Thickness dependence of LRS impedance. The LRS are obtained by using identical (negative) voltage procedure: 0V⟶ -2V⟶ 0V. (a) LRS capacitance (*C*) *vs.* reciprocal thickness (1/δ). (b) LRS resistance (*R*) thickness (δ). (c) LRS resonance frequency (*f*$_{max}$) *vs.* thickness (δ). (Nanometallic film: Si$_3$N$_4$:4%Pt, device size: 512×512 μm$^2$)

### 4.4.3 Composition Dependence

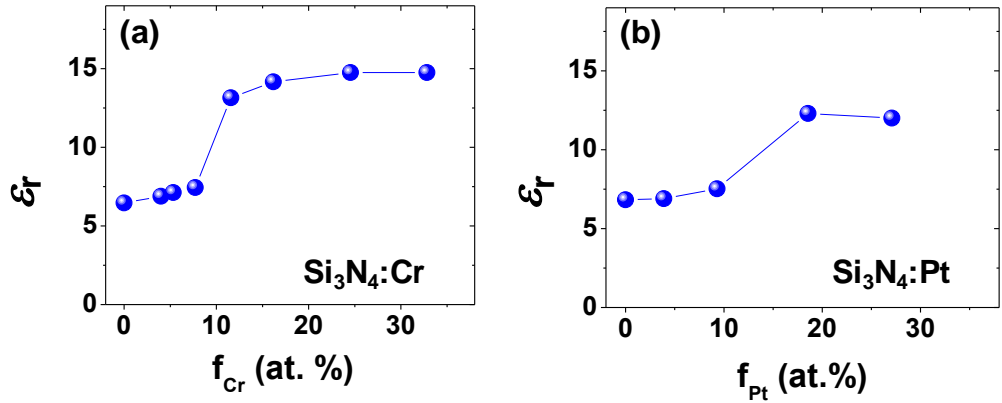

**Figure 4.12.** Relative dielectric constant $\varepsilon_r$ of (a) (1-*f*)Si$_3$N$_4$:*f*Cr films *vs.* Cr composition. Data are from capacitance values $C=\varepsilon_o\varepsilon_r A/\delta$ ($\varepsilon_o$ is the permittivity of vacuum) collected in **Figure 4.9**; (b) (1-*f*)Si$_3$N$_4$:*f*Pt films *vs.* Pt composition.

Supported by the above area and thickness dependence, which is consistent with the behavior of a uniform linear dielectric capacitor, we will from now on extract the relative dielectric constant from the capacitance data using the relation $C=\varepsilon_r\varepsilon_0 A/\delta$. These values are already listed in **Figure 4.9** for the HRS of various Cr concentrations in Si$_3$N$_4$. **Figure 4.12a** summarizes computed relative dielectric constants $\varepsilon_r$ of Si$_3$N$_4$:Cr films in the HRS (same data as in **Figure 4.9**). As *f* increases, $\varepsilon_r$ initially rises linearly, then abruptly increases at a certain critical $f$ (~10%) beyond which $\varepsilon_r$ apparently saturates. We also computed the relative dielectric constant of the LRS capacitance for the sample described in the last section, made of 90% $Si_3N_4$:10% Pt. Assuming the capacitance value of HRS and LRS differs by ~20%, the relative dielectric constant of the LRS is ~5.6 for this low-metal-concentration device and ~12 for a high- metal-concentration device. Very similarly behaving data are shown in **Figure 4.12b** for $Si_3N_4$:Pt films in their HRS states, again showing a sudden rise, at $f$~15%. Both figures share the same $\varepsilon_r$ value of 6.5 at $f$=0, which is reasonably close to the data in the literature[16-17]: $\varepsilon_r$=7.1 for a 97% dense $\alpha$-$Si_3N_4$ ceramic[16], and $\varepsilon_r$=7.0 for a PEPVD amorphous film of unknown density[17]. (Our film most likely contains some porosity, which decreases the relative dielectric constant.) Therefore, without metal dopant, the amorphous $Si_3N_4$ film in our RRAM has a similar dielectric property of a dense amorphous $Si_3N_4$ film.

### 4.4.4 Voltage/Field Dependence

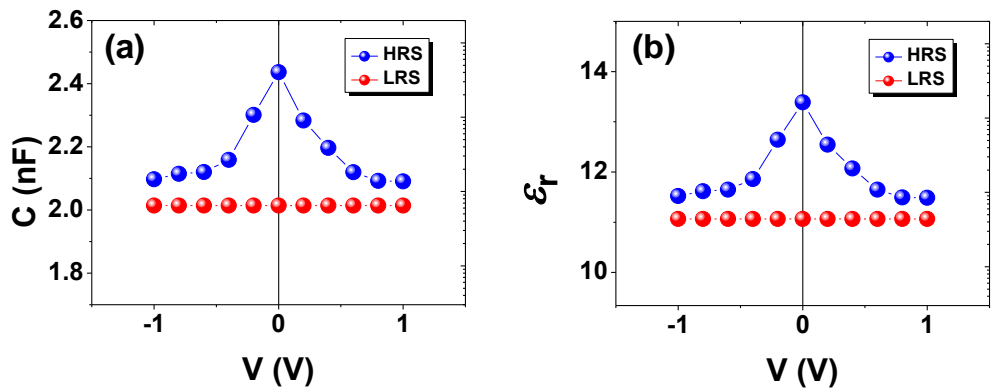

**Figure 4.13.** (a) Capacitance *vs.* bias voltage for HRS and LRS. (b) Converted dielectric constant *vs.* bias voltage using nominal geometry (nanometallic film: $Si_3N_4$:12%Cr, $\delta$=10 nm, $d$=512 μm)

**Figure 4.13** shows the voltage dependence of capacitance and calculated apparent dielectric constant (using nominal geometry) in a $Mo/Si_3N_4$:Cr/Pt nanometallic RRAM. The HRS dielectric constant exhibits a weak but obvious voltage dependence: it displays a maximum at 0 V, and monotonically and symmetrically decreases as the DC bias increases in either polarity. In contrast, the LRS dielectric constant is voltage independent. A similar voltage dependence was also obtained for $Mo/Si_3N_4$:Al/Pt, $Mo/Si_3N_4$:Cr/Pt, $Mo/Si_3N_4$:Ta/Pt, $Mo/SiO_2$:Pt/Pt, $Mo/SiO_xN_y$:Pt/Pt, $Mo/Al_2O_3$:Pt/Pt as shown in **Appendix**.

### 4.4.5 Temperature Dependence

The Cole-Cole plot of zero-bias HRS and LRS impedance of a $Mo/Si_3N_4$:10%Pt/Pt device determined at various temperature are shown in **Figure 4.14a & b**. Similar to the room temperature result, all complex impedance shows a capacitive behaviors with two well-distinguished $R//C$ elements corresponding to two separate arcs in two different frequency regions. Again, the high frequency arc is independent of the resistance state; moreover, it seems to be independent of temperature. Therefore, it most likely corresponds to some unchanged element arising from the electrodes, electrode/film interfaces or a serial part of the nanometallic film that is not affected by resistance switching or temperature (see **0**). As before, the low frequency arcs obviously correspond to the switchable nanometallic film since they are state (HRS & LRS) dependent. Moreover, they are temperature dependent. By repeating the same fitting procedure for all curves, we obtained the capacitance (**Figure 4.14c**), resistance (**Figure 4.14d**) and associated ideality factor $n$ (**Figure 4.14c inset**) of the HRS and LRS at different temperatures. The capacitance results in terms of relative dielectric constant are presented in **Figure 4.14c inset**.

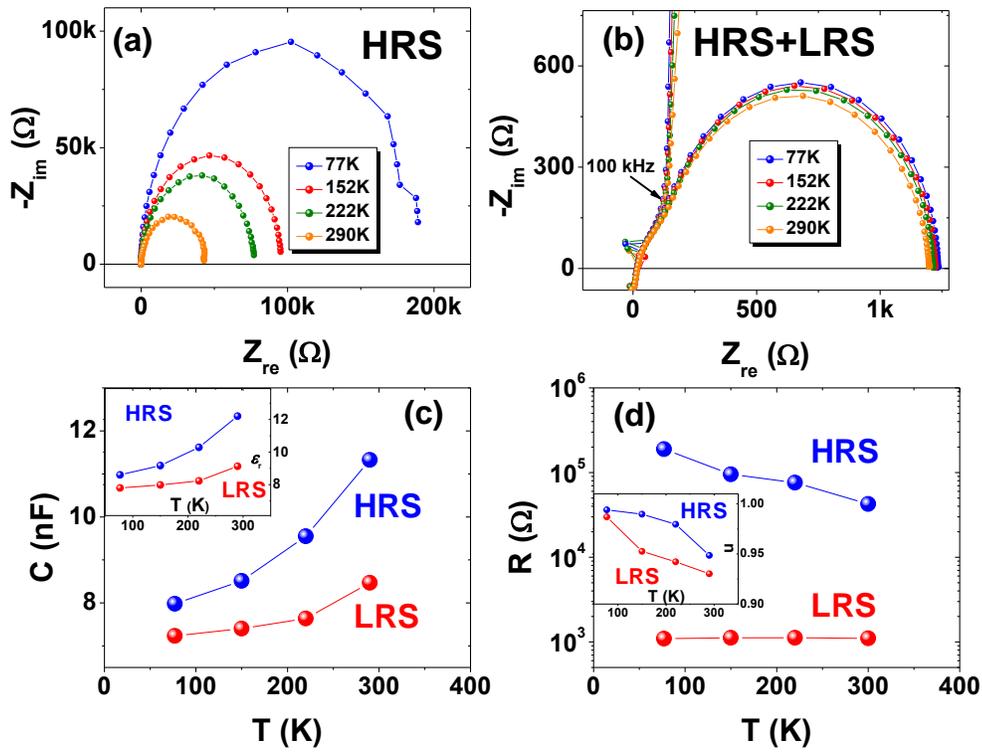

**Figure 4.14.** Impedance data at 0 V DC bias (device size: $d$=1100 μm). (a) Cole-Cole plot of HRS under various temperatures. (b) Cole-Cole plot of HRS (same as (a)) and LRS under various temperature. (c) Calculated capacitance (low frequency arc) after fitting data in (b). Inset: converted relative dielectric constant $\varepsilon_r$ *vs.* $T$. (d) Calculated resistance (low

frequency arc) after fitting data in (b). Inset: ideality factor *n vs. T.* ($f_{Pt}$=10%)

Impedance data under a nonzero bias were also collected at various temperatures, as shown in **Figure 4.15**. Again, by repeating the same fitting procedure, we calculated the device capacitance (**Figure 4.15a**) and resistance (**Figure 4.15b**) at each voltage bias and temperature. The capacitance of the HRS again shows a voltage dependence similar to the one seen in **Figure 4.13** at room temperature, but the nonlinearity is much smaller at lower temperature. In contrast, although the resistance also decreases with voltage (*i.e.*, it is non-Ohmic) as usual, the non-linearity is much stronger at lower temperature. The resonance frequency, which is related to the reciprocal product $1/2\pi RC$, is also temperature dependent (**Figure 4.15c**).

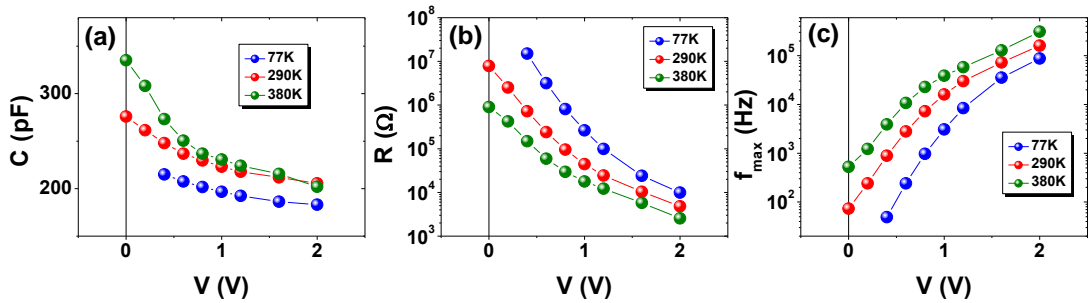

**Figure 4.15.** Impedance data under various temperatures at nonzero DC bias (device size: $200\times200$ μm$^2$). (a) Calculated *C-V* dependence from Cole-Cole plot under various temperatures. (b) Calculated *R-V* dependence from Cole-Cole plot under various temperatures. (c) Resonance frequency $f_{max}$ under various temperatures (nanometallic film: Si$_3$N$_4$:12%Cr).

From the complete impedance data in **Figure 4.14b**, we found frequency $f$=100 kHz is especially useful for our measurements: it is high enough for the low frequency $R_1//C_1$ arc (thus dominated by $C_1$) but low enough for the high frequency $R_2//C_2$ arc (thus dominated by $R_2$). Therefore, at $f$=100 kHz the equivalent circuit can be approximated as ($R_0$+ $R_2$)-$C_1$, which is a serial $R$-$C$ that allows the use of Method 1 (see **Section 4.2.1**) for a quick estimation of $C_1$. **Figure 4.16** shows the temperature dependent capacitance (dielectric constant) for different device sizes determined at $f$=100 kHz. After renormalizing capacitances by their nominal geometry, all sizes exhibit identical $\varepsilon_r(T)$ behavior (**Figure 4.16b**), which confirms that the nanometallic device is dielectrically uniform. It also demonstrates that at 100 kHz the dielectric constant increases with temperature.

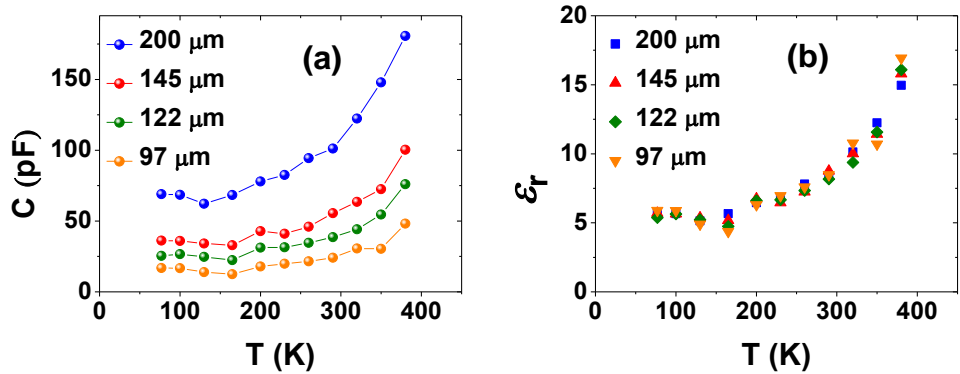

**Figure 4.16.** (a) Capacitance *vs.* temperature measured by method 1(see **0**) assuming $R$-$C$ circuit. (b) Dielectric constant *vs.* temperature by converting (a) with nominal geometry of devices.

Dielectric properties measured at 100 kHz for other nanometallic films of different metal compositions are illustrated by a set of dispersion curves $\varepsilon'(T)$ and tan $\delta$ $(T)$ in **Figure 4.17**. At low $f_{metal}$ ($f_{Cr}$=1%, 5%, 10%), the dielectric constant is weakly dependent on temperature (with a slightly positive slope). At high $f_{metal}$ ($f_{Cr}$=16%, 30%), it exhibits a much stronger temperature dependence: it initially increases with temperature (similar to the one shown in **Figure 4.14c**), then reaches a maximum around 400 K, beyond which there is a tendency for it to decrease. At ~400 K, the dielectric constant of these high $f_{metal}$ films is several times of that of the low-temperature data and the low $f_{metal}$ films.

A limited study was also made on the frequency dependence of dielectric response. As shown in **Figure 4.18** for $f_{Cr}$=16% sample, both the dielectric constant and its temperature dependence decreases with frequency, being rather flat giving $\varepsilon_r$~10 at 1 MHz. This strongly suggests that (electron) conduction plays a central role in the dielectric response. This is also evident from the tan$\delta$ which rapidly increases at less than 10 kHz, indicating that only data at higher than 10 kHz reflect capacitative behavior.

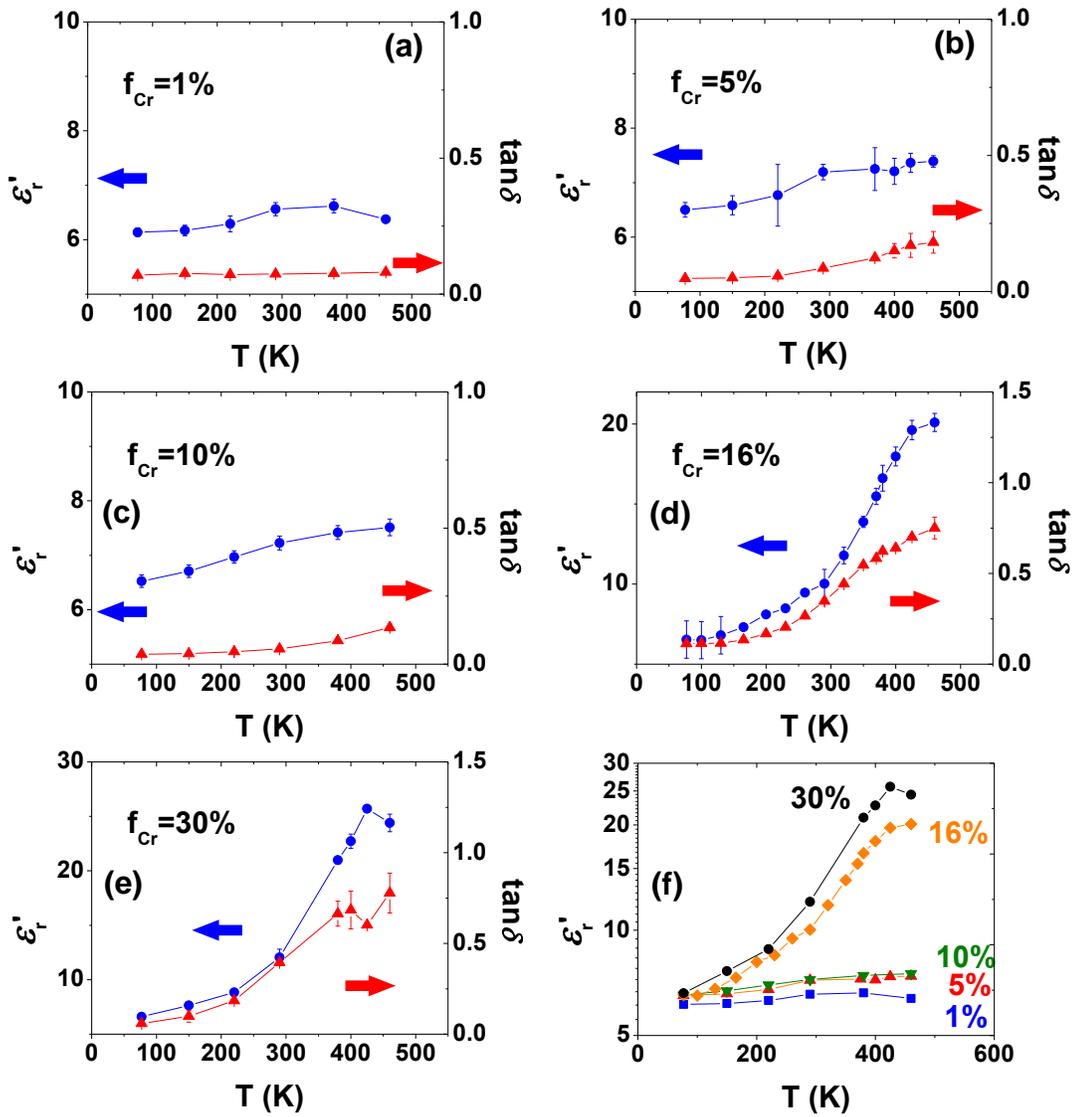

**Figure 4.17.** Dielectric constant ($\varepsilon_r$') and tangent loss (tan $\delta$) measured at $f$=100 kHz for (a) $f_{Cr}$=1%, (b) $f_{Cr}$=5%, (c) $f_{Cr}$=10%, (d) $f_{Cr}$=16%, (e) $f_{Cr}$=30%. (f) A summary of dielectric constant in (a)-(e).

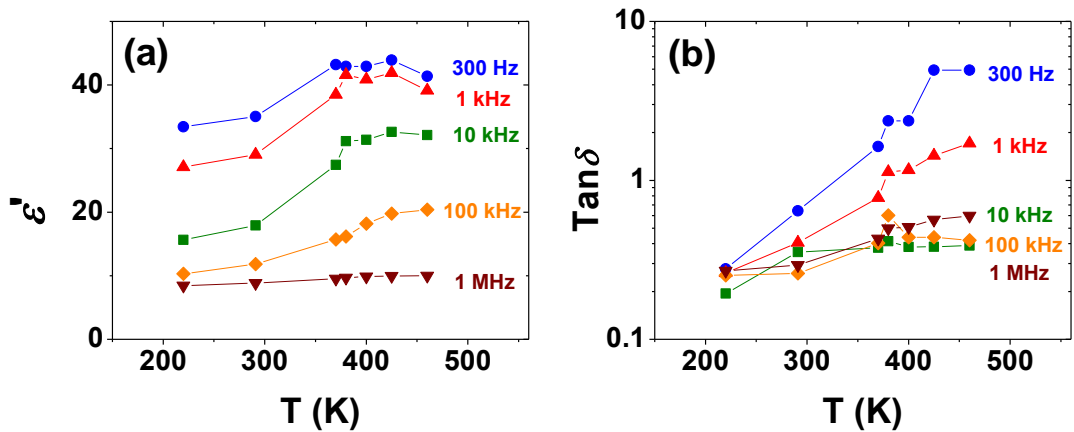

**Figure 4.18.** (a) Dielectric constant ($\varepsilon$') and (b) tangent loss (tan $\delta$) measured for $f_{Cr}$=16% sample at various frequencies.

### 4.4.6 A Comparison with Filamentary HfO₂ RRAM

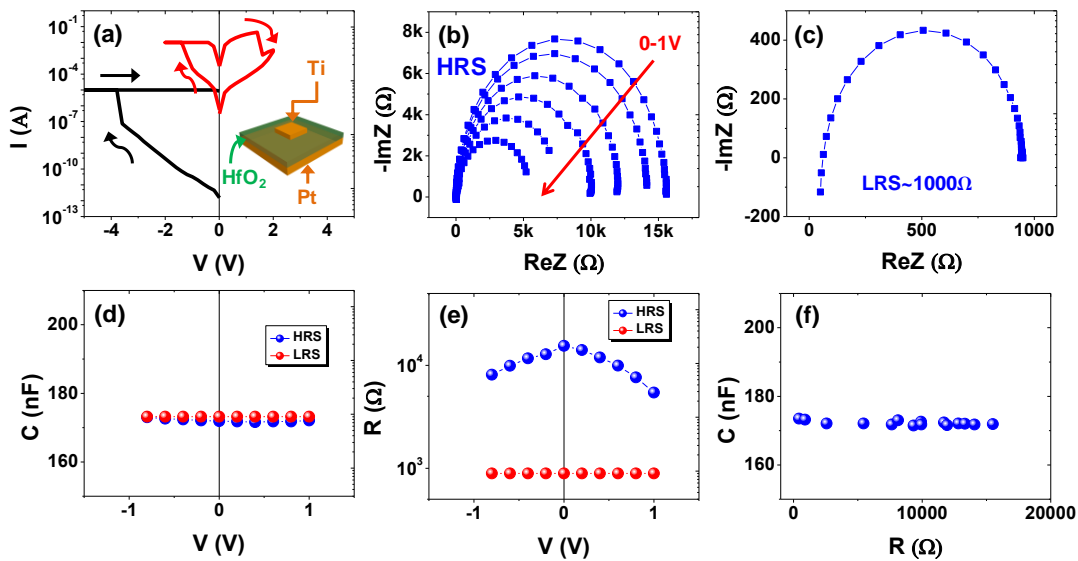

**Figure 4.19.** (a) Electroforming process and subsequent *I-V* switching curve of Pt/HfO₂/Ti filamentary RRAM. (b) Cole-Cole plot of HRS under various bias. (c) Cole-Cole plot of LRS. (d) Fitted capacitances of different states as a function of voltage. (e) Fitted resistances of different states as a function of voltage. (f) Capacitance *vs.* resistance.

Conventional filamentary RRAMs such as NiO (ref.[2]), TiO$_2$ (ref.[4]) and HfO$_x$ (ref.[6]) feature the same capacitance for different resistance states and under different DC voltage bias. This is the case found in our RRAM as described above. In order to exclude the possibility that our different findings might stem from any instrument/method-related artifact, we also fabricated and characterized Pt (BE) /HfO$_2$/ Ti (TE) RRAM, which is a valence change memory (VCM) type of filamentary RRAM. (Here, the 10 nm HfO$_2$ layer was deposited by atomic layer deposition (ALD) at 150 $^o$C.) Virgin devices were found to be in a very high resistance state (>1 G$\Omega$), which required a standard electroforming process (the black curve in **Figure 4.19a**) to form conducting filament(s). After electroforming, the device operated as a bipolar RRAM with a much higher current value (the red curve in **Figure 4.19a**). As shown in **Figure 4.19b & c**, all states exhibit a capacitive behavior, with the HRS having low frequency arcs that are voltage dependent. The capacitance and resistance values shown in **Figure 4.19d, e & f** are independent of the resistance state and the voltage bias (**Figure 4.19d & f**), in agreement with the literature[6]. Therefore, the different behavior of nanometallic RRAM shown in **Figure 4.13 & Figure 4.25** is not an artifact.

## 4.5 Discussion

### 4.5.1 Origin of Nanometallic Dielectric Behavior

In this section, we will discuss possible mechanisms for nanometallic dielectric behavior and conclude that it arises from a combination of factors due to changes of morphology/shape, effective geometry and metallicity.

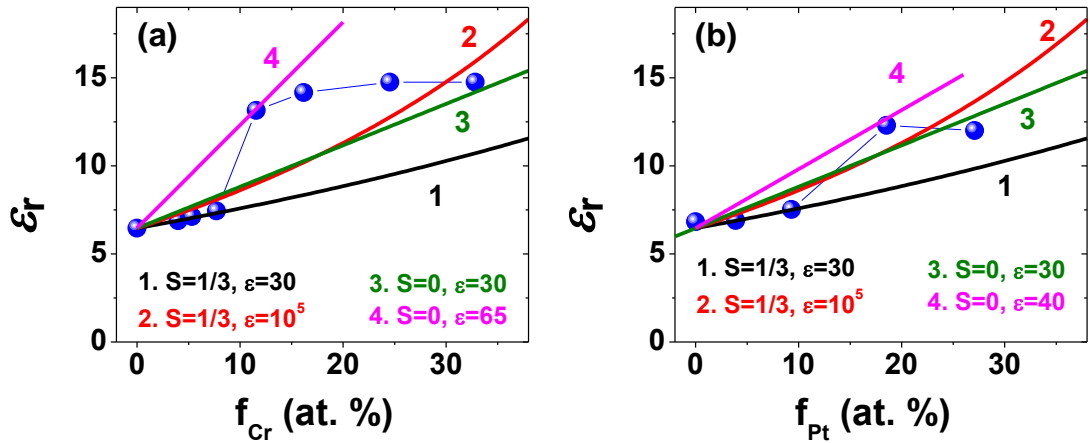

**Figure 4.20.** Fitting results for **Figure 4.12** using effective medium theory for (a) $Si_3N_4$:Cr and (b) $Si_3N_4$:Pt nanometallic films. Curves (1-4) are predictions of Maxwell-Garnett effective medium theory using following parameters: (1) spherical particles ($S$=1/3) with $\varepsilon_r$=30; (2) spherical particles with $\varepsilon_r = \infty$ (metallic); (3) needle-like particles ($S$=0) with $\varepsilon_r$=30; (4) needle-like particle ($S$=0) with $\varepsilon_r$=65 ($Si_3N_4$:Cr) or $\varepsilon_r$=40 ($Si_3N_4$:Pt). Curve (1) fits the data of $f$<0.1. Data jump at $f$~0.1 requires transition from Curve (1) to Curve (4).

The effective dielectric constant of a two-phase composite containing second-phase particles embedded in a dielectric matrix is described by the Maxwell-Garnett equation for effective medium[18-19]:

$$\frac{\varepsilon_r - \varepsilon_i}{(\varepsilon_r - \varepsilon_i)S + \varepsilon_i} = f \frac{\varepsilon_c - \varepsilon_i}{(\varepsilon_c - \varepsilon_i)S + \varepsilon_i}$$

In the above, $\varepsilon_i$ is the relative dielectric constant of the matrix, $\varepsilon_c$ is that of the second phase, $\varepsilon_r$ is that of the composite, and $S$ is a shape factor (also called the Lorentz factor) which is reasonably bounded between $S=1/3$ (for spheres) and $S=0$ (for cylinders or needles aligned along the applied field). In general, for second-phase particles that have a higher dielectric constant than that of the matrix, the smaller the shape factor, the stronger their effect on enhancing $\varepsilon_r$.

Using the above equation and assuming $S=1/3$ at low $f$, we found the best-fit $\varepsilon_c$ is 30 for our data of both Si$_3$N$_4$:Cr and Si$_3$N$_4$:Pt (curve 1 in **Figure 4.20**); metallic spheres ($\varepsilon_c=\infty$) are ruled out because they lead to a slope (curve 2) twice that of the data plot. On the other hand, the large abrupt rise in $\varepsilon_r$ observed at $f\sim0.1$ cannot be accounted for by a shape change or a property change alone. The most extreme shape change from $S=1/3$ to $S=0$ while keeping $\varepsilon_c=30$ would predict a jump from curve 1 to curve 3 giving a rise at $f=0.1$ that is much smaller than observed; the most extreme property change from $\varepsilon_c=30$ to $\varepsilon_c=\infty$ while keeping $S=1/3$ would predict a jump from curve 1 to curve 2 that is equally inadequate to explain the observation. Only a simultaneous change in shape *and* property (for example, from $\varepsilon_c=30$ and $S=1/3$, curve 1, to $\varepsilon_c=65$ and $S=0$, curve 4 in **Figure 4.20a**) can account for the sharp $\varepsilon_r$ rise at $f\sim0.1$. This implies that there are at least two apparently coupled $f$-triggered morphological and insulator/metal transitions happening to the metal-rich clusters. In the literature, a similarly coupled morphological (also involving size) and property transition (also implicating insulator/metal transition) was observed for supported catalyst of Au nanoparticles[20-21]. This postulation is indeed supported by the nanostructures of high $f$ nanometallic films shown in **Figure 4.21**: metallic clusters partially form needle-like entities of a high aspect ratio. Note, however, that when the metal composition is rich, it is not possible to have all the metal atoms behaving like being included in metallic particles. This is because even the most conservative estimate of spherical metallic particles (curve) would give an overestimate of the dielectric constant at $f_{Cr}$ and $f_{Pt} > 0.3$.

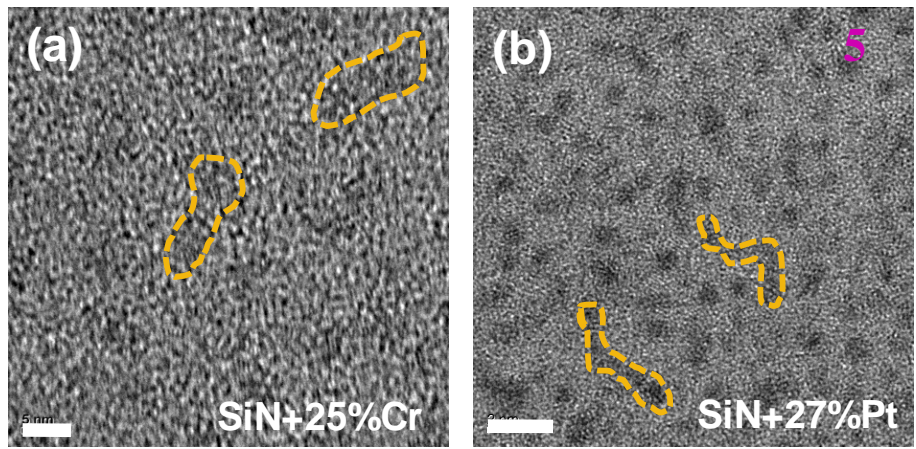

**Figure 4.21.** TEM image of (a) $SiN_{4/3}$:25%Cr and (b) $SiN_{4/3}$:27%Pt films showing high aspect ratio metal "needles" (enclosed yellow regions). Scale bar: 5 nm.

The above results are consistent with our picture of atomically dispersed deposition of metal atoms in the insulating dielectric during ambient-temperature sputtering. Since a single metal atom cannot be metallic (in the sense that its electron wave function is still localized to itself instead of being an extended Bloch wave function as seen in metallic crystals), metal-lean nanometallic films with atomically dispersed metal atoms are not expected to show metallicity either. In fact, even in films of large metal compositions in which chance encounters of metal atoms may have resulted in many metal-atom clusters that do have metallicity, some atomically dispersed, hence nonmetallic metal atoms must still exist. Our fitting results are consistent with this picture.

On the other hand, since nanometallic films contain thermodynamically immiscible constituents (an oxide/nitride insulator and a metal), clustering is inevitable and metal-rich clusters will form to some extent, by diffusion or by chance encounter. During ambient temperature sputter-deposition, diffusion is limited to the surface (diffusion time being the interval for sputter-growing one monolayer, about 2-10 seconds in our experiments) and it probably reaches to the nearest-neighbor sites or thereabout at most. Therefore, a multi-atom metal cluster cannot form unless $f$ is rich enough to statistically place metal atoms one or two atomic distances away from each other. Under this circumstance, the growth mechanism—by chance encounter aided by very nearest neighbor diffusion—does not allow metal clusters to achieve compact packing. So the morphology of clusters is likely to be ramified. In particular, when two small clusters are joined by a newly arrived atom that forms an inter-cluster "bridge," the joined cluster is likely to remain in a "dumb-bell" configuration almost indefinitely. It is interesting to note that such occurrence—the joining of two clusters— instantly gives rise to a new cluster of twice the size and aspect ratio. This mechanism provides a plausible explanation why, at a critical $f$ (10% in **Figure 4.20**) when cluster-joining become statistically frequent events, a simultaneous size and morphology transition can occur to account for the sharp $\varepsilon_r$ rise. Later, as $f$ further increases, more sputtered metal atoms will arrive, by chance, to join the clusters, so "belly-filling" of the dumbbell begins making the linked clusters appearing more rounded in their overall shape. This has the effect of decreasing the aspect ratio and increasing $S$ (from 0 for needle in curve 4, to 1/3 for sphere in curve 2, in **Figure 4.20**). Such an evolution can explain the apparent saturation of $\varepsilon_r$ at high $f$ in **Figure 4.12a** (**Figure 4.20**) or even a slight decrease of $\varepsilon_r$ at the highest $f$ shown in **Figure 4.12b**. It is also consistent with the microstructure evolution observed in **Figure 4.22** for $Si_3N_4$:Pt films.

The discussion above focuses on static dielectric response of nanometallic films of various metal contents. We will return to these films in a later section to consider the temperature and frequency effect which influence the dynamic dielectric response.

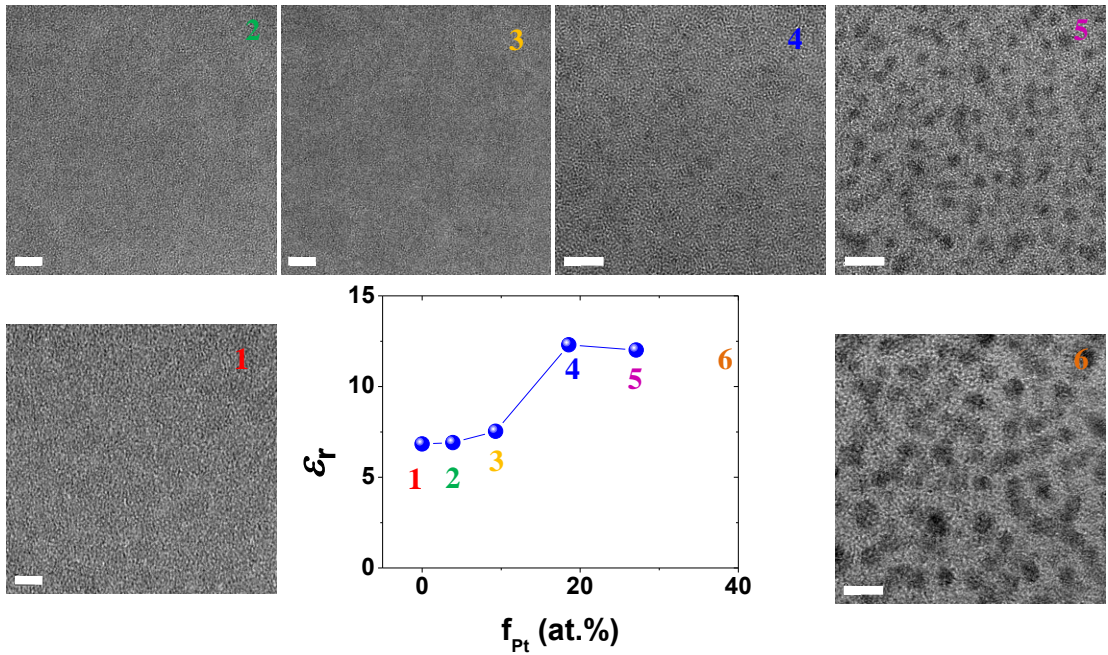

### 4.5.2 Conductivity and Capacitance

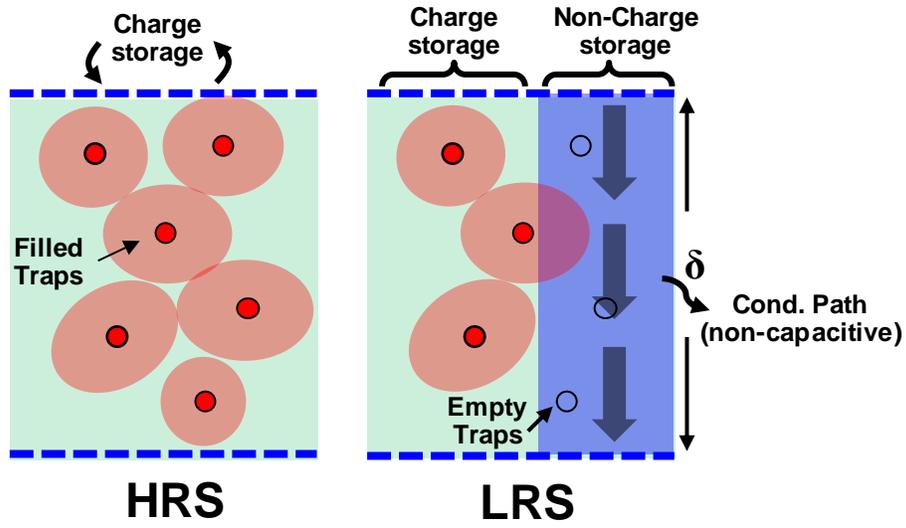

**Figure 4.23.** Microscopic picture of HRS and LRS for charge storage.

The capacitance difference $C_{HRS} > C_{LRS}$ is not seen in filamentary RRAM such as NiO (ref.[2]), TiO$_2$ (ref.[4]) and HfO$_x$ (ref.[6]). In the literature, this is rationalized by the fact of the very low (volume and area) fraction occupied by the filaments (~1 nm filament per ~1 μm device, giving an area ratio of 10$^{-6}$), making filament's contribution to the overall capacitance negligible despite its pivotal role in resistance transition. If we adopt the same argument to account for the 20% lower capacitance of the LRS relative to the HRS, we would expect the LRS in the nanometallic memory to have a conducting cross section of about 20% more than that of the HRS, as schematically illustrated in **Figure 4.23**. This percentage is huge compared to the filamentary structure. The picture is supported by the study of the capacitance of the intermediate resistance states (IRS). (See **Chapter VII** on how IRS is programmed.) As shown in **Figure 4.25**, the capacitance and resistance of the IRS, HRS, and LRS are "correlated": a lower $C$ correlates to a lower $R$. This clearly demonstrates that as more non-capacitive conducting paths are formed, the effective capacitance is reduced along with the overall resistance. To build a quantitative model, we employ the circuit model in **Chapter VII** again and envision the device as a parallel connection of a high-resistance cross section ($r_H$ and $c_H$ per area, area fraction=1-$F$) and a low-resistance cross section ($r_L$ and $c_L$ per area, area fraction=$F$). As shown in **Figure 4.24** inset, given a resistance state: $\frac{1}{R} = \frac{1-F}{r_H} + \frac{F}{r_L}$, the corresponding capacitance can be written as: $C = (1-F)c_H + Fc_L$ . Therefore, the following capacitance-resistance relation is obtained:

$$C = -A/R + C_0 \text{ or } \varepsilon_r = -A'/R + C_0^{'}$$

where $A = \frac{c_H - c_L}{\frac{1}{r_L} - \frac{1}{r_H}}$ and $C_0 = \frac{c_H/r_L - c_L/r_H}{\frac{1}{r_L} - \frac{1}{r_H}}$. The red line in **Figure 4.24** shows the fitting results using the above equation, which seems to be in reasonable agreement with the experimental data.

One might expect that nanometallicity can reduce the effective thickness, thus affecting the capacitance and apparent dielectric constant. This is because, even in the HRS, short range conducting paths may extend inside the nanometallic film within an electron diffusion distance $\sim\zeta$, allowing free electrons to be stored within this layer inside the film. However, if this is the case, one would expect the LRS to have a much thinner effective thickness, thus a much higher capacitance, which is contrary to our experimental observation. Therefore, we can rule out the effect of nanometallicity on the effective dielectric thickness of the RRAM capacitor.

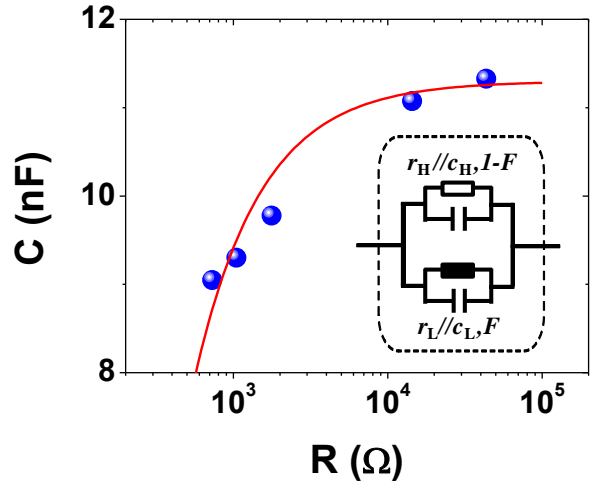

**Figure 4.24.** State dependence of capacitance. $C(\text{nF})=11.3-1883/R(\Omega)$.

### 4.5.3 *C-V* Dependence

The polarity-symmetric *C-V* curve immediately rules out the contribution of the interfacial (Schottky) barrier. This is because the capacitance of the Schottky barrier depends on a space-charge layer, which can be asymmetrically modulated by an external electrical field following $C \propto \frac{1}{\sqrt{V_{bi}-V}}$ ($V_{bi}$: built-in potential)[22].

Regarding the *C-V* dependence (*C* decreases at a larger *V*), there are two possible explanations:

(i) <u>Intrinsic non-linear field dependence.</u> In ferroelectric and highly polar paraelectric

oxides (*e.g.*, BaTiO$_3$, SrTiO$_3$, and their solid solutions), polarization becomes incrementally harder under a larger electrical field (~10 MV/m). Such field dependence can be approximated by[23]:

$$\varepsilon_r(T, E) = \frac{b(T)}{\sqrt{a(T) + E^2}}$$

where $a(T)$ and $b(T)$ are temperature–dependent coefficients. In our experiment, the DC bias is on the order of ~0.1 V corresponding to a field of ~10 MV/m. Therefore, this could be a viable mechanism. However, if such effect were to dominate, we should expect a similar field (voltage) dependence for the capacitance of the LRS, which is absent (**Figure 4.13**). Therefore, this explanation can be ruled out.

(ii) "Volatile" conducting paths. Following the same reasoning used to rationalize the state dependence in **Section 4.5.2**, one may argue that the decrease of capacitance could stem from the increasing number of conducting paths in nanometallic films. These conducting paths shrink the effective area of charge storage and therefore lead to a smaller capacitance. Using the capacitance of 2.4 nF at 0 V and 2.1 nF at ±1 V, we can estimate that the voltage-induced conducting paths may take up ~13% of the total effective area of the dielectric in the RRAM capacitor in **Figure 4.13**. Again, this percentage is huge compared to the percentage of filaments. It is worth noting that conducting paths here are different from the ones described earlier for the LRS, in the sense of volatility, "volatile" here but "non-volatile" in **Section 4.5.2**. The origin of the voltage dependence of conductivity path will be discussed in the chapter on transport properties. For example, at room temperature, the voltage dependence of conductivity paths is caused by field-dominated variable range hopping.

Regardless of whether the conducting paths are volatile or not, from the viewpoint of the parallel circuit model, one should expect a similar correlation between the effective area and the fraction of conducting paths $F$. This seems to be confirmed in **Figure 4.25**: capacitance and resistance are strongly correlated to each other and all data points lie on the same curve irrespective to the nature of resistances change—whether it originates from the different resistance states (HRS, IRS, or LRS), or from a DC voltage bias. Such correlation is seen in both $Si_3N_4$:Pt and $Si_3N_4$:Cr systems. A similar $C$-$R$ dependence is also observed in the $SiO_2$:Pt system (**Figure 4.25b**), which is even stronger. This is not unexpected: the conducting path effect should be more pronounced when the metal atoms are embedded in a background of a lower dielectric constant ($\varepsilon_{r,\,SiO2}=3.9 < \varepsilon_{r,\,Si3N4}=7$). In addition, the $SiO_2$ film is nanoporous, thus the conducting paths that are localized near the nanoporous region are also expected to have a stronger effect.

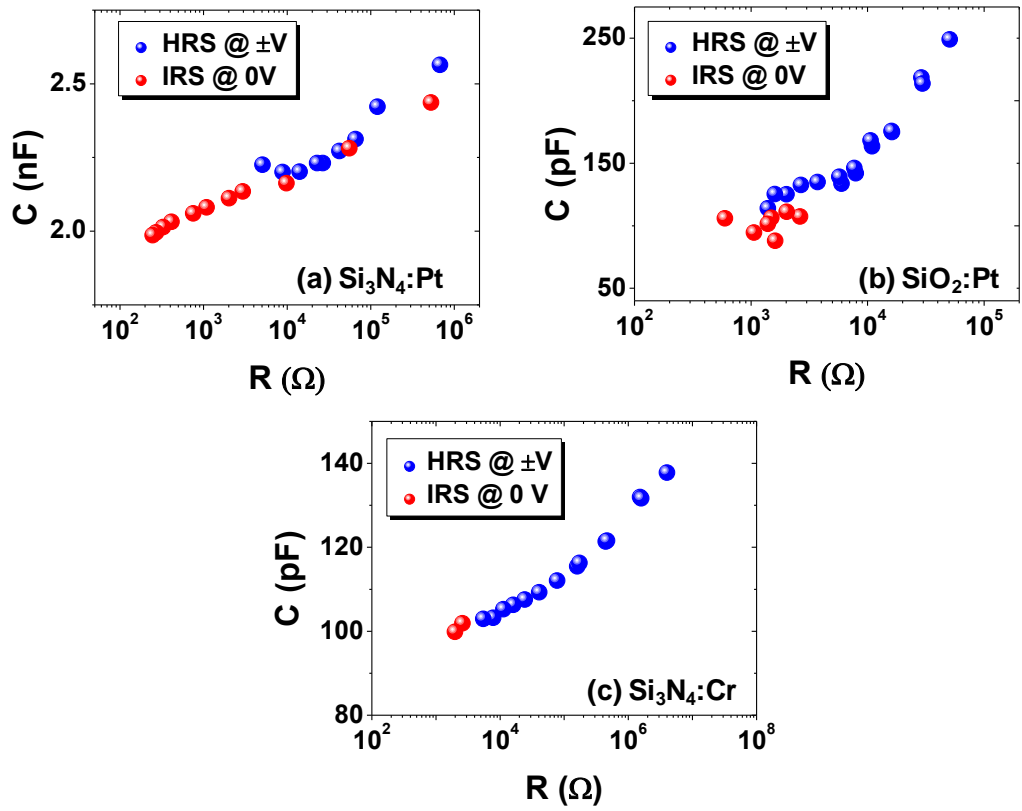

**Figure 4.25.** Capacitance *vs.* resistance for all states (acquired by either IRS at 0 V or HRS at various bias) in (a) Si₃N₄:Pt (*d*=512 μm, *δ*=10 nm), (b) SiO₂:Pt (*d*=200 μm, *δ*=20 nm) and (c) Si₃N₄:Cr (*d*=122 μm, *δ*=10 nm) nanometallic devices.

## 4.5.4 Temperature Effect

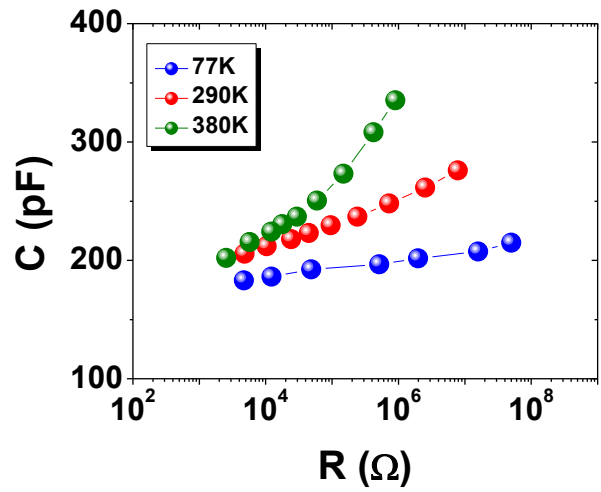

**Figure 4.26.** Capacitance *vs.* resistance for various states (re-plotted from **Figure 4.15**) under different temperatures for a $Si_3N_4$:12%Cr nanometallic film.

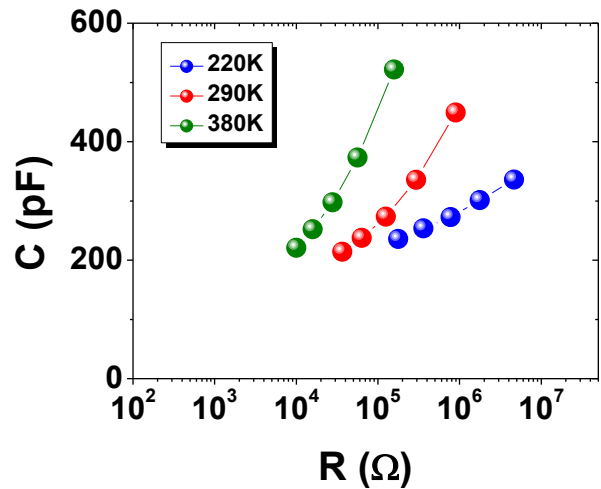

**Figure 4.27.** Capacitance *vs.* resistance for various states for a $Si_3N_4$:16%Cr nanometallic film.

By applying the same procedure, we replot the *C-V* dependence of **Figure 4.15** into the *C-R* dependence at various temperatures in **Figure 4.26**. For each temperature, the capacitance increases with *R*, which is the same correlation found above between the capacitance and the effective dielectric cross section in the RRAM capacitor. However, the positive *C-R* correlation should not be taken too far. At zero bias, as the temperature increases, resistance decreases for the HRS but the capacitance increases (**Figure 4.16**, or **Figure 4.15** at 0 V.) Likewise, at zero bias, as the metal concentration increases, resistance decreases for the HRS but the capacitance increases (**Figure 4.12**) Therefore, higher temperature and higher metal content can both increase the capacitance despite the fact that they also decreases the resistance hence the dielectric cross section. If we use voltage bias to tune resistance, and compare capacitance in **Figure 4.26** at the same resistance, its increase with temperature is obvious: it becomes more pronounced at higher resistance. The same holds for other compositions, see **Figure 4.27**. Regarding the combined effect of composition, the capacitance increase with temperature is gradual in films of low metal concentrations (which is more resistive) but much steeper at higher metal concentrations (which is more conductive). Therefore, there appear to be an additive effect of these two factors: higher temperature and higher metal composition, in combination, cause the most increase in capacitance. This suggests that the two effects may have the same or similar origin.

It is known that dielectric constant can be affected by the response of both induced dipoles and permanent dipoles. Dielectric constant due to permanent dipoles typically shows two types of temperature responses. If electric dipoles are free to rotate, then thermodynamics dictates that the dielectric constant increases with decreasing temperature. This may result in the Curie law in which the dielectric constant diverges at 0 K, or in the Curie-Weiss law in which the peak dielectric constant occurs at the Curie temperature due to an internal field arising from dipole-dipole interactions. In contrast, if dipoles are not free to rotate, *i.e.*, they are frozen, then they make no contribution to the dielectric constant. The latter situation typically changes at higher temperature when dipole rotation becomes easier, then some contribution to the dielectric constant is made and the dielectric constant gradually increases. This is often seen in amorphous materials such as polymers. Similar dielectric response also widely exist in compound semiconductors which lack inversion symmetry, including GaAs[24], CdTe[24], ZnSe[24], CdS[25], ZnO[25], GaP[26] and InP[27], where a universal linear expression $\varepsilon(T) = \varepsilon(0)(1+\lambda T)$ are employed for data fitting ($\lambda > 0$, $\lambda \sim 10^{-4}$ K$^{-1}$). It is generally believed that electron-phonon and phonon-phonon interaction contributing to such behavior[26].

In our amorphous dielectric films, dipoles due to various defects possibly exist. They may originate from covalently-bonded cation-anion pairs, electrochemically controlled metal (inclusion)-insulator (matrix) interfaces and compositional/structural disorders. Their rotation is likely to be very difficult at lower temperatures considering the refractory nature of $Si_3N_4$ and $SiO_2$. The broad but gradual rise with temperature is likely due to the contributions from these "hard" defects, which gradually become free to rotate. This seems to be responsible for the dielectric response in nanometallic films of low metal concentrations.

When the metal concentration exceeds a critical value, however, the temperature dependence increases rapidly. Adopting the Curie-Weiss law and use $\varepsilon_r = 1 + C/(T_c - T)$ to analyze our data, we can see from **Figure 4.28** that the data of $f_{Pt}$=16% and $f_{Pt}$=30% nanometallic films follow the Curie-Weiss law, with a critical temperature ~500 K and a similar Curie-Weiss constant. Films with leaner metal composition do not show such behavior. Since the Curie-Weiss behavior, which emerges at high metal composition, does cause a large increase of dielectric response, it now becomes clear that temperature-caused increase of capacitance is correlated to metal-composition-caused increase of capacitance.

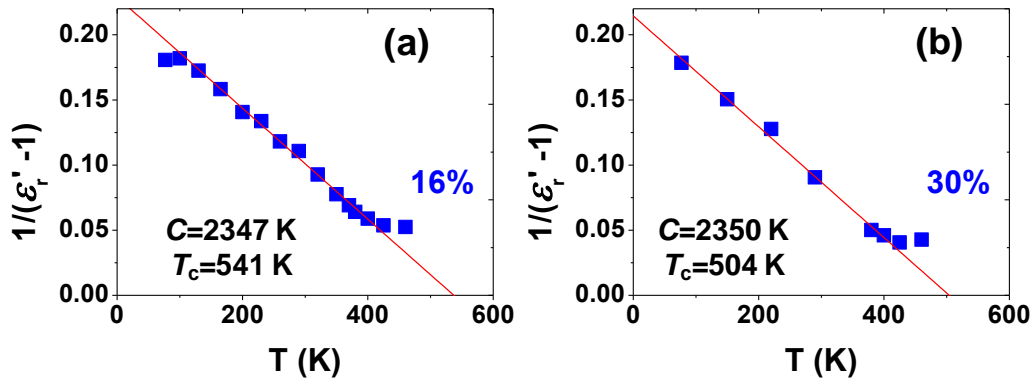

**Figure 4.28.** Curie-Weiss plot for (a) $f_{Pt}$=16% and (b) $f_{Pt}$=30% nanometallic films.

Earlier, we found that the deposited metal atoms acquire metallicity above a critical concentration of ~10%. Therefore, the apparently abrupt emergence of the Curie-Weiss behavior is probably related to the emergence of metallic nanoparticles. This may be understood as follow. Metallic nanoparticles have a large dipole, moreover, the dipole can reorient without a configurational change (*i.e.*, a physical rotation of the nanoparticle) by redistribution of conducting electrons. Thus, these are strong and soft dipoles unlike the weak and hard atomic defects in amorphous $Si_3N_4$ or $SiO_2$ films. Meanwhile, because of their high concentration at high metal contents, dipole-dipole interaction between metallic nanoparticles is relatively strong so that an internal field also arises along with metallicity. As a result, there is an apparent Curie-Weiss law that describes the dielectric behavior of these films containing metal nanoparticles, which explain both the temperature dependence and the compositional dependence. Lastly, redistribution of conducting electrons in metallic nanoparticles is obviously a conduction process, which explains the strong frequency dependence and large tanδ shown in **Figure 4.18** for $f_{Cr}=16\%$ data, especially below 10 kHz.

To more quantitatively examine the compositional effect, we need to compare the Curie-Weiss coefficient $C$ as a function of composition. Note that the dielectric constant of the "dielectric" of the RRAM is probably underestimated at high metal concentrations because of the large portion of the conducting paths. Using the room temperature resistance as a reference, we see that the resistance of the 0 V resistance decreases by a factor of ~30× from $f_{Pt}=16\%$ to $f_{Pt}=30\%$. Taking this into account, we believe that the $f_{Pt}=30\%$ dielectric response is probably stronger by the same factor than that of the $f_{Pt}=16\%$ composition, even if the nominal Curie constant $C$ is essentially the same for both in **Figure 4.28**.

We finally note that the fitting factor $n$ describing the temperature response of the dielectric constant in **Figure 4.14** is closer to 1 at lower temperature; it reduces to <0.95 at room temperature. In an ideal Debye dielectric, $n=1$ corresponding to a single relaxation time ($\tau=RC=\rho\varepsilon$); a larger deviation from this ideal situation implies a wider $\tau$-dispersion, which is apparently the case at a higher temperature. Although this is illustrated in **Figure 4.14** for one composition only, it is likely to be the case in films of other metal compositions. This may be interpreted in at least two ways. (i) At higher temperatures, more modes of dipole rotations become feasible, leading to an increase of relaxation mechanisms and dispersion. (ii) Lattice vibration (phonon) introduces electrical randomness through electron-phonon and phonon-phonon scattering, creating local inhomogeneity and thus time constant spreading $\Delta\tau\sim\Delta\rho\varepsilon+\rho\Delta\varepsilon$.

## 4.6 Conclusions

(1) Nanometallic devices can be described by an equivalent circuit composed of $R_0$-$R_1//C_1$-$R_2//C_2$, where $R_0$ and $R_1//C_1$ stem from spreading/line resistance, bottom electrode and interface, whereas $R_2//C_2$ is attributed to the nanometallic film itself. The value of $R_2$ can change by orders of magnitude because of resistance switching and non-Ohmic behavior.

(2) The capacitance of a nanometallic RRAM follows the standard scaling law: $C\propto A/\delta$ for both the HRS and LRS, indicating that it is essentially made of a linear dielectric that is spatially uniform over the length scale of the capacitor (such as the thickness of a few nm).

(3) The effective dielectric constant slowly increases with $f_{metal}$ at $f_{metal}<10\%$, but abrupt increases at $f_{metal}\sim10\%$ and then saturates afterwards. This variation results from a combination of changes of morphology, shape/geometry and metallicity of metal atoms and their clusters.

(4) Both HRS and LRS capacitance are symmetric with respect to voltage. The HRS capacitance decreases with voltage, whereas the LRS capacitance is insensitive to voltage. Meanwhile $C_{HRS} > C_{LRS}$ by up to ~20%. A strong correlation between $C$ and $R$ was also found. These results suggest that the nominal capacitance depends on the effective cross sectional area of the non-conducting dielectric, which is significantly reduced in the LRS and by a DC voltage bias, as pictured by the parallel circuit model.

(5) The capacitance of filamentary RRAMs is the same for the HRS and the LRS, and it shows no voltage dependence and has no $C$-$R$ correlation. This suggests that, unlike nanometallic RRAM, the cross sectional area of the insulating dielectric remains the same in filamentary RRAM at all states and voltages.

(6) Dielectric constant increases with temperature between 77 K and ~400 K. At high metal compositions, the rise is especially steep allowing the dielectric constant to increase significantly, with a peak at $T > 400$ K. The latter behavior is consistent with the Curie-Weiss law with a Curie temperature ~500 K. This is suggested to arise from dipole-dipole interactions between dipoles associated with metallic nanoparticles. This behavior, however, disappears at high frequency (*e.g.*, 1 MHz) indicating that electron redistribution in metallic nanoparticles can be suppressed.

# Chapter V. Transport in Nanometallilc Memory

## 5.1 Introduction

Almost all the studies in the RRAM literature were performed at room temperature. As a result, the information is limited, and such limitation has especially affected the interpretation of the conduction mechanism. Most insulators exhibit a non-linear *I-V* curve, which can sometimes be simplified to an exponential form: $I \sim \exp(V/V_0)$. At a constant temperature or a very narrow temperature range near the room temperature, it can be quantitatively fitted by various models based on Poole-Frenkel emission[1-2], Schottky emission[3-4], space charge limited conduction[5-7], or trap assisted tunneling[8], which have all been proposed for RRAM, sometimes for the same material (*e.g.*, HfO$_2$). Likewise, most literature claims a metallic LRS based on its linear *I-V* behavior, which is not convincing since many conduction mechanisms do yield a linear *I-V* behavior at small voltage. In principle, temperature dependence of conductivity can help distinguish different mechanisms and perhaps establish a unique conduction mechanism, and still more information can be supplied by magnetic field dependence. Such a study is undertaken here from 2 K to 300 K at a magnetic field up to 9 T to interrogate the conduction mechanism of nanometallic RRAM.

The study is of particular interest to nanometallic resistive switching devices. As discussed in **Chapter II**, at room temperature ultra-thin nanometallic films do not follow Ohm's law in thickness scaling and instead exhibit an exponential dependence that can be attributed to electron localization. However, to truly manifest the localization behavior of electron's wave function in random material, a low temperature study is mandatory since at room temperature thermally activated hopping is expected to overwhelm tunneling, thus obscuring the genuine localization feature. In addition, nanometallic RRAM is believed to rely on electronic switching and not ion or atom migration. This implies that switching requires little thermal assistance and can occur at all temperatures, including low temperature when the thermal energy ($\sim k_B T$) is considerably less than $eV_{switching}$, *i.e.*, the switching voltage should be independent of temperature. Magnetic study is also relevant in view of the random nature of nanometallic material and the predominance of electron tunneling because the interference of electron wave function can be easily altered by a magnetic field.

In the next sections, a brief summary of conduction mechanisms will be presented first, followed by the presentation of data and their quantitative fitting as a function of temperature and voltage using a set of self-consistent microscopic model parameters for elastic and inelastic tunneling. A variety of states of the nanometallic devices, including different resistance, thickness, and metal content, will be investigated to provide a comprehensive picture of the energy landscape for electron conduction in these devices.

## 5.2 Mechanisms for Electron Conduction: A Brief Summary

### 5.2.1 Conduction in Metals

The simplest model of electron conduction is the Drude model, which explains the Ohm law as a constitutive equation: current density is proportional to the applied field, $J=\sigma E$. Electrons are thought to diffuse via random walk, in each walk (called mean free path) which last a time $\tau$ (called mean free time, which is the time between collisions), the electron movement is monotonically accelerated by the electric force provided by the electric field. The conductivity $\sigma$, or its reciprocal, resistivity, is expressed by[9-10]

$$\sigma = \frac{ne^2\tau}{m}, \text{ or } \rho = \frac{m}{ne^2\tau}$$

where $e$ and $m$ are the charge and mass of electron, $n$ is the density of electrons. Although $\tau$ in classical physics is associated with the mean free time between adjacent collisions, it can also be associated with electron relaxation/decoherence time in quantum mechanics, so that the diffusion model is also adopted in mesoscopic theory of electron conduction. Generally, the temperature dependence of metal in this model can be divided into four regions[9-10]:

(1) At high temperature ($T > \Theta_D$, the Debye temperature), electrons experience scattering from lattice vibration with a scattering probability ($P \sim 1/\tau$) proportional to the average squared atomic displacement, which is proportional to $kT$ ($P \propto \overline{\Delta x^2} \propto k_B T$). This leads to a linear law: $\rho(T) \sim 1/\tau \sim T$.

(2) Below the Debye temperature, the number of phonons available is proportional to $(T/\Theta_D)^3$, each contributing a scattering cross section proportional to $(\mathbf{Q} \cdot \mathbf{u})^2$ (where $Q$ is the momentum transfer during electron-phonon collision/scattering), which is proportional to $kT$, and the fraction of electrons participating in the $N$-process electron-phonon collision (without changing the Brillouin zone) is from a section of the size $kT/E_f$ on the Fermi surface. Thus, the overall temperature dependence is $T^5/\Theta_D^3$ due to electron-phonon interaction.

(3) Electron-electron scattering contributes to a resistivity that follows a square-law

$\rho(T) \sim 1/\tau \sim T^2$. This $T^2$ dependence dominates over the Debye law at low temperature in systems where electrons form a Fermi liquid.

(4) At 0 K, impurity and defect scattering results in a finite resistivity instead of zero resistivity as in a perfect metal.

### 5.2.2 Conduction in Insulators

Electrons are localized in insulators. To move to other sites, a certain energy barrier needs to be overcome either by tunneling or hopping, the latter being inelastic tunneling substantially assisted by phonons (*i.e.*, thermal energy). The nature of the barrier depends on the type of defects or dopants present. **Figure 5.1** lists some possibilities for an electron that attempts to travel from a cathode, through an insulator film to an anode[11]. The possibility that provides the largest electron transition rate is the dominant conduction mechanism.

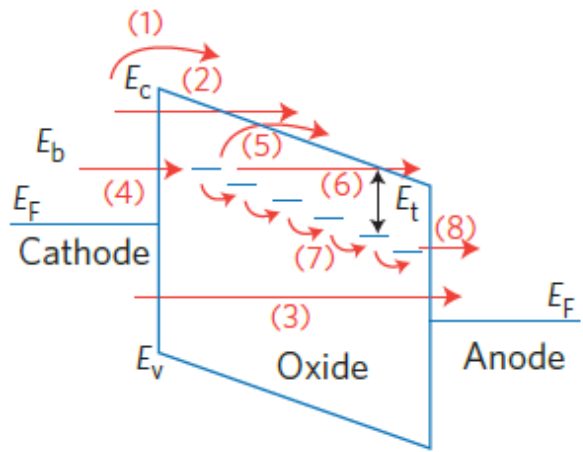

**Figure 5.1.** Schematic of some possible electron conduction paths through a MIM stack. (1) Schottky emission (SE). (2) Fowler-Nordheim tunneling

tunneling (FNT). (3) Direct tunneling (DT). If the oxide has substantial number of traps, trap-assisted tunneling contributes to additional conduction, including the following steps: (4) tunneling from cathode to traps; (5) emission from trap to conduction band (Poole-Frenkel emission); (6) F-N-like tunneling from trap to conduction band; (7) trap to trap hopping or tunneling; (8) tunneling from traps to anode. Adapted from ref. [11].

### 5.2.2.1 Schottky emission (SE) or thermionic emission (TE)

Under a field assistance electrons from one electrode may be thermally activated and injected into the conduction band of the insulator, surmounting the barrier. Due to imaging force between the electrode and the electron, the barrier is lowered and the process is made slightly easier. The SE/TE conduction follows a linear relation between $\ln(J/T^2)$ and $1/T$ (Ref. [12]), which can be analytically expressed by

$$J \propto AT^2 \exp\left(-\frac{e\left(\Phi_b - \sqrt{eE/4\pi\varepsilon_i}\right)}{k_B T}\right)$$

In which the field dependence is non-linear because of image charge consideration. One distinguishing feature of this mechanism is an asymmetry when two electrodes with different work functions are involved, giving different barriers.

### 5.2.2.2 Frenkel-Poole emission (FPE)

Electrons from one electrode may first enter a trap inside the insulator, where it sees a lower barrier because of field-shearing. The trap electron then undergoes field-assisted thermal emission into the conduction band and travel to the other electrode in a similar manner as SE/TE. When overall barrier lowering due to field shearing and image charge is considered, the following expression is obtained:

$$J \propto E \exp\left(-\frac{e\left(\Phi_b - \sqrt{eE/\pi\varepsilon_i}\right)}{k_B T}\right)$$

Mathematically, FPE turns out to have the same voltage dependence as SE, except for a larger amount barrier lowering[13].

### 5.2.2.3  Space Charge Limited Conduction (SCLC)

Space charge refers to the electric charges which may be treated not as distinct point-like entities but as a charge continuum distributed over a region of space[14]. In a capacitor-type device which contains a trap-free insulator gap, the total space charge in the insulator is given by $Q=CV=\varepsilon AV/\delta$, accumulated over a time of the order of $t=\delta/\mu E= \delta^2/\mu V$ ($\mu$ is electron mobility, $\mu E$ with $E=V/\delta$ is electron velocity). This leads to the following simple estimate for the current which follows a square $V$ dependence:

$$I \propto \frac{Q}{t} = \frac{\varepsilon\mu A}{\delta^3} V^2$$

Since the mobility may have a strong temperature dependence, space charge limited conduction also varies with temperature in an Arrhenius manner $\sim\exp(-A/k_B T)$. Furthermore, this mechanism is very sensitive to the presence of traps inside the insulator. If the insulator only contains shallow traps close to the conduction band, the square $V$ law is still valid but is modified with an additional constant prefactor accounting for the "trapped electrons" fraction. A more realistic circumstance is to have traps that are distributed in energy, and to allow the external voltage to elevate the Fermi level and thus effectively modulate the electron density. Consequently, for the case of uniformly distributed traps, SCLC becomes an exponential form:

$$I \propto \left( \frac{V q \mu_0}{\delta^2} \right) \left( \frac{e n_{c0}}{C} \right) \exp\left( \alpha V \right)$$

where $n_{e0}$ is the initial (thermally-equilibrated)) equilibrium concentration of free carriers. Another commonly used SCLC form assumes the trap density follows the Boltzmann distribution, $n_t \sim \exp(-E/k_B T_c)$ ($E$ is measured from the bottom of the conduction band, $T_c$ is a characteristic temperature indicating how rapidly the trap distributions varies), which leads to a stronger voltage dependence for $T < T_c$:

$$I \propto V^{(T_c/T)+1}$$

For $T > T_c$, this expression reduces to the same square-$V$ law as for shallow traps (square law). A good reference for SCLC is in ref.[15]

### 5.2.2.4 Tunneling

Tunneling refers to a process by which an electron wave function maintaining a definite phase and energy moves through a barrier that has a barrier height higher than the energy of the wave function. Tunneling probability is determined by the ratio of the wave function amplitude of the final state to that of the initial sites. Simmons derived a generalized formula for electron tunnelling between similar electrodes separated by an arbitrary, spatially varying tunnel barrier $\Phi_b(\boldsymbol{r})$ (Ref.[16]). For a trapezoidal barrier and when the applied bias is less than the barrier height, the Simmons formula reduces to an analytical form:

$$I = \frac{e^2 A}{4\pi^2 \hbar d^2}\left\{\left(\Phi_b - \frac{V}{2}\right)\exp\left(-\frac{2d\sqrt{2m^*e}}{\hbar}\sqrt{\Phi_b - \frac{V}{2}}\right) - \left(\Phi_b + \frac{V}{2}\right)\exp\left(-\frac{2d\sqrt{2m^*e}}{\hbar}\sqrt{\Phi_b + \frac{V}{2}}\right)\right\}$$

Here $A$ is the junction area, $d$ is the barrier width, $m^*$ is the effective mass of electron, $\Phi_b$ is the barrier height and $e$ is electron charge. Several simpler forms have also been proposed in the literature based on simplified models that carry more straightforward physical meaning.

**A. Direct tunneling (DT):** DT occurs in a very thin insulator through which an electron can elastically tunnel without losing any energy[17]. For a rectangle barrier, DT leads to:

$$I \propto V \exp\left(-\frac{2d\sqrt{2m^*e\Phi_b}}{\hbar}\right)$$

This is exactly the low field limit ($V\rightarrow 0$) of Simmons' formula for the same rectangular barrier. Thus, it applies at very low field only. An important feature in this limit is the linear $I$-$V$ relation, *i.e.*, there is a finite conductance at zero field.

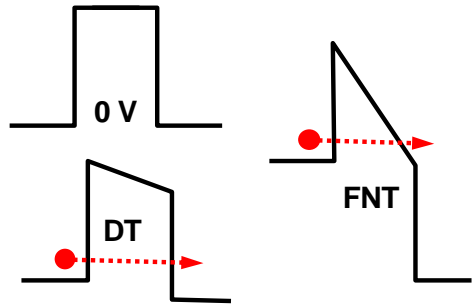

**Figure 5.2.** Direct tunneling (DT) and Fowler-Nordheim tunneling (FNT).

**B. Fowler-Nordheim tunneling (FNT, field emission):** In the presence of a high

electrical field, a rectangular energy barrier may be so severely sheared to reduce to a triangular barrier. In this scenario, electrons see a "thinner" barrier and thus tunneling becomes much easier. This is the FNT situation, which leads to[18]:

$$I \propto V^2 \exp\left(-\frac{4d\sqrt{2m^* e\Phi_b^3}}{3\hbar V}\right)$$

It can be represented as a simple relation between $\ln(I/V^2)$ and $1/V$.

**C. Trap assisted tunneling (TAT):** In a way somewhat similar to PFE, tunneling becomes much easier when a thick insulator contains traps. These traps serve as staging stations for electrons, which now see a reduced barrier width between stations (traps).

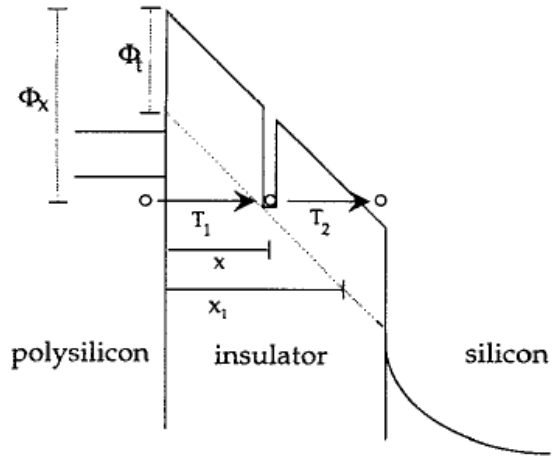

**Figure 5.3.** Trap assisted tunneling (TAT). Adapted from ref.[19].

Cheng *et al.*[19] proposed a method to calculate TFT by assuming a simple Fowler-Nordheim-type potential profile (**Figure 5.3**). In this picture, an electron from one electrode first tunnels to an intermediate state (a trap) with a rate $T_1$, it next tunnels to the conduction band of the insulator (which is responsible for the initially rectangular barrier) with a rate $T_2$, thus reaching the other electrode therefrom. The two rates can be expressed as $T_1=CN_t(1\text{-}f)P_1$ and $T_2=CN_tfP_2$, where $C$ is a constant, $N_t$ is the trapping site concentration, $f$ is the fraction of the trap states that is occupied, and $P_1$ and $P_2$ are tunneling probability of these two steps (calculated using the WKB approximation for tunneling assuming a triangular potential):

$$P_1 = \exp\left(-\frac{4\sqrt{2m*e}}{3\hbar E}\left(\Phi_x^{3/2} - \Phi_t^{3/2}\right)\right), \; P_2 = \exp\left(-\frac{4\sqrt{2m*e}}{3\hbar E}\Phi_t^{3/2}\right)$$

By imposing detailed balance of the trap state occupancy, $f$ can be expressed in terms of $P_1$ and $P_2$, and the overall tunneling current can be expressed as:

$$J = \int_0^{x_t} \frac{eC_t N_t P_1 P_2}{P_1 + P_2}\, dx$$

which is linearly dependent on the trap concentration. Unfortunately, even for this oversimplified case, an analytical form is difficult to obtain. Thus, this relation is difficult to use. However, the above result has a simple interpretation: if the two regions between the trap and the two electrodes can be considered as two separate insulators with two resistances, then the total resistance between the electrodes is the sum of the two separate resistances.

**D. Fluctuation induced tunneling (FIT):** FIT describes elastic tunneling across weak barriers between metallic patches separated by a nano-gap, facilitated by the Johnson fluctuation of the gap voltage. Since the fluctuation increases with temperature and the tunneling probability is non-linear and asymmetric with respect to voltage (*i.e.*, $P(V+\Delta V)+P(V-\Delta V)>2P(V)$), net tunneling rate gradually increases with temperature despite the fact that the process is entirely elastic (**Figure 5.4**). Since the voltage fluctuation across the nano-gap capacitor is $\pm(k_B T/C)^{1/2}$, which can be significant when $C$ is small[20], this is a nano effect not seen in macroscopic capacitors.

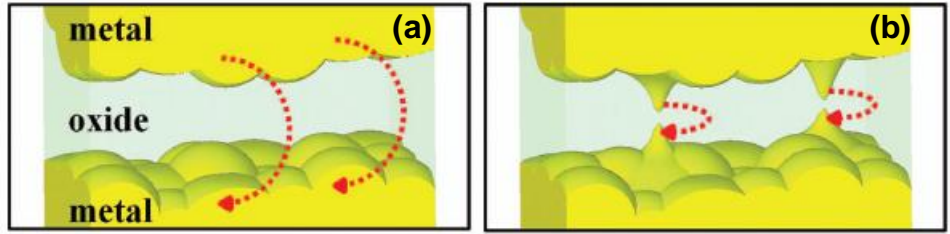

**Figure 5.4.** (a) Tunneling through the entire insulator. (b) Tunneling through small gaps. Adapted from ref.[21].

An analytical form of FIT is expressed by[20-21]

$$G(T) = G_0 \exp\left(-\frac{T_1}{T + T_0}\right)$$

$$T_1 = \frac{8\varepsilon_r A \Phi_b^2}{k_B e^2 w}, \quad T_0 = \frac{16\varepsilon_r \hbar A \Phi_b^{3/2}}{\pi (2m^*)^{1/2} k_B e^2 w^2}$$

Here, $A$ and $w$ are the area and width of the gap, respectively, $\Phi_b$ is the potential barrier height, $\varepsilon_r$ is the dielectric permittivity of the insulator, and $e$ and $m^*$ are the electron charge and mass, respectively. As temperature varies, the expression reduces to two asymptotes: $G_0\exp(-T_1/T_0)$ in the low temperature limit, and $G_0$ in the high temperature limit. In transport literature, there is also an empirical formula for voltage dependence of FIT, expressed by

$$G(V) = G_{0,T} \frac{\exp(V/V_0)}{1 + h_v\left[\exp(V/V_0) - 1\right]}$$

where $G_{0,T}$ is the temperature-dependent small-voltage conductance, $V_0$ is a voltage scale factor and $h_v$ is the ratio of conductance at low and high voltages. Since the form is empirical, the fitting parameters carry little physical meaning.

### 5.2.2.5  Inelastic Tunneling (Hopping)

Hopping is an incoherent, thermally activated process in which an electron loses all its phase information as it moves from one site to another. In detail, the electron still advance by extending its wave function, but it also interacts with phonon so that its energy and momentum are no longer constant. Since there is no coherence between the amplitudes for finding an electron at different sites, hopping is akin to diffusive transport in which there is no explicit need for prescribing a wave function to carry the information of amplitude and phase: energy information alone will suffice.

**Variable range hopping (VRH):** VRH was originally developed for doped/ disordered semiconductors. Electrons are localized in these systems. In addition, statistically, the probability of an electron to find its nearest available electron states to have an identical energy is very low. Therefore, elastic tunneling is highly improbable. Since electrons may gain enough thermal energy $k_BT$, they can also inelastically tunnel if they can find some nearby states that have an energy within $k_BT$ of themselves. The probability of finding such states is much higher at a higher temperature and with the electrons reaching out to probe a larger zone, although as they do so their own wave function decays so the tunneling probability also diminishes. The overall probability, which provides the conductance, can be expressed by

$$G \propto P \propto \exp\left(-\frac{2r}{\zeta} - \frac{\Delta E}{k_B T}\right)$$

where $\zeta$ is the localization length of the wave function, $r$ is the hopping distance and $\Delta E$ is the energy difference between the initial and final states. Mott[22] pointed out the energy difference $\Delta E$ available is related to the density of states, $N_u$, since the probability of finding a state of such energy within all the states in a space of a linear dimension $r$ should be about unity, thus $\Delta E \sim 1/N_u r^d$, where $d$ is the dimension of the system. By maximizing the probability, Mott obtained the following relation (Mott's law)[23]:

$$G \propto \exp\left(-\left(\frac{T_0}{T}\right)^{\frac{1}{d+1}}\right)$$

which provides a constant conductance in the low field limit ($F \rightarrow 0$, where $F$ is the electric field). In particular, the 3D VRH model predicts $\ln G$ is linearly proportional to $T^{1/4}$ and:

$$T_{0,3D} = \frac{24}{\pi} \frac{1}{k_B N_u \zeta^3}$$

In this picture, the most probably range $r$ is temperature dependent: it is much longer at low temperature than at high temperature, and most of the time it is longer than the distance between the nearest neighbor states, hence the name of variable range hopping. In the high field limit, reached when the electric work $\sim erF$ is larger than the site energy difference $\Delta E$, hopping becomes much easier and exponentially dependent on the electrical field $F$:

$$I \propto k_B T \nu_{ph} \exp\left[-\left(\frac{F_0}{F}\right)^{\frac{1}{d+1}}\right]$$

Here, the field constant is, for 3D conduction,

$$F_0 = \frac{81}{16\pi} \frac{1}{e N_u \zeta^4}$$

This is essentially an athermal hopping (AH) process where conduction is weakly dependent on temperature. The transition field between the "low field" regime and the "high field" regime can be estimated as $F_c \approx k_B T / e\zeta$ (ref. [24]).

**Coulomb-gap hopping (CG-VRH):** Another type of VRH was proposed by Efros and Shklovkii, who considered the Coulomb interaction between localized electrons. They found such interaction inevitably opens a "soft" gap in the density of states near the Fermi level. This gap of density of states obviously affects the statistics of available states for variable range hopping, leading to[25]

$$G \propto \exp\left(-\left(\frac{T_{ES}}{T}\right)^{1/2}\right)$$

which is formally the same as the temperature dependence in conventional VRH in 1D. In the above, $T_{ES}$ is given by:

$$T_{ES} = \frac{\beta e^2}{\varepsilon k_B \zeta}$$

where $\varepsilon$ is a macroscopic dielectric constant and $\beta$ is a numerical constant of the order of unity ($\beta = 2.8$ for 3D)[26]. The Coulomb gap width $\Delta$ is related to the unperturbed density of states $N_{uo}$: $\Delta = e^3 \sqrt{N_{u0}} / \varepsilon^{3/2}$. Therefore, if $k_B T \gg \Delta$, the Coulomb gap is thermally smeared so the conduction mechanism crosses over from ES VRH to Mott VRH. In 3D, the ES-Mott crossover temperature is[27]

$$T_{cross} = 16 T_{ES}^2 / T_M$$

## 5.3 Experimental Setup

Two experimental stations were employed to interrogate temperature related electrical properties. One was a cryogenic probe station (Lakeshore), which has the advantage of accommodating the device size (smallest $d \sim 25$ μm) and providing better statistics (probing many cells on the same chip). It also has a lower parasitic impedance for AC signal detection. The other was a Physical Property Measurement System (PPMS, Quantum Design), which requires permanent wiring connections thus limiting the device size and number and causing a higher parasitic impedance. However, it provides a wider temperature range (2 K-350 K) with a much better temperature control and the possibility of additional property (*e.g.* magnetic, optical, *etc.*) measurements. It also exerts little mechanical contact stress/wear during testing and temperature excursions, which is always a problem with a cryogenic probe station due to the temperature drift. (In addition to the possibility of damaging the cells, mechanical stress can cause switching, see **Chapter III**.) Devices of the Mo/$Si_3N_4$:Pt/Pt type fabricated using methods described in previous chapters were used. Data collected in both stations are comparable. Briefly, the probe station was used for initial studies and to ascertain the size dependence and sample statistics. Further testing was performed in the PPMS to determine properties below 77K and the magnetic effects.

### 5.3.1 Cryogenic Probe Stations

The Lakeshore cryogenic probe station provides $10^{-7}$ Torr level vacuum and is equipped with a heating/cooling stage, allowing temperature control from 77K to 400 K. This was implemented by purging the sample stage with liquid nitrogen (LN) while electrically heating the stage simultaneously (**Figure 5.5**). The testing setup is outlined in **Figure 5.5a & c**, and a customized LabView program was developed to automatically control/ monitor temperature and collect data. (This is the same setup previously used for impedance study, in **Chapter IV**.)

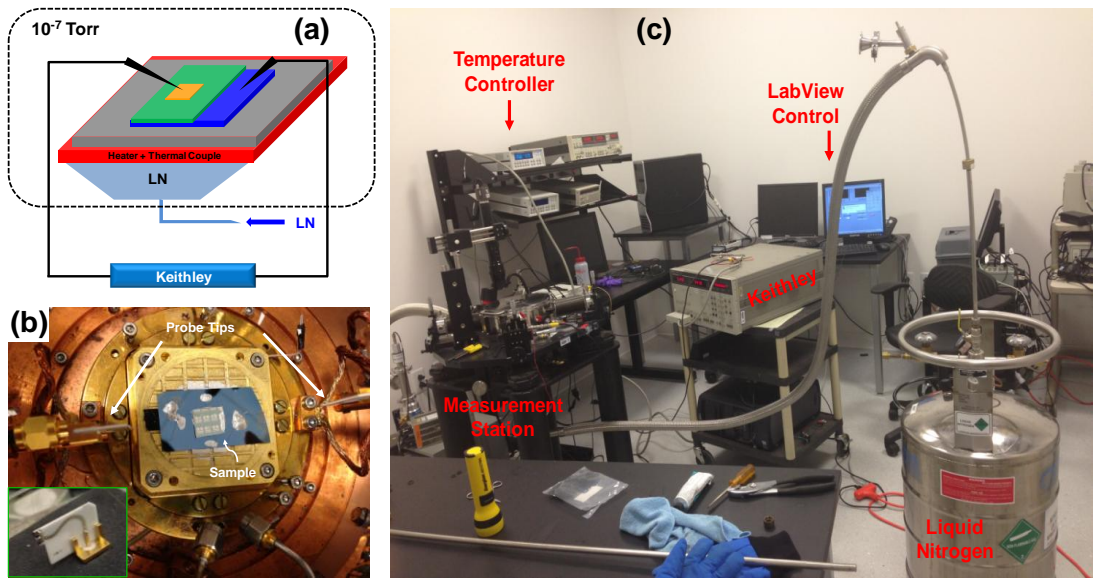

**Figure 5.5.** (a) Schematic of experimental setup using Lakeshore Cryogenic Probe Station. (b) Sample configuration and probe tips inside vacuum chamber. Inset: Probe tip. (c) Complete experimental setup.

### 5.3.2 Physical Property Measurement System (PPMS)

The PPMS was equipped with an Ever-Cool helium cooling system and a heater for temperature control. It also supplies a magnetic field (-9 T to 9 T) in the vertical direction, which in most cases is perpendicular to the film of our devices (**Figure 5.6**). Samples (≤1×1 cm$^2$) were mounted on a special chip holder with a heat conducting vacuum grease (**Figure 5.6b**). Silver paint was used to bond Gold wires ($d$=2 mil) to device electrodes and to connect to the pins on the sample holder.

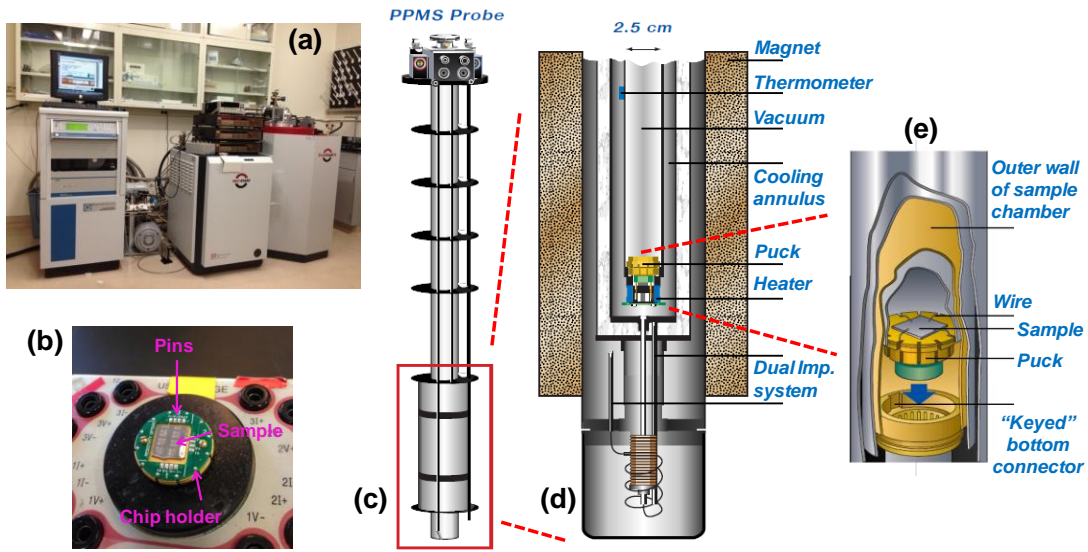

**Figure 5.6.** (a) Physical Property Measurement System (PPMS). (b) Sample and sample holder. (c, d, e) PPMS probe configuration. (c-e) are adapted from PPMS operational manual (Quantum Design, Inc.).

## 5.4 Results

### 5.4.1 Resistance Switching at Various Temperatures

Resistance switching characteristics were first investigated. The results are illustrated for two samples measured in the two systems, demonstrating very similar trends. As shown in **Figure 5.7a** & **c**, switching voltages show a weak or no temperature dependence from 2 K to 300 K; at high temperature (>300 K) the reset (off→on) voltage $V_{reset}$ appears to increase slightly. The insulator-like HRS resistance is more non-linear (non-Ohmic) at lower temperatures (but Ohmic at small voltage, see **Figure 5.7e**). The small-voltage resistance (near 0 V) increases by ~5 orders of magnitude as the temperature decreases from 380K to 77 K, while the high-voltage resistance (*e.g.*, at 2 V) only changes by ~1 order of magnitude (**Figure 5.7a**) over the same temperature range. However, at very low temperatures, there is no further increase of the resistance (overlapped 2K & 20K curves in **Figure 5.7c**). In contrast, the LRS shows little temperature dependence, its set (on→off) voltage ($V_{set}$) again depends on the maximum negative voltage used in the previous scan (**Figure 5.7a** *vs.* **c**) as described in **Chapter VII**. Furthermore, *R-V* (*I-V*) curves are symmetric at all temperatures for both resistance states.

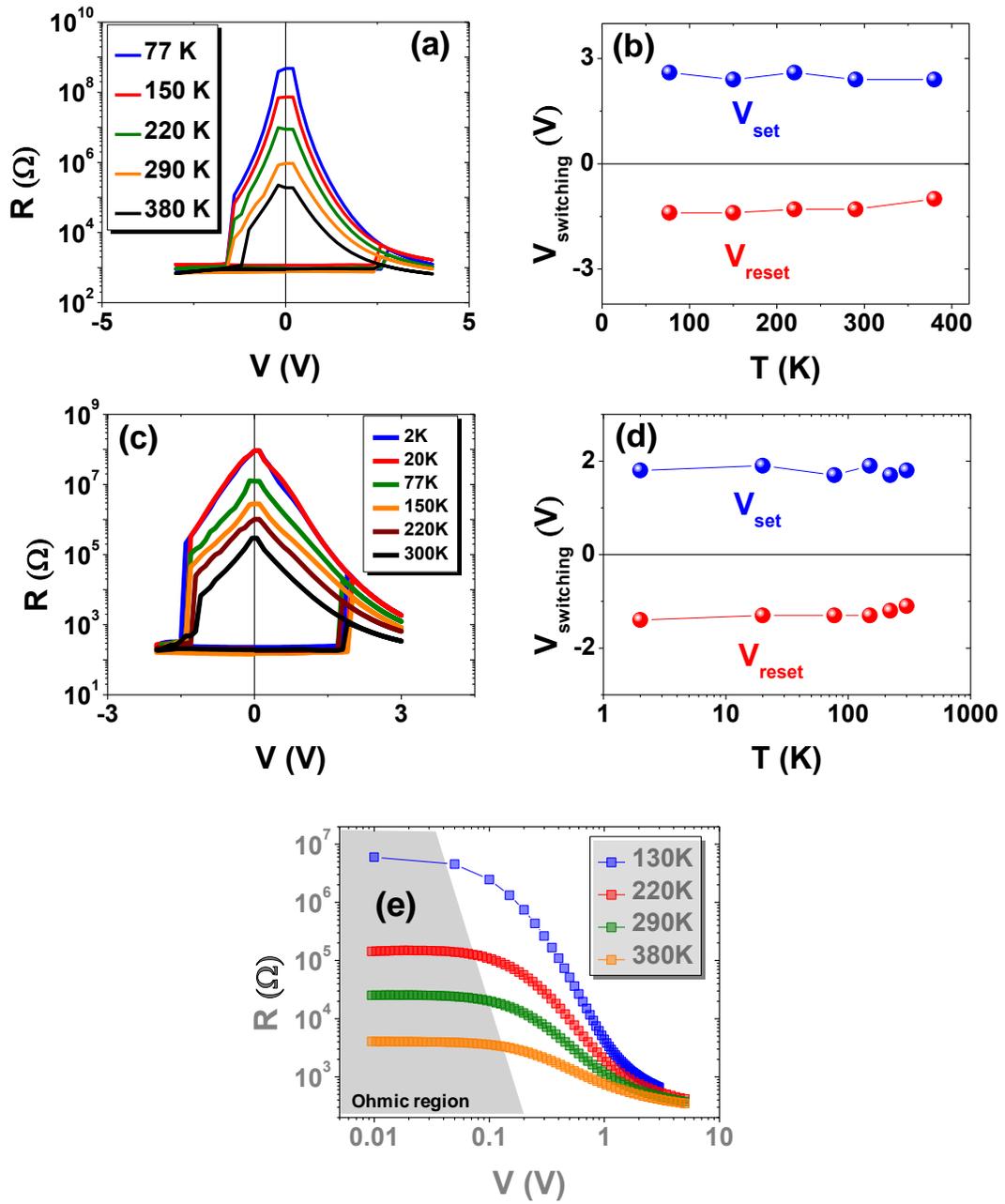

**Figure 5.7.** (a) Switching *R-V* characteristics from 77 K to 380 K, measured by cryogenic probe station (Sample: Mo/Si₃N₄:5%Pt (10 nm)/Pt, diameter *d*=342 μm). (b) Switching voltages in (a). (c) Switching *R-V* characteristics from 2 K to 300 K, measured by PPMS (Hard-wired

sample: Mo/Si$_3$N$_4$:10%Pt (9 nm)/Pt, diameter $d$=512 µm). (d) Switching voltages in (c). (e) HRS $R$-$V$ curve under log-log scale showing an Ohmic region as $V\to0$.

The size dependence was determined using the cryogenic probe station. At a low voltage (0.2 V), the HRS resistance follows Ohm's law giving a set of parallel Resistance-Area ($R\sim1/A$) curves at different temperatures (**Figure 5.8a**), suggesting uniform resistivity independent of sample size. The Ohm law is demonstrated in **Figure 5.8b** by plotting the product of resistances and area ($R\times A$). Therefore, any suitable cell size within the range of 100-512 µm may be used to determine the temperature effect. All the data shown in the next section were collected using cells of 512 µm.

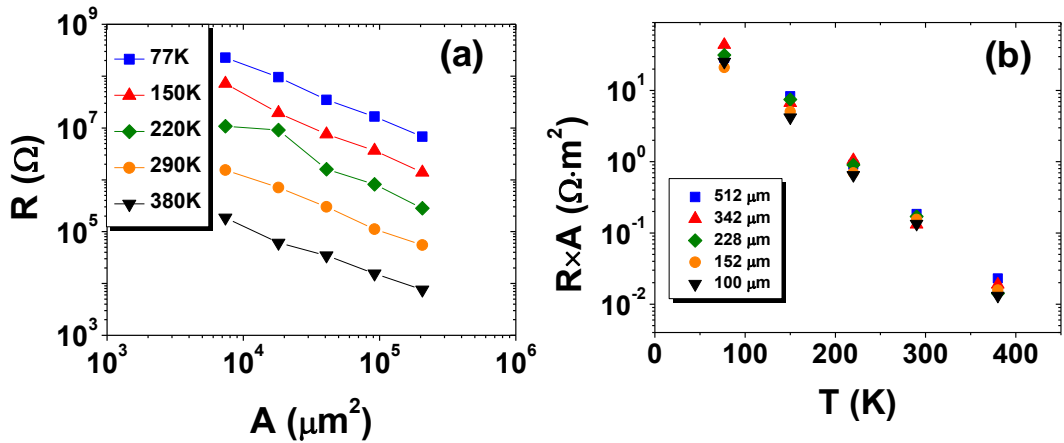

**Figure 5.8.** (a) HRS (read at 0.2 V) *vs.* device area $A$. (b) Renormalized resistance ($R\times A$ product) *vs.* temperature ($T$). (Sample: Mo/Si$_3$N$_4$:5%Pt (10 nm)/Pt)

### 5.4.2 No Magnetic Effect on Resistance Switching

Resistance switching characteristics under a 9 T magnetic field was examined at 2 K. As shown in **Figure 5.9**, there is no apparent magnetic effect on the switching curves, which are plotted in both linear form and semi-logarithmic form. Essentially identical switching voltages and resistance values are shown in these curves regardless of magnetic fields. This excludes any possible magnetic effect, which has been reported for nano-clusters of nominally non-magnetic element such as Pt[28].

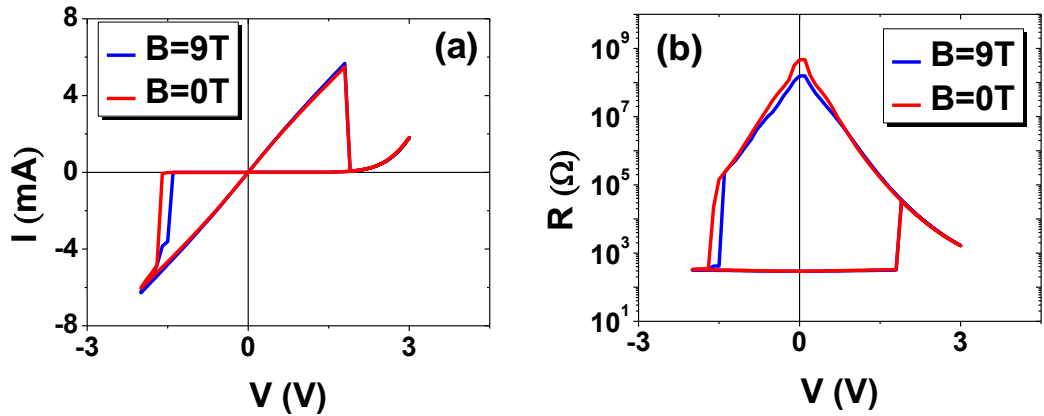

**Figure 5.9.** (a) Switching *I-V* characteristics under magnetic field (9 T *vs.* 0 T) at 2 K. (b) *R-V* characteristics for (a).

### 5.4.3 Resistance Behavior

#### 5.4.3.1 The HRS and the IRS

Different resistance states were probed in the same cell by first arresting several intermediate states during the HRS-to-LRS transition as described in **Chapter VII**. Small-voltage (0.05 V) transport data were then collected for each state during heating and cooling cycles; they generally coincided showing little hysteresis. These results are plotted as resistance ($R$)-temperature ($T$) curve, where the resistance is simply defined as $R=V/I$. To refer to these states, we designate the HRS as the state of the highest resistance $R_{HRS}$, which is also the HRS state obtained in usual resistance switching from the LRS with a typical voltage sweeping window ($\pm4$ V). Intermediate resistance states (IRS) have $R_{IRS}<R_{HRS}$ but their resistance must show a negative $R$-$T$ slope (the insulator type) to be designated as such. The LRS is a metallic state exhibiting a positive $R$-$T$ slope (the metal type), which will be described in the next subsection. **Figure 5.10** shows temperature plots of $R_{HRS}$ and several $R_{IRS}$ in a Mo/ Si$_3$N$_4$:4%Pt (10 nm)/ Pt device. All curves can be separated into three regions:

(1) A flat $R$-$T$ curve at 2 K to ~50K, to be referred to as the low temperature behavior (the resistance in this limit may be used to characterize the state);

(2) A gradually decreasing $R$-$T$ curve at ~100K, to be referred to as the intermediate temperature behavior;

(3) A rapidly decreasing $R$-$T$ curve at >200K, to be referred to as the high temperature behavior.

It is clear from **Figure 5.10** that the intermediate and high temperature $R$-$T$ dependence becomes progressively weaker as the HRS changes to various IRS of a lower resistance. (Note that **Figure 5.10d-g** are linear plots, not semi-logarithm plots as in **Figure 5.10a-c**.) At the same time, the upper temperature limit of the low temperature regime progressively decreases in the progression from the HRS to IRS of decreasing resistance.

This strongly suggests that thermally activated processes are more likely to proceed in an IRS of a lower resistance.

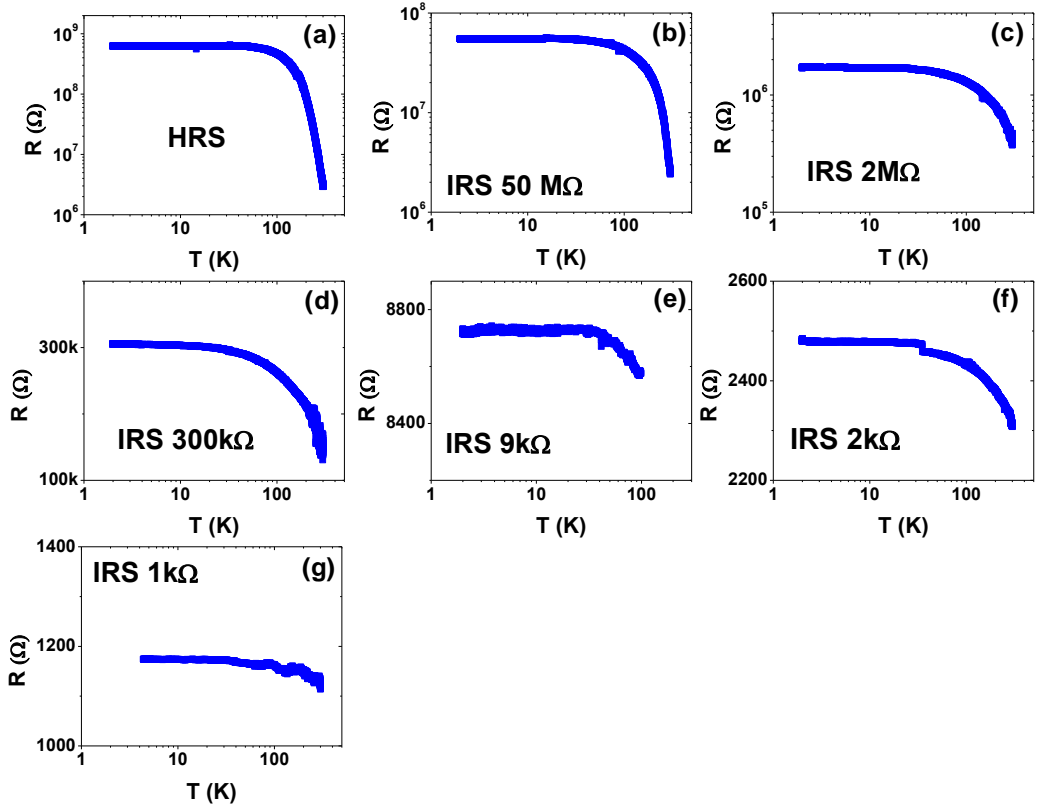

**Figure 5.10.** Resistance ($R$)-temperature ($T$) for various resistance states: (a) HRS; (b) $R_{IRS}$~50 MΩ; (c) $R_{IRS}$~2 MΩ; (d) $R_{IRS}$~300 kΩ; (e) $R_{IRS}$~9 kΩ; (f) $R_{IRS}$~2 kΩ; (g) $R_{IRS}$~1.2 kΩ. $R$ values here are "saturated" resistance at 2 K. Nanometallic film: Si$_3$N$_4$: Pt, $f_{Pt}$=4%, δ=10 nm.

### 5.4.3.2  The LRS

The LRS is distinct from the IRS and HRS because it has an entirely different temperature dependence featuring a rising $R$-$T$ curve. However, it is not unique since several LRS states can be arrested following the procedure described in **Chapter VII**.

We studied several of them (with a low-temperature resistance from 80 Ω to 600 Ω) at a small voltage (0.05 V). Similar to the HRS and IRS data, the resistance saturates at low temperature; it then rises slightly (by a few percent) starting at the intermediate temperature. Since the LRS resistance (~ a few hundred Ω) is comparable to the bottom electrode resistance (serial load), the load resistance need to be subtracted from the measured resistance to allow a more accurate analysis.

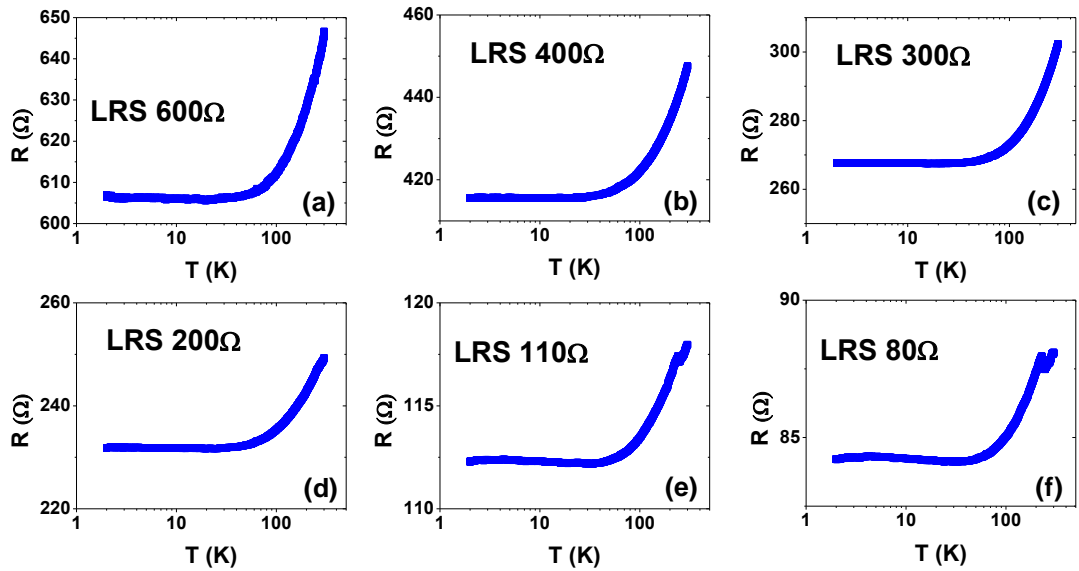

**Figure 5.11.** Resistance ($R$)-temperature ($T$) for various LRS: (a) $R$~600 Ω; (b) $R$~400 Ω; (c) $R$~300 Ω; (d) $R$~200 Ω; (e) $R$~110 Ω; (f) $R$~80 Ω. $R$ values here are "saturated" resistance at 2 K. Nanometallic film: $Si_3N_4$: Pt, $f_{Pt}$=4%, $\delta$=10 nm.

### 5.4.3.3 Thickness Dependence of HRS

We studied the thickness dependence of resistance of the HRS. Since the resistances of the IRS and LRS states are dependent on the arresting procedure, which is dependent on the load resistance and parasitic capacitance of the circuit, it is not possible to ascertain the same "state" is attained in samples of different thickness. On the other hand, from our past experience (**Chapter II & VII**), there is reasonable certainty that the HRS is unique, and can be reliably reached after a normal switching cycle after the LRS-to-HRS transition if a typical voltage sweeping window (±4 V) is used. Therefore, the data below pertains to the HRS only. As shown in **Figure 5.12**, the room temperature resistance already exhibits a strong (exponential) thickness dependence: $R_{300K}$ increases by ~2 orders of magnitude as the film thickness increases by merely two-fold, from 7 nm to 14 nm. The already strong exponential dependence becomes even stronger at lower temperatures: $R_{2K}$ increases by ~4 orders of magnitude for the same two-fold thickness increase. Three temperature regions described in **Section 5.4.3.1** can again be identified. Interestingly, the limiting temperature of these regimes seems to be independent of thickness.

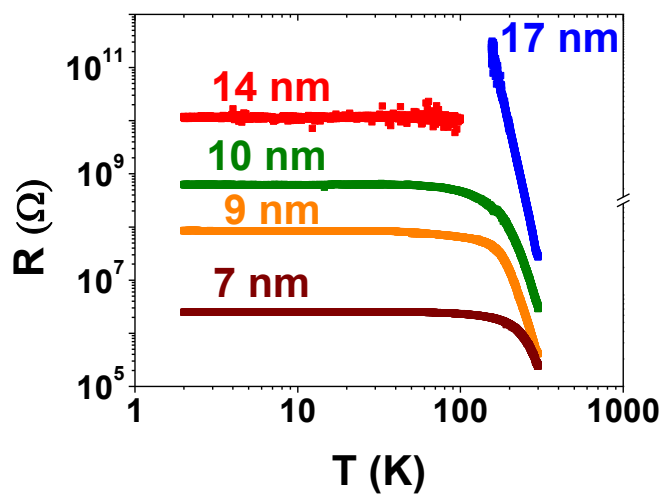

**Figure 5.12.** HRS transport characteristics for $\delta$=7 nm, 9 nm, 10 nm, 14 nm (incomplete data) and 17 nm Mo/Si$_3$N$_4$:10%Pt/Pt devices. Reading voltage: 0.05 V. Data collection for 17 nm sample was stopped at 150 K because of instrument limit (low current).

### 5.4.3.4 Metal Concentration Dependence of HRS

The data presented above were obtained from samples of 4% Pt. Samples of several other Pt concentrations with an identical thickness of 10 nm were studied to determine the compositional effect on the HRS. As shown in **Figure 5.13**, the *R-T* curve of each composition qualitatively follows the three-region behavior in **Section 5.4.3.1**. The room temperature resistance monotonically decreases as the metal amount increases, consistent with the common experience of "more metal, more conducting". However, while the low temperature resistance eventually increases with the metal concentration, it initially increases with Pt concentration reaching a maximum at around 20% Pt. It is also evident that the upper temperature limit of the low-temperature regime progressively decreases with the metal concentration. This strongly suggest that thermally activated process may become operational at a lower temperature in Pt-rich samples, in a similar manner as in IRS of a lower resistance value. Moreover, even at a very high metal concentration of 39% Pt, when the nanometallic films have already approached/reached its metal percolation limit (typically ~40% metal), the *R-T* curve, while quite flat, is still of the insulator type. Therefore, all of the samples in this figure are indeed in the HRS.

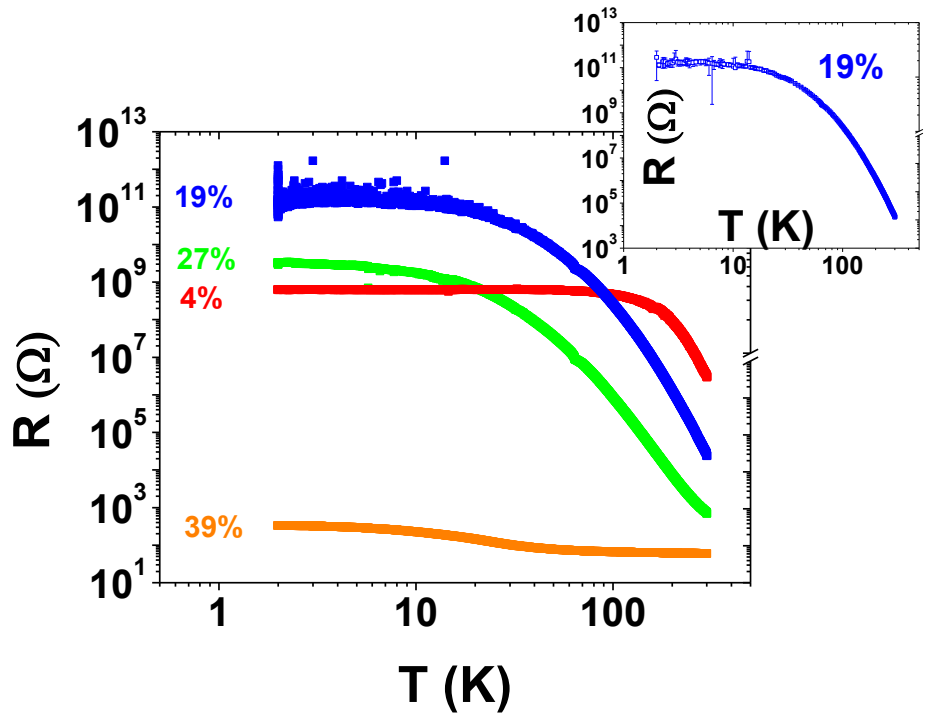

**Figure 5.13.** HRS transport characteristics for $f_{Pt}$=4%, 19%, 27%, 39% devices (Mo/Si$_3$N$_4$:Pt (10 nm)/Pt). Reading voltage: 0.05 V. Inset: $f_{Pt}$=4% data with explicit error bar.

## 5.5  Model Fitting: LRS

### 5.5.1  General Behavior and Empirical Fitting

As shown in **Figure 5.11**, all LRS show metallic behavior with a positive *R-T* slope. However, based on the discussion in **Chapter II** and **VII**, the underlying bottom electrode resistance significantly contributes to the apparent resistance. Therefore, it must be subtracted before further analysis of the LRS data.

We first use a polynomial to mathematically fit the bottom electrode *R-T* characteristics. The device resistance at the most negative bias, which has the lowest resistance, *e.g.*, **Figure 5.11f**, is taken as the bottom electrode resistance. We next subtract this resistance from all the LRS data in **Figure 5.11a-e**, and one of the results is replotted in **Figure 5.14a**. The remaining (corrected) resistance still shows a positive *R-T* slope, indicating the residual resistance after removing the contribution of the bottom electrode is still metallic. The data can be fit by an empirical formula:

$$\rho_{LRS,empirical}(T) = \rho_0 + AT + BT^2$$

(similar empirical formula can also be found in ref [29]) as shown in **Figure 5.14b**. Applying the same procedure, we can correct all the LRS data in **Figure 5.11** and replot them in **Figure 5.14c**. Remarkably, all such curves scale similarly: they collapse to the same curve after normalization with respect to the residual resistance at 0 K. Likewise, we have corrected the LRS data for devices of other thickness and presented the normalized data in **Figure 5.14d**. Again, the normalized data collapse into one curve. Therefore, there is apparently no fundamental difference in different LRS, differing in either residual resistance (at 0K) or film thickness. This allows us to use one representative curve to interrogate the LRS transport below.

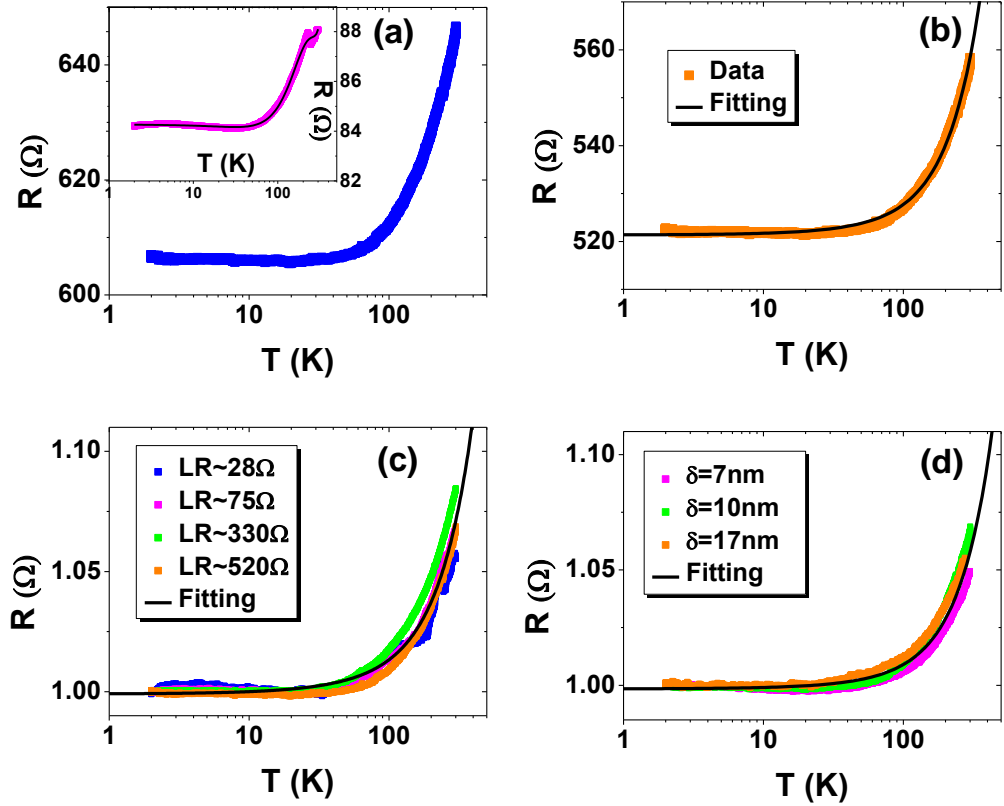

**Figure 5.14.** (a) LRS in **Figure 5.11a**. Inset: LRS (using -10V reset voltage) in **Figure 5.11f** and a polynomial fitting (black curve). (b) *R-T* relation after data subtraction in (a) and an empirical fitting using $\rho(T) = \rho_0 + AT + BT^2$. (c) Renormalized *R-T* data for various LRS and empirical fitting. (Lowest LRS is not shown because of high noise to signal ratio) (d) Renormalized *R-T* data for various thickness and empirical fitting.

### 5.5.2 Electron-phonon and Electron-electron Scattering

The LRS resistance can be fit with the following formula, $a+b(T-T_0)+cT^2$, in which $a$ is the most significant term at all temperature and $cT^2$ is noticeable only in the low temperature range. (In the above, we obviously cannot distinguish $a$ and $bT_0$ at low temperature, although at high temperatures when there is an apparent bilinear behavior it is possible to determine $T_0$ and $b$ independently.) According to the conduction theory of electrons in metals (see introduction part), the high temperature resistivity originates from electron-phonon scattering, leading to a linear law. Such linear law is verified in **Figure 5.15a** in the temperature range of 100 K-300K:

$$R_{LRS,100K-300K}(T) = R_0(1+\alpha T) = 511\Omega \times (1+2.96\times10^{-4}T)$$

The temperature coefficient $\alpha$ is about one order of magnitude smaller than that of a typical good metal (*e.g.* $\alpha_{\text{Cu}}$=6.8×10$^{-3}$ K$^{-1}$, $\alpha_{\text{Al}}$=3.9×10$^{-3}$ K$^{-1}$, $\alpha_{\text{W}}$=4.5×10$^{-3}$ K$^{-1}$, $\alpha_{\text{Pt}}$=3.9×10$^{-3}$ K$^{-1}$)[30], but is similar to that of some metal alloys (*e.g.* $\alpha_{\text{Ni-Cr}}$=4.0×10$^{-4}$ K$^{-1}$, $\alpha_{\text{Cr-Al}}$=5.8×10$^{-4}$ K$^{-1}$)[30], where a larger disorder exists.

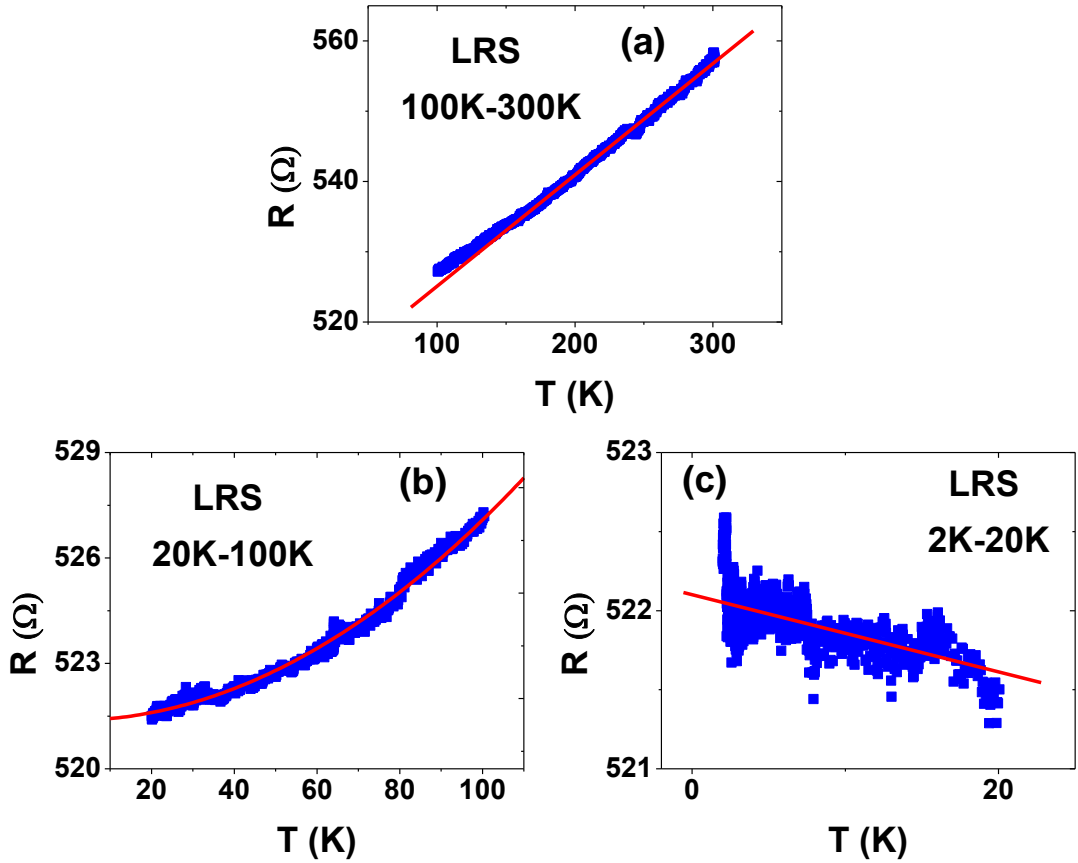

**Figure 5.15.** (a) Linear-law fitting: $R(T)= R_0(1+\alpha T)$ in temperature range of 100 K-300 K. (b) Parabolic-law fitting: $R(T)= R_0'(1+\beta T^2)$ in temperature range of 20 K-100 K. (c) Remaining resistance below 20 K (red curve is a trend line only).

The linear law breaks down at low temperature, where a curvature is clearly seen in the data between 20 K and 100 K (**Figure 5.15b**). In this regime, resistance can be fitted by a parabolic law:

$$R_{LRS,20K-100K}(T) = R_0'(1+\beta T^2) = 521\Omega\times(1+1.09\times10^{-6}T^2)$$

with the parabolic term arising from the electron-electron interaction/scattering. This parabolic form is typically observed in strongly correlated materials[29,31-33], in which strong electron-electron interaction exists. Based on our discussion in **Chapter II & VII**, different LRS correspond to different de-trapping states, in which trapped electrons still exist but do not fully block free-electron passage. Therefore, electron-electron interaction/scattering is expected. The temperature coefficient $\beta$ above is similar to or smaller than the ones shown in the literature (*e.g.*, $\beta_{\text{Ni-Nanowire}}=2.8\times10^{-6}$ K$^{-2}$ (ref.[33]), $\beta_{\text{LaCeCuO4}}=7.1\times10^{-5}$ K$^{-2}$ (ref.[32]), $\beta_{\text{W}}=5.1\times10^{-5}$ K$^{-2}$ (ref.[31])), indicating a weaker electron-electron interaction than in standard strongly correlated systems.

At less than 20 K, resistance saturates or even slightly increases (**Figure 5.15c**). The nonzero resistance at 0 K in metals usually implies impurities/defects serving as scattering centers. In our case, scattering may arise from two origins: one is from structural disorder inherent in the amorphous film or induced by metal-atom doping, the other is from charge disorder of trapped charges. Structural disorder is expected to remain unchanged for different states while charge disorder is expected to vary with states. Based on the experimental observation that residual resistance at 0 K is highly dependent on the resistance states, we can infer that charge disorder is the dominant origin of scattering.

### 5.6   Model Fitting: HRS

### 5.6.1   Preliminary Considerations

#### 5.6.1.1   HRS energy barrier

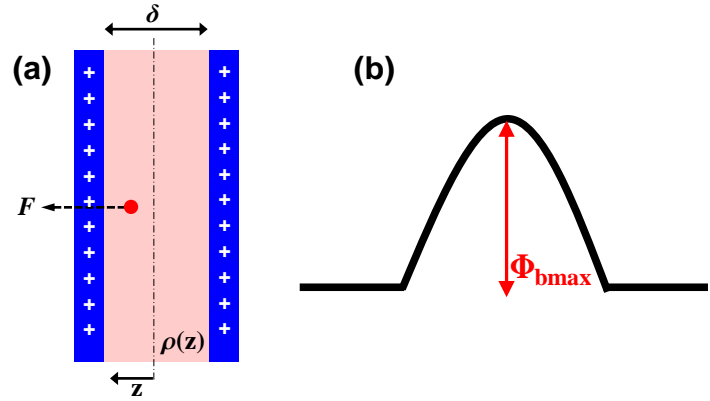

**Figure 5.16.** (a) Simplified picture of charge distribution. Red region represents uniform distributed trap (negative) charge, compensated by positive charge on electrodes (blue). (b) Parabolic Coulombic potential barrier induced by charge distribution in (a).

HRS is a metastable state, in which negatively charged electrons are trapped. One might hope to model the trapped charge effect on the energy landscape by assuming a uniform distribution of such charge. This will establish a long range field in the device, but not outside the device, because of the screening effect (compensating charges) of the electrodes. The solution of the electrostatic problem is standard: for a uniform charge density $\rho(z)=\rho$, it has a parabolic potential

$$\Phi_b(z) = \frac{e\rho}{2\varepsilon}\left(\frac{\delta^2}{4} - z^2\right), \quad -\frac{\delta}{2} \leq z \leq \frac{\delta}{2}$$

with the maximum being

$$\Phi_{b,\max} = \frac{e\rho\delta^2}{8\varepsilon}$$

at the center of the film. For an estimate, a 10 nm thick film with a dielectric constant $\varepsilon=10\varepsilon_0$ and a charge density of $\rho=4.4\times10^{25}$/m$^3$ (*i.e.*, 1 electron per 3 nm$\times$3 nm$\times$3 nm cube) will have a 1 eV maximum at the center.

While the above solution is applicable to an insulating film, our cells all manifest various degrees of electron conduction even at 2 K. Therefore, the long-range field provided by the parabolic potential cannot be sustained: it will be compensated by electron redistribution which, over time, will completely wipe out the field.

A more realistic picture of the energy landscape in our film is obtained by considering the short range potential of the trapped charge. For two (negative e) point charges located at $(x, y)=(\pm d/2, 0)$, there is a saddle point at $(0, 0)$. The computed Coulomb potential profile is shown in **Figure 5.17**, in which the maximum along $(0, y)$ at $(0, 0)$ is the saddle point and the energy profile along $(0, y)$ is

$$\Phi_{\min}(0, y) = \frac{e^2}{2\pi\varepsilon} \frac{1}{\sqrt{(d/2)^2 + y^2}}$$

The saddle pont defines the "effective barrier height" (**Figure 5.17**) for electrons passing between $x=\pm d/2$. In our picture, these electrons are in the channel, and in the HRS they need to overcome the above barrier to communicate from channel to channel, eventually from one electrode to another electrode, thus providing current. This barrier reaches 1 eV when the two charges are separated by 6 nm. Note that the barrier profile is smooth and does not have the square-wall shape commonly assumed in tunneling models. This may have implications on the transport properties as a smooth barrier always provides a stronger voltage dependence for tunneling current than a square-wall shaped barrier.

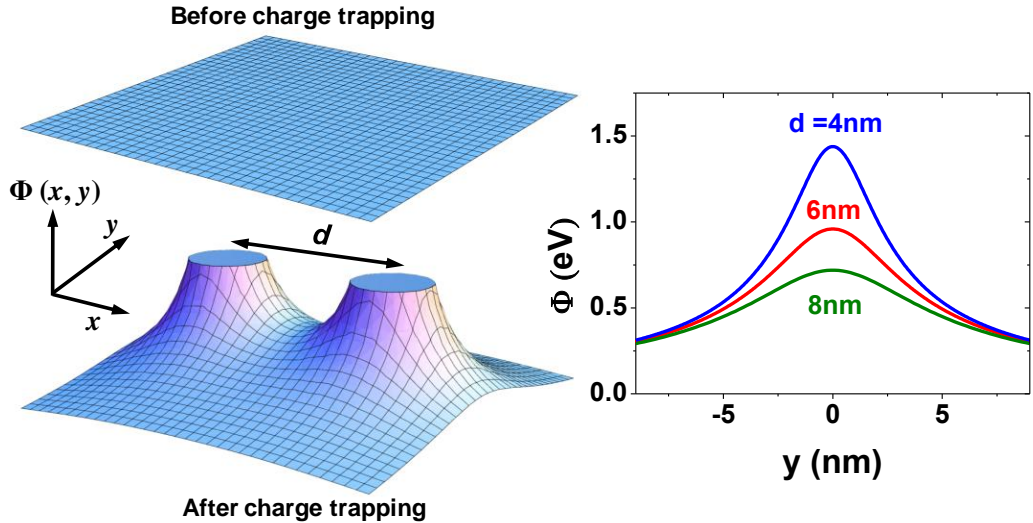

**Figure 5.17.** Point charge model: energy potential before and after two symmetric charges are trapped. Inset: minimum energy profile along $y$ axis assuming two charges are $d$=4 nm, 6 nm, 8 nm apart.

### 5.6.1.2 Overview of fitting strategy

In the following, we will systematically model-fit the HRS transport data using various transport models. Since qualitatively the same temperature dependence is seen in samples of various composition and thickness, in the HRS and various IRS, the procedure will be repeated to obtain pertinent microscopic parameters that govern electron transport in various devices. Below we describe the overall strategy.

(i) Low temperature: We believe elastic tunneling is the dominant transport mechanism here. There are three microscopic model parameters: the barrier height, the barrier width, and the area fraction of the device that participates in transport. Our picture is that the barrier is the patch between neighboring but disconnected channels, and it must be overcome by tunneling with the assistance of an applied field. Since in its simplest form the $R$-$V$ curve has at least three characteristic features, $R(0)$, $R'(0)$, and the high-voltage/field dependence, it is mathematically possible to solve the problem.

(ii) High temperature: We believe thermally assisted (inelastic) and field assisted tunneling is the dominant mechanism. We will rely on variable range hopping, taking into consideration of field assistance, to model-fit the data. Again, there are three microscopic model parameters: the density of states, the extent of the wave function (a localization length), and the area fraction of the device that participates in transport. As in low temperature, it is possible to determine these three parameters mathematically using the data at $V$=0 and at high fields.

(iii) Intermediate temperature: It turns out that the data of the intermediate temperature cannot be satisfactorily explained by the superposition of the low-temperature transport and high-temperature transport. Therefore, another transport mechanism is needed to explain the relatively gradual temperature dependence of resistance. We will choose the fluctuation-induced tunneling mechanism, which describes elastic tunneling albeit with the consideration of the Johnson noise, which is temperature dependent and can cause fluctuation in the barrier height. Again, we will determine the three microscopic parameters, two for the model, the third one for the area fraction of the device.

Lastly, we will present a simple mathematical formula that turns out to provide good numerical fitting to all the data. The physical meaning of this formula, however, is less certain, which will be discussed in **Appendix**.

### 5.6.2 The Low Temperature Region: Direct Tunneling and Fowler-Nordheim Tunneling

Resistance saturation at low temperature is a strong indication of a purely tunneling process. We will extract its physical parameters by assuming the simplest energy barrier profile with a rectangular shape. Generally speaking, direct tunneling (DT) occurs at low voltage bias when the entire width of the barrier affects the decay of electron wave function. As voltage or electrical field increases, barrier shearing gradually happens and gives rise to a higher tunneling current. When the voltage reaches the barrier height, a rectangular barrier is severely sheared into a triangle shape, which triggers the Fowler-Nordheim Tunneling (FNT).

The probability $T(E)$ that an electron can penetrate, in the $x$ direction, a potential barrier of height $\Phi_b(x)$ is given by the Wigner-Kramers-Brillouin (WKB) approximation[34]:

$$T(E) = \exp\left\{ -\frac{4\pi}{h} \int_{x1}^{x2} \sqrt{2m^* e\left(\Phi_b(x) - E\right)} dx \right\}$$

where $E = mv_x^2/2$ is the energy component of the incident electron in the $x$ direction. Under a bias $V$, the net flow of electrons $J$ is determined by the difference of forward and backward electron flow:

$$J = \int_0^E T(E) dE \times \left\{ \frac{4\pi m^{*2} e}{h^3} \int_0^\infty \left[ f(E) - f(E + eV) \right] dE \right\}$$

where $f(E)$ is the Fermi-Dirac distribution function.

Over a rectangular barrier ($\Phi_b(x) = \Phi_b$), direct tunneling under a small bias (DT, $V \rightarrow 0$) and the Fowler-Nordham tunneling (FNT) at large $V$ have the following analytical forms of current density:

$$I_{DT}(V) = A \frac{3\sqrt{2m^* e\Phi_b}}{2d} \frac{e^2}{h^2} V \exp\left(-\frac{2d\sqrt{2m^* e\Phi_b}}{\hbar}\right)$$

$$I_{FNT}(V) = A \frac{e^2}{16\pi^2 \hbar \Phi_b d^2} V^2 \exp\left(-\frac{4d\sqrt{2m^* e}}{3\hbar V} \Phi_b^{3/2}\right)$$

Here, $m^*$ is electron effective mass ($m^*$=0.5$m_0$ is assumed), $d$ is tunneling distance, $\Phi_b$ is barrier height and $A$ (less than $A^*$, the area of the top electrode) is effective tunneling area. It is worth noting that tunneling is not necessarily a one-step process across the entire film, $i.e.$, the tunneling distance is uncertain and need to be determined by data fitting. If the same set of tunneling barriers ($i.e.$, if $d$, $\Phi_b$ and $A$ are unique) controls the DT and FNT, then we can obtain these barrier parameters using the following procedure:

(1) Fitting the large-voltage current in the $\ln(I_{FNT}/V^2)$ $vs.$ $1/V$ plot, to obtain $d\Phi_b^{3/2}$ from the slope (**EQ1**) and $A/\Phi_b d^2$ from the intercept (**EQ2**).

(2) Fitting the low-voltage current in the $I_{DT}$ $vs.$ $V$ plot, to obtain another equation for ($d$, $\Phi_b$, $A$) from the slope (**EQ3**).

**EQ1 & EQ2**: $\ln(I_{FNT}/V^2) = \ln(\frac{Ae^2}{8\pi h \Phi_b d^2}) - \frac{4d\Phi_b^{3/2}\sqrt{2m^* e}}{3h} \times \frac{1}{V}$

**EQ3**: $I_{DT} = A \frac{3e^2\sqrt{2me\Phi_b}}{2h^2 d} V \exp\left(-\frac{4\pi d}{h}\sqrt{2m^* e\Phi_b}\right)$

Therefore, the three barrier parameters can be uniquely determined by the three equations (**EQ1 & 2 & 3**). As shown in **Figure 5.18a-e**, devices with different thickness all follow FNT in the high voltage limit. Their slopes of $\ln(I_{\text{FNT}}/V^2)$ *vs.* $1/V$ uniquely determine $d\Phi_b^{3/2}$ listed in **Table 5.1**. Since both tunneling distance and barrier height positively contributes to the tunneling resistance, we can define:

$$\text{“Barrier hardness”} = d\Phi_b^{3/2}$$

to characterize how difficult the tunneling process is. A thicker film is found to have a higher $d\Phi_b^{3/2}$ (**Table 5.1**) and more difficult to tunnel.

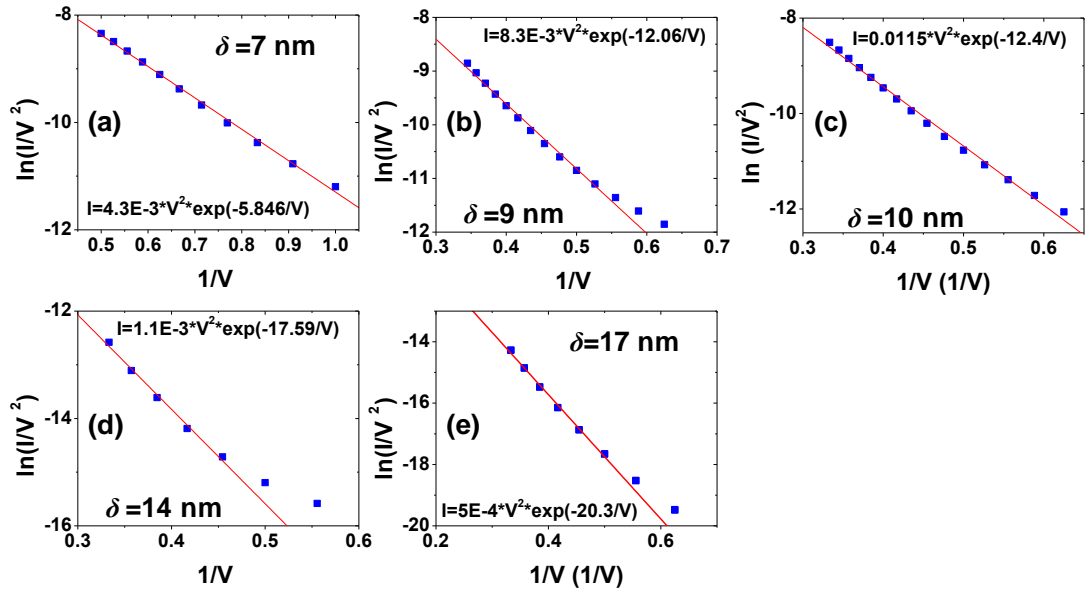

**Figure 5.18.** (a)-(e) FNT plot for various film thickness, from standard *I-V* curves at 2 K.

| Thickness | $\delta$=7 nm | $\delta$=9 nm | $\delta$=10 nm | $\delta$=14 nm | $\delta$=17 nm |
|---|---|---|---|---|---|
| $d\sqrt{\Phi_b^3}$ (nm·eV$^{3/2}$) | 1.22 | 2.51 | 2.58 | 3.67 | 4.23 |

| $A/\Phi_b d^2$ (1/eV) | 2798 | 5400 | 7807 | 716 | 351 |
|---|---|---|---|---|---|

**Table 5.1.** Barrier hardness $d\sqrt{\Phi_b^3}$ for samples with various thickness, extracted from slope of FNT plot.

Next, data in the small voltage limit are considered in **Figure 5.19**, where low bias *I-V* curves are fitted with a parabolic law ($I = A_1 V + A_2 V^2$) from which the linear DT term ($A_1 V$) can be unambiguously extracted. After applying the aforementioned fitting procedure and incorporating the FNT fit, barrier parameters obtained are listed in **Table 5.2**. It is found that the barrier area is indeed very small (~100×100 nm$^2$ out of device area 500×500 μm$^2$). This "localized" nature suggests the tunneling current is dominated by the regions with the weakest barrier. These weakest barriers are not lumped elements (unlike a single filament or a few of them); they are uniformly distributed throughout the film cross section because the HRS has a strong size dependence ~1/$A^*$, meaning $A$ is proportional to $A^*$. Indeed, 100×100 nm$^2$ is relatively large and unlikely to come from one filament (~a few atoms wide) or a few of them. The barrier width $d$ obtained in **Table 5.2** is sensible in that it is smaller than the sample thickness and is within the reachable range of tunneling (~1 nm). The barrier height $\Phi_b$ ~ 1 eV seems reasonable for a Coulombic potential: a trapped electron at a distance of 1.4 nm away can elevate the electrostatic energy by 1 eV. (Screening effects are not considered here, which might lower the Coulombic potential).

In **Table 5.2**, there is a general trend of increasing barrier height and width with increasing thickness. This suggests that, statistically, a thicker film has a higher probability to encounter a harder barrier. Such statistical fluctuation is expected to become stronger as the film thickness $\delta$ is close to the tunneling distance $d$, which may explain why the barrier height increases by almost two-fold when the thickness increases from 7 nm to 9 nm. The change of these barrier parameters (more precisely, "barrier hardness"=$d\Phi_b^{3/2}$) is partly responsible for the "exponential thickness dependence" that we observed extensively in nanometallic films (**Chapter II**).

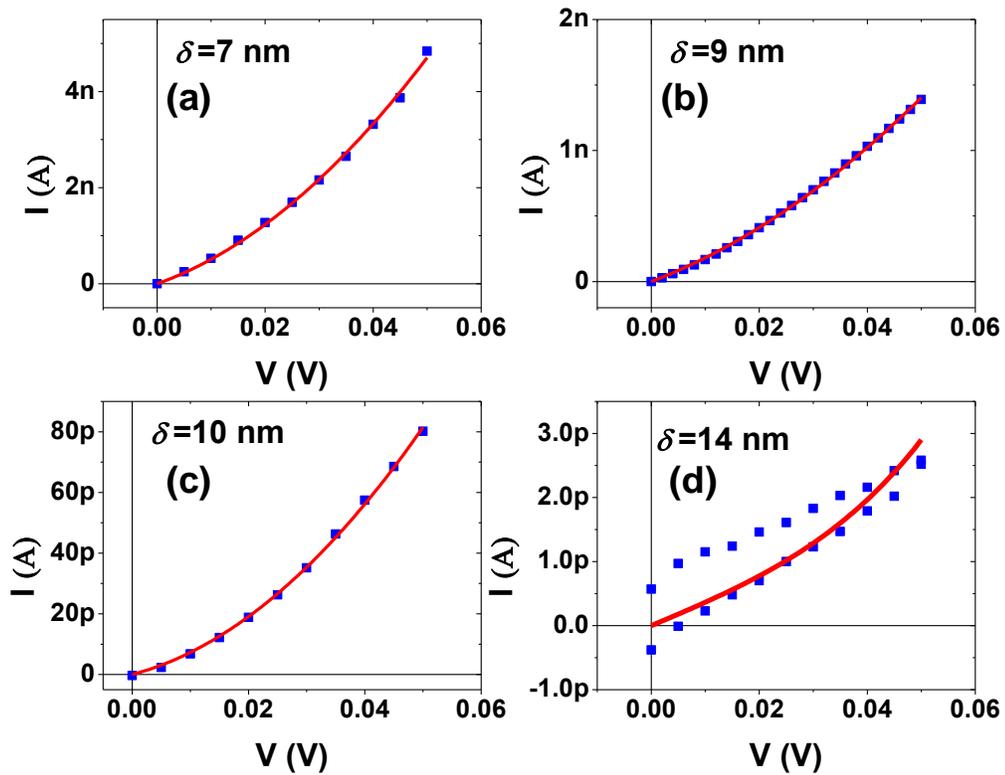

**Figure 5.19.** (a)-(d) Small voltage data fitted with parabolic law: $I = A_1 V + A_2 V^2$ from which linear DT term ($A_1 V$) can be extracted. Sample with $\delta$=17 nm is too resistive for low-bias $I$-$V$ to be measured. The hysteresis in (d) is due to low signal-to-noise signal ratio at the pA current level.

| Thickness | $\delta$=7 nm | $\delta$=9 nm | $\delta$=10 nm | $\delta$=14 nm |
|---|---|---|---|---|
| $d\sqrt{\Phi_b^3}$ (nm·eV$^{3/2}$) | 1.22 | 2.51 | 2.58 | 3.67 |
| $a$ (nm) | 109 | 176 | 252 | 70 |
| $d$ (nm) | 2.68 | 2.33 | 3.09 | 2.23 |
| $\Phi_b$ (eV) | 0.59 | 1.05 | 0.89 | 1.39 |

**Table 5.2.** Calculated barrier parameter ($A=a^2$, $d$, $\Phi_b$) by assuming identical barrier for DT and FNT.

By applying a similar procedure to other nanometallic films, we can obtain their physical parameters, listed in **Table 5.3**. In metal-rich films, barrier hardness $d\Phi_b^{3/2}$ becomes smaller, which is primarily due to the lower barrier height $\Phi_b$ that decreases from 0.89 eV to 0.25 eV as the Pt concentration changes from 4% to 27%. Such barrier lowering effect might arise from different distribution of trapped electrons in different nanometallic film. For a higher $f_{Pt}$ film, trapped electrons tend to be more sparsely distributed near the conduction paths thus leading to a lower barrier (saddle point). Interestingly, the tunneling distance $d$ appears to slightly increase with metal concentration, negatively contributing to the conduction. This suggests that, at high Pt concentrations, the majority of Pt atoms is already in Pt nanoparticles thus belonging to the channel, and the channel-to-channel separation may actually increase because of Pt-clustering, resulting in an increase of the tunneling distance. Alternatively, tunneling distance may scale with the separation between trapped electrons, whose concentration decreases at a higher $f_{Pt}$ according to the $\Phi_b$-$f_{Pt}$ dependence, again leading to a decrease of $d$ at higher $f_{Pt}$. Since direct tunneling is more sensitive to the barrier width (~exp($d$)) than the barrier height ($\sim\exp(\sqrt{\Phi_b}\,)$), the increase of tunneling distance has a more pronounced effect on DT than the increase of barrier height. This is in contrast to the case of FNT, which is more strongly affected by the decrease of "barrier hardness". This contrast results in a crossover: $R$(high $f_{Pt}$)$> R$(low $f_{Pt}$) at low bias where DT dominates but $R$(high $f_{Pt}$)$< R$(low $f_{Pt}$) at high bias where FNT dominates, which explains the "voltage crossover" illustrated in **Figure 5.20**. The same observation, that $R$(high $f_{Pt}$)$> R$(low $f_{Pt}$) at low bias where DT dominates, is also the origin of the temperature crossover shown in **Figure 5.13**, since as will be shown in the next section, the high temperature transport is dominated by VRH which gives $R$(high $f_{Pt}$)$< R$(low $f_{Pt}$).

| Thickness | $f_{Pt}$ =4 at.% | $f_{Pt}$ =19 at.% | $f_{Pt}$ =27 at.% |
|---|---|---|---|
| $d\sqrt{\Phi_b^3}$ (nm·eV$^{3/2}$) | 2.58 | 1.72 | 0.828 |
| $a$ (nm) | 252 | 760 | 575 |
| $d$ (nm) | 3.09 | 5.40 | 6.75 |
| $\Phi_b$ (eV) | 0.89 | 0.47 | 0.25 |

**Table 5.3.** Calculated barrier parameter ($A=a^2$, $d$, $\Phi_b$), faced by both DT and FNT, for various metal composition $f_{Pt}$.

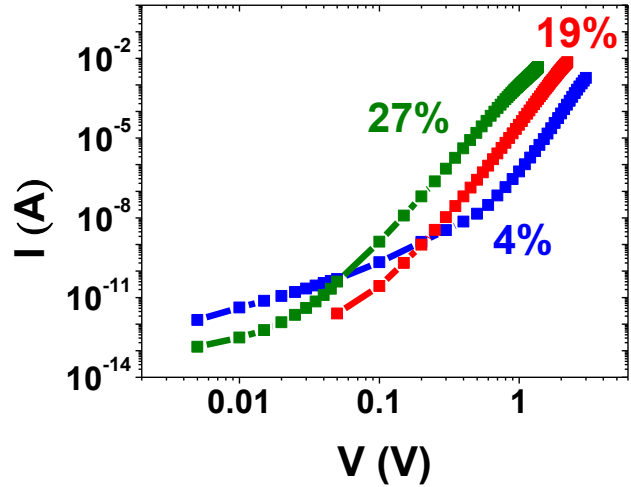

**Figure 5.20.** *I-V* curves for various metal composition films at 2K. Current is higher at a larger *V* but lower at smaller *V* for the sample with a higher $f_{Pt}$.

To estimate the effective barrier height for various (intermediate) resistance states, a simplified formula for direct tunneling $R = A \exp\left(2d\sqrt{2m^{*}e\Phi_{b}}/\hbar\right)$ is used. Since there is no microscopic structure change for various resistance states in the same device, we first assume barrier height $\Phi_{b}$ is the only variable causing different resistances while keeping *d*, the "channel-to-channel" distance, constant. (If *d* is related to the separation between trapped electrons, then both *d* and $\Phi_{b}$ might change in principle. However, high voltage IRS cannot be measured because switching will occur ~1 V. This makes it difficult to uniquely determine *d* and $\Phi_{b}$. Therefore, we let *d* remain constant, for simplicity). As shown in **Table 5.4**, a lower resistance state (defined by the "flat" resistance at 2 K) has a lower energy barrier. In our picture, this may be explained by a weaker Coulombic repulsion potential because of a larger spacing between (fewer) trap electrons.

| RS ($R_{2K}$) | 600MΩ | 55MΩ | 2MΩ | 300kΩ | 8.7kΩ | 2.5kΩ |
|---|---|---|---|---|---|---|
| $\Phi_b$(eV) | 0.89 | 0.70 | 0.47 | 0.36 | 0.20 | 0.15 |

**Table 5.4.** Tunneling barrier for various resistance states by assuming $\Phi_{b,600MΩ}$ =0.89 eV for $\delta$=10 nm sample. The tunneling distance $d$ =3.09 nm and area $A$=252×252 nm$^2$ are fixed.

## An Alternative Interpretation

There is an alternative way to interpret the tunneling data if we allow the possibility that DT and FNT shown in different voltage/field regimes are governed by two different sets of barrier. This extra degree of freedom makes it impossible to determine the barrier parameters ($d_{DT}$, $\Phi_{b,DT}$, $A_{DT}$, $d_{FNT}$, $\Phi_{b,FNT}$, $A_{FNT}$) using the above procedure, so we will need other assumptions to make the problem tractable.

One possibility is to assume that the tunneling distance for FNT is the apparent film thickness ($d=\delta$). Then the calculated FNT barrier height is on the order of $\Phi_{b,FNT}$~0.4 eV, and the effective area (~a few hundred nanometers) is not much different from the previous case (**Table 5.5**). The DT contribution is assumed to come from the entire device area and tunneling is through trap-assisted tunneling (TAT, a simplified $n$-step serial DT model is assumed: $R_{TAT}=nR_{DT}$) in which the tunneling barrier height is the energy offset between the Fermi-level of Pt and the conduction band of the Si$_3$N$_4$ dielectric (3.5 eV). This gives a calculated DT tunneling distance (*i.e.*, the trap distance) of $d_{DT}$~3 nm. Thus deduced parameters do not seem unreasonable, so this alternative interpretation cannot be excluded.

| Thickness | $\delta$=7 nm | $\delta$=9 nm | $\delta$=10 nm | $\delta$=14 nm |
|---|---|---|---|---|
| $d\sqrt{\Phi_b^3}$ (nm·eV$^{3/2}$) | 1.22 | 2.51 | 2.58 | 3.67 |
| $d_{FNT}$ (nm) | 7 | 9 | 10 | 14 |
| $\Phi_{b,FNT}$ (eV) | 0.32 | 0.43 | 0.41 | 0.41 |
| $a_{FNT}$ (nm) | 207 | 434 | 550 | 239 |
| $a_{DT}$ (µm) | 512 | 512 | 512 | 512 |
| $\Phi_{b,DT}$ (eV) | 3.5 | 3.5 | 3.5 | 3.5 |
| $d_{DT}$ (nm) | 2.43 | 2.5 | 2.75 | 2.97 |

**Table 5.5.** Calculated barrier parameter ($A=a^2$, $d$, $\Phi_b$) by assuming, in DT, $A_{DT}$=device area, and in FNT, $d_{FNT}=\delta$. $\Phi_{b,DT}$ is barrier between Pt Fermi-level to Si$_3$N$_4$ conduction band (DT is TAT). Yellow shaded regions are assumed values.

### 5.6.3 The High Temperature Region: Variable Range Hopping (VRH)

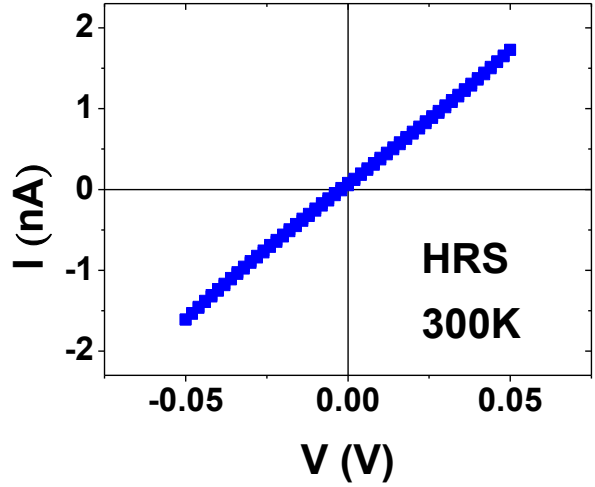

**Figure 5.21.** Small voltage *I-V* characteristics at room temperature 300 K.

As suggested by the small-voltage transport data in **Figure 5.10a**, a thermally activated process is involved in the high temperature region (~200 K to 300 K), over which there is a rapid change (by several orders of magnitude) of resistance. Up to 0.05 V, the *I-V* curve obeys Ohm's law, shown as a linear relation in **Figure 5.21**. As will be shown below, such data can be well explained by the variable range hopping (VRH) model, which is widely observed in disordered systems[35]:

$$\rho = \rho_0 \exp\left[\left(\frac{T_0}{T}\right)^{1/d+1}\right]$$

Here *d* is the dimensionality of the system, and $T_0$ is a scaling parameter. VRH describes inelastic tunneling (with additional energy benefit ~ $k_B T$) of localized electrons between a series of distributed localized electronic states near the Fermi level. These distributed states are expected in an amorphous structure, where disorders may be further enhanced by metal doping. After renormalizing the data in **Figure 5.12** taking into consideration of device geometry (using Ohm's law), we obtain resistivity which follows a straight line in the $\ln(\sigma)$-$T^{-1/4}$ plot for the high temperature part (**Figure 5.22a**), consistent with 3-D variable range hopping. The extracted slope $T_0 = 151^4$ K may be related to physical quantities through the following relation:

$$T_0 = \frac{24}{\pi} \frac{1}{k_B N_u \zeta^3}$$

where $k_B$ is the Boltzmann constant, $N_u$ is the density of states (DOS) at the Fermi level and $\zeta$ is the localization radius of states near the Fermi level. In principle, the two parameters $N_u$ and $\zeta$ cannot be uniquely determined from one equation, the one shown above. This problem can be solved by further considering the high field limit, as suggested by J. J. van Hapert[23].

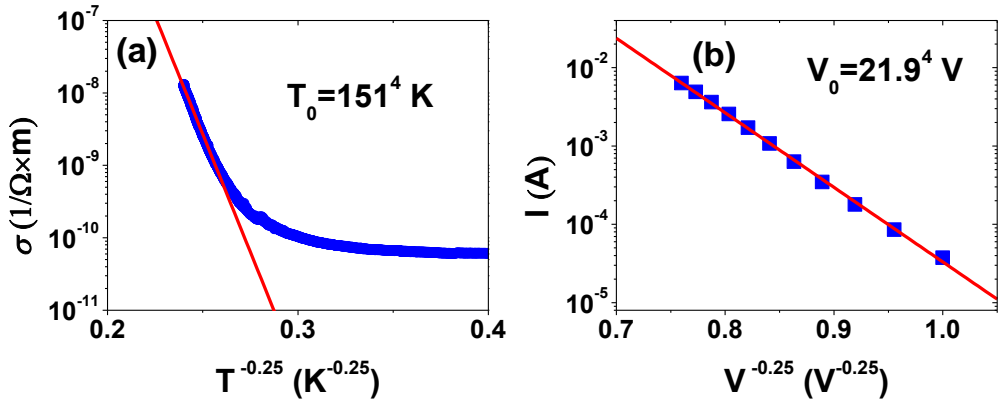

**Figure 5.22.** (a) Conductivity (log $\sigma$) *vs.* $T^{1/4}$ with a linear region at high temperature. (b) Current (log $I$) *vs.* $V^{1/4}$ with a linear region at high voltage/field. Nanometallic film: $Si_3N_4$: Pt, $f_{Pt}=4\%$, $\delta=10$ nm.

Under the high-field limit where electrical work exceeds the site energy difference, variable range hopping becomes much easier and exponentially dependent on the electrical field $F$:

$$I \propto \exp\left[-\left(\frac{F_0}{F}\right)^{1/4}\right],$$

with

$$F_0 = \frac{81}{16\pi}\frac{1}{eN_u\zeta^4}$$

Therefore, combining the low-field and the high-field data, we can unambiguously determine $N_u$ and $\zeta$. As shown in **Figure 5.22b**, the high-field $I$-$V$ can be fitted by the field-dominant VRH equation with a slope $V_0=21.9^4$ V (thus $F_0=2.2\times10^{11}$ V/cm). These $T_0$ and $F_0$ are close to data reported in the literature: $T_0=4.2\times10^8$ K, $F_0=5.1\times10^{11}$ V/cm[23] for amorphous $SiO_x$ (ref. [23]); and $T_0=10^7$-$10^9$ K, $F_0=10^{10}$ V/cm for $SrTiO_3$ (ref.[36-37]). Also, $T_0$ is compatible with my colleagues' data, $T_0=1.5\times10^7$ K in a perovskite nanometallic film[38] and $T_0=4.0\times10^8$ K in a $SiO_2$:Pt nanometallic film[39]. From $T_0$ and $F_0$, we calculate $N_u=2.57\times10^{18}$ cm$^{-3}$eV$^{-1}$ and $\zeta=0.41$ nm. The obtained $\zeta$ is smaller than the sample thickness ($\delta=10$ nm), confirming that the HRS is a "localized" state. The above parameters are compared with the reported VRH values in oxides ($N_u=10^{19}$~$10^{20}$ cm$^{-3}$eV$^{-1}$ and $\zeta=0.07$~$0.7$ nm in amorphous $SiO_x$ (ref.[23]), $N_u=10^{15}$~$10^{18}$ cm$^{-3}$eV$^{-1}$ and $\zeta=2$ nm in $SrTiO_3$ (ref.[37])) and in a quantum-dot system ($N_u=10^{18}$~$10^{20}$ cm$^{-3}$eV$^{-1}$, $\zeta=0.7$~$1.1$ nm in CdSe (ref.[26])).

To further validate the VRH mechanism, we performed the following consistency check:

(1) The optimized hopping distance $r_0$ should be longer than the localization length $\zeta$ to guarantee the validity of the VRH model (otherwise it will be nearest neighbor hopping). This is confirmed by

$$r_0 = \frac{3}{4}\left(\frac{3\zeta}{2\pi N_u k_B T}\right)^{1/4} = 5.5\,nm \gg \zeta$$

at 300 K.

(2) The low-field condition for VRH requires:

$$F < \frac{4}{3e}\left(\frac{2\pi N_u}{3\zeta}\right)^{1/4}\left(k_B T\right)^{5/4} = 5\times10^6\,V/m$$

which is equivalent to $V<0.05$ V for a 10 nm nanometallic film. This is satisfied by our voltage bias of 0.05V in **Figure 5.22a**.

(3) The high-field condition for VRH requires:

$$F > \frac{k_B T}{2e\zeta} = 3\times10^7\,V/m$$

which is equivalent to $V>0.3$ V for a 10 nm nanometallic film. This is satisfied by our voltage bias (1 V~3 V) in **Figure 5.22b**.

Therefore, VRH is a reasonable process that may account for high temperature transport in the HRS.

| Thickness | $\delta$=7 nm | $\delta$=9 nm | $\delta$=10 nm | $\delta$=17 nm |
|---|---|---|---|---|
| $T_0^{1/4}$ (K$^{1/4}$) | 112 | 144 | 151 | 180 |
| $V_0^{1/4}$ (K$^{1/4}$) | 15.4 | 19.2 | 21.9 | 28.1 |
| $N_u$ ($10^{18}$ cm$^{-3}$eV$^{-1}$) | 13.0 | 1.55 | 2.57 | 0.55 |

| ζ (nm) | 0.353 | 0.513 | 0.406 | 0.530 |
|---|---|---|---|---|
| $r_0$ (nm) @ 300 K | 3.6 | 6.6 | 5.5 | 8.6 |

**Table 5.6.** Values of density of states $N_u$, localization length ζ and optimized hopping distance $r_0$ (300 K) for various thickness ($f_{Pt}$=4%).

By applying the same procedure to other nanometallic films of different thickness, we can obtain physical parameters as shown in **Table 5.6**. The localization length ζ is almost unchanged for all samples. This is expected since at the same metal concentration, electron localization should be within a similar range independent of thickness. However, the calculated density of states $N_u$ seems higher in thinner films. Such dependence becomes reasonable if we consider the rate-limiting transport process to occur at some critical patches. As all the critical patches are connected in series, the apparent resistance is mainly determined by the "most difficult" VRH patch. Statistically, a thicker film has a higher probability of involving a "more difficult" VRH patch, corresponding to a lower effective DOS in the patch. This statistical behavior is reminiscent of the higher tunneling barrier in thicker films seen in low-temperature tunneling. (See the previous subsection.) In addition, surface states near the electrodes may also contribute to the higher DOS in thinner films.

The $N_u$ and ζ calculated for films of other metal concentrations are shown in **Table 5.7**. We find ζ increases with Pt concentration, which is reasonable because Pt atoms can extend the extent of electron's wave-functions thus delocalizing the electron. For the Pt-lean sample ($f_{Pt}$ =4 at.%), ζ is comparable to the size of a $SiN_4$ tetrahedron (0.4 nm), meaning at this composition the material is essentially an insulator. From such value, $\zeta$ then rapidly increases with the Pt content, reaching $\zeta$=6.67 nm at $f_{Pt}$ =27 at.%, which is close to the sample thickness. This provides a direct confirmation of our material design: the localization length can be extended by simply doping an insulator with metal atoms. (At this composition, the film is still an insulator with a negative temperature dependence of resistance since $\zeta$ is still less than $\delta$ (10 nm).) On the other hand, **Table 5.7** also shows that $N_u$ decreases, rather significantly, with high metal concentration. This suggests that at higher Pt concentrations, an increasing fraction of the metal atoms actually belongs to the Pt nanocrystals not participating in VRH, and only an increasingly smaller number of isolated Pt atoms remains and provides localized electron states involved in VRH. This is reminiscent of the increasing tunneling distance seen at higher Pt concentrations in the low-temperature region (see the previous subsection.) The trend is not contradictory to the increased $\zeta$: as the localization length increases, the channel regions expand and fewer isolated states are left for VRH to hop between. Because of these changes, the optimized hopping distance increases with increasing $f_{Pt}$. Nevertheless, the decreasing $T_0$ and $F_0$ give rise to faster VRH with Pt doping.

| Composition | $f_{Pt}$ =4 at.% | $f_{Pt}$ =19 at.% | $f_{Pt}$ =27 at.% |
|---|---|---|---|
| $T_0^{1/4}$ ($K^{1/4}$) | 151 | 157 | 112 |
| $V_0^{1/4}$ ($K^{1/4}$) | 21.9 | 16.0 | 16.8 |
| $N_u$ ($cm^{-3}eV^{-1}$) | $2.57 \times 10^{18}$ | $3.04 \times 10^{16}$ | $2.87 \times 10^{15}$ |
| $\zeta$ (nm) | 0.406 | 1.69 | 6.67 |
| $r_0$ (nm) @ 300 K | 5.5 | 24 | 61 |

**Table 5.7.** Values of density of states $N_u$ localization length $\zeta$ and optimized hopping distance $r_0$ (300 K) for various concentration $f_{Pt}$ ($\delta$=10 nm).

For other resistance states (IRS) in **Figure 5.10**, the associated density of state $N_u$ can also be derived. Since these IRS are from the same sample, we expect the localization lengths to be independent of the state (a large voltage will switch the IRS so high field data cannot be obtained experimentally). By fixing $\zeta$=0.406 nm from the HRS data, the calculated $N_u$ are shown in **Table 5.8**. The DOS decreases rapidly as the resistance increases, indicating the effect of trapped charge is to remove electron states from the vicinity of the Fermi level, thus making them unaccessible by hopping. (Another equivalent interpretation is that the energy range of the electron state distribution is expanded by the addition of the Coulombic energy due to the trapped charge. As a result, the effective density of state decreases. However, this interpretation is quantitatively tenuous since $N_u$ in **Table 5.8** varies by many orders of magnitude, while the energy range is at most expected to vary by a few times, between 0.1 eV and 10 eV.) Note that the DOS values for 8.7 k$\Omega$ and 2.5 k$\Omega$ are probably unreasonable (too large for a typical VRH). In such case, nearest neighbor hopping (NNH) takes over VRH, as there are always nearest-neighbor states available for each conduction electron to hop into. Indeed, in **Table 5.8**, we find the calculated optimized hopping distance to fall below the size of a SiN$_4$ tetrahedron, confirming that VRH is not needed and NHH must take over in the 8.7 k$\Omega$ and 2.5 k$\Omega$ states. An Arrhenius plot ($R$ *vs.* $1/T$) is given in **Figure 5.23** for a 2.5 k$\Omega$ state.

| RS ($R_{2K}$) | 600MΩ | 55MΩ | 2MΩ | 300kΩ | 8.7kΩ | 2.5kΩ |
|---|---|---|---|---|---|---|
| $T_0^{1/4}$ ($K^{1/4}$) | 151 | 129 | 27.7 | 19.7 | 6.12 | 1.03 |
| $N_u$ (cm$^{-3}$eV$^{-1}$) | $2.6 \times 10^{18}$ | $4.7 \times 10^{18}$ | $2.2 \times 10^{21}$ | $8.7 \times 10^{21}$ | $1.0 \times 10^{24}$ | $1.2 \times 10^{27}$ |
| $\zeta$ (nm)* | 0.406 | 0.406 | 0.406 | 0.406 | 0.406 | 0.406 |
| $r_0$ (nm)@300K | 5.5 | 4.7 | 1.0 | 0.72 | 0.22 | 0.037 |

**Table 5.8.** Values of density of states $N_u$ and optimized hopping distance $r_0$ (300 K) for various resistance states (HRS and IRS), listed by their resistance $R_{2K}$ at 2K. ($f_{Pt}$=4%, $\delta$=10 nm). *Localization length $\zeta$ is fixed at $\zeta$=0.406 nm according to the HRS ($R_{2K}$=600MΩ) data.

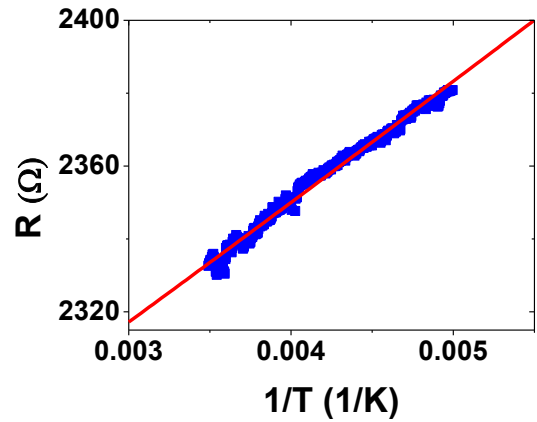

**Figure 5.23.** Fitting high temperature $R$-$T$ curve on Arrhenius plot for 2.5 kΩ state. $R$=2221 Ω ×exp(1.21 meV/$k_B T$).

### 5.6.4  Intermediate Temperature: Fluctuation Induced Tunneling (FIT)

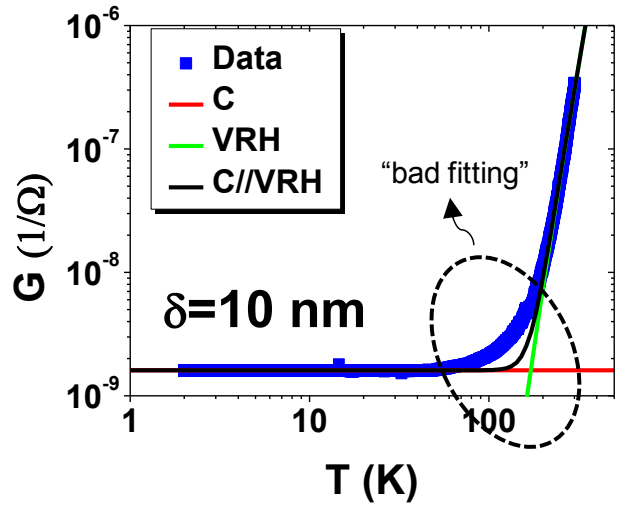

**Figure 5.24.** Fitting *G-T* curve (*V*=0.05 V) using only two parallel mechanism.

While the data in the high temperature region and the low temperature region can be well understood in terms of tunneling (DT and FNT) and VRH, we found their superposition cannot account for the conductance data in the intermediate region. An example is shown in **Figure 5.24**, where the fitted conductance is several times smaller than the experimental conductance. Specifically, since the low temperature conductance $G_1$ is constant, while the high temperature VRH conductance is $G_2 \sim \exp(-AT^{-1/4})$, their sum $G_1+G_2$ would appear as a sharp transition in **Figure 5.24**, which is at variance with the rather smooth variation of the data.

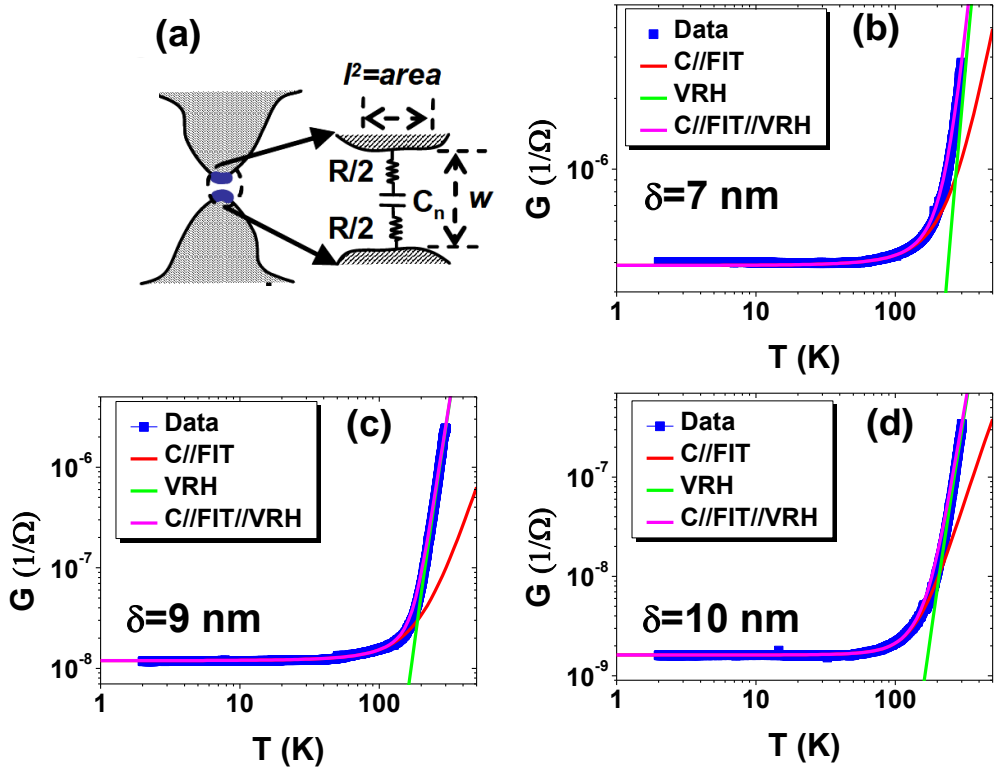

**Figure 5.25.** (a) Schematic of critical tunneling junction involved in FIT and its circuit simplification. (b)-(d) Experimental data and fitted curves in the entire temperature range. $f_{Pt}=4\%$.

Adding fluctuation induced tunneling (FIT) to the transport mechanisms provides a reasonable explanation of the smooth transition between the athermal tunneling region and the thermally activated hopping region. For a single parabolic barrier, temperature dependent conductance of FIT can be expressed by[20-21]:

$$G(T) = G_0 \exp\left(-\frac{T_1}{T + T_0}\right)$$

$$T_1 = \frac{8\varepsilon_r A'\Phi_b^2}{k_B e^2 w} \, , \ T_0 = \frac{16\varepsilon_r \hbar A'\Phi_b^{3/2}}{\pi(2m^*)^{1/2} k_B e^2 w^2}$$

In the above, $k_B T_1$ is approximately the energy of the tunneling barrier, *i.e.*, the energy required for an electron to cross the gap between the conductive patches, when expressed in terms of the equivalent thermal energy. Therefore, above $T_1$, thermal fluctuation is more significant than FIT. The critical junction in the above patch is described by two geometrical parameters $A'$ and $w$, which are the area and the width of the gap, respectively. More specifically, $\Phi_b$ is the potential barrier height of the gap and $\varepsilon_r$ is the dielectric permittivity of the material. As usual, $e$ and $m^*$ are the electron charge and mass, respectively.

As shown in **Figure 5.25b-d**, with the aid of FIT, conductance can be satisfactorily fitted over the entire temperature range (2 K-300 K). The fitting parameters of FIT for the above data and other data at different film thickness are summarized in **Table 5.9**: $T_1$ is on the order of $10^3$ K-$10^4$ K and $T_0$ is on the order of $10^2$ K-$10^3$ K. By further assuming a gap area $A'$=0.5×0.5 nm$^2$, a potential barrier height $\Phi_b$ (~1 eV), a gap $w$ of ~a few nm can be obtained. The above barrier height is reasonable for a saddle point barrier arising from the Coulombic potential of trapped charge (see discussion in **Section 5.6.1.1**). The gap size $w$ is also reasonable because it is larger than the localization length (0.4 nm, see **Section 5.6.3**), but smaller than the film thickness.

| Thickness | $\delta$=7 nm | $\delta$=9 nm | $\delta$=10 nm | $\delta$=14 nm |
|-----------|-----------|-----------|-----------|-----------|
| $T_0$ (K) | 443 | 558 | 267 | 196 |
| $T_1$ (K) | 5585 | 8899 | 4654 | 6837 |

| $\Phi_b$ (eV) | 0.8 | 1.1 | 0.9 | 1.4 |
|---|---|---|---|---|
| $w$ (nm) | 1.7 | 1.9 | 2.3 | 3.7 |

**Table 5.9.** FIT fitting parameters and calculated barrier height and width (Junction area $A' = 0.5 \times 0.5$ nm$^2$ is assumed).

In the above, the observation that the gap size $w$ (and perhaps, the barrier height) increases with thickness can again be rationalized by invoking a statistical argument. Because the rate limiting transport process is determined by the "worst junction", we expect a thicker film to more likely contain a "more difficult" tunneling junction.

### 5.6.5 An Empirical Fit

As discussed above, transport physics in different temperature regimes are different and different mathematical forms are needed for their description. Yet I found a single empirical formula $R(T) = R_0 \exp[-(T/T_0)^n]$ can surprisingly fit the conduction data over the entire temperature range. This empirical form is physically motivated by the general observation of three temperature regions of the resistance. At the low temperature limit, it gives $R(T \to 0) = R_0$, in agreement to the temperature-independent tunneling process. At the intermediate temperature ($T \sim T_0$), the resistance decreases but at a much slower rate than in the high-temperature region ($T > T_0$) if an exponent $n > 1$ is chosen. Here, the empirical $T_0$ corresponds to the characteristic temperature dividing the low-temperature region and the high-temperature region. In general, both $T_0$ and $n$ may depend on the voltage, resistance (state), thickness, and composition.

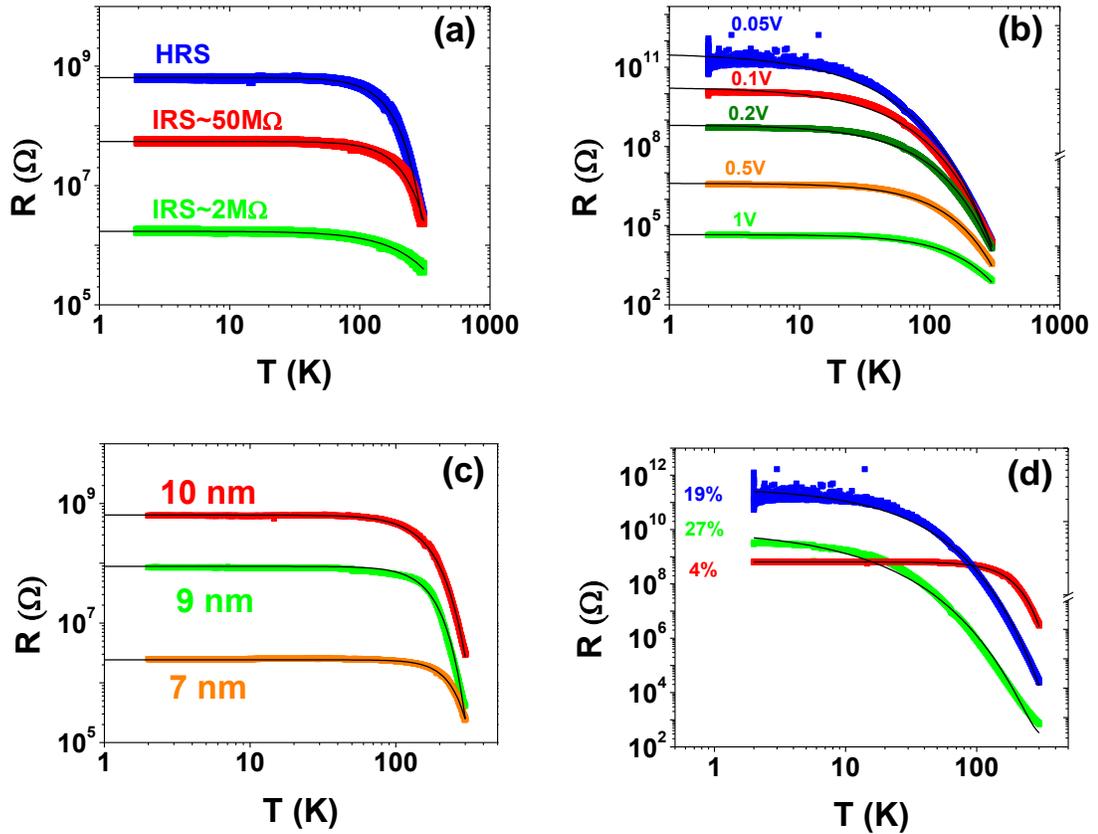

**Figure 5.26.** Data fitting (black curves) using empirical formula $R(T)=R_0\exp[-(T/T_0)^n]$ for devices of (a) different states, (b) different voltages, (c) different thickness and (d) different metal concentration.

The above formula allows excellent fitting for the experimental data on the log-log scale as shown in **Figure 5.26**. Fitting parameters summarized in **Figure 5.27** follow systematic trends for the resistance states, voltage, thickness, and composition, which in turn allows easy interpretation: a lower $T_0$ represents a more prominent role of thermal activation, a larger $n$ represents a stronger temperature dependence at higher temperatures. For example, for the Resistance State dependence (**Figure 5.27a**): $T_0$ increases with resistance, and n decreases with resistance. Other dependences are self-explanatory in **Figure 5.27b-d.**

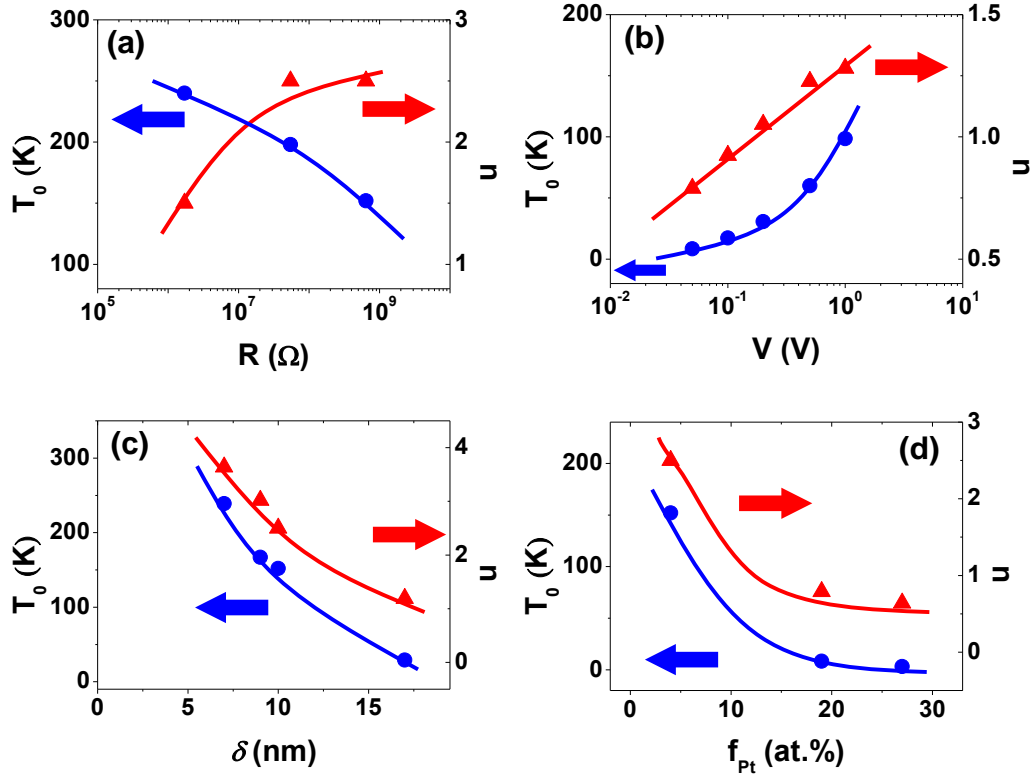

**Figure 5.27.** Fitting parameters $T_0$ and $n$ in **Figure 5.26** for different (a) resistance states, (b) voltages, (c) thickness and (d) metal concentration.

## 5.7 Discussion

### 5.7.1 Conduction Map

Our fitting results clearly indicate that different conduction mechanisms dominate in different temperature and field domains. We can group them into two broad types, which may be considered as two parallel conduction paths: a tunneling path with no (or weak, if considering FIT) temperature dependence, and a thermally activated path with a very strong temperature dependence ($\exp(-1/T^n)$ type. (The latter is sometimes called the stretched exponential type, which reduces to the Arrhenius type if $n$=1.) Their combined conductance is $G_{tunnel}+G_{hop}$; in terms of resistance, it is $1/(1/R_{tunnel}+1/R_{hop})$, as schematically shown in **Figure 5.28**. The schematic may be interpreted as a $2\times2$ matrix, with the two rows corresponding to the two conducting paths/channels, and the two columns corresponding to the two LRS and HRS (including various IRS) states. For the LRS (column), where there is no trap charge or insufficient trapped charge to block the conducting channels, the hopping path is inactive at all temperatures since electrons can tunnel through the entire film along the conducting channel with a transmission probability $T$=1 (barrier-less). In this case, conduction is Ohmic throughout the entire voltage range, and the temperature dependence of conduction may be attributed to scattering, between electrons at low temperatures and between electrons and phonons at high temperatures. For the HRS (and various IRS) column, the Coulombic repulsion from trapped charge modifies the energy landscape and the conduction mechanisms in the two paths (rows). (i) In the tunneling path, energy barriers of the order of ~1 eV are created, which breaks down the conducting channels into separate patches between them tunneling is required with a transmission probability $T$<1. This row (path) dominates at low temperature. (ii) In the hopping path, the energy levels of electron states are dispersed due to random Coulombic potential; in particular, it removes DOS from the Fermi level in a way that is similar to the Coulomb gap. This row (path) dominates at high temperature.

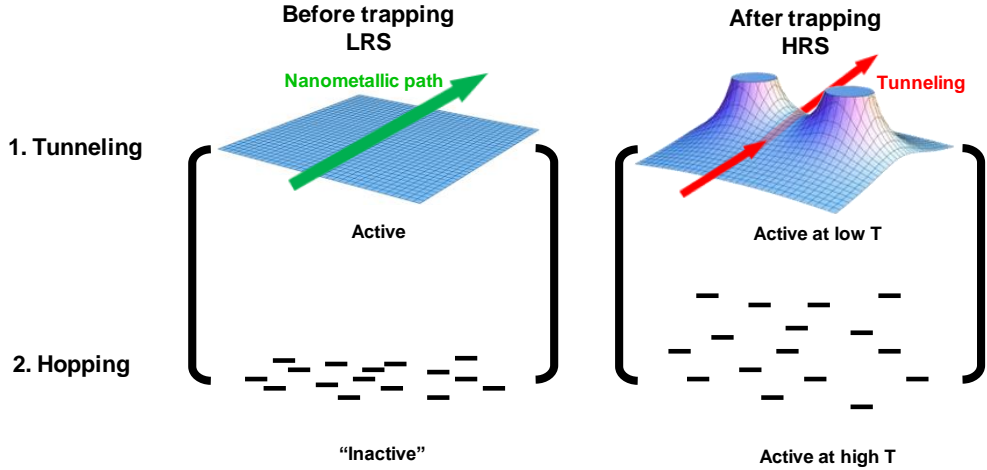

**Figure 5.28.** Two (parallel) channel conduction model.

Crossover between different paths in the HRS and various IRS can be quantified by finding the conditions when $G_{tunnel}=G_{hop}$. This procedure can be further refined if individual tunneling (DT, FNT, and FIT) and hopping (low-field VRH, high-field VRH, and NNH (or Coulomb gap hopping if appropriate)) mechanisms are distinguished as $G_i$. Then the conditions (boundaries) between mechanisms are computed from $G_i=G_j$ for the most dominant mechanisms, $i$ and $j$, as appropriate. For a given film thickness, composition, and resistance (trapped charge density) states, the condition can be delineated in terms of voltage and temperature, as in **Figure 5.29**, which depicts a mechanism map showing various regions where a single mechanism dominates. (Here, the FIT mechanism is omitted since we have no information of its voltage dependence.) In addition to illustrating different regions for one set of film thickness, composition and resistance (trapped charge density) state, **Figure 5.29** also presents a set of iso-conductance contours, calculated from the combined conductance, to indicate the systematic variation of conductance as a function of temperature and voltage. In addition, arrows at the domain boundaries in **Figure 5.29** indicate the direction of the boundary shift when the film thickness, composition or trapped charge (which increases the resistance) increases.

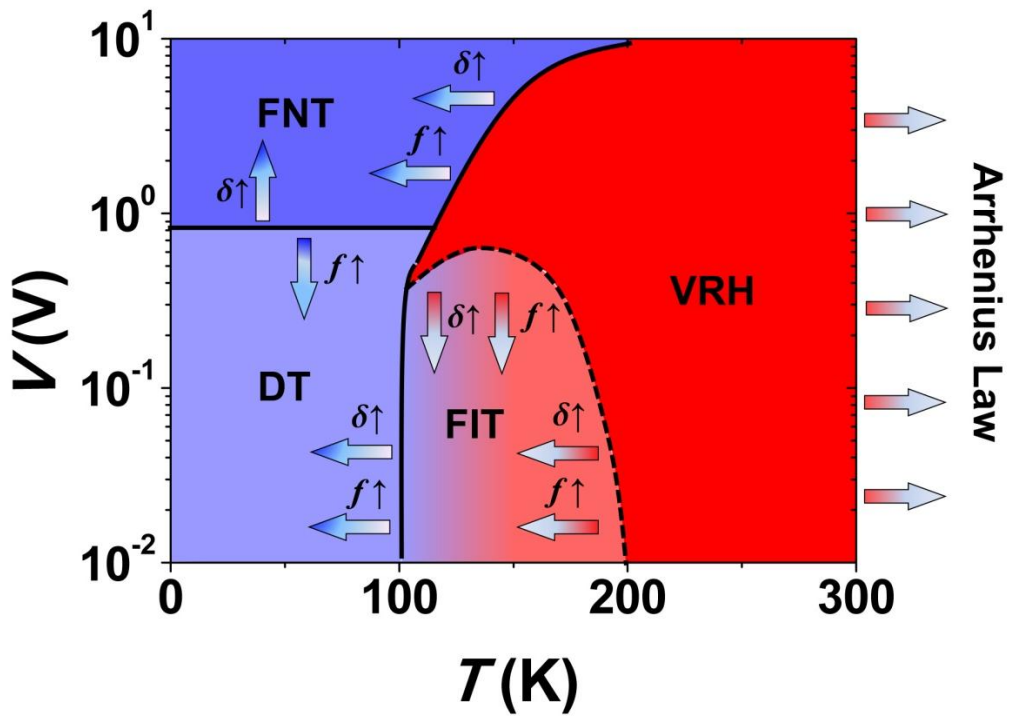

**Figure 5.29.** HRS transport mechanism at different temperature and voltage domains. Arrows indicate the direction of boundary shift when film thickness, metal concentration, or resistance increases. (Nanometallic film: $f_{Pt}$=4%, $\delta$=10 nm).

A comprehensive view of the conduction mechanisms is made transparent by such map. In the low temperature region, tunneling dominates, by DT at low voltage and FNT at high voltage. In the high temperature region, VRH dominates, again separated into the low-field mechanism and the high-field mechanism. For a thicker film, tunneling becomes more difficult, thus the tunneling-hopping boundary in **Figure 5.29** moves to lower temperature. Meanwhile, the DT-FNT boundary also moves to a higher field because the tunneling height tends to increase with the film thickness. For metal-richer film, there is a larger tunneling distance $d$ but so is the localization length $\zeta$. This suppresses tunneling (due to a larger $d$) but facilitates hopping process (due to a larger $\zeta$), thus the tunneling/hopping boundary moves towards a low temperature. Meanwhile, the lower tunneling barrier height in the metal-rich film favors FNT over DT, therefore the DT-FNT boundary move towards a lower field. For a lower resistance IRS, tunneling barrier is lower and thus DT-FNT boundary move towards a lower field (Although this may not be observed in experiment because resistance switching will not allow detecting stable FNT behavior). Meanwhile, density of states rapidly increases which facilitate VRH dominating at a lower temperature.

### 5.7.2 Other Mechanisms

In RRAM and the broader thin film literature, other putative conduction mechanisms are often reported for various systems based on the data at or near the room temperature. They include Frenkel-Poole emission, Schottky emission, space charge limited conduction, and trap assisted tunneling. In the following, these possible mechanisms are examine to see if they are appropriate for our materials and devices.

### Frenkel-Poole Emission

As described in the Introduction of this Chapter, Frenkel-Poole emission describes electron emission from localized traps to the conduction band in the insulator/semiconductor film, which has a mathematical form:

$$J \propto E \exp\left( -\frac{e\left(\Phi_b - \sqrt{eE/\pi\varepsilon_i}\right)}{k_B T} \right)$$

**Figure 5.30** shows the fitting results for Frenkel-Poole emission, having a good fit in the intermediate regime (0.2 V~2 V). An activation energy of ~0.3 eV and a dielectric constant of ~22 can be extracted. However, this dielectric constant is too large, given the typical static dielectric constant of Pt-poor $Si_3N_4$:Pt is close to $\varepsilon_{0,SiN}$~7 (**Chapter IV**). (The dielectric constant involved in the FP process is the high frequency dielectric constant, which is lower: at $\lambda$~500 nm, it is 4 (ref.[40]).) Therefore, this mechanism does not seem to apply at least in the metal-poor nanometallic films. However, FPE may still be possible in nanometallic films with a higher $f$ where a higher dielectric constant is feasible.

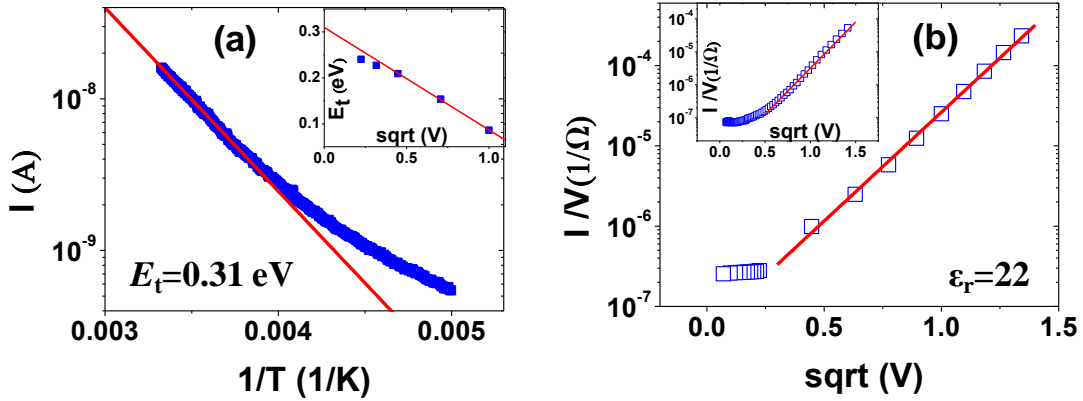

**Figure 5.30.** Frenkel-Poole emission model fitting of the experimental results: (a) Arrhenius plot *I vs.* 1/*T* to extract the activation energy $E_t$. Inset: Extrapolating $E_t$ to zero bias. The trap energy (~0.3 eV) is obtained from the intersection at *V*=0. (b) ln(*I/V*) *vs.* sqrt(*V*). Dielectric constant (~22) is extracted from the slope. Inset: another sample showing similar results.

### Schottky Emission

Schottky emission describes electron emission from the electrode to the conduction band of the insulator/semiconductor film. By fitting our conduction data with:

$$J \propto AT^2 \exp\left(-\frac{e\left(\Phi_b - \sqrt{eE/4\pi\varepsilon_i}\right)}{k_B T}\right)$$

we obtain the Schottky emission barrier $\Phi_b$~0.2 eV and the dielectric constant ~2.9. Since pure $Si_3N_4$ already has a dielectric constant ~7 (static) or ~4 (high frequency), this extracted dielectric constant is somewhat low. More importantly, Schottky emission typically implies an asymmetric *I-V* transport behavior when two dissimilar electrodes with different work functions are used. The absence of asymmetry in our transport data thus excludes the possibility of Schottky emission.

## Space Charge Limited Conduction

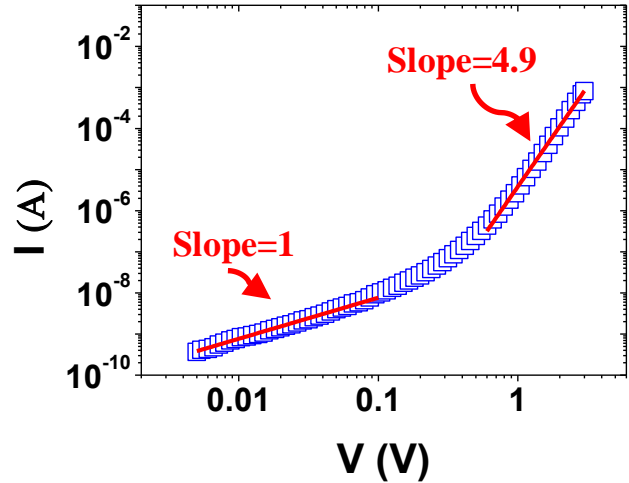

**Figure 5.31.** *I-V* on a log-log plot at 300 K. Slope=1 and slope=4.9 are found for low field and high field limit.

Space charge limited conduction mechanism features itself by parabolic law ($I{\sim}V^2$) at high field. However, experimental data shows a much stronger voltage dependence ($I{\sim}V^{4.9}$) in **Figure 5.31**, which immediately excludes the possibility of space charge limited conduction. (There is a transition region of "slope 2" between "slope 1" (low field) and "slope 4.9" (high field), but the range is too limited.)

## Trap Assisted Tunneling

Since tunneling is a coherent transport process with very weak temperature dependence, the activation energy ~0.3 eV (see discussion in Frenkel-Poole emission) at high temperature immediately exclude such possibility. However, it is the dominating conduction at low temperature.

### Other VRH-like Hopping Models

The possibility of other hopping models, including 1D Mott VRH, Efros-Shklovskii VRH and Nearest Neighbor Hopping, are discussed in **Appendix**.

## 5.8  Conclusions

(1) Nanometallic resistive switching occurs over a broad range of temperature (2 K to 300 K) and magnetic field (-9 T to 9 T) with the same switching voltages. This strongly suggests an electronic switching mechanism.

(2) The LRS is a metallic state, although showing considerable impurity scattering near 0 K. Scattering from 20 K to ~100 K is dominated by electron-electron interaction judging from the $T^2$ law for resistance, and above ~100 K by electron-phonon interaction judging from the linear temperature dependence of resistance.

(3) The HRS is an insulating state. From 2 K to ~100 K, tunneling dominates conduction evolving from direct tunneling to FN tunneling as the voltage increases. Tunneling becomes easier for a thinner film, a higher metal composition film and a less resistive film.

(4) Above ~200 K, VRH is the dominant conduction mechanism in the HRS. It is

facilitated Electron localization length increases With increasing metal contents, VRH is facilitated by increasing electron localization length, albeit somewhat counteracted by a decreasing density of states.

(5) At intermediate temperature, from ~100 K to ~200 K, FIT contributes significantly to conduction and is primarily responsible for the temperature dependence.

(6) The IRS are also insulators and share the same conduction mechanisms as the HRS. The LRS may also have different resistance states without losing the main metallic character.

(7) The tunneling barrier height, the tunneling distance, and the density of states for VRH all show a systematic variation with film thickness. This suggests that the overall resistance is controlled by the most difficult barrier or hopping patch, which becomes statistically more difficult in thicker films.

(8) While the tunneling/hopping height tends to decrease with the metal content, the tunneling distance and density of states also tend to decrease. This suggests that concentration of isolated metal atoms, which are most pivotal for tunneling/hopping, are fewer in metal-richer films.

(9) The trapped charge provides Coulombic repulsion forming barriers to electron tunneling/hopping. The magnitude of the barrier is of the order of 1 eV, which may be accounted for by a single trapped electron at 1.4 nm away, or by two separate trapped electrons at 6 nm apart.

# Chapter VI. Environmental Effect [1]

Dielectric thin films in nanodevices may absorb moisture leading to physical changes and property/performance degradation, such as altered data storage and readout in resistance-random-access-memory (RRAM). Here we demonstrate using a nanometallic memory that such degradation proceeds *via* nanoporosity, which facilitates water wetting in otherwise non-wetting dielectrics. Electric degradation only occurs when the device is in the charge-storage state, which provides a nanoscale dielectrophoretic force directing $H_2O$ to internal field centers (sites of trapped charge) to enable bond rupture and charged hydroxyl formation. They also cause an offset in the current-voltage curve indicating the presence of an internal field due to electrode oxidation. While these processes are dramatically enhanced by an external DC or AC field and electron-donating electrodes, they can be completely prevented by eliminating nanoporosity, depositing a barrier layer or using an oxidation resistant electrode. These findings provide insight for understanding high performance memory and field-assisted degradation of nanodevices.

## 6.1 Introduction

Thin films of dielectrics and their hybrids, including ones containing metals, are widely used in electronic and optical nanodevices. Dielectric degradation under ambient conditions has long been recognized as a reliability issue, and it becomes especially important in nanodevices because of nanodimensions. Thin films deposited by physical vapor deposition methods such as sputtering typically are nanoporous with densities lower than theoretical[2-5]; nanoporosity can absorb moisture in the ambient air. This may have a profound influence on the device properties. For example, water with a refractive index $n = 1.3$ can introduce optical mismatch hence scattering in the film. They can also cause chemical reactions (*e.g.*, forming hydroxides that affect permittivity) or change physical dimensions (*e.g.*, swelling that roughens surfaces)[6]. Entrapped moisture can lower the electric resistance of insulating films, even causing electric shorts[7]. It can also create charge trapping sites thus affecting device performance[8-9]. Moreover, moisture penetration may extend to the buried layers (*e.g.*, bottom electrodes and substrates) leading to more extended degradation beyond films.

Among dielectric nanodevices, of special recent technological interest are ones that exhibit resistive switching properties, thus having the potential of becoming a new class of nonvolatile resistance random access memory (RRAM) materials[10]. Most of these devices switch by forming (soft-breakdown) conducting nanofilaments, undergoing redox reactions, or both[11-17]. Naturally, redox reactions are often moisture sensitive[18-20]. But nanoscale transport may also be environment sensitive, since gas absorption may affect the energetic of ion migration, and oxygen activity may affect defect populations that in turn affect $n/p$-type conduction in nanodevices[21-22]. Indeed, for a $Pt/SiO_2/Cu$ RRAM, Tsuruoka *et al.* reported residual water to affect both redox-reaction thermodynamics and ion-migration kinetics[19]. No such study has been reported for charge-trapping RRAM, however. More broadly speaking, among the very large number of RRAM papers published in recent years, only very few reports on this subject[19,23-27]. This is not entirely surprising: many RRAM suffer from large variability in their resistance and switching voltage characteristics[28-30], which can easily mask the time-dependent effects of environment.

In this respect, our nanometallic RRAM based on nanometallic films may provide an excellent platform for studying the moisture effect[31]. As described already in previous chapters, these films (5-30 nm thick) are atomic mixtures of random insulators ($SiO_2$ or $Si_3N_4$) and metals (Pt, Cr, etc.), with wavefunctions of itinerant electrons reaching the full thickness of the films[31-34]. In the as-fabricated form, these films are conductors. Subsequently, under a voltage trigger, Coulomb barriers arise from trapped charge which blocks the wavefunctions, so the film switches from the conducting state (the low-resistance state, LRS, being the "on state") to a high-resistance (insulating) state (HRS, being the "off state"). This switching process is reversible, and it occurs by an electronic (charge trapping/detrapping) mechanism without forming any filament or Schottky barrier. Our previous work on nanometallic RRAM (made of $SiO_2$:Pt with a top coat of ~4 nm $Al_2O_3$) demonstrated outstanding uniformity of switching characteristics (resistances of HRS and LRS, and on-switching and off-switching voltages)[32-34]. In this work, we have systematically investigated the performance of several nanometallic RRAM under various environmental conditions to shed light on the possible degradation mechanisms. In this setting, static and dynamic electric testing is used both to tune degradation kinetics and to probe degradation mechanisms. Solutions to mitigate moisture-induced degradation in these nanodevices have also been demonstrated. Electrical characteristics related to the movement of moisture-introduced ionic species and electrode oxidation are also measured to contrast with those of "dry" films in which only electron movement is possible.

## 6.2 Experimental Methods

**Device fabrication:** All devices were fabricated on $SiO_2/Si$ substrates (**Figure 6.1a**). Substrates were cleaned by acetone and isopropyl-alcohol followed by heating at 150 °C in vacuum for 30 min in a sputter chamber (vacuum level: $3 \times 10^{-7}$ Torr). After cooling down to room temperature, a bottom Mo electrode was deposited by DC sputtering before a nanometallic film was deposited by RF magnetron co-sputtering using targets of a metal (Pt or Cr) and an insulator ($SiO_2$ or $Si_3N_4$). (DC Power: 300 W; RF Power: 200 W for insulator $SiO_2$ or $Si_3N_4$; 10 W- 30 W for Pt or Cr; no vacuum break between these steps.) Next, top Pt electrodes were RF sputter-deposited (RF Power: 30 W) and patterned (having a circular area of various sizes) lithographically or by a shadow mask. On some devices, a capping layer of 4 nm-thick $Al_2O_3$ was deposited after device fabrication, by atomic layer deposition (ALD) under the following condition, precursor: trimethylaluminum; oxidant: $H_2O$; wafer temperature: 150 °C. For a comparative study, another set of samples used in a previous research[31-32], with the same $SiO_2$:Pt composition and Pt top electrode but $SrRuO_3$ (SRO) bottom electrode deposited by pulse laser deposition, was also investigated.

**Characterization:** For density determination, nanometallic films on MgO substrates were studied by Rutherford Backscattered Spectrometry (RBS, NEC Minitandem Ion Accelerator), and the composition and density distributions were obtained using the simulation software (SIMNRA). Density was also confirmed using X-ray-reflectometry (Bruker D8 Discover) using films on Si wafers. DC electric measurements were conducted using a semiconductor parameter analyzer (SPA, Keithley 237), while AC impedance spectroscopy was studied using an impedance analyzer (Gamry G750). For these studies, samples were placed on a probe station (Signatone S1160) and a voltage was applied between the top and the bottom electrodes. (Current flowing from the top electrode to the bottom electrode is considered positively biased.) A homemade electric hotplate was used to heat the sample with the temperature monitored by a thermocouple nearby. For IR spectroscopy, nanometallic films were deposited on IR-transparent KBr substrates, and measured using a Nicolet Nexus 470 FTIR spectrometer. Static contact angles of water were measured on various films/materials, without patterned top electrodes, by a rame-hart goniometer (Model 200) using 2.0 μL droplets and averaged over several locations. Unless otherwise noted, characterization was conducted in air at room temperature.

### 6.3  Results

### 6.3.1  Physical Appearance

In the as-fabricated device (**Figure 6.1a**), films of ~20 nm thick 80% $SiO_2$: 20%  Pt film on Mo (bottom electrode) /$SiO_2$/Si (substrate) had a light tan appearance when photographed under natural light, with the Pt top-electrode areas appearing white (**Figure 6.1b**). After kept at room ambient conditions for 1 year, the appearance changed to that of **Figure 6.1c** showing dark green films with ring-like features around cells. The evolution of appearance was accelerated by placing the device in a chamber of saturated humidity: the green appearance emerged followed by a violet one after a few hours. This also occurred to films already exposed to air for 1 year: moisture in the chamber of saturated humidity produced the appearance shown in **Figure 6.1d**. Such violet color is consistent with the color of $MoO_2$ (ref.[35]) which like many other non-stoichiometric $MoO_x$ compounds is a conductor ($x<3$)[35-36]. In contrast, devices made of $Si_3N_4$:Pt showed no change in appearance when left at room ambient conditions: in **Figure 6.1e** is a one-year-old device which looks just like a new one. The $SiO_2$:Pt device can also become immune to changes if protected by a ~4 nm $Al_2O_3$ capping layer, as shown in **Figure 6.1f**) after 1-year storage. Likewise, when $SrRuO_3$ was used as the bottom electrode, the $SiO_2$:Pt device too became immune to change during storage over several years (figure not shown.) During electric testing, the (uncapped) $SiO_2$:Pt device exhibits additional changes in appearance. **Figure 6.1g** shows the outer edge of a tested cell (lower left) turning dark after one voltage-sweep cycle (0 V to −5 to 0 V to 5 V to 0 V); meanwhile nearby untested cells were unaffected. Such darkening typically occurred during the positive voltage sweep. In addition, these changes sometimes varied from device to device and from cell to cell. For example, some cells experienced "bubble formation" during testing (**Figure 6.1h**); these cells could not be switched anymore. Interestingly, these "bubbles" seemed to remain indefinitely: there is no change in appearance after many weeks (**Figure 6.2**). This is as if the water bubbles are so strongly attracted on the device that evaporation is completely suppressed. If a cell survived the test without change in appearance, during storage it could still change if left in the HRS, which contains trapped charge. One example is shown in **Figure 6.1i** in which the HRS cells are surrounded by a dark ring whereas the LRS cells are not. In contrast, SiO$_2$:Pt films with the same Pt top electrodes but with SrRuO$_3$ bottom electrodes showed no changes during testing (see ref.[31-32]).

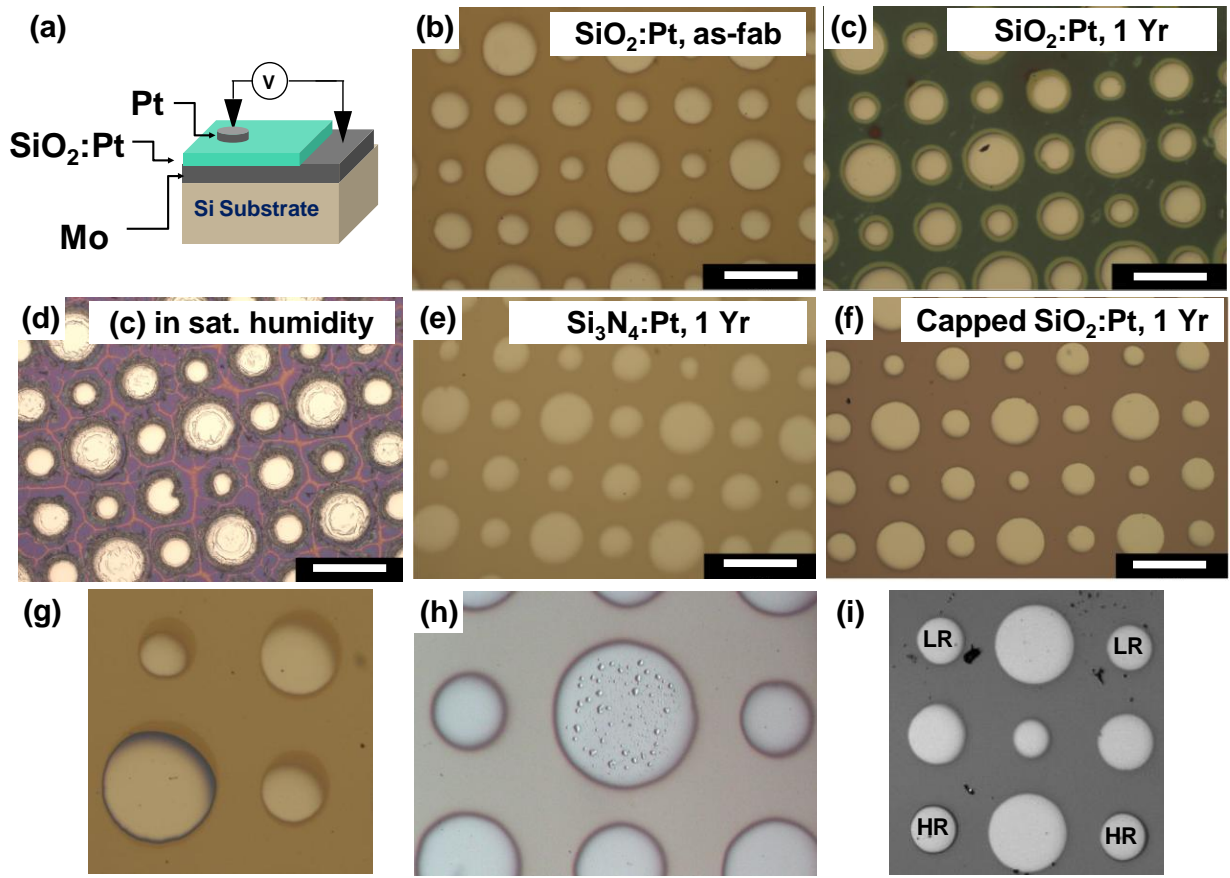

**Figure 6.1.** (a) Schematic of nanometallic RRAM device. Active layer (SiO$_2$:Pt, Si$_3$N$_4$:Pt or Si$_3$N$_4$:Cr) sandwiched between top electrode Pt and bottom electrode Mo. Light microscopy image of (b) as-fabricated 80% SiO$_2$: 20% Pt device; (c) device (b) after 1 yr in ambient air; (d) device (c) after 3 hr in saturated humidity; (e) 95% Si$_3$N$_4$:5%Pt device after 1 yr in ambient air; (f) Al$_2$O$_3$-capped 80% SiO$_2$: 20% Pt device after 1 yr in ambient air (Scale bar: 250 μm); (g) voltage tested

cell (lower left) in 80% $SiO_2$: 20% Pt device showing a dark ring at the cell edge; (h) voltage tested cell (center) in 80% $SiO_2$: 20% Pt device showing bubbles on top electrode; and (i) 80% $SiO_2$: 20% Pt device in different resistance states; dark rings around HRS (lower two, marked) but not LRS (upper two, marked) cells.

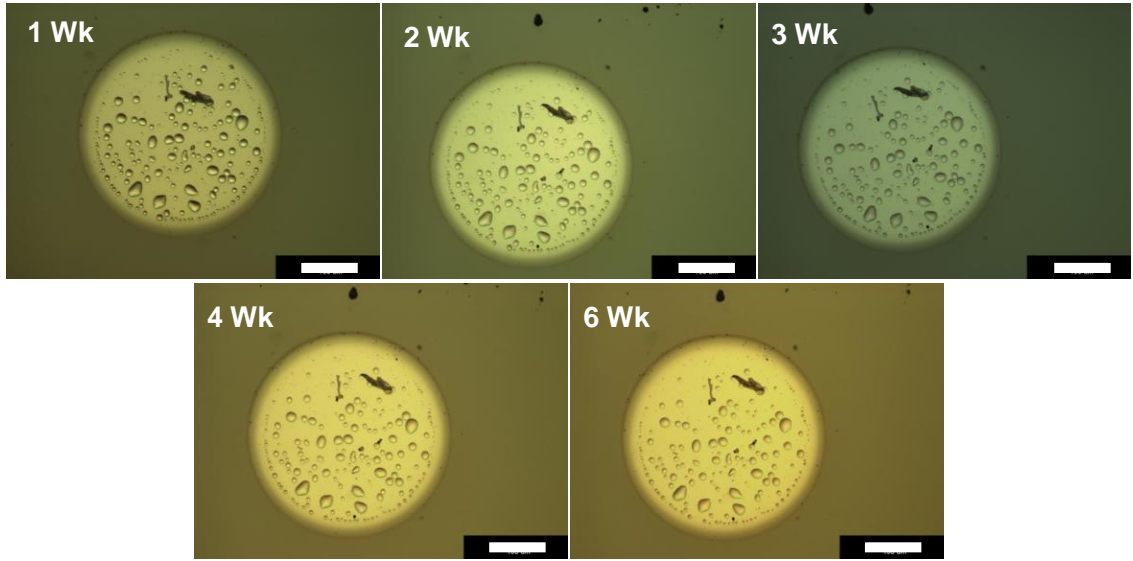

**Figure 6.2.** Voltage tested cell stored in ambient air after 1 week, 2 weeks, 3 weeks, 4 weeks and 6 weeks showing negligible evaporation of "bubbles". Scale bar: 100 μm.

### 6.3.2 Density

According to transmission electron microscopy[31], all the films studied here are amorphous showing no evidence of porosity (low-electron density regions). However, density data of $SiO_2$:Pt films measured by both RBS and X-ray reflectometry, which are consistent with each other as shown in **Figure 6.3**, are less than the theoretical ones suggesting nanoporosity. Without incorporating any Pt, sputtered $SiO_2$ films have a density ~2.16 g/cm$^3$, which is 82% of the theoretical silica density of 2.64 g/cm$^3$ (ref.[37]).

This confirms similar reports of low-density for sputtered $SiO_2$ films[38]. Hybrid $SiO_2$:Pt films also have a lower density than the theoretical one, taken as

$$\rho_{Theoretical} = f_{Pt}\rho_{Pt} + (1 - f_{Pt})\rho_{SiO_2}$$

where $f$ is the atomic fraction of metal, and theoretical densities of pure substances ($\rho_{SiO_2} = 2.64 g/cm^3$ for silica and $\rho_{Pt} = 21.46 g/cm^3$ for Pt metal) are used. The relative densities of all the Pt-containing films in **Figure 6.3** (the same data as in **Figure 6.1h**) are around 0.5, indicating a highly porous structure, which was also reported for other oxide-metal hybrids[39]. On the other hand, sputtered $Si_3N_4$ films had a density of 3.48 $g/cm^3$, close to that of fully dense $Si_3N_4$ ceramic in the literature[40]. The density of 93% $Si_3N_4$: 7%Cr films, to be used for device study later in this work, had a density of 3.62 $g/cm^3$, which is only 3% lower than $\rho_{Theoretical}$ ($\rho_{Si_3N_4} = 3.44 g/cm^3$, $\rho_{Cr} = 7.19 g/cm^3$). This means $Si_3N_4$: Cr films are almost free of pores. Other evidence of different nanoporosity in different films came from atomic force microscopy (**Figure 6.4**). Comparing area (10 × 5 μm²) scans of surface topography, $SiO_2$:Pt film exhibits much larger roughness (1.8 nm rms, consistent with the data in the literature[31]) than $Si_3N_4$: Cr film (0.29 nm rms).

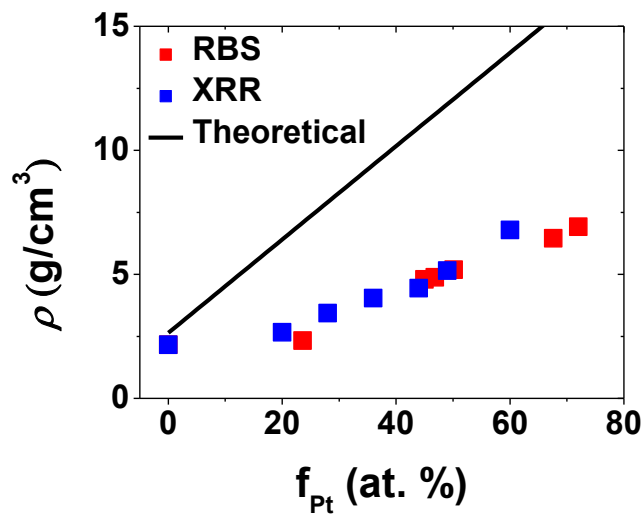

**Figure 6.3**. Density of SiO$_2$:Pt film at various Pt concentration (same data as in **Figure 6.1h**). Data from RBS (red) and XRR (blue). Line: theoretical density $f_{Pt}\rho_{Pt} + (1-f_{Pt})\rho_{SiO_2}$ ($\rho_{Pt} = 21.46 g/cm^3$, $\rho_{SiO_2} = 2.64 g/cm^3$).

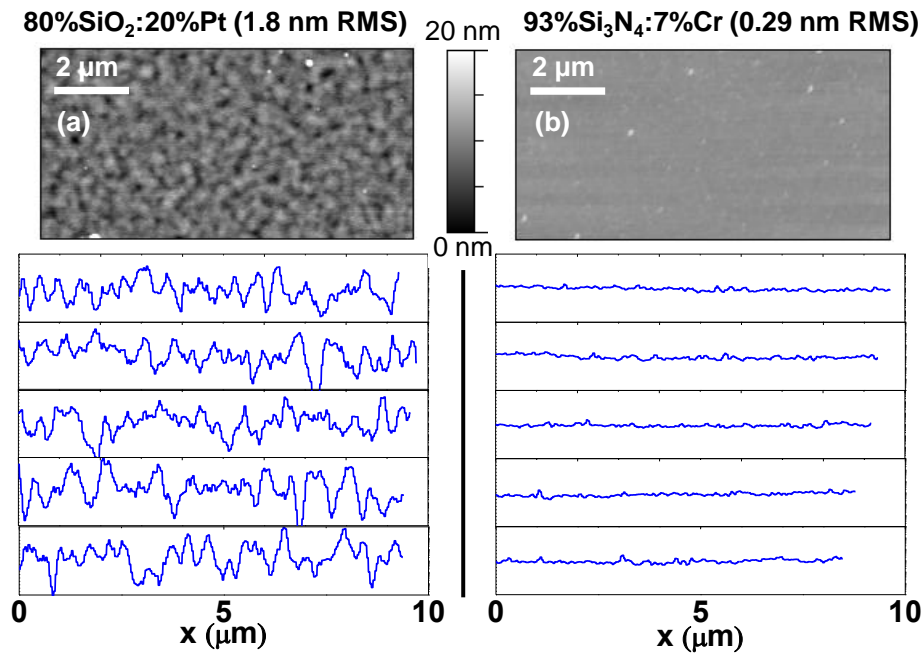

**Figure 6.4.** (Top) Topography of (a) 80%SiO$_2$:20%Pt and (b) 93%Si$_3$N$_4$:7%Cr film. (Bottom) Line scan profiles at several randomly selected locations. (Range of displayed height: 10 nm).

### 6.3.3 IR Spectroscopy

Transmission FTIR spectra of ~100 nm films deposited on KBr substrates are shown in **Figure 6.5**. (These films were left at room ambient conditions for a few days after removal from the sputtering chamber.) The spectra of 80% SiO$_2$:20% Pt from 400 cm$^{-1}$ to 4,000 cm$^{-1}$ contain peaks listed in **Table 6.1**, in which the three lowest-wavenumber ones are Si-O-Si vibration modes while the rest are all environment (H$_2$O, CO$_2$) related[41-42]. Similarly, in the 95% Si$_3$N$_4$: 5% Pt spectra two characteristic Si-N-Si peaks are identified at lower wavenumbers, while all the higher wavenumber peaks are probably environment related. (Very similar data were obtained for 93% Si$_3$N$_4$: 7% Cr, confirming that the peaks are due to Si-N-Si and environmental species, and not affected by atomically dispersed metals.) Relative to the strongest characteristic peak (1,057 cm$^{-1}$ for Si-O-Si stretching and 870 cm$^{-1}$ for Si-N-Si stretching), most of the environment-related peaks are more prominent in the SiO$_2$:Pt film, suggesting more H$_2$O absorption.

| Location (cm$^{-1}$) | Assignment |
| --- | --- |
| 463 | Si-O-Si rocking |
| 795 | Si-O-Si bending |
| 1,057 | Si-O-Si stretching |
| 1,632 | bending of free H$_2$O and/or H$_2$O H-bonded to silanol |
| 2,341 | CO$_2$ vapor |
| 3,225 | free/H-bonded H$_2$O |
| 3,450 | stretching of H$_2$O bound to silica network |
| 3,600 | OH stretching of H-bonded silanol groups |

**Table 6.1.** IR absorption bands of 80% SiO$_2$:20% Pt.

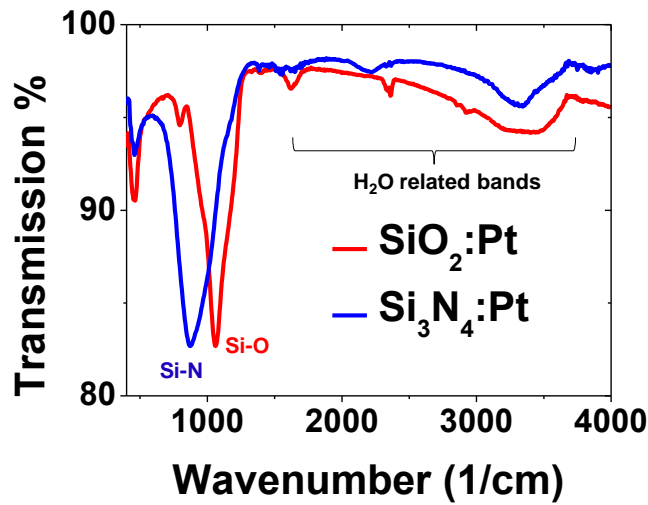

**Figure 6.5.** IR spectroscopy of 80% SiO$_2$: 20% Pt and 95% Si$_3$N$_4$: 5% Pt film from 400 cm$^{-1}$ to 4,000 cm$^{-1}$. Strongest Si-N and Si-O stretching bands are labeled. See other band assignments in **Table 6.1**.

### 6.3.4 Wetting

On sputtered 80% $SiO_2$:20% Pt nanometallic films (~200 nm, without the top electrode) water wetting is remarkably complete (wetting angle <3°) as shown in **Figure 6.6a**: as soon as the water droplet touched the film, it immediately spread out. This is surprising since $SiO_2$ is not known to be wetting. Indeed, a presumably dense thermal oxide $SiO_2$ (~200 nm, formed by oxidation of a Si wafer) had a contact angle ~50° (**Figure 6.6b**), which is consistent with the reported data for silica in the literature[43]. Such huge difference in the wetting behavior between sputtered $SiO_2$:Pt films and dense $SiO_2$ films is known to occur: a highly porous surface of a hydrophobic substance can become super-hydrophilic.[44] Since no visible porosity was found in our films according to microscopy, it must be nanoporosity that causes both lower density and apparent wetting. Similarly, we compared the wetting angles of our sputtered 95% $Si_3N_4$: 5% Pt film (47° in **Figure 6.6c**) and a dense hot-pressed SiAlON (in which Al and O partially substitute Si and N in the $Si_3N_4$ network) ceramic (60° in **Figure 6.6d**)[45]. observing a similar, though less drastic, increase in wettability in the film, presumably also due to nanoporosity. In contrast, when 80% $SiO_2$:20% Pt is capped with a $Al_2O_3$ ALD film (~4 nm), it shows a wetting angle (52°, **Figure 6.6e**) close to that of a dense sintered alumina ceramic (50°, **Figure 6.6f**), suggesting that the ALD $Al_2O_3$ film is rather dense. Lastly, 93% $Si_3N_4$: 7% Cr films have the same wetting property (~49° as $Si_3N_4$: Pt films. These results are summarized in **Table 6.2**.

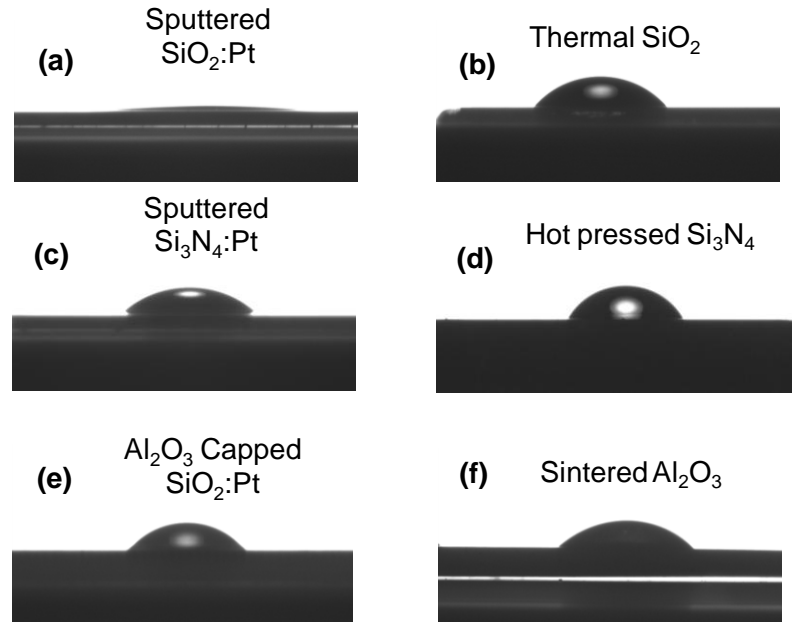

**Figure 6.6.** Wetting by static sessile water drop on (a) sputtered 80% $SiO_2$:20% Pt film; (b) thermally oxidized $SiO_2$ film; (c) sputtered 95% $Si_3N_4$:5% Pt film; (d) hot pressed $Si_3N_4$ ceramic; (e) ALD $Al_2O_3$-capped 80% $SiO_2$:20% Pt film; and (f) sintered $Al_2O_3$ ceramic. Calculated wetting angles listed in **Table 6.2**.

| Sample | Wetting Angle |
|---|---|
| Sputtered $SiO_2$:Pt | <3° |
| Thermal $SiO_2$ | 50° |
| Sputtered $Si_3N_4$:Pt | 47° |
| Hot pressed $Si_3N_4$ | 60° |
| $Al_2O_3$ Capped $SiO_2$:Pt | 52° |
| Sintered $Al_2O_3$ | 50° |

**Table 6.2**. Wetting angles of various films and ceramics.

### 6.3.5 Electric Characterization of 80% $SiO_2$:20%Pt Films

As-fabricated devices were all in a conducting state with an initial resistance $R$ between $10^2$ and $10^3$ Ω. A typical current ($I$)-voltage ($V$) and $R$-$V$ cycle is shown in **Figure 6.7a** having the following characteristics. When a forward bias voltage is applied, the current suddenly drops at a certain voltage (~4 V), which transitions the cell to the HRS (the "off" state in RRAM). The cell is "non-volatile" and "bipolar" in that the HRS is maintained at 0 V and that a reverse bias is required to switch it back to the LRS (the "on" state) through a series of current increases indicating a multi-step process. This is in agreement with our previous findings[31-34]: when Pt is used as top electrode and Mo or $SrRuO_3$ as bottom electrode, the switching direction is counterclockwise in the $R$-$V$ hysteresis, changing from LRS to HRS at a positive bias and vice versa at a negative bias.

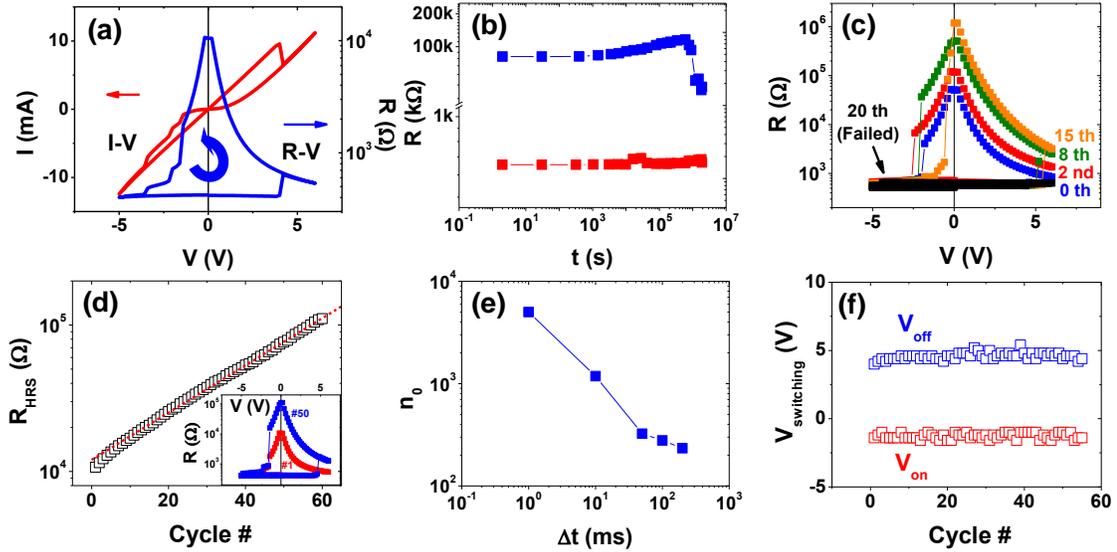

**Figure 6.7.** (a) Current ($I$)-voltage ($V$) and resistance ($R$)-voltage ($V$) characteristic of 80% $SiO_2$:20% Pt nanometallic device. (b) Resistance retention (read at 0.2 V) in ambient air for HRS (blue) and LRS (red); HRS rolls-off at around $10^6$ s. (c) Cell held in HRS and switched 3 times (switching $R$-$V$ curves shown) during first 15 days; failure after 20 days. (d) Exponential increase of HRS resistance during consecutive 60 switching cycles (0 V→5 V→6 V→-5 V→0 V, cycle duration $\Delta t$~1s). Inset: $R$-$V$

curves of $1^{st}$ and $50^{th}$ cycle, respectively. (e) Degradation factor $n_0$ *vs.* pulse duration $\Delta t$. Negative slope ~1 indicates degradation depends on total time under electric loading. (f) Switching voltages ($V_{off}$ and $V_{on}$) remain constant during consecutive switching cycles.

Devices made of $SiO_2$:Pt with Mo bottom electrode and without capping is not stable at the room ambient conditions in the sense that its resistance (especially HRS) changes with time and external voltage. **Figure 6.7b** shows the results of a retention test for a device at the HRS (kept at room ambient conditions with a relative humidity 10%-15%.) In this experiment, several fresh cells were first switched to the HRS with $R_{HRS}$~81 kΩ. The resistance of one cell (named cell A) was periodically monitored using a small voltage (0.2 V), while other cells (cells B, C, *etc.*) were used as surrogates for evaluating switchability. (We assume cell A and its surrogates behave similarly as long as they are not disturbed by a voltage sweep cycle required for the switchability test.) After $6 \times 10^5$ s (~7 days), the cell A resistance had gradually increased to 116 kΩ but the device remained switchable (according to the test on surrogate cells B, C, *etc.*). After that the cell A resistance dropped rapidly indicating a "loss of memory". Eventually, failure occurred in the sense that the cell could no longer be switched to the LRS (again, according to the switchability test on another virgin surrogate cell D) after $2 \times 10^6$ s (~23 days). It should be noted that the resistance change only occurred when the cell was kept in the HRS; a cell kept in the LRS showed no change in resistance when held at room ambient conditions (see the lower branch data in **Figure 6.7b**).

As mentioned already, during the above retention test the cell A was not disturbed by the voltage sweep cycle; the switchability was established through testing the surrogates. If instead the same cell is periodically disturbed and tested for switchability, then its $R_{HRS}$ increase is much faster, and the failure comes sooner. **Figure 6.7c** shows such a cell having an initial $R_{HRS} \sim 51$ k$\Omega$ at 0.2 V, which increased to 1.2 M$\Omega$ after 15 days during which it was switched merely 3 times. When checked on day 20, the cell already lost its memory and was not able to switch to the HRS after on-switching. This indicates that repeated switching, even just a few times, can cause much more degradation if the cell is held in the HRS. To further investigate the effect of repeated switching, we conducted the voltage-sweep cycle without interruption in the so-called fatigue or endurance test. As shown by the $R$-$V$ curves in the inset of **Figure 6.7d**, the HRS of the 1$^{st}$ cycle differs significantly from that of the 50$^{th}$ cycle. (The switching voltages are relatively constant: off-switching at 4 V and on-switching at $-1$ V, as is the LRS remaining at 420 $\Omega$). The $R_{HRS}$ increased (**Figure 6.7d**) exponentially with the number of cycle $n$: $R_{HRS} = R_0 \times \exp(n/n_0)$ with $n_0 = 27$, meaning doubling $R_{HRS}$ every 19 cycles. To investigate whether the HRS degradation is dependent on the voltage-sweep period $\Delta t$, we repeated the fatigue test at other $\Delta t$ and analyzed the results in terms of $n_0$ used in the exponential "law." As shown in **Figure 6.7e**, a faster pulse yielded a larger $n_0$, and vice versa. However, since the initial slope of log $n_0$ *versus* log $\Delta t$ is close to $-1$, meaning $n_0$ $\Delta t \sim$ constant, the HRS degradation (at least during fast switching) is dependent on the total time ($n\Delta t$) only, following $R_{HRS} = R_0 \times \exp(t/t_0)$ with $t_0 \sim 10^4$ s. This result may be regarded as representative of "dynamic" degradation caused by both an internal field (due to the trapped charge in the HRS) and an external field (due to the voltage sweep). Compared to "static" degradation in **Figure 6.7b** ($t_o \sim 1.8 \times 10^6$ s) caused by only an internal field without an external field, dynamic degradation is much faster. Lastly, in contrast to the exponential rise of $R_{HRS}$ (**Figure 6.7d**), switching voltages ($V_{off}$ and $V_{on}$) remained unchanged during sweeping cycles (**Figure 6.7f**). This is unlike the case of Tsuruoka et al in which large variations in switching voltage with environment were found for the $Cu/SiO_2/Pt$ and $Cu/Ta_2O_5/Pt$ cells[19].

### 6.3.6 Temperature Dependence of $R$, $\varepsilon_r^{'}$ and $\tan\delta$

The DC resistance and lower-frequency dielectric spectra of the HRS of a $SiO_2$:Pt film shown in **Figure 6.8** all display a pronounced temperature dependence. As-fabricated devices in this study had a $R_{HRS} \sim 200$ k$\Omega$, which degraded to $\sim 1$ M$\Omega$ after three months. This change is reversible: as shown in **Figure 6.8a**, it decreased to 35 k$\Omega$ at just above 100°C. $R_{HRS}$ continued to gradually decrease when the cell was heated above 100°C, reaching $\sim 20$ k$\Omega$ at 175°C. After cooling, it recovered the as-fabricated resistance. The above behavior suggests a two-stage process: the first stage (the low temperature part) is related to the loss of water, the second stage (the high temperature part) reflects the negative $dR/dT$ property (*i.e.*, increased conductivity) common to all insulators. Relative dielectric constant (the real part, $\varepsilon_r^{'}$, **Figure 6.8b**) also appears to undergo a large decrease above 100°C as water evaporates, especially at low frequency. The much stronger frequency dependence below 100°C signals a contribution of a polar species, presumably $H_2O$ or OH, that undergoes relatively slow dielectric relaxations compared to intrinsic ionic and electronic contributions. The influence of water is too implicated in **Figure 6.8c** by the large tanδ, which can arise from both dielectric loss and conductivity loss. At low temperature and high frequency (100 kHz), it already reaches 0.5 even though the DC resistance is still high (>1 MΩ, **Figure 6.8a**); most likely, the dominant contribution here comes from dielectric relaxation of water. Remarkably, as the temperature increases, tanδ continues to increase at least up to 100°C despite the expected evaporative loss of water which should reduce tanδ. This suggests that the thermally activated dielectric relaxation of $H_2O$ molecule is much more temperature sensitive than water evaporation. (A drop in tanδ reflecting the water loss eventually occurs at ~100°C in the spectra of 10 and 100 kHz in **Figure 6.8c**). Lastly, above 120°C, tanδ increases again at all frequencies in coincidence with the decreased DC resistance, which is consistent with our earlier suggestion of the negative $dR/dT$ property (*i.e.*, increasing conductivity) common to all insulators. The "intrinsic" background dielectric constant may be taken as $\varepsilon_r'$ ~10 seen in the high-frequency, high-temperature limit, which is considerably higher than that of porous $SiO_2$ (3.8 for fused $SiO_2$) because of the metal content. In the literature, other water-containing $SiO_2$ films are also known to have a higher dielectric constant than pristine silica; this is generally attributed to the easily polarizable OH groups (such as Si-OH) that populate the films[46].

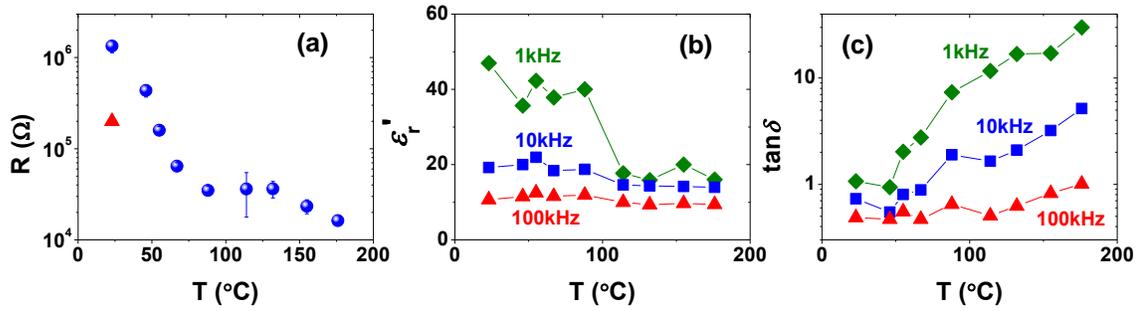

**Figure 6.8**. Temperature dependence of (a) HRS DC resistance, (b) real part of dielectric constant $\varepsilon_r'$ and (c) tangent loss ($\tan\delta$) of 80% SiO$_2$:20% Pt device during heating; before test, device was in ambient air for 3 months. Red triangle in (a): data of as-fabricated device, which coincide with the extrapolation of resistance data above 100$^{\circ}$C.

### 6.3.7 Robust Devices

Robust electric characteristics were observed using 93% Si$_3$N$_4$: 7% Cr films, 95% Si$_3$N$_4$: 5% Pt films, Al$_2$O$_3$-capped 80% SiO$_2$: 20% Pt films, and 80% SiO$_2$: 20% Pt films with SrRuO$_3$ bottom electrodes. Retention data in **Figure 6.9a** of a 93% Si$_3$N$_4$:7% Cr cell show that HRS degradation is largely avoided. (If the data are extrapolated to 10 years, they can easily fulfill the requirement for data storage of a good memory.) Supporting evidence also came from *R-V* loops shown in **Figure 6.9b**: the 50[th] loop is indistinguishable from the 1[st] loop. Indeed, even after 1 year the *R-V* curve hardly changed despite the fact that the cell had been periodically tested as shown in **Figure 6.9c**. Likewise, a capped SiO$_2$:Pt device showed the same robust behavior: no HRS change in retention (**Figure 6.9d**), and no change in the *R-V* loops in either cycle tests (**Figure 6.9e**) or during storage (**Figure 6.9f**). (The Al$_2$O$_3$ capping layer was deposited over the film and the top electrode. Before electric testing, the capping layer was poked by the probe tip under a small voltage to establish electric contact to the top electrode.) Similar data for 95% Si₃N₄: 5% Pt with Mo bottom electrode and for SiO₂:Pt with SrRuO₃ bottom electrode showing no degradation are provided in **Figure 6.10** and **Figure 6.11**.

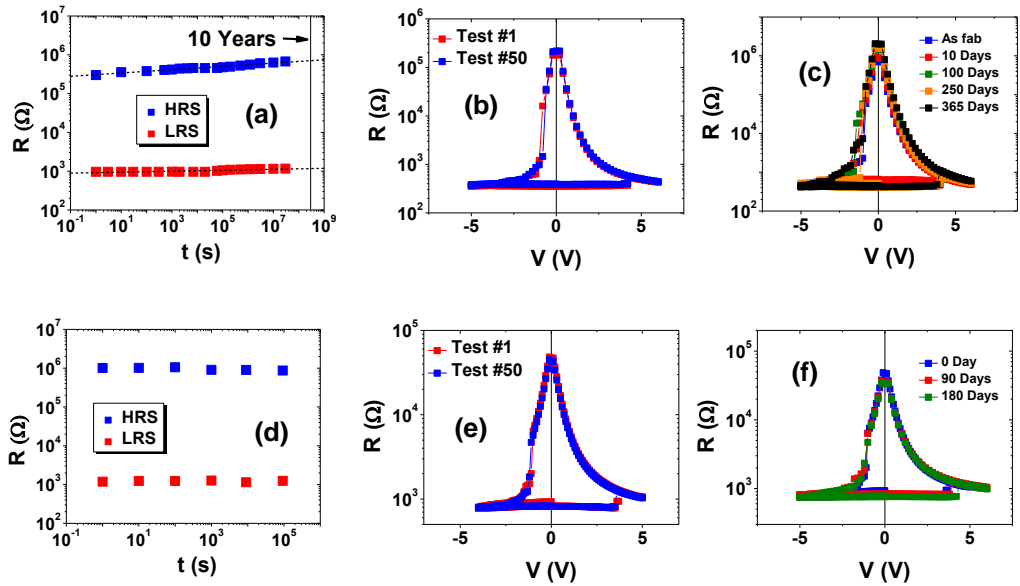

**Figure 6.9**. Robust devices made of 93% Si$_3$N$_4$: 7% Cr showing (a) resistance retention with extrapolated HRS and LRS resistance to 10 yr; (b) *R-V* curves of 1$^{st}$ and 50$^{th}$ cycles over consecutive 50 cycles, and (c) *R-V* curves taken over 365 days, with almost identical curves in (b) and (c). Likewise, robust devices made of Al$_2$O$_3$-capped 80% SiO$_2$: 20% Pt showing (d) resistance retention over 10$^5$ s; (e) *R-V* curves of 1$^{st}$ and 50$^{th}$ cycles over consecutive 50 cycles, and (f) *R-V* curves taken over 180 days, with almost identical curves in (d) and (e).

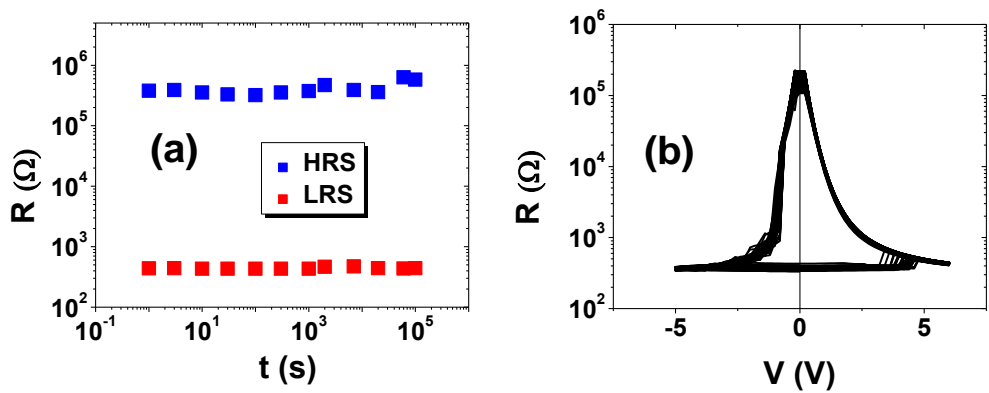

**Figure 6.10.** Device properties of 95% Si$_3$N$_4$: 5%Pt device in ambient air (Mo bottom electrode). (a) Retention test with 0.2V read voltage for HRS and LRS. No variation or degradation was found for 10$^5$s. (b) Consecutive 88 cycles, showing almost identical *R-V* curves without any variation/degradation.

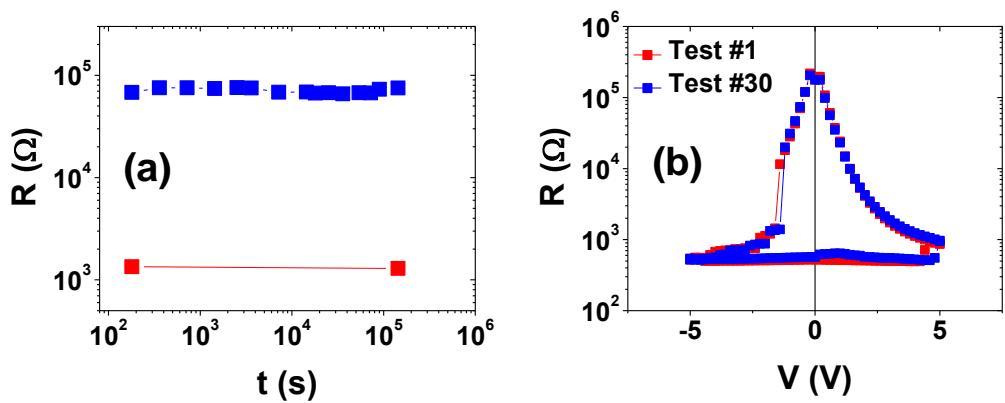

**Figure 6.11.** Device properties of 80% SiO$_2$: 20%Pt device (SrRuO$_3$ bottom electrode) in ambient air. (a) Retention test with 0.2 V read voltage for HRS and LRS. No variation or degradation was found for 10$^5$ s. (b) 1st and 30th cycles, showing almost identical *R-V* curves without any variation/degradation.

## 6.4 Discussion

It is obvious that (a) moisture is connected to the film and device degradation, (b) the severity is influenced by the intrinsic physicochemical nature of the film and/or nanoporosity, (c) the moisture effect can be "blocked" by having a dense capping layer, (d) electric fields, both internal and external, accelerate degradation, and (e) bottom electrode, being Mo or $SrRuO_3$, has a decisive influence on the degradation effect. In the above, (a)-(c) are strongly correlated to wetting: all three non-degrading devices described in **Figure 6.9** have relatively high wetting angles on the films, whereas rapidly degrading $SiO_2$:Pt films are nearly completely wetting. The measured "intrinsic" water wetting angles $\theta_e$ of dense dielectrics studied here are only slightly different, increasing from 50° to 60° in the order of $SiO_2$, $Al_2O_3$ and $Si_3N_4$. Intrinsic wetting properties reflect the physical chemistry of the materials pair (water and dielectric) in contact[47-48]. In our case, ionic characters of both Si-O and Al-O bonds make $SiO_2$ and $Al_2O_3$ relatively hydrophilic; meanwhile, Si-O bonds very likely are also present on the surface of $Si_3N_4$ making it too similarly hydrophilic. This suggests that intrinsic physicochemical nature is similar for all the films in the present study, thus it is not a differentiating factor for degradation.

On the other hand, the relatively small difference in $\theta_e$ can be overwhelmed by the influence of surface roughness, caused by nanoporosity, which varies greatly between films. Such roughness factor may be related to the bulk porosity assuming spherical pores of a radius $a$. According to quantitative stereography, on any random cross-section the surface fraction of intersected pores equals the volume fraction of pores, $P$. Therefore, 1-*P* fraction of the surface is flat, while *P* fraction of the surface has depressions made of exposed pores. The average surface area of an exposed pore can be found by the following consideration. Divide a sphere randomly into two parts; the two parts may differ in size but their total surface areas always add up to $4\pi a^2$. Thus their average surface area is $2\pi a^2$. The average cross section of the exposed pore can be found by the following consideration. Take a sphere and slice it into 2a slices of identical thickness 1. Since their total volume is $4/3\pi a^2$ the average area of each slice is $2/3\pi a^2$. Therefore, the roughness factor, which is total surface area divided by the projected area, is

$$r = (1-P) + P\left(2\pi a^2 / \frac{2}{3}\pi a^2\right) = 1 + 2P$$

We consequently conclude that porosity (volume fraction *P*) increases the effective contact area by a (roughness) factor $r = 1+2P>1$. Furthermore, according to the Wenzel equation

$$\cos\theta_r^w = r\cos\theta_e$$

$r>1$ leads to a smaller wetting angle $\theta_r^w$ on a porous rough surface than that on a dense smooth surface. For $SiO_2$:Pt films with *P*=0.5, we estimate $r = 2$, which predicts complete wetting. Nearly complete wetting was indeed observed in **Figure 6.6a**. For other films, nanoporosity is too small to significantly affect wetting, which is also in agreement with our observations in **Figure 6.6(b-f)**. Therefore, we may conclude that in our study, it is nanoporosity that is the direct cause for degradation. Moreover, since some nanoporosity always exists even in non-degrading films, we believe it is interconnected nanoporosity which penetrates deep inside the films that causes degradation. In addition, films with a small $P$ probably do not have such penetrating nanoporosity, hence not suffering from moisture-induced degradation. This scenario is also consistent with the outcome of surface capping: once the surface is capped by a dense film, moisture penetration apparently stops.

We next examine the possible mechanisms *via* which moisture aided by an internal or external electric field can cause $R_{HRS}$ to increase when the bottom electrode is Mo, but not when the bottom electrode is SrRuO₃. Note that this occurs without any significant change in $R_{LRS}$ and switching voltages. The latter observation rules out any effect on the interface, since any rise in the interface resistance must be reflected in a corresponding rise in the switching voltage, as the actual voltage across the film to trigger switching must become less due to voltage sharing by the interface resistance[32]. This leaves distributed degradation throughout the film the more likely cause for the $R_{HRS}$ increase. We propose this occurs in two steps: (1) $H_2O$ attraction to the film interior *via* dielectrophoresis[49], and (2) $H_2O$- mediated Mo oxidation and silicic acid formation, creating negative charge centers $(OH)^-$. Concerning step (1), in dielectrophoresis, an object (such as $H_2O$ molecule) that has a higher dielectric constant than the surrounding experiences a force in the direction of increasing field (absolute) magnitude. If there is only an internal field $E_{int}$, the force is proportional to the gradient of $E_{int}^2$, so $H_2O$ should migrate toward the centers of this field, which are located inside the film at trapped charges (electrons) at the so-called negative-$U$ centers[33]. (In nanometallic memory, filling these negative-$U$ centers by trapped charge is responsible for the LRS-to-HRS transition. Conversely, emptying these centers by detrapping is responsible for the HRS-to-LRS transition.) When a uniform AC/DC external field $E_{ext}$ is added, the net force is proportional to the gradient of $E^2 = (E_{int} + E_{ext})^2$, i.e, the gradient of $E_{int}^2 + 2E_{int}E_{ext}$.[49] So it always enhances dielectrophoresis along some direction of $E_{int}$, pointing to the center of $E_{int}$: the trapped charge. Therefore, an external field merely amplifies the dielectrophoretic force and accelerates moisture-caused degradation without otherwise altering the direction.

Concerning step (2), we envision a $H_2O$-mediated redox reaction resulting in electron transfer from Mo to $SiO_2$ in the following reaction: $Mo + SiO_2 + 2H_2O = Mo^+ + [Si(OH)_4]^-$. Here, Mo oxidation with attendant $SiO_2$ reduction is aided by hydroxyl formation. Moreover, we believe it is further aided by either a mechanical stress due to hydrolysis strain, or $E_{int} + E_{ext}$, both of which deform the $SiO_2$ network thus facilitating Si-O-Si bond rupture [50-51] followed by hydroxyl formation. (Obviously, $E_{ext}$ accelerates this reaction. Also, the stress/strain/field gradient is generally larger at the rim of the cell, which may be the reason for the dark-ring appearance in **Figure 6.1i**.) Indeed, the negative-$U$ center itself is often a site of strained bond, which relaxes upon capturing an electron *via* electron-phonon interaction. The net result of step (1) and (2) is an additional trapped electron at the center of $E_{int}$, which is already the location of a previously trapped electron as mentioned above. The combined $E_{int}$ of these two trapped electrons increases the Coulomb barrier that impedes the movement of free electrons, hence a higher $R_{HRS}$. Note that the "half-cell reactions" (using the term in electrochemistry) in the above overall redox reaction generally do not occur at the same site: the $Mo/Mo^+$ reaction occurs at the electrode whereas the $SiO_2$-$H_2O$/$[Si(OH)_4]^-$ reaction occurs at the strained Si-O-Si site or

negative-$U$ center. The overall reaction involves both half-cell reactions and electron transport between the two sites, which are separated at a distance no more than the film thickness $\sim 10$ nm.

The above two-step mechanism is consistent with the observation that a positive voltage polarity causes cell darkening (Fig. 1g), since it corresponds to a negative voltage on Mo bottom electrode, which favors electron donation by Mo. The mechanism is also consistent with the observation of no degradation with $SrRuO_3$ as the bottom electrode. This is because $Ru^{4+}$ having the low-spin $4d^4$ electron configuration is relatively stable[52-53]; therefore, even though $H_2O$ may still enter the $SiO_2$:Pt film, there is no redox reaction without electron donation from $SrRuO_3$.

Further insight may be gained from scrutinizing the evolution of $R_{HRS}$ and switching voltages in light of the above picture. To begin with, we note that increase of $R_{HRS}$ with time was also observed in amorphous chalcogenide phase-change memory due to structural relaxation[54]. Characteristically, such relaxation decelerates with time and asymptotically approaches saturation. This is in contrast with the behavior of **Figure 6.7b**, which shows acceleration in the log-log plot with no evidence of saturation before a relative sudden drop leading to failure. According to our mechanism, the $R_{HRS}$ increases because of water-facilitated increase in trapped charge and possibly trap sites (negative $U$-centers). However, water is also known to cause loss of insulating properties of dielectrics, even creating electric shorts[7]. Such water-induced electric degradation is expected to worsen as the internal electric field increases. In our picture, a higher $R_{HRS}$ corresponds to a state of higher internal field, so degradation of the $SiO_2$ matrix can explain the eventual loss of $R_{HRS}$ after an initial increase. (Another possibility is increased leakage of trapped electrons through tunneling or Frenkel-Poole hopping, which is aided by the increasing density of trapped charge and facilitated by the matrix degradation. Such leakage does not occur in the absence of water infiltration—see **Figure 6.9d** and the retention data elsewhere[32].) After failure, the HRS cell may not be switchable anymore if trapped electrons are stabilized by the water-altered environment, where local bonding could have changed dramatically. (This process can again be aided by the internal electric/mechanical field). So the cell becomes "stuck at the HRS." Alternatively, a failed cell may be returned to the LRS, then becomes "stuck at the LRS" because the dielectric matrix is too leaky to withstand the switching voltage, making it impossible to switch. (The HRS-to-LRS transition requires very little current, so it could precede the above event.) These failure modes have both been described in the previous section. Lastly, the fact that switching voltages in **Figure 6.7f** are not changed (until the cell fails) despite the progression of $R_{HRS}$ implies that the additional trapped charge/trapping sites created by water-induced reactions still largely fall into the same energy-level distribution as for the original population. This is because switching in nanometallic RRAM is voltage-controlled depending on the actual voltage in the film and independent of film thickness, area and temperature[31-34]. For trapped charge/trapping sites of the same energy level, the critical voltage for detrapping/trapping is the same.

In filament-conducting RRAM, moisture was also shown to cause degradation by altering the redox state of the filament-forming species (*e.g.*, $Cu/Cu^+$): the additional electrochemical driving force in the $Pt/H_2O/Cu/Cu^+$ "cell" lowers the threshold voltage for filament formation and dissolution, hence the switching voltages[19]. This mechanism is not applicable to our study, which witnessed no change in switching voltages when altering environment or temperature. Indeed, the present study supports our claim that nanometallic RRAM is a charge-trapping-controlled switching system[31-34], unrelated to redox-controlled switching mechanisms and/or filamentary mechanisms that operate in many RRAM[10-16]. On the other hand, moisture-related reliability issues no doubt can arise in other dielectrics which serve as the active layer in most filamentary RRAM: by affecting breakdown, changing carrier concentration and altering defect thermodynamics[18-22], and moisture-induced charge trapping occurring at resistance-critical junctions of broken filaments could cause a drift of $R_{HRS}$. However, since charge trapping is not considered important in filamentary RRAM, to the extent that the switching layer indeed has no stored charge, the lack of any internal field could imply less severe moisture-induced problems than observed in our study.

One consequence of moisture-enabled (Mo) electrode oxidation is the redox current and internal field associated with the reaction. This reaction requires anion flow toward the bottom electrode, meaning an electric current toward the top electrode in the absence of an external voltage. As a result, the zero-current condition is obtained only when (a) oxidation is complete and the device (an electrochemical cell) reaches an equilibrium state, or (b) a positive voltage is applied to the top electrode to compensate for the negative "oxidation" current. Because of the slow ion mobility and oxidation kinetics at ambient temperature, equilibrium is difficult to attain and non-zero ion currents and internal fields do rise during testing. (In this respect, it is worth noting that the voltage sweeping rate used for RRAM characterization/operating is many orders of magnitude faster than typically employed in electrochemical cells such as batteries and electrolysis cells.) This non-zero voltage offset is a common feature in the so-called "redox" RRAMs (see **Figure 6.12a**), and has been named the "nanobattery effect" in the literature (ref.[55]).

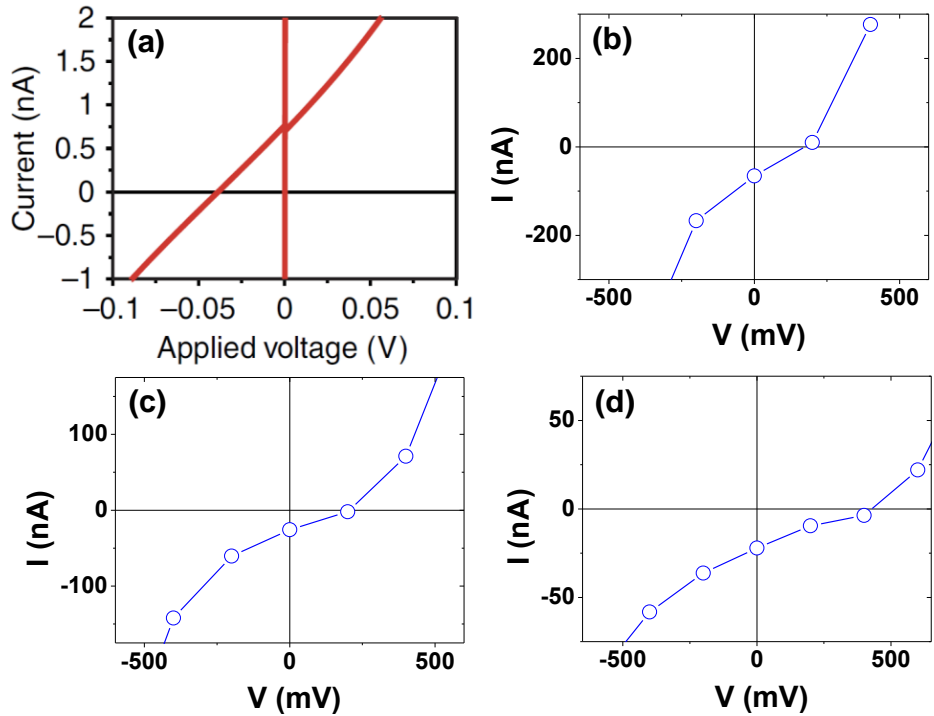

**Figure 6.12.** (a) Non-zero crossing near 0 V for *I-V* curve in ref.[55] (voltage polarity is different from ours). (b)-(d) Three uncapped $SiO_2$:Pt samples ("moisturized") showing large voltage-offset near 0 V.

We indeed observed such voltage offset in testing the HRS in "moist" (uncapped) $SiO_2$:Pt devices. As shown in **Figure 6.12b-d** (also reported in ref.[56]), typical non-zero voltage offsets are a few hundred mV, which is a rather large compared to the one reported in the literature (tens of mV). The sign of the voltage offset is consistent with Mo oxidation, and is opposite to that in **Figure 6.12a** due to the opposite polarity convention used in ref.[55] In contrast, in "capped" or "dense" films, such offset was not observed as shown in **Figure 6.13**. Using the voltage offset ($V_{offset}$) as a measure of the internal field, we find it monotonically increasing with the number of cycles in the uncapped device (**Figure 6.14**). This is in sharp contrast with the capped sample where $V_{offset}\sim0$ V irrespective of switching cycles. The increasing voltage offset indicates that the "active" electrode area increases with cycling, since the electrochemical potential should be the same but a larger area facilitates a larger "oxidation" current which demands a larger positive voltage offset to provide a larger compensation current. These results are all consistent with the moisture attack. Furthermore, if a vanishing $V_{offset}$ is considered as evidence of the absence of redox reaction, then the nanometallic device is indeed an intrinsically electronic device without ionic contribution. Therefore, $V_{offset}$ can serve as an important metric to differentiate ionic and electronic devices.

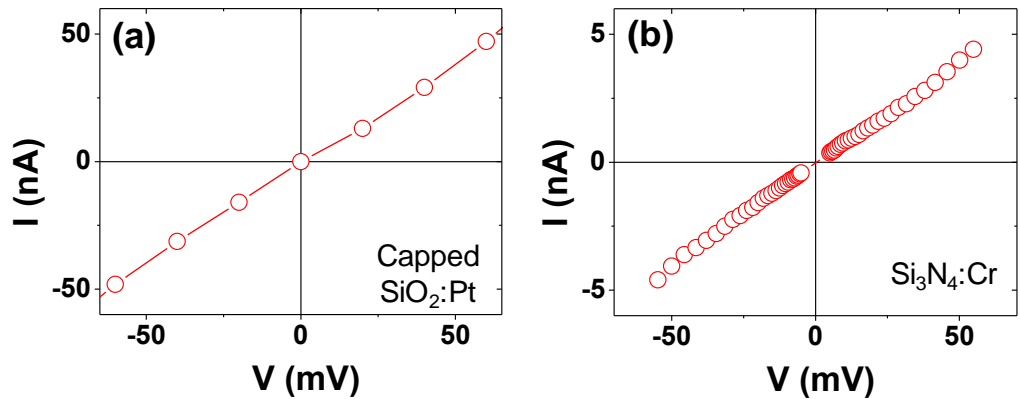

**Figure 6.13.** *I-V* curve near 0 V for (a) capped $SiO_2$:Pt device and (b) dense $Si_3N_4$:Cr device.

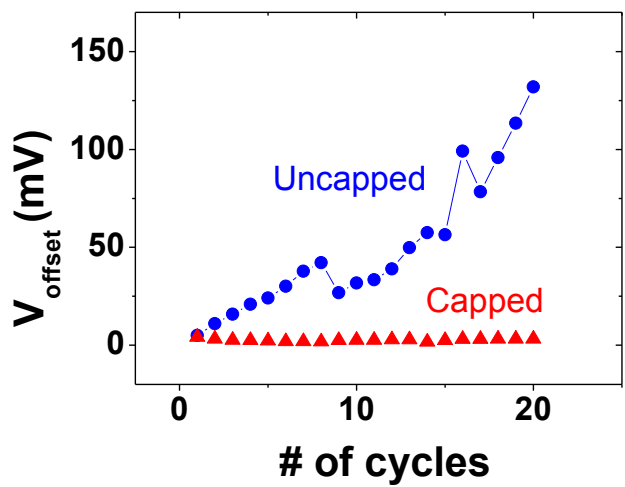

**Figure 6.14.** $V_{offset}$ (V-axis intercept) increases with the number of cycles in uncapped device, but is absent in capped device.

Finally, we return to the observation of nanoporosity formation in these films. As commonly reported in the literature[5,19,38-39], sputtered dielectric films deposited on unheated substrate are nanoporous because of sluggish surface diffusion. This situation is exacerbated by co-sputtering of metal species, which may have faster kinetics but are insoluble in dielectrics. Such metal atoms are likely located on the surface of nano-pores, like metal catalyst situated on a nanoporous support. As a result, the probability of metal clustering increases with $P$. Since nanometallicity arises from delocalized electrons placed in an environment of random potential[32-33,57], metal clustering compared to atomic dispersion is actually detrimental to nanometallicity because it increases the spacing between electron providers (atomically dispersed metal atoms or metal clusters.) This is consistent with our experience. As the amount of nanoporosity increases, the required metal content for resistance switching increases (See **Figure 6.15**). In this respect, the most "efficient" nanometallic films having the lowest metal-content requirement should also be the most moisture resistant—they have the lowest nanoporosity.

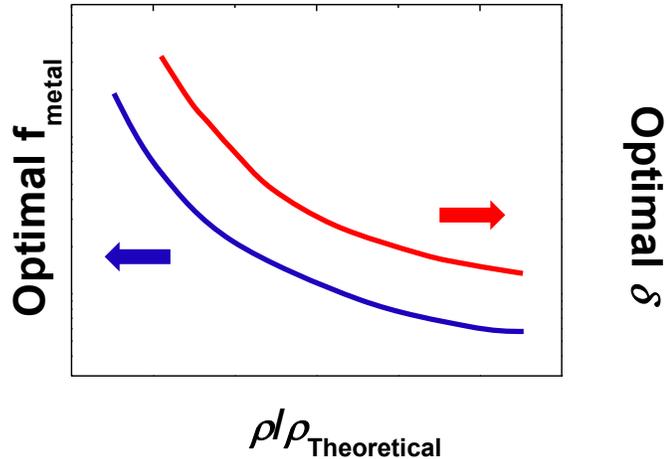

**Figure 6.15**. Schematic relationship between film density and (blue) optimal metal concentration $f_{metal}$, and (red) film thickness $\delta$, for RRAM device. As relative density increases, metallic clusters are avoided, allowing more metal atom dispersion thus reducing the required metal content and film thickness.

## 6.5 Conclusions

1. When electron-donating Mo is used as the bottom electrode, moisture induces physical and electric/dielectric changes in $SiO_2$:Pt nanometallic RRAM, including de-coloration and increases in dielectric constant and $R_{HRS}$, but not $R_{LRS}$. However, unlike filamentary RRAM, switching voltages remain unchanged in these devices.

2. Moisture penetrates *via* nanoporosity in the films, which can become almost completely water wetting due to nanoporosity-caused surface roughening.

3. Internal electric fields attract $H_2O$ molecules *via* dielectrophoresis, and motivate redox reaction causing Mo oxidation and $SiO_2$ reduction, which traps electrons at Si-OH, providing increasingly stronger Coulomb repulsion, thus a higher $R_{HRS}$. This process is greatly enhanced by an external field.

4. With lower nanoporosity or a dense capping layer, wetting, de-coloration and device degradation can be avoided, rendering robust switching and memory retention readily feasible in nanometallic RRAM. The same can also be achieved by replacing Mo with an oxidation-resistant electrode such as $SrRuO_3$.

5. Non-zero transient ion current and voltage offset at zero current exist in moist nanometallic device, indicative of an electrochemical potential caused by the on-going oxidation of the metal electrode. Such offset is completely eliminated in capped devices or dense devices.

# Part B: Nanometallic Devices and Circuit

# Chapter VII. Circuit Model

## 7.1 Introduction

As discussed in **Chapter I**, resistive switching is a universal phenomenon covering myriad of simple and advanced materials. Depending on specific material properties (*e.g.*, morphology, carrier diffusivity/mobility, temperature, *etc.*), RRAMs may have very different switching mechanisms. However, their macroscopic characteristics (*e.g.*, $I$-$V$ and $R$-$V$ switching curves) are remarkably similar regardless of switching mechanisms. For example, in a bipolar switching RRAM, the following features are commonly observed. (i) The HRS goes through one or a few gradual steps (indicating intermediate states) of decreasing resistance after an on-switching voltage $V_{on}$ is applied. (ii) The LRS can be controlled by current compliance or the maximum stress voltage, namely $R_{LRS}=R_{LRS}(I_c, V_{max})$. (iii) OFF switching proceeds at an off-switching voltage $V_{off}$, which varies as a function of historic current compliance or maximum stress voltage, *i.e.*, $V_{off}=V_{off}(I_c, V_{max})$. (iv) Both HRS and LRS can present multiple sub-states (intermediate states). These common characteristics suggest the possibility of a phenomenological description of all the RRAMs using a set of phenomenological circuit parameters. Such description could provide the critical link between the device-level picture (for "material engineer") and the architecture/system-level picture (for "electrical engineer"). In this chapter, I will introduce a simple circuit model which is in remarkable agreement with experiment data. The model can take either a discrete or continuum form. An earlier version of this model and its experimental justification have been published elsewhere[1].

## 7.2 The Discrete Model [1]

The model is intended to capture the following features of a typical RRAM: a sharp transition or a multi-step transition for off-switching, and a gradual transition (with typically only one or two or sharp steps) for on-switching. It attributes the "steps" to intermediate states realized by different resistance states in a circuit. The model takes voltage as the control parameter. Simulations will be used to verify the model by comparing the simulation results with the experimental data. In the following, the basic assumptions are stated.

### 7.2.1 Assumptions

**Assumption 1**: The device is composed of $N$ resistors in a parallel connection. Each resistor can take two resistance values, which define two states: $R_{lrs}$ and $R_{hrs}$. Here we let $R_{lrs}$ be a constant while treating $R_{hrs}$ as voltage dependent, taking a form of $R_{hrs} = R_0 \exp\left(-V/V_0\right)$.

**Assumption 2**: Any lrs (low resistance state) resistor under a positive bias $V \geq V_m^+$ will switch to an hrs state, *i.e.*, off-switching. Conversely, any hrs (high resistance state) resistor under a negative bias $V \leq -|V_m^-|$ will switch to an lrs state, *i.e.*, on-switching. For simplicity, we let $V_m^+ = |V_m^-| = V_m$, but allow certain distribution. such as a logarithmic normalized distribution, to govern $V_m$, *i.e.*, $V_m \sim LN\left(\mu, \sigma^2\right)$.

**Assumption 3**: The device also contains a voltage-independent load resistor $R_{load}$ in a serial connection with the parallel network. It may originate from the electrode, interfaces, connecting wires or other parasitic elements in the device circuit.

### 7.2.2 Simulation Methods

In **Figure 7.1**, lrs resistors are in green, and hrs resistors are in red. If all the resistors are lrs resistors, then the device is least resistive. If they are all hrs resistor, then the device is most resistive. The former may be termed the LRS itself, and the latter may be termed the HRS itself. Intermediate resistance states (IRS) are ones that contain $n$ lrs states and $N$-$n$ hrs states. The resistance of the device is

$$R = \frac{1}{n/R_{LRS} + (N-n)/R_{HRS}} + R_{BE}$$

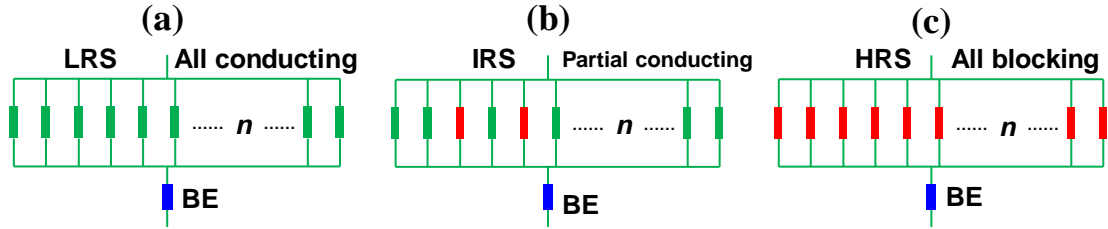

**Figure 7.1.** Parallel circuit representation of (a) low resistance state (LRS), (b) intermediate resistance state (IRS) and (c) high resistance state (HRS). Each resistor (conducting path) has two states, conducting (green, linear) or blocking state (red, non-linear)

Let the voltage on the entire circuit be $V_{total}$, and the voltage on individual hrs and lrs resistor be $V_{sample}$. Clearly, $V_{sample} < V_{total}$. Starting with the LRS, as the positive applied voltage reaches the minimum value required for switching one lrs resistor, such resistor will switch to an hrs resistor (green to red in **Figure 7.1**). As a result, $R_{sample}$ increases, and $V_{sample}$ increases. Such increased $V_{sample}$ may satisfy the switching voltage for one or more resistors, causing them to switch and $R_{sample}$ and $V_{sample}$ to further increase. This sets up a positive feedback process which may lead to avalanche switching. On the opposite side, starting from the HRS, if the negative applied voltage reaches the minimal value required for switching an hrs resistor, such resistor will switch to a lrs resistance. Thus, $R_{sample}$ decreases, and $V_{sample}$ decreases. Any further switching must wait for $V_{total}$ to increase so much that $V_{sample}$ exceeds the required switching voltage of another hrs resistor. This negative feedback process leads to a stepwise on-switching.

The simulation is done in Wolfram Mathematica 7.0.

**Pseudo Code** (off-switching):

```
Construct(N resistors);
Vm[i] = LogNormalDistribution(<Vm>, σ²);
Initialize R[i] = RHRS or RLRS;
For (V = 0, V →Vmax→0)
    {Calculate Rsample;
              Calculate Vsample;
              Iteration:
              If (Vsample ≥ Vm[i]) then R[i] = RHRS;
              Update Rsample;
              Update Vsample;
              }
```

### 7.2.3 Simulation Results

The parameters used in simulation are shown in **Figure 7.2**. We used 100 resistors with various $V_m$ distributed around 1.16 V, ranging from 0.8 V to 1.5 V. The initial state is assumed as 100% $R_{LRS}$, *i.e.*, an LRS. The voltage sweep cycle chosen is 0 V→+8 V→-6 V→0 V.

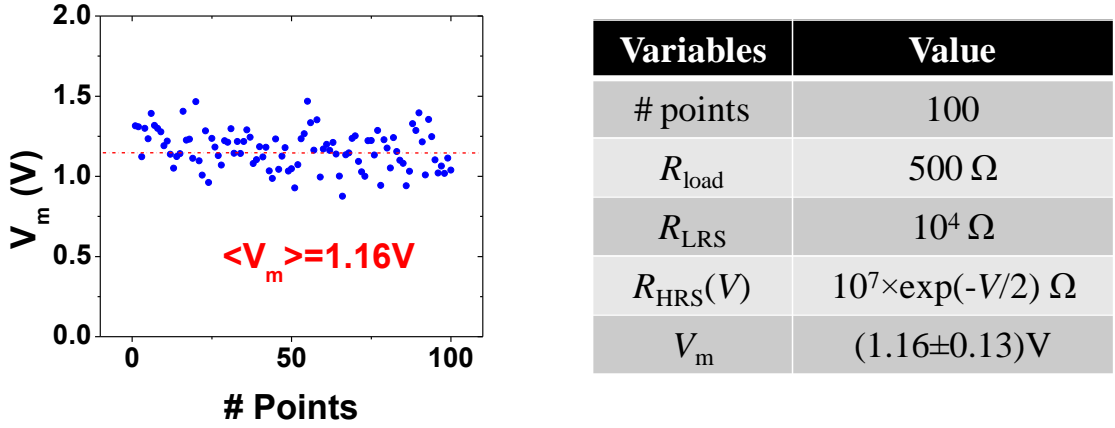

| Variables | Value |
|-----------|-------|
| # points | 100 |
| $R_{load}$ | 500 Ω |
| $R_{LRS}$ | $10^4$ Ω |
| $R_{HRS}(V)$ | $10^7 \times \exp(-V/2)$ Ω |
| $V_m$ | $(1.16 \pm 0.13)$V |

**Figure 7.2.** Generated $V_m$ distribution and simulation parameters

The simulated *R-V* curve is shown in **Figure 7.3**. Remarkably, it captures all the key features of switching dynamics. The LRS holds until ~+6 V, followed by a sudden jump to the HRS with all 100 lrs resistors switched to hrs resistors. After that, the resistance follows a highly nonlinear curve up and down the voltage, but there is no further switching. The off-switching voltage ~+6 V is expected because in this simulation, $R_{sample}$=100 Ω and $R_{load}$=500 Ω, so at $V_{total}$ =6 V, $V_{sample}$=1 V, which is what is required to trigger off-switching. On the side of negative $V_{total}$, a sharp drop occurs at ~-1 V, after that the sample undergoes a gradual reset process. A more negative $V_{total}$ allows more hrs resistors to switch to lrs resistors, but at -6 V only 77% of the hrs resistors have switched. Thus, it corresponds to an intermediate resistance state (IRS), whose resistance is maintained as the voltage returns to 0 V.

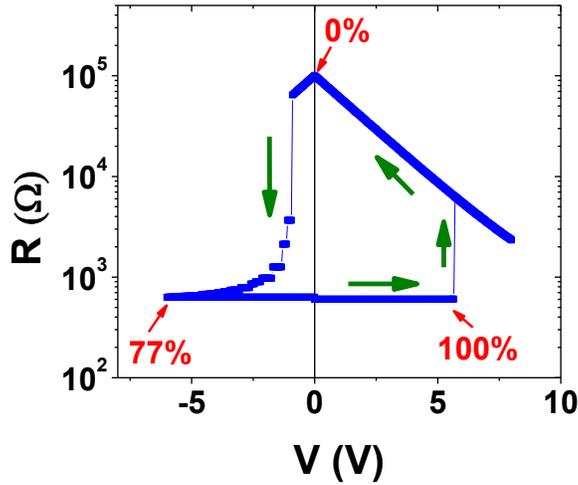

**Figure 7.3**. Simulated DC sweeping cycle.

### 7.2.3.1 True sample voltage ($V_{sample}$)

Voltage partitioning due to $R_{load}$ generally leads to $V_{sample}$ deviating from $V_{total}$, especially for the LRS. **Figure 7.4** explicitly shows the true voltage on the sample during off-switching. Before off-switching, $V_{sample}$ is linearly proportional to the total voltage $V$ with a slope 100 $\Omega$/600 $\Omega$=1/6 (**Figure 7.4a**). Such linear relation holds until the switching condition $V \sim <V_m>$ is met, then followed by a sharp jump to 6 V on a straight line with a slope 1. In this case, voltage sharing by $R_{load}$ is negligible as the device is turned off, making $R_{load} << R_{HRS}$. On the other hand, during on-switching (**Figure 7.4b**), negligible $R_{load}$ initially leads to a slope 1 straight line until the first hrs resistor reaching its on-switching voltage $V_m[i]$. Then the negative feedback holds the sample voltage around $<V_m>$ with small fluctuations. However, the $R_{load}$-mediated negative feedback also makes it difficult to reach LRS (100% lrs) again, unless an infinitely large negative bias is applied. For an estimation of sample resistance, we assume $V_{sample}=<V_m>$, then

$$R_{sample} = \frac{<V_m>}{V - <V_m>} R_{BE}$$

with an asymptote $\sim 1/V$.

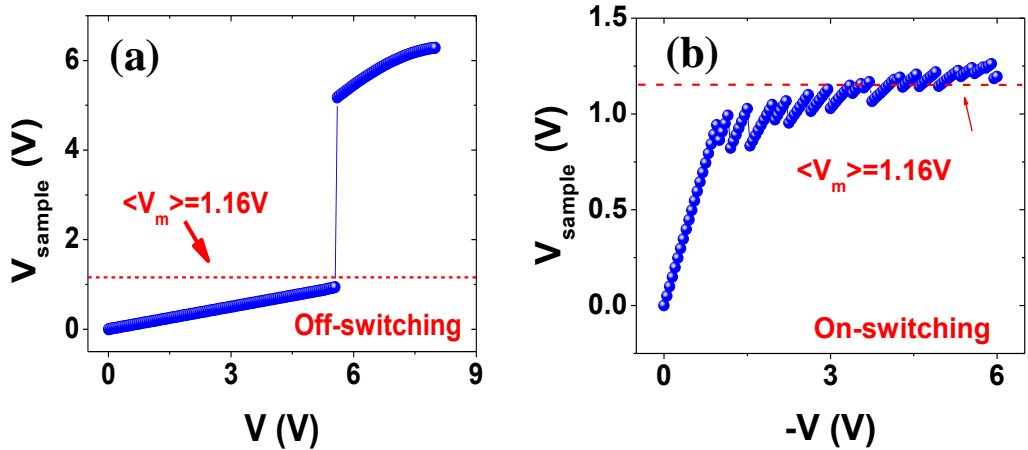

**Figure 7.4.** True voltage on sample during (a) off- and (b) on-switching.

### 7.2.3.2 On-switching to IRS

The simulation satisfactorily captures stepwise on-switching characteristics. On the *R-V* plot, a set of intermediate states (IRS) are presented and actual values depend on the percentage of hrs and lrs resistors, which can be physically controlled by a negative voltage (**Figure 7.5a**). Simulated *R-V* curves (**Figure 7.5b**) follow such behavior: a higher negative bias leads to a lower final resistance (a higher ratio of *n/N*), which is verified in **Figure 7.5c**

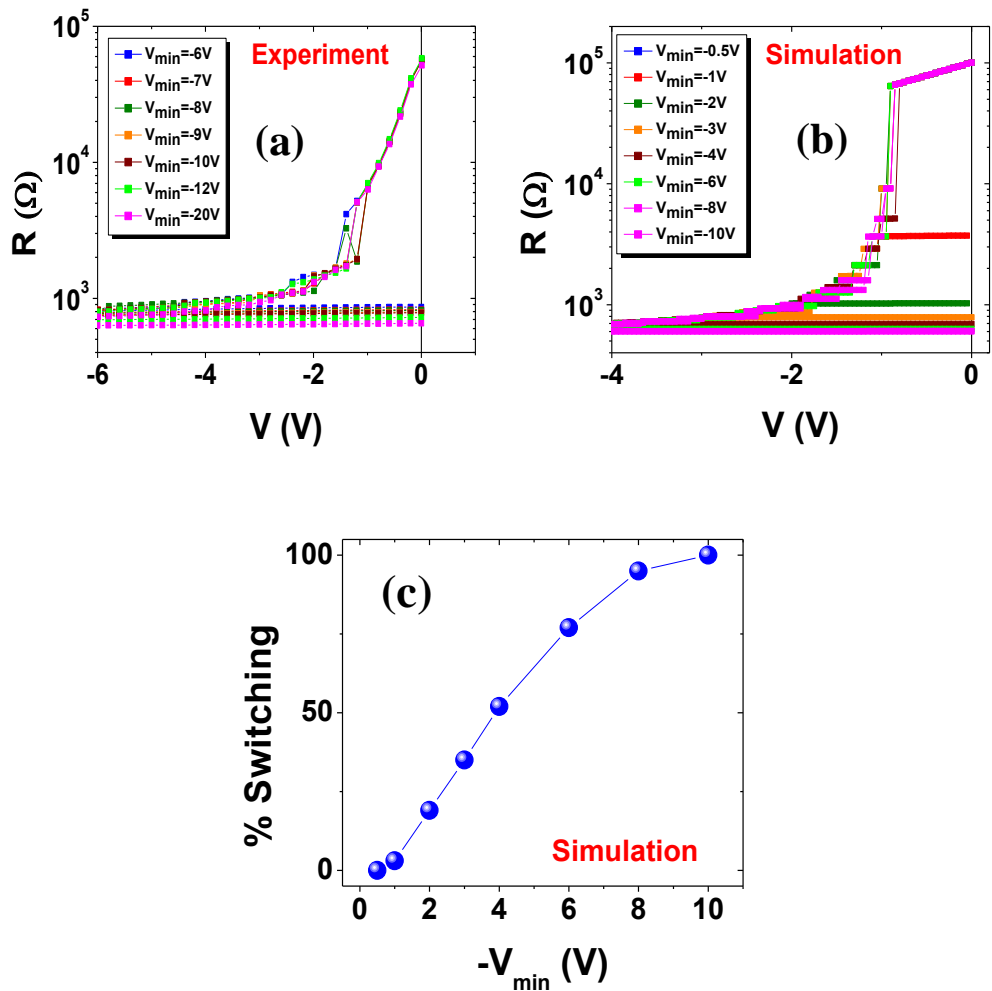

**Figure 7.5**. *R-V* characteristics of (a) experimental and (b) simulated on-switching process. (c) Switching percentage *vs.* negative bias.

### 7.2.3.3 Off-switching from IRS

The resultant IRS with certain conducting path percentage forms a new initial state for subsequent cycles. A more resistive IRS, with a lower lrs percentage, allows a higher amount of voltage partitioning, thus easier switching. This is verified by both experiment (**Figure 7.6a**) and simulation (**Figure 7.6b**). Here, the positive feedback ensures a complete and sharp switching all the time (IRS→100% HRS).

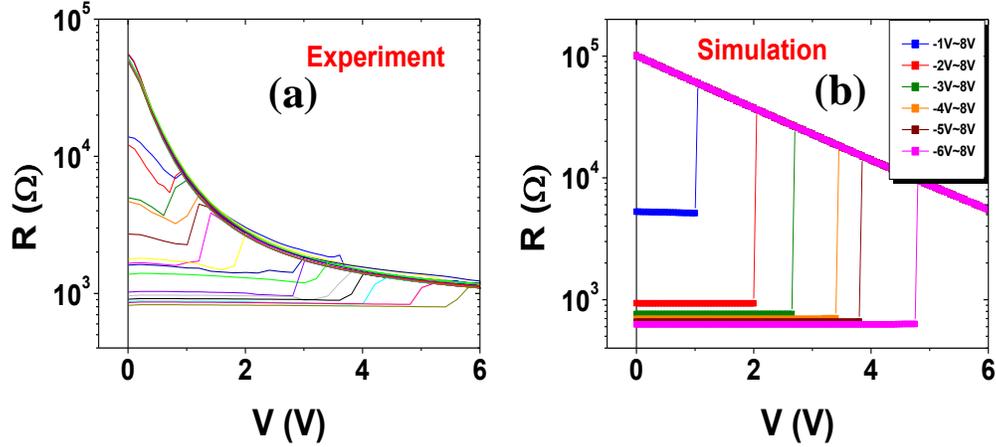

**Figure 7.6.** *R-V* characteristics of (a) experimental and (b) simulated off-switching process.

### 7.2.3.4 Intermediate HRS

The state variable in the parallel circuit model is the percentage ($n/N$). Although in most cases, the apparent HRS has nearly $n=0$, other $n>0$ states can be accessed by weakening the positive feedback (*e.g.*, lowering the $R_{sample}/R_{load}$ ratio). A more resistive IRS has a more gradual $V_{sample}$ increase during switching and hence, off-switching tends to follow the intrinsic $V_m$ distribution rather than an avalanche behavior. In such case an apparent HRS with $n>0$ can be easily captured by voltage control during gradual switching. Such idea is verified both in experiment (**Figure 7.7a**) and simulation (**Figure 7.7b**). These intermediate states can be used for building multi-level cells, which will be elaborated in a later section.

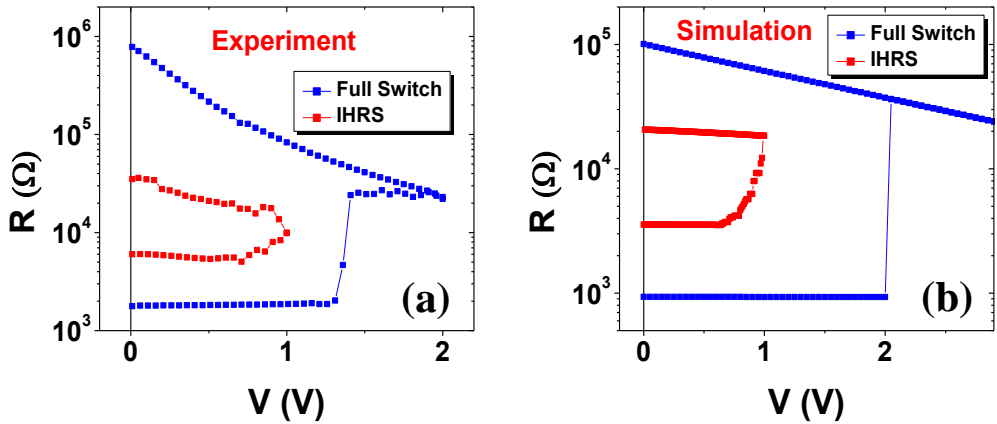

**Figure 7.7**. *R-V* characteristics of (a) experimental and (b) simulated off-switching process for capturing intermediate HRS.

### 7.2.3.5 Distribution of switching voltages ($V_m$)

The form of the total resistance

$$R = R_{BE} + R_{sample} = R_{BE} + \frac{V_m}{I}$$

suggests a plot of $R$ against $1/I$ may reveal the characteristics of $R_{BE}$ and $V_m$. The plot yields a straight line if $V_m$ is a constant or a narrowly distributed variable around a characteristic value. Meanwhile, the intercept is $R_{BE}$. From the on-switching data plotted in **Figure 7.8**, we obtain $R_{BE}$=407 Ω.

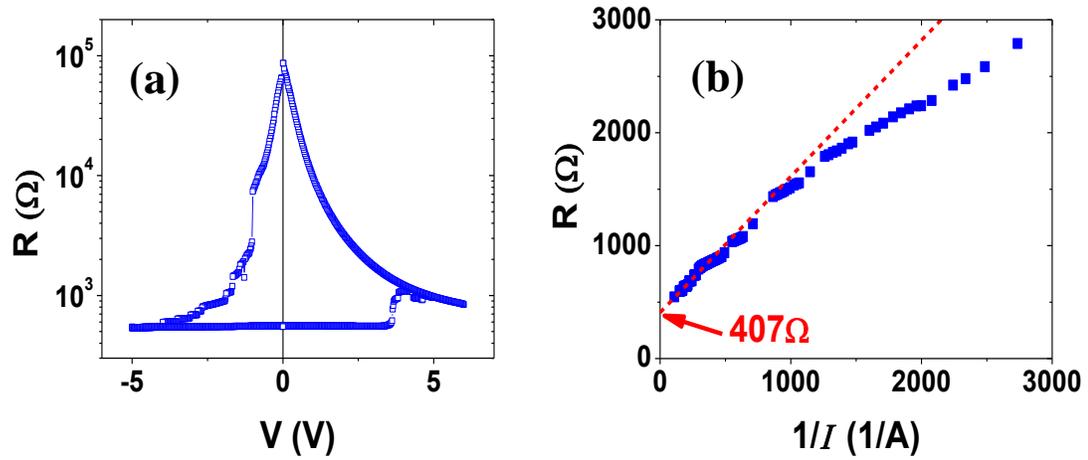

**Figure 7.8**. (a) Experimental *R-V* curve. (b) *R vs.* 1/*I* plot, the extrapolated intercept gives $R_{BE}$=407 Ω.

Knowing $R_{BE}$, we can back-calculate $V_m$ using $V_m=(R-R_{BE})\times I$. The obtained $V_m$ distribution is shown in **Figure 7.9**. The distribution (1.16±0.19) V is quite close to 1 V, which is the basis of our assumption in the previous simulations.

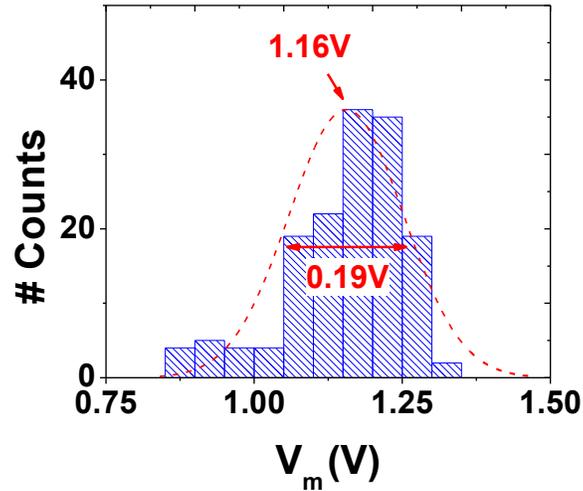

**Figure 7.9**. Calculated $V_m$ distribution.

### 7.2.3.6 Effect of $N$

To investigate the effect of discretness, $R_{hrs}$ and $R_{lrs}$ are set so that constant $R_{hrs}/N$ and $R_{lrs}/N$ are maintained for different $N$. (This maintains the same device resistance in the HRS and LRS state.) Assuming the same form of a lognormal distribution (1.16±0.13V), we generated $V_m$ distributions for different $N$ as shown in **Figure 7.10**. (All distribution here are assumed, not derived from experimental data). Using these distributions, the simulated $R$-$V$ curves become smoother as $N$ increases (**Figure 7.11**). Specifically, the $R$-$V$ of the $N$=10 case shows a set of sharp jumps and flat IRS in **Figure 7.11a**, but the $N$=10000 case resembles a completely continuous process in **Figure 7.11d**. This behavior is easy to understand if we realize that the first resistance discontinuity is roughly of the order of $R_{lrs}$, which may approach $R_{hrs}/N$ (*i.e.*, the resistance of the HRS) if $N$ is sufficiently large. Therefore, if sharp discontinuities are indeed observed experimentally, an optimized *N* can be used to simulate it. In this respect, *N*=100 shows the best agreement (**Figure 7.11b**) with the experimentally observed *R-V*; thus it will be used for subsequent simulation. It is worthwhile to note that "discreteness" could also stem from the voltage step $\Delta V$ used in simulation or experiment, simply because a large $\Delta V$ might "group" a large number of resistors and yield collective switching.

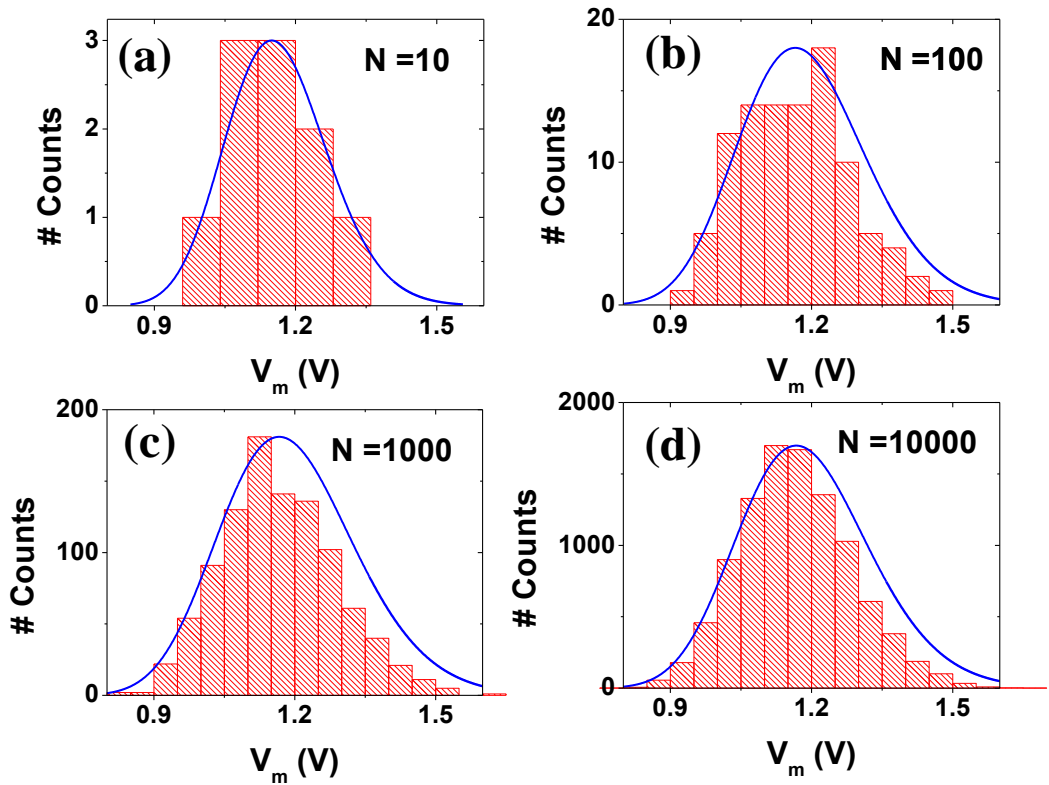

**Figure 7.10.** Identical $V_m$ distribution generated for simulation using (a) *N*=10, (b) *N*=100, (c) *N*=1000, (d) *N*=10000.

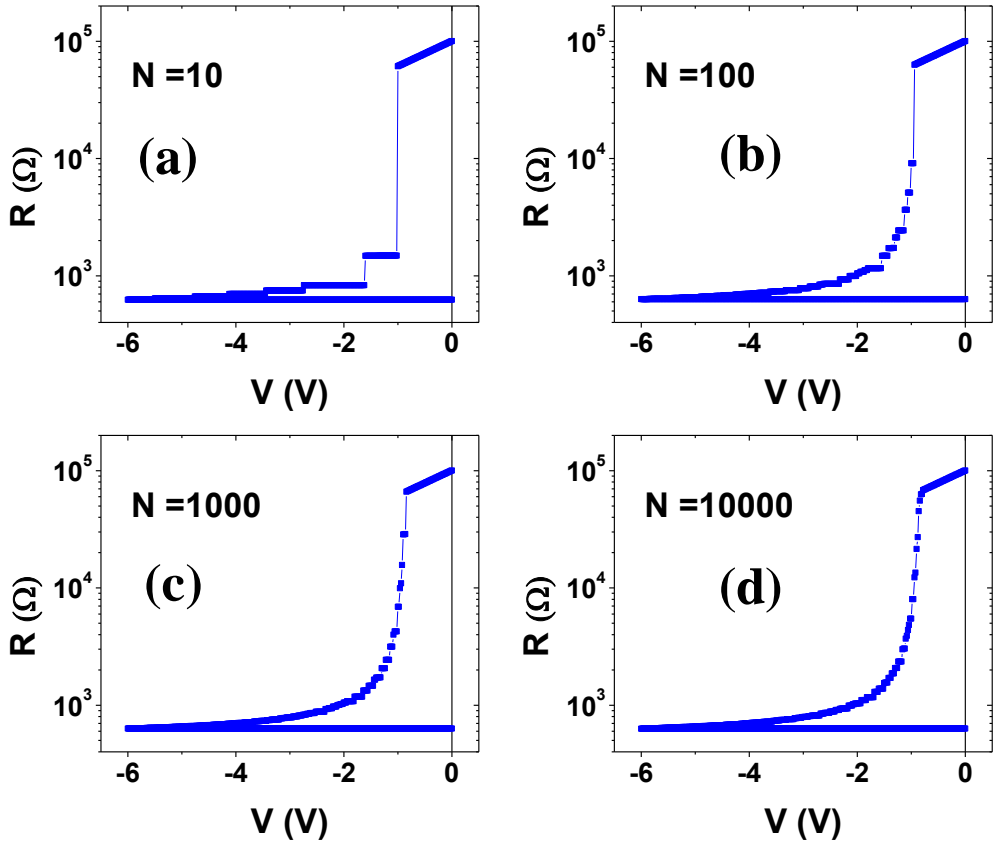

**Figure 7.11.** Simulated *R-V* curves for (a) *N*=10, (b) *N*=100, (c) *N*=1000, (d) *N*=10000.

### 7.2.4   Fitting of Experimental Data

In **Figure 7.12**, the fitting curves for Mo/SiO$_2$:Pt/Pt nanometallic RRAM exhibiting good agreement with the experimental data were obtained with following simulation parameters:

$N$=100; $R_{load}$=$R_{BE}$=407 Ω; $R_{lrs}$=14.4 kΩ; $<V_m>$=1.16 V, $\sigma$ =0.11 V;

$R_{hrs} = \exp(11.41 - 3.21\,|V| + 1.12\,|V|^2 - 0.25642\,|V|^3 + 0.032\,|V|^4 - 0.0016\,|V|^5)$

Here, $R_{hrs}$ is obtained by fitting the HRS resistance with an exponential-polynomial function, which is purely empirical without any physical meaning ascribed to it.

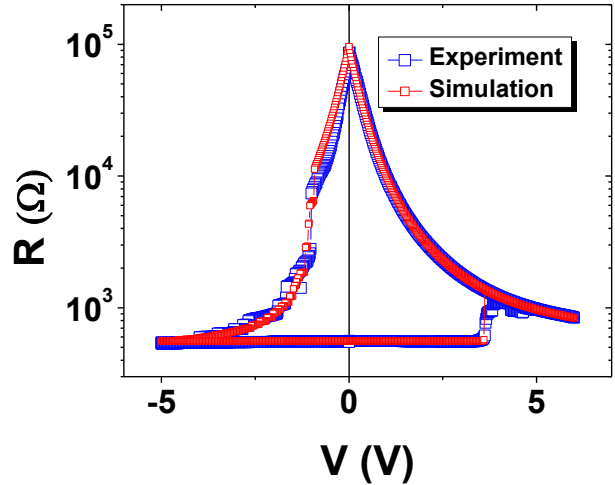

**Figure 7.12.** Fitted *vs.* experimental *R*-*V* curves for Mo/SiO$_2$:Pt/Pt RRAM.

## 7.3 A Continuum Model [2-3]

Treating the RRAM device as a serial connection of a cell resistance $R_c$ and a load resistance $R_l$ in **Figure 7.13a**, we let $R_c$ be a parallel connection of a low-resistance cross section ($r_L$ per area, area fraction=$F$) and a high-resistance cross section ($r_H$ per area, area fraction=1-$F$), see **Figure 7.13b**. In this picture, $F$ is the state variable varying between 0 (the most resistive state) and 1 (the most conducting state). Specifically, on-switching corresponds to the transition from the $F$=0 state to the $F$=1 state, off-switching corresponds to the transition from the $F$=1 state to the $F$=0 state, and intermediate states corresponds to intermediate $F$ values between 0 and 1. On-switching is $R_l$ dependent: the larger the $R_l$, the smaller the current in the cell, thus the smaller the $F$ of the intermediate state, and the higher the $R_c$. In general,

$$R = \frac{1}{\dfrac{1}{AFr_L} + \dfrac{1}{A(1-F)r_H}} + R_l$$

where $A$ is the sample area. $F$ and $r_H$ have hidden voltage dependence, $i.e.$, $F=F(V)$ and $r_H = r_H(V)$.

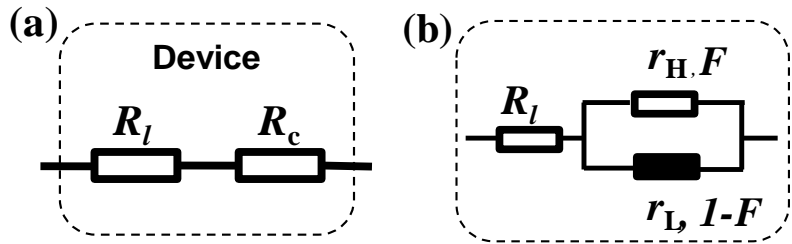

**Figure 7.13.** Equivalent circuits of (a) RRAM device consisting of cell resistor $R_c$ and load resistor $R_l$. (b) cell resistor consisting of high-resistance cross section ($r_H$ per area, area fraction $F$) and low resistance cross section ($r_L$ per area, area fraction 1-$F$).

### 7.3.1   $V_m$ Generation

We use a logarithmic normal distribution with two parameters

$$F(V_m) = 1 - 0.5 \times erfc\left[ -\frac{\ln V_m - \mu}{\sqrt{2}\sigma} \right]$$

$$dF/dV_m = -\frac{1}{\sqrt{2\pi}V_m\sigma} \times erfc\left[ -\frac{(\ln V_m - \mu)^2}{\sqrt{2}\sigma^2} \right]$$

for $V_m>0$. For simplicity, the distribution at $V_m<0$ part is assumed to be the symmetric counterpart of the above distribution. The distribution and its derivative are shown in **Figure 7.14**, which also delineates two voltages at which switching is concentrated.

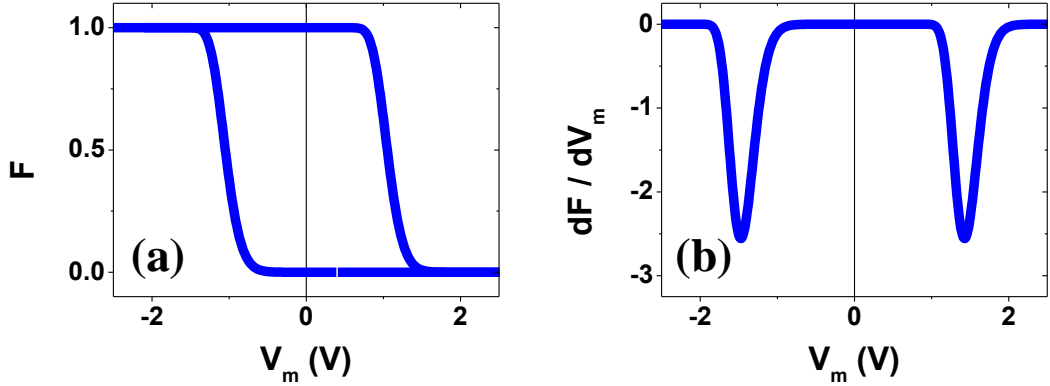

**Figure 7.14.** (a) Distribution function $F(V_m)$ and (b) its derivative $dF/dV_m$. with $\mu=0.05$ ($<V_m>=1.05V$) and $\sigma=0.15$.

### 7.3.2 Simulation of *R-V* Curves

Mo/Si$_3$N$_4$:Cr/Pt nanometallic RRAM is simulated by this model with the following parameters:

$V_c*$(V)=$\pm$(1.2$\pm$0.2), $R_l(\Omega)$=330, $r_L$/A($\Omega$)=90,

$r_H / A(\Omega) = \exp(14.74 - 5.45|V| + 1.56|V|^2 - 0.25|V|^3 + 0.019|V|^4 - 0.00059|V|^5)$,

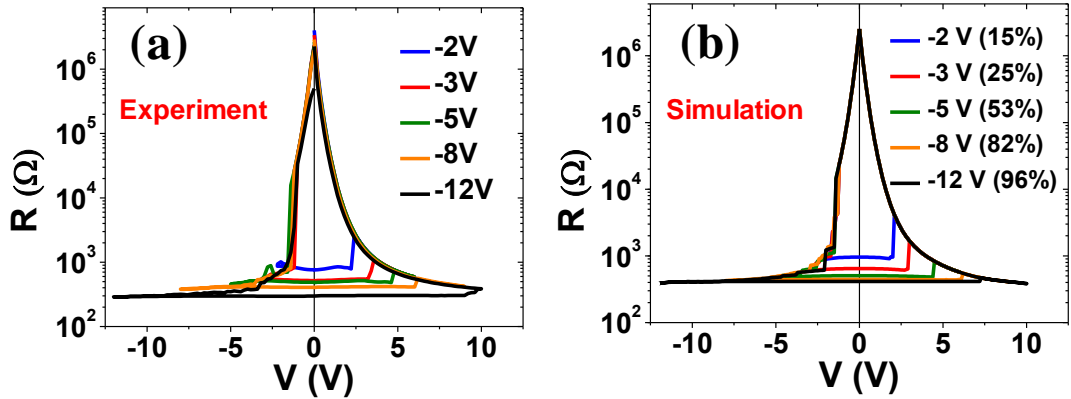

**Figure 7.15.** (a) Experimental *vs.* (b) simulated *R-V* curves for Mo/Si$_3$N$_4$:Cr/Pt RRAM. The percentage denotes the fraction *F*.

Simulated curves shown in **Figure 7.15b** satisfactorily reproduce the experimental data in **Figure 7.15a**. The simulation also identified the fraction (*F*) for each low-resistance plateau, which ranges from 15% to 96% as the negative bias increases from -2V to -12V.

### 7.3.3 Area Dependence

Area dependence was investigated using the continuous model (**Figure 7.16**, parameters given in caption). The results are summarized as follows.

1) *V*$_\text{off}$ *vs. A*: Off-switching voltages are initially around the intrinsic *V*$_\text{m}$. At this stage, for small area the voltage sharing effect is weak since *R*$_\text{LRS}$~$r_L$/A>$R_l$. Above a critical area where *R*$_\text{LRS}$~$r_L$/A~$R_l$, voltage sharing becomes significant such that the apparent off-switching voltage increases.

2) *F*$_\text{on}$ *vs. A*: The on-state fraction *F*$_\text{on}$ in a small device can reach *F*$_\text{on}$~1. However, as the load resistor starts to share voltage, negative feedback prevents complete switch from happening and only $F_{on} \ll 1$ can be achieved during on-switching.

3) $R_{off}$ vs. $A$: $R_{off}$ is inversely proportional to $A$, which obeys Ohm's law as it must.

4) $R_{on}$ vs. $A$: $R_{on}$ is inversely proportional to $A$ for small $A$ when $R_{LRS} \sim r_L/A \gg R_l$. However, as the voltage sharing effect becomes more prominent for larger $A$, the area dependence of LRS is masked by $R_l$. This is further confirmed in a loading effect study in **Figure 7.17**. In fact, as the $r_L/R_l$ ratio increases (*e.g.* vary $r_L$ while fix $R_l$), the onset of area-dependent $R_{on}$ is shifted to a larger $A$. This provides a theoretical explanation for the apparent size-independent on-state resistance in most experiments.

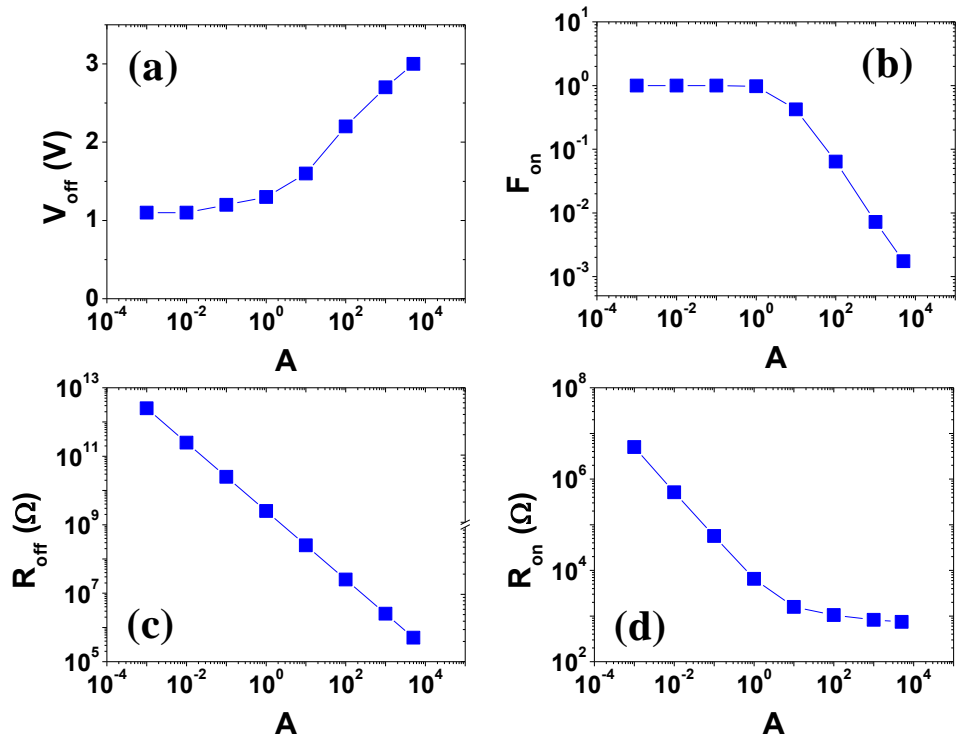

**Figure 7.16.** Area dependence of (a) off-switching voltage $V_{off}$, (b) fraction of LRS $F_{on}$, (c) HRS, (d) LRS. In the simulation, initial states are set as $F=0$ (HRS),

voltage scans are identically performed as 0V→-2V→+5V→0V. $r_L$=5000Ω, $R_l$=300 Ω. $r_H = \exp(14.74 - 5.45\,|\,V\,| + 1.56\,|\,V\,|^2 - 0.25\,|\,V\,|^3 + 0.019\,|\,V\,|^4 - 0.00059\,|\,V\,|^5)$.

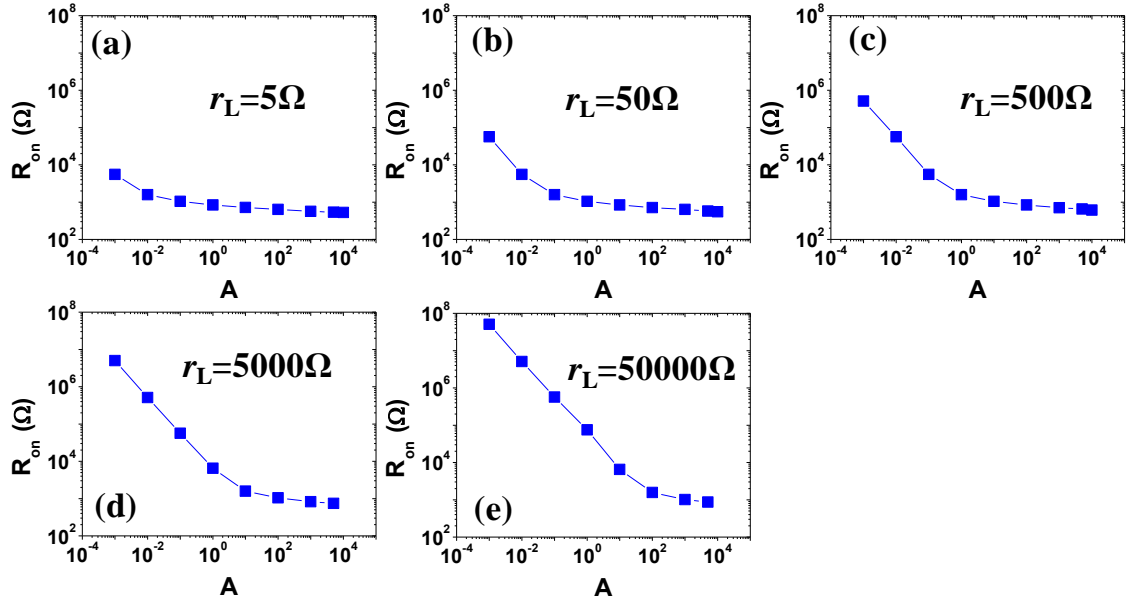

**Figure 7.17.** $R_{on}$ *vs. A* for different load, (a) $r_L$=5Ω, (b) $r_L$=50Ω, (c) $r_L$=500Ω, (d) $r_L$=5000Ω, (d) $r_L$=50000Ω. Other simulation parameters are identical to ones shown in **Figure 7.16** caption.

Motivated by the observation of intermediate states and the load resistance effect above, studies of practical RRAM devices motivated will be conducted in the next few chapters. The results will be interpreted in terms of the circuit model. These devices serve to further verify the model. In addition, they provide additional functionalities that go beyond standard RRAMs.

## 7.4 Conclusions

6. RRAM can be treated as an external load in series with a parallel network of adjustable resistors, the latter representing the film that undergoes resistance switching. Discrete and continuum modeling based on the above equivalent circuit can satisfactorily simulate experimentally measured $I$-$V$ and $R$-$V$ characteristics. This model is applicable to all RRAMs and may easily be integrated into SPICE simulation software.

7. Voltage-sharing between the load and the parallel network introduces asymmetry for on- and off- switching, exerting negative and positive feedback on switching dynamics. Using voltage/current compliance techniques, it is possible to control the voltage across the active film and access a rich population of intermediate states of intermediate resistance.

# Chapter VIII. Multilevel Switching [1]

## 8.1 Introduction

RRAM that exhibit more than 2 resistance states may provide multi-bit storage[2-8], in an analogous way to the multi-level cell (MLC) flash memory. In principle, if each single cell can display $2^N$ distinguished states, then for the same storage capacity the required die area will scale with $1/N$. Stated differently, for the same die area, each $2^N$-state layer of a 2D memory has same storage capacity of $N$ 2-state layers of a 3D memory. Obviously, this will greatly increase the storage density and/or reduce the integration complexity.

In the literature, multilevel states of RRAM are usually revealed by imposing a current compliance[3-4] or through voltage programming[2,7]. The fact that these approaches seem to be applicable to all types of RRAM, irrespective of their underlying switching/conduction mechanisms, suggests a common understanding and control may be possible. However, no such understanding has been provided to-date. Moreover, despite the relatively common observation of multilevel states, there has been no report of two-way switching between all the $2^N$ states, which will be required in order for them to function properly as reprogrammable memory. Here we will demonstrate two-way switching for the $N$=2 case (4 multiple states: 00, 01, 10, 11) using a protocol guided by the circuit model developed in **Chapter VII**.

## 8.2  Experimental Methods

We demonstrate these ideas using nanometallic RRAM, which is based on a hybrid amorphous structure with a distributed electronic energy profiles, thus naturally allowing multiple states[7-10]. Such RRAM exhibits several outstanding properties including excellent uniformity with small variations in switching voltages and resistance values, thus possibly providing highly reproducible multilevel states[10]. Although nanometallic RRAM can be implemented using a large variety of insulator:metal pairing (**Chapter II**), here we focus on $Si_3N_4$:Cr films (10 nm thick) with a Mo bottom electrode and a Pt top electrode (**Figure 8.1a**, left inset). The fabrication methods are identical to the ones introduced in **Chapter II**.

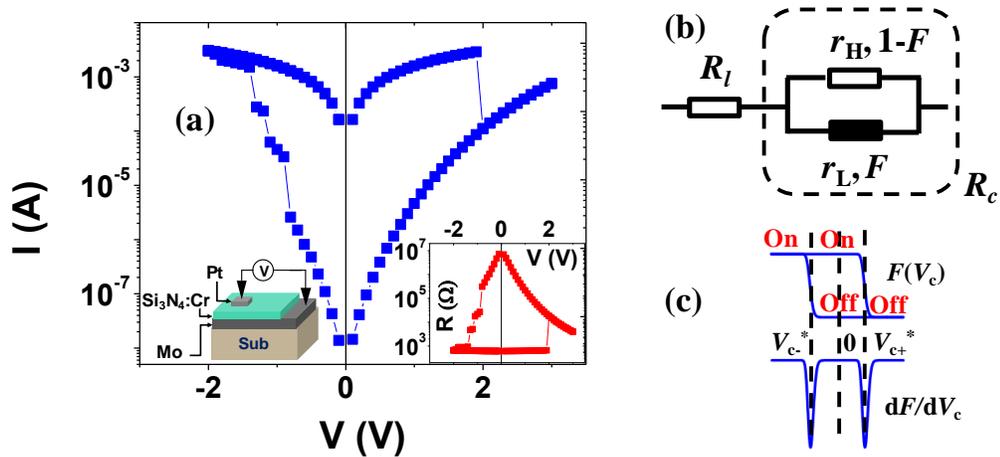

**Figure 8.1.** (a) Characteristic *I-V* curve of nanometallic bipolar RRAM: On switching progresses in multiple steps, off switching displays one step. Left inset: schematic of device. Right inset: *R-V* curve. (b) Equivalent circuit of RRAM device. Cell resistor consists of high-resistance cross

section ($r_H$ per area, area fraction 1-$F$) and low-resistance cross section ($r_L$ per area, area fraction $F$). (c) Schematic $F(V_c)$ and d$F$/d$V_c$ depicting on-switching and off-switching.

## 8.3 Results and Discussion

As-fabricated RRAM devices exhibit bipolar switching behavior as shown in the *I-V* curve in **Figure 8.1a** obtained using the following voltage sweep sequence: 0 V, to 3 V, to -2 V, and to 0 V. Here a positive bias means current flowing from top to bottom as usual. Initially conducting, the device shows a linear *I-V* curve corresponding to a flat resistance in the *R-V* curve (**Figure 8.1a**, right inset). With an increasingly positive voltage, the device is sharply "turned off" at ~2 V. Next, it stays at an insulating, high-resistance state (HRS) which exhibits a nonlinear *I-V* and *R-V* behavior. Under a negative voltage, the HRS passes through several intermediate states before eventually returning to the initial low-resistance state (LRS). From the shape of the *I-V* and *R-V* curves, it is obvious that there is easy access to the intermediate states from the HRS but not from the LRS. Similar problems (with similar *I-V* and *R-V* curves) often exist for other types of RRAM according to the literature[11-13].

As described in our previous model, *I-V/R-V* curves of the above kind can be very satisfactorily and quantitatively modeled by treating the device as a series connection of a load resistance $R_l$ and a cell resistance $R_c$, **Figure 8.1b.** Here, $R_l$ is the sum of all the non-film resistances in the device (electrodes, interface, line and compliance resistance), whereas $R_c$ is the resistance of the film, which has a low-resistance cross section (area fraction=$F$, with a constant resistance= $r_L$ per area) and a high-resistance cross section (area fraction=1-$F$, with a non-linear resistance= $r_H$ per area). As shown in **Figure 8.1c**, on-switching corresponds to the transition from the initial $F$=0 state to various larger $F$ (intermediate) states at increasing $|V_c|$, around a characteristic $|V_c.*|$. During such transition, the cell resistance $R_c$ decreases, which causes $|V_c|$ on the cell to decrease and the voltage on $R_l$ to increase. Therefore, to compensate for the drop in $|V_c|$, the partially switched device will need more external (negative) voltage before continuation of on-switching. This negative feedback during on-switching ensures a gradual *I-V/R-V* curve with many (indeed, infinite) intermediate states. Conversely, off-switching corresponds to the transition from the initial $F$=1 state to a lower-$F$ state when $V_c$ rises past a characteristic $V_{c+}*$. But since any decrease in $F$ leads to an increase in $R_c$, hence a higher $V_c$, there is a positive feedback: it results in a self-propelling transition to HRS, a transition that is completed as soon as it is started.

Because asymmetric feedback, which is rooted in $R_l$, is the reason why intermediate states are not accessible during off-switching, tuning the $R_c/R_l$ ratio should provide a means to adjust feedback to allow access. **Figure 8.2a** shows simulated *R-V* curves for a cell with a log-normally distributed d$F$/d$V_c$ ($V_{c+}*$ ~ 1.05 V at $F$=0.5, $\Delta V_{c+}*$=$\pm$0.23 V at $F$=0.1 and 0.9, respectively.) When simulation is run with an applied voltage increasing at 0.1 V increment, for a device starting at 0 V with a very low resistance (small $R_c/R_l$) corresponding to a large initial $F$, the off-switching is sharp and completed in one voltage increment (see $F$=0.9 and 0.5 in **Figure 8.2a**). This is the case of large positive feedback: during the one-step transition, the voltage spent on the film rises from ~$V_{c+}*$ to ~$V$, the latter well exceeding $V_{c+}*$+$\Delta V_{c+}*$, even though the applied voltage $V$ merely increases by 0.1 V. After the transition, the final state already has $F$=0 (HRS). On the other hand, if the device starts from a higher resistance (a smaller initial $F$, a larger $R_c/R_l$), then the required off-switching voltage is lower (see $F$=0.2 and 0.05 in **Figure 8.2a**) because the cell now shares a higher fraction of the applied voltage. Meanwhile, intermediate states begin to appear on the switching curve. This corresponds to the case of limited feedback: even though $V_c$ also rises from ~$V_{c+}$* to approach $V$ in one voltage increment, it has not exceeded $V_{c+}$*+$\Delta V_{c+}$*, thus not triggered avalanche switching.

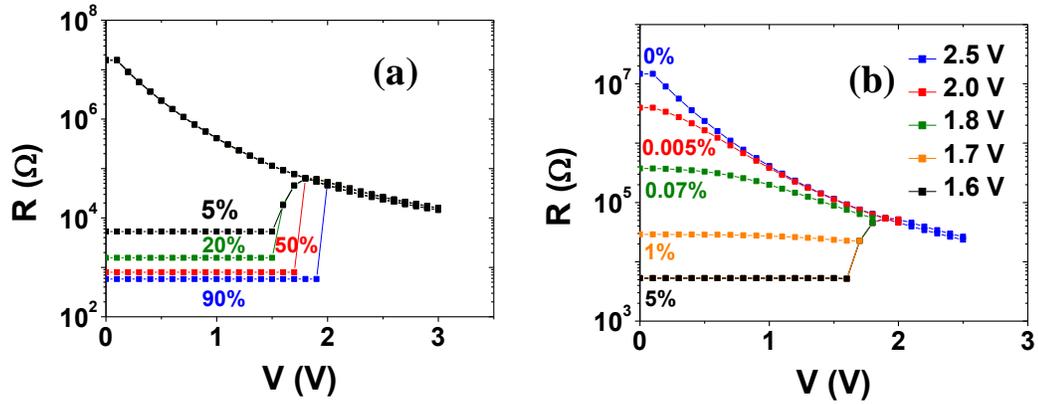

**Figure 8.2.** Simulated $R$-$V$ curves for off-switching using parallel circuit model in **Figure 8.1b**. (a) $R$-$V$ curves starting from different LRS, showing one step switching ($F$=0.9 and 0.5) and multi-step switching ($F$=0.2 and 0.05). (b) $R$-$V$ curves for off-switching from one resistance state ($F$=0.05) to four other resistance states ($F$=1%, 0.07%, 0.005%, 0%) by using different off-switching voltage. Simulation parameters: $V_{c+}^*(V) = 1.05$ (at $F$=0.5), $\Delta V_{c+}^*(V) = \pm 0.23$ (+ at $F$=0.1, − at $F$=0.9), $R_l(\Omega)$=300, $r_L(\Omega)$=250, $r_H(\Omega) =$ exp ( 17.05 − 5.45 × |V| + 1.56 × |V|² − 0.25 × |V|³ + 0.0193 × |V|⁴ − 0.0005913 × |V|⁵ ), where $V$ is voltage in volt.

The simulation also predicted the "unloading" $R$-$V$/$I$-$V$ curves of the intermediate states, **Figure 8.2b**. For a device with an initial $F$=0.05, three intermediate states of $F$=1%, 0.07% and 0.005% are obtained after three successive $V$ increments. Decreasing the applied voltage afterwards causes "unloading" of $V_c$, so these $F$ values for the intermediate states remain unchanged. However, because $r_H$ is non-linear, the "unloading" $R$-$V$ curve is also non-linear, the more so the smaller $F$ (**Figure 8.2b**). At $V$=0, these intermediate states have distinctly different resistance values well separated from each other and from HRS ($F$=0) and the nominal LRS ($F$=0.05). If these memory states can be realized in practice, they should be rather easy to distinguish and to read.

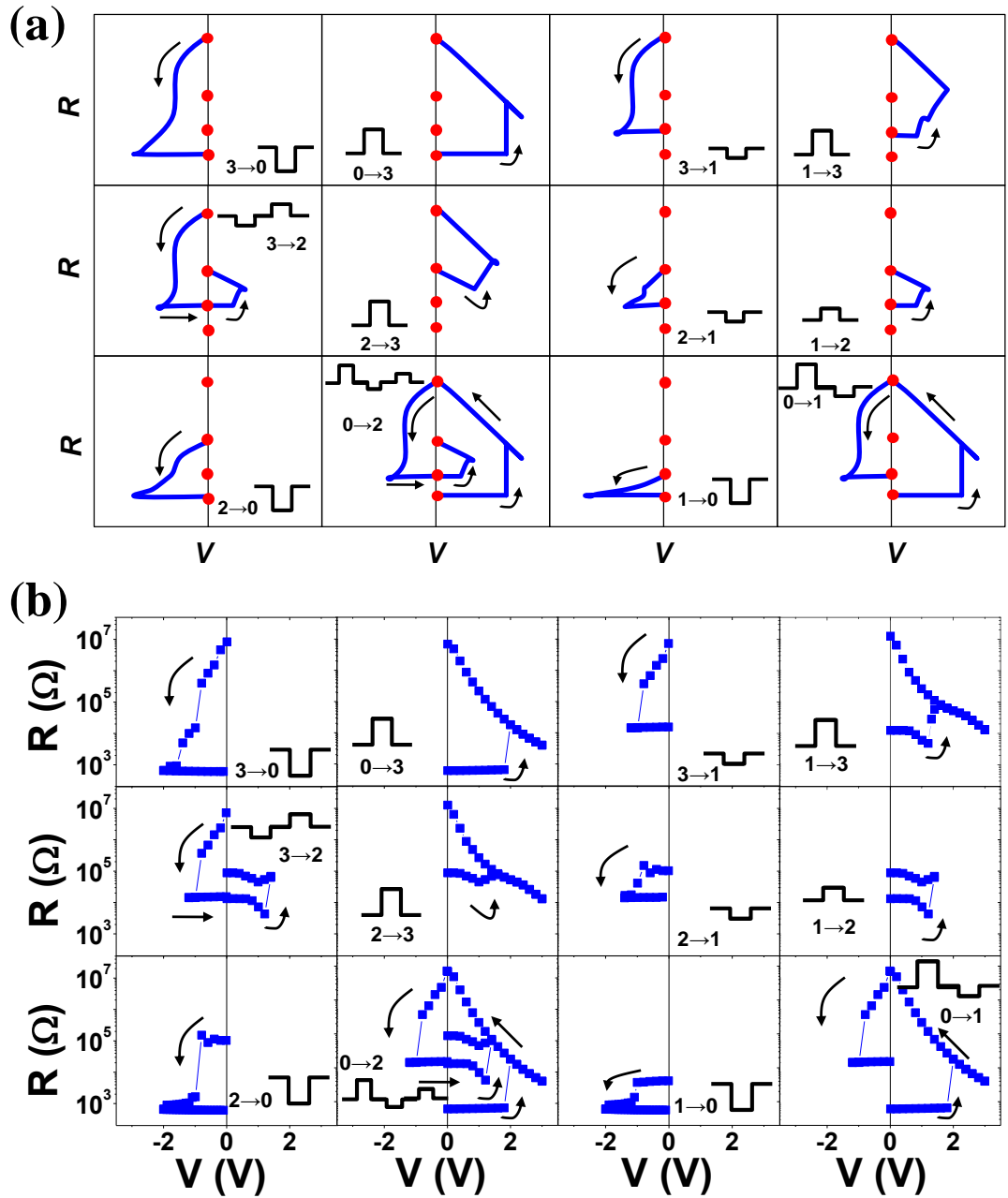

**Figure 8.3.** (a). Schematic *R-V* curves of two-way switching in 2-bit memory between any two resistance states from 0 to 3 (resistance at 0 V in red). Inset: Pulse trains of switching voltage. (b) Experimental

R-V curves verifying (a). Memory cells constructed using nanometallic Mo/Si$_3$N$_4$:Cr/Pt film.

For these states we next illustrate a scheme for two-way switching. Here we use a 2-bit (N=2) memory, having four states that are ranked as 0-3 by their increasing resistance. (Note that this is different from the standard notation of calling the LRS the "1" state.) The schematic switching R-V curves and triggering voltage-pulse trains are shown in **Figure 8.3**. Most (9 out of 12) switches are straightforward requiring only a one-step pulse. However, to access intermediate states (state 1 and 2) from the LRS (state 0), a multi-step pulse with a small negative voltage step is required to raise the cell resistance to limit off-switching avalanche. This is the case of 0→2 and 0→1, in which a detour *via* state 3 is made before applying the negative voltage step. For 3→2 switching, a two-step pulse is illustrated in **Figure 8.3a**, but a one-step pulse for direct transition is also feasible if the device is under a compliance control that prevents 3→1 transition. The above scheme has been experimentally verified in nanometallic RRAM, its data provided in **Figure 8.3b**.

Next, we employ voltage pulses (single pulse width: 100 ns) to implement the above scheme at a realistic write/rewrite speed. The blue curve in **Figure 8.4a** shows that the initial state 0 holds its resistance until $V_{pulse}$>1.8 V, then transitions to state 3. On the other hand, if the device starts from state 1, a 1 V pulse will transition it to state 2 (shown as the green curve), while a 2 V pulse will switch it to state 3. On a negative pulse, the state 3 can be reset back to either state 1 (with a -1 V pulse) or state 0 (with a -2 V pulse).

These transitions are the essential ones that ensure the success implementation of the scheme in **Figure 8.3**, but other switches had all been verified using one-step or multi-step pulses. These states were stable: they maintained their resistance values without roll off during retention tests lasting over $10^5$ s. As shown in **Figure 8.4b**, states 0, 1, and 3 can all hold constant resistance. State 2 does show some resistance scatter, which may indicate exchanges between similar intermediate resistance states when subject to small perturbations. However, the scatter is small and will not affect distinguishing state 2 from neighboring state 1 and 3.

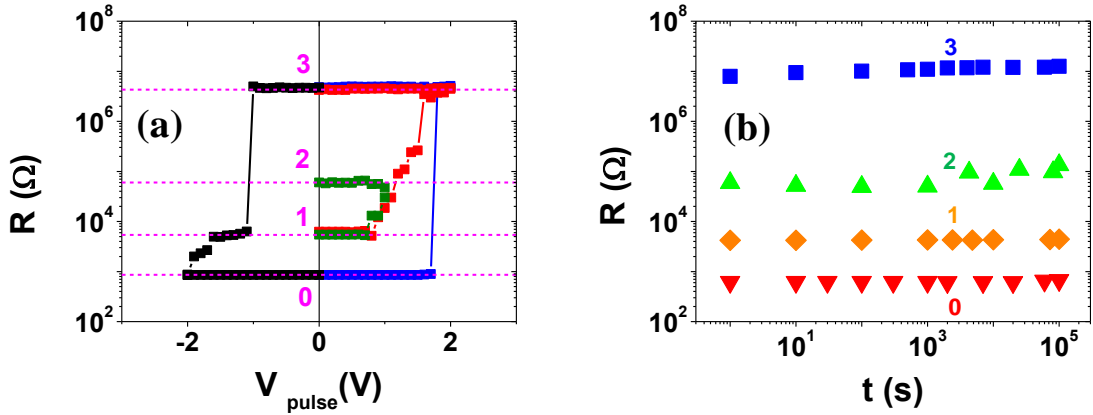

**Figure 8.4.** (a) Resistance-pulse-voltage traces (pulse width = 100 ns) used to define four resistance states 0-3. After each voltage pulse, resistance is read at 0.2 V. (b) Resistance retention test (read at 0.2V) for four states, each maintaining starting resistance over tested $10^5$ s without roll-off.

The two-bit memory above should not suffer from long $RC$ time or slow switching speed. Concerning the $RC$ time, we refer to **Figure 8.1b** and envision a cell capacitance $C_c$ in parallel to $R_c$. It is then trivial to show that the $RC$ time of the device is $R_l C_c \left( \frac{R_c}{R_c + R_l} \right)$, which is bounded by $R_l C_c$ and essentially independent of the bit resistance. Typical values measured for our RRAM are $R_l \sim 1$ k$\Omega$ and $C_c \sim 100$ pF for a $100 \times 100$ μm$^2$ cell, which gives $R_l C_c \sim 10^{-7}$ s. Since $R_l$ in our device is mainly due to spreading resistance and is relatively area-independent but the capacitance scales linearly with the area, the projected $R_l C_c$ is $\sim 10^{-13}$ s for a $100 \times 100$ nm$^2$ cell, which is more than satisfactory. Using the state-of-the-art CMOS technology, for which the typical sheet resistance for metal conductor layers is 0.05 $\Omega$/sq, we also estimate for a 10 Gbit storage unit ($10^5 \times 10^5$) a line resistance of the order of 0.05 $\Omega$/sq $\times 10^5$ sq or 5 k$\Omega$, which is 5× the value for $R_l$ above. Thus, the delay time (5× longer) is still extremely short for a $100 \times 100$ nm$^2$ cell. In addition, since nanometallic films can easily exhibit intermediate states with $R_{intermediate} \gg R_l$, $R_l$ of the above type would not affect the readability/detectability of intermediate states.

Concerning the switching time, we refer to **Figure 8.3a** to compute an average for a 2-bit memory array, assuming all 4 states are equally populated and all 12 transitions are equally executed. Since 9 transitions need 1-step pulses, 2 transition need 2-step pulses and 1 transition ($0 \rightarrow 2$) needs a 3-step pulse, the average switching time of this 2-bit-4-state memory is 1.33 time that of a 1-bit-2-state memory. Therefore, the tradeoff between a higher storage density (2×) and a slower writing speed seems favorable.

We now return to the constitutive basis of multistate memory. Referring to **Figure 8.1c**, we see that multiple state transition is the result of the gradual inter-conversion curve ($F(V_c)$) between $r_H$ and $r_L$ elements. In nanometallic memory, the $r_L$ state is the metallic state of a random conductor, and the $r_H$ state is the insulating state in which charge-filled negative-$U$ centers have blocked electron passage in their vicinity. Trapping and detrapping are thus responsible for the $r_L \leftrightarrow r_H$ conversion, which is voltage-driven governed by the energy landscape in the random material. Since the landscape in such material is inherently diverse, by nature this conversion must be energetically dispersive and can be triggered by a range of voltages as schematically illustrated in **Figure 8.1c**. However, although a multitude of intermediate states naturally exist, they may be masked by voltage overload because of the positive feedback induced by the load resistance $R_l$. This can lead to a cell-voltage overshoot above the critical voltage $\Delta V_{c+}*$ and even beyond the energy dispersion ($\Delta V_{c+}*$), which then creates a switching avalanche bypassing all intermediate states. The condition for this to occur can be obtained from the following simple analysis. (i) Because of voltage sharing, off-switching cannot be initiated until $V = \frac{R_l + R_c}{R_c} V_{c+}^*$. (ii) Once initiated, with a positive feedback, the entire applied voltage is soon spent essentially on the cell, giving $V_c \sim V$. (iii) If $V_c > V_{c+}* + \Delta V_{c+}*$, then overshoot will occur: transition will complete as soon as it is initiated. Combining (i-iii), we obtain $R_l/R_c > \Delta V_{c+}*/V_{c+}*$ as the criterion for switching avalanche. In **Figure 8.2**, which uses $R_l$=300 Ω and $\Delta V_{c+}*/V_{c+}*$=0.22, the condition separating sharp and continuous switching should be $R_c$= 1363 Ω. Indeed, continuous switching in **Figure 8.2a** begins with $F$=0.2, corresponding to $R_c$=1250 Ω.

Our circuit model can be easily used to explain the switching $I$-$V$/$R$-$V$ behavior of other RRAM irrespective of the underlying switching mechanisms. Multilevel states in other RRAM have been reported, and are typically accessed by voltage programming or current compliance. Although it may seem reasonable to attempt more storage bits by expanding the resistance range (e.g., lowering the LRS by using a larger on-switching voltage) or increasing the current compliance through a larger $R_l$, the above analysis points to the shortcomings of these approaches: a large $R_l/R_c$ is inherently unstable for off-switching, and a large $R_l$ will increase the $RC$ time. On the other hand, RRAM with a negligible $R_l$ has some advantage: for example, in **Figure 8.3** state 0 can directly switch to state 2 if no positive feedback is provided. Systems with a highly dispersive $dF/dV_c$, i.e., a large $\Delta V_{c+}*/V_{c+}*$, are obviously desirable from this perspective.

Lastly, we address two practical issues in implementing the current scheme. The first issue/concern for multilevel RRAMs is about their large $R$ contrast, which could cover several orders of magnitude making it difficult to differentiate different states without using a complicated sensing circuit. Such concern is particularly valid for conventional (filamentary or/and ionic) multilevel RRAM, in which it is very difficult to tune the resistance values by fine-tuning the device composition and/or configuration. However, this is not the case in nanometallic RRAM. Indeed, a competitive advantage of nanometallic RRAM over conventional RRAM is its ability to tune the resistance values through either thickness (HR resistance increasing with thickness following an exponential dependence) or metal concentration (HR resistance decreasing with concentration spanning several orders of magnitude)[8-10]. Therefore, it is entirely feasible to adjust the HR resistance to "squeeze" all the resistance states into a certain range so that they are all readable by the standard sensing circuit. The other concern is the complexity demanded on the drive circuit to generate the multi-impulse pulse trains in

**Figure 8.3**. However, the complexity level of our pulse trains is fundamentally the same as that used in writing multilevel NAND memory. Moreover, since only one drive circuit is needed for each memory array, the complexity will not increase with storage size and therefore not significantly affect the space/cost saving consideration for using multi-bit cells.

## 8.4  Conclusions

We have demonstrated and analyzed a stable 2-bit-4-state nanometallic memory which can be read, written and rewritten using voltage pulses. Even with only 2 bits, such storage memory already enjoys an advantage ($2\times$) in space/area saving at a modest increase ($1.33\times$) of average programming time. These results are applicable to other RRAM systems, and further advances in developing multi-bit memory may accelerate the adoption of highly integrated RRAM in future generations of digital memory.

# Chapter IX. Dynamic-Load-Enabled Ultra-low Power Multiple-State RRAM Devices[1]

## 9.1 Introduction

Power consumption is a key issue for electron devices including resistive random access memory (RRAM), which has attributes of high density, fast write/read speed, fatigue endurance and long retention[2]. In a RRAM, on-switching (also called set-switching) consumes relatively little power because the current is limited by the relatively high (off) resistance. So the power consumption is dictated by off-switching (also called reset-switching) which has a relatively low (on) resistance. Off-switching power should be proportional to the area of the resistance cell if the voltage/current density required to trigger switching is independent of the area. Indeed, literature data of off-switching power of some 20 RRAMs shown in **Figure 9.1** (also see **Table 9.1** for details) support such a "scaling law": they vary from the mW range for micrometer-sized devices to the μW range for nanometer-sized devices[3-23]. Recognizing such a trend, our goal here is to systematically seek scalable strategies to further lower the power for RRAM off-switching. Our power data and the scaling prediction are summarized in **Table 9.1**.

| Off-switching power | Materials (size) |
|---|---|
| >10 mW | $TiO_2$ (300 μm)[3], $Sr_2TiO_4$ (~100 μm)[4], polymer:Au[5], NiO (120 μm)[6], $Al_2O_3$ (80 μm)[7], LSMO (800 μm)[8] |
| 1-10 mW | $SiO_2$:Pt (80 μm)[9], Fe:STO[10], $CoSiO_x$ (8 μm)[11], $Cu_xO$ (300 μm)[12], PCMO (200 μm)[13], V:$SrZrO_3$ (220 μm)[14], Nb:$SrTiO_3$ (100μm)[15], ZnO (50 μm)[16] |
| 100-1000 μW | $ZrO_2$ (6 μm)[17], $HfO_x$ (<60μm)[18], $Ge_2Sb_2Te_5$[19] |
| 10-100 μW | $TaO_x$ (30 nm)[20], LSMO (AFM Tip)[21], $TiO_2$ (50 nm)[22], a-Si (50 nm)[23] |

**Table 9.1.** Off-switching power consumption in reported ReRAM systems (Off-switching power is defined as product of *I* and *V* at the onset of off-switching.)

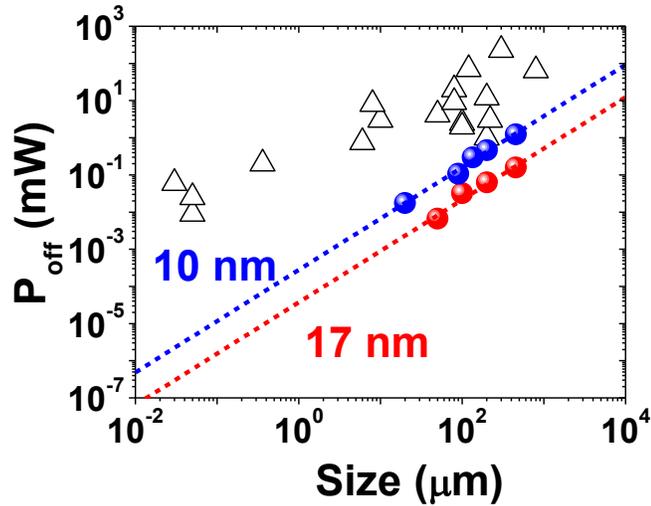

**Figure 9.1.** Scaling behavior of off-switching power consumption in literature (triangles) and in this work (filled circles) using asymmetric load for devices of two thickness, 10 nm (blue) and 17 nm (red). Extrapolation (dash line) gives 12 nW and 500 pW for $100 \times 100$ nm$^2$ device and $10 \times 10$ nm$^2$ device, respectively (10 nm thick), and 1.5 nW and 60 pW for their 17 nm counterparts. See **Table 9.1** for details of literature data. All power data are calculated from $P_{off} = V_{off}^2 / R_{on}$, where $V_{off}$ is off-switching voltage and $R_{on}$ is off-switching resistance at $V_{off}$.

We begin by treating a RRAM device as a serial connection of a cell resistance $R_c$ and a load resistance $R_l$, see **Figure 9.2a**. The latter may come from word/bit lines, electrodes and interfaces. Depending on the configuration $R_l$ may or may not be inversely proportional to the cell size or area. (For example, the spreading resistance of a very thin bottom-electrode substrate is logarithmically dependent on the reciprocal cell size.) We next designate the critical cell-switching voltage $V_c*$ as a characteristic of cell material. The intrinsic switching power is thus $P_c^* = V_c^{*2} / R_c$ per cell. However, since the device-switching voltage $V*$ equals $(1+\Delta)V_c*$, where $\Delta = R_l / R_c$, the device-switching power $P$ must exceed $P_c^*$. Indeed, $P = (1+\Delta)P_c^*$.

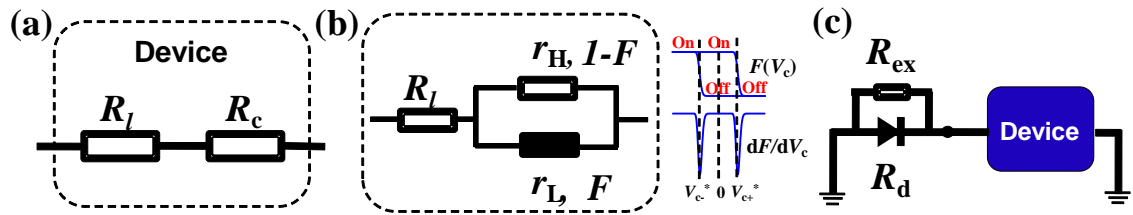

**Figure 9.2.** Equivalent circuits of (a) RRAM device consisting of cell resistor $R_c$ and load resistor $R_l$, (b) cell resistor consisting of low-resistance cross section ($r_L$ per area, area fraction $F$) and high resistance cross section ($r_H$ per area, area fraction 1-$F$), (c) dynamic load consisting of parallel diode $R_d$ and external resistor $R_{ex}$. Inset of (b): schematic $F(V_c)$ and d$F$/d$V_c$ depicting on-switching and off-switching.

The above consideration suggests that power minimization requires maximizing $R_c$ and minimizing $R_l$. However, $R_c$ and $R_l$ are interrelated in many RRAM that contains multiple intermediate states: if a larger $R_l$ is used during on-switching, it provides a current compliance often causing switching to an intermediate state of a higher (intermediate) $R_c$ than would otherwise. (Compliance control by the source meter is widely used for this purpose.[24]) The interplay between $R_l$ and $R_c$ for multi-state RRAM can be understood by viewing the cell as a parallel connection of a low-resistance cross section ($r_L$ per area, area fraction=$F$) and a high-resistance cross section ($r_H$ per area, area fraction=1-$F$), see **Figure 9.2b**. In this picture, $F$ is a state variable that characterizes the cell: it lies between 0 (the most resistive state) and 1 (the most conducting state). As schematically shown in the inset of **Figure 9.2b**, on-switching corresponds to the transition from the $F$=0 state to the $F$=1 state, and off-switching the transition from the $F$=1 state to the $F$=0 state. However, various intermediate $F$ states can also result during on-switching because the transition path is $R_l$ dependent: the larger the $R_l$, the smaller the current in the cell, thus the smaller the $F$ of the intermediate state, and the higher the $R_c$. We believe that this $R_l$-$F$ relationship can be exploited in many RRAM to lower $P$ further by including an asymmetric dynamic load: the load is large during on-switching to minimize $F$ hence maximize $R_c$, but small during off-switching to minimize $\Delta$. In the following, we demonstrate this design using a new nitride-based nanometallic RRAM made of amorphous $Si_3N_4$ with atomically dispersed Cr.

## 9.2  Experimental Methods

Fabrication procedure are identical to the ones used in **Chapter II**. Electric tests were conducted using the following convention: a positive polarity is defined by having electric current flowing from the top electrode to the bottom electrode. Testing cycles follow a pre-described loop starting from 0 V to a negative voltage limit, to a positive voltage limit, to the negative voltage limit again, and back to 0 V. To determine the state without the test cycle, the resistance was read at 0.2 V.

The *R-V* hysteresis under a constant $R_l$ was calculated using the parallel circuit model which gives $R_c^{-1} = (F/r_L + (1-F)/r_H)A$. Since the resistance ratio is $\Delta = R_l/R_c = (F/r_L + (1-F)/r_H)AR_l$, we can immediately obtain $V=V_c(1+\Delta)$ for any point on the $F(V_c)$ transformation curve in **Figure 9.2b inset**. (Clearly, $AR_l$ emerges as the control parameter for the switching behavior.) Specifically, we start at an initial point (the initial state of the cell) on one branch of the $F(V_c)$ curve, we then follow the curve and continuously record $F$ and compute $R_c$, $\Delta$ and $V$ until the limit for $V$ (specified by the range of the voltage cycle) is reached. We then reverse the direction and follow the other branch of the $F(V_c)$ curve until the other limit for $V$ is reached before reversing again. During off-switching at larger $AR_l$ under a positive $V_c$, it is possible to encounter a region of negative slope, $dV/dV_c<0$, which implies a jump in $V_c$ without changing $V$. This corresponds to an abrupt drop in $F$, hence a first order transition in the resistance state.

## 9.4   Results

### 9.4.1   Bipolar Switching Involving Intermediate States

As-fabricated devices (without forming) were nearly Ohmic-conducting. Under a voltage sweep, they exhibited bipolar switching behavior as shown in the *I-V* curve in **Figure 9.3a** obtained using the following voltage sweep: 0 V, to −12 V, to 10 V, to −12 V, and to 0 V. The initial sweep from 0 V to −12 V does not result in any sharp transition. Positive-voltage off-switching occurs at 8 V consuming ~250 mW, after that a non-Ohmic high resistance state (HRS), corresponding to $F=0$, is reached. The HRS returns to the low resistance at −1 V, consuming ~30 μW during on-switching. In the above, power to operate the device was equated to the product of the applied voltage and current at the onset of switching. (This convention will be used throughout our work.) As shown in the inset of **Figure 9.3a**, from −12 V to 8 V the resistance ($R_c$+$R_l$) is flat, which will be referred to as plateau resistance to indicate no transition. This plateau-resistance state is actually one of the intermediate states with $F$ <1. Note that the off-switching voltage in **Figure 9.3a** is relatively high signifying a relatively large $\Delta$. This is caused by the very small $R_c$ and large $F$, which was made possible according to **Figure 9.3c** by using a very large negative voltage limit (−12 V) during the negative sweep. Such high $\Delta$ and low $R_c$ in turn raise $P$.

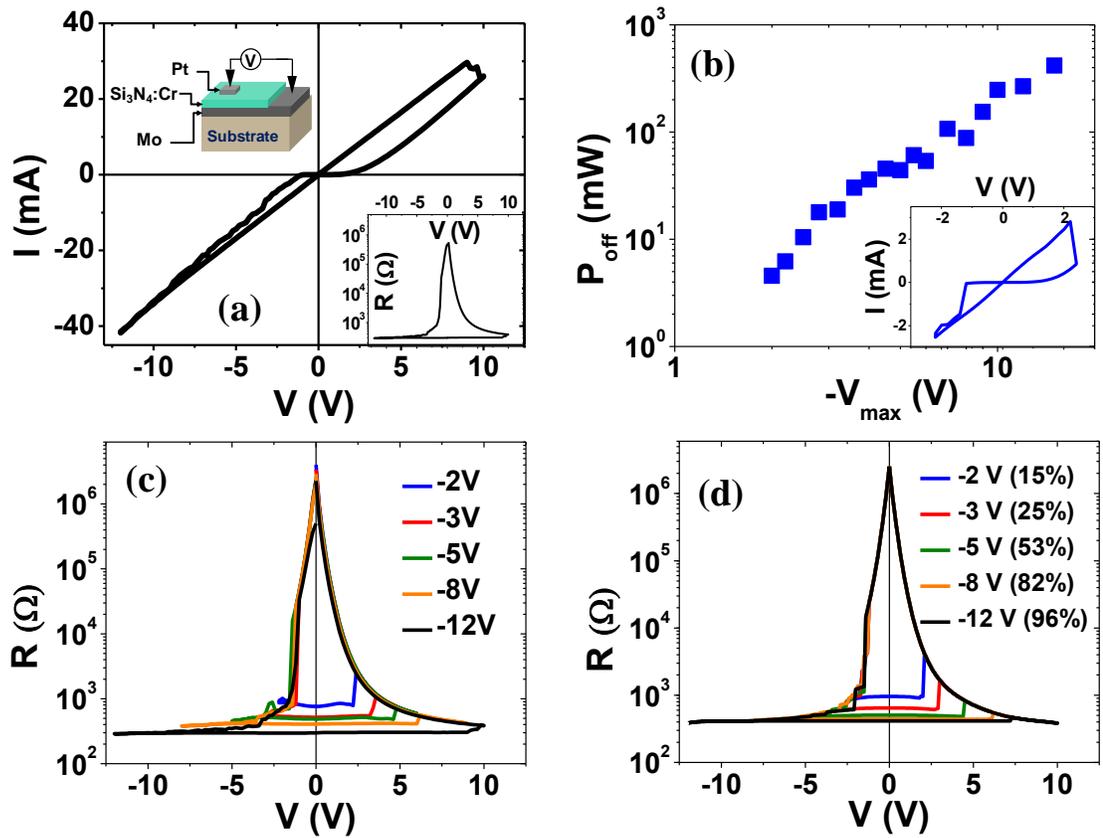

**Figure 9.3** (a) Characteristic *I-V* curve of nanometallic bipolar RRAM: on-switching under negative voltage, off-switching under positive voltage. On-switching progresses in multiple steps, suggesting possibility of multi-bit memory. Cell size: $100 \times 100\ \mu m^2$. Upper left inset: schematic of device: Lower right inset: *R-V* curve. (b) Off-switching power *vs.* negative voltage limit, $-V_{max}$, showing $\sim 60\times$ power reduction as $-V_{max}$ decreases from 12 V to 2 V. Inset: *I-V* curve for $V_{max}=-2$ V. (c) *R-V* curves for various negative voltagelimits from $-12$ V to $-2$ V. Plateau resistance increases as negative voltagelimitreduces, causing off-switching voltage to decrease. (d) Simulated *R-V* curves under different $-V_{max}$ using parallel circuit model in **Figure 9.2b**. Percentage in the bracket shows different $F$ at plateau resistance. Simulation parameters: $V_c{}^*(\text{V})=\pm(1.2\pm0.2)$, $R_l(\Omega)=330$, $r_L/\text{A}(\Omega)=90$,

$$r_H/\text{A}(\Omega)=\exp(14.74-5.45\,|\,V\,|+1.56\,|\,V\,|^2-0.25\,|\,V\,|^3+0.019\,|\,V\,|^4-0.00059\,|\,V\,|^5)$$

, where $V$ is voltage in volt.

The plateau resistance can be substantially increased without altering $R_l$. As shown in **Figure 9.3c**, a progressively smaller voltage for the negative sweep leaves a progressively higher plateau resistance, corresponding to a progressively smaller $F$ value for the intermediate state. As the range of the sweep voltage decreases from −12 V to −2 V, the plateau resistance increases from 300 Ω to 1 kΩ. Correspondingly, the off-switching voltage decreases from 8 V to 2 V. Meanwhile, the power decreases (**Figure 9.3b**) from 250 mW to 4 mW. The $I$-$V$ curve for the 4 mW case is shown in **Figure 9.3b inset** to illustrate the ∼ 10× reduction in current compared to **Figure 9.3a**. This behavior can be understood in terms of the parallel circuit model (**Figure 9.2b**). As shown in **Figure 9.3d**, all the $R$-$V$ curves can be satisfactorily reproduced by numerical calculation using the model taking $R_l$∼300 Ω and $V_c{}^*$∼ ±1 V. This also allows us to identify the $F$ value for each plateau resistance, which ranges from 0.96 to 0.15 as marked in **Figure 9.3d**.

### 9.4.2 Asymmetric Load

We next demonstrate that the plateau resistance can be increased and the off-switching power drastically reduced by introducing a dynamic load that has an asymmetric response to voltage. This was achieved using a diode in parallel with another external load $R_{ex}$, as schematically shown in **Figure 9.2c.** Under a positive bias which includes off-switching, the diode is in the forward direction ($R_d\sim0$) allowing $R_{ex}$ to be short-circuited. So the net load is nearly $R_l$, still equal to 300 $\Omega$. Under a negative bias which includes on-switching, the diode is in the reverse direction and is almost open-circuited. Thus the net load is the sum of 300 $\Omega$ and $R_{ex}$. In reality, under a positive bias, a typical diode also introduces a positive voltage drift due to its threshold voltage $V_{th}$. ($V_{th}$ can be as low as 0.2 V in a Schottky diode, but is 0.6 V in the silicon diode used in the experiment described here.) In addition, the diode in the reverse direction has a characteristic resistance $R_d$ which is 100 M$\Omega$ in our experiment. As shown in **Figure 9.4a** for a 100×100 μm$^2$ device, such a diode results in a switching curve with ~10× reduction in the on-switching current and ~10× increase in plateau resistance (**Figure 9.4a inset**). Under a positive bias, current increase starts near $V_{th}\sim0.6$ V and off-switching occurs at $V^*\sim1.4$V, corresponding to a maximum in current and a minimum in resistance (being $R_c$+300 $\Omega$). For this (on) resistance and $V^*$, the off-switching $P$ is 0.25 mW.

We systematically examined whether off-switching $P$ can be further reduced by varying $R_{ex}$. With $R_{ex}$ increasing from 10 $\Omega$ to 100 k$\Omega$, **Figure 9.4b** (the 100 μm curve) shows $P$ to decrease from 2.5 mW to ~110 μW for the same 100×100 μm$^2$ device. The decrease essentially begins when $R_{ex}$ is comparable to $R_l$, which was again 300 $\Omega$ in this experiment. Beyond this point the asymmetric load starts to cause the plateau resistance and the on-resistance ($R_c$+$R_l$) to increase (the 100 μm branch in **Figure 9.4c**) by arresting the intermediate state at a progressively higher $R_c$. Meanwhile, the off-switching voltage $V^*$ also decreases (**Figure 9.4c**) signifying transition initiating at the low-voltage tail of the $V_c^*$ distribution (**Figure 9.2b inset**), resulting in an abrupt off-transition to some intermediate $F$ state. (Later, full transition to the $F=0$ state occurs during the remainder of the positive voltage sweep, but such subsequent transition consumes much less power because of the much higher $R_c$.) Eventually, $P$ reaches a lower limit when $R_{ex}$ becomes comparable to the resistance of the HRS; any further increase of $R_{ex}$ will postpone on-switching to an impractically large (negative) voltage, again because $V^*$ is much larger than $V_c^*$ when $\Delta$ is too high. This limits the minimum $P$ for a $100 \times 100$ $\mu m^2$ device to ~110 $\mu W$. For all the $R_{ex}$, a large on-off ratio of resistance (read at 0.2 V) exceeding $10\times$ is maintained as shown in **Figure 9.4b inset**.

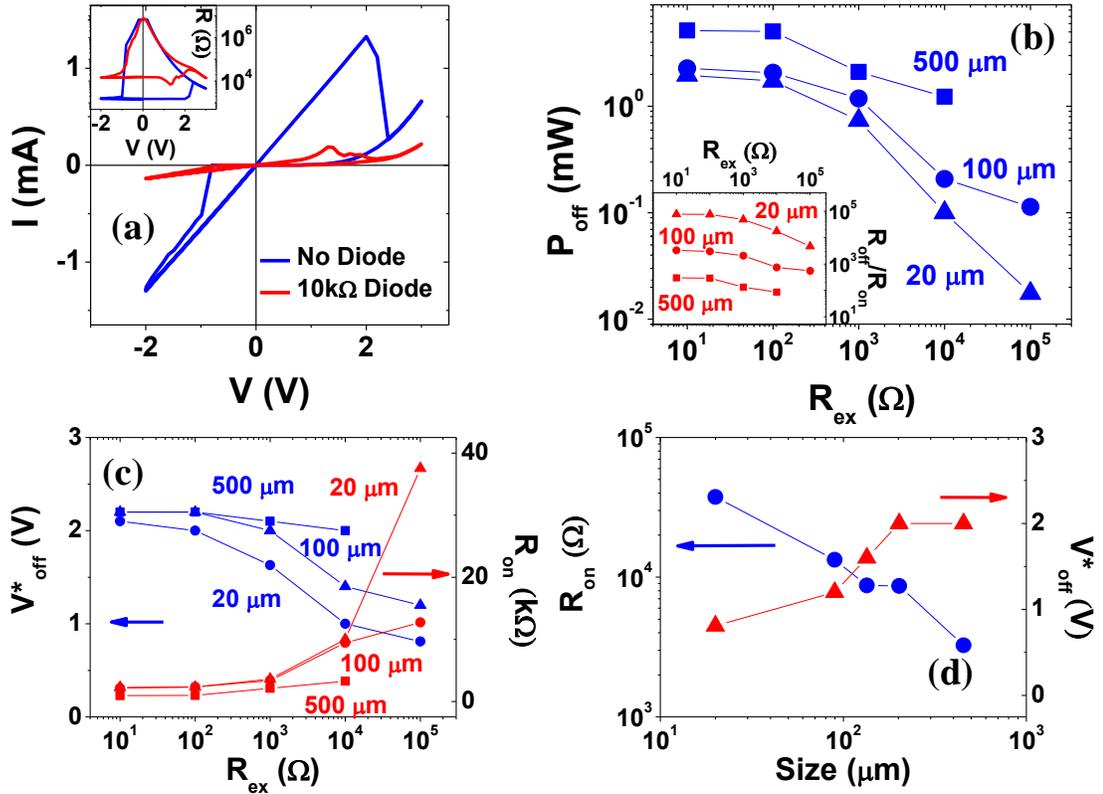

**Figure 9.4** (a) *I-V* curves for RRAM device with and without asymmetric load, which reduces current and off-switching voltage. Inset: corresponding *R-V* curves. Cell size: 100×100 μm². Off-switching (b) power $P_{off}$ and (c) voltage $V_{off}*$ and on-resistance $R_{on}$ *vs.* $R_{ex}$ for three cells of different sizes, showing systematic size-dependent $P_{off}$ and $V_{off}$ decreases and $R_{on}$ increases. Inset of (b): $R_{off}/R_{on}$ *vs* $R_{ex}$ for cells of different sizes. (d) Scaling behavior of $R_{on}$ and $V_{off}*$. See **Figure 9.1** for scaling behavior of $P_{off}$.

### 9.4.3 Scalability

The above approach is scalable. This is illustrated in **Figure 9.4b-c** for two other cells ~25× larger/smaller in cell area. Here, we used the same diode but extended the range of $R_{ex}$. They depict a systematic shift of off-switching $P$ (**Figure 9.4b**), $V*$ and on-resistance (**Figure 9.4c**) with cell size: in the smaller cells the significant reduction in $P$ and $V*$, along with the significant increase of on-resistance, starts at a higher $R_{ex}$ because the cell resistance of smaller cells is higher. Since the same trend is obeyed for all cell areas, we may assign the limiting $P$, $V*$ and plateau resistance values as the ones obtained at the highest $R_{ex}$ before on-switching becomes impractical. These assigned values follow an apparent scaling "law" in **Figure 9.4d** for $V*$ and plateau resistance and in **Figure 9.1** for $P$. (Data of two additional cells of intermediate areas have also been included in these plots.) Although the data on the scaling plot **Figure 9.1** are somewhat scattered because only a few $R_{ex}$ were used, we have tentatively extrapolated the scaling line to smaller area. For a $100 \times 100$ nm$^2$ device, which is readily manufacturable today, the projected off-switching power is 12 nW. For a $10 \times 10$ nm$^2$ device, the projected power is 500 pW.

While the validity of the above projection is not known until future experimental verification, we can nevertheless examine the basis for the projection to identify any potential causes for its breakdown. Our power data in **Figure 9.1** apparently follow a cell area ($A$) scaling behavior of $A^{0.7}$. Since the off-switching voltage is only very weakly dependent on $A$, most of the above scaling may be attributed to on-resistance. As mentioned above, the maximum $R_{ex}$ usable is determined by the resistance of HRS. This, in turn, determines $R_c$ and on-resistance (the relation between $R_c$, $R_{ex}$ and HRS is non-linear because of the non-linearity in the $V_c*$ distribution in **Figure 9.2b**.) In the insulating state, HRS should scale with $A^{-1}$ (ref.[9]). So the slightly weaker area dependence of power and plateau resistance is understandable in terms of mostly HRS and partly the interplay between the diode, $R_{ex}$, and the $F$-transition curve (**Figure 9.2b inset**). It also follows that whether such scaling behavior can continue at small cell areas depends on whether (a) HRS continues to scale with $A^{-1}$ and (b) there is a large spectrum of intermediate states of wide ranging resistance between HRS and $R_c+R_l$ to interact with the dynamic load. Data for $A^{-1}$ scaling of HRS are presented in previous chapters, which seem quite robust. In the following, we examine (b) with the aid of modeling.

To clarify (b), we simulated the $R$-$V$ hysteresis under a constant $R_l$ for $A$ spanning over 4 orders of magnitude using the model in **Figure 9.2b**. As shown in **Figure 9.5**, as the area decreases, the $R$-$V$ curves develop an expanding gap (to about 7 orders of magnitude) between the plateau resistance and the HRS resistance, the latter indeed scales with $A^{-1}$. Meanwhile, off-switching continues to occur between 1 V and 2 V even though the transition is no longer abrupt at small $AR_l$. (The abrupt transition is due to the negative slope $dV/dV_c$, which becomes positive definite at small $AR_l$.) While the detailed outcome of the simulated results (e.g., the $F$ value of the plateau state) obviously depends on the parameter used, such as $R_l$ which we assumed to be area-independent, these findings do suggest that item (b) should not be a concern, thus lending support to our scaling hypothesis under a dynamic load. Moreover, since the plateau resistance does increase at smaller $A$, meaning that $R_c>R_l$ in such case, for a sufficiently small $A$ there will be less need for compliance control rendered by $R_{ex}$. As a result, $R_{on}$ should increase less rapidly at small $A$ than indicated by **Figure 9.4d**.

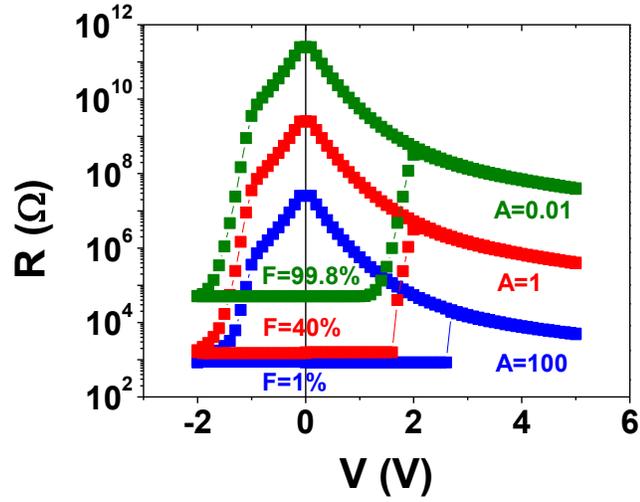

**Figure 9.5.** Simulated *R-V* curves for different cell area *A* using parallel circuit model in **Figure 9.2(b)**. Percentage in the bracket shows *F* at plateau resistance. Simulation parameters: $V_c^*(V)=\pm(1.35\pm0.15)$, $R_l(\Omega)=300$, $r_L(\Omega)=500$,

$$r_H(\Omega)=\exp(21.65-5.45\,|V|+1.56\,|V|^2-0.25\,|V|^3+0.019\,|V|^4-0.00059\,|V|^5)\,, \text{ } V \text{ is voltage.}$$

### 9.4.4 Increasing the Dynamic Range Using Nanometallic Feature

Although the approach of employing an asymmetric dynamic load to reduce *P* was demonstrated above using a nanometallic RRAM, it is applicable to other bipolar RRAM that satisfies two requirements: (i) intermediate states are accessible using compliance control, and (ii) switching is triggered by a critical cell voltage independent of cell area. Nevertheless, nanometallic RRAM does have two important advantages. First, since it switches by a purely electronic mechanism, fast switching speed should be possible (<50 ns as already measured in our laboratory, much faster also likely), assuring a very small energy for switching per bit. Second, reflecting the elastic tunneling nature of itinerant electrons in random materials, the HRS of nanometallic thin films follows a unique exponential dependence on thickness, $R_c \sim \exp(\delta/\zeta_{HR})$, where $\delta$ is the thickness and $\zeta_{HR}$ (of the order of a few nm) is the localization length in the HRS[9]. ($\zeta_{HR}$ essentially defines the spatial extent of electron's wave-function, which decays exponentially in a random material.) Meanwhile, $V_c^*$ is thickness independent in the nanometallic regime[9,25]. These unique attributes allow additional freedom to increase $R_c$ and decrease $P$ by many orders of magnitude by using thicker films to take advantage of their higher HRS (see **Figure 9.6a**). This is demonstrated by the data (red) in **Figure 9.1**, which were collected for a set of thicker (17 nm) film devices following the same procedure described above. The on-resistance for each of the 17 nm device in **Figure 9.1** is shown in **Figure 9.6b** along with the $R_{ex}$ it contains and the $V_{off}^*$ it exhibits. Comparing these data with those of similar cell areas in **Figure 9.4d**, it is clear that $V_{off}^*$ is maintained at the same value but the on-resistance is raised in the 17 nm film devices.

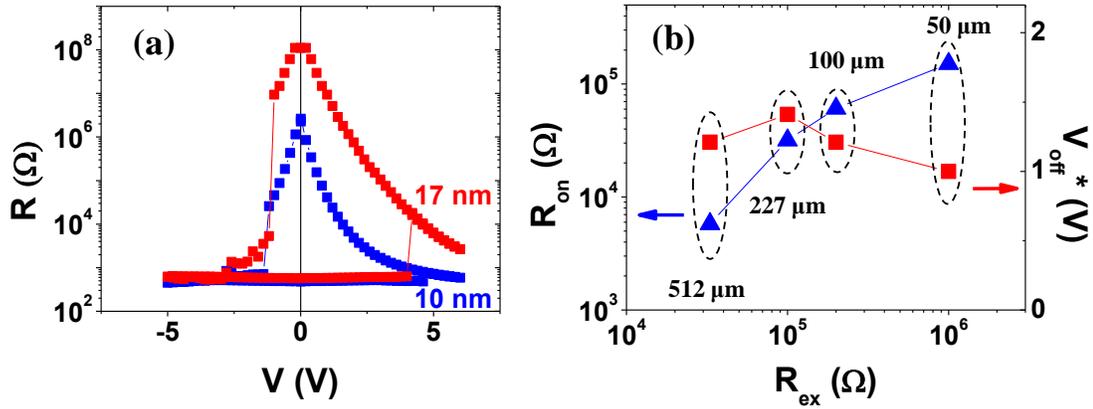

**Figure 9.6.** (a) *R-V* curves of 100×100 µm² cells of 10 nm and 17 nm thickness. (b) On-resistance $R_{on}$ and off-switching voltage $V_{off}$* for the 17 nm film cells of various areas at their minimum $P_{off}$ configuration (optimal $R_{ex}$).

## 9.5 Discussion

In the RRAM literature, diodes have been introduced to the unipolar devices to lower the leakage current at zero bias, thus lowering power dissipation. However, during switching the diode is "on" so it does not necessarily lead to a reduction of switching current or switching power. This is apparent from **Figure 9.4b**: *P* decreases only when an appropriately large $R_{ex}$ is also present. For bipolar RRAM, selectors having very large resistance below a bipolar threshold voltage have been introduced to eliminate the so-called "sneak-path" problem.[26-27] Once again, they can lower the leakage current at zero bias, but not the switching current since switching occurs when the selector resistance is low. Unlike these modifications, our asymmetric dynamic load can reduce not only the leakage current but also switching power. Moreover, unlike the selector used to eliminate sneak paths, our asymmetric dynamic load need not be integrated into the device stack at each cell. Indeed, only one such load is required for each write/read test-circuit module that has access to an array of $N \times N$ device stacks. Therefore, the approach can be implemented without additional increase in fabrication complexity.

Nanometallic memory switches by an electronic mechanism, so the switching speed is ultimately limited by the $RC$ delay time in the circuit. To compute the delay time, the pertinent capacitance is that of the memory cell, which scales with $A$ and is ~100 pF for a $100 \times 100$ μm$^2$ cell in our experiment. The pertinent resistance is that of the total load, including $R_l$ and the asymmetric external load. During off-switching, the diode is in the forward direction and contributes to little load. Likewise, during reading, the diode is not needed and the total load is again small. Therefore, the longest delay time is the one encountered during on-switching when the total load is ~$R_{ex}$. Since $R_{ex}$ is bounded by HRS but has a weaker area dependence, say of $A^{-1+s}$, where $s \sim 1/3$, we expect $R_{HRS}C > R_{ex}C \sim A^s$. For a $100 \times 100$ μm$^2$ cell, $R_{HRS} = 100$ kΩ at on-switching, giving an $RC$ product of 10 μs if we use an asymmetric load. For a $100 \times 100$ nm$^2$ cell, the on-switching time should be $100 \times$ smaller, reaching 100 ns. At even smaller $A$, we expect $R_c$ to rise which lessens the need for $R_{ex}$ to increase (see discussion on **Figure 9.5**), so the $R_{ex}C$ should also rise less rapidly. Moreover, the exponential dependence on thickness can be utilized to further increase $R_c$ without increasing $R_{ex}$, thus limiting the $RC$ delay to a reasonable value.

Our work has demonstrated that an ultra-low power solution for multiple-state nanometallic RRAM devices can be devised using two strategies. First, an asymmetric dynamic load involving a diode and a linear resistance can be used. This approach is applicable to other RRAM, and the scaling results presented here have likely set the lower limit of power consumption for most devices. Second, by increasing the film thickness by merely a few nm, the HRS resistance of nanometalic RRAM can be dramatically increased to broaden the dynamic range of the asymmetric load, thereby further lowering the power consumption by several orders of magnitude. Both strategies are scalable: power consumption using a 17 nm film is $P_{on}$=460 nW and $P_{off}$=30 μW for a 100×100 μm$^2$ device, and is projected to be $P_{on} < $ 460 fW and $P_{off} < $ 1.5 nW for a 100×100 nm$^2$ device. For a 10×10 nm$^2$ device with thickness/composition optimized nanometallic film, we believe 1 pW $P_{off}$ is ultimately feasible. To realize the anticipated low power, however, an improved ability for current readout will be required, since ultimately it is the product of current readout and $V^*_{off}$ (~1 V for the best RRAM devices today) that sets the power limit. Such advances may accelerate the adoption of RRAM technology.

# Chapter X. High Tunability of Resistance States Based on an Electroforming-free Nanometallic Complementary RRAM

## 10.1  Introduction

Resistive random access memory (RRAM) is promising for future digital memory because of its superior properties of nano-second speed, >10 year retention, and >$10^7$ cycle endurance, along with good scalability (<100 nm)[1]. Typical RRAMs present at least two distinguished states, in which information is stored and electrical stimulus can be employed to trigger inter-state conversion. One of the difficulties for RRAM engineering lies in its low resistance state (LRS), which is often less controllable. In most RRAMs, the LRS is typically a linear resistor with resistance $R$<100 $\Omega$ and such resistance is weakly dependent on device dimensions. One issue associated with such low resistance is power ($V^2/R_{LRS}$) consumed when using a voltage $V$ during read/write operation. This becomes more problematic as the memory array scales down, which implies a higher power density. In addition, parasitic resistance (*e.g.*, resistance of interconnection line) could easily overwhelm LRS and thus cause read/write inaccuracy.

In the RRAM literature, there are reports of increasing the LRS resistance through either a voltage control[2-3] or a current compliance control[4-5] of filamentary growth. However, the increase is small and not statistically reliable because an accurate control of filament growth is difficult. On the other hand, the LRS resistance should not be too high either since a smaller $R_{HRS}/R_{LRS}$ leads to a smaller read margin. In fact, there is an optimal $R_{LRS}$ value, dictated by compromising parasitic resistance and the $R_{HRS}/R_{LRS}$ ratio, which varies with the array size[6]. A simple method for LRS tuning is needed to achieve such optimal LRS.

In **Chapter II**, we showed the HRS resistance of nanometallic RRAM is highly tunable by adjusting film thickness ($R \sim \exp(\delta/\zeta)$) or metal concentration. In the Chapter, I will demonstrate how such tunable HRS may achieve a highly controllable low-power RRAM with a high $R_{HRS}/R_{LRS}$ ratio when used in a complementary structure. (The complementary structure has two anti-serially connected RRAMs with different thickness.) In such configuration, the HRS of one of the constituent RRAMs serves as the LRS of the combined devices, while keeping its bipolar switching characteristics. A schematic of the device is shown in **Figure 10.1a**.

## 10.2 Device Fabrication

To fabricate a complementary RRAM, we first coated a Si/SiO$_2$ substrate with a Pt electrode by RF sputtering. Second, square shaped holes with various sizes were photolithographically patterned by removing the photoresist. Afterwards, five layers in the following sequence, Pt/Si$_3$N$_4$:Pt/Mo/Si$_3$N$_4$:Pt/Pt, were consecutively sputter-deposited using different target materials (Pt, Si$_3$N$_4$ or Mo) without breaking vacuum. In the above, the thickness of the two Si$_3$N$_4$:Pt layers are different, being 5.4 nm and 10 nm. Finally, a conventional lift-off procedure was applied to form the final square shaped devices (**Figure 10.1b**). To confirm the multi-layer structure, we employed Focus Ion Beam (FEI Strata DB235) to cut a cross section from top electrode to substrate. As shown in **Figure**

**10.1c**, five distinct layers (Pt/Si$_3$N$_4$:Pt/Mo/Si$_3$N$_4$:Pt/Pt) can be readily seen from their different brightness, which is mainly determined by the atomic weight ($Z_{Pt} > Z_{Mo} > Z_{Si}$).

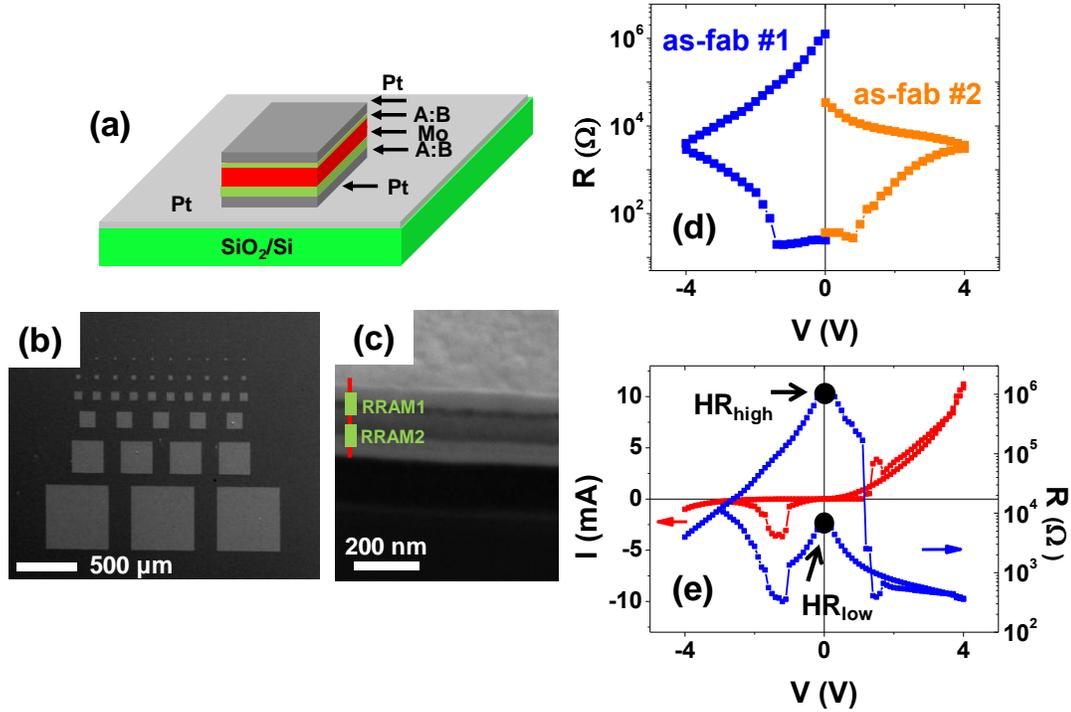

**Figure 10.1**. (a) Schematic of complementary RRAM device. (b) SEM top view of fabricated device. (c) SEM cross section view of fabricated device cut by FIB. (d) Initial *R-V* curve for negative and positive polarity. (e) *I-V* & *R-V* switching curve.

## 10.3   Results and Discussion

The as-fabricated complementary device shows a fairly low initial resistance state (~20 $\Omega$), without forming. Such state exhibits a linear *I-V* or constant *R-V* behavior up to a critical voltage (~±1 V, **Figure 10.1d**), indicating a metallic state, which is further verified by the positive d*R*/d*T* slope in **Figure 10.2** (the metallic nature was extensively studied in **Chapter V**). In a serial connection, this characteristic is possible only if both constituent RRAM devices (top Pt/Si$_3$N$_4$:Pt/Mo ($\delta_1$=5.4 nm) and bottom Mo/Si$_3$N$_4$:Pt/Pt ($\delta_2$=10 nm)) are both metallic in their as fabricated state and each has a resistance on the order of 10 $\Omega$.

Next, at ~1 V with either a positive or a negative polarity the device switches to the HRS (**Figure 10.1d**), which has a non-Ohmic behavior with decreasing resistance with increasing voltage. However, because of different thickness of the active layers in the two constituents RRAMs, meaning that they have very different resistance due to the exponential thickness dependence of resistance of a nanometallic film, the resistance value of the HRS is polarity dependent: when a negative polarity is used (0 V→-4 V→0 V), the transition ends at ~2 M$\Omega$, which will be referred to as HR$_{high}$ hereafter, compared to ~30 k$\Omega$ when a positive polarity is used (0 V→+4 V→0 V), which will be referred to as HR$_{low}$ hereafter. The negative cycle turns the bottom constituent RRAM to the HRS, while maintaining the top constituent RRAM at the LRS. In contrast, the positive cycle turns the top constituent RRAM to the HRS while maintaining the bottom constituent RRAM at the LRS. The above result is understandable within the framework of nanometallic RRAM, which is bipolar with the switching polarity determined by the relative work functions of the electrodes. The as-fabricated active layers are thin enough ($\delta$~10 nm) to allow electrons to freely tunnel through, *i.e.*, they are at the LRS with $\delta<\zeta$. When a critical voltage (~1 V) applied, overcoming the energy barrier for electron trapping, it triggers the transition to the HRS state in one of the two constituent RRAMs.

(It is the one that experiences a current flowing from the Pt electrode to the Mo electrode.)

After the initial cycle which sets one of the constituent RRAMs into the HRS, the complementary structure exhibits regular switching cycles as shown in **Figure 10.1e**. It is convenient to view the entire switching process as two standard $R$-$V$ curves with head-to-head polarity. If the device starts from 2 M$\Omega$ (corresponding to LRS/HRS from top to down or $HR_{high}$) and is electrically stressed with a voltage cycle 0 V$\rightarrow$+4 V$\rightarrow$0 V$\rightarrow$-4 V$\rightarrow$0 V, it follows a nonlinear $R$-$V$ curve until +1 V, then abruptly switches to a low resistance state (corresponding to LRS/LRS). Further positive voltage leads to another resistance transition around +2 V (corresponding to HRS/LRS or $HR_{low}$). As the voltage returns to 0 V, resistance reaches 10 k$\Omega$. Similarly, at the negative voltage side, the device transitions to the low resistance state (corresponding to LRS/LRS) at -1 V and it further transitions to the initial high resistance state (corresponding to LRS/HRS or $HR_{high}$) near +2 V. It is obvious that if we treat the entire complementary structure as one device, and use the $HR_{high}$ as the HRS and $HR_{low}$ as the LRS, we have indeed created a new bipolar RRAM with a much more resistive LRS.

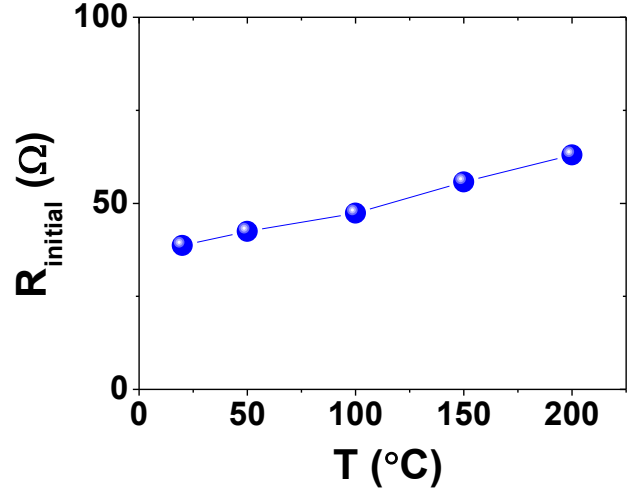

**Figure 10.2**. Temperature dependence of initial resistance state revealing its metallic nature.

Since both HR$_{high}$ and HR$_{low}$ contain one constituent RRAM at the HRS, and the resistance value of the HRS of a nanometallic material can be easily tuned by adjusting the film thickness ($R \sim \exp(\delta/\zeta)$), we are able to tune the HR$_{low}$, hence the resistance of the LRS in a complementary device. We demonstrate this by changing the top device thickness $\delta_1$ from 2 nm to 12 nm, while fixing the bottom device thickness $\delta_2$ at 10 nm. A series of $R$-$V$ curves in **Figure 10.3** shows the HR$_{low}$ value changes by several orders of magnitude as the top RRAM thickness varies. With a top RRAM thickness $\delta_1$=2 nm, HR$_{low}$ is 70 $\Omega$, way below HR$_{high}$. It increases and approaches HR$_{high}$ as $\delta_1$ increases. Eventually, as the top RRAM thickness ($\delta_1$=12 nm) exceeds that of the bottom RRAM, the HR$_{low}$ value exceeds the HR$_{high}$ value. Then the situation is reversed and the two states exchange roles. As illustrated in **Figure 10.4**, the HR$_{low}$-$\delta_1$ dependence follows the scaling law for nanometallic film, *i.e.* $R \sim \exp(\delta/\zeta)$. Meanwhile, the ratio of $HR_{high}/HR_{low}$ spans over 5 orders of magnitude (from $10^4$ to $10^{-1}$) as the bottom device thickness changes from 2 nm to 12 nm. Therefore, the exponential $R$-$\delta$ relation has provided a wide range of possible LRS resistance in a complementary RRAM, allowing arbitrary HRS and LRS combinations.

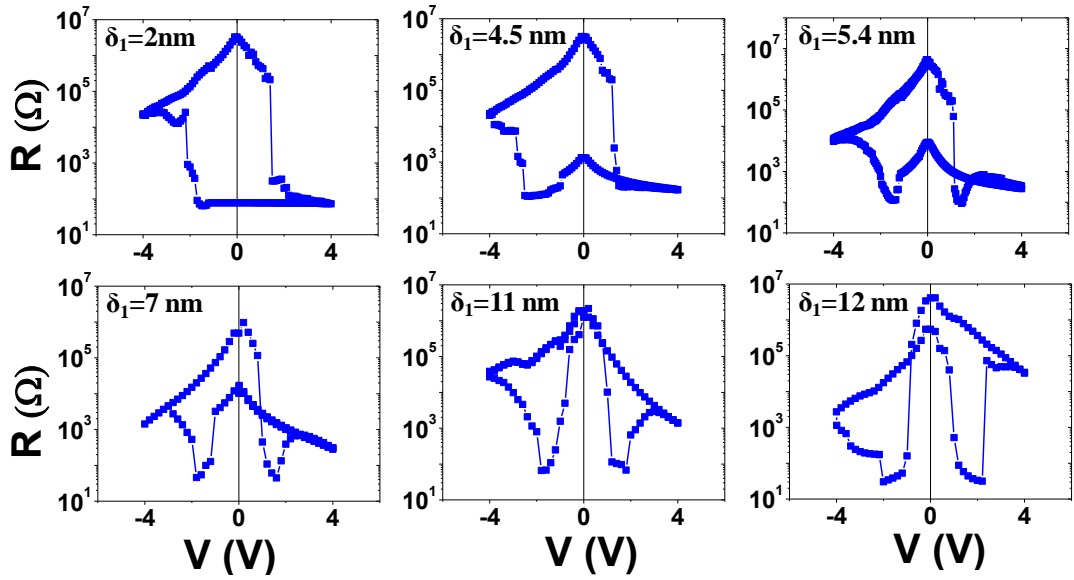

**Figure 10.3**. *R-V* curve of complementary device shown in **Figure 10.1a**. Thickness of RRAM2 $\delta_2$ was fixed at 10 nm and thickness of RRAM1 $\delta_1$ varied from 2 nm to 12 nm.

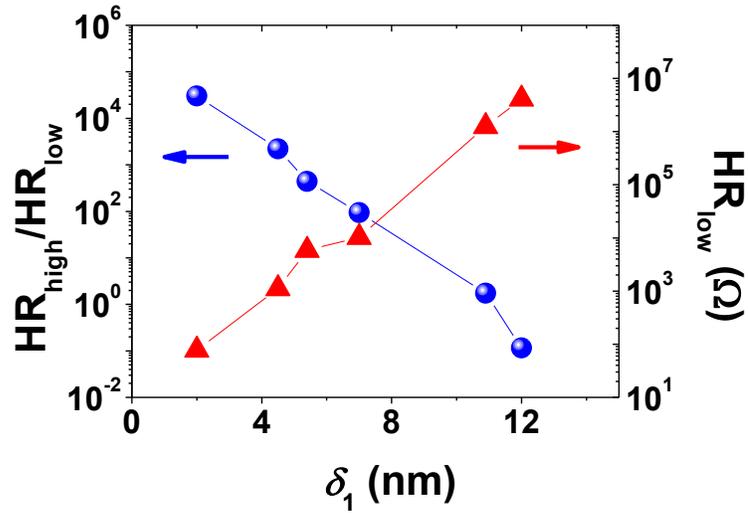

**Figure 10.4**. Summary of HR *vs.* $\delta_1$ information in **Figure 10.3**.

In **Chapter II**, we showed that switching at 1 μs is possible for nanometallic RRAM. Our previous work[7] also demonstrated 100 ns switching. The complementary device can switch at about the same speed, as shown in **Figure 10.5a**. It depicts transient *V-t* behaviors of a complementary RRAM under 1 μs voltage pulsing. The device was pre-set to the LRS/HRS state ($HR_{high}$) before pulsing. Under +3.6 V, the device undergoes the LRS/HRS-to-HRS/LRS ($HR_{high}$-to-$HR_{low}$) transition, which involves an abrupt (current) peak which subsequently decays to ~3.4 mA within 50 ns. Such peak is indicative of an *RC*-circuit charging/de-charging behavior, with the steady state reached by the exponential decay corresponding to a static resistance of ~1 kΩ. This value is consistent with the DC resistance value of $HR_{low}$ state at +3.6 V in **Figure 10.1e**. Afterwards, under a negative -3.6 V pulse, the resistance states switches to the LRS/HRS ($HR_{high}$) state, which completely suppresses the current, giving $I \approx 0$ mA. Again, an *RC* circuit charging/de-charging behavior ($\tau_{RC}$<50 ns) provides a negative peak near the edge of the voltage pulse.

To confirm such peaks and their associated decays originate from the circuit itself, we performed a quantitative SPICE simulation, assuming instant switching of the film at the pulse edges but the overall device performance is dictated by the *RC* device response to such abrupt switching. The simulation (**Figure 10.6**) agrees well with the experimental data and thus verifies our argument. Regarding the real switching time, we can only conclude from the simulation that the switching is completed well within the *RC* delay time, *i.e.*, $\tau_{switch}$<<$\tau_{RC}$<50 ns. This applies to both transitions, $HR_{high} \rightarrow HR_{low}$ and $HR_{low} \rightarrow HR_{high}$. The device can continue to switch many times between the $HR_{high}$ and the $HR_{low}$ without suffering degradation, as illustrated in the endurance data in **Figure 10.5b**.

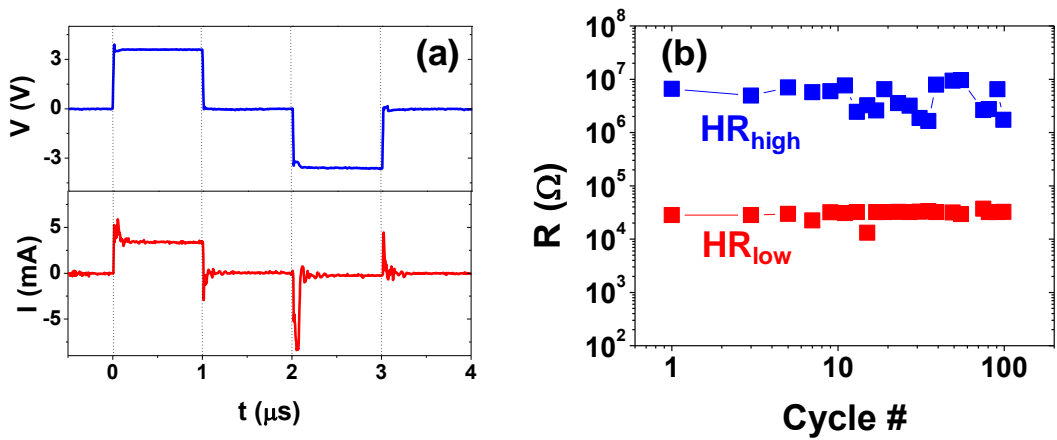

**Figure 10.5**. (a) Dynamic *V-t* and *I-t* during switching. (b) Endurance test of device with complementary structure.

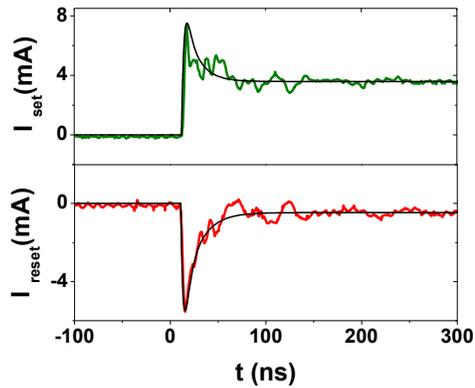

| | R | C |
|---|---|---|
| HRS 1 | 10 MΩ | 24 pF |
| LRS 1 | 940 Ω | 29 pF |
| HRS 2 | 43.7 kΩ | 51 pF |
| LRS 2 | 800 Ω | 67 pF |

**Figure 10.6**. SPICE simulation of transient current during switching. The table is the parameters used for simulation.

Nanometallic complementary RRAM also exhibits multi-bit capability. **Figure 10.7** shows that if voltage is withdrawn at some intermediate values (6 states are used for demonstration), six different resistance states are obtained. Each state is non-volatile and quite stable, with no roll-off over $10^4$ s (**Figure 10.7b**). Their resistance windows are wide enough for low-error-rate reading. Theoretically, if we treat the complementary device as a collection of two constituent RRAMs, each with $N$ distinguishable states, then such complementary structure potentially has $2N$ states for data storage. Manipulation of inter-state conversion can be implemented based on a two-way switching method, as we will explore in **Chapter VIII**.

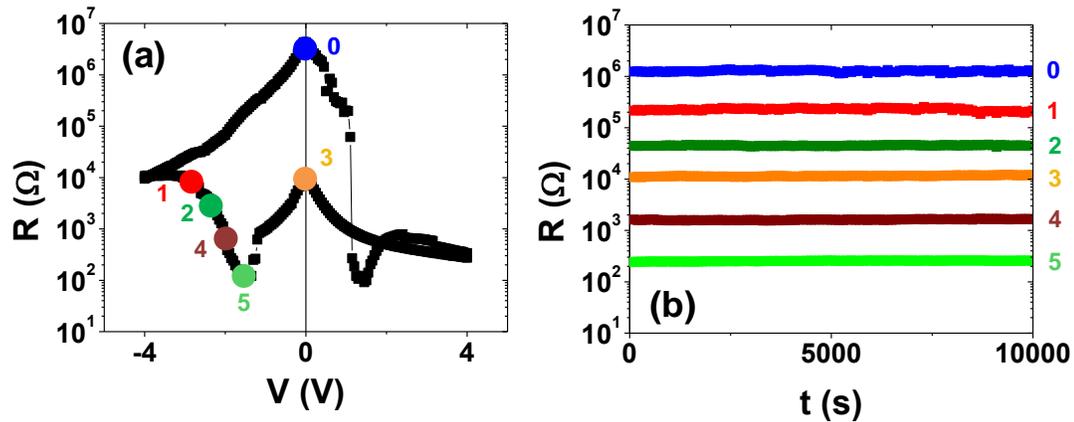

**Figure 10.7**. (a) *R-V* curve labeled with intermediate resistance states. (b) Retention test for each state.

One known feature for complementary device made of two antiserially connected RRAM is its built-in compliance control. For a standard nanometallic RRAM, a current compliance is typically needed at off→on switching to avoid excess current damaging the device. However, complementary structure already has a built-in compliance since the constituent RRAM of the opposite polarity will be turned off before the current becomes too high.

Meanwhile, the switching voltage and switching windows (corresponds to intermediate state LRS/LRS) depend on the intrinsic off→on and on→off voltage. For example, in **Figure 10.1e**, LRS/HRS→LRS/LRS and HRS/LRS→LRS/LRS transitions occurs at ±1 V, close to the off→on voltage of a stand-alone nanometallic RRAM (∼±1 V). The switching voltage for LRS/LRS→LRS/HRS or LRS/LRS→HRS/LRS needs to consider the voltage sharing between the two LRS RRAMs. Therefore, a higher external voltage is required for switching. In our case, the off→on voltage of a stand-alone RRAM is ~1 V, and the apparent voltage is >1 V. Moreover, if a line or load resistance $R_0$ is considered, then its voltage sharing effect also need to be considered in determining the switching windows. The apparent off-switching (LRS/LRS→HRS/LRS) voltage now becomes $\left(\frac{R_0 + 2R_{LRS}}{R_{LRS}}\right) V_{off}$, which increases with the load $R_0$.

The equivalent circuit of a complementary device includes a line resistance $R_0$ in series with $R_1//C_1$ and $R_2//C_2$. To simplify the problem, we ignore the RRAM resistance (letting them be infinite) and consider only a $R_0$, $C_1$ and $C_2$, which overestimates the $RC$ delay. With this approximation, the delay is $R_0 \times C_1 C_2 /(C_1 + C_2)$. As we showed in **Chapter IV**, $C$ is on the order of 100 pF for a $100 \times 100$ μm$^2$ cell, which linearly scales with size. Thus, for a $100 \times 100$ nm$^2$ cell and a 50 kΩ line resistance, the $RC$ delay is less than 5 ps. (In the above, we refer to the CMOS technology in which the typical sheet resistance is 0.05 Ω/sq for the metal conductor layer. For a 1 Tbit storage unit ($10^6 \times 10^6$), a line resistance of the order of 0.05 Ω/sq $\times 10^6$ sq or 50 kΩ is thus estimated.)

**Merits of Complementary Nanometallic RRAM**

Complementary nanometallic RRAM have several advantages. First, Complementary RRAM structures may be easier to fabricate because they contain the same materials as stand-alone RRAM, and their constituent layers can be consecutively deposited without breaking the vacuum. Second, using nanometallic RRAM, the exponential thickness dependence ($R \sim \exp(\delta/\zeta)$) provides the possibility of tuning all the characteristics of the complementary device including resistance and power. This may be further supplemented by using the metal composition as another tuning parameter since, unlike the case of filamentary RRAM, the resistance of a nanometallic RRAM can be varied by several orders of magnitude by varying the composition. Both adds the flexibility to the selection of the current sensing circuit and lowers the power consumption during reading, which is controlled by $HR_{low}$. Third, nanometallic supplementary RRAM may achieve a larger memory array size. This is because the memory array size/density can be effectively increased by taking advantage of the high LRS ($HR_{low}$) and its non-linearity, which implies a high interconnect line electrode resistance, which in turn requires an even higher LRS ($HR_{low}$) to avoid reading error—this can be easily achieved in nanometallic RRAM by adjusting the thickness or composition, making it easier to achieve the optimal $R_{LRS}$. Fourth, nanometallic RRAM can more readily be designed to satisfy the following requirements: $R_{LRS}$ should be high enough to mitigate the sneak path leakage and to sense the voltage to switch, but low enough to guarantee a large $R_H/R_L$ ratio or a good reading contrast. Lastly, nanometallic complementary RRAM has a highly non-linear LRS, which is beneficial for solving the sneak path problem. In summary, nanometallic complementary RRAM may be a promising memory for practical applications.

# APPENDICES

# Appendix I. Fabrication and Characterization

## I.1  Thin Film Deposition

Thin film deposition technology has been extensively investigated in the past few decades, prompted by the demanding VLSI technology. A variety of chemical vapor deposition (CVD) and physical vapor deposition (PVD) methods have been developed in industry to meet specific "quality" criterions, such as composition, morphology, uniformity, contamination levels, defect density, mechanical and electrical properties. Furthermore, the shrinking dimension of feature size and increasing numbers of stacking layers eventually pose stringent demands of "conformal coverage" for thin film deposition. Generally speaking, CVD, relying on chemical reactions between the introduced reactant gases and substrate surface, provides a better film quality, step coverage and more accurate stoichiometric control than PVD, in which physical methods are employed to produce the constituent atoms. CVD has historically been used in IC industry primarily for silicon and dielectric deposition. However, PVD offers much versatile choices (almost all materials can be deposited by PVD methods), and hence, it has been used extensively for metal and alloy deposition[1].

### I.1.1  Chemical Vapor Deposition (CVD)

For CVD, reactant gases are introduced into deposition chamber, in which chemical reactions take place, and consequently form the desired film on the surface of the substrate. Such reactant gases can be either single gas which will decompose under certain condition to supply the necessary components for the film, or multiple gases which will interact to form the film. Sometimes desired reactants are not available in gas phase and in such case, liquid source might be used. Due to its chemical reaction feature, CVD is generally considered as an isotropic deposition.

CVD can result in various forms, including single crystalline, polycrystalline, amorphous and epitaxial, depending on detailed reaction condition (gas, temperature, pressure, *etc.*) and substrate type. Major CVD includes:

1) Atmospheric pressure CVD (APCVD): CVD at atmospheric pressure ~760 Torr.

2) Low pressure CVD (LPCVD): CVD at sub-atmospheric pressure ~1 Torr.

3) Ultrahigh vacuum CVD (UHVCVD): CVD at $<\sim10^{-8}$ Torr.

4) Atomic layer CVD (ALCVD or ALD): precursors react with substrate surface on at a time in a sequential, self-limiting and repeating manner. ALD provides a very accurate control of film density, chemical stoichiometry, thickness and homogeneity, and thus has large area capability, excellent reproducibility and conformality. The major limitation is its slowness, typically ~one monolayer per cycle. An improved version is plasma-enhanced ALD (PEALD).

5) Plasma-enhanced CVD (PECVD): Highly ionized plasma supplies extra energy to the reactant gases and thus typically requires much lower temperature. Combining PECVD with bias sputtering leads to a new version: high-density plasma CVD (HDPCVD), featuring itself with very good fillings of narrow gaps.

## I.1.2 Physical Vapor Deposition (PVD)

PVD techniques are generally more flexible than CVD methods. In PVD, individual atoms or molecules or clusters, are produced by either evaporation of a solid source (thermal/e-beam evaporator), or by bombardment of energetic gaseous ions with a solid target in plasma (sputter). These atoms or molecules or clusters then travel through a vacuum or low pressure gas, impinge on the substrate, and condense on the sample surface to form a new layer. During such physical process, chemical reaction can also occur (*e.g.* reactive sputtering) by properly introducing secondary reactive gas. Since deposition occurs in a low pressure system, it is believed very few target atoms colliding with working gas and thus PVD is typically a highly directional or anisotropic approach. As a result, PVD arises several critical issues such as poor thickness uniformity, step coverage, and shadowing effect.

### Evaporator

In evaporator, the source material is heated by heater (*e.g.*, tungsten filament heated up by electrical current) or a high electron beam (controlled by magnetic field) in a high vacuum chamber. As temperature reaches its melting point, materials start to melt, then evaporate towards all directions isotropically and condense on sample surface. In microelectronics, e-beam evaporation is more popular, which can achieve higher temperature so that a broader range of materials can be evaporated. In addition, e-beam evaporation typically results in purer films than thermal evaporation, in which contamination was often found because sodium and potassium were used in the production of the tungsten filament. On the other hand, e-beam systems produce high energy X-ray, which can create trap charges in gate oxides, and thus require extra post-annealing step.

**Sputter**

Sputter deposition usually requires looser vacuum conditions than evaporation (sputter: 1-100 mTorr *vs*. evaporation: $<10^{-5}$ Torr). In addition, it is capable of large area coating and covering almost every kind of condensed material. For such reasons, sputter deposition was the preferred method in semiconductor technology. It is also the main deposition tool for my material development.

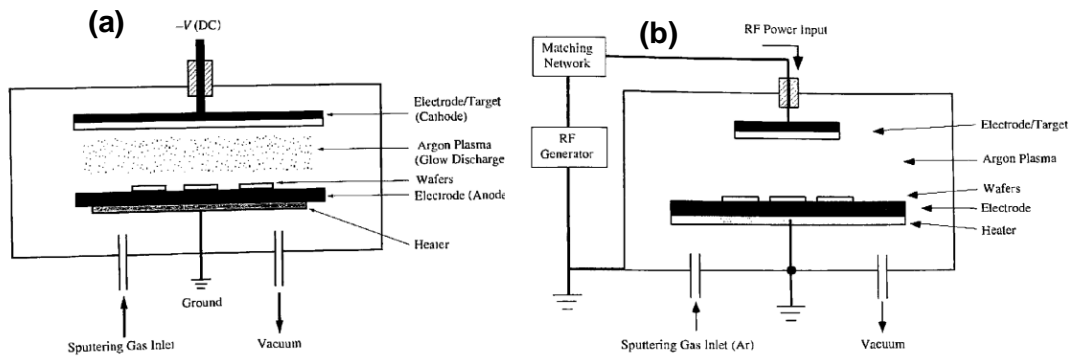

Figure 1. Schematic diagram of (a) DC (b) RF powered sputter deposition system. An extra matching network is required to match the impedance for RF sputter[1].

A basic configuration of sputter system is shown in **Figure 1**. A cathode mounted with depositing source (target) provides a high voltage bias (either DC or RF) with respect to grounded sample stage. Such voltage supplies sufficiently high energy to working gas (*e.g.* argon), and ionize Ar atoms making a conducting medium (plasma) between target and sample. Then under such high bias, the energetic positive ions ($Ar^+$) in the plasma are accelerated to negatively biased cathode, strike the target and dislodge, or sputter, the target atoms. These atoms then freely travel through the plasma (or collide/react with reactant gas) and impinge on the surface of the wafer. DC sputter is commonly used for conducting material deposition (*e.g.* metal). However, DC cathode cannot effectively neutralize charge if target is an insulator, and hence arises difficulties to sustain. To avoid such charging issue, RF sputter deposition is typically employed for non-conducting materials (*e.g.* dielectric).

Sputter is a non-linear kinetic process, which qualitatively obeys the following rules:

1) The higher the supplied power, the higher deposition rate, but larger surface roughness.

2) The higher working pressure, the more atoms sputtered from target materials, also the more chances these atoms scattered by working gas.

3) The closer distance from target to substrate, the faster the deposition rate.

4) RF ignition is easier than DC ignition and thus suitable for slower rate deposition.

5) Negative substrate bias helps planarized films as well as a better filling and higher density[1-2].

## I.2  Thin Film Device Fabrication/Lithography

Lithography technology is the cornerstone for modern IC industry, which allows submicron functional features accurately "printed" on chips. Mainstream of manufacturing ICs are currently replying on photolithography, which is also the main tool for device patterning in this thesis. The concept is simple and straightforward. A light sensitive polymer, namely photoresist, is spun onto sample surface first and then selectively exposed by shining light through a mask which contains sophisticated pattern information. Finally, the photoresist is developed which completes pattern transfer from mask to sample. Such fabricated pattern essentially forms a new mask, which can be used for later deposition, ion implantation and etching. Since the feature size is restricted by wavelength of light source (typically 100 nm~1000 nm), conventional photolithography has difficulties for submicron patterns. One way to solve such problem is adopting immersion technique, allowing light passing through liquid with higher refractive index ($n>1$ virtually shortens the wavelength). The other solution is replacing illumination light with shorter wavelength source, such as extreme ultra-violet (EUV), X-ray or electron beam. Among these nano-size techniques, electron beam lithography (EBL) is my major tool for developing nano-device. EBL is a maskless method, which can theoretically achieve ~10 nm accuracy without diffraction limit. However, the critical downside of EBL is extremely low throughput, because of sequential point-by-point exposure. Also, charging effect from electron poses a requirement of conducting substrate/resist, as backscattering or secondary electrons may lead to unwanted exposure / pattern in resist. A comparison of photolithography and e-beam lithography is summarized in **Table 1**. Alternative lithography techniques include scanning beam lithography, nanoimprinting, colloidal/polymer assembly, holographic lithography, two-photon lithography, *etc*.

| | Photolithography | E-beam Lithography |
|---|---|---|
| Size | Micron-scale | Nano-scale |
| Speed | Parallel process (fast) | Serial process (slow) |
| Mask | Photomask required | Mask-less |
| Efficiency | High throughput | Low throughput |
| Issue | contact / proximity effect | Backscattering/secondary electrons, charging effect |

Table 1. Photolithography *vs.* E-beam lithography.

## I.3  Thin Film Device Characterization

Thin film characterization in this thesis mainly focuses on materials characterization and electrical/optical characterization. As a brief summary, morphology and structure information are acquired by combination of microscope (optical, electron, scanning probe) and X-ray techniques; chemical composition and bonding information are obtained by analyzing various spectroscopies. Electrical characterizations are mainly done on probe stations for micron-device, or a customized conducting AFM (CAFM) for nano-device. Temperature related measurement is performed in a Physical Property Measurement System (PPMS) with electrical and magnetic capability. Electrical automation is realized through NI LabView or Agilent VEE Pro programming, which connects various instruments of interest to central computer. **Table 2** summarizes characterization techniques used in my work. Detailed characterization methods will be introduced in subsequent chapters.

| Tool/Instrument | Purpose |
|---|---|
| **Materials Characterization** | |
| Optical Microscope | Morphology |
| SEM (FEI Quanta FEG ESEM) | Morphology, spectroscopy, electron flux |
| TEM (JEOL 2010F TEM/STEM) | Morphology, spectroscopy |
| AFM (Asylum MFP-3D) | Morphology, thickness, CAFM, uniaxial stress |
| XRD (Rigaku Miniflex) | Crystallization, material identification |
| XRR (Bruker D8 Discover) | Density, thickness, roughness |
| EDX (FEI Quanta FEG ESEM, JEOL 2010F TEM/STEM) | Composition, material identification |
| EELS (JEOL 2010F TEM/STEM) | Composition, material identification, Electronic structures |
| FTIR (Nicolet Nexus 470 spectrometer) | Ionic bonding |
| RBS (NEC Minitandem Ion Accelerator) | Composition, material identification |
| UV-Vis (Varian Cary 5000 spectrometer) | Optical properties |
| Goniometer (Rame-hart model 200) | Wetting properties |
| **Electrical Characterization** | |
| Probe Station (Signatone S1160, Lakeshore) | Electrical measurement testbench (vacuum and low temperature test for Lakeshore) |
| PPMS (Quantum Design) | Electrical/magnetic properties (2K-350K) |
| Source Meter (Keithley 237, 2400) | 2pt/4pt $I$-$V$ ($R$-$V$), FET transistor |
| Function/pulse generator (Agilent 81104A) | Endurance, transient properties, $V$-$t$ properties |
| Oscilloscope (HP Infinium 54825A) | Transient properties |
| Impedance analyzer (HP4192A, Gamry G750) | AC impedance analysis |
| **Programming/Simulation** | |
| NI LabView, Agilent VEE Pro | Electronic automation |
| Mathematica, Matlab | Circuit simulation |
| LTSpice | Circuit simulation (pulse related) |
| Comsol 4.3 | Electromagnetic simulation |

Table 2. Summary of characterization tools/instrument used in this thesis and their purposes.

# Appendix II. Electrode Dependence and Endurances for Nanometallic RRAM

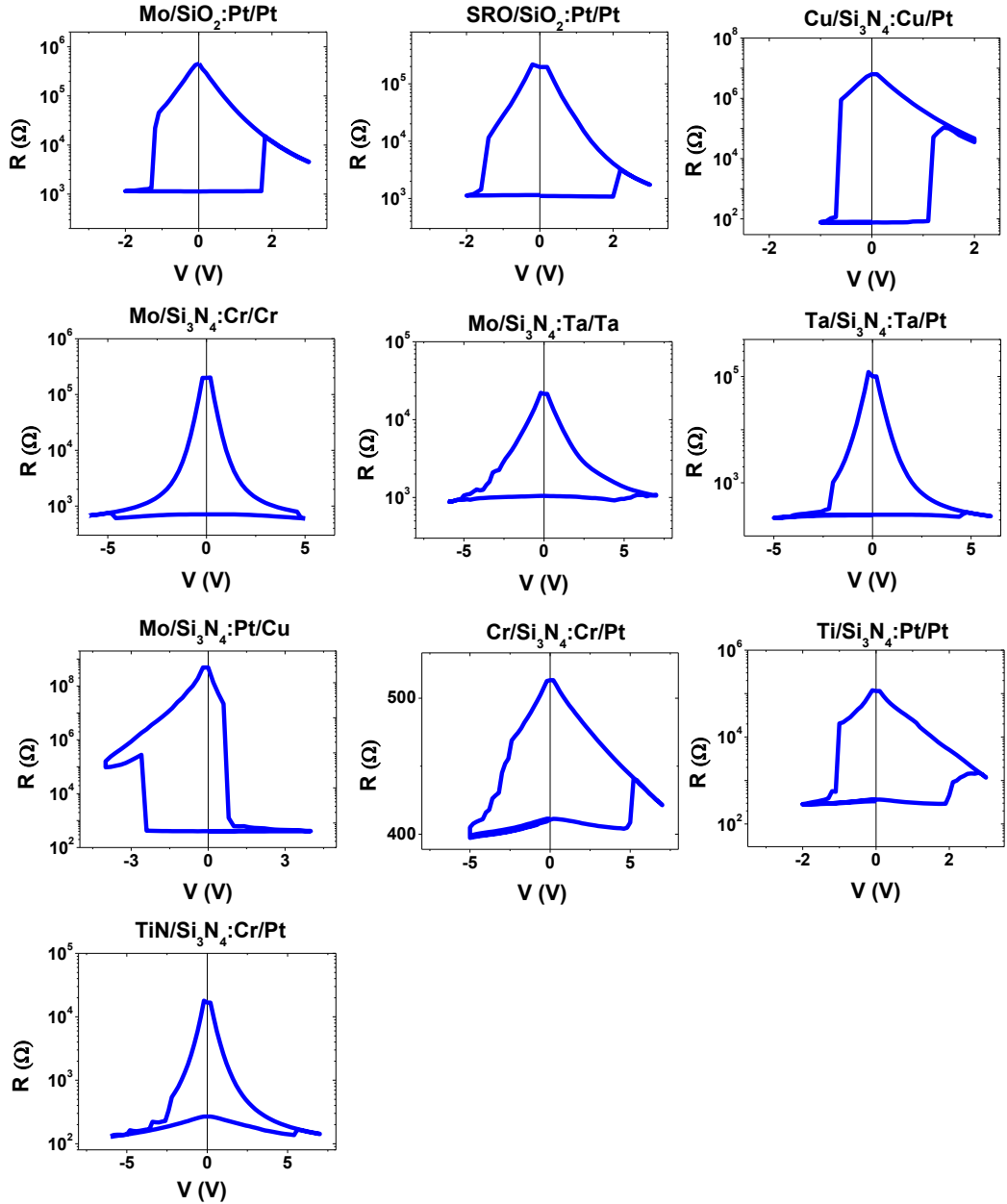

Figure 1. Various electrode combinations for nanometallic RRAMs (Bottom electrode / nanometallic films / top electrode).

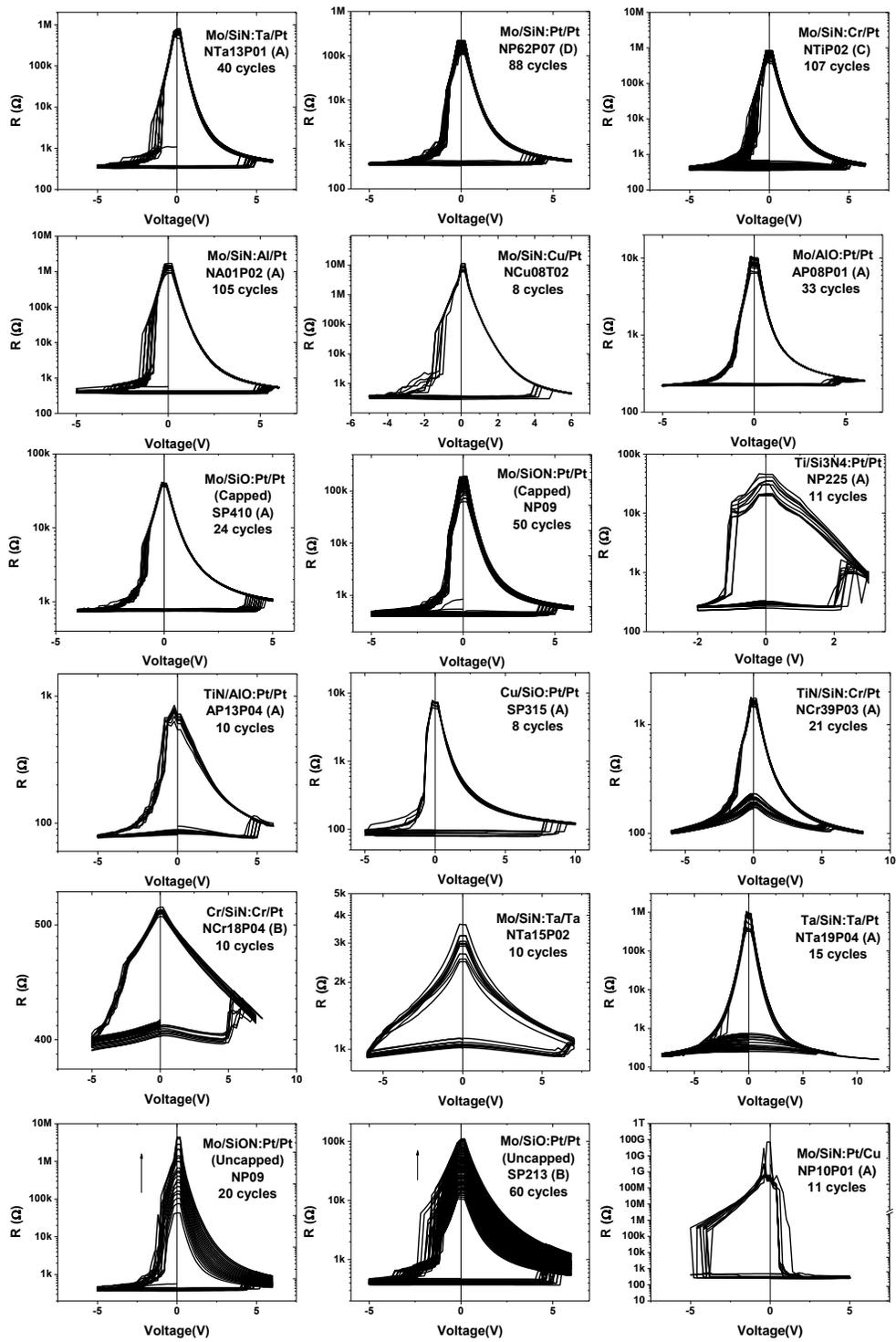

Figure 2. DC endurance data for various nanometallic RRAMs

# Appendix III. Dynamics of Pulse Switching

Digital oscilloscope provides the most straightforward method to monitor switching dynamics. Here we employed two schemes:

1. **Pump probe method**. An excitation pulse (pump) was sent into device, followed by a small probing voltage. Current levels associated with probing signals reflect the resultant resistance states. This is similar to the method used in **Chapter II**, but with a further improvement that probing signal can be controlled closer to original excitation pulses. This method cannot provide *in-situ* switching information, but is a convenient approach to check switchability, especially useful as circuit LCR response severely suppressed real switching signals.

2. **Lead time dynamic method**. A trapezoidal voltage pulse with certain lead time was sent into device and switching dynamics was directly monitored by oscilloscope. The change of resistance states are typically associated with abrupt variations of current, thus switching voltage can be easily derived from current discontinuity points. This method provides *in-situ* switching information, but becomes ineffective as circuit LCR response severely suppressed real switching signals.

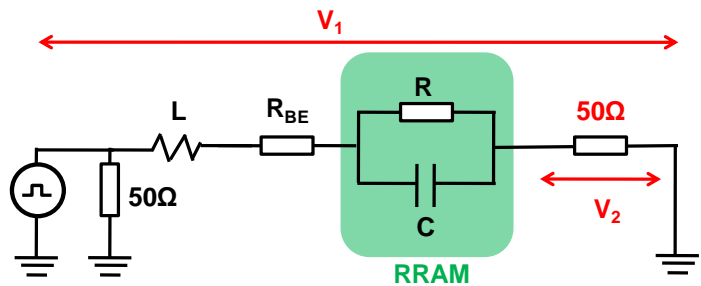

Figure 1. Testing circuit.

Testing circuit is shown in **Figure 1**. The pulses were provided by a pulse generator (Agilent 81104A). A multi-channel digital oscilloscope (HP Infinium 54520) was used to monitor voltage ($V_1$) and current ($I=V_2/50$ Ω). Standard 50 Ω resistors were used for impedance matching.

➤ **Pump Probe Method.**

**ON→OFF switching**

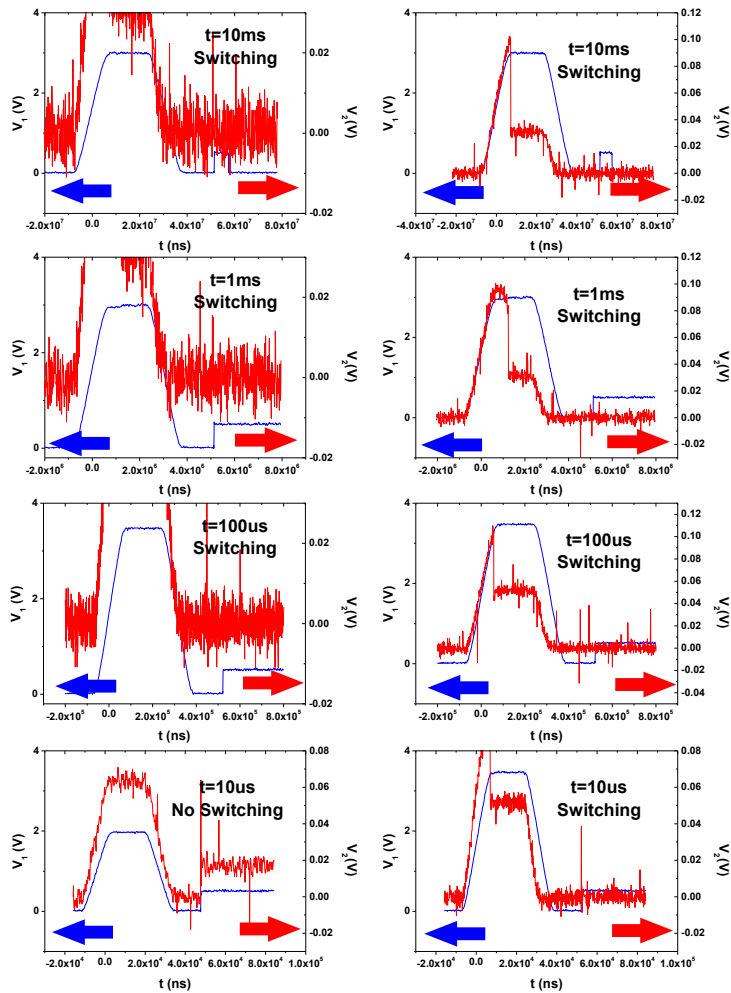

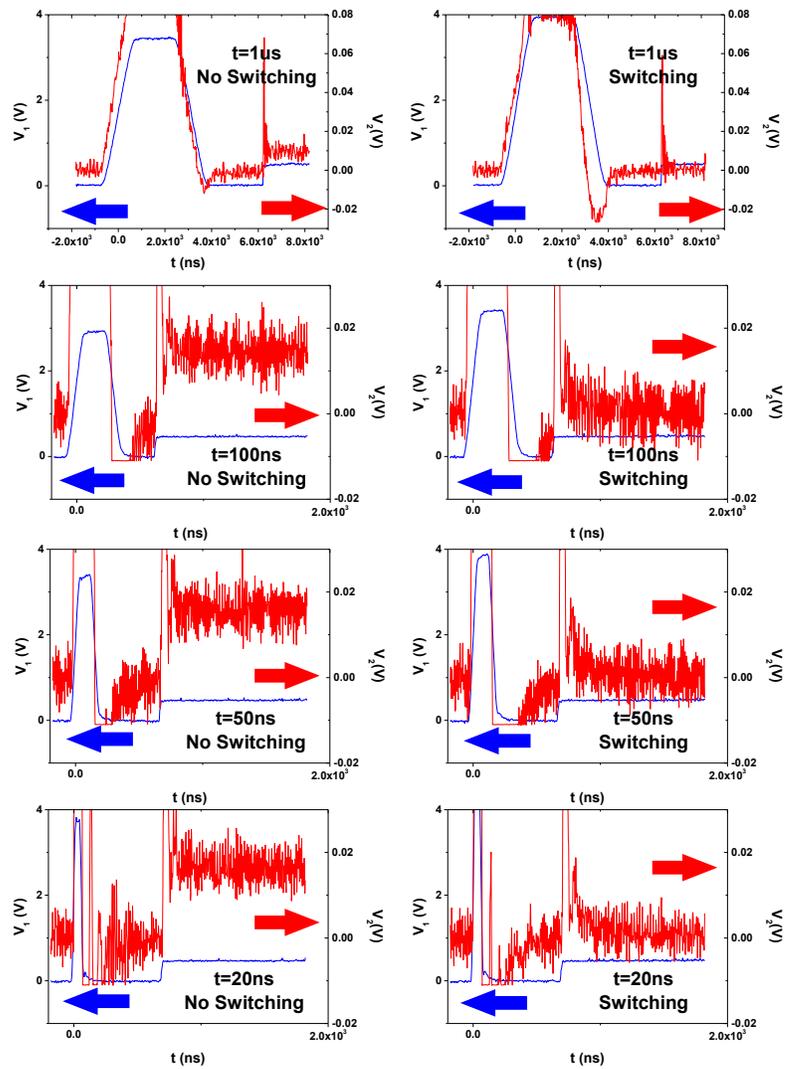

Figure 2. ON→OFF switching dynamics using pump probe method. Blue curves are excitation pulses ($V_1$), red curves are currents ($I = V_2/50\ \Omega$). A successful ON→OFF switching is identified as a low current level (HRS) in probing signals.

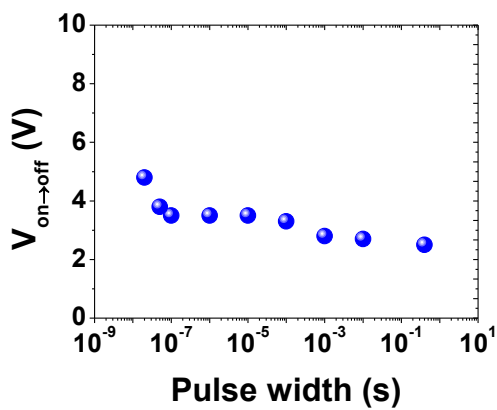

Figure 3. A summary of ON→OFF switching voltages *vs.* pulse widths using pump probe method.

**OFF→ON switching**

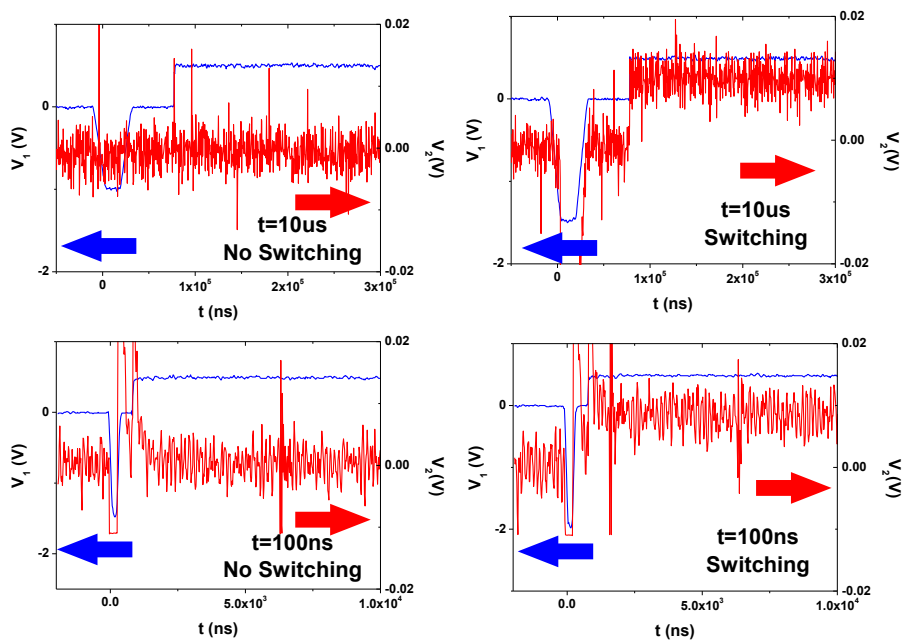

Figure 4. OFF→ON switching dynamics using pump probe method. Blue curves are excitation pulses ($V_1$), red curves are currents

($I=V_2/50\ \Omega$). A successful OFF→ON switching is identified as a high current level (LRS) in probing signals.

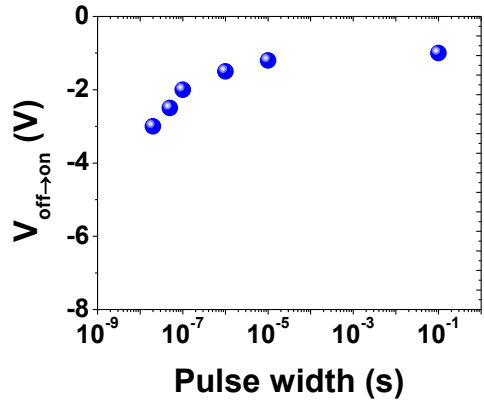

Figure 5. A summary of OFF→ON switching voltages *vs.* pulse widths using pump probe method.

➢ **Lead Time Dynamic Method.**

**Pulse Width**

Interval between leading and trailing edge medians:

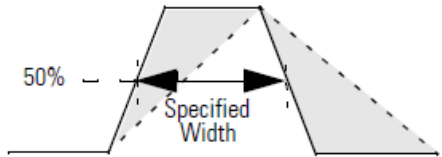

**Transition Time (Lead Time)**

Interval between the 10% and 90% amplitude points on the leading/ trailing edge:

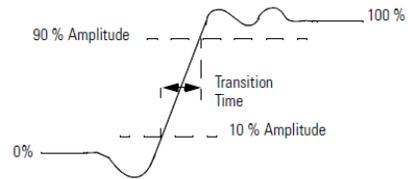

Figure 6. Definitions of pulse parameters.

**ON→OFF switching**

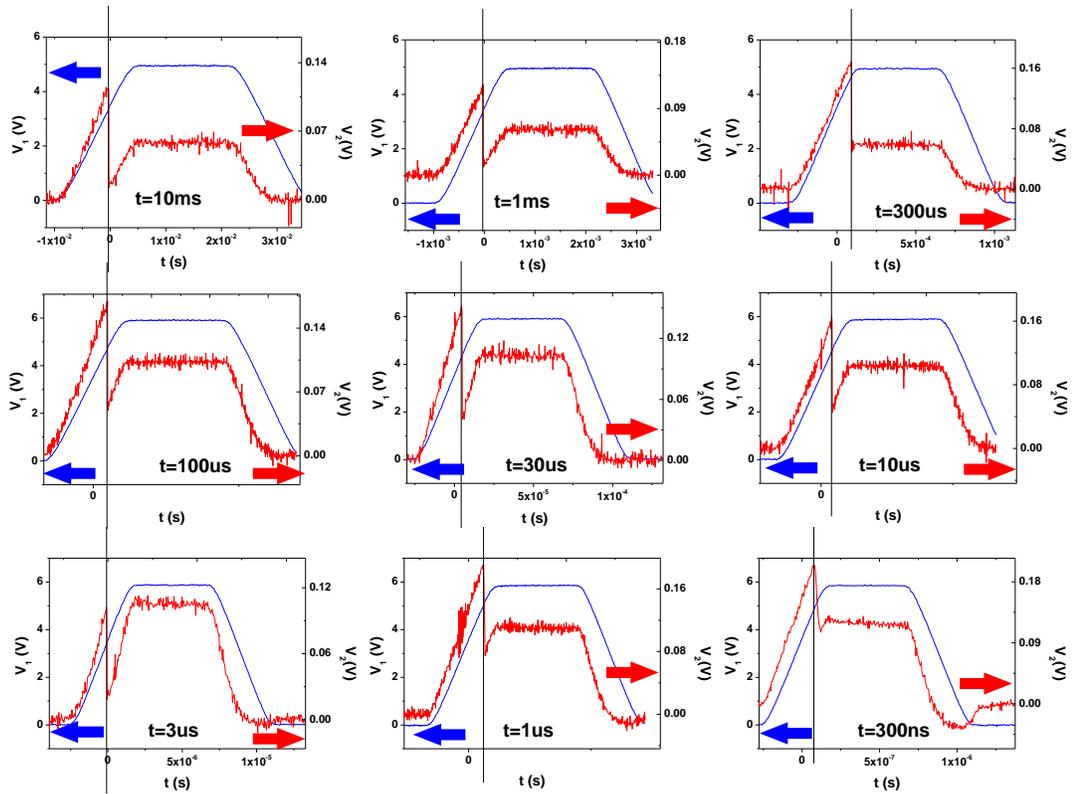

Figure 7. ON→OFF switching dynamics using lead time dynamic method. Blue curves are excitation pulses ($V_1$), red curves are currents ($I=V_2/50\ \Omega$). ON→OFF switching occurs at a critical voltage, reflected as a discontinuity point on current curves.

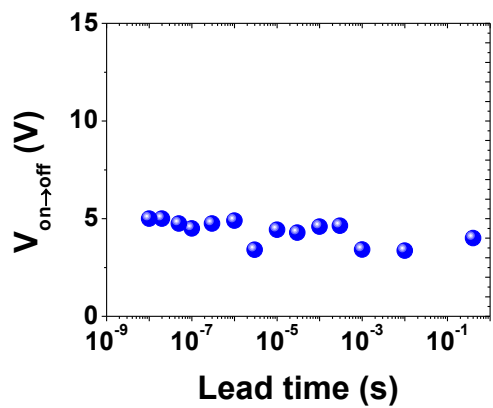

Figure 8. A summary of ON→OFF switching voltages *vs.* pulse widths
using lead time dynamic method.

**OFF→ON switching**

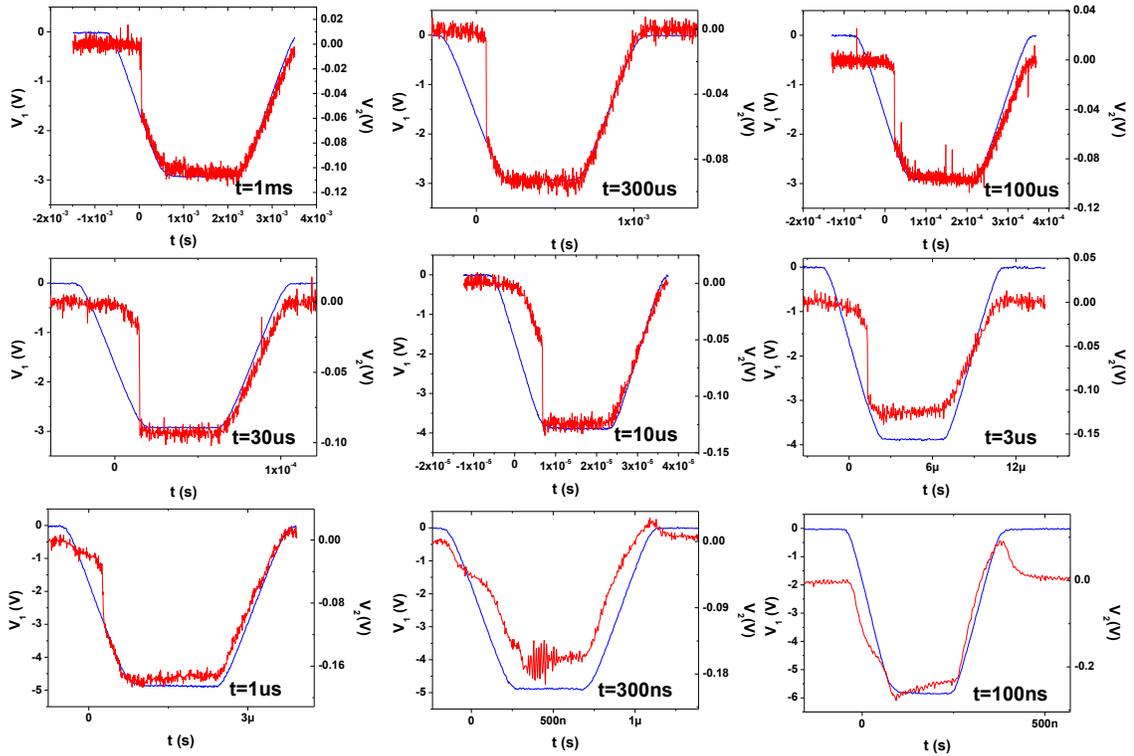

Figure 9. OFF→ON switching dynamics using lead time dynamic
method. Blue curves are excitation pulses ($V_1$), red curves are
currents ($I=V_2/50$ Ω). OFF→ON switching occurs at a critical
voltage, reflected as a discontinuity point on current curves.

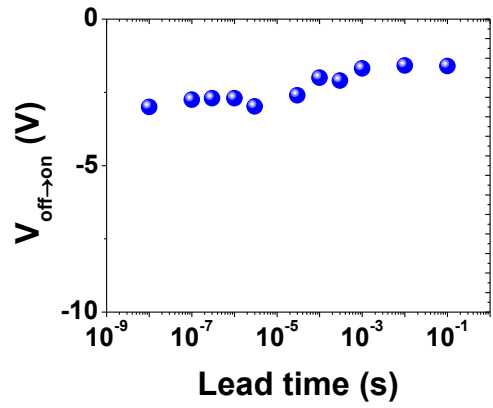

Figure 10. A summary of OFF→ON switching voltages *vs.* pulse widths using lead time dynamic method.

# Appendix IV. Four Point Measurement

Random access memory based on resistive switching materials have attracted considerable interests for future memory, as they stem from easily fabricated structure and exhibit outstanding memory properties such as fast speed (~ns), long retention (>10 years) and high scalability (~10 nm). Resistive switching exists in a large material community, including binary oxide (*e.g.* NiO, TiO$_2$, Al$_2$O$_3$, ZrO$_2$, HfO$_2$), colossal magnetoresistance materials (*e.g.* Pr$_{0.3}$Ca$_{0.7}$MnO$_3$, La$_{1-x}$Sr$_x$MnO$_3$), doped perovskites (e.g. SrZrO$_3$, SrTiO$_3$ doped with Cr, V, or Mo), polymer materials with embedded metallic nanoparticles / nanowires and ionic conductors (*e.g.* Ag-Ge-S). These materials typically require two metallic electrodes on both sides to provide good contacts or redox source. Such electrodes and associated interfaces in fact serve as series load resistors ($R_{load}$) and therefore contribute certain resistance to the intrinsic cell. This is especially an important issue for low resistance state (LRS) that can show a resistance <10Ω: apparent $R_{LRS}$ can be completely suppressed if $R_{LRS} << R_{load}$. In this case, care should be taken to separate intrinsic cell resistance and parasitic resistance. However, few works have been done on this significant topic. In this work, I will introduce a universal four point measurement method to separate different resistance components, which eventually assists us obtain more accurate understanding of resistive switching phenomenon.

The tested device within this work is based on the simplest but most widely used fabrication structure and sequence: planar bottom electrode covered with planar active layer, followed by top electrode patterning using shadow mask or photolithography. Specifically, a Mo film (10 nm thick) was first deposited to cover the entire fused silica substrate by DC sputtering. Such insulating substrate was chosen to avoid current bypass paths along a conducting substrate like Si or Si/SiO$_2$. Next hybrid Si$_3$N$_4$:Cr film (10 nm thick) was fabricated by co-sputtering Si$_3$N$_4$ and Cr using separate Si$_3$N$_4$ and Cr targets in a magnetron RF sputtering system at room temperature. A top Pt electrode (40 nm thick) was then deposited using RF sputtering through a shadow mask. Entire device schematic is illustrated in **Figure 1a**.

Test was performed by scratching one side of film with a diamond pen and then filling such scratching area with conducting silver paste. To create a four point configuration, the other side of sample was also treated similarly. Equivalent test configuration is shown in **Figure 1b**, where current source connects top electrode and one scratching area while voltage sense connects top electrode and the different scratching area. Electrical current flows through only one side of sample (1→2→3) but not the other because of high input impedance of voltage meter (2 and 4 are equal potential). Therefore, voltage sense meter can exclude voltage partition within bottom electrode, allowing us to separate voltage contribution from $R_{TE}$ (top electrode), $R_{cell}$, $R_{int}$ (interface) and $R_{BE}$. Based on this idea, LabView program of four point measurement was developed (see **Appendix XIII**). An equivalent circuit can be simplified as **Figure 1c**. Source meter provides voltage $V_s$ and current $I_s$, giving a total two-point resistance $R_{2pt}=V_s/I_s$. The sensing loop detects a lower voltage $V_{4pt}$ (between "1" and "2" in **Figure 1b & c**) and 4-point resistance is thus calculated as $V_{4pt}/I_s$. Bottom electrode resistance $R_{BE}$ is calculated as $R_{BE}= R_{2pt} - R_{4pt}$.

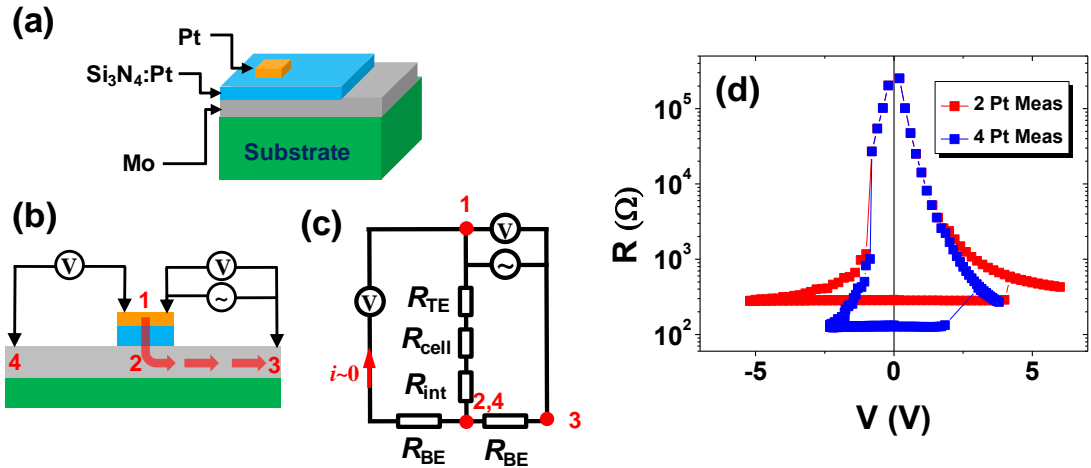

Figure 1. (a) Schematic of device configuration. (b) Schematic of four point measurement setup and current flow. (c) Equivalent circuit. (d) Experimental results $R_{2pt}$ & $R_{4pt}$ vs. $V$ during switching.

The testing results are shown in **Figure 1d**. For HRS, $R_{4pt}$ and $R_{2pt}$ show negligible difference, which is easy to understand because $R_{BE} << R_{HRS}$. On a sharp contrast, LRS shows a significant difference between $R_{4pt}$ and $R_{2pt}$, indicating $R_{BE} \sim R_{HRS}$ ($R_{BE} \sim 200\ \Omega$ and $R_{LRS} \sim 100\ \Omega$ in **Figure 1d**). Therefore, a higher apparent switching voltage is required to compensate $R_{BE}$ effect ($V_{2pt,\ switching} > V_{4pt,\ switching}$). These results immediately provide solid evidences for nontrivial voltage partition that we assumed in circuit model (**Chapter VII**).

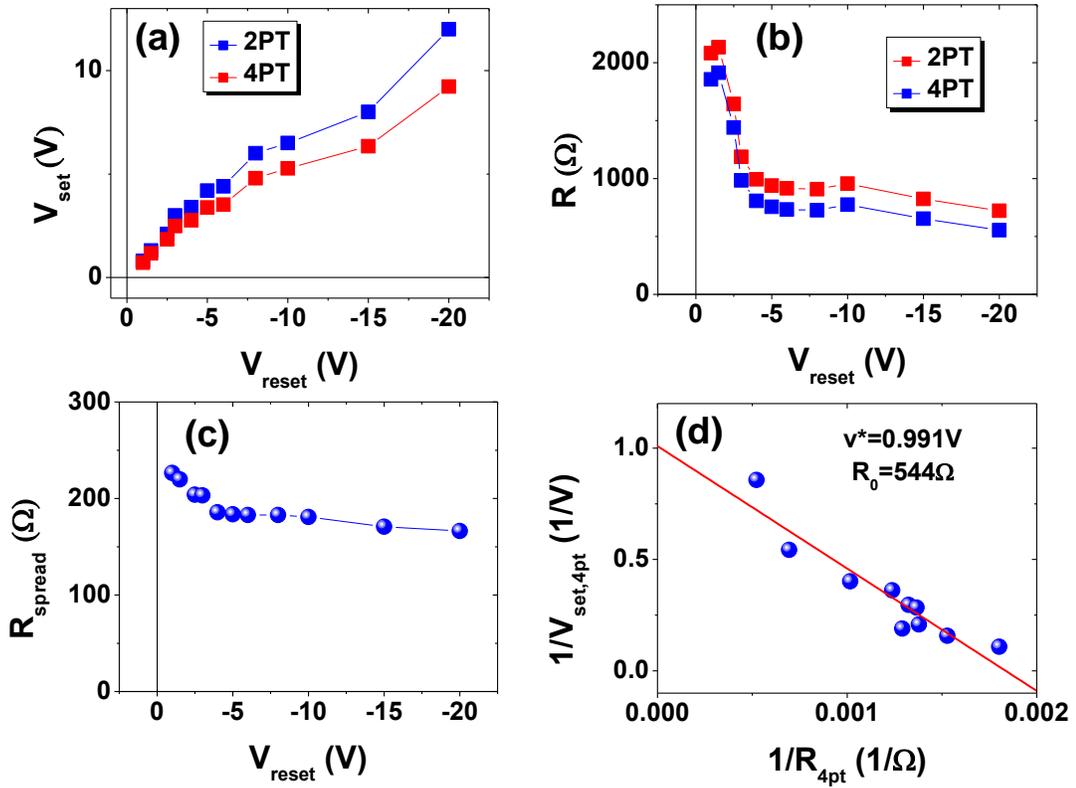

Figure 2. (a) $V_{set}$ (off-switching voltage) *vs.* $V_{reset}$ (maximum negative voltage used for on-switching). (b) $R$ *vs.* $V_{reset}$. (c) Calculated $R_{BE} = R_{2pt} - R_{4pt}$ (d) $1/V_{4pt}$ *vs.* $1/R_{4pt}$ during off-switching, where intrinsic switching voltage $V_m^* \sim 1$ V can be extracted.

Switching voltage dependence was also explored. As shown in **Figure 2a**, as maximum negative voltage increases ($V_{reset}$), off switching voltage ($V_{set}$) also increases for both 4-point and 2-point cases. This implies other than $R_{BE}$, there exists an intrinsic constant resistance which is not switchable (otherwise, 4-point $V_{set}$ should be a constant according to our model in **Chapter VII**). Such non-switchable element can originate from interface ($R_{int}$ in **Figure 1c**) or impurity resistance in nanometallic film. Resistance decreases as $V_{reset}$ increases (**Figure 2b**), revealing a higher negative bias can trigger more parallel conducting paths and reduce the film resistance. By using $R_{BE} = R_{2pt} - R_{4pt}$, bottom electrode resistance can be obtained (**Figure 2c**), which shows almost a constant value ~200 $\Omega$ irrespective of operation voltages. Since this 4-point measurement cannot completely exclude the constant resistance, the following methods were developed to extract the real switching voltage.

$$R_{4pt} = \frac{V_{4pt}}{I_{4pt}}, \ I_{4pt} = \frac{V_0}{R_0} = \frac{V_{4pt} - V_m^*}{R_0}$$

where $V_0$ and $R_0$ are constant remainder resistance in film and $V_m^*$ is the real voltage for switching. We can then rewrite it as:

$$\frac{1}{V_{4pt}} = \frac{1}{V_m^*} - \frac{R_0}{V_m^*}\frac{1}{R_{4pt}}$$

, indicating a linear relation between $1/V_{4pt}$ and $1/R_{4pt}$ in which $R_0$ and $V_m^*$ can be uniquely determined from slope and intercept. Applying this technique to experimental data, $V_m^*$=0.99 V and $R_0$=544 $\Omega$ can be obtained (**Figure 2d**). These values are highly consistent with our circuit model in **Chapter VII**, revealing an intrinsic $V^*$~1 V is required for switching.

Impedance analysis was used to confirm this 4-point method. As shown in **Figure 3b**, two semi-circles are visible revealing two $R$-$C$ elements. The one at low frequency has various diameters for different states, corresponding to switchable elements with variable resistances. The one at high frequency exhibits a constant diameter (thus constant $R$-$C$) irrespective of states, corresponding to constant remainder resistance ($R_0 \sim 544\ \Omega$). The intercept represents the spreading/line/BE resistance, which is about $\sim 167\Omega$. These values quantitative agree with the above DC 4-point measurement results.

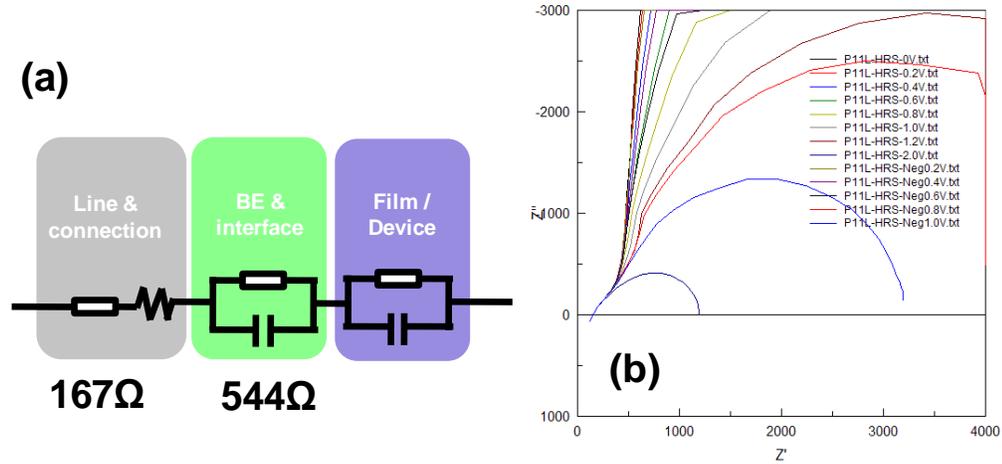

Figure 3. (a) Equivalent circuit from impedance analysis. (b) Cole-Cole plot revealing two semi-circles (one is with constant diameter (resistance) at high frequency part irrespective of states and voltages) and a constant intercept.

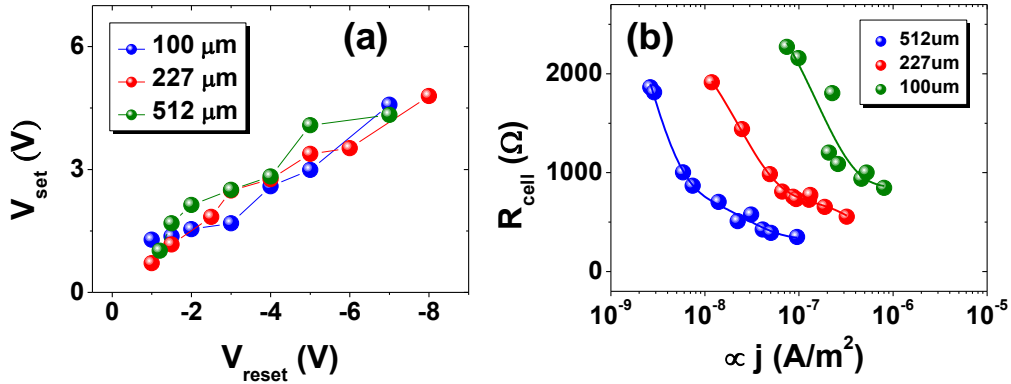

Figure 4. (a) $V_{set}$ (off-switching voltage) *vs.* $V_{reset}$ (maximum negative voltage used for on-switching) for various sizes, showing no size effect. (b) $R_{4pt}$ *vs.* $j$ (current density = $I/A$) for various sizes. All the values are defined at off-switching event ($V_{set}$).

This 4-point method was also used to explore size dependence of switching voltage and LRS (HRS dependence is already shown in **Chapter II**). As shown in **Figure 4a**, $V_{set}$ increases with $V_{\text{reset}}$, and the quantitative relation is identical for all sizes. This reveals switching voltage is irrelevant to device lateral size and thus device switching properties are uniform. Furthermore, we define current density as $j=I/A$ (A is device area) and use this parameter to investigate size effect of LRS. Differently from most works in the RRAM literature which generally believe a fair comparison is drawn under "same voltage", we believe the fair comparison should be better performed under "same current density". Under this spirit, resistance can be rewritten as:

$$R_{4pt} = R_0 + \frac{V_m^*}{I} = R_0 + \frac{1}{j}\frac{V_m^*}{A}$$

**Figure 4b** shows the plot of $R_{4\text{pt}}$ *vs.* $j$, where a higher current density naturally corresponds to a lower resistance for all sizes. This is equivalent to the fact that a lower LRS needs more current (voltage) to be switched off. For a constant current density, a larger device size has a lower resistance, following $R\sim1/A$ Ohm's law scaling.

# Appendix V. Additional Impedance Data for Nanometallic Films

➤ **Mo/Si₃N₄:Cr/Pt Nanometallic Device**

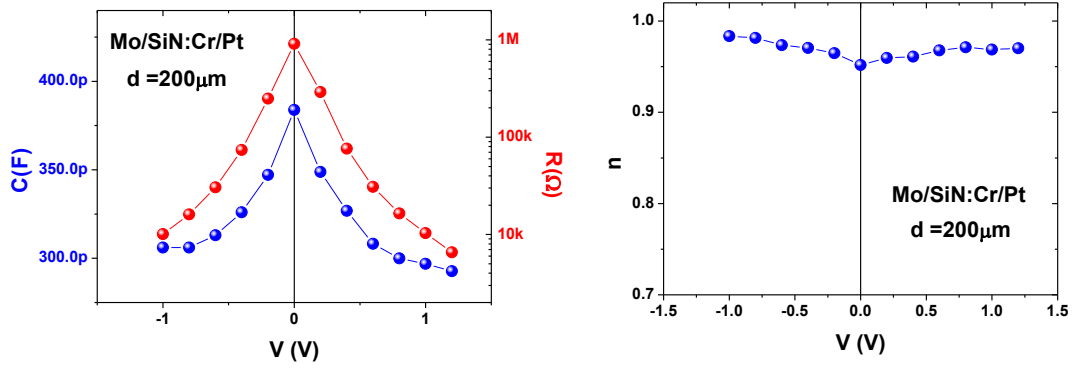

Figure 1. Mo/Si₃N₄:Cr/Pt device ($f_{Cr}$=4%, $d$=200 μm, $\delta$=10 nm)

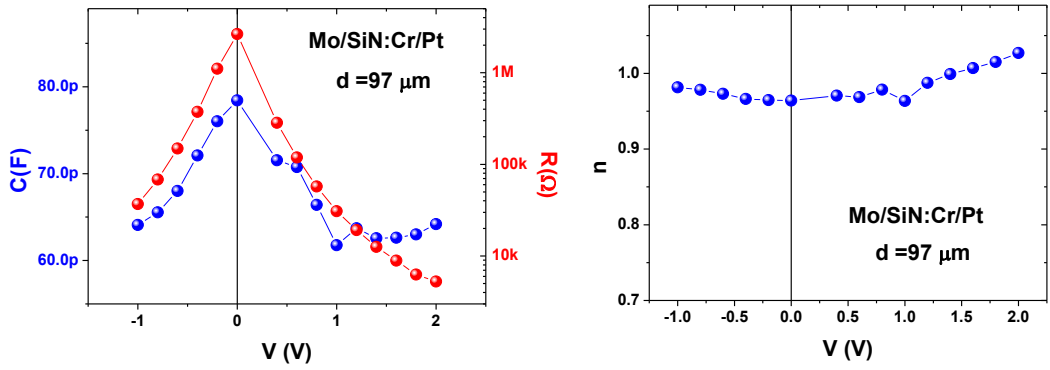

Figure 2. Mo/Si₃N₄:Cr/Pt device ($f_{Cr}$=4%, $d$=97 μm, $\delta$=10 nm)

➤ **Mo/Si₃N₄:Pt/Pt Nanometallic Device**

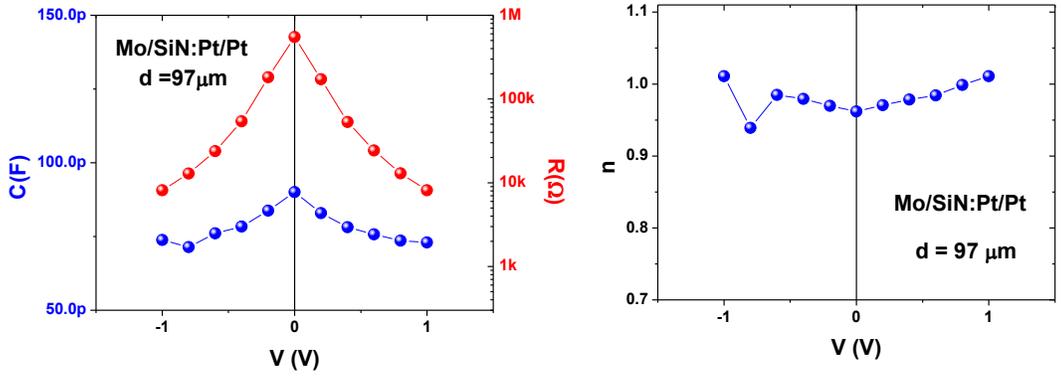

Figure 3. Mo/Si₃N₄:Pt/Pt device ($f_{Pt}$=5%, $d$=97 μm, $\delta$=10 nm)

➢ **Mo/Si₃N₄:Al/Pt Nanometallic Device**

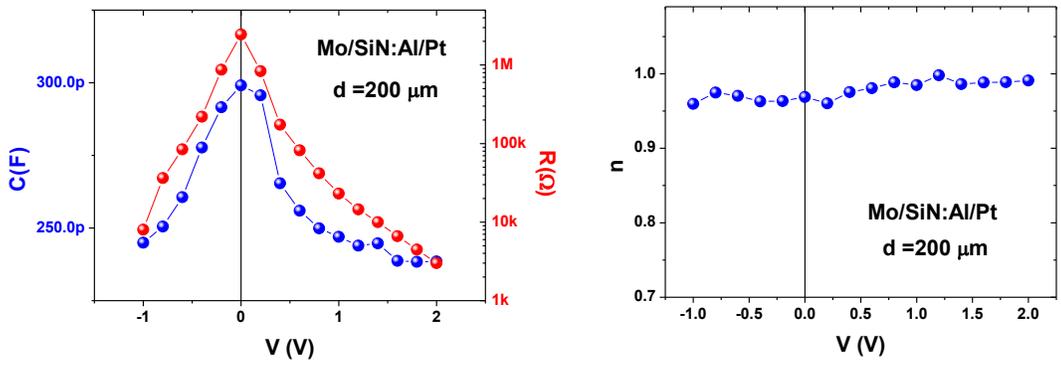

Figure 4. Mo/Si₃N₄:Al/Pt device ($d$=200 μm, $\delta$=10 nm)

➢ **Mo/Si₃N₄:Ta/Pt & Ta/Si₃N₄:Ta/Pt Nanometallic Device**

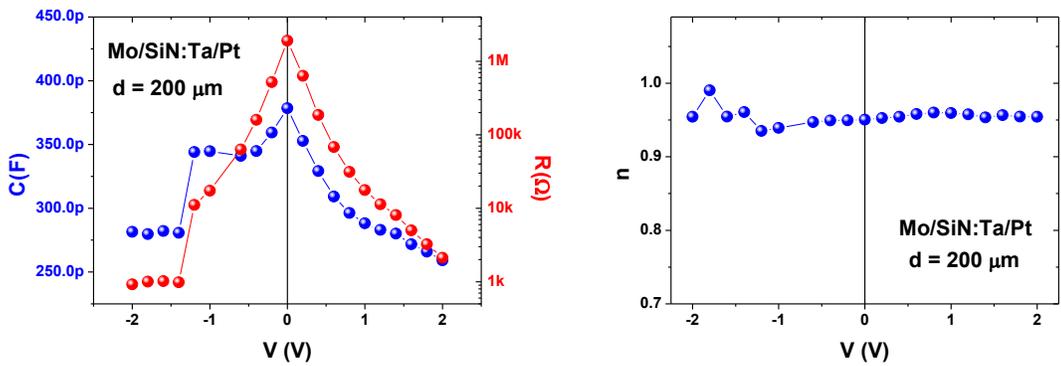

Figure 5. Mo/Si₃N₄:Ta/Pt device ($d$=200 μm, $\delta$=10 nm)

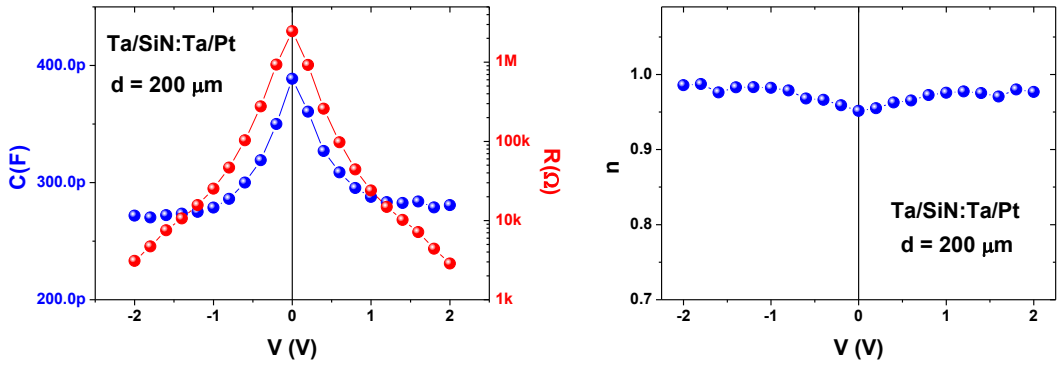

Figure 6. Ta/Si$_3$N$_4$:Ta/Pt device ($d$=200 μm, $\delta$=10 nm)

➤ **Mo/SiO$_2$:Pt/Pt Nanometallic Device**

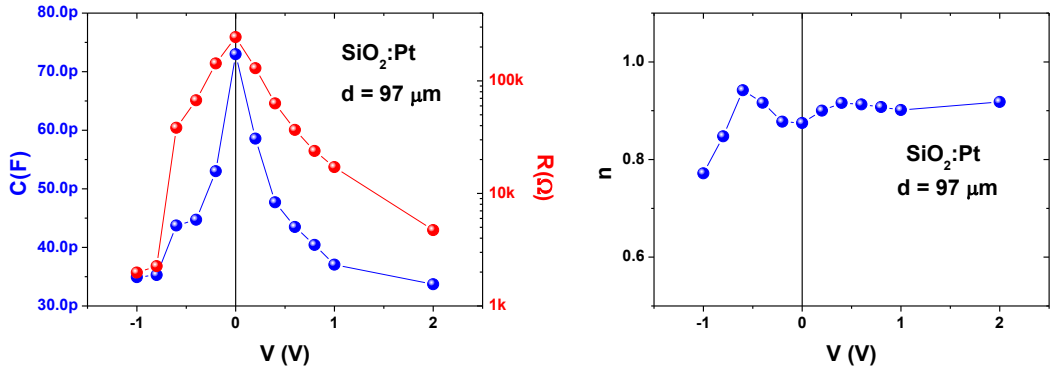

Figure 7. Mo/SiO$_2$:Pt/Pt device ($f_{Pt}$=20%, $d$=97 μm, $\delta$=20 nm)

➤ **Mo/Al$_2$O$_3$:Pt/Pt Nanometallic Device**

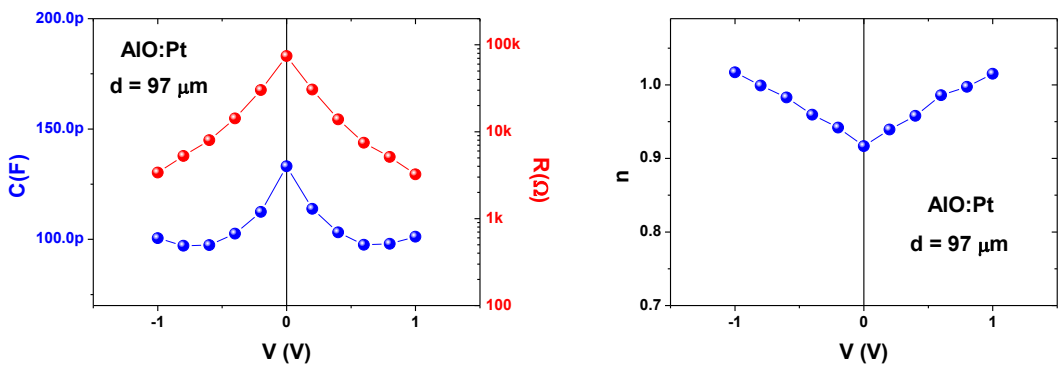

Figure 8. Mo/Al$_2$O$_3$:Pt/Pt device ($d$=97 μm, $\delta$=10 nm)

# Appendix VI. Optical Data

To extend the interrogation of nanometallic film to higher frequency domains, we measured IR, visible and UV spectra of our films (30 nm thick) from $10^{13}$ Hz to $10^{15}$ Hz. At these frequencies, the "capacitative" element in the equivalent circuit dominates, allowing these experiments to again probe the metal-rich clusters through their influence on the effective dielectric functions, which are now complex, $\omega$-dependent. As shown in **Figure 1a**, the IR spectra mostly reflect various vibrational modes of $Si_3N_4$ with little information concerning the clusters. In the UV-Vis range (**Figure 1b**), however, we see evidence of metallic clusters in a broad peak at 470 nm in the $Si_3N_4$:Cr films of higher $f$, which may be reasonably assigned to the plasmon resonance of metallic Cr. (Plasmon resonance is a light-induced cooperative oscillation of conduction electrons in a metallic particle. For Cr, $E_p$~2.5 eV or 496 nm[1].) This peak appears to be washed out below $f$=0.25, leaving nearly identical reflectance as that of $Si_3N_4$ ($f$=0). Note that the reflectance of $Si_3N_4$ is not flat in this region because of interfering reflections from the substrate (fused silica, which has a lower refractive index). The interference peak should lie at about 200 nm, which is difficult to avoid except in films (below 15 nm) too thin to give measurable signals. Therefore, a comparison with model calculations is needed to better understand these data.

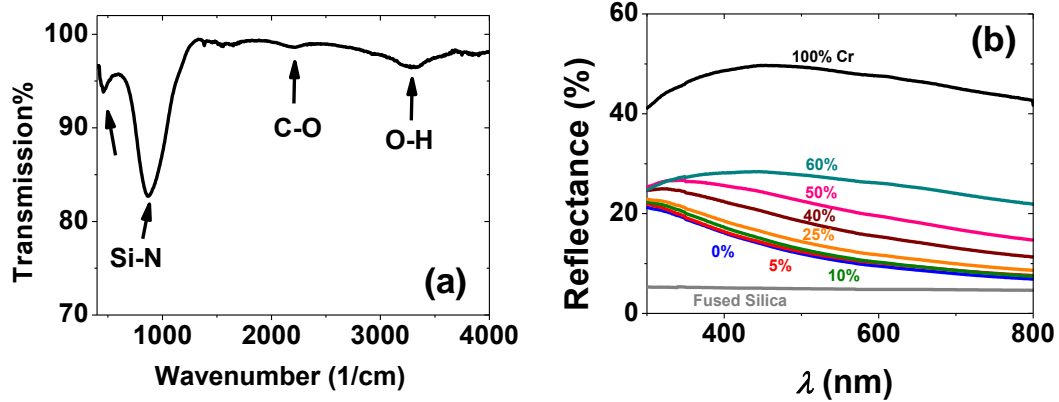

Figure 1. (a) IR transmission spectrum of $Si_3N_4$: 0.25Cr on KBr substrate. Two low frequency characteristic peaks corresponding to Si-N vibration mode comes from $Si_3N_4$ host; C-O and O-H peaks come from background absorption. (b) UV reflectance *vs.* wavelength of $(1-f)Si_3N_4$:$f$Cr films for various $f$.

As before, we employed Maxwell-Garnett's effective media theory[2] to calculate the dielectric functions and reflectance assuming a $Si_3N_4$ matrix with metallic Cr particles. (We used metallic dielectric functions from the reference library in TFCompanion software. The model prediction is an underestimate since no consideration is given to the particle size, which leads to underestimates of interparticle interactions and local field contributions at the particle/matrix interface.)[3] In agreement with our data, the calculation found a broad maximum at ~500 nm that does not emerge until $f$>0.2 (See **Figure 2a**). However, it also revealed a significant increase in the reflectance even at $f$<0.2, suggesting that the lack of such increase in our data is caused by the lack of metallicity of Cr-rich clusters in our films at $f$<0.2. To quantify this, we plot the integrated difference reflectance (the difference between $f$>0 and $f$=0 spectra) from 300 nm to 800 nm in **Figure 2b**. The calculated spectra portray a difference reflectance that linearly increases with $f$, in contrast to our data which display little increase until after $f$=0.25. (We also tried several other wavelength ranges for integration, which all showed the same trends as above.)

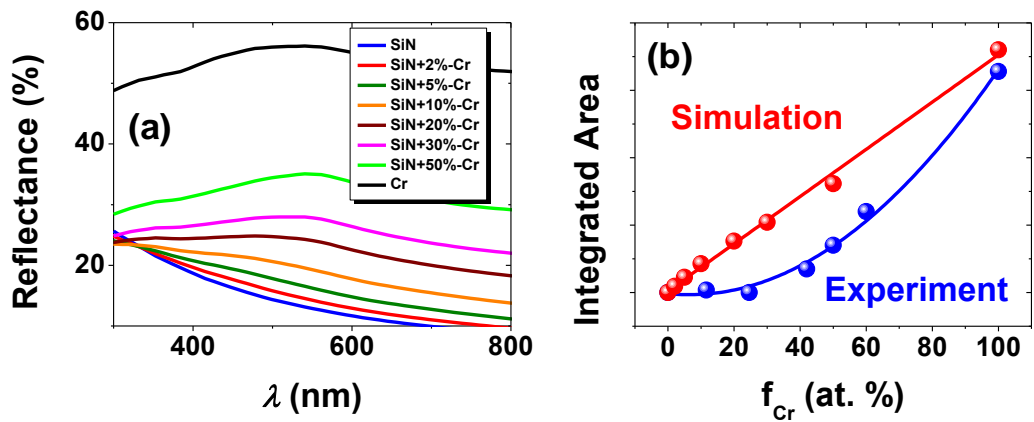

Figure 2. (a) Simulated reflection *vs.* wavelength for $(1-f)Si_3N_4$:$f$Cr films of various Cr concentration. (b) Integrated difference reflectance (area between $f>0$ reflectance and $f=0$ reflectance from 300 nm to 800 nm) *vs.* $f$. Up to $f=0.25$, measured difference reflectance is much smaller than simulated one (based on Maxwell-Garnett effective medium theory assuming metallic Cr particles) indicating absence of optically metallic inclusions at low $f$.

A parallel study of $SiO_2$:Pt films made previously[4] was also similarly analyzed and again revealed the same difference in the calculated *vs.* measured difference reflectance below $f=0.25$ (**Figure 3**). (No interference occurred in $SiO_2$:Pt films because the fused-silica substrate and $SiO_2$ have essentially the same refractive index, making it easy to discern the plasmon peak in the calculated spectra (**Figure 3a**), which is lacking in the measured spectra until after $f=0.3$ (**Figure 3c**)). Therefore, both studies indicate that optical metallicity of metal-rich clusters does not emerge until $f\sim0.25$.

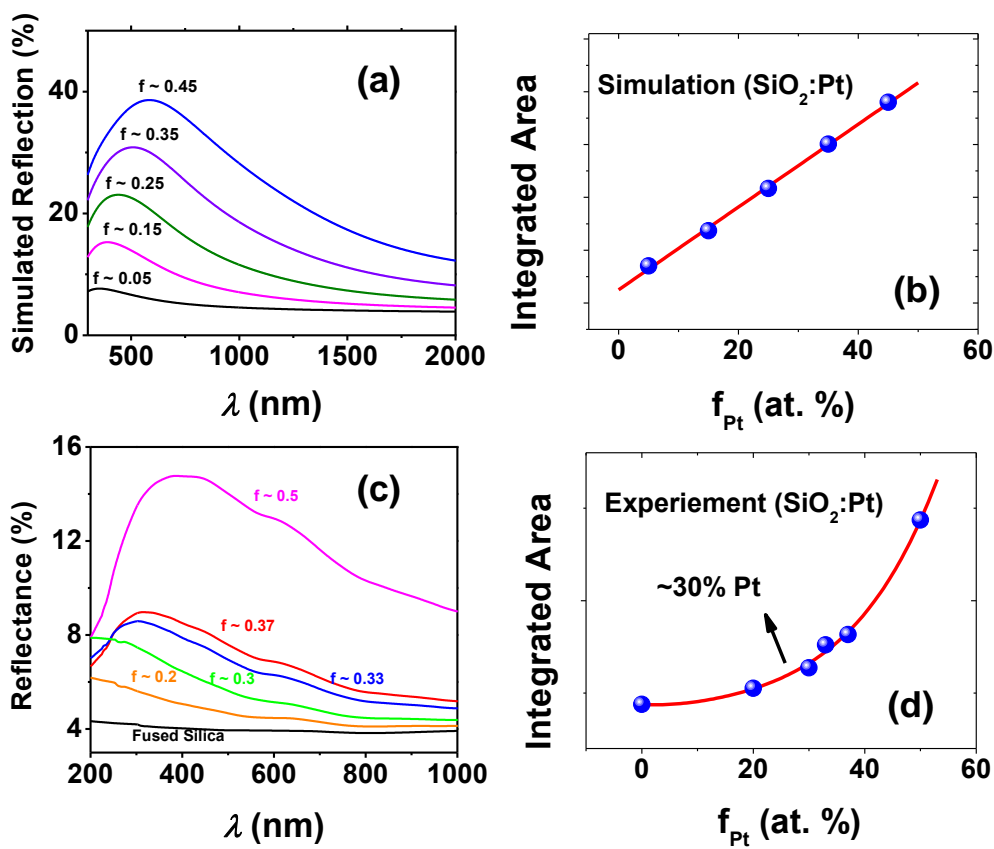

**Figure 3.** (a) Simulated reflection *vs.* wavelength for (1-*f*)SiO₂:*f*Pt films of various Pt concentration. (b) Integrated area between *f* and *f*=0 curves in (a), from 300 nm to 800 nm (c) Experimental data of (1-*f*)SiO₂:*f*Pt reflection.[4] (d) Integrated area between *f* and *f*=0 curves in (c), from 300 nm to 800 nm.

# Appendix VII. Additional Materials Characterization

## ➤ Low energy electron energy loss spectroscopy

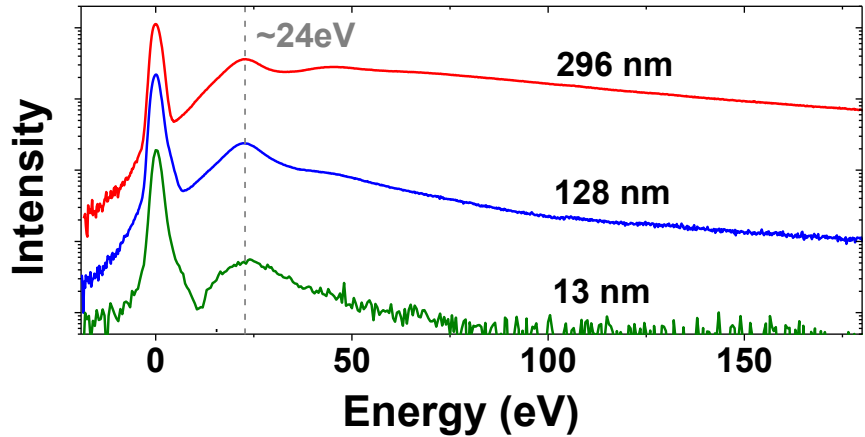

**Figure 1.** Low energy EELS data for pure $Si_3N_4$ film with various thickness ($\delta$=13 nm, $\delta$=128 nm, $\delta$=296 nm). A thin film ($\delta$=13 nm) only has first harmonic Plasmon resonance peak at 24 eV, corresponding to valence electron excitation of $Si_3N_4$. A thick film ($\delta$=296 nm) has multiple peaks, corresponding to higher harmonic Plasmon resonance peak at $24 \times N$ eV ($N$=integer).

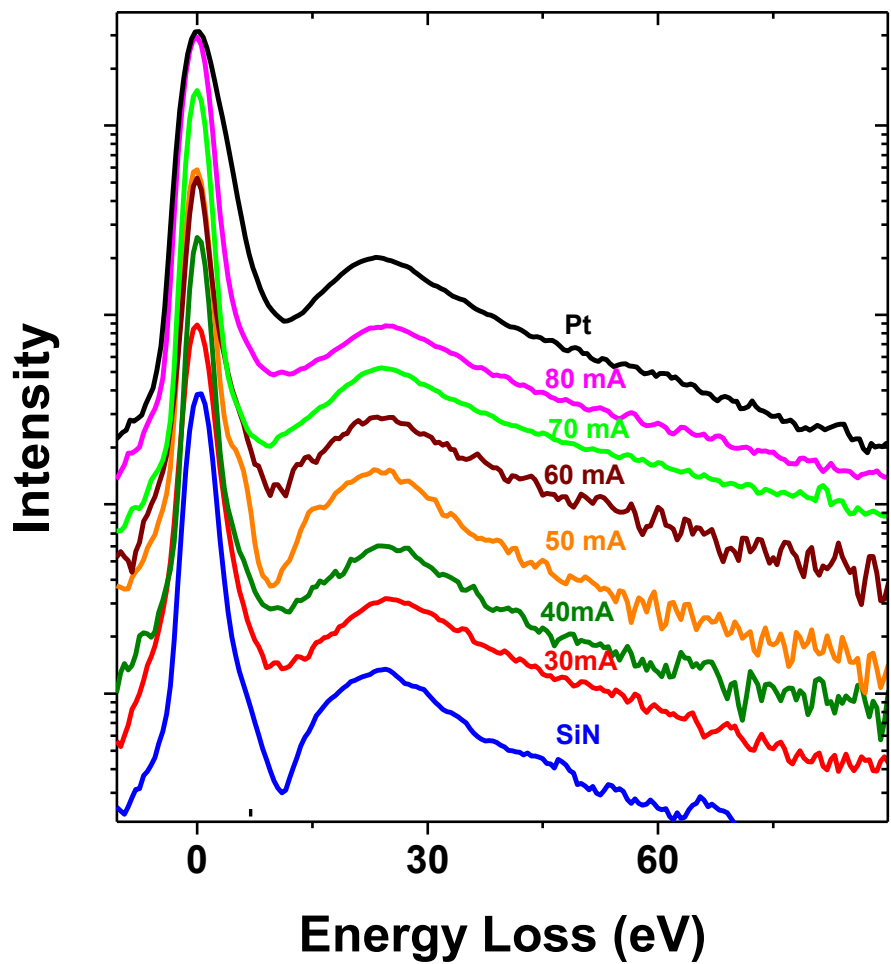

**Figure 2.** Low energy EELS data for nanometallic $Si_3N_4$:Pt film ($\delta \approx 10$ nm) with various $f_{Pt}$ (labeled as sputtering power). All Plasmon peaks are located at ~24 eV. Low energy EELS is thus not a good method to detect free electron (plasmon) information in $Si_3N_4$:Pt film.

➢ **TEM of SiO₂:Pt films**

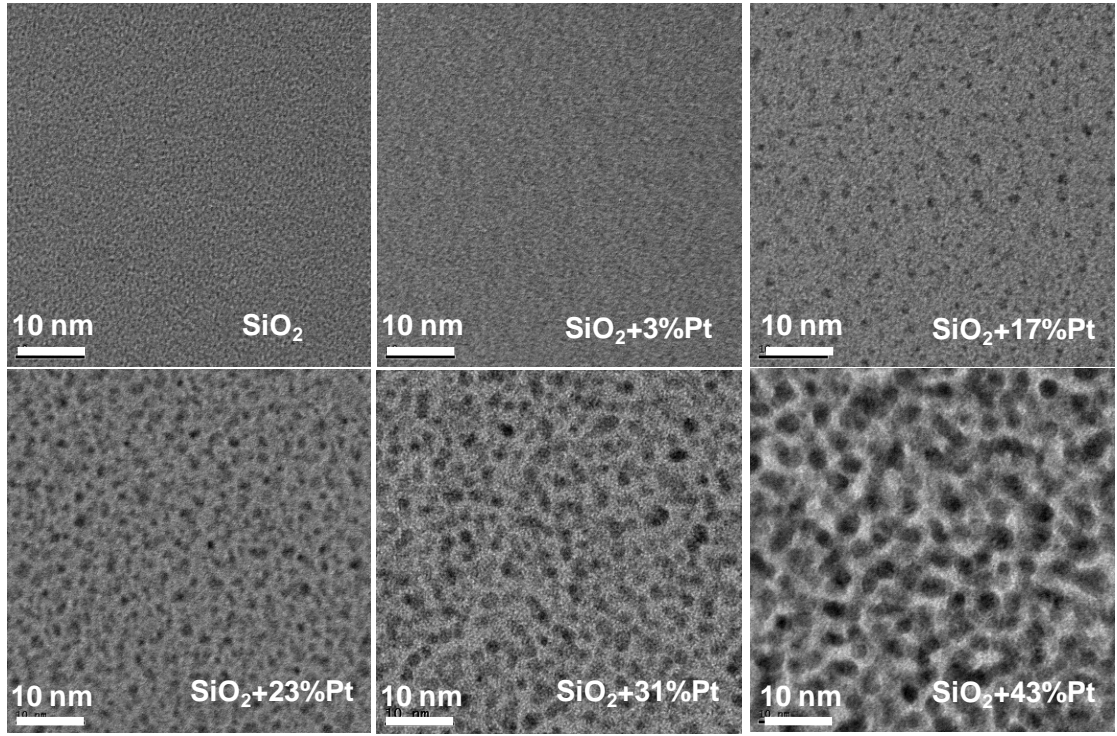

**Figure 3.** TEM for various SiO₂:Pt films.

## ➢ FIB Cutting for TEM (Lamella Preparation)

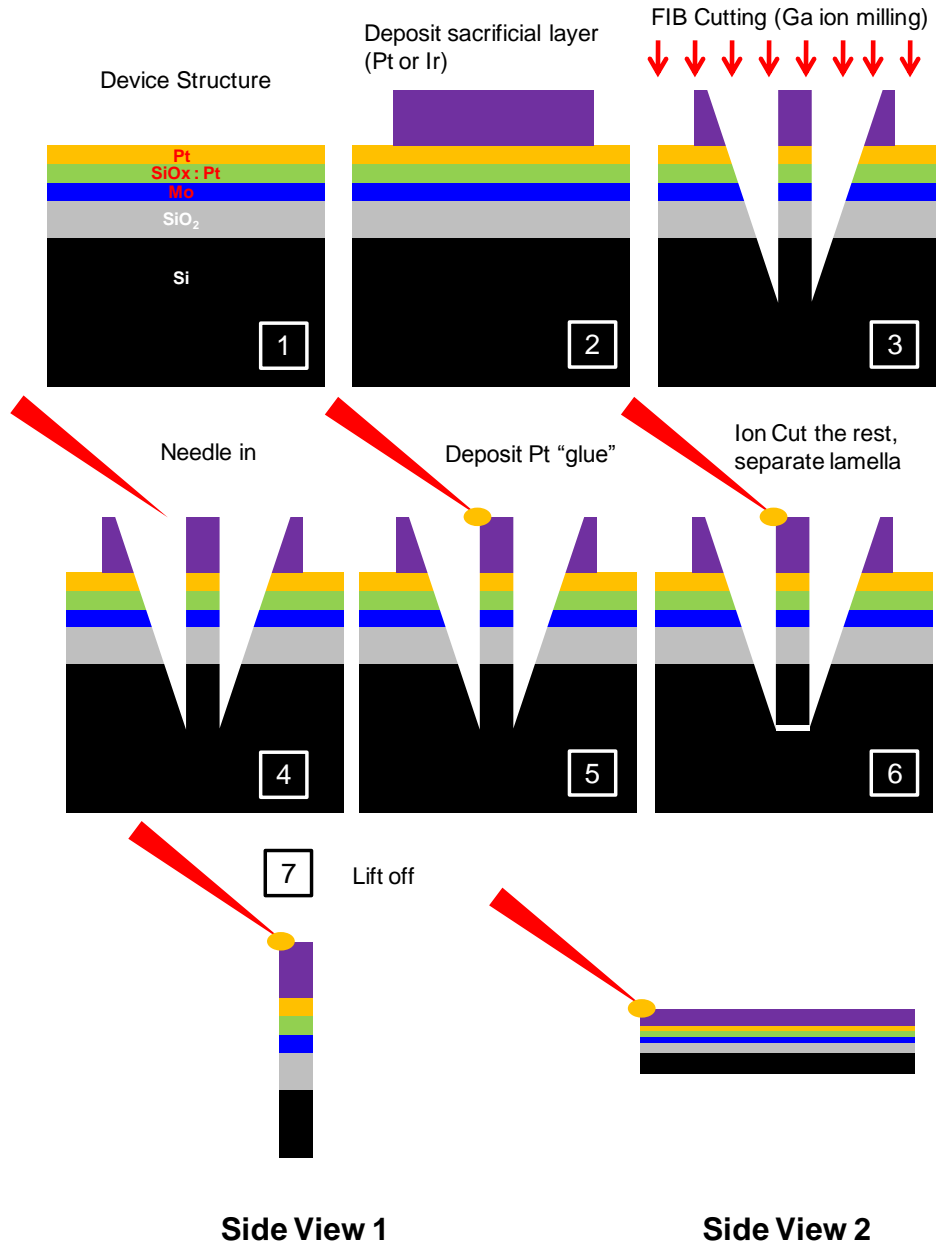

Device Structure

Deposit sacrificial layer
(Pt or Ir)

FIB Cutting (Ga ion milling)

Pt
SiOx : Pt
Mo
SiO₂
Si

Needle in

Deposit Pt "glue"

Ion Cut the rest,
separate lamella

7  Lift off

**Side View 1**

**Side View 2**

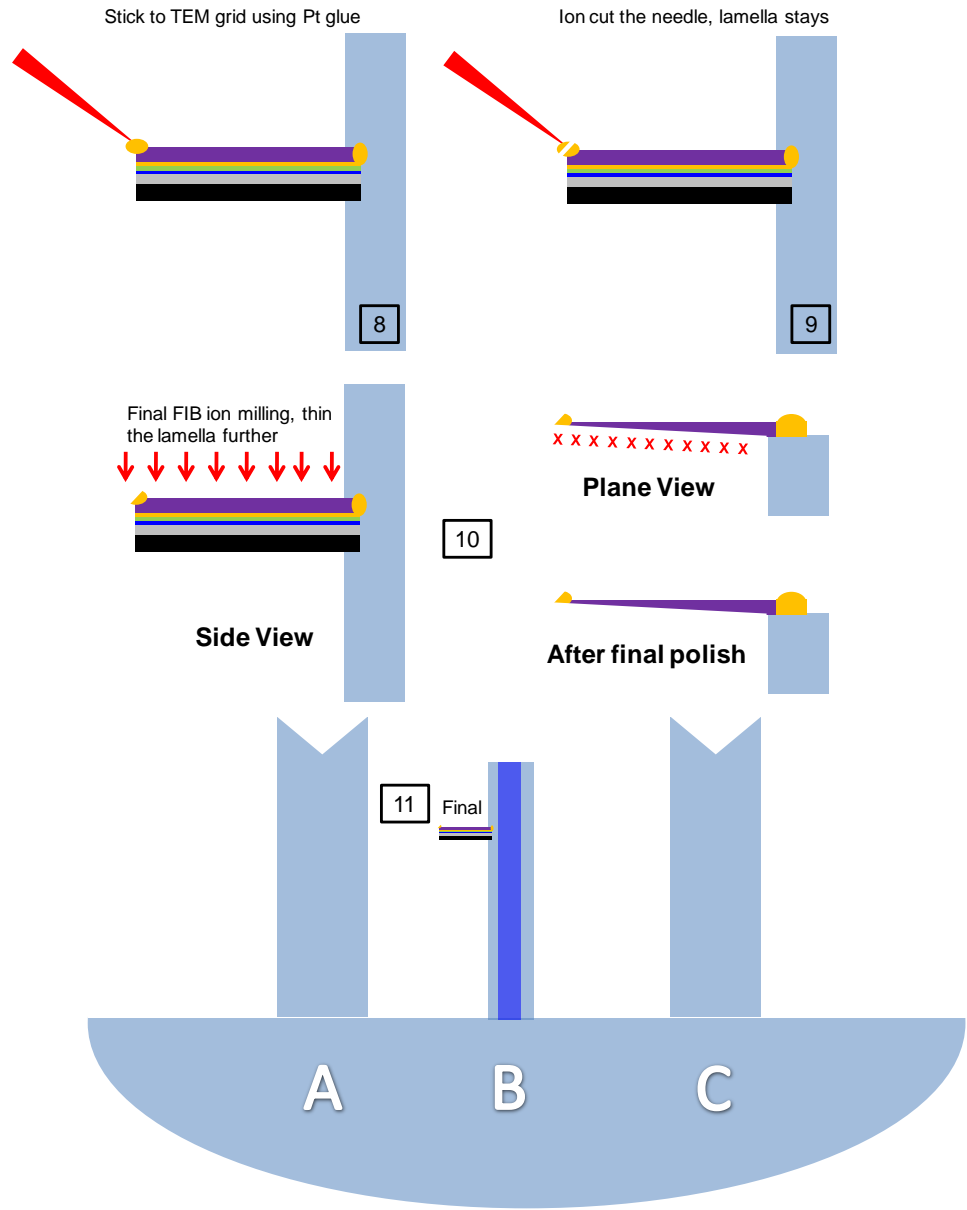

**Figure 4.** FIB cutting procedure for cross sectional TEM (Lamella preparation).

**E-Beam View**

**I-Beam View**

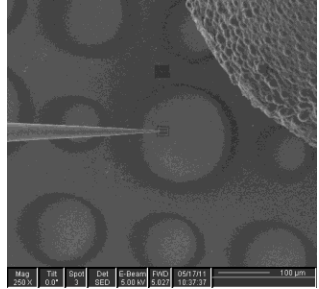
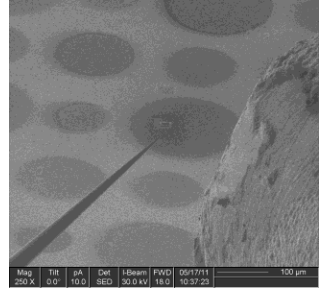
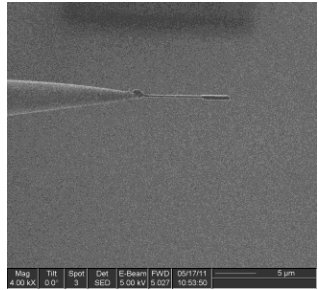
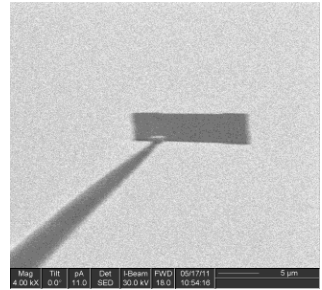

**Step 6&7**

**E-Beam View**

**I-Beam View**

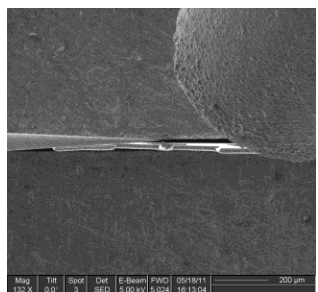
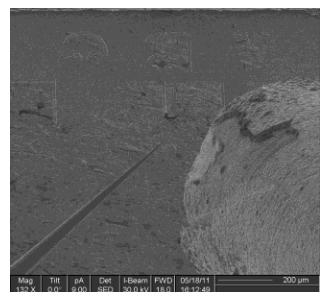
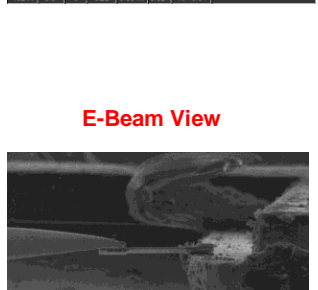
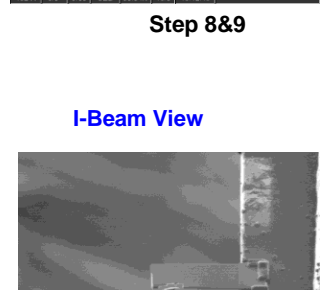

**Step 8&9**

**E-Beam View**

**I-Beam View**

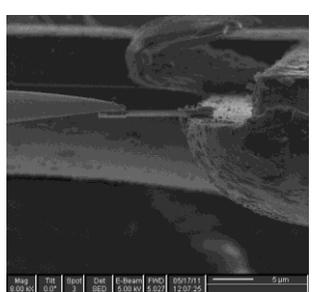
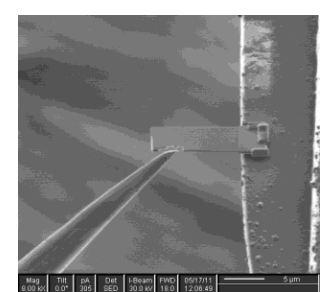
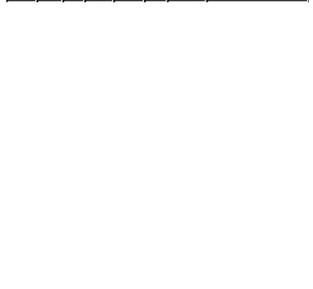
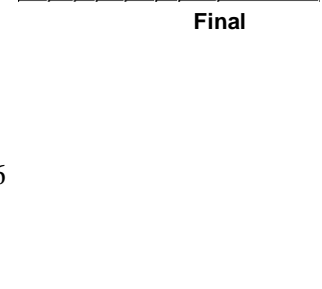

**Final**

**Figure 5.** E-beam view and I-beam view during Lamella preparation.

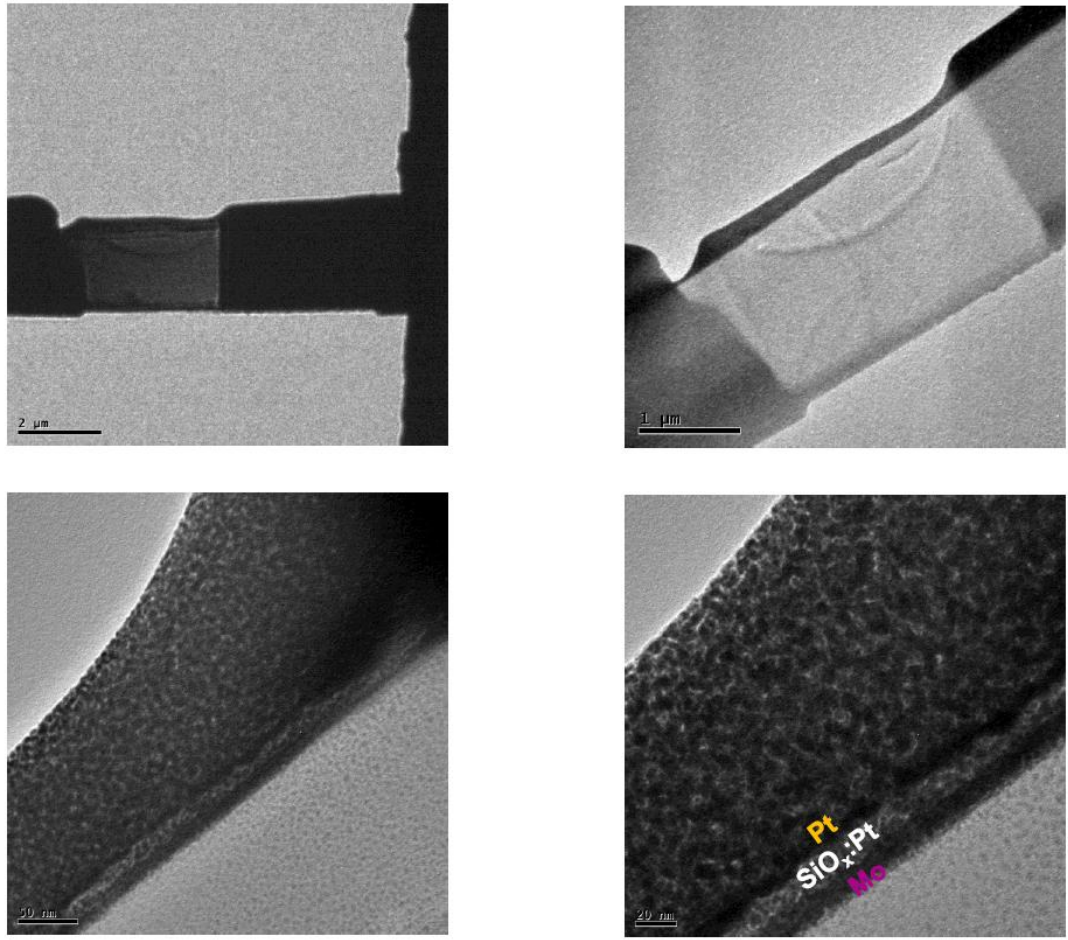

**Figure 6.** Final cross sectional TEM

# Appendix VIII. E-bunch images

## Induce current at TE edges

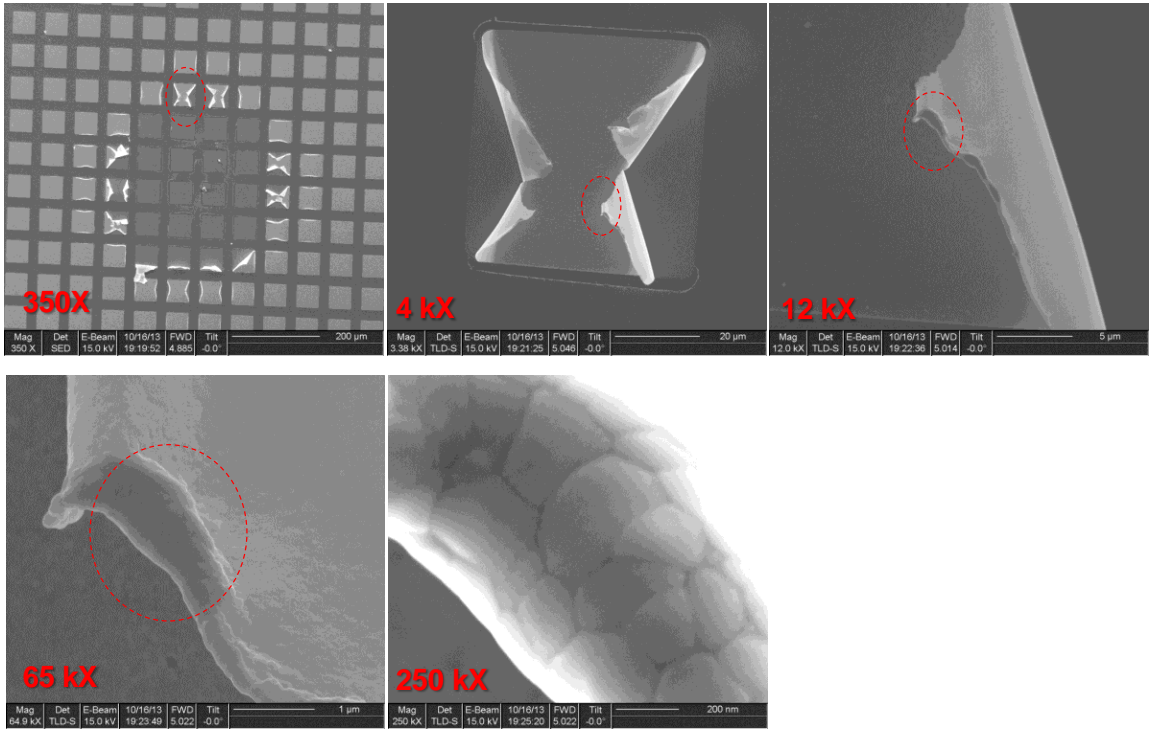

**Figure 1.** SEM images of top electrode morphology (at the edge near crack region) under different magnifications. There is a clear evidence that Pt grain grows near the edge (along **E**-field direction of ebunch), which implies a large current concentrated near the electrode edge. Red circles are the spots magnified in the next figures.

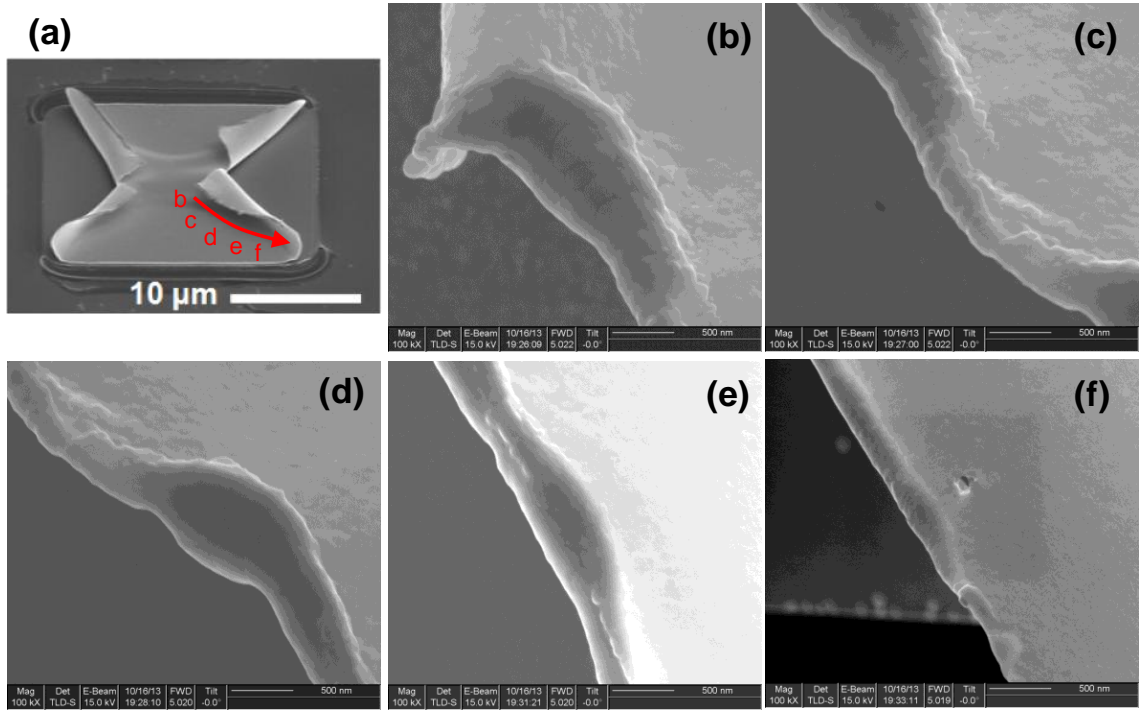

**Figure 2.** SEM images of top electrode morphology along the edge. (a) Overview of Pt TE. (b)-(f) Sequential images along edge indicated by red arrow in (a). Pt grain growth reaches a maximum near the center (b), implying current is maximum near the center.

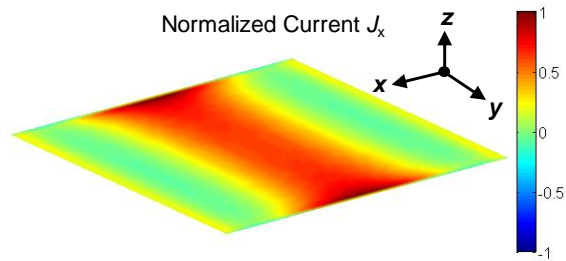

**Figure 3.** Simulated current distribution on top electrode. A linearly polarized excitation along $x$ direction was used.

# Appendix IX. Derivation of Variable Range Hopping (VRH)

## Xiang Yang, I-Wei Chen

> **Conductivity in the absence of electrical field**

Conductivity is proportional to the probability of tunneling between sites separated by a distance $r$:

$$P \propto \exp(-2\alpha r - \Delta E / kT)$$

This formula is based on the fact that in disordered system, wave functions are localized on a single site, the probability of the jump, either by phonon-assisted transition over the barrier or tunneling through the barrier, is proportional to the overlap of the wave function on the two neighboring sites, which falls off exponentially as $\exp(-\alpha r)$, where $r$ is the hopping distance and $\alpha$ is a decay constant ($1/\alpha \sim$ localization length $\xi$). This formula is quite meaningful: usually sites with identical energy ($\Delta E = 0$) are spatially far apart, therefore tunneling probability coming from elastic transport is quite small; on the other hand, electrons can sacrifice a small energy $\Delta E$ but reach more abundant states within smaller $r$, thus keep a larger probability. Therefore, such probability can be maximized by finding the optimized energy $\Delta E_{max}$.

Assume a density of state (DOS) as $N(E)$, then the number of states within sphere of hopping distance is: $4/3\pi r^3 N(E)\Delta E$. Then let it be constant, we obtain the relation between $r$ and $\Delta E$, given same available conduction states, as:

$$r = \frac{C}{\left(N(E)\Delta E\right)^{1/3}}$$

Then make partial derivative on probability to obtain the optimized $\Delta E$:

$$\frac{\partial P}{\partial \Delta E} = 0, \text{ or } \frac{\partial \exp(-2\alpha r - \Delta E / kT)}{\partial \Delta E} = 0$$

We can find:

$$\Delta E_{max} = \left(\frac{2\alpha CkT}{3N(E)^{1/3}}\right)^{3/4}$$

Plug it back to original form, we can find the maximum probability as:

$$P \propto \exp\left[ -\frac{4}{3^{3/4}} \frac{(2C\alpha)^{3/4}}{N(E)^{1/4}(kT)^{1/4}} \right] = \exp\left[ -\left(\frac{T_0}{T}\right)^{1/4} \right], \quad T_0 = \frac{8\sqrt{2}C^3\alpha^3}{27N(E)}$$

> **Conductivity under electrical field**

Consider a general case in which tunneling probability in forward and backward directions are:

$$P_+ = \exp\left( -2\alpha r - \frac{\Delta E - \beta e F r}{kT} \right)$$

$$P_- = \exp\left( -2\alpha r - \frac{\Delta E + \beta e F r}{kT} \right) = P_+ \exp\left( -\frac{2\beta e F r}{kT} \right)$$

This relation will be valid as electrical field $F$ is small, *i.e.*

$$\beta e F r \leq \Delta E$$

Under high field circumstance, the jumping probability reduced to:

$$P_+ = \exp(-2\alpha r)$$

$$P_- = 0$$

Before optimization of tunneling probability, we still have:

$$4/3 \pi r^3 N(E) \Delta E = const.$$

To simplify the derivation, we define the following dimensionless energy, distance, and field as:

$$\Delta \tilde{E} = \frac{\Delta E}{kT}, \quad \tilde{R} = 2\alpha r, \quad \tilde{F} = \frac{\beta e F}{2\alpha kT}$$

Therefore, we can rewrite the above equation as:

$$P_+ = \exp\left( -\tilde{R} - \Delta \tilde{E} + \tilde{F}\tilde{R} \right)$$

$$P_- = P_+ \exp\left( -2\tilde{F}\tilde{R} \right), \text{ if } \tilde{F}\tilde{R} \leq \Delta \tilde{E}$$

Or high field condition:

$$P_+ = \exp\left( -\tilde{R} \right)$$

$$P_- = 0, \text{ if } \tilde{F}\tilde{R} > \Delta \tilde{E}$$

With the constraint relation:

$$\Delta \tilde{E} \tilde{R}^3 = const. \frac{6\alpha^3}{\pi k T N(E)} = \frac{A}{T}$$

To optimize the forward tunneling probability, let $\partial \ln P_+ / \partial \tilde{R} = 0$, given the constraint differential equation: $\partial \left( \Delta \tilde{E} \tilde{R}^3 \right) / \partial R = 0$, we have a clean relation:

- $\Delta \tilde{E} = \frac{1}{3} \tilde{R}$, if $\tilde{F} \rightarrow 0$ (low field limit)

- $\Delta \tilde{E} = \frac{1}{3} \left( 1 - \tilde{F} \right) \tilde{R}$, if $0 \leq \tilde{F} \leq 1/4$ (moderate field)

- $\Delta \tilde{E} = \tilde{F} \tilde{R}$, if $\tilde{F} \geq 1/4$ (high field)

By solving these relation with constraint condition for $\Delta \tilde{E}$ and $\tilde{R}$, we can obtain the following:

- $\tilde{R} = \left( \frac{3A}{T} \right)^{1/4}$, if $\tilde{F} \rightarrow 0$ (low field limit)

- $\tilde{R} = \left( \frac{3A}{T \left( 1 - \tilde{F} \right)} \right)^{1/4}$, if $0 \leq \tilde{F} \leq 1/4$ (moderate field)

- $\tilde{R} = \left( \frac{A}{T \tilde{F}} \right)^{1/4}$, if $\tilde{F} \geq 1/4$ (high field)

Then plug these relation back to original probability equation, we can obtain:

- $P_+ = \exp \left[ -\frac{4}{3} \left( \frac{3A}{T} \right)^{1/4} \right]$, if $\tilde{F} \rightarrow 0$ (low field limit)

- $P_+ = \exp \left[ -\frac{4}{3} \left( 1 - \tilde{F} \right) \left( \frac{3A}{T \left( 1 - \tilde{F} \right)} \right)^{1/4} \right]$, if $0 \leq \tilde{F} \leq 1/4$ (moderate field)

- $P_+ = \exp \left[ -\left( \frac{A}{T \tilde{F}} \right)^{1/4} \right]$, if $\tilde{F} \geq 1/4$ (high field)

Next, let's consider the conductivity. Current density can be written as:

$$j = e N(E) v k T r \left( P_+ - P_- \right)$$

Therefore conventional conductivity is simply:

$$\sigma = \frac{j}{F} = \frac{e N(E) v k T r}{F} \left( P_+ - P_- \right)$$

1. For zero field limit ($\tilde{F} \to 0$), noticing that $\left(1 - e^{-x}\right)/x \to 1$

$$\sigma = \frac{e^2 N(E)\beta\nu}{(2\alpha)^2} \frac{\tilde{R}}{\tilde{F}} P_+ \left(1 - \exp\left(-2\tilde{F}\tilde{R}\right)\right) = 2BP_+ \tilde{R}^2 = 2B\left(\frac{3A}{T}\right)^{1/2} \exp\left[-\frac{4}{3}\left(\frac{3A}{T}\right)^{1/4}\right]$$

where $B = \dfrac{e^2 N(E)\beta\nu}{(2\alpha)^2}$

2. For moderate field ($0 \le \tilde{F} \le 1/4$), results becomes complicated but still completely analytical:

$$\sigma = \frac{B}{\tilde{F}}\left(\frac{3A}{T\left(1-\tilde{F}\right)}\right)^{1/4} \times \exp\left[-\frac{4}{3}\left(1-\tilde{F}\right)\left(\frac{3A}{T\left(1-\tilde{F}\right)}\right)^{1/4}\right] \times \left[1 - \exp\left(-2\tilde{F}\left(\frac{3A}{T(1-\tilde{F})}\right)^{1/4}\right)\right]$$

$$= \frac{2\alpha kTB}{\beta eF}\left(\frac{3A}{T-\dfrac{\beta eF}{2\alpha k}}\right)^{1/4} \times \exp\left[-\frac{4}{3}\left(1-\frac{\beta eF}{2\alpha kT}\right)\left(\frac{3A}{T-\dfrac{\beta eF}{2\alpha k}}\right)^{1/4}\right] \times \left[1 - \exp\left(-\frac{\beta eF}{\alpha kT}\left(\frac{3A}{T-\dfrac{\beta eF}{2\alpha k}}\right)^{1/4}\right)\right]$$

3. For high field ($\tilde{F} \ge 1/4$), we can obtain a weakly temperature dependent conductivity as:

$$\sigma = \frac{B}{\tilde{F}}\left(\frac{A}{T\tilde{F}}\right)^{1/4} \times \exp\left[-\left(\frac{A}{T\tilde{F}}\right)^{1/4}\right]$$

$$= \frac{2\alpha kTB}{\beta eF}\left(\frac{2\alpha kA}{\beta eF}\right)^{1/4} \times \exp\left[-\left(\frac{2\alpha kA}{\beta eF}\right)^{1/4}\right]$$

# Appendix X. Justification of the Semi-empirical Parametric Model

Here, we provide a quantitative argument on the proposed semi-empirical model for HRS in **Chapter V**:

$$R(T) = R_0 \exp\left(-\left(\frac{T}{T_0}\right)^n\right)$$

Intuitively, transport in an insulator can be visualized as a process that electron passes through a "barrier". Here the "barrier" might be a physical energy barrier or merely a synonym of a terminology describing how an electron experiences resistance in the presence of external stimuli ($E$-field, $k_B T$, $etc.$). Generally speaking, "barrier" affects the transport in an exponential manner. Therefore, we have

$$I \sim \exp\left(-C\Phi_b^m\right)$$

At an elevated temperature, electron transport inside insulator is assisted by $k_B T$ and we incorporate such effect into an equivalent "barrier lowering" term, namely

$$\Phi_b = \Phi_{b0} - AT^n$$

The coefficient $A$ should be very small because in reality, "barrier" modification should be tiny ($AT^n \rightarrow 0$). Therefore, we obtain the following empirical form:

$$I \sim \exp\left[-C\left(\Phi_b - AT^n\right)^m\right]$$

Keeping the leading term ($AT^n \rightarrow 0$), we can get:

$$I \sim \exp\left[-C\Phi_b^m\left(1 - \frac{mAT^n}{\Phi_b}\right)\right] = I_0\left(V, \Phi_b, m\right)\exp\left[\left(\frac{T}{T_0}\right)^n\right]$$

or

$$\rho(T) = \rho_0 \exp\left(-\left(\frac{T}{T_0}\right)^n\right)$$

As one physical example, such barrier lowering effect could stem from thermal motion induced barrier oscillation. Mathematically, the product of barrier factor $C\Phi_b^m$ can be approximated with a first-order Taylor expansion around atom equilibrium positions:

$$C\Phi_B^m = \left(C_0 - \sum_i \chi_i \Delta x_i\right)\left(\Phi_{B0} - \sum_i \gamma_i \Delta x_i\right)^m$$

$$\approx C_0 \Phi_{B0}^m - \sum_i \Delta x_i \left(C_0 m \gamma_i \Phi_{b0}^{m-1} + \chi_i \Phi_{b0}^m\right)$$

A Boltzmann distribution of the vibration probability density can be used,

$$p(\Delta x_i) = \sqrt{\frac{A_i}{2\pi T}} \exp\left(-\frac{A_i}{2T}\Delta x_i^2\right)$$

We can then translate macroscopic current is $I(T)\sim\exp[-C\Phi_b(T)]$ to average current of microscopic current:

$$I(T) \sim \left\langle \exp(-C\Phi_B^m)\right\rangle_T$$

$$= \prod_i \int_{-\infty}^{\infty} \exp\left[-C_0 \Phi_{B0}^m + \sum_i \Delta x_i \left(C_0 m \gamma_i \Phi_{b0}^{m-1} + \chi_i \Phi_{b0}^m\right)\right]$$

$$\times \sqrt{\frac{A_i}{2\pi T}} \exp\left(-\frac{A_i}{2T}\Delta x_i^2\right) d^3 \Delta x_i$$

The integral can be solved as

$$\sqrt{\frac{A_i}{2\pi T}} \int_{-\infty}^{\infty} \exp\left[\Delta x_i \left(C_0 m \gamma_i \Phi_{b0}^{m-1} + \chi_i \Phi_{b0}^m\right) - \frac{A_i}{2T}\Delta x_i^2\right] d^3 \Delta x$$

$$= \exp\left(\frac{C_0^2 \left(m \gamma_i \Phi_{b0}^{m-1}\right)^2 T}{2 A_i} + \frac{\chi_i^2 \Phi_{b0}^{2m} T}{2 A_i}\right)$$

After integral is carried out in all three dimensions, the final value of the current is

$$I(T) \sim \exp\left[-C_0 \Phi_{B0}^m + 3T \sum_i \frac{C_0^2 \left(m \gamma_i \Phi_{b0}^{m-1}\right)^2 + \chi_i^2 \Phi_{b0}^{2m}}{2 A_i}\right]$$

leading to a $n=1$ law $\rho(T) = \rho_0 \exp\left(-AT\right)$.

# Appendix XI. Other VRH-like Hopping Models

In this part, we discuss some possible crossovers between (other) different hopping models, which may possibly exist in **Chapter V**.

## 1D Mott Variable Range Hopping

We have shown that the 3D Mott-VRH can describe the high temperature conduction data very well. However, we found the 1D Mott-VRH seems to work equally well. The 1D VRH is described by the following equations, one under zero bias and the other under a large bias, respectively:

$$I \propto \exp\left[-\sqrt{\frac{T_{1D}}{T}}\right], \text{ where } T_{1D} = \frac{1}{k_B N_u \zeta}$$

$$I \propto \exp\left[-\sqrt{\frac{F_{1D}}{F}}\right], \text{ where } F_{1D} = \frac{1}{e N_u \zeta^2}$$

Using a similar fitting procedure as described previously for the 3D Mott-VRH, we can obtain the DOS and the localization length for the 1D Mott-VRH, which are summarized in **Table 1**, **Table 2** & **Table 3**.

| Thickness | $\delta$=7 nm | $\delta$=9 nm | $\delta$=10 nm | $\delta$=17 nm |
|---|---|---|---|---|
| $T_0^{1/2}$ (K$^{1/2}$) | 229 | 290 | 294 | 358 |
| $V_0^{1/2}$ (V$^{1/2}$) | 8.1 | 11.4 | 13.5 | 16.2 |
| $N_{u,1D}$ ($10^6$ cm$^{-1}$eV$^{-1}$) | 4.6 | 2.8 | 3.2 | 1.2 |
| $\zeta_{1D}$ (nm) | 0.48 | 0.49 | 0.41 | 0.71 |

**Table 1.** Values of localization length $\zeta$ for various thickness ($f_{Pt}$=4%) from 1D Mott-VRH.

| Composition | $f_{Pt}$ =4 at.% | $f_{Pt}$ =19 at.% | $f_{Pt}$ =27 at.% |
|---|---|---|---|
| $T_0^{1/2}$ (K$^{1/2}$) | 294 | 303 | 199 |
| $V_0^{1/2}$ (V$^{1/2}$) | 13.5 | 8.43 | 3.4 |
| $N_{u,1D}$ ($10^6$ cm$^{-1}$eV$^{-1}$) | 3.2 | 1.1 | 0.97 |
| $\zeta_{1D}$ (nm) | 0.41 | 1.11 | 3.03 |

**Table 2.** Values of density of states $N_u$ localization length $\zeta$ and optimized hopping distance $r_0$ for various concentration $f_{Pt}$ ($\delta$=10 nm) from 1D Mott-VRH.

| RS ($R_{2K}$) | 600MΩ | 55MΩ | 2MΩ | 300kΩ | 8.7kΩ | 2.5kΩ |
|---|---|---|---|---|---|---|
| $T_0^{1/2}$ (K$^{1/2}$) | 294 | 277 | 56 | 39 | 14 | 2.1 |
| $N_{u,1D}$ (cm$^{-1}$eV$^{-1}$) | $3.2\times10^6$ | $3.7\times10^6$ | $9.1\times10^7$ | $1.9\times10^8$ | $1.5\times10^9$ | $6.6\times10^{10}$ |
| $\zeta_{1D}$ (nm)* | 0.41 | 0.41 | 0.41 | 0.41 | 0.41 | 0.41 |

**Table 3.** Values of density of states $N_u$ for various resistance state (RS) at 2 K. ($f_{Pt}$=4%, $\delta$=10 nm) from 1D Mott-VRH. *Localization length $\zeta$ is fixed at $\zeta_{1D}$=0.41 nm from previous case.

Comparing these calculated density of states and the localization length with those for the 3D Mott-VRH (**Table 5.6, Table 5.7 & Table 5.8**), we can arrive at the common conclusions that apply to both mechanisms: (i) The DOS decreases with the thickness but $\zeta$ remians at ~0.4 nm. (ii) For a higher $f_{metal}$ film, the DOS decreases but the $\zeta$ increases. (iii) The DOS increases rapidly as the resistance states decreases. Therefore, all the interpretations in **Section 5.6.3** (**Chapter V**) can be restated if the 1D Mott-VRH is used for data fitting, confirming that the physics of hopping remains unchanged.

**Efros-Shklovskii Variable Range Hopping (ES-VRH)**

Coulombic interaction generated by trapped electrons could in principle open a soft gap in the DOS, therefore the possibility of ES-VRH exist. ES-VRH follows: $G \propto \exp\left(-\sqrt{T_{ES}/T}\right)$, where $T_{ES}$ is given by $T_{ES} = 2.8e^2/\varepsilon k_B \zeta$. The crossover temperature of ES-VRH to Mott-VRH can be written as: $T_{cross} = 16 T_{ES}^2/T_M$ and Coulomb gap can be estimated as $\Delta = e^3 \sqrt{N_{u0}}/\varepsilon^{3/2}$ ($N_{u0}$ is the unperturbed DOS). For a typical nanometallic film ($f_{Pt}$=4%, $\delta$=10 nm), we can estimate the crossover temperature as $T_{cross}$=230 K ($T_{ES}$=294$^2$ K, $T_M$=151$^4$ K). This means Mott-VRH dominates above 230 K. If we further assume the calculated DOS is unperturbed, an equivalent Coulomb gap of $\Delta$=0.12 eV can be estimated. As metal doping increases, a stronger Columbic interaction involves and thus we expect such $T_{cross}$ might move to a higher temperature. This means Mott-VRH might be suppressed by ES-VRH for a high $f_{Pt}$ film in our temperature range (200 K-300 K). Nevertheless, the fitting results for localization lengths of Mott-VRH and ES-VRH are quantitatively consistent within the same order of magnitude, as shown in **Table 4**.

| Composition | $f_{Pt}$ =4 at.% | $f_{Pt}$ =19 at.% | $f_{Pt}$ =27 at.% |
|---|---|---|---|
| $T_{ES}^{1/2}$ (K$^{1/2}$) | 294 | 303 | 199 |
| $\zeta_{ES}$ (nm) | 0.679 | 0.639 | 1.48 |
| $\zeta_{Mott}$ (nm) | 0.406 | 1.69 | 6.67 |

**Table 4.** Values of localization length $\zeta$ for various concentration $f_{Pt}$ ($\delta$=10 nm) from ES-VRH. $\zeta_{Mott}$ is from **Table 5.7**.

**Nearest Neighbor Hopping (NNH)**

If the DOS is large enough to provide nearest hopping states, NNH could dominate. This may possibly happen for intermediate resistance states, which have a higher $N_u$. We can estimate the onset of NNH by comparing the optimized hopping distance with the smallest possible hopping distance (in $SiN_4$, it is the tetrahedron size, $a$=0.4 nm):

$$r_0 = \frac{3}{4}\left(\frac{3\zeta}{2\pi N_u k_B T}\right)^{1/4} = a$$

For our standard nanometallic film ($f_{Pt}$=4%, $\delta$=10 nm) with $\zeta$=0.41 nm, NNH starts to appear for $N_u$=$10^{23}$ eV$^{-1}$cm$^{-3}$ at $T$>300K. However, for a typical HRS, $N_u$ (HRS) =2.6×$10^{18}$ eV$^{-1}$cm$^{-3}$, NNH (requiring $T$>$10^7$ K) is unlikely to occur. NNH is also unimportant in the metal-rich films, since as the metal doping increases, $N_u$ decreases rapidly, making NNH even harder to achieve. Therefore, NNH can be ignored except for a very low resistance IRS.

# Appendix XII. Magnetoresistance (MR) Data

## Xiang Yang, Jay Kikkawa, I-Wei Chen

Magnetoresistance (MR) measurement was performed for various resistance states (HRS, IRS, LRS, *etc.*) in PPMS under T=2 K, with magnetic field ranging from -9 T to 9 T. MR effect is typically very small (<1% change for resistance at 0 T and ±9 T). As shown in **Figure 1**, HRS follows no definite MR laws (sometimes positive parabolic MR (**Figure 1b&c&f**), sometimes negative parabolic MR (**Figure 1d**), sometimes very flat MR (**Figure 1a&e**)). Finer multi-level states can be easily found (**Figure 1c&d**) under MR testing when signal to noise ratio is high. On the other hand, LRS (**Figure 1g&h**) always shows a positive MR following linear law ($\Delta R = \alpha |B|$).

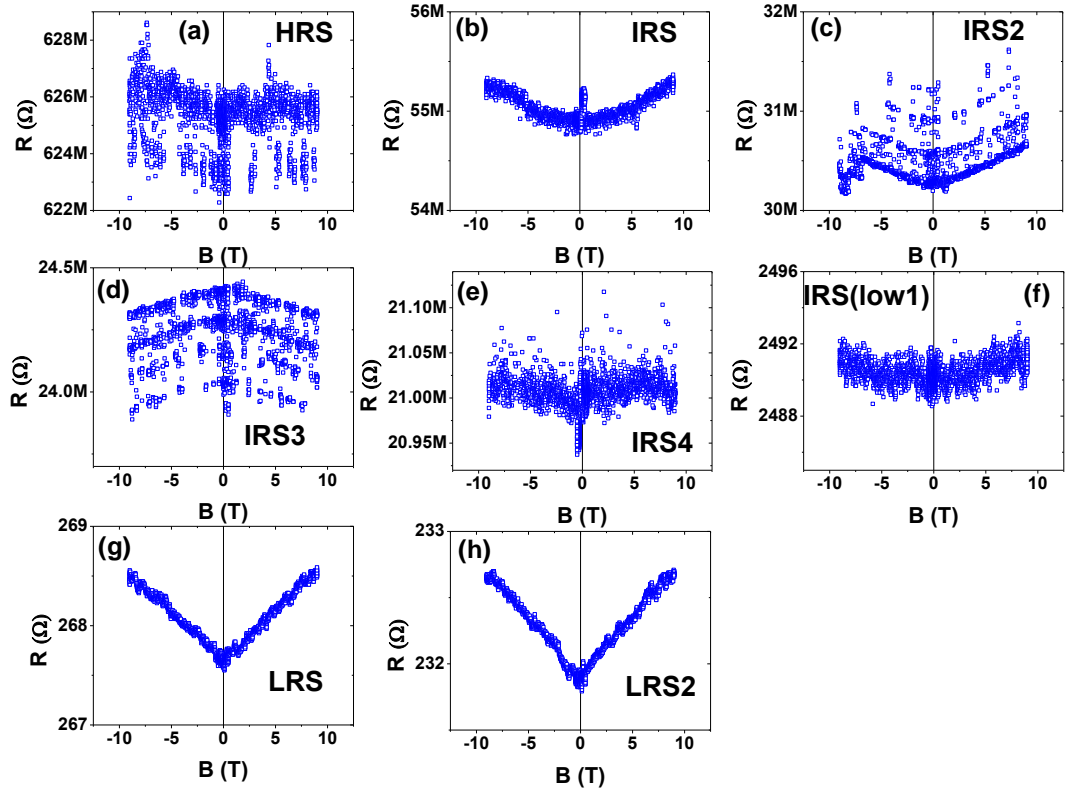

Figure 1. MR data for standard nanometallic device (Mo/Si$_3$N$_4$:4%Pt/Pt, $d$=512 μm, $\delta$=10 nm). The test was performed at 2K under 0.05V DC bias.

Positive linear law was also shown in **Figure 2** with more LR states. Since parasitic resistance also could contribute significantly to MR signal, a differential method was used:

$$R_{\text{film}} = R_{\text{total}} - R_{\text{lowest LR}}$$

where $R_{\text{film}}$ is real resistance of nanometallic film, $R_{\text{total}}$ is experimentally measured resistance and $R_{\text{lowest LR}}$ is the lowest LRS under very high negative reset voltage (assuming only parasitic resistance was left under such extreme, **Figure 2h**).

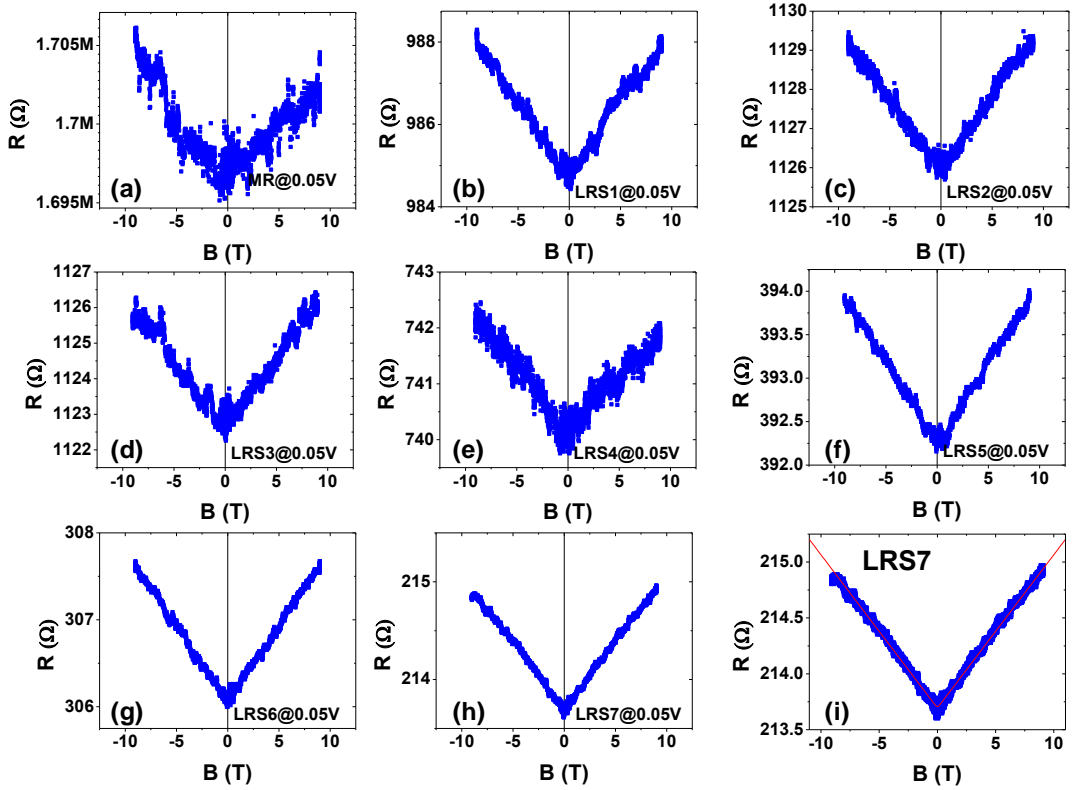

Figure 2. (a)-(h) A set of MR curve for IRS or LRS, showing positive linear law. (i) Lowest LRS (BE) fitted by $214\Omega \times (1+6.4 \times 10^{-8}|B|)$, $B$ is in unit of Oe.

After calibration of raw data with differential method, real film MR data are shown in **Figure 3**. Positive linear law $R(B)=R(B=0)\times(1+\alpha|B|)$ is still valid and MR coefficient $\alpha \sim 10^{-8}$/Oe.

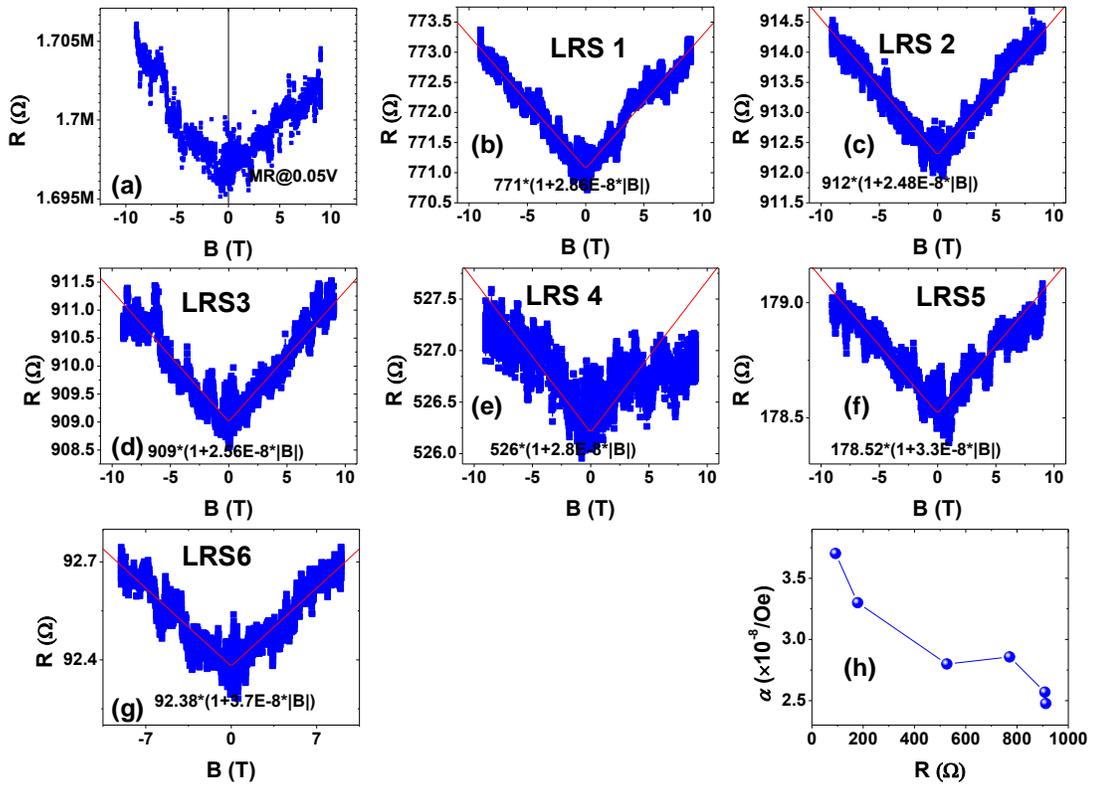

Figure 3. (a)-(g) MR data after subtraction of BE contribution, following $R(B)=R(B=0)\times(1+\alpha|B|)$. (h) MR coefficient $\alpha$.

# Appendix XIII. LabView Program: Four Point Measurement

## Xiang Yang

## Front Panel

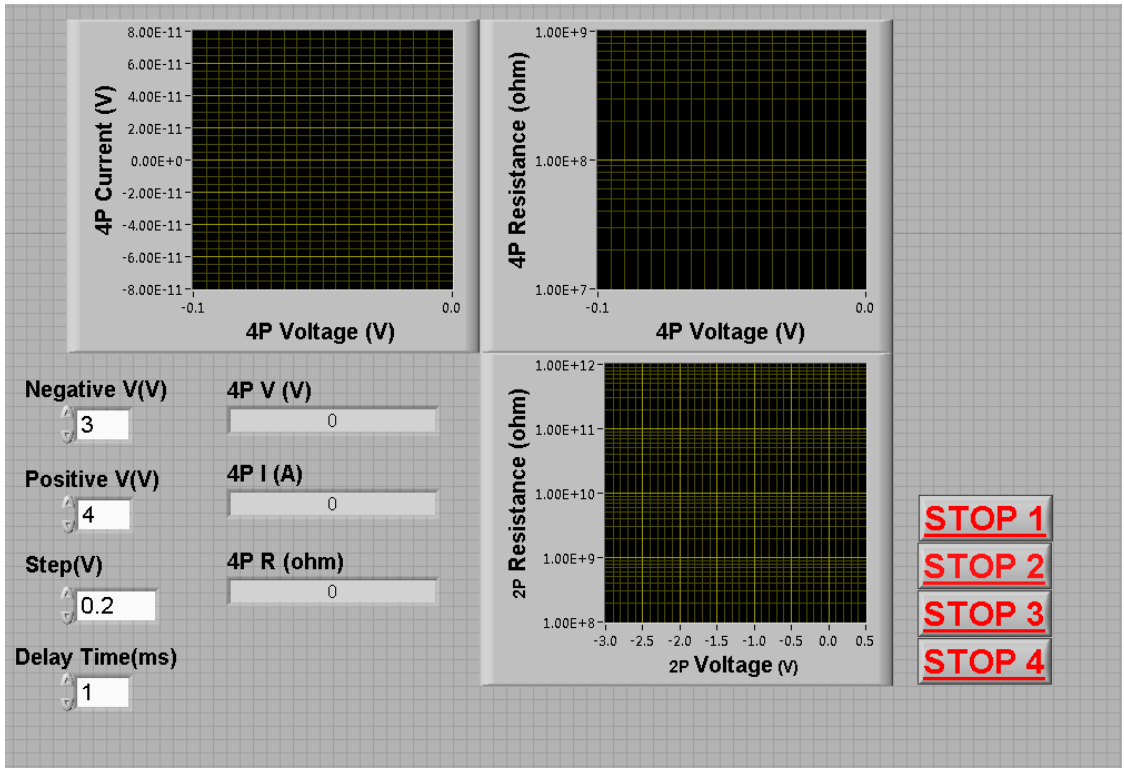

# Block Diagram

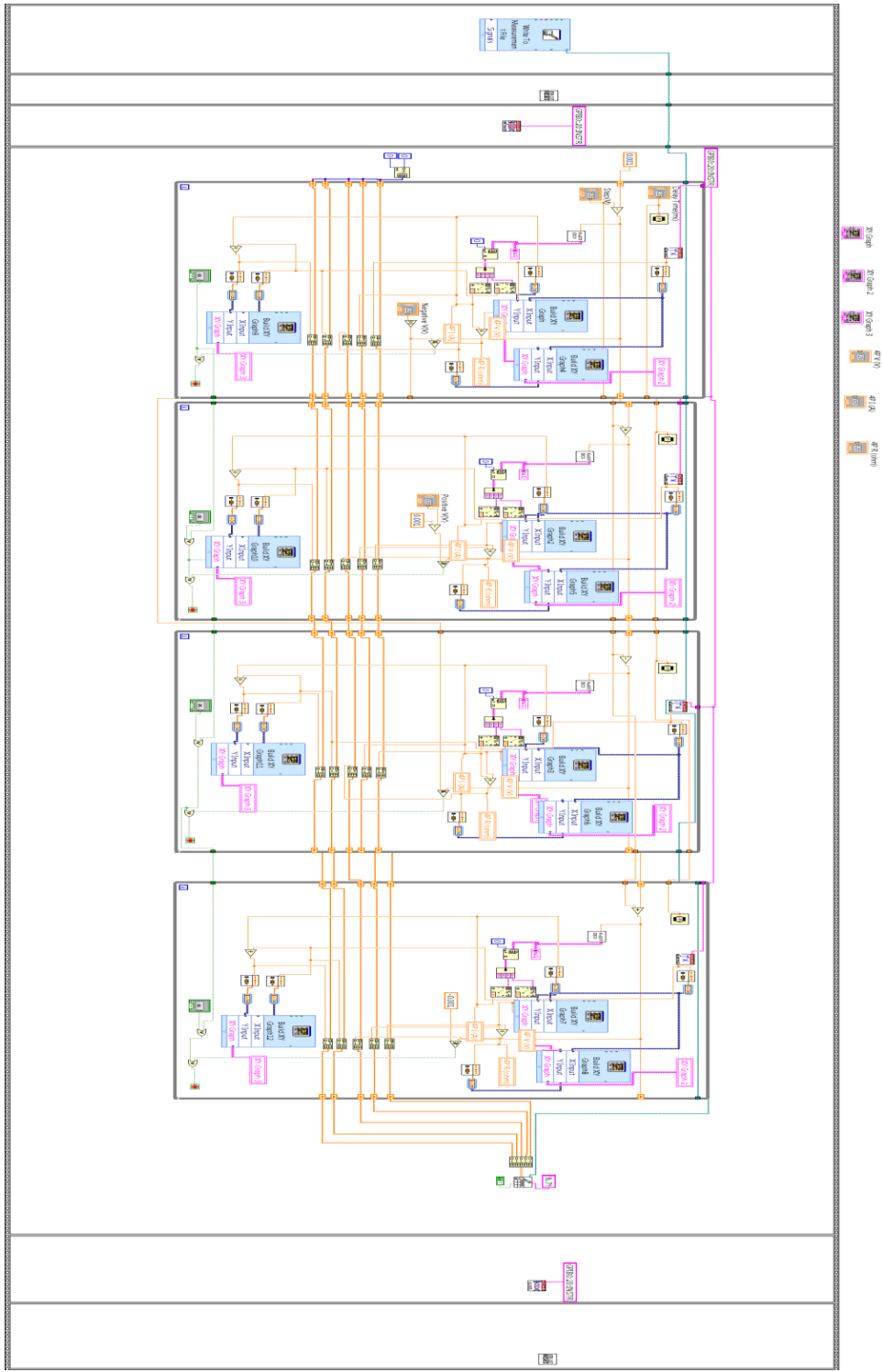

# Appendix XIV. LabView Program: Transistor Testing ($I_{ds}$-$V_g$)

## Xiang Yang

**Front Panel**

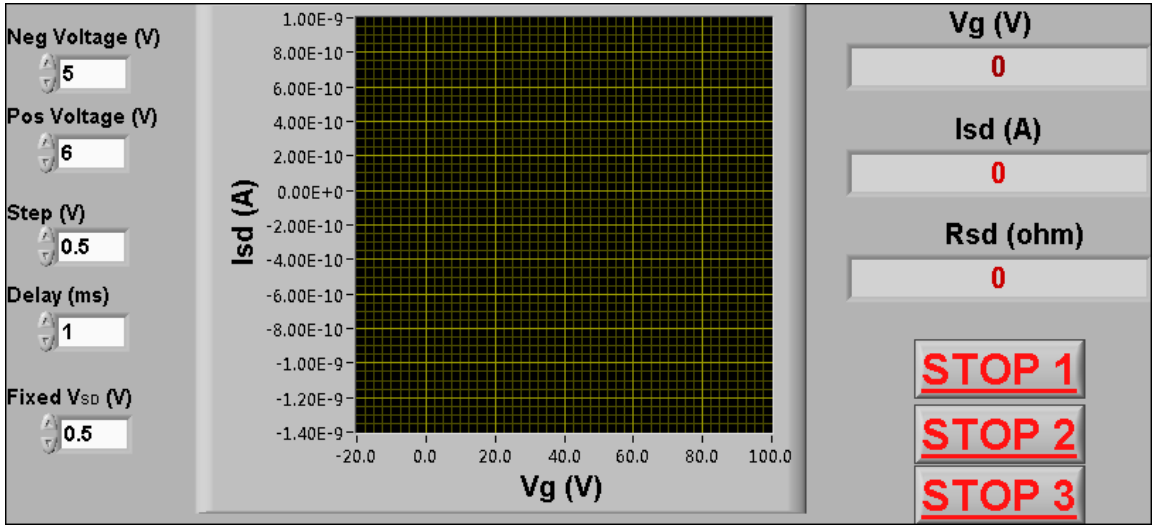

# Block Diagram

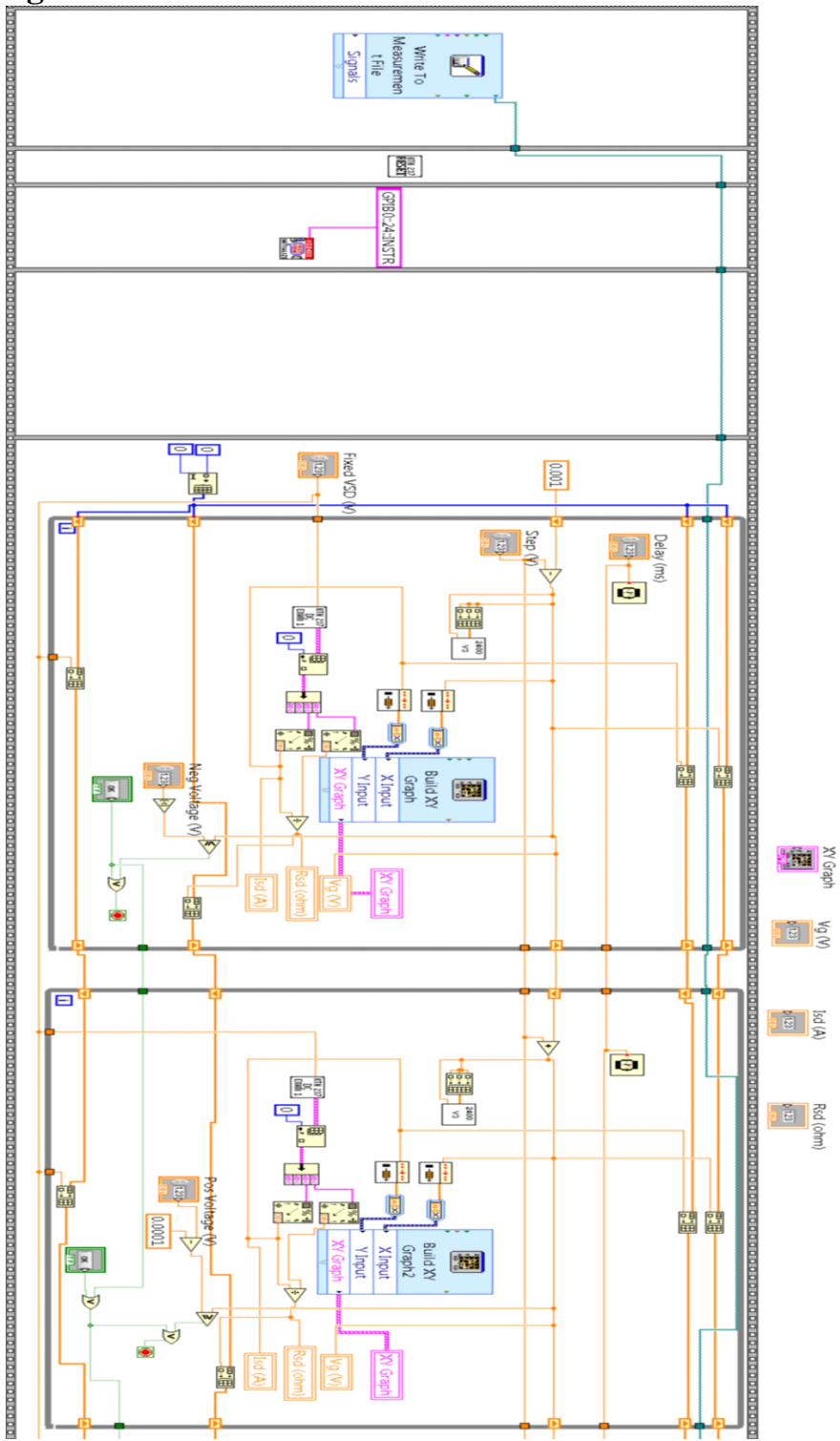

# Appendix XV. LabView Program: Transistor Testing ($I_{ds}$-$V_{ds}$)

**Xiang Yang**

## Front Panel

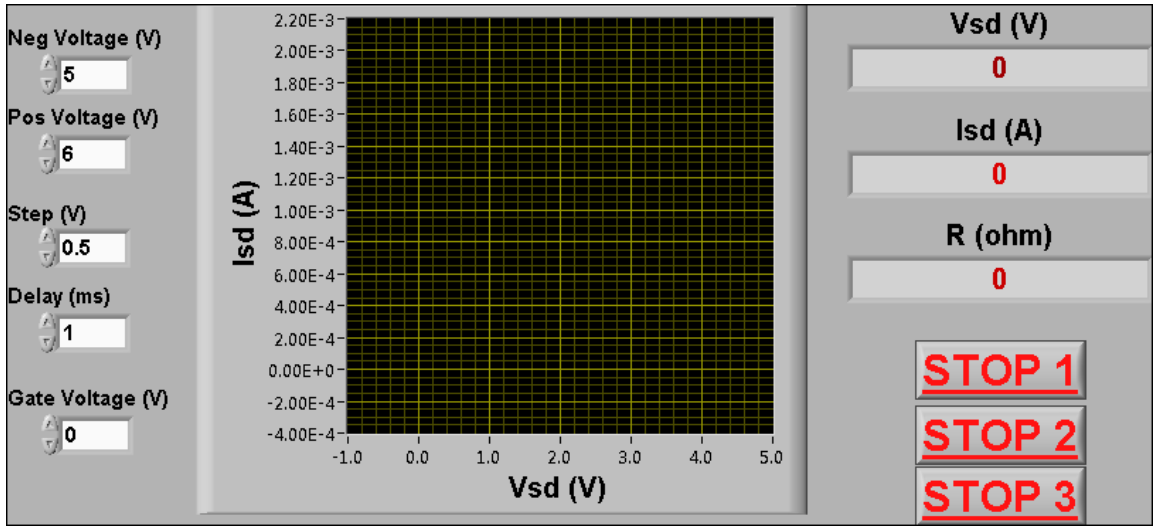

# Block Diagram

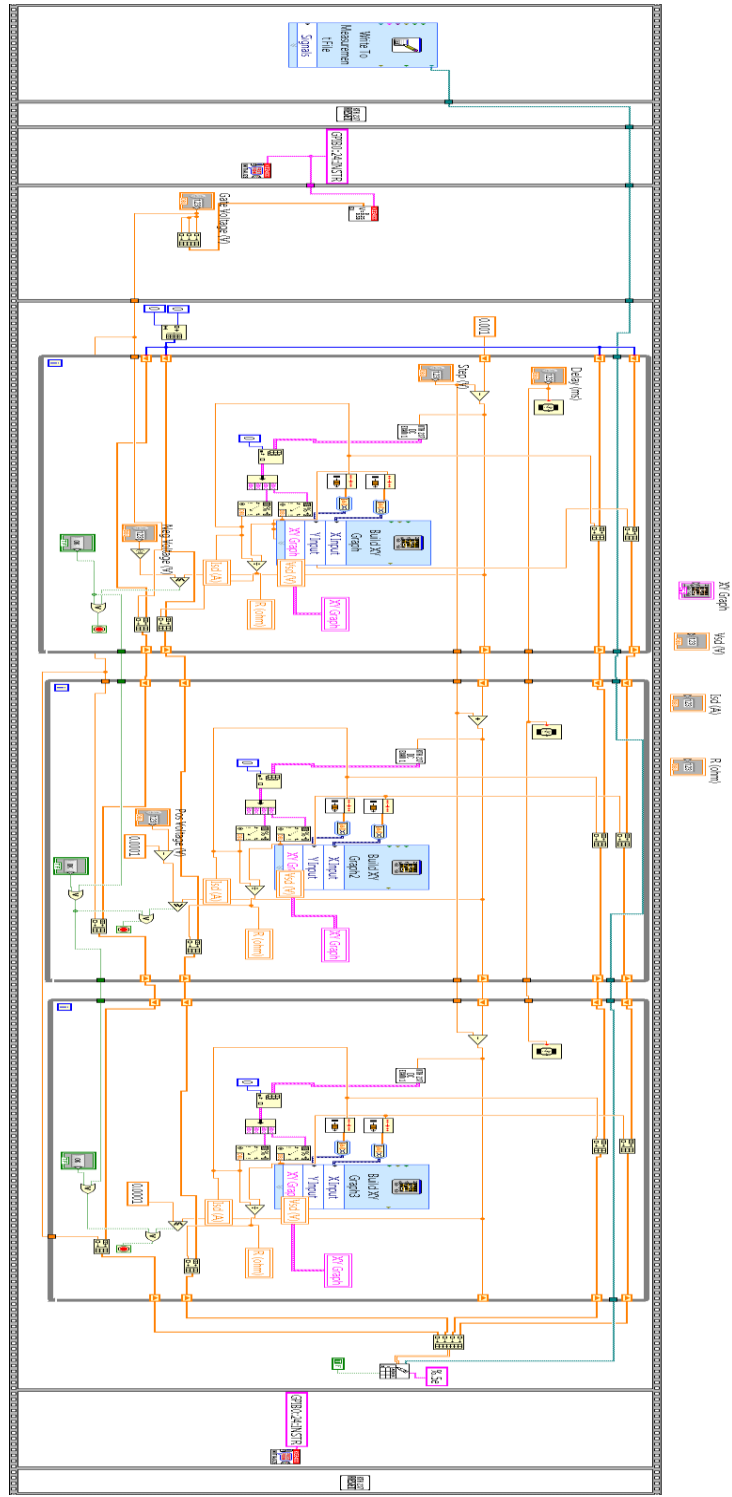

# Appendix XVI. LabView Program: Temperature

# Controlled Impedance

## Xiang Yang

## Front Panel

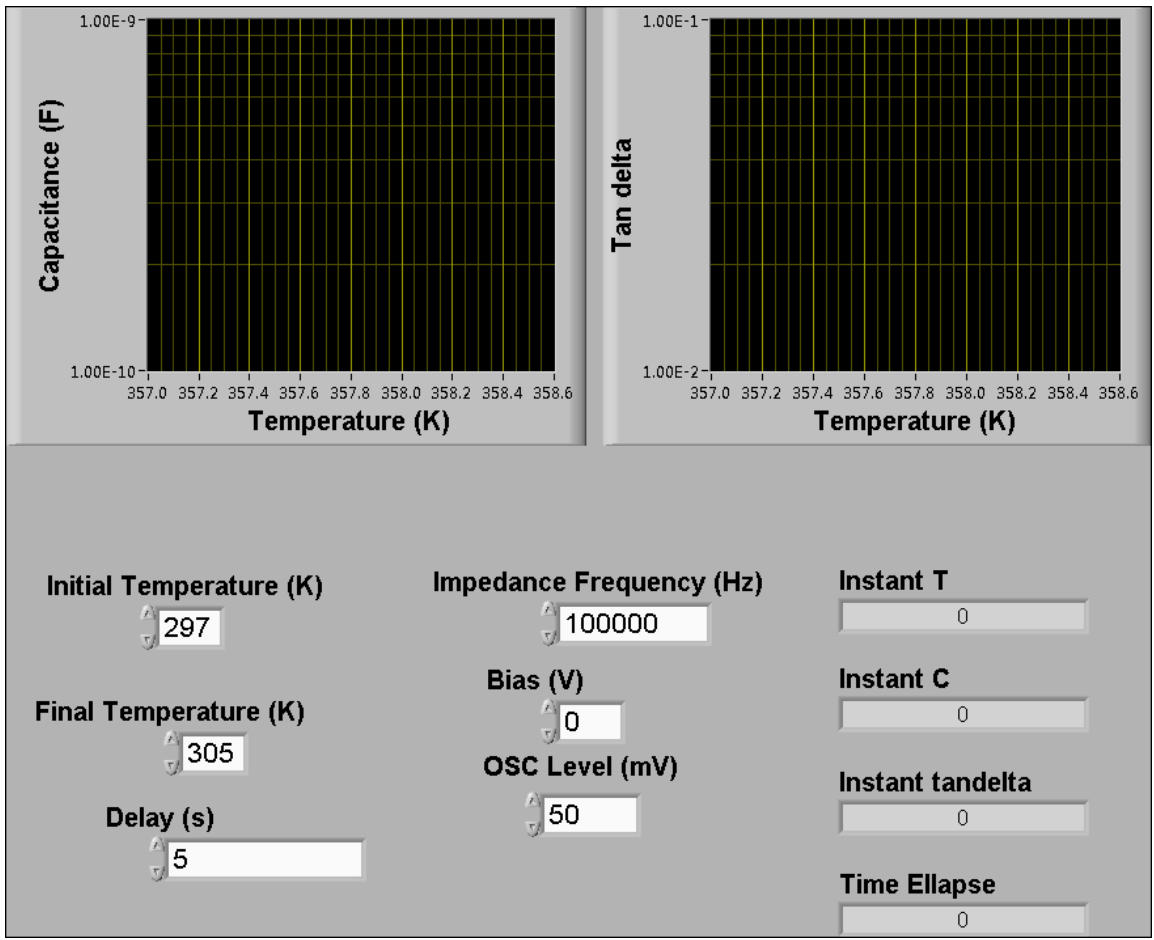

# Block Diagram

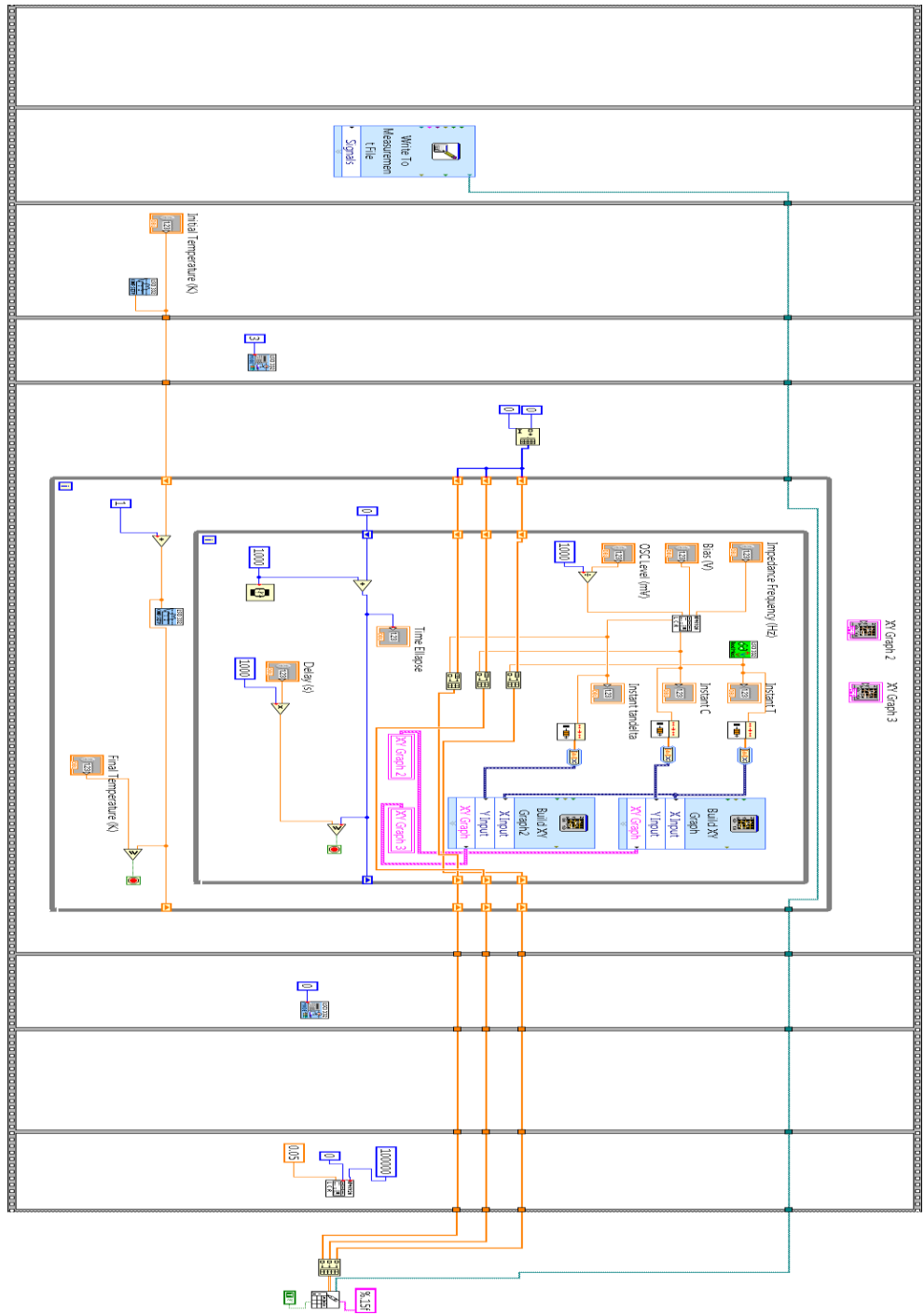

# VITA

## Xiang Yang

Chen group, Materials Science and Engineering         Tel: (858)-349-0216

University of Pennsylvania         Email: betterfutureyx@gmail.com

**Education**

2009-2014: University of Pennsylvania, Philadelphia, PA, U.S.A

- Ph. D. candidate in *Material Science and Engineering*. Advisor: I-Wei Chen

2012-2014: University of Pennsylvania, Philadelphia, PA, U.S.A

- M. S. in *Electrical Engineering*.

2008-2009: University of California, San Diego, La Jolla, CA, U.S.A

- M. S. in *Physics*.

2004-2008: Tsinghua University, Beijing, P.R.China

- B. S. in *Physics*. Advisor: Bangfen Zhu

**Patent**

I-Wei Chen & **X. Yang**. "Non-Volatile Resistance Switching Thin Film Devices".
International Patent PCT Application No. PCT/US2013/030178 (Filed 3/11/2013).

**Publication**

1.    **X. Yang**, I. Tudosa, B. J. Choi, A. B. K. Chen and I-Wei Chen, "Resolving Voltage–Time Dilemma Using an Atomic-Scale Lever of Subpicosecond Electron–Phonon Interaction", ***Nano Lett.*** **14**, 5058-5067 (2014).

2.    **X. Yang**, B. J. Choi, A. B. K. Chen & I-Wei Chen, "Cause and Prevention of Moisture Induced Degradation of Resistance Random Access Memory Nanodevices", ***ACS Nano*** **7**, 2302 (2013).

3.    **X. Yang**, A. B. K. Chen, B. J. Choi & I-Wei Chen, "Demonstration and Modeling of Multi-bit Resistance Random Access Memory", ***Appl. Phys. Lett.*** **102**, 043502 (2013).

4.    **X. Yang** & I-Wei Chen, "Dynamic-Load-Enabled Ultra-low Power Multiple-State RRAM Devices" ***Sci. Rep.*** **2**, 744 (2012).

5.    A. B. K. Chen, B. J. Choi, **X. Yang** & I-Wei Chen, "A Parallel Circuit Model for Multi-State Resistive Switching Random Access Memory" ***Adv. Funct. Mater.*** **22**, 546 (2012).

6.    B. J. Choi, A. B. K. Chen, **X. Yang** & I-Wei Chen, "Purely Electronic Switching with High Uniformity, Resistance Tunability and Good Retention in Pt-dispersed $SiO_2$ Thin Films for ReRAM" ***Adv. Mater.*** **23**, 3847 (2011).